\documentclass{dune}
\pdfoutput=1

\usepackage[pdftex,bookmarks,hidelinks]{hyperref}

\graphicspath{ {graphics/} }

\newif\ifdp
\newif\ifsp

\def\expshort{DUNE\xspace}

\def\explong{The Deep Underground Neutrino Experiment\xspace}

\def\thedocsubtitle{Deep Underground Neutrino Experiment (DUNE)} 
\def\tdrtitle{Technical Design Report}
\def\voltitleexec{Introduction to DUNE\xspace}
\def\volnumberexec{I}

\def\introchnd{Volume~\volnumberexec{}, \voltitleexec{}, Chapter~5\xspace}

\def\voltitlephysics{DUNE Physics\xspace}
\def\volnumberphysics{II}

\def\voltitletc{DUNE Far Detector Technical Coordination\xspace}
\def\volnumbertc{III}

\def\voltitlesp{The DUNE Far Detector Single-Phase Technology\xspace}
\def\volnumbersp{IV}

\def\spchdaq{Volume~\volnumbersp{}, \voltitlesp{}, Chapter~7\xspace}

\def\voltitledp{The DUNE Far Detector Dual-Phase Technology\xspace}
\def\volnumberdp{V}

\newcommand{\refsec}[2]{Volume~\csname volnumber#1\endcsname \xspace Section~#2}
\newcommand{\refch}[2]{Volume~\csname volnumber#1\endcsname \xspace Chapter~#2}
\newcommand{\refinch}[2]{#2 in Volume~\csname volnumber#1\endcsname \xspace}

\newcommand{\numu}{\ensuremath{\nu_\mu}\xspace}
\newcommand{\nue}{\ensuremath{\nu_e}\xspace}
\newcommand{\nutau}{\ensuremath{\nu_\tau}\xspace}

\newcommand{\anumu}{\ensuremath{\bar\nu_\mu}\xspace}
\newcommand{\anue}{\ensuremath{\bar\nu_e}\xspace}

\newcommand{\dm}[1]{\ensuremath{\Delta m^2_{#1}}\xspace} 

\newcommand{\sinst}[1]{\ensuremath{\sin^2\theta_{#1}}\xspace} 
\newcommand{\sinstt}[1]{\ensuremath{\sin^22\theta_{#1}}\xspace}  

\newcommand{\deltacp}{\ensuremath{\delta_{\rm CP}}\xspace}   
\newcommand{\mdeltacp}{\ensuremath{\delta_{\rm CP}}}   

\newcommand{\nuxtonux}[2]{\ensuremath{\nu_{#1} \to \nu_{#2}}\xspace}  
\newcommand{\numutonumu}{\nuxtonux{\mu}{\mu}}
\newcommand{\numutonue}{\nuxtonux{\mu}{e}}

\newcommand{\numubartonumubar}{
\ensuremath{\overline{\numu}\rightarrow\overline{\numu}}\xspace
}

\newcommand{\numubartonuebar}{
\ensuremath{\overline{\numu}\rightarrow\overline{\nue}}\xspace
}
\newcommand{\ptoknubar}{\ensuremath{p\rightarrow K^+ \overline{\nu}}\xspace}
\newcommand{\ptoepizero}{\ensuremath{p \rightarrow e^+ \pi^0}\xspace}
\newcommand{\ntoek}{\ensuremath{n\rightarrow e^{-}K^{+}}\xspace}
\newcommand{\nnbar}{\ensuremath{n-\bar{n}}\xspace}

\def\argon40{${}^{40}$Ar}       
\def\Ar39{$^{39}$Ar}
\def\Cl40{$^{40}$Cl}
\def\K40{$^{40}$K}
\def\B8{$^{8}$B}
\newcommand\isotope[2]{\textsuperscript{#2}#1} 

\def\ndfromtarget{\SI{574}{\meter}\xspace} 
\def\fdfiducialmass{\SI{40}{\kt}\xspace}

\def\larmass{\SI{17.5}{\kt}\xspace} 

\def\cryostatht{\SI{17.8}{\meter}\xspace} 
\def\cryostatlen{\SI{65.8}{\meter}\xspace} 
\def\cryostatwdth{\SI{18.9}{\meter}\xspace} 

\def\nominalmodsize{\SI{10}{kt}\xspace} 
 
\def\spmaxfield{\SI{500}{\volt/\centi\meter}\xspace} 
\def\spmaxdrift{\SI{3.5}{\m}\xspace}
\def\tpcheight{\SI{12.0}{\meter}\xspace} 

\def\sptargetdriftvoltpos{\SI{180}{\kilo\volt}\xspace} 

\def\dpmaxdrift{\SI{12.0}{\m}\xspace} 
\def\dpnominaldriftfield{\SI{500}{\volt/\cm}\xspace} 
\def\dptargetdriftvoltpos{\SI{600}{\kV}\xspace} 

\newcommand{\efield}{E field\xspace}

\newcommand{\threed}{3D\xspace}
\newcommand{\twod}{2D\xspace}
\newcommand{\phel}{photoelectron\xspace} 
\newcommand{\frfour}{FR-4\xspace} 

\newcommand{\lsim}{{\;\raise0.3ex\hbox{$<$\kern-0.75em\raise-1.1ex\hbox{$\sim$}}\;}}
\newcommand{\gsim}{{\;\raise0.3ex\hbox{$>$\kern-0.75em\raise-1.1ex\hbox{$\sim$}}\;}}
\newcommand{\beq}{\begin{equation}}
\newcommand{\eeq}{\end{equation}}
\newcommand{\bea}{\begin{eqnarray}}
\newcommand{\eea}{\end{eqnarray}}

\mathchardef\minus="002D

\newcommand{\rrt}[1]{}

\newcommand{\cherenkov}{Cherenkov\xspace}  
\newcommand{\superk}{Super--Kamiokande\xspace} 
\newcommand{\hyperk}{Hyper--Kamiokande\xspace} 
\newcommand{\microboone}{MicroBooNE\xspace} 
\newcommand{\minerva}{MINERvA\xspace} 
\newcommand{\nova}{NOvA\xspace} 
\newcommand{\lariat}{LArIAT\xspace} 
\newcommand{\argoneut}{ArgoNeuT\xspace} 
  
\newcommand{\miniboone}{MiniBooNE\xspace}

\newcommand{\lartpc}{LArTPC\xspace}

\newcommand{\larsoft}{LArSoft\xspace}

\newcommand{\fnal}{Fermilab\xspace} 
\newcommand{\surf}{SURF\xspace}

\newcommand{\detmodule}{detector module\xspace}
\newcommand{\dual}{DP\xspace}

\newcommand{\single}{SP\xspace}

\newcommand{\lar}{LAr\xspace}

\DeclareSIUnit \s {\second}
\DeclareSIUnit \MB {\mega\byte}
\DeclareSIUnit \GB {\giga\byte}
\DeclareSIUnit \TB {\tera\byte}
\DeclareSIUnit \PB {\peta\byte}
\DeclareSIUnit \Mbps {\mega\bit/\s}
\DeclareSIUnit \Gbps {\giga\bit/\s}
\DeclareSIUnit \Tbps {\tera\bit/\s}
\DeclareSIUnit \Pbps {\peta\bit/\s}
\DeclareSIUnit \kton {\kilo\tonne} 
\DeclareSIUnit \kt {\kilo\tonne}
\DeclareSIUnit \Mt {\mega\tonne}
\DeclareSIUnit \eV {\electronvolt}
\DeclareSIUnit \keV {\kilo\electronvolt}
\DeclareSIUnit \MeV {\mega\electronvolt}
\DeclareSIUnit \GeV {\giga\electronvolt}
\DeclareSIUnit \m {\meter}
\DeclareSIUnit \cm {\centi\meter}
\DeclareSIUnit \in {\inchcommand}
\DeclareSIUnit \km {\kilo\meter}
\DeclareSIUnit \kV {\kilo\volt}
\DeclareSIUnit \kW {\kilo\watt}
\DeclareSIUnit \MW {\mega\watt}
\DeclareSIUnit \MHz {\mega\hertz}
\DeclareSIUnit \mrad {\milli\radian}
\DeclareSIUnit \year {year}
\DeclareSIUnit \POT {POT}
\DeclareSIUnit \sig {$\sigma$}
\DeclareSIUnit\parsec{pc}
\DeclareSIUnit\lightyear{ly}
\DeclareSIUnit\foot{ft}
\DeclareSIUnit\ft{ft}
\DeclareSIUnit \ppb{ppb}
\DeclareSIUnit \ppt{ppt}
\DeclareSIUnit \samples{S}

\newcommand{\ktyr}{\si{\kt\year}\xspace}
\newcommand{\Mtyr}{\si{\Mt\year}\xspace}
\newcommand{\ktMWyr}{\si{\kt\MW\year}\xspace}
\newcommand{\SIadj}[2]{\SI{#1}{#2}}

\newcommand{\ktadj}[1]{\SIadj{#1}{\kt}}

\newcommand{\MeVadj}[1]{\SIadj{#1}{\MeV}}

\newcommand{\ktmwyr}[1]{\SI[inter-unit-product=\ensuremath{{}\cdot{}}]{#1}{\kt\MW\year}}

\usepackage[toc]{glossaries}
\makeglossaries

\newcommand{\dshort}[1]{\glsentrytext{#1}}  
\newcommand{\dlong}[1]{\glsentrylong{#1}}  

\newcommand{\dword}[1]{\gls{#1}}
\newcommand{\dwords}[1]{\glspl{#1}}
\newcommand{\Dword}[1]{\Gls{#1}}
\newcommand{\Dwords}[1]{\Glspl{#1}}

\newcommand{\newduneword}[3]{
    \newglossaryentry{#1}{
        text={#2},
        long={#2},
        name={\glsentrylong{#1}},
        first={\glsentryname{#1}},
        firstplural={\glsentrylong{#1}\glspluralsuffix},
        description={#3},
        sort={#2}
    }
}

\newcommand{\newduneabbrev}[4]{
  \newglossaryentry{#1}{
    text={#2},
    long={#3},
    shortplural={{#2}\glspluralsuffix},
    longplural={{#3}\glspluralsuffix{}},
    name={\glsentrylong{#1}{} (\glsentrytext{#1}{})},
    first={#3 (#2)},
    firstplural={#3\glspluralsuffix{} (\glsentrytext{#1}\glspluralsuffix{})},
    description={#4},
    sort={#2}
  }
}

\newcommand{\newduneabbrevs}[5]{
  \newglossaryentry{#1}{
    text={#2},
    long={#3},
    plural={#4},
    shortplural={{#2}\glspluralsuffix},
    longplural={#4},
    name={\glsentrylong{#1}{} (\glsentrytext{#1}{})},
    first={#3 (#2)},
    firstplural={#4 (\glsentrytext{#1}\glspluralsuffix{})},
    description={#5},
    sort={#2}    
  }
}

\newduneword{dword}{DUNE Word}{A term in the DUNE lexicon}

\newduneword{nasa}{NASA}{U.S. National Aereonautics and Space Administration}

\newduneabbrev{nd}{ND}{near detector}{Refers to the detector(s) 
 installed close to the neutrino source at \fnal }

\newduneabbrev{fd}{FD}{far detector}{The \SI{70}{kt} total (\fdfiducialmass fiducial) mass \gls{lartpc} DUNE detector, composed of four \larmass total (\nominalmodsize fiducial) mass modules,  
  to be installed at the far site at \surf in
  Lead, SD, USA}

\newduneabbrev{sp}{SP}{single-phase}{Distinguishes one of the DUNE far detector technologies by the fact that it operates using argon in its liquid phase only}

\newduneabbrev{dp}{DP}{dual-phase}{Distinguishes one of the DUNE far detector technologies by the fact that it operates using argon 
 in both gas and liquid phases}

\newduneabbrev{pds}{PD system}{photon detection system}{The detector 
  subsystem sensitive to light produced in the \lar }

\newduneabbrev{hvs}{HVS}{high voltage system}{The detector 
  subsystem that provides the \gls{tpc} drift field}

\newduneabbrev{tpc}{TPC}{time projection chamber}{A type of particle detector that uses an \efield together with a sensitive volume of gas or liquid, e.g., \gls{lar}, to perform a \threed reconstruction of a particle trajectory or interaction. The activity is recorded by digitizing the waveforms of current
  induced on the anode as the distribution of ionization charge passes by
  or is collected on the electrode} 

\newduneabbrev{lartpc}{LArTPC}{liquid argon time-projection chamber}{A \gls{tpc} filled with liquid argon; 
the basis for the \gls{dune} \gls{fd} modules} 

\newduneabbrevs{apa}{APA}{anode plane assembly}{anode plane assemblies}{A unit of the \single
  detector module containing the elements sensitive to ionization in the \lar. 
  It contains two faces each of three planes of wires, and interfaces to the cold
  electronics and photon detection system} 

\newduneabbrev{awg}{AWG}{American wire gauge} {U.S. standard set of non-ferrous wire conductor sizes}

\newduneabbrev{ufer}{Ufer}{concrete encased electrode} {U.S. National Electrical Code grounding method refered to as Concrete Encased Electrode}

\newduneabbrev{cro}{CRO}{charge readout}{The system for detecting
  ionization charge distributions in a \dual detector module}

\newduneabbrev{lro}{LRO}{light readout}{The system for detecting
  scintillation photons in a \dual  detector module}

\newduneabbrev{shv}{SHV}{safe high voltage}{Type of bayonet mount
connector used on coaxial cables that has additional insulation 
compared to standard BNC and MHV connectors that makes it safer
for handling \gls{hv} by preventing accidental contact with the
live wire connector in an unmated connector or plug}

\newduneabbrev{fe}{FE}{front-end}{The front-end refers a point that is
  ``upstream'' of the data flow for a particular subsystem. 
  For example the \gls{sp} front-end electronics is where the cold electronics
  meet the sense wires of the TPC and the front-end \gls{daq} is where the
  \gls{daq} meets the output of the electronics}

\newduneabbrev{daqrou}{DAQ RU}{DAQ readout unit}{The first element in the data flow of the \gls{daq}}

\newduneabbrev{cots}{COTS}{commercial off-the-shelf}{Items, typically hardware such as 
computers, that may be purchased whole, without any custom design or fabrication and 
thus at normal consumer prices and availability}

\newduneabbrev{i2c}{I2C}{Inter-Integrated Circuit}{I$^2$C or I2C is a synchronous, 
multi-master, multi-slave, packet switched, single-ended, serial computer bus widely used 
for attaching lower-speed peripheral ICs to processors and microcontrollers in short-distance, 
intra-board communication} 

\newduneabbrev{spi}{SPI}{Serial Peripheral Interface}{The Serial Peripheral Interface is a 
synchronous serial communication interface specification used for short distance 
communication, primarily in embedded systems}

\newduneabbrev{miso}{MISO}{master in slave out}{The Master In Slave Out is a logic
signal on the \gls{spi} bus on which the data from the slave are transmitted once
a request from the master is received} 

\newduneabbrev{mosi}{MOSI}{master out slave in}{The Master Out Slave In is a logic
signal on the \gls{spi} bus on which the data from the master is transmitted} 

\newduneabbrev{uart}{UART}{Universal Asynchrous Receiver/Transmitter}{A universal 
asynchronous receiver-transmitter is a computer hardware device for asynchronous 
serial communication in which the data format and transmission speeds are configurable}

\newduneword{cr}{CR}{Capacitance-Resistance} 

\newduneword{dc}{DC}{direct coupling} 

\newduneword{ac}{AC}{capacitive coupling}  

\newduneabbrev{pll}{PLL}{Phase-Locked Loop}{A control system that generates an
output signal whose phase is related to the phase of an input signal}  

\newduneword{fifo}{FIFO}{First-In-First-Out} 

\newduneword{tsmc}{TSMC}{Taiwan Semiconductor Manufacturing Company}

\newduneword{saci}{SACI}{\gls{slac} \gls{asic} Control Interface}

\newduneword{om3}{OM3}{Type of multi-mode fiber optic cable, typically capable of \SI{10}{Gbps} data transmission at lengths up to \SI{300}{m}}

\newduneword{om4}{OM4}{Type of multi-mode fiber optic cable, typically capable of \SI{10}{Gbps} data transmission at lengths up to \SI{550}{m}}

\newduneword{qfp}{QFP}{Quad Flat Package} 

\newduneabbrev{ams}{AMS}{analog and mixed signal}{Verilog-AMS is a derivative of the Verilog hardware description language that includes analog and mixed-signal extensions (AMS) in order to define the behavior of analog and mixed-signal systems}

\newduneabbrev{hepa}{HEPA}{High Efficiency Particulate Air}{The High Efficiency Particulate Air filters are a type of air filter that remove \num{99.97}\% of particles that have a size greater han or equal to \SI{0.3}{$\mu$m}}  

\newduneabbrev{uvm}{UVM}{universal verification methodology}{The Universal Verification Methodology is a standardized methodology for verifying integrated circuit designs}   

\newduneword{lhc}{LHC}{Large Hadron Collider}

\newduneabbrev{lsb}{LSB}{least significant bit}{The bit with the lowest numerical value in a binary number}

\newduneabbrev{ldo}{LDO}{low-dropout regulator}{A low-dropout or LDO regulator is a \gls{dc} linear voltage regulator that can regulate the output voltage even when the supply voltage is very close to the output voltage}

\newduneabbrev{adc}{ADC}{analog-to-digital converter}{A sampling of a voltage
  resulting in a discrete integer count corresponding in some way to
  the input}

\newduneabbrev{inl}{INL}{integral non-linearity}{A commonly used measure of performance in \glspl{adc}. It is the deviation between the ideal input threshold value and the measured threshold level of a certain output code}

\newduneabbrev{dnl}{DNL}{differential non-linearity}{A commonly used measure of performance in \glspl{adc}. The DNL error is defined as the difference between an actual step width and the ideal value of one \gls{lsb}}

\newduneword{pnp}{PNP}{Type of bipolar junction transistor consistning of a
layer of N-doped semiconductor sandwiched between two layers of P-doped material}

\newduneword{spice}{SPICE}{SPICE
(``Simulation Program with Integrated Circuit Emphasis'') is a general-purpose, 
open-source analog electronic circuit simulator. It is a program used in integrated 
circuit and board-level design to check the integrity of circuit designs and to 
predict circuit behavior}

\newduneabbrev{daq}{DAQ}{data acquisition}{The data acquisition system
  accepts data from the detector \gls{fe} electronics, buffers
  the data, performs a \gls{trigdecision}, builds events from the selected
  data and delivers the result to the offline \gls{diskbuffer}}

\newduneword{detmodule}{detector module}{The entire DUNE far detector is
  segmented into four modules, each with a nominal \SI{10}{\kton}
  fiducial mass}

\newduneword{detunit}{detector unit}{A 
portion of a \gls{detmodule} may be further partitioned into a number of similar parts.   For example the \gls{sp} \gls{tpc} 
is made up of \gls{apa}  units (and other elements)}

\newduneword{diskbuffer}{secondary DAQ buffer}{A secondary
  \dshort{daq} buffer holds a small subset of the full rate as
  selected by a \gls{trigcommand}. 
  This buffer also marks the interface with the DUNE Offline}

\newduneabbrev{om}{OM}{online monitoring}{Processes that run inside
  the \gls{daq} on data ``in flight,'' specifically before landing on the
  offline disk buffer, and that provide feedback on the operation of
  the \gls{daq} itself and the general health of the data it is marshalling}

\newduneabbrev{dqm}{DQM}{data quality monitoring}{Analysis of the raw
  data to monitor the integrity of the data and the performance of the
  detectors and their electronics. This type of monitoring may be
  performed in real time, within the \gls{daq} system, or in later
  stages of processing, using disk files as input}

\newduneword{dumpbuffer}{DAQ dump buffer}{This \gls{daq} buffer
  accepts a high-rate data stream, in aggregate, from an associated
  portion of a \gls{detmodule} sufficient to collect all data likely relevant to
  a potential \gls{snb}}

\newduneabbrev{etl}{ETL}{external trigger logic}{Trigger processing
  that consumes \gls{detmodule} level \gls{trignote} information
  and other global sources of trigger input and emits
  \gls{trigcommand} information back to the \glspl{mtl}}
\newduneabbrev{daqeti}{ETI}{external trigger interface}{Interface between \glspl{mtl} and external source and sinks of relevant trigger information}

\newduneword{trignote}{trigger notification}{Information provided by
  \gls{mtl} to \gls{etl} about \gls{trigdecision} 
  processing}

\newduneword{trigprimitive}{trigger primitive}{Information derived by
  the \gls{daq} \gls{fe} hardware that describes a region of space (e.g.,
  one or several neighboring channels) and time (e.g., a contiguous set
  of \gls{adc} sample ticks) associated with some activity}

\newduneword{externtrigger}{external trigger candidate}{Information
  provided to the \gls{mtl} about events external to a
  \gls{detmodule} so that it may be considered in forming
  \glspl{trigcommand}}

\newduneabbrev{daqoob}{OOB dispatcher}{out-of-band trigger command
  dispatcher}{This component is responsible for dispatching a \gls{snb} dump
  command to all \glspl{daqfer} in the \gls{detmodule}}

\newduneabbrev{mtl}{MTL}{module trigger logic}{Trigger processing
  that consumes \gls{detunit} level \gls{trigcommand} information
  and emits \glspl{trigcommand}. 
  It provides the \gls{etl} with \glspl{trignote} and receives back any
  \glspl{externtrigger}}

\newduneword{octant}{octant}{Any of the eight parts into which 4$\pi$
  is divided by three mutually perpendicular axes. 
  In particular in referencing the value for the mixing angle
  $\theta_{23}$}

\newduneword{trigcandidate}{trigger candidate}{Summary information derived
  from the full data stream and representing a contribution toward
  forming a \gls{trigdecision}}

\newduneword{trigcommand}{trigger command}{Information derived from
  one or more \glspl{trigcandidate}  that directs elements of the
  \gls{detmodule} to read out a portion of the data stream}

\newduneabbrev{tcm}{TCM}{trigger command message}{A message flowing
  down the trigger hierarchy from global to local context.  Also see \gls{tpm}}

\newduneabbrev{mlt}{MLT}{module level trigger}{The \gls{daq} component responsible for producing a \gls{trigdecision} that will be used to command the readout of a detector module}

\newduneword{trigdecision}{trigger decision}{The process by which
  \glspl{trigcandidate} are converted into \glspl{trigcommand}}

\newduneabbrev{tpm}{TPM}{trigger primitive message}{A message flowing
  up the trigger hierarchy from local to global context.  Also see \gls{tcm}}

\newduneabbrev{ipc}{IPC}{inter-process communication}{A system for software elements to exchange information between threads, local processes or across a data network.  An IPC system is typically specified in terms of protocols  composed of message types and their associated data schema}

\newduneword{daqdispre}{discovery and presence}{As used in the context of the \gls{ipc}, a system that provides mechanisms for a node on a communication network to learn of the existence of peers and their identity (discovery) as well as determine if they are currently operational or have become unresponsive (presence)}

\newduneabbrev{pubsub}{PUB/SUB}{publish-subscribe communication pattern}{An \gls{ipc} communication pattern where one element, the publisher, sends data to all connected elements, the subscribers.  Each subscriber may connect to multiple publishers.  A variant is PUB/SUB with topics where a subscriber may register an identifier, the topic, to limit the information received to just an associated subset}

\newduneabbrev{eb}{EB}{event builder}{A software agent that executes \glspl{trigcommand}  for one  \gls{detmodule} by reading out the requested data}

\newduneabbrev{daqdfo}{DFO}{data flow orchestrator}{The process by which trigger commands are executed in parallel and asynchronous manner by the back-end output subsystem of the \gls{daq}}

\newduneabbrev{daqubi}{UBI}{upstream DAQ buffer interface}{The process which provides read-only access to data residing in the upstream \gls{daq} buffers to processes on the network}

\newduneabbrev{cob}{COB}{cluster on board}{An ATCA motherboard housing four RCEs}

\newduneabbrev{rce}{RCE}{reconfigurable computing element}{Data processor located outside of the cryostat on a \gls{cob} that contains \gls{fpga}, RAM and \gls{ssd} resources, responsible for buffering data, producing trigger primitives, responding to triggered requests for data and synching \gls{snb} dumps}

\newduneabbrev{bow}{BOW}{Bump On Wire}{A working name for the front-end readout computing elements used in the nominal \gls{daq} design to interface the \dual  crates to the \gls{daq} front-end computers}

\newduneabbrev{atca}{ATCA}{Advanced Telecommunications Computing
  Architecture}{An advanced computer architecture specification developed for the telecommunications, military, and aerospace industries that incorporates the latest trends in  high-speed interconnect technologies, next-generation processors, and improved reliability, availability and serviceability} 

\newduneabbrev{utca}{$\mu$TCA}{Micro Telecommunications Computing Architecture}{The computer architecture specification followed by the crates that house charge and light readout electronics in the \gls{dpmod}} 

\newduneabbrev{udp}{UDP}{user datagram protocol}{A simple,
  connectionless Internet protocol that supports data integrity
  checksums, requires no handshaking, and does not guarantee packet delivery}

\newduneabbrev{amc}{AMC}{advanced mezzanine card}{Holds digitizing
  electronics and lives in \gls{utca} crates}

\newduneabbrev{rf}{RF}{radio frequency}{Electromagnetic emissions
  that are within the (radio) frequency band of sensitivity of the detector
  electronics}

\newduneabbrev{fpga}{FPGA}{field programmable gate array}{An
integrated circuit technology that allows the hardware to be reconfigured to
execute different algorithms after its manufacture and deployment}

\newduneabbrev{fmc}{FMC}{FPGA mezzanine card}{Boards holding \glspl{fpga} and other integrated circuitry that attach to a motherboard}

\newduneabbrev{felix}{FELIX}{Front-End Link eXchange}{A
  high-throughput interface between \gls{fe} and trigger electronics
  and the standard PCIe computer bus}

\newduneword{daqpart}{DAQ partition}{A cohesive and
 coherent collection of \gls{daq} hardware and software working together to trigger and read out some portion of one detector module; it consists of an integral number of \glspl{daqfrag}. 
 Multiple \gls{daq} partitions may operate simultaneously, but each instance operates independently}

\newduneabbrev{fec}{DAQ FEC}{DAQ front-end computer}{The portion of one
  \gls{daqpart} that hosts the \gls{daqdr}, \gls{daqbuf} and
  \gls{daqds}.  It hosts the \gls{daqfer} and corresponding portion of the \gls{daqbuf}}

\newduneword{daqfrag}{DAQ front-end fragment}{The portion of one
  \gls{daqpart} relating to a single \gls{fec} and corresponding to an
  integral number of \glspl{detunit}.  See also \gls{datafrag}}

\newduneword{datafrag}{data fragment}{A block of data read out from a single \gls{daqfrag} that
span a contiguous period of time as requested by a \gls{trigcommand}}

\newduneabbrev{daqfer}{FER}{DAQ front-end readout}{The portion of a
  \gls{daqfrag} that accepts data from the detector electronics and
  provides it to the \gls{fec}}

\newduneabbrev{daqdr}{DDR}{DAQ data receiver}{The portion of the
  \gls{daqfrag} that accepts data from the \gls{daqfer}, emits
  trigger candidates produced from the input trigger primitives, and
  forwards the full data stream to the \gls{daqbuf}}

\newduneword{daqbuf}{DAQ primary buffer}{The portion
  of the \gls{daqfrag} that accepts full data stream from the
  corresponding \gls{detunit} and retains it sufficiently long for it
  to be available to produce a \gls{datafrag}}

\newduneword{daqds}{data selector}{The portion of the \gls{daqfrag}
  that accepts \glspl{trigcommand} and returns the corresponding
  \gls{datafrag}.  Not to be confused with \gls{daqdsn}}

\newduneword{daqdsn}{data selection}{The process of forming a trigger decision for selecting a subset of detector data for output by the \gls{daq} from the content of the detector data itself.  Not to be confused with \gls{daqds}}

\newduneabbrev{daqros}{DAQ RO}{DAQ readout subsystem}{The subsystem of the \gls{daq} for accepting and buffering data input from detector electronics}

\newduneabbrev{daqdss}{DAQ DS}{DAQ data selection subsystem}{The subsystem of the \gls{daq} responsible for forming a trigger decision based on a portion of the input data stream.  The majority subset of the \gls{daqtrs}}

\newduneabbrev{daqtrs}{DAQ TS}{DAQ trigger subsystem}{The subsystem of the \gls{daq} responsible for forming a trigger decision}

\newduneabbrev{daqbes}{DAQ BE}{DAQ back-end subsystem}{The portion of the \gls{daq} that is generally toward its output end.  It is responsible for accepting and executing trigger commands and marshaling the data they address to output storage buffers}

\newduneabbrev{daqtss}{DAQ TSS}{DAQ timing and synchronization subsystem}{The portion of the \gls{daq} that provides for timing and synchronization to various components}

\newduneabbrev{femb}{FEMB}{front-end mother board}{Refers a unit of
  the \gls{sp} \gls{ce} that contains the \gls{fe} amplifier
  and \gls{adc} \glspl{asic} covering 128 channels}

\newduneword{asic}{ASIC}{application-specific integrated circuit}

\newduneword{lv}{LV}{low voltage}

\newduneabbrev{iceberg}{ICEBERG}{ICEBERG R\&D cryostat and electronics}{Integrated Cryostat and Electronics Built for Experimental Research Goals:
a new double-walled cryostat built and installed at \gls{fnal} 
for liquid argon detector R\&D and for testing of DUNE detector components}

\newduneword{coldadc}{ColdADC}{A newly developed 16-channels \gls{asic} providing analog to digital conversion}

\newduneword{coldata}{COLDATA}{A 64-channel control and communications \gls{asic}}

\newduneword{cryo}{CRYO}{Integrated ASIC including \gls{fe} circuitry providing signal amplification and pulse shaping, analog to digital conversion, and control and communication functionalities for 64 channels}

\newduneword{larasic}{LArASIC}{A 16-channel \gls{fe} \gls{asic} that provides signal amplification and pulse shaping}

\newduneword{cmos}{CMOS}{Complementary metal-oxide-semiconductor}

\newduneabbrev{enc}{ENC}{equivalent noise charge}{The equivalent noise charge is the input charge that corresponds to a 
$\gls{snr}=1$}

\newduneword{sar}{SAR}{successive approximation register}

\newduneword{protodune}{ProtoDUNE}{Either of the two DUNE prototype detectors constructed at \gls{cern}. 
  One prototype implements \gls{sp} technology and the other \gls{dp}}
  
\newduneword{protodune2}{ProtoDUNE-2}{The second run of a \gls{protodune} detector}

\newduneword{pdsp}{ProtoDUNE-SP}{The \gls{sp} \gls{protodune} detector at \gls{cern}}

\newduneword{pddp}{ProtoDUNE-DP}{The \gls{dp} \gls{protodune} detector at \gls{cern}}

\newduneword{wa105}{WA105 DP demonstrator}{The \SI[product-units=power]{3x1x1}{m} WA105 \gls{dp} prototype detector at \gls{cern}}

\newduneword{rawevent}{DAQ event block}{The unit of data output by the
  \gls{daq}.  
  It contains trigger and detector data spanning a unique, contiguous
  time period and a subset of the detector channels}

\newduneabbrev{ssd}{SSD}{solid-state disk}{Any storage device that
  may provide sufficient write throughput to receive, both collectively and
  distributed, the sustained full rate of data from a \gls{detmodule}
  for many seconds}
\newduneabbrev{nvme}{NVMe}{Non-volatile memory express}{A specification for an interface to storage media attached via PCIe}

\newduneabbrev{hlt}{HLT}{high-level trigger}{This is actually a filter applied to data that has been triggered and aggregated in order to further reduce or characterize it}

\newduneabbrev{pid}{PID}{particle ID}{Particle identification}

\newduneword{readout window}{readout window}{A fixed, atomic and
  continuous period of time over which data from a \gls{detmodule}, in
  whole or in part, is recorded. 
  This period may differ based on the trigger that initiated the
  readout}

\newduneabbrev{zs}{ZS}{zero-suppression}{Used to delete some portion of a
  data stream that does not significantly deviate from zero or
  intrinsic noise levels. 
  It may be applied at different granularity from per-channel to per
  \gls{detunit}}

\newduneabbrev{rc}{RC}{run control}{The system for configuring,
  starting and terminating the \gls{daq}}

\newduneword{r-c}{RC}{resistive-capacitive (circuit)}

\newduneabbrev{daqccm}{CCM}{DAQ control, configuration and monitoring subsystem}{A system for controlling, configuring and monitoring other systems in particular those that make up the \gls{daq} where the CCM encompasses \gls{rc}}

\newduneword{daqrun}{DAQ run}{A period of time over which relevant data taking conditions and \gls{daq} configuration are asserted to be unchanged. 
  Multiple \gls{daq} runs may occur simultaneously when multiple \glspl{daqpart} are active. 
  This term should not be confused with DUNE experiment or beam ``runs'' that typically span many \gls{daq} runs}
\newduneword{daqrunnum}{DAQ run number}{A monotonically increasing count that uniquely and globally identifies a \gls{daqrun}}

\newduneabbrev{snb}{SNB}{supernova neutrino burst}{A prompt 
  increase in the flux of low-energy neutrinos emitted in the first few seconds of a core-collapse supernova.  It can also refer to a trigger command type that may be due to this phenomenon,
  or detector conditions that mimic its interaction signature}

\newduneabbrev{snble}{SNB/LE}{supernova neutrino burst and low
  energy}{Supernova neutrino burst and low-energy physics program}

\newduneabbrev{snews}{SNEWS}{SuperNova Early Warning System}{A global
  supernova neutrino burst trigger formed by a coincidence of \gls{snb} 
  triggers collected from participating experiments}

\newduneabbrev{pps}{1PPS signal}{one-pulse-per-second signal}{An
  electrical signal with a fast rise time and that arrives in real
  time with a precise period of one second}

\newduneabbrev{sls}{SLS}{spill location system}{A system residing at
  the DUNE far detector site that provides information, possibly
  predictive, indicating periods of time when neutrinos are being
  produced by the \fnal Main Injector beam spills}

\newduneabbrev{wib}{WIB}{warm interface board}{Digital electronics
  situated just outside the \gls{sp} cryostat that receives digital data
  from the \glspl{femb} over cold copper connections and sends it to the \gls{rce}
  \gls{fe} readout hardware}

\newduneabbrev{gps}{GPS}{Global Positioning System}{A satellite-based system that provides a highly accurate \gls{pps} that may be used to synchronize clocks and determine location}

\newduneabbrev{ntp}{NTP}{Network Time Protocol}{A networking protocol that allows synchronizing of clocks to within a few \si{\milli\second} of a time standard on a local network and within a few tens of \si{\milli\second} over the Internet} 

\newduneabbrev{ptproto}{PTP}{Precision Time Protocol}{A networking protocol that allows synchronizing of clocks to within a few \si{\micro\second} of a time standard on a local network} 

\newduneabbrev{irig}{IRIG}{inter-range instrumentation group}{A standards body that defined a time-code standard for transferring timing information}

\newduneabbrev{nic}{NIC}{network interface controller}{Hardware for controlling the interface to a communication network.  Typically, one that obeys the Ethernet protocol}

\newduneabbrev{wiec}{WIEC}{warm interface electronics crate}{Crates mounted on the signal flanges that contain the \glspl{wib}}

\newduneabbrev{ptc}{PTC}{power and timing card}{Cards that provide further processing and distribution of the signals entering and exiting the \gls{sp} cryostat}

\newduneabbrev{ptb}{PTB}{power and timing backplane}{Backplane used to connect the \gls{wib}s and the \gls{ptc}s on the \gls{wiec}. Also connects the \gls{ce} flange on the cryostat penetration}

\newduneabbrev{sipm}{SiPM}{silicon photomultiplier}{A solid-state
  avalanche photodiode sensitive to single \phel signals}

\newduneabbrev{cisc}{CISC}{cryogenic instrumentation and slow controls}{Includes equipment to monitor all detector  components and  \gls{lar} quality and behavior, and provides a control system for many of the detector components}

\newduneword{fte}{FTE}{full-time equivalent. A unit of labor
  for the project. One year of work from one person}

\newduneword{art}{art}{A software framework implementing an
  event-based execution paradigm} 

\newduneabbrev{sam}{SAM}{sequential
  access via metadata}{A data-handling system to store and retrieve
  files and associated metadata, including a complete record of the
  processing that has used the files}

\newduneword{artdaq}{artdaq}{A data acquisition toolkit for data transfer, aggregation and processing}

\newduneword{beamline}{beamline}{A sequence of control and monitoring devices used for the formation of a directed collection of particles}
\newduneabbrev{cdr}{CDR}{conceptual design report}{A formal project
  document 
   that describes the experiment
  at a conceptual level}

\newduneabbrev{cf}{CF}{conventional facilities}{Pertaining to
  construction and operation of buildings and conventional infrastructure, and for \gls{lbnf-dune}, CF includes the excavation caverns}

\newduneabbrev{cp}{CP}{charge parity}{Product of charge and parity
  transformations}

\newduneabbrev{cpt}{CPT}{charge, parity, and time reversal symmetry}{product of charge, parity
  and time-reversal transformations}

\newduneabbrev{cpv}{CPV}{charge-parity symmetry violation}{Lack of
  symmetry in a system before and after charge and parity
  transformations are applied. 
  For CP symmetry to hold,  a particle turns into its
 corresponding antiparticle under a charge transformation, and a parity
transformation inverts its space coordinates, i.e., 
produces the mirror image}

\newduneword{doe}{DOE}{U.S. Department of Energy}

\newduneabbrev{fra}{FRA}{Fermi Research Alliance}{A joint partnership of the University of Chicago and the Universities Research Association (URA) that manages and operates Fermilab on behalf of the \gls{doe}}

\newduneabbrev{dune}{DUNE}{Deep Underground Neutrino Experiment}{A leading-edge, international experiment for neutrino science and proton decay studies}

\newduneabbrev{esh}{ES\&H}{environment, safety and health}{A discipline and specialty that studies and implements practical aspects of environmental protection and safety at work} 

\newduneabbrev{ppe}{PPE}{personnel protective equipment}{Equipment worn to minimize exposure to hazards that cause serious workplace injuries and illnesses}

\newduneabbrev{odh}{ODH}{oxygen deficiency hazard}{a hazard that occurs when inert gases such as nitrogen, helium, or argon displace room air and thus reduce the percentage of oxygen below the level required for human life}

\newduneabbrev{feshm}{FESHM}{Fermilab Environment, Safety and Health Manual}{The document that contains Fermilab's policies and procedures designed to manage environment, safety, and health in all its programs}

\newduneabbrev{fscf}{FSCF}{far site conventional facilities}{The
  \gls{cf} at the DUNE far detector site, \surf}
  
\newduneabbrev{nscf}{NSCF}{near site conventional facilities}{The
  \gls{cf} at the DUNE near detector site, \fnal}

\newduneabbrevs{gut}{GUT}{grand unified theory}{grand unified theories}{A class of theories that unifies the electro-weak and strong forces}

\newduneabbrev{lar}{LAr}{liquid argon}{Argon in its liquid phase; it is a cryogenic liquid with a boiling point of $\SI{-90}{^\circ{C}}$ (\SI{87}{K}) and density of \SI{1.4}{g/ml}}

\newduneabbrev{lbl}{LBL}{long-baseline}{Refers to the distance between the 
  neutrino source  and the \gls{fd}.  It can also refer to the distance between the near and far detectors. 
  The ``long'' designation is an approximate and relative distinction. For DUNE, this distance  (between \gls{fnal} and \gls{surf}) is approximately \SI{1300}{km}}

\newduneabbrev{lbnf}{LBNF}{Long-Baseline Neutrino Facility}{The
  organizational entity responsible for developing the neutrino beam, the cryostats
  and cryogenics systems, and the conventional facilities for DUNE}
  
\newduneabbrev{lbnf-dune}{LBNF/DUNE}{LBNF and DUNE project}{The overall global project, including \gls{lbnf} and \gls{dune}}

\newduneabbrev{lbnc}{LBNC}{Long-Baseline Neutrino Committee}{The committee, composed of internationally prominent scientists with relevant expertise, charged by the \gls{fnal} director to review the scientific, technical, and managerial progress, plans and decisions associated with \gls{dune}}

\newduneabbrev{ncg}{NCG}{Neutrino Cost Group}{A group of internationally prominent scientists with relevant experience that is charged by the \gls{fnal} director to review the cost, schedule, and associated risks for the \gls{dune} experiment}

\newduneabbrev{mh}{MH}{mass hierarchy}{Describes the separation
  between the mass squared differences related to the solar and
  atmospheric neutrino problems}

\newduneabbrev{mi}{MI}{Fermilab Main Injector}{An accelerator at
  \fnal that provides a beam of high-energy protons that upon
  striking a target produce secondaries that decay to provide the
  neutrinos directed toward the DUNE far detector}

\newduneabbrev{pot}{POT}{protons on target}{Typically used as a unit
  of normalization for the number of protons striking the neutrino
  production target}

\newduneabbrev{qa}{QA}{quality assurance}{The set of actions taken to provide confidence that quality requirements are fulfilled, and to detect and correct poor results}

\newduneabbrev{qc}{QC}{quality control}{An aggregate of activities (such as design analysis and inspection for defects) performed to ensure adequate quality in manufactured products}

\newduneabbrev{sm}{SM}{standard model}{Refers to a theory describing
  the interaction of elementary particles}

\newduneabbrev{tdr}{TDR}{technical design report}{A formal project
  document 
  that describes the experiment at a technical level}

\newduneabbrev{prelimdr}{PDR}{preliminary design report}{A formal project
  document 
  that describes the experiment at a preliminary design level}

\newduneabbrev{tp}{IDR}{interim design report}{An intermediate
milestone on the path to a full \gls{tdr}} 

\newduneabbrev{ckm}{CKM matrix}{Cabibbo-Kobayashi-Maskawa
  matrix}{Refers to the matrix describing the mixing between mass and
  weak eigenstates of quarks}

\newduneabbrev{cl}{CL}{confidence level}{Refers to a probability
  used to determine the value of a random variable given its
  distribution}

\newduneabbrev{pmns}{PMNS}{Pontecorvo-Maki-Nakagawa-Sakata}{A type of matrix that describes the mixing between mass and weak eigenstates of
  the neutrino}

\newduneabbrevs{cpa}{CPA}{cathode plane assembly}{cathode plane assemblies}{The component of the \single detector module that provides the drift HV cathode}

\newduneabbrev{fc}{FC}{field cage}{The component of a \gls{lartpc} that contains and shapes the applied \efield}

\newduneword{cpafc}{CPA/FC}{A pair of \gls{cpa} panels and the top and bottom \gls{fc} portions that attach to the pair; an intermediate assembly for installation into the \gls{spmod} }

\newduneabbrev{topfc}{top FC}{top field cage}{The horizontal portions of the \gls{sp} \gls{fc}   on the top of the \gls{tpc}}

\newduneabbrev{botfc}{bottom FC}{bottom field cage}{The horizontal portions of the \gls{sp} \gls{fc} on the bottom of the \gls{tpc}}

\newduneabbrev{ewfc}{endwall FC}{endwall field cage}{The vertical portions of the \gls{sp} \gls{fc} near the wall}

\newduneabbrev{gp}{GP}{ground plane}{An electrode held electrically neutral relative to Earth ground voltage; it is mounted on the \gls{fc} in a \gls{spmod} to protect the cryostat wall}

  \newduneword{gg}{ground grid}{An electrode held electrically neutral relative to Earth ground voltage; it is installed between the cathode and the \glspl{pd} in a \gls{dpmod} to protect the \glspl{pmt}, maintaining high transparency to light}

\newduneabbrev{alara}{ALARA}{as low as reasonably
  achievable}{Typically used with regard management of radiation
  exposure but may be used more generally. It means making every
  reasonable effort to maintain e.g., exposures, to as far below the
  limits as practical, consistent with the purpose for that the
  activity is undertaken}

\newduneabbrev{ecal}{ECAL}{electromagnetic calorimeter}{A detector
  component that measures energy deposition of traversing particles (in the near detector conceptual design)}

\newduneabbrev{hv}{HV}{high voltage}{Generally describes a voltage
  applied to drive the motion of free electrons through some media, e.g., LAr}

\newduneword{spmod}{SP module}{single-phase DUNE \gls{fd} module}
\newduneword{dpmod}{DP module}{dual-phase DUNE \gls{fd} module}

\newduneabbrev{tcoord}{TC}{technical coordinator}{A member of the \gls{dune} management team responsible for organizing the technical aspects of the project effort; is head of \gls{tc}}

\newduneabbrev{rcoord}{RC}{resource coordinator}{A member of the \gls{dune} management team responsible for coordinating the financial resources of the project effort}

\newduneword{tc}{technical coordination}{The DUNE organization responsible for overall integration 
of the detector elements and successful execution of the detector
construction project; areas of responsibility include 
general project oversight, systems engineering, \gls{qa} 
and safety}

\newduneabbrev{exb}{EB}{executive board}{The highest level DUNE
  decision-making body for the collaboration}

\newduneabbrev{tb}{TB}{technical board}{The DUNE organization responsible for
  evaluating technical decisions}

\newduneabbrev{rrb}{RRB}{Resources Review Board}{A part of \gls{dune}'s international project governance structure, composed of representatives of all funding agencies that sponsor the project, and of  \gls{fnal} management, established to provide coordination among funding partners and oversight of \gls{dune}}

\newduneabbrev{inc}{INC}{International Neutrino Council}{A highest-level international advisory body to the U.S. \gls{doe} and the  \gls{fnal} directorate on matters related to the  \gls{lbnf} and the  \gls{pip2} projects. This council is composed of representatives from the international funding agencies and  \gls{cern} that make major contributions the infrastructure}

\newduneabbrev{cc}{CC}{charged current}{Refers to an interaction
  between elementary particles where a charged weak force carrier
  ($W^+$ or $W^-$) is exchanged}

\newduneabbrev{dis}{DIS}{deep inelastic scattering}{Refers to the 
  interaction of an elementary charged particle with a nucleus in an
  energy range where the interaction can be modeled as taking place with
  individual nucleons}

\newduneabbrev{fsi}{FSI}{final-state interactions}{Refers to
  interactions between elementary or composite particles subsequent to
  the initial, fundamental particle interaction, such as may occur as
  the products exit a nucleus}

\newduneword{geant4}{Geant4}{A
  software toolkit for the simulation of the passage of particles
  through matter using \gls{mc} methods}

\newduneabbrev{genie}{GENIE}{Generates Events for Neutrino Interaction
  Experiments}{Software providing an object-oriented neutrino
  interaction simulation resulting in kinematics of the products of
  the interaction}

\newduneabbrev{mc}{MC}{Monte Carlo}{Refers to a method of numerical
  integration that entails the statistical sampling of the integrand
  function. 
  Forms the basis for some types of detector and physics simulations}

\newduneabbrev{qe}{QE}{quasi-elastic}{Refers to interaction between
  elementary particles and a nucleus in an energy range where the
  interaction can be modeled as occurring between constituent quarks
  of one nucleon and resulting in no bulk recoil of the resulting
  nucleus}

\newduneabbrev{mou}{MoU}{memorandum of understanding}{A document
  summarizing an agreement between two or more parties}

\newduneabbrev{pip2}{PIP-II}{Proton Improvement Plan II}{A \gls{fnal} project for
  improving the protons on target delivered delivered by the \gls{lbnf} neutrino production beam. 
  This is version two of this plan and it is planned to be followed by a PIP-III}
  
\newduneabbrev{sdsta}{SDSTA}{South Dakota Science and Technology
  Authority}{The legal entity that manages \gls{surf}, in Lead, S.D}
  
\newduneabbrev{sdsd}{SDSD}{Fermilab South Dakota Services Division}{A Fermilab division responsible providing host laboratory functions at SURF in South Dakota}

\newduneabbrev{firus}{FIRUS}{Facility Information Reporting Utility System}
 {The safety system at \surf}

\newduneabbrev{bsi}{BSI}{building and site infrastructure}
 {The work package for outfitting of the \gls{lbnf} underground infrastructure}

\newduneabbrev{wbs}{WBS}{work breakdown structure}{An organizational
  project management tool by which the tasks to be performed are
  partitioned in a hierarchical manner}

\newduneabbrev{br}{BR}{branching ratio}{A fractional probability for a
  decay of a composite particle to occur into some specified set or
  sets of products}
\newduneword{bsm}{BSM}{beyond the standard model}

\newduneabbrev{dm}{DM}{dark matter}{The term given to the unknown
  matter or force that explains measurements of galaxy motion 
  that are otherwise inconsistent with the amount of mass associated
  with the observed amount of photon production}
  
  \newduneabbrev{bdm}{BDM}{boosted dark matter}{A new model that describes a relativistic dark matter particle boosted by the annihilation of heavier dark matter participles in the galactic center or the sun}

\newduneabbrev{cern}{CERN}{European Organization for Nuclear
Research}{The leading particle physics laboratory in Europe and home to the ProtoDUNEs. (In French, the Organisation Europ\'{e}enne pour la Recherche Nucl\'{e}aire, derived from Conseil Europ\'{e}en pour la Recherche Nucl\'{e}aire}

\newduneabbrev{dsnb}{DSNB}{diffuse supernova neutrino background}{The
  term describing the pervasive, constant flux of neutrinos due to all
  past supernova neutrino bursts}

\newduneabbrev{espp}{ESPP}{European Strategy for Particle Physics}{The
cornerstone of Europe's
decision-making process for the long-term future of the
field. Mandated by the \gls{cern} Council, it is formed through a broad
consultation of the grass-roots particle physics community, it
actively solicits the opinions of physicists from around the world,
and it is developed in close coordination with similar processes in
the USA and Japan in order to ensure coordination between regions and
optimal use of resources globally}

\newduneabbrev{gar}{GAr}{gaseous argon}{argon in its gas phase}
\newduneabbrev{gartpc}{GArTPC}{gaseous argon time-projection chamber}{A \gls{tpc} filled with gaseous argon; a possible technology choice for the \gls{nd}}

\newduneabbrev{globes}{GLoBES}{General Long-Baseline Experiment
  Simulator}{A software package for simulating energy spectra of
  neutrino flux, interactions, and energy spectra measured after application of some
  model of a detector response)}

\newduneabbrev{snowglobes}{SNOwGLoBES}{SuperNova
Observatories with GLoBES} {From the official description~\cite{snowglobes}: 
SNOwGLoBES is public software for computing interaction rates and distributions of observed quantities for \gls{snb} neutrinos in common detector materials} 

\newduneword{l/e}{L/E}{length-to-energy ratio}
\newduneword{lri}{LRI}{long-range interactions}

\newduneabbrev{nc}{NC}{neutral current}{Refers to an interaction
  between elementary particles where a neutrally charged weak force carrier
  ($Z^0$) is exchanged}

\newduneabbrev{nh}{NH}{normal hierarchy}{Refers to the neutrino mass
  eigenstate ordering whereby the sign of the mass squared difference
  associated with the atmospheric neutrino problem is positive}

\newduneabbrev{ih}{IH}{inverted hierarchy}{Refers to the neutrino mass
  eigenstate ordering whereby the sign of the mass squared difference
  associated with the atmospheric neutrino problem is negative}

\newduneabbrev{no}{NO}{normal ordering}{Refers to the neutrino mass
  eigenstate ordering whereby the sign of the mass squared difference
  associated with the atmospheric neutrino problem is positive}

\newduneabbrev{io}{IO}{inverted ordering}{Refers to the neutrino mass
  eigenstate ordering whereby the sign of the mass squared difference
  associated with the atmospheric neutrino problem is negative}

\newduneabbrev{msw}{MSW}{Mikheyev-Smirnov-Wolfenstein effect}{Explains
  the oscillatory behavior of neutrinos produced inside the sun as
  they traverse the solar matter}

\newduneabbrev{nsi}{NSI}{nonstandard interaction}{A general class of
  theory of elementary particles other than the Standard Model}

\newduneabbrev{pfive}{P5}{Particle Physics Project Prioritization
Panel}{The Particle Physics Project Prioritization Panel (P5) was a
subpanel of the High Energy Physics Advisory Panel (HEPAP). It completed
its Report, a ten-year strategic plan for high energy physics in the
U.S., in 2014. This report included a recommendation that ``host a world-leading neutrino
program that will have an optimized set of short- and long-baseline neutrino oscillation experiments, and its long-term focus
is a reformulated venture referred to here as the Long Baseline
Neutrino Facility (LBNF)''}

\newduneabbrev{sme}{SME}{standard-model extension}{an effective field theory that contains the \gls{sm}, general relativity, and all possible operators that break Lorentz symmetry (Wikipedia)}

\newduneabbrev{susy}{SUSY}{supersymmetry}{Theoretical symmetry between a fermion and a boson}

\newduneabbrev{wimp}{WIMP}{weakly-interacting massive particle}{A
  hypothesized particle that may be a component of dark matter}

\newduneabbrev{ce}{CE}{cold electronics}{Analog and digital readout electronics that operate at cryogenic temperatures}

\newduneabbrev{crp}{CRP}{charge-readout plane}{In the \gls{dp} technology, a  collection of
  electrodes in a planar arrangement placed at a particular voltage
  relative to some applied \efield such that drifting electrons
  may be collected and their number and time may be measured}

\newduneabbrev{dram}{DRAM}{dynamic random access memory}{A computer memory technology}

\newduneabbrev{fnal}{Fermilab}
{Fermi National Accelerator Laboratory}{U.S. national laboratory in Batavia, IL. It is the laboratory that hosts \gls{dune} and serves as its near site}

\newduneabbrev{bnl}{BNL}{Brookhaven National Laboratory}{US national laboratory in Upton, NY}

\newduneabbrev{slac}{SLAC}{SLAC National Accelerator Laboratory}{US national laboratory in Menlo Park, CA}

\newduneabbrev{lbnl}{LBNL}{Lawrence Berkeley National Laboratory}{US national laboratory in Berkeley, CA}

\newduneabbrev{anl}{ANL}{Argonne National Laboratory}{US national laboratory in Lemont, IL}

\newduneabbrev{lanl}{LANL}{Los Alamos National Laboratory}{US national laboratory in Los Alamos, NM}

\newduneabbrev{fs}{FS}{full stream}{Relates to a data stream that has not undergone selection, compression or other form of reduction}

\newduneabbrev{lem}{LEM}{large electron multiplier}{A micro-pattern detector suitable for use in ultra-pure argon vapor; LEMs consist of copper-clad PCB boards with sub-millimeter-size holes through which electrons undergo amplification}

\newduneabbrev{lng}{LNG}{liquefied natural gas}{Pertaining to natural gas in its liquid phase}

\newduneabbrev{mip}{MIP}{minimum ionizing particle}{Refers to a
  particle traversing some medium such that the particle's mean energy loss is  
  near the minimum}

\newduneabbrev{pd}{PD}{photon detector}{The detector
  elements involved in measurement of the number and arrival times of
  optical photons produced in a detector module} 

\newduneabbrev{pmt}{PMT}{photomultiplier tube}{A device that makes use
  of the photoelectric effect to produce an electrical signal from the
  arrival of optical photons}

\newduneabbrev{ppm}{ppm}{parts per million}{A concentration equal to one part in $10^{-6}$}
\newduneabbrev{ppb}{ppb}{parts per billion}{A concentration equal to one part in $10^{-9}$}
\newduneabbrev{ppt}{ppt}{parts per trillion}{A concentration equal to one part in $10^{-12}$}

\newduneword{rio}{RIO}{reconfigurable input output}

\newduneabbrev{s/n}{S/N}{signal-to-noise}{signal-to-noise ratio}
\newduneword{snr}{\mbox{S/N}}{signal-to-noise ratio}

\newduneword{ssp}{SSP}{SiPM signal processor}

\newduneabbrev{sbn}{SBN}{Short-Baseline Neutrino}{A \gls{fnal} program consisting of three collaborations, \gls{microboone}, \gls{sbnd}, and \gls{icarus}, to perform sensitive searches for $\nue$ appearance and $\numu$ disappearance in the Booster Neutrino Beam}

\newduneword{stt}{STT}{straw tube tracker}

\newduneword{wire board}{wire board}{At the head end of the APA in the \single TPC, stacks of electronics boards referred to as ``wire boards'' are arrayed to anchor the wires.  They also provide the connection between the wires and the cold electronics} 

\newduneabbrev{wls}{WLS}{wavelength-shifting}{A material or process by
  which incident photons are absorbed by a material and photons are
  emitted at a different, typically longer, wavelength}
  
\newduneabbrev{tpb}{TPB}{tetra-phenyl butadiene}{A 
\gls{wls} material}

\newduneabbrev{ptp}{PTP}{p-terphenyl}{A 
\gls{wls} material}

\newduneabbrev{sft}{SFT}{signal feedthrough}{A cryostat penetration allowing for the passage of cables or other extended parts}
\newduneabbrev{sftchimney}{SFT chimney}{signal feedthrough chimney}{In the \dual technology, a volume above the cryostat penetration used for a signal feedthrough}

\newduneabbrev{catiroc}{CATIROC}{charge and time integrated readout chip}{A complete read-out chip manufactured in AustriaMicroSystem designed to read arrays of 16 photomultipliers}

\newduneabbrev{wr}{WR}{White Rabbit}{A component of the timing system that forwards clock signal and time-of-day reference data to the master timing unit}

\newduneabbrev{mch}{MCH}{MicroTCA Carrier Hub}{An network switching device}

\newduneabbrev{wrmch}{WR-MCH}{White Rabbit \gls{utca} Carrier Hub}{A card mounted in \gls{utca} crate that recieves time syncronization information and trigger data packets over \gls{wr} network and disributes them to the \gls{amc} over \gls{utca} backplane} 

\newduneabbrev{wrtsn}{WR-TSN}{White Rabbit TimeStamping Node}{A unit on the \gls{wr} network that timestamps the trigger signals and sends out trigger data packets to \gls{wrmch}}

\newduneword{cmp}{CMP}{configuration management plan}
\newduneword{qap}{QAP}{quality assurance plan} 
\newduneword{ieshp}{IESHP}{integrated environmental, safety and health plan}
\newduneword{dmp}{DMP}{data management plan} 
\newduneword{qam}{QAM}{quality assurance manager} 

\newduneabbrev{dss}{DSS}{detector support system}{The system used to support a \gls{sp} \gls{detmodule} within its cryostat}

\newduneabbrev{ddss}{DDSS}{DUNE detector safety system}{The system used to manage key aspects of detector safety}

\newduneabbrev{lc}{LC}{logistics center}{A facility where \gls{lbnf} and \gls{dune} components will be received and transhipped to \gls{surf}}

\newduneabbrev{tco}{TCO}{temporary construction opening}{An opening in the side of a cryostat through which detector elements are brought into the cryostat; utilized during construction and installation}

\newduneabbrev{surf}{SURF}{Sanford Underground Research Facility}{The laboratory in South Dakota where the \gls{lbnf} \gls{fscf} will be constructed and the \gls{dune} \gls{fd} will be installed and operated}

\newduneabbrev{sit}{SIT}{surface installation team}{An organizational unit responsible for logistics and integration in South Dakota}

\newduneabbrev{uit}{UIT}{underground installation team}{An organizational unit responsible for installation in the underground area at the \gls{surf} site}

\newduneabbrev{cmgc}{CMGC}{construction manager/general contractor}{The organizational unit responsible for management of the construction of conventional facilities at the underground area at the \surf site}

\newduneword{cdrev}{conceptual design review}{A project management device by which a conceptual design is reviewed} 
\newduneword{pdr}{preliminary design review}{A project management device by which an early design is reviewed} 
\newduneword{fdr}{final design review}{A project management device by which a final design is reviewed}
\newduneword{prr}{production readiness review}{A project management device by which the production readiness is reviewed}
\newduneword{irr}{installation readiness review}{A project management device by which the plan for installation is reviewed}
\newduneword{orr}{operational readiness review}{A project management device by which the operational readiness is reviewed}
\newduneword{ppr}{production progress review}{A project management device by which the progress of production is reviewed} 
\newduneabbrev{edms}{EDMS}{engineering document management system}{A computerized document management system developed and supported at \gls{cern} in which some DUNE documents, drawings and engineering models are managed}
\newduneabbrev{ecr}{ECR}{engineering change request}{The first step in the change control process in which a proposed change is described}
\newduneabbrev{docdb}{DocDB}{Document DataBase}{A computerized document management system developed and supported at \gls{fnal} in which virtually all LBNF and most DUNE documents are managed}

\newduneword{wrgm}{WR grandmaster}{White Rabbit grandmaster}

\newduneabbrev{larsoft}{LArSoft}{Liquid Argon Software}{A shared base of physics software across \lartpc experiments}
\newduneword{nova}{NOvA}{The \nova off-axis neutrino oscillation experiment at \gls{fnal}}
\newduneword{minerva}{MINERvA}{The \minerva neutrino cross sections experiment at  \gls{fnal}}
\newduneword{microboone}{MicroBooNE}{The \lartpc-based \microboone neutrino oscillation experiment at  \gls{fnal}}
\newduneword{sbnd}{SBND}{The Short-Baseline Near Detector experiment at  \gls{fnal}}
\newduneabbrev{nexo}{nEXO}{Enriched Xenon Observatory}{Experiment at Lawrence Livermore National Laboratory (U.S. national lab in Livermore, CA)searching for new physics with neutrinoless double-beta decay}
\newduneword{argoneut}{ArgoNeuT}{The ArgoNeuT test-beam experiment and \gls{lartpc} prototype at  \gls{fnal}}
\newduneword{icarus}{ICARUS}{A neutrino experiment that was located at the Laboratori Nazionali del Gran Sasso (LNGS) in Italy, then refurbished at \gls{cern} for re-use in the same neutrino beam from \gls{fnal} used by the MiniBooNE, \gls{microboone} and \gls{sbnd} experiments. The ICARUS detector is being reassembled at \gls{fnal}}
\newduneword{atlas}{ATLAS}{One of two general-purpose detectors at the \gls{lhc}. It investigates a wide range of physics, from the search for the Higgs boson to extra dimensions and particles that could make up \gls{dm}}

\newduneword{lbne}{LBNE}{Long Baseline Neutrino Experiment (a terminated US project that was reformulated in 2014 under the auspices of the new \gls{dune} collaboration, an internationally coordinated and internationally funded program, with \gls{fnal} as host)}

\newduneabbrev{lbno}{LBNO}{Long Baseline Neutrino Observatory} {A terminated European project that, during its six-year duration, assessed the feasibility of a next-generation deep underground neutrino observatory in Europe)}

\newduneword{wirecell}{Wire-Cell}{A tomographic automated \threed neutrino event reconstruction method for \lartpc{}s}
\newduneabbrev{wct}{WCT}{Wire-Cell Toolkit}{A software toolkit with data flow processing components for \lartpc noise and signal simulation, noise filtering, signal processing, and tomographic \threed ionization activity imaging}
\newduneword{ftslite}{F-FTS-lite}{Light-weight version of the \fnal File Transfer system used for rapid data transfers out of the online systems}
\newduneabbrev{fts}{FTS}{File Transfer System}{A file transfer system developed at \fnal to catalog and move data to permanent storage}

\newduneword{35t}{35 ton prototype}{A prototype cryostat and \gls{sp} detector built at \fnal before the \gls{protodune} detectors}

\newduneabbrev{mcr}{MCR}{main communications room}{Space at the \gls{fd} site for cyber infrastructure}

\newduneabbrev{cuc}{CUC}{central utility cavern}{The utility cavern at the 4850L of \gls{surf} located between the two detector caverns. It contains utilities such as central cryogenics and other systems, and the underground data center and control room}

\newduneabbrev{cfd}{CFD}{computational fluid dynamics}{High performance computer-assisted modeling of fluid dynamical systems}
\newduneword{vuv}{VUV}{vacuum ultra-violet}
\newduneword{tallbo}{TallBo}{A cylindrical cryostat at \gls{fnal} primarily used for developing scintillation light collection technologies for \gls{lartpc} detectors}

\newduneword{root}{ROOT}{A modular scientific software toolkit. It provides all the functionalities needed to deal with big data processing, statistical analysis, visualisation and storage. It is mainly written in C++ but integrated with other languages such as Python and R}

\newduneabbrev{eos}{EOS}{EOS}{The XRootD-based distributed file system developed by CERN}
\newduneabbrev{ehn1}{EHN1}{Experiment Hall North One}{Location at CERN of the ProtoDUNE experiments}
\newduneword{led}{LED}{Light-emitting diode}
\newduneabbrev{rtd}{RTD}{resistance temperature detector}{A temperature sensor consisting of a material with an accurate and reproducible resistance/temperature relationship}
\newduneword{swc}{SWC}{Software \& Computing}
\newduneabbrev{las}{LAS}{LEM-anode Sandwich}{In the \dual technology, a \gls{lem} and its corresponding anode are mounted together in a module called a LEM-anode sandwich}

\newduneword{roi}{ROI}{region of interest}
\newduneabbrev{hpc}{HPC}{high-performance computing}{high-performance computing facilities; generally computing facilities emphasizing parallel computing with aggregate power of more than a teraflop}

\newduneword{comfund}{common fund}{The shared resources of the collaboration}
\newduneabbrev{ims}{IMS}{integrated master schedule}{A project management device consisting of linked tasks and milestones}

\newduneword{hvdb}{HVDB}{HV divider board}
\newduneword{sas}{SAS}{Another term for the materials airlock; a pass-through chamber used to ensure safe transfer of materials into a clean room, avoiding contamination in both directions}

\newduneabbrev{fea}{FEA}{finite element analysis}{Simulation of a physical phenomenon using the numerical technique called Finite Element Method (FEM), a numerical method for solving problems of engineering and mathematical physics}

\newduneword{fss}{FSS}{field shaping strips}
\newduneword{lvds}{LVDS}{low-voltage differential signaling}

\newduneword{esd}{ESD}{electrostatic discharge}

\newduneabbrev{rp}{RP}{resistive panel}{Resistive panels form the constant potential surfaces for a \gls{spmod} \gls{cpa}; they are composed of a thin layer of carbon-impregnated Kapton and laminated to both sides of a \frfour sheet}

\newduneword{uhmwpe}{UHMWPE}{ultra-high molecular weight polyethylene}

\newduneword{cts}{CTS}{Cryogenic Test System}
\newduneword{plc}{PLC}{programmable logic controller}

\newduneword{mppc}{MPPC}{\SI{6}{mm}$\times$\SI{6}{mm} Multi-Pixel Photon Counters produced by Hamamatsu\texttrademark{} Photonics K.K}

\newduneabbrev{sfp}{SFP}{small form-factor pluggable}{a particular standard for optical transceivers}

\newduneabbrev{minipod}{MiniPOD}{miniature parallel optical device}{a family of types of multi-channel optical transceivers}

\newduneword{ccc}{CCC}{configuration change command}
\newduneword{act}{ACT}{activation time stamp}
\newduneword{lcm}{LCM}{light calibration module}
\newduneword{lpm}{LPM}{light pulser module}
\newduneword{dac}{DAC}{digital-to-analog converter}
\newduneword{arapuca}{ARAPUCA}{A \gls{pds} design that consists of a light trap that captures wavelength-shifted photons inside boxes with highly reflective internal surfaces until they are eventually detected by \gls{sipm} detectors or are lost}
\newduneword{sarapu}{S-ARAPUCA}{Standard \gls{arapuca} design with different \gls{wls} coatings on both faces of the dichroic filter window(s) of the cell}
\newduneword{xarapu}{X-ARAPUCA}{Extended \gls{arapuca} design with \gls{wls} coating on only the external face of the dichroic filter window(s) but with a \gls{wls} doped plate inside the cell}
\newduneword{feb}{FEB}{front-end board}

\newduneabbrev{lsnd}{LSND}{Liquid Scintilator Neutrino Detector}{A scintillation detector and associated experiment located at Los Alamos National Laboratory}

\newduneabbrev{cvn}{CVN}{convolutional visual network}{An algorithm for identifying neutrino interactions based on their topology and without the need for detailed reconstruction algorithms}

\newduneword{pandora}{Pandora}{The Pandora multi-algorithm approach to pattern recognition} 

\newduneabbrev{pma}{PMA}{Projection Matching Algorithm}{A reconstruction algorithm that combines \twod reconstructed objects to form a \threed representation}
\newduneabbrev{bdt}{BDT}{boosted decision tree}{A method of multivariate analysis}
\newduneabbrev{cnn}{CNN}{convolutional neural network}{A deep learning technique most commonly applied to analyzing visual imagery}
\newduneword{pdg}{PDG}{Particle Data Group}

\newduneword{pci}{PCI}{Peripheral Component Interconnect}

\newduneword{labview}{LabVIEW}{Laboratory Virtual Instrument Engineering Workbench is a system-design platform and development environment for a visual programming language from National Instruments}

\newduneword{pcb}{PCB}{printed circuit board}

\newduneword{crio}{cRIO}{Compact Reconfigurable Input Output}

\newduneword{dcs}{DCS}{Distributed Communications System}

\newduneword{opc-ua}{OPC-UA}{OPC  Unified Architecture is a machine to machine communication protocol for industrial automation developed by the OPC Foundation. OPC stands for Object Linking and Embedding for Process Control}

\newduneword{cabangle}{Cabibbo angle}{A quark mixing parameter that governs the coupling of up quarks to strange quarks}
\newduneword{valor}{VALOR}{A neutrino oscillation fitting framework that is used by \gls{t2k}; the name stands for VALencia-Oxford-Rutherford, the original three institutions that developed it}
\newduneword{cafana}{CAFAna}{Common Analysis File Analysis}
\newduneabbrev{pca}{PCA}{principal component analysis}{A statistical procedure that uses an orthogonal transformation to convert a set of observations of possibly correlated variables into a set of values of linearly uncorrelated variables called principal components (Wikipedia)}
\newduneword{numi}{NuMI}{a set of facilities at \fnal, collectively called ``Neutrinos at the Main Injector.''  The NuMI neutrino beamline target system converts an intense proton beam into a focused neutrino beam}
\newduneword{gibuu}{GiBUU}{Giessen Boltzmann-Uehling-Uhlenback Project; a unified theory and transport framework in the MeV and GeV energy regimes for elementary reactions on nuclei }
\newduneabbrev{rpa}{RPA}{random phase approximation} {an approximation method commonly used for describing the dynamic linear electronic response of electron systems (Wikipedia)}
\newduneword{t2k}{T2K}{T2K (Tokai to Kamioka) is a long-baseline neutrino experiment in Japan studying neutrino oscillations}
\newduneword{mptdet}{MPT detector}{multipurpose tracking detector}

\newduneword{lariat}{LArIAT}{The repurposed ArgoNeuT \gls{lartpc}, modified for use in a charged particle beam, dedicated to the calibration and precise characterization of the output response of these detectors}

\newduneword{captain}{CAPTAIN}{Experimental program sited at \gls{lanl} that is designed to make measurements of scientific importance to \gls{lbl} neutrino physics and physics topics that will be explored by large underground detectors}

\newduneword{dayabay}{Daya Bay}{a neutrino-oscillation experiment in Daya Bay, China, designed to measure the mixing angle $\Theta_{13}$  using antineutrinos produced by the reactors of the Daya Bay and Ling Ao nuclear power plants}

\newduneword{nuwro}{NuWro}{neutrino interaction generator}

\newduneabbrev{neut}{NEUT}{neutrino interaction generator}{A neutrino interaction simulation program library for the studies of atmospheric accelerator neutrinos}

\newduneword{minos}{MINOS}{A long-baseline neutrino experiment, with a near detector at \gls{fnal} and a far detector in the Soudan mine in Minnesota, designed to observe the phenomena of neutrino oscillations (ended data runs in 2012)}

\newduneabbrev{efig}{EFIG}{Experimental Facilities Interface Group}{The body responsible for the required high-level coordination between the \gls{lbnf} and \gls{dune} projects}
\newduneword{ashriver}{Ash River}{The Ash River, Minnesota, USA \gls{nova} experiment far site, used as an assembly test site for \gls{dune}} 

\newduneword{ipd}{project integration director}{Responsible for integration and installation of \gls{lbnf} and \gls{dune} deliverables in South Dakota. Manages the \gls{integoff}}

\newduneabbrev{jpo}{JPO}{Joint Project Office}{The framework through which team members from the LBNF project office, \gls{integoff}, and DUNE \gls{tc} work together to provide coherence in project support functions across the global enterprise. 
Its functions include global project configuration and integration, installation planning and coordination, scheduling, safety assurance, technical review planning and oversight, development of partner agreements, and financial reporting}

\newduneword{ifbeam}{IFbeam}{Database that stores beamline information 
indexed by timestamp}

\newduneabbrev{marley}{MARLEY}{Model of Argon Reaction Low Energy
Yields}{Developed at UC Davis, MARLEY is the first realistic model of
neutrino electron interactions on argon for enegies less than \SI{50}{MeV}. This includes the energy range important for \gls{snb}
neutrinos and also solar 8--boron neutrinos}

\newduneabbrev{es}{ES}{elastic scattering}{Events in which a neutrino
elastically scatters off of another particle}

\newduneabbrev{cno}{CNO}{carbon nitrogen oxygen}{The CNO cycle (for carbon-nitrogen-oxygen) is one of the two known sets of fusion reactions by which stars convert
hydrogen to helium, the other being the proton-proton chain reaction
(pp-chain reaction). In the CNO cycle, four protons fuse, using
carbon, nitrogen, and oxygen isotopes as catalysts, to produce one
alpha particle, two positrons and two electron neutrinos}

\newduneabbrev{sdwf}{SDWF}{South Dakota Warehouse Facility}{Warehousing operations in South Dakota responsible for receiving LBNF and DUNE goods and coordinating shipments to the Ross shaft at \gls{surf}}

\newduneabbrev{wms}{WMS}{warehouse management system}{Commercial software package used to track shipments and interface to freight forwarders. This includes a database for shipping}

\newduneabbrev{dcdb}{DCDB}{DUNE construction database}{Database used by DUNE to track the history and testing of all parts of each \gls{detmodule}}

\newduneabbrev{aup}{AUP}{acceptance for use and possession}{Required for beneficial occupancy of the underground areas at SURF for LBNF and DUNE}

\newduneabbrev{bms}{BMS}{building management system}{Part of the safety system at \gls{surf} that includes the fire and life safety system}
\newduneabbrev{fls}{FLS}{fire and life safety system}{Part of the safety system at \gls{surf}}

\newduneabbrev{sno}{SNO}{Sudbury Neutrino Observatory}{The Sudbury
Neutrino Observatory was a detector built 6800 feet under ground, in
INCO's Creighton mine near Sudbury, Ontario, Canada. SNO was a
heavy-water Cherenkov detector designed to detect neutrinos produced
by fusion reactions in the sun}

\newduneword{sk}{Super-Kamiokande}{Experiment sited in the Kamioka-mine, Hida-city, Gifu, Japan that uses a large water Cherenkov detector to study neutrino properties through the observation of solar neutrinos, atmospheric neutrinos and man-made neutrinos}

\newduneabbrev{id}{ID}{inner diameter}{Inner diameter of a tube}

\newduneabbrev{od}{OD}{outer diameter}{Outer diameter of a tube}

\newduneabbrev{rms}{RMS}{root mean square}{The square root of the arithmetic mean of the squares of a set of values, used as a measure of the typical magnitude of a set of numbers, regardless of their sign}

\newduneabbrev{orc}{ORC}{operational readiness clearance}{Final safety approval prior to the start of operation}

\newduneabbrev{gsc}{GSC group}{global safety coordination group}{DUNE group that evaluates applicable codes and standards, including international code equivalency, for the design, assembly, and installation of the \gls{fd}}

\newduneabbrev{ha}{HA}{hazard analysis}{A first step in a process to assess risk; the result of hazard analysis is the identification of the hazards present for a task or process}
\newduneword{har}{HAR}{hazard analysis report}

\newduneabbrev{tap}{TAP}{trip action plan}{A document required for any trip by a worker to the underground area at \gls{surf}, per that site's access control program; 
it describes the work to be accomplished during the trip} 

\newduneword{em}{EM}{emergency management}
\newduneword{ert}{ERT}{emergency response team}

\newduneabbrev{ndk}{NDK}{nucleon decay}{The hypothetical, baryon number violating decay of a proton or a bound neutron into lighter particles}

\newduneabbrev{emi}{EMI}{electromagnetic interference}{Disturbance generated by an external source that affects an electrical circuit by electromagnetic induction, electrostatic coupling, or conduction}

\newduneabbrev{pe}{PE}{photoelectron}{An electron ejected from the surface of a material by the photoelectric effect}

\newduneabbrev{spe}{SPE}{single photoelectron}{A single photoelectron}

\newduneabbrev{fwhm}{FWHM}{full width at half maximum}{Width of a distribution measured between those points at which the distribution is equal to half of its maximum amplitude}

\newduneabbrev{gdml}{GDML}{geometry description markup language}{An application-indepedent, geometry-description format based on XML}

\newduneabbrev{xml}{XML}{extensible markup language}{A markup language that defines a set of rules for encoding documents in a format that is both human-readable and machine-readable}

\newduneabbrev{crt}{CRT}{cosmic ray tagger}{Detector external to the TPC designed to tag TPC-traversing cosmic ray particles}

\newduneabbrev{sn}{SN}{supernova}{Event that occurs upon the death of certain types of stars}

\newduneabbrev{wg}{WG}{working group}{A group of persons working together to achieve specified goals}

\newduneabbrev{ctsf}{CTSF}{coating, testing and storage facility}{A facility where the the \dual photon detectors will be coated, tested, and stored}

\newduneword{rucio}{Rucio}{Data management system originally developed
by \gls{atlas} but now open-source and shared across HEP}
\newduneabbrev{doma}{DOMA}{data organization, management, and
access}{data organization, management, and access efforts through the
HEP Software Foundation}

\newduneabbrev{hsf}{HSC}{HEP Software Foundation Collaboration}{A foundation that facilitates cooperation and common efforts in high energy physics software and computing internationally}

\newduneabbrev{wlcg}{WLCG}{Worldwide LHC Computing Grid}{Worldwide LHC
Computing Grid}
\newduneabbrev{osg}{OSG}{Open Science Grid}{Open Science Grid}
\newduneabbrev{sci}{SCI}{Scientific Computing Infrastructure}{Proposed
extension of the infrastructure component of \gls{wlcg} to other
experiments}
\newduneabbrev{csc}{CSC}{computing and software consortium}{DUNE
computing and software consortium}

\newduneword{dirac}{DIRAC}{Computing workflow management designed for
LHCb and now used by many HEP experiments}

\newduneword{frp}{FRP}{fiber-reinforced plastic}
\newduneabbrev{hdpe}{HDPE}{high-density polyethylene}{High-density polyethylene plastic}
\newduneword{hvps}{HVPS}{\gls{hv} power supply}
\newduneword{aisi}{AISI}{American Iron and Steel Institute}
\newduneword{ific}{IFIC}{Instituto de Fisica Corpuscular (in Valencia, Spain)}
\newduneabbrev{rsds}{RSDS}{radioactive source deployment system}{Proposed calibration system based on the deployment of
radioactive sources inside the \gls{dune} cryostat}
\newduneword{2p2h}{2p2h}{two particle, two hole}
\newduneabbrev{duneprism}{DUNE-PRISM}{\gls{dune} Precision Reaction-Independent Spectrum Measurement}{a mobile near detector that can perform measurements over a range of angles off-axis from the neutrino beam direction in order to sample many different neutrino energy distributions}
\newduneword{arcube}{ArgonCube}{The name of the core part of the \gls{dune} \gls{nd}, a \gls{lartpc}}

\newduneabbrev{citf}{CITF}{cryogenic instrumentation test facility}{A facility at \fnal with small ($<\,\SI{1}{ton}$) to intermediate ($\sim\,\SI{1}{ton}$) volumes of instrumented, purified TPC-grade \lar, used for testing devices intended for use in \gls{dune}}

\newduneabbrev{3dst}{3DST}{3D scintillator tracker}{The core part of the \threed projection scintillator tracker spectrometer in the near detector conceptual design}
\newduneabbrev{3dsts}{3DST-S}{3D scintillator tracker spectrometer}{The \threed projection scintillator tracker spectrometer  in the near detector conceptual design}
\newduneabbrev{mpd}{MPD}{multi-purpose detector}{A component of the near detector conceptual design; it is a magnetized system consisting of a \gls{hpgtpc} and a surrounding \gls{ecal}}
\newduneabbrev{hpg}{HPG}{high-pressure gas}{gas at high pressure to be used in a \gls{hpgtpc}} 
\newduneabbrev{hpgtpc}{HPgTPC}{high-pressure gaseous argon TPC}{A \gls{tpc} filled with gaseous argon; a possible component of the \gls{dune} \gls{nd}}

\newduneword{src}{SRC}{short-range correlated nucleon-nucleon interactions}
\newduneword{larpix}{LArPix}{ \gls{asic} pixelated charge readout for a \gls{tpc} }
\newduneword{arclt}{ArCLight}{a light detector \gls{arcube} effort}
\newduneword{fhc}{FHC}{forward horn current ($\numu$ mode)}
\newduneword{rhc}{RHC}{reverse horn current ($\overline{\nu}_{\mu}$ mode)}
\newduneword{mwpc}{MWPC}{multi-wire proportional chamber}
\newduneword{na61}{NA61}{CERN hadron production experiment}
\newduneword{pdnd}{ProtoDUNE-ND}{a prototype \gls{dune} \gls{nd}}
\newduneword{ccqe}{CCQE}{charged current quasielastic interaction} 
\newduneabbrev{roc}{ROC}{readout chamber}{readout chamber for gaseous argon \gls{tpc}}
\newduneabbrev{iroc}{IROC}{inner readout chamber}{inner (radial) readout chamber for gaseous argon \gls{tpc}}
\newduneabbrev{oroc}{OROC}{outer readout chamber}{outer (radial) readout chamber for gaseous argon \gls{tpc}}

\newduneword{lux}{LUX}{Large Underground Xenon (LUX) dark matter detector at \gls{surf} }

\newduneword{mjdemo}{Majorana Demonstrator}{Experiment sited at \gls{surf} that  seeks to determine whether neutrinos are their own antiparticles}

\newduneword{lz}{LZ}{Experiment sited at \gls{surf} that  seeks to detect faint interactions between galactic dark matter and regular matter}

\newduneword{mu2e}{Mu2e}{An experiment sited at \gls{fnal} that searches for charged-lepton flavor violation and seeks to discover physics beyond the \gls{sm}}

\newduneword{pdsp2}{ProtoDUNE-SP-2}{A second test run in the singe-phase
ProtoDUNE test stand at CERN, acting as a validation of the final
single-phase detector design}

\newduneword{osha}{OSHA}{Occupational Safety and Health Administration (USA Department of Labor) formed by the Occupational Safety and Health Act of 1970}
\newduneabbrev{pns}{PNS}{pulsed neutron source}{Calibration system based
on neutron capture gamma showers spread out in the whole detector}

\newduneabbrev{fv}{FV}{fiducial volume}{The detector volume within the \gls{tpc} 
that is selected for physics analysis through cuts on reconstructed event position}

\newduneword{p6}{P6}{framework used to plan and status the resource-loaded schedule of activities associated with the USA contributions to \gls{lbnf} and \gls{dune} }
\newduneabbrev{evms}{EVMS}{earned value management system}{Earned Value Management is a systematic approach to the integration and measurement of cost, schedule, and technical (scope) accomplishments on a project or task. It provides both the government and contractors the ability to examine detailed schedule information, critical program and technical milestones, and cost data (text from the US DOE); the EVMS is a system that implements this approach}

\newduneword{core}{CORE}{CORE contributions are in either monetary units or labor hours. They can be technical components for the facility or experiment and the effort of the staff needed to produce, install, and test them;  major facilities for the experiment; or other products and services relevant for the completion of the facility or experiment} 

\newduneabbrev{ahj}{AHJ}{Authority Having Jurisdiction}{An organization, office, or individual responsible for enforcing the requirements of a code or standard, or for approving equipment, materials, an installation, or a procedure (OSHA)}
\newduneword{cte}{CTE}{coefficient of thermal expansion}

\newduneabbrev{opc}{OPC}{open platform communications}{Open platform communications is a series of standards and specifications for industrial telecommunication} 
\newduneword{scada}{SCADA}{supervisory control and data acquisition}
\newduneword{ln}{LN$_2$}{liquid nitrogen}
\newduneabbrev{lapd}{LAPD}{Liquid Argon Purity Demonstrator}{Cryostat at Fermilab for long-term studies requiring a large volume of argon}

\newduneabbrev{pab}{PAB}{Proton Assembly Building}{Home of several \gls{lar} facilities at Fermilab}
\newduneword{hep}{HEP}{high energy physics}
\newduneword{sc}{SC}{scientific computing}  
\newduneword{cms}{CMS}{Compact Muon Solenoid experiment at CERN}
\newduneword{alice}{ALICE}{A Large Ion Collider Experiment, at CERN}
\newduneword{gpib}{GPIB}{general purpose interface bus}

\newduneabbrev{pfparticle}{PFParticle}{particle flow particle}{Each of the individual reconstructed particles in the hierarchy (or particle flow) describing the reconstructed event interaction}

\newduneabbrev{mcparticle}{MCParticle}{Monte Carlo Particle}{Individual true simulated particle}
\newduneword{au}{AU}{astronomical unit}
\newduneword{nufit}{NuFIT 4.0}{The NuFIT 4.0 global fit to neutrino oscillation data}

\newduneabbrev{sgft}{SGFT}{term}{add def (DP install)}
\newduneword{uhv}{UHV}{ultra high vacuum}
\newduneword{lps}{LPS}{laser positioning system}

\newduneword{unicamp}{UNICAMP}{University of Campinas, Sao Paulo, Brazil}
 
\newduneabbrev{fbk}{FBK}{Fondazione Bruno Kessler}{FBK is a research non-profit entity in Trento, Italy that partners in the development of technology with applications in various fields including High Energy Physics}

\newduneword{fft}{FFT}{fast Fourier transform}
\newduneabbrev{enob}{ENOB}{effective number of bits}{The effective number of bits is a measure of the dynamic range of an \gls{adc} and its associated circuitry. The resolution of an \gls{adc} is specified by the number of bits used to represent the analog value, in principle giving 2N signal levels for an N-bit signal. However, all real \gls{adc} circuits introduce noise and distortion. ENOB specifies the resolution of an ideal \gls{adc} circuit that would have the same resolution as the circuit under consideration}
\newduneabbrev{sndr}{SNDR}{signal to noise and distortion ratio}{Also known as SINAD. Ratio of the \gls{rms} signal amplitude to the mean value of the root-sum-square of all other spectral components, including harmonics, but excluding \gls{dc} levels. It is a good indication of the overall dynamic performance of an \gls{adc} because it includes all components which make up noise and distortion}
\newduneabbrev{sfdr}{SFDR}{spurious free dynamic range}{Spurious free dynamic range is the ratio of the \gls{rms} value of the signal to the \gls{rms} value of the worst spurious signal regardless of where it falls in the frequency spectrum. The worst spur may or may not be a harmonic of the original signal}
\newduneabbrev{thd}{THD}{total harmonic distortion}{Total harmonic distortion is the ratio of the \gls{rms} value of the fundamental signal to the mean value of the root-sum-square of its harmonics} 
\newduneword{tvs}{TVS}{transient voltage suppression}

\newduneword{riskprob}{risk probabilities}{The risk probability, after taking into account the planned mitigation activities, is ranked as 
L (low $<\,$\SI{10}{\%}), 
M (medium \SIrange{10}{25}{\%}), or 
H (high $>\,$\SI{25}{\%}). 
The cost and schedule impacts are ranked as 
L (cost increase $<\,$\SI{5}{\%}, schedule delay $<\,$\num{2} months), 
M (\SIrange{5}{25}{\%} and 2--6 months, respectively) and 
H ($>\,$\SI{20}{\%} and $>\,$2 months, respectively)}

\newduneabbrev{lbls}{LBLS}{laser beam location system}
{Auxiliary calibration system providing an independent location measurement of the ionization laser beams direction}

\newduneabbrev{lsst}{LSST}{Large Synoptic Survey Telescope}{8.4 m telescope with 3.2G-pixel camera that will start taking data in 2023}
\newduneabbrev{ska}{SKA}{Square Kilometer Array}{International radio telescope array planned to start data-taking in 2027}
\newduneabbrev{hyperk}{HyperK}{Hyper Kamiokande}{260 kt water Cerenkov neutrino detector to begin construction at Kamiokande in 2020}
\newduneword{lhcb}{LHCb}{LHC experiment dedicated to forward physics}
\newduneword{belleii}{Belle II}{B-factory experiment now running at KEK}

 \newduneabbrev{ldm}{LDM}{light-mass dark matter}{Refers to dark matter particles with mass values much lower than the electroweak scale, specifically below the 1~GeV level}
 
\newduneabbrev{bnv}{BNV}{baryon-number violating}{Describing an interaction where \gls{baryonnumber} is not conserved}

\newduneword{bugey}{Bugey}{Neutrino experiment that operated at the Bugey nuclear power plant in France}

\newduneword{minosplus}{MINOS$+$}{The successor to the \gls{minos} experiment, utilizing the same detectors and beam line, but operating at higher beam energy tune than \gls{minos}, parasitic with \gls{nova}}

\newduneword{baryonnumber}{baryon number}{A quantity expressing the total number of baryons in a system minus the number of antibaryons}

\newduneword{np04}{NP04}{CERN North Area hadron beamline used for the \gls{sp} test beam run}

\newduneword{ua1}{UA1}{UA1 (Underground Area 1) was a particle detector at \gls{cern}'s  Super Proton Synchrotron (SPS). It ran from 1981 until 1990, when the SPS was used as a proton-antiproton collider, searching for traces of W and Z particles in collisions. (CERN) The UA1 dipole magnet was reused in the NOMAD experiment and currently provides the magnetic field for the \gls{t2k} ND280 detector}

\newduneword{ssc}{SSC}{The Superconducting Super Collider was to be a huge underground ring complex beneath the area near Waxahachie, Texas, USA, that would have been the world’s most energetic particle accelerator. It was begun in 1990, but canceled by the U.S. Congress in 1993 (scientificamerican.com Oct 2013)}

\newduneword{daphne}{DAPHNE}{Detector electronics for Acquiring PHotons from NEutrinos is a custom-developed warm front-end waveform digitizing electronics module derived from the readout system developed at Fermilab for the Mu2e experiment}
 
\newduneword{nersc}{NERSC}{National Energy Research Computing Facility at \gls{lbnl}}

  \newduneword{integoff}{integration office}{The office that incorporates the onsite team responsible for coordinating integration and installation activities at SURF}

\newduneabbrev{sma}{SMA}{SubMiniature version A}{Connector interface for coaxial cables
with a screw-type coupling mechanism}

\newduneword{kloe}{KLOE}{KLOE is a $e^+ e^-$ collider detector spectrometer operated at DAFNE, the $\phi$-meson factory at Frascati, Rome. In DUNE it will consist of a \SI{26}{cm} Pb+scintillating fiber ECAL surrounding a cylindrical open detector region that is  \SI{4.00}{m} in diameter and \SI{4.30}{m} long. The ECAL and detector region are embedded in a \SI{0.6}{T} magnetic field created by a \SI{4.86}{m} diameter superconducting coil and a \SI{475}{tonne} iron yoke}

\newduneword{ro}{review office}{An office within the \gls{integoff} that organizes reviews }

\newduneabbrev{doecd}{CD}{critical decision}{The U.S. DOE's Order 413.3B outlines a series of staged project approvals, each of which is referred to as a critical decision (CD)}

\newduneabbrev{lbnfspac}{LBNF SPAC}{LBNF Strategic Project Advisory Committee}{A committee charged by the host laboratory director to provide expert, independent advice on significant issues and strategies related to LBNF project organization, management, and risks}

\newduneabbrev{sand}{SAND}{System for on-Axis Neutrino Detection}{The beam monitor component of the near detector that remains on-axis at all times and serves as a dedicated neutrino spectrum monitor}

\newduneword{4850l}{4850L}{The depth in feet (1480 m) of the top of the cryostats underground at SURF; used more generally to refer to the DUNE underground area. Called the ``4850 level'' or ``4850L''}

\hypersetup{
    pdftitle={\expshort TDR \thedocsubtitle},
    pdfauthor={\expshort Collaboration},
    final=true,
    colorlinks=false,
    linktocpage=true,
    linkbordercolor=blue,
    citebordercolor=green,
    urlbordercolor=magenta,
    filecolor=black,
    pdfpagemode=UseOutlines,
    pdfborderstyle={/S/U},  
}

\renewcommand\thedoctitle{\voltitlephysics} 
\newcommand\thevolumenumber{\volnumberphysics} 

\begin{document}

\pagestyle{titlepage}
\includepdf[pages={-}]{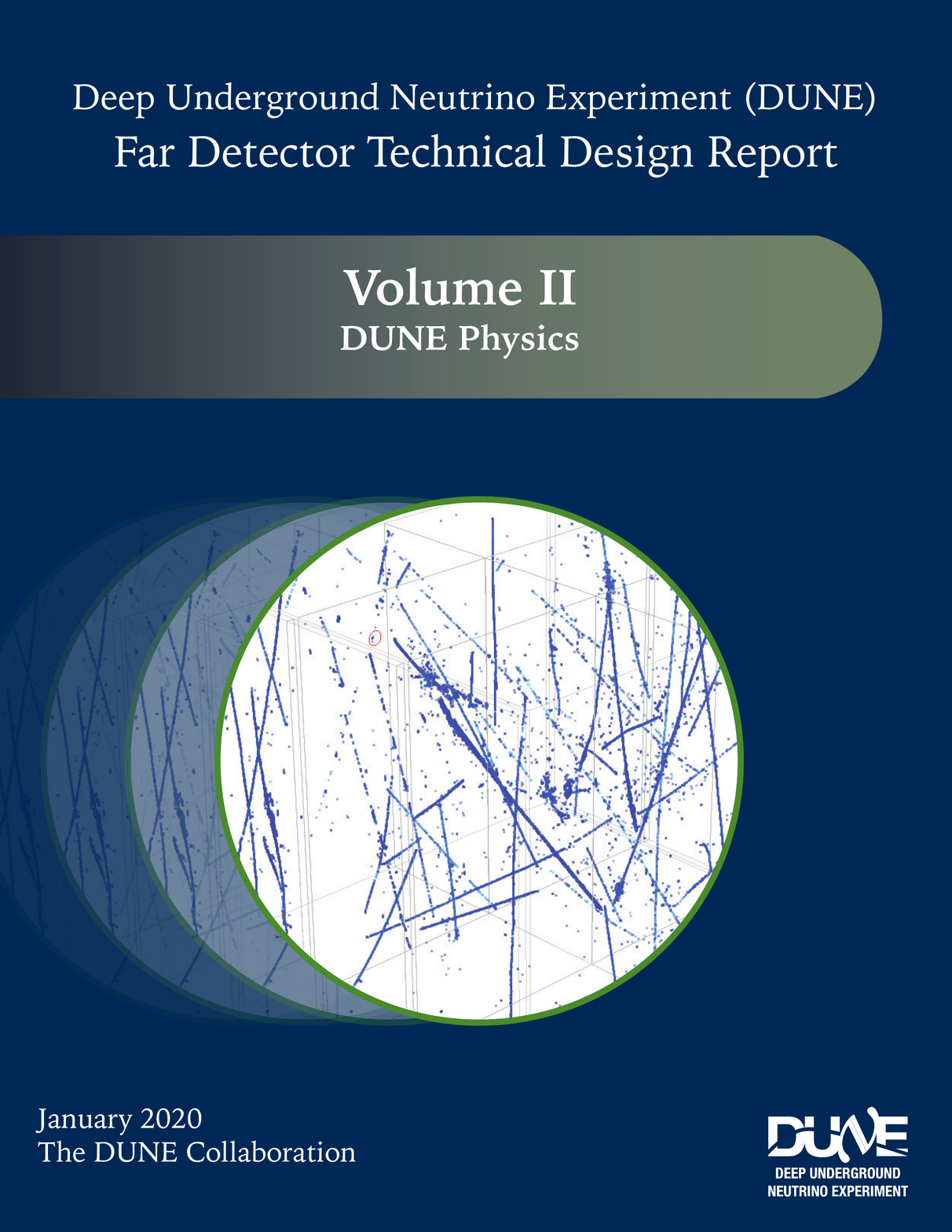}
\cleardoublepage

\cleardoublepage
\vspace*{16cm} 
  {\small  This document was prepared by the DUNE collaboration using the resources of the Fermi National Accelerator Laboratory (Fermilab), a U.S. Department of Energy, Office of Science, HEP User Facility. Fermilab is managed by Fermi Research Alliance, LLC (FRA), acting under Contract No. DE-AC02-07CH11359.
  
The DUNE collaboration also acknowledges the international, national, and regional funding agencies supporting the institutions who have contributed to completing this Technical Design Report.  
  }
\includepdf[pages={-}]{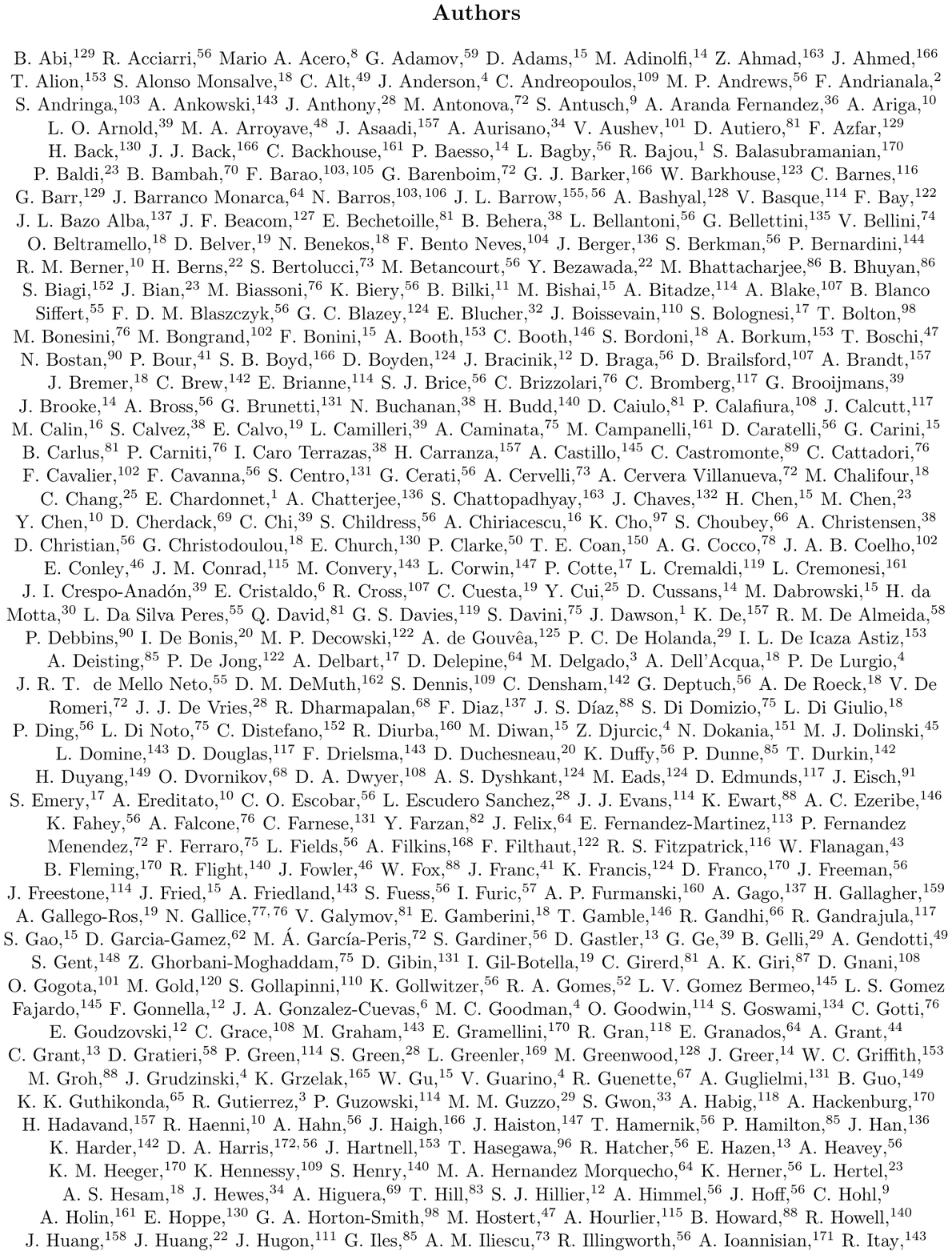}

\renewcommand{\familydefault}{\sfdefault}
\renewcommand{\thepage}{\roman{page}}
\setcounter{page}{0}

\pagestyle{plain}

\textsf{\tableofcontents}

\textsf{\listoffigures}

\textsf{\listoftables}
  \vspace{4mm}
  \addcontentsline{toc}{chapter}{A Roadmap of the DUNE Technical Design Report}

\iffinal\else
\textsf{\listoftodos}
\clearpage
\fi

\renewcommand{\thepage}{\arabic{page}}
\setcounter{page}{1}

\pagestyle{fancy}

\renewcommand{\chaptermark}[1]{%
\markboth{Chapter \thechapter:\ #1}{}}
\fancyhead{}
\fancyhead[RO,LE]{\textsf{\footnotesize \thechapter--\thepage}}
\fancyhead[LO,RE]{\textsf{\footnotesize \leftmark}}

\fancyfoot{}
\fancyfoot[RO]{\textsf{\footnotesize The DUNE Technical Design Report}}
\fancyfoot[LO]{\textsf{\footnotesize \thedoctitle}}
\fancypagestyle{plain}{}

\renewcommand{\headrule}{\vspace{-4mm}\color[gray]{0.5}{\rule{\headwidth}{0.5pt}}}



\cleardoublepage
\chapter*{A Roadmap of the DUNE Technical Design Report}

The \dword{dune} \dword{fd} \dword{tdr} describes the proposed physics program,  
detector designs, and management structures and procedures at the technical design stage.  

The TDR is composed of five volumes, as follows:

\begin{itemize}
\item Volume~\volnumberexec{} (\voltitleexec{}) provides an overview of all of DUNE for science policy professionals.

\item Volume~\volnumberphysics{} (\voltitlephysics{}) describes the DUNE physics program.

\item Volume~\volnumbertc{} (\voltitletc{}) outlines DUNE management structures, methodologies, procedures, requirements, and risks. 

\item Volume~\volnumbersp{} (\voltitlesp{}) and Volume~\volnumberdp{} (\voltitledp{}) describe the two \dword{fd} \dword{lartpc} technologies.

\end{itemize}

The text includes terms that hyperlink to definitions in a volume-specific glossary. These terms  appear underlined in some online browsers, if enabled in the browser's settings.

\cleardoublepage

\cleardoublepage

\chapter{Introduction and Executive Summary}
\label{ch:exec-summ}

The Physics volume of the DUNE  \dword{fd} Technical Design Report
(TDR) presents the science program of DUNE.
Within, we describe the array of 
identified scientific opportunities and key goals.  
Crucially, we also report our best current understanding
of the capability of DUNE to realize these goals,
along with the detailed arguments and investigations 
on which this understanding is based.

In the context of the complete set of DUNE TDR volumes,
a central role for this volume is to document the scientific
basis underlying the conception and design of the LBNF/DUNE
experimental configurations.  As a result, it is the
description of DUNE's experimental capabilities that constitutes
the bulk of the document.  Key linkages between requirements for
successful execution of the physics program and
primary specifications of the experimental configurations
are drawn and summarized.

This document also serves a wider purpose as a statement on the
scientific potential of DUNE as a central component within
a global program of frontier theoretical and experimental
particle physics research. Thus, the presentation also
aims to serve as a resource for
the particle physics community at large.

In this chapter,
the scientific goals, the methodologies utilized to obtain
sensitivity projections, the corresponding results
for selected elements of the scientific program, and
the demands placed on the experiment design and performance
are presented in summary form.  
Together with the two chapters that follow,
this summary establishes the context for the
detailed descriptions specific to each area of
research that comprise the remaining chapters.


\section{Overview of DUNE and its Science Program}
\label{sec:exec-program-overview}

The Deep Underground Neutrino Experiment (DUNE) will be a
world-class neutrino observatory and nucleon decay detector
designed to answer
fundamental questions about the nature of elementary particles
and their role in the universe. The international DUNE
experiment, hosted by the U.S. Department of Energy's \fnal{},
will consist of a far detector to be located about \SI{1.5}{km}
underground at the Sanford Underground Research Facility
(\surf) in South Dakota, USA, at a distance of  \SI{1300}{\km}
from \fnal{}, and a near detector to be located at \fnal in
Illinois. The far detector will be a very large, modular
liquid argon time-projection chamber (\lartpc) with a total mass of nearly \SI{70}{kt} 
(fiducial mass of at least  
\fdfiducialmass{}). This \lar technology will make
it possible to reconstruct neutrino interactions
with image-like precision.

The far detector will be exposed to the world's most intense 
neutrino beam originating at \fnal{}. A high-precision near 
detector, located in a hall \SI{574}{m} from the neutrino source on 
the \fnal site, will be used to characterize
the intensity and energy spectrum of this wide-band 
beam in real time.  Over the long term, the near detector 
will also enable many strategies for mitigating systematic 
errors, both through direct cancellation of errors common to both near and far detectors, as well as through dedicated studies
of exclusive neutrino interaction channels, beam line characteristics, 
and reconstructed neutrino energy uncertainties, to name a few.

In this section, the goals of the DUNE science program are
presented.  Assumptions and methods utilized in determining
DUNE's capabilities to meet these goals are summarized,
with more detail appearing in Chapter~\ref{ch:tools}.
Finally, experimental sensitivities for selected
physics measurements are shown to illustrate the achieved level of 
performance demonstrated.

\subsection{Key Goals of the DUNE Science Program}
\label{sec:exec-key-goals}

The LBNF/DUNE strategy has been developed to meet the
requirements set by the U.S. Particle Physics Project Prioritization Panel
(P5) in 2014. It also takes into account the recommendations
of the European Strategy for Particle Physics (ESPP) adopted
by the CERN Council in 2013, which classified the \dword{lbl}
neutrino program as one of the four scientific objectives requiring significant resources, sizable collaborations, and sustained commitment.  

As a benchmark, the P5 report~\cite{p5report2014} set the goal of
reaching a sensitivity to \dword{cpv} of better than three
standard deviations (\num{3}$\sigma$) over more than $75\%$
of the range of possible values of the unknown
\dshort{cp}-violating phase \deltacp.
Based partly on this goal, it stated that ``the
minimum requirements to proceed are the identified capability
to reach an exposure of \num{120}~\ktMWyr{} by the 2035 time
frame, the far detector situated underground with cavern space
for expansion to at least \fdfiducialmass \lar fiducial volume,
and \SI{1.2}{MW} beam power upgradeable to multi-megawatt power.
The experiment should have the demonstrated capability to
search for \dwords{snb} and for proton decay, providing a
significant improvement in discovery sensitivity over current
searches for the proton lifetime.''
These requirements are discussed below and in the sections
that follow.

To summarize, the DUNE experiment will combine the world's most
intense neutrino beam, a deep underground site, and massive \lar
detectors to enable a broad science program addressing some of
the most fundamental questions in particle physics.
This program is articulated in brief form below.

The primary science goals of DUNE are to:
\begin{itemize}

\item Carry out a comprehensive program of neutrino oscillation measurements using
      \numu and \anumu beams from \fnal. This program includes measurements of
      the \dword{cp} phase, determination of the neutrino mass ordering
      (the sign of \dm{31}$ \equiv m_3^2-m_1^2$), measurement of the mixing
      angle $\theta_{23}$ and the determination of the octant in which this
      angle lies, and sensitive tests of the three-neutrino paradigm.
      Paramount among these is the search for \dword{cpv} in neutrino oscillations, potentially offering 
      insight into the origin of the matter-antimatter asymmetry,
      one of the fundamental questions in particle physics and cosmology.

\item Search for proton decay in several 
decay modes.
      The observation of proton decay would represent a ground-breaking discovery
      in physics, satisfying a key requirement of the grand unification of the forces.

\item Detect and measure the $\nu_\text{e}$ flux from a core-collapse
      supernova within our galaxy, should one occur during the lifetime
      of the DUNE experiment. Such a measurement would provide a wealth
      of unique information about the early stages of core-collapse, and
      could even signal the birth of a black hole.

\end{itemize}

The intense neutrino beam from LBNF, the massive DUNE \lartpc
far detector, and the high-resolution DUNE near detector will also
provide a rich ancillary science program, beyond the primary goals
of the experiment. The ancillary science program includes:
\begin{itemize}

\item Other accelerator-based neutrino flavor transition measurements with
      sensitivity to beyond the standard model (BSM) physics, such as non-standard
      interactions (NSIs), Lorentz invariance violation, \dword{cpt} violation,
      sterile neutrinos, large extra dimensions, heavy neutral leptons,
      and tests with measurements of tau neutrino appearance;

     \item Measurements of neutrino oscillation phenomena using atmospheric neutrinos;

     \item Searches for dark matter utilizing a variety of
           signatures in both
           near and far detectors, as well as  
           non-accelerator searches for BSM physics 
           such as neutron-antineutron oscillation.

     \item A rich neutrino interaction physics program utilizing the DUNE near detector,
           including a wide-range of measurements of neutrino cross sections and studies of
           nuclear effects.

\end{itemize}

Further advancements in the \lartpc 
technology during the course of the far detector construction
may enhance DUNE's capability to observe very low-energy
phenomena such as solar neutrinos or even the diffuse
supernova neutrino flux.

\subsection{Summary of Assumptions and Methods Employed}
\label{sec:exec-assm-meth}

Scientific capabilities are determined assuming DUNE
is configured according to the general parameters described above.
Further assumptions regarding the neutrino beam and detector 
systems, and their deployment, are stated here in
Sections~\ref{sec:exec-assm-meth-beamdetector} and
\ref{sec:exec-assm-meth-deployment}.  More detail is given
in later chapters as appropriate.

Determination of experimental sensitivities relies on the
modeling of the underlying physics and background processes,
as well as the detector response, including calibration and
event reconstruction performance and the utilization of data
analysis techniques and tools.
While a brief discussion of the strategies employed is given below
in Sec.~\ref{sec:exec-assm-meth-simreco}, a dedicated chapter
(\ref{ch:tools}) is devoted to the presentation of this material.
Considerations specific to individual elements of the science
program are presented in detail in the corresponding chapters.

\subsubsection{Beam and Detector}
\label{sec:exec-assm-meth-beamdetector}

This document presents physics sensitivities using
the optimized design of the 1.2~MW neutrino beam and
corresponding \dword{pot} per year assumed to
be 1.1 $\times 10^{21}$ \dword{pot}.  These numbers assume a combined
uptime and efficiency of the \fnal accelerator complex and the
LBNF beamline of 56\%.\footnote{This projection, from which one  
year of LBNF beam operations can be expressed as \SI{1.7e7}{seconds}, 
is based on extensive 
experience with intense neutrino beams at Fermilab, and in particular 
the NuMI beam line, which incorporates elements like those in the  
proposed LBNF beamline design and faces similar operating conditions.} 
The beam design, simulation and associated
uncertainties are described in
Sec.~\ref{sec:physics-lbnosc-flux}.

For the neutrino oscillation physics program, it is assumed that
equal exposures (time-integrated beam power times fiducial mass) are obtained with both horn current polarities,
and therefore with the corresponding mix of primarily \numu
and \anumu data samples (see Sec.~\ref{sec:physics-lbnosc-osc}).

It is assumed that the DUNE far detector will include some
combination of the different \nominalmodsize fiducial volume
implementations (single or dual-phase) of the \lartpc concept
for which technical designs have been developed.
For much of the science program, it is expected that the
capabilities of the two proposed far detector module 
implementations will be comparable.  As a result of the
current state of reconstruction and analysis software development
(see Sec.~\ref{sec:exec-assm-meth-simreco}), the
physics sensitivity studies reported in this TDR are based on
the single-phase \lartpc implementation,
documented in full in Volume~\volnumbersp{}.

It is also assumed that validation of the DUNE far detector 
designs will come from data and operational experience acquired 
with the large-scale ProtoDUNE detectors staged at CERN, 
including single-particle studies of data obtained 
in test-beam running.  Although this program is in early stages, 
beam data has already been collected with the ProtoDUNE-SP detector.  
Where possible, preliminary results from initial analyses of these 
data are presented in this document (see Chapter~\ref{ch:tools}).

The near detector for DUNE has been under active development,
and a Conceptual Design Report is in preparation.
Correspondingly, the descriptions utilized in this TDR
are consistent with this level of development.  As the
beam-based neutrino oscillation program depends strongly
on the capabilities of the near detector systems, a brief
summary of these systems and their expected performance is
given in Chapter~\ref{ch:osc}.

\subsubsection{Deployment Scenario}
\label{sec:exec-assm-meth-deployment}

Where presented as a function of calendar year,
sensitivities are calculated with the following
assumed deployment plan, which is based on a
technically limited schedule.
\begin{itemize}
    \item Start of beam run: Two \dword{fd} module 
    volumes for total fiducial mass of 20 kt, 1.2 MW beam
    \item After one year: Add one \dword{fd} module  volume for total fiducial mass of 30 kt
    \item After three years: Add one \dword{fd} module  volume for total fiducial mass of \fdfiducialmass
    \item After six years: Upgrade to 2.4 MW beam
\end{itemize}

\subsubsection{Simulation, Reconstruction and Data Analysis Tools}
\label{sec:exec-assm-meth-simreco}

The development of algorithms and software infrastructure needed
to carry out physics sensitivity studies has been an active 
effort within DUNE and the associated scientific community.  
As demonstrated in Chapter~\ref{ch:tools}, significant progress 
has been made: event reconstruction 
codes can be run on fully simulated neutrino interaction events 
in DUNE far detector modules; the DUNE computing infrastructure 
allows high-statistics production runs; and end-user interfaces 
are functioning.  Robust end-to-end analyses not 
possible a year ago have now been done and are being 
reported in this document.

For some aspects -- for example, beamline modeling
and GeV-scale neutrino interaction simulations --
well-developed and validated (with data) software packages have
been available throughout much of DUNE's design phase.
For others, corresponding tools did not exist and needed to be
either developed from scratch or adapted with substantial
modifications from other experimental programs.  Concurrent
with these development efforts, interim descriptions such
as parametric detector response modeling, necessarily simple
but based on reasonable extrapolation from experience and
dedicated studies, were employed to assess physics capabilities.
Even for the case of the better-developed tools -- again, neutrino 
interaction modeling is a good example -- significant incremental
improvements have been made as data from neutrino experiments
and other sources have become available and as theoretical
understandings have advanced.

As a result of the rapid pace of development as well as 
practical considerations including human 
resource availability, different levels
of rigor have been applied in the evaluation of physics capabilities for different elements of the program.  
The strategy adopted for
this TDR has been to hold the primary elements of the program
to the highest standard of rigor, involving direct analysis
of fully simulated data, utilizing actual event reconstruction
codes and analysis tools that could be applied to real data
from DUNE far detector modules.  For other elements of the
program, sensitivities utilize realistic beam and
physics simulations, but employ parametric detector
response models in place of full reconstruction.

The implementation of this strategy comes with caveats
and clarifications that are discussed in the corresponding
chapters.  Some of these are mentioned here.
\begin{itemize}
\item In the case of the long-baseline oscillation physics
      program, this approach requires a combination of the 
      fully end-to-end analysis of simulated far detector data
      with the concurrent analysis of simulated data from
      near detector systems to capture in a realistic way 
      the level of 
      control over systematic errors.  Given the 
      current state of development in the DUNE near detector design and 
      corresponding analysis tools,  it has been necessary to 
      employ parametric detector response modeling for near 
      detector components, as described in
      Sec.~\ref{sec:nu-osc-06}.

\item In the case of the nucleon decay searches
      (Sec.~\ref{sec:nonaccel-ndk}),
      reconstruction and analysis tools dedicated toward
      addressing the particular challenges presented are 
      not as well developed as in the case of the 
      beam-based oscillation physics program. Effort is 
      ongoing to improve the performance of these tools. 

\item The supernova neutrino burst program (Sec.~\ref{sec:snb-lowe-snb}) relies on reconstruction of event signatures from LArTPC signals generated by low-energy (MeV-scale) particles (electrons and de-excitation gammas).  Full simulation and reconstruction is used for some studies, such as for the directionality study described in Sec~\ref{sec:pointing}.
For other studies, a modified strategy is employed in order to efficiently explore model space:  reconstruction metrics (resolution smearing matrices, for example) are derived from analysis of fully simulated and reconstructed low-energy particles and events in the far detector, and applied to understand mean detector response over a range of signal predictions.


\item It should be noted that for scientific program elements where
      analysis of fully reconstructed simulated data has 
      not yet been performed, the parametric response models used for the analyses presented here have
      been well characterized with dedicated studies
      and incorporation of results from other experiments.
      The demonstration of sensitivities for the long-baseline
      oscillation physics program (with full reconstruction) 
      that are comparable to those
      previously obtained based on parametric response
      provides validation for this approach.
\end{itemize}

\subsection{Selected Results from Sensitivity Studies}
\label{sec:exec-sensitiv-results}

In this section, selected sensitivity projections from the 
central elements of the DUNE science program are presented.  
This selection is intended to convey just the headlines from 
what is an extensive and diverse program of frontier science.

\subsubsection{DUNE can discover \dword{cpv} in the neutrino 
sector and precisely measure oscillation parameters}

The key strength of the DUNE design concept is its ability to 
robustly measure the oscillation patterns of \numu and \anumu 
over a range of energies spanning the first and second 
oscillation maxima (see, {\sl e.g.},  Fig.~\ref{fig:oscprob} 
in Chapter~\ref{ch:osc}). 
This is accomplished by a coordinated analysis of the 
reconstructed \numu, \anumu, \nue, and \anue energy spectra 
in near and far detectors, 
incorporating data collected with forward (neutrino-dominated) 
and reverse (antineutrino-dominated) horn current polarities.  

The statistical power of DUNE relative to the current 
generation of long-baseline oscillation experiments 
is a result of many factors including  
(1) on-axis operations, (2) the LBNF beam power, 
(3) long baseline and correspondingly high energy 
oscillation maxima and strong separation of 
normal and inverted neutrino mass ordering scenarios, 
(4) detector mass, and (5) event 
reconstruction and selection capabilities. 
Tables~\ref{tab:execsumm-apprates} and 
\ref{tab:execsumm-disrates} 
(reproduced later as Tables~\ref{tab:apprates} and 
\ref{tab:disrates}) 
give the expected event 
yields for the appearance (\nue and \anue) 
and disappearance (\numu and \anumu) channels, respectively, 
after seven years of operation, assuming $\mdeltacp = 0$ and
\dword{nufit}~\cite{Esteban:2018azc,nufitweb} 
values (given in Table~\ref{tab:oscpar_nufit}) for other parameters.
\begin{dunetable}
[\nue and \anue appearance rates]
{lrr}
{tab:execsumm-apprates}
{\nue and \anue appearance rates: Integrated rate of selected $\nu_e$ \dword{cc}-like events between 0.5 and 8.0~GeV assuming \num{3.5}-year (staged) exposures in the neutrino-beam and antineutrino-beam modes.  The signal rates are shown for both normal mass ordering (NO) and inverted mass ordering (IO), and all the background rates assume normal mass ordering.  All the rates assume $\mdeltacp = 0$, and \dword{nufit}~\cite{Esteban:2018azc,nufitweb} 
values for other parameters.}
& \multicolumn{2}{c}{Expected Events (3.5 years staged per mode)} \\ \toprowrule
 & $\nu$ mode & $\bar{\nu}$ mode  \\
 \colhline 
 \nue Signal NO (IO) & 1092 (497) & 76 (36) \\
 \anue Signal NO (IO) & 18 (31)   & 224 (470) \\
  \colhline
 Total Signal NO (IO) & 1110 (528) & 300 (506) \\
  \colhline 
 Beam $\nu_{e}+\bar{\nu}_{e}$ \dword{cc} background & 190 & 117 \\
 \dword{nc} background & 81  & 38\\
 $\nu_{\tau}+\bar{\nu}_{\tau}$ \dword{cc} background & 32 & 20 \\
 $\nu_{\mu}+\bar{\nu}_{\mu}$ \dword{cc} background & 14 & 5 \\
  \colhline
 Total background & 317 & 180\\
\end{dunetable}

\begin{dunetable}
[\numu and \anumu disappearance rates]
{lr}
{tab:execsumm-disrates}
{\numu and \anumu disappearance rates: Integrated rate of selected $\nu_{\mu}$ \dword{cc}-like events between 0.5 and 8.0~GeV assuming a \num{3.5}-year (staged) exposure in the neutrino-beam mode and antineutrino-beam mode.  The rates are shown for normal mass ordering and $\mdeltacp = 0$.}
& Expected Events (3.5 years)\\ \toprowrule
 &  $\nu$ mode  \\
 \colhline 
 \numu Signal & 6200  \\
 \colhline 
  \anumu \dword{cc} background & 389 \\
 \dword{nc} background & 200 \\
 $\nu_{\tau}+\bar{\nu}_{\tau}$ \dword{cc} background & 46 \\
 $\nu_e+\bar{\nu}_e$ \dword{cc} background & 8 \\
 \toprowrule
 $\bar{\nu}$ mode  & \\
\colhline 
 \anumu Signal & 2303 \\
\colhline 
  \numu \dword{cc} background & 1129 \\
 \dword{nc} background & 101 \\
 $\nu_{\tau}+\bar{\nu}_{\tau}$ \dword{cc} background & 27 \\
 $\nu_e+\bar{\nu}_e$ \dword{cc} background & 2 \\
\end{dunetable}

\begin{dunefigure}[Significance of the DUNE determination of 
CP-violation]{fig:cpv_staging_execsum}
{Significance of the DUNE determination of CP-violation (i.e.: \deltacp 
$\neq 0$ or $\pi$) for the case when \deltacp=$-\pi/2$, and for 50\% and 
75\% of possible true \deltacp values, as a function of time in calendar 
years. True normal ordering is assumed. The width of the band shows the 
impact of applying an external constraint on \sinstt{13}.}
\includegraphics[width=0.95\linewidth]{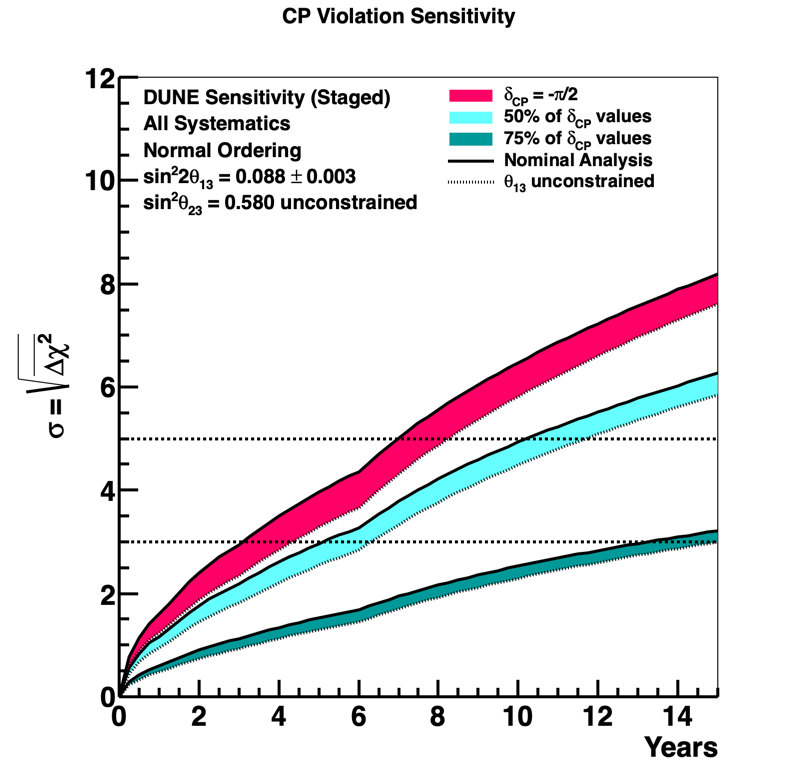}
\end{dunefigure}

Fig.~\ref{fig:cpv_staging_execsum} (reproduced later as 
Fig.~\ref{fig:cpv_staging}) illustrates DUNE's ability to distinguish 
the value of the \dword{cp} phase \deltacp from \dword{cp}-conserving 
values (0 or $\pi$) as a function of time in calendar year.  
These projections incorporate a sophisticated treatment of systematic 
error, as described in detail in Chapter~\ref{ch:osc}.  Strong evidence ($>3\sigma$) for \dword{cpv} is obtained for favorable values (half of the phase space) of \deltacp after five 
years of running, leading to a $>5\sigma$ determination after ten years.

A summary of representative sensitivity milestones for neutrino 
mass ordering and \dword{cpv} discovery, as well as precision on 
\deltacp and \sinstt{13} is given in 
Table~\ref{tab:milestones_execsumm}.  The ultimate level of 
precision that can be obtained on oscillation parameters 
highlights the point that DUNE will provide crucial input for  
flavor physics:  Patterns required by particular symmetries 
underlying fermion masses and mixing angles may appear.  The 
unitarity of the neutrino mixing matrix can be tested directly 
through comparisons of \sinstt{13} with the value obtained from 
reactor experiments.  
In conjunction with \sinstt{13} and 
other parameters, the precise value of \deltacp can  
constrain models of leptogenesis that are leading 
candidates for explanation of the baryon asymmetry of the Universe.

\begin{dunetable}[Projected DUNE oscillation physics milestones]
{lcc}
{tab:milestones_execsumm}
{Exposure in years, assuming true normal ordering and equal 
running in neutrino and antineutrino mode, required to reach 
selected physics milestones in the nominal analysis, using the 
NuFIT 4.0~\cite{Esteban:2018azc,nufitweb} best-fit values for the oscillation parameters. As 
discussed in Section~\ref{sec:physics-lbnosc-oscvar}, there are 
significant variations in sensitivity with the value of
\sinst{23}, so the exact values quoted here 
(using \sinst{23} = 0.580) are strongly dependent on that choice. 
The staging scenario described in 
Section~\ref{sec:exec-assm-meth-deployment} is assumed. Exposures 
are rounded to the nearest year.}
 Physics Milestone & Exposure (staged years) \\
 5$\sigma$ Mass Ordering & 1 \\
 \phantom{xxx}(\deltacp = -$\pi/2$) & \\ \colhline
 5$\sigma$ Mass Ordering & 2 \\
 \phantom{xxx}(100\% of \deltacp values) & \\ \colhline
 3$\sigma$ CP Violation & 3 \\
 \phantom{xxx}(\deltacp = -$\pi/2$) & \\ \colhline
 3$\sigma$ CP Violation & 5 \\
 \phantom{xxx}(50\% of \deltacp values) & \\ \colhline
 5$\sigma$ CP Violation & 7 \\
 \phantom{xxx}(\deltacp = $-\pi/2$) & \\ \colhline
 5$\sigma$ CP Violation & 10 \\
 \phantom{xxx}(50\% of \deltacp values) & \\ \colhline
 3$\sigma$ CP Violation & 13 \\
 \phantom{xxx}(75\% of \deltacp values) & \\ \colhline
 \deltacp Resolution of 10 degrees & 8 \\
 \phantom{xxx}(\deltacp = 0) & \\ \colhline
 \deltacp Resolution of 20 degrees & 12 \\
 \phantom{xxx}(\deltacp = -$\pi/2$) & \\ \colhline
 \sinstt{13} Resolution of 0.004 & 15 \\ 
\end{dunetable}

\subsubsection{DUNE can discover proton decay and other 
baryon-number violating processes}

By virtue of its deep underground location and large fiducial 
mass, as well as its excellent event imaging, particle 
identification and 
calorimetric capabilities, the DUNE far detector will be 
a powerful instrument for discovery of baryon-number violation.
As described in Chapter~\ref{ch:nonaccel}, DUNE will be able 
to observe signatures of decays of protons and neutrons, 
as well as the phenomenon of neutron-antineutron mixing, at 
rates below the limits placed by the current generation of 
experiments.

Many nucleon decay modes are accessible to DUNE.  
As a benchmark, a particularly compelling discovery channel 
is the decay of a proton to a positive kaon and a neutrino, 
\ptoknubar.  In this channel, the kaon and its decay products 
can be imaged, identified, and tested for kinematic consistency 
with the full decay chain, together with precision sufficient to 
reject backgrounds due to atmospheric muon and neutrino 
interactions. 
Preliminary analysis of single-particle beam and cosmic ray tracks 
in the \dword{pdsp} \lartpc is already demonstrating the particle 
identification capability of DUNE, as illustrated in 
Fig.~\ref{fig:pdsp_dedx_execsum}.  
The signature of the kaon track and its observable decay particles is 
sufficiently rich that a credible claim of evidence for 
proton decay could be made on the basis of just 
one or two sufficiently well-imaged events, for the case 
where background sources are expected to contribute much less 
than one event (see Chapter~\ref{ch:nonaccel} for a more complete 
discussion). 

\begin{dunefigure}[Reconstructed $dE/dx$ of protons and muons in 
\dshort{pdsp}]{fig:pdsp_dedx_execsum}
{Energy loss of protons (left) and muons (right) in 1-GeV  
running with the \dword{pdsp} \lartpc at CERN, as a function of 
residual range.  The protons are beam particles identified from 
beamline instrumentation; the muons are reconstructed stopping 
cosmic rays collected concurrently.  
The red curves represent the mean of the 
corresponding expected signature.  Note the difference in 
the vertical scale of the two plots.  The kaon $dE/dx$ curve 
will lie between the two curves shown.}
\includegraphics[width=0.45\linewidth]{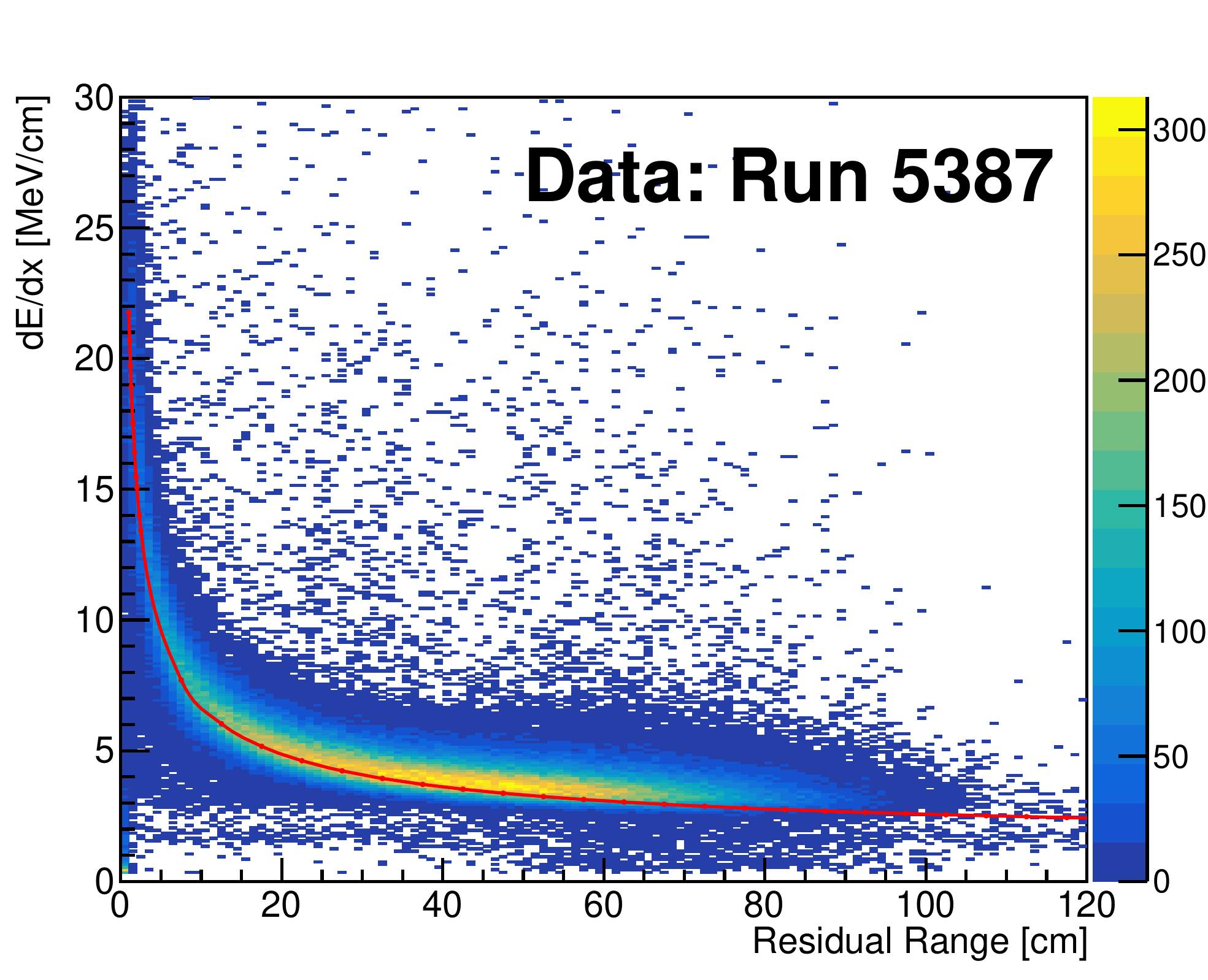}\hspace{0.05\linewidth}
\includegraphics[width=0.45\linewidth]{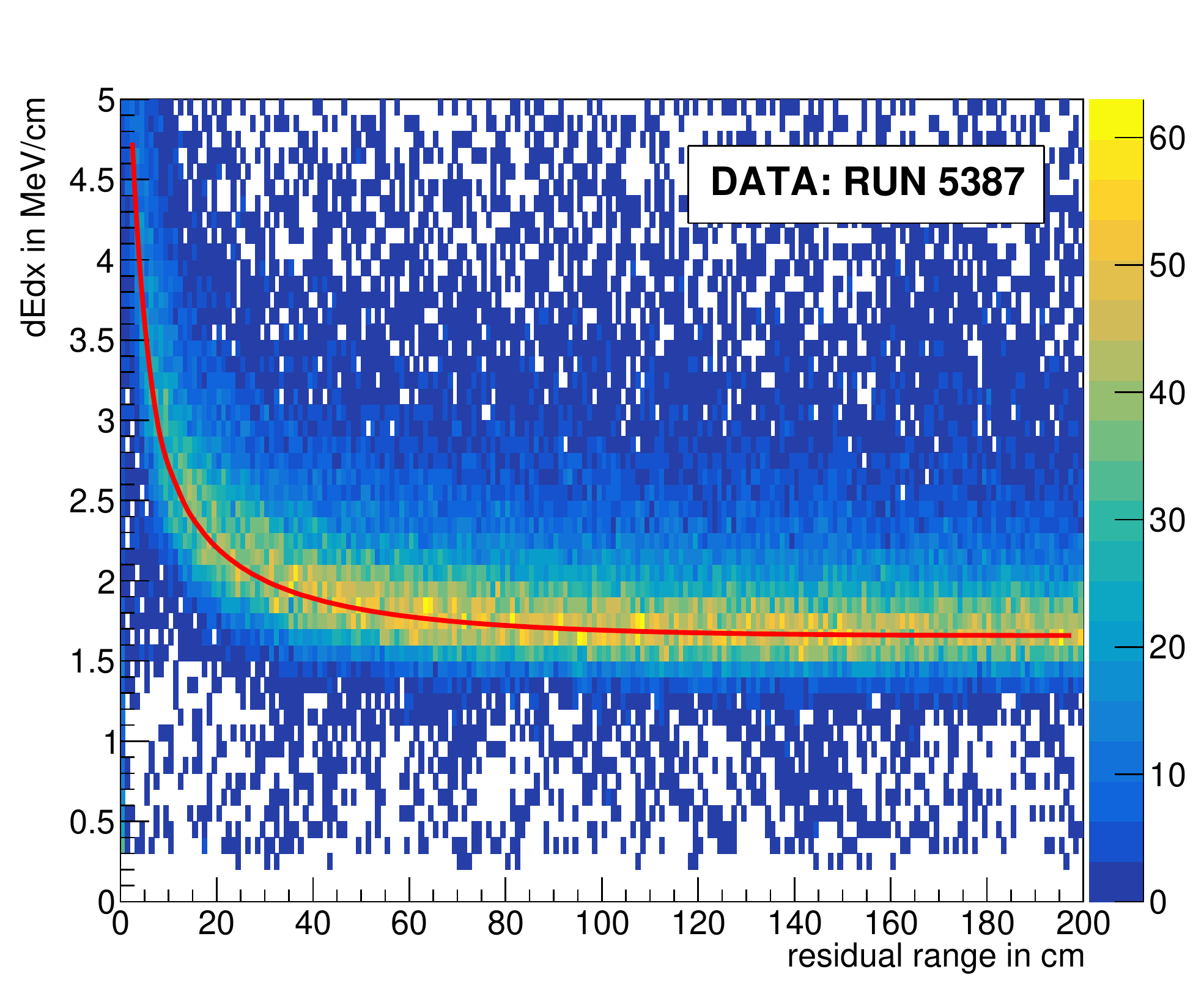}
\end{dunefigure}

Projecting from the current analysis of \ptoknubar in the DUNE 
far detector, with a detection efficiency of \num{30}\% as described in Chapter~\ref{ch:nonaccel}, the 
expected 90\% CL lower limit on lifetime divided by branching 
fraction is \SI{1.3e34}{years} for a 
\num{400}-\SI{}{\ktyr} 
exposure, assuming no candidate events are observed.  This 
is roughly twice the current limit of 
\SI{5.9e33}{years} from \superk~\cite{Abe:2014mwa}, 
based on an exposure of \SI{260}{\ktyr}.  Thus, should the rate 
for this decay be at the current \superk limit, five candidate 
events would be expected in DUNE within ten years 
of running with four far detector modules.  Ongoing work is aimed 
at improving the efficiency in this and other channels.

\subsubsection{DUNE can probe galactic supernovae via measurements of neutrino bursts}

As has been demonstrated with SN1987a, the observation 
of neutrinos~\cite{Bionta:1987qt,Hirata:1987hu} from a 
core-collapse supernova can reveal much about these  
phenomena that is not accessible in its  
electromagnetic signature.  Correspondingly, there is a 
wide range of predictions from supernova models for even 
very basic characteristics of the neutrino bursts.  Typical  
models predict that a supernova explosion in the 
center of the Milky Way will result in several thousand 
detectable neutrino interactions in the DUNE far detector 
occurring over an interval of up to a few tens of seconds.
The neutrino energy spectrum peaks around \SI{10}{MeV}, 
with appreciable flux up to about \SI{30}{MeV}.

\lar based detectors are sensitive to the \nue 
component of the flux, while water Cherenkov and organic 
scintillator detectors are most sensitive to the \anue 
component.  Thus DUNE is uniquely well-positioned to study the 
neutronization burst, in which \nue's are produced during the 
first few tens of milliseconds.  More generally,  
measurements of the (flavor-dependent) neutrino flux and energy 
spectrum as a function of time over the entirety of the burst 
can be sensitive to astrophysical properties of the supernova 
and its progenitor, and distortions relative to nominal 
expectations can serve as signatures for phenomena such 
as shock wave and turbulence effects, or even black hole 
formation.  

The sensitivity of the DUNE far detector to these 
phenomena is discussed in Chapter~\ref{ch:snb-lowe}. An illustration 
of one element of the program is given in Fig.~\ref{fig:fullSN_execsum}, 
which indicates a pointing resolution of better than $5^\circ$ that 
can be obtained by analysis of both subdominant highly-directional $\nu$-$e$ elastic scattering 
events and dominant weakly-directional $\nu_e$CC events within a supernova burst, based 
on full reconstruction and analysis. The DUNE results can be 
combined with corresponding measurements in other neutrino detectors to 
provide supernova localization from neutrinos alone in real time.
\begin{dunefigure}[Supernova direction determination from $\nu-e$ elastic scattering events]{fig:fullSN_execsum}{Left: Log
    likelihood values as a function of direction for a
    supernova sample with 260 $\nu$-$e$ elastic scattering (ES) events.  Right: Distribution of angular differences for
    directions to 10-kpc supernova using a maximum likelihood
    method.}
  \includegraphics[width=0.44\textwidth]{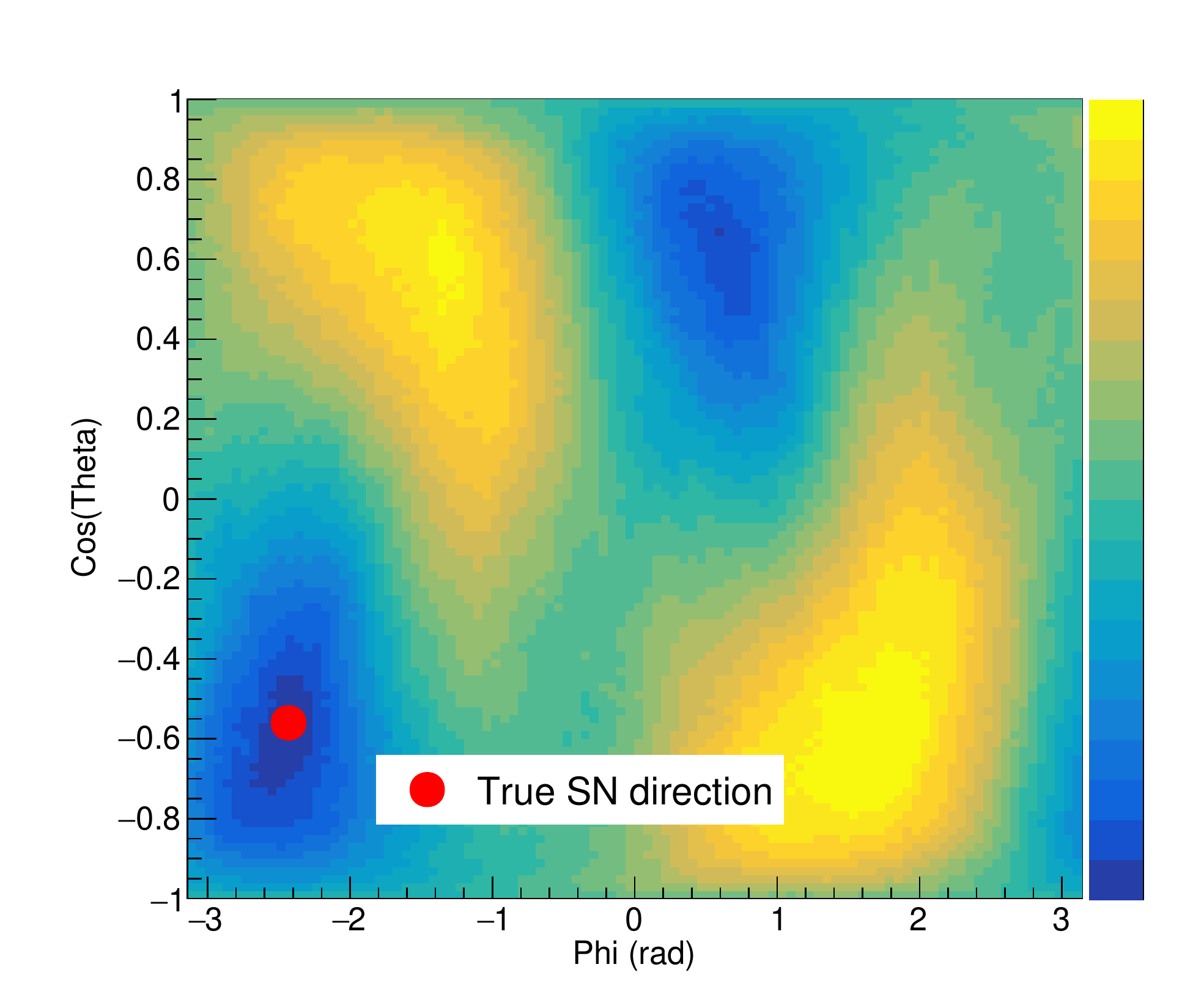}
  \includegraphics[width=0.48\textwidth]{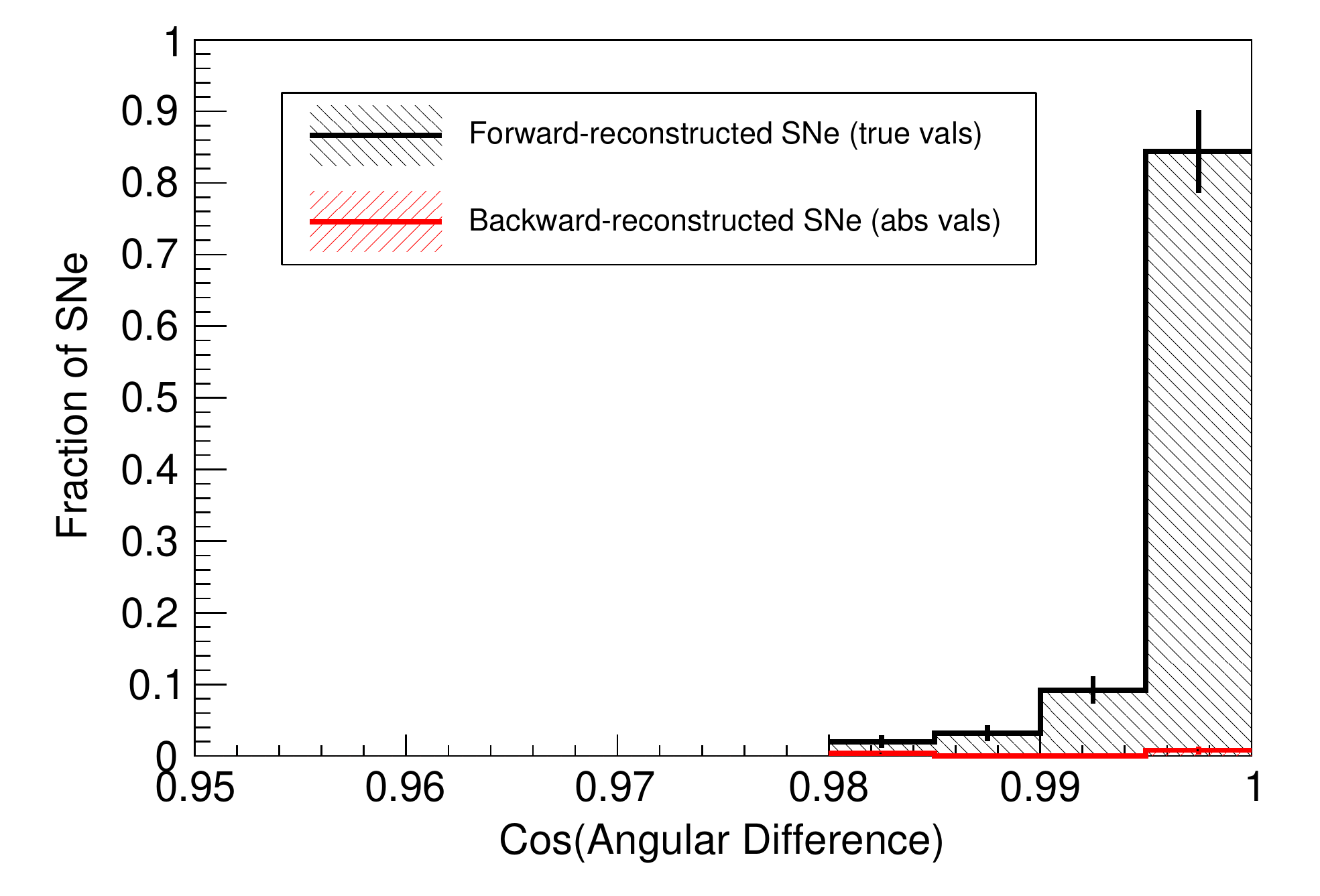}
\end{dunefigure}

\section{Science Drivers for LBNF/DUNE Design Specifications}
\label{sec:exec-key-reqs}

The scientific case summarized in the sections above
is predicated on the suitability of the experimental
configuration of LBNF/DUNE.  In this section we summarize
the ways in which the primary physics goals drive key features
of this configuration.  Further elaboration of the impacts of
physics goals on specific high-level performance needs of the
experimental systems is presented in the corresponding chapters
that follow.  Translation of performance needs into the
corresponding LBNF/DUNE design specifications is addressed within
the appropriate TDR volume.

\subsection{General LBNF/DUNE Operating Principles}

Worldwide scientific and technical planning for the
ambitious next-generation deep underground long-baseline
neutrino oscillation experiment that LBNF/DUNE now represents
has been under way for more than a decade.  Much
development preceded the formation of the DUNE science
collaboration in 2015 (see, for example,
Refs.~\cite{Adams:2013qkq} and \cite{LAGUNA-LBNO-deliv}).

Extensive study and discussion within the community
have led to the principal elements of the LBNF/DUNE
configuration:
\begin{itemize}
  \item {\bf\boldmath High-intensity conventional wide-band
  $\nu_\mu$ beam}\\
	The current generation of long-baseline neutrino experiments
	have benefited from narrow-band beam characteristics 
	associated with off-axis detector deployment. The principal 
	advantage is a low background rate in both \nue appearance 
	and \numu disappearance channels from misidentified neutral 
	current interactions of high energy neutrinos.  
	However, this advantage comes at a cost of flux and 
	spectral information relative to an on-axis detector 
    configuration~\cite{Adams:2013qkq,Agarwalla:2014tca}.
    The DUNE concept
    builds on the notion that a highly-performant detector
    technology with excellent neutrino energy reconstruction  and background rejection capabilities can
    optimize sensitivity and cost with an on-axis exposure to
    an intense conventional (magnetic horn-focused) beam.

  \item {\bf Far detector site selection for long baseline}\\
    The \SI{1300}{\km} baseline offered by locating the DUNE far detector
    at the Sanford Underground Research Facility in Lead, 
    South Dakota is well-optimized for the neutrino oscillation 
    physics goals of the program~\cite{Bass:2013vcg}.

  \item {\bf Deep underground location for far detector modules}\\
    Early studies (see, {\sl e.g.}, Ref.~\cite{homestake:depth}) 
    demonstrated that to realize the non-accelerator based elements
    of the DUNE science program, a deep underground far detector
    location is required.  These studies also indicate that the 
    4850 Level of Sanford Lab provides sufficient attenuation of 
    cosmic rays in the rock above, conclusions that have been 
    supported by more recent studies (see,
    {\sl e.g.}, Refs.~\cite{bib:docdb3384,bib:docdb1752}).

  \item {\bf \lartpc technology for far detector modules}\\
    Combining intrinsic scalability with high-performance event 
    imaging, calorimetry and particle identification capabilities, 
    the concept of large \dword{lartpc} detectors was developed 
    for the broad-based underground science program of DUNE.  This design choice integrates well with the
    other basic design elements described above.
    For example, the
    excellent neutrino energy reconstruction capability
    of \dword{lartpc}'s 
    is especially important for the long-baseline program with a 
    wide-band neutrino beam.
    Additionally, the \dword{lartpc} technology choice provides 
    valuable complementarity to other 
    existing and planned detectors pursuing many
    of the same goals.  As an example of this complementarity,
    the sensitivity of DUNE to the \nue component of supernova 
    neutrino flux, prevalent in the neutronization phase of the 
    explosion, provides distinct information relative to that 
    provided by water or organic scintillator-based detectors in 
    which \anue interactions dominate.  
\end{itemize}

The scientific basis for the above foundational experimental
design choices has been examined and validated through extensive
review, undertaken at all stages of DUNE concept development.
Recent experimental and theoretical developments have only
strengthened the scientific case for DUNE and its
basic configuration.  The technical underpinnings for
these choices have also been strengthened over time through a worldwide
program of R\&D and engineering development, as described in a suite
of LBNF/DUNE project documents including this \dword{tdr}, as
well as in sources describing independent experiments and development
activities.

\subsection{Far Detector Performance Requirements}
\label{sec:exec-summ-fd-requirements}


The number of detector design parameters that have direct
or indirect impact on performance is large.  These design
parameters have been studied over the years by past LArTPC
experiments, by DUNE during early detector optimization work,
through the successful construction and now operation of
ProtoDUNE-SP, and through continuing studies within the
DUNE consortia and physics groups.

\subsubsection{High-level Observables in the Far Detector}

DUNE's suite of physics measurements relies on a relatively small
number of event observables, through which the physics of interest
can be accessed.  Foremost are:
\begin{itemize}
\item {\bf Particle energies}  \\
Examples include the total visible energy in a supernova
neutrino interaction; the reconstructed energy of a beam
neutrino for oscillation measurements; the reconstructed
energy of a muon track in a nucleon decay candidate event.

\item {\bf Particle identification}  \\
This comes from spatial patterns and energy depositions.
Examples include photon/electron separation in the \nue{}
appearance analysis; proton/kaon separation in certain
nucleon decay channels.

\item {\bf Event time} \\
This allows for fiducialization
in the TPC, drift corrections, and macroscopic timing for
beam neutrinos and \dword{snb} physics.
\end{itemize}

\subsubsection{Physics Case Studies}

\paragraph{\bf CP violation search}
The primary demonstration that the detector design meets
the physics needs is the full simulation and analysis
being documented in this TDR volume.  As described in 
Sec.~\ref{sec:exec-sensitiv-results}, 
Figure~\ref{fig:cpv_staging_execsum}
presents the time-evolution of the \dword{cpv} sensitivity 
obtained in this way.

To break the measurement apart into the three
observables above, we start with neutrino energy.
The charged lepton and hadronic shower energies are
reconstructed separately and then summed.  In electron
neutrino events, the leptonic energy resolution
(spectrum averaged) is 8\%, the hadronic energy resolution
is 49\%, and the neutrino energy resolution is 13\%.
Note that a significant portion of the energy smearing
comes from the physics of neutrino-nucleus scattering
and hadronic shower production rather than from detector
performance.  In the impossible case that the lepton
energy could be perfectly reconstructed, the electron
neutrino (muon neutrino) energy resolution would only
change by approximately $13\%\rightarrow 10\%$ 
($18\%\rightarrow 17\%$).
Equivalently, small degradations in detector response
have minimal leverage to affect the final neutrino energy
resolution.

Particle identification is critical for the oscillation analysis 
in that it enables neutrino flavor identification.  
For \nue{} appearance in particular, one
must positively identify the presence of a high-energy
electron while avoiding misclassification of high energy
photons as electrons. The LArTPC design meets this challenge
by having spatial resolution that is much smaller than the
radiation length ($0.5~\rm{cm} \ll 14~\rm{cm}$) to make
visible the gaps between an event's reconstructed vertex
and any photon conversions, and charge resolution that
provides additional $dE/dx$ separation based on pre-EM-shower
depositions, as demonstrated in an operating detector by
ArgoNeuT~\cite{Acciarri:2016sli} and with DUNE simulation
in Figure~\ref{fig:emdedx}.  The DUNE study~\cite{bib:docdb3384}
also shows alternative detector designs for the 
single-phase \lartpc implementation.  As long as
the signal-to-noise ratio is high on the readout wires,
minor adjustments to the wire angle and pitch have negligible
impact on $dE/dx$ separation power.
\begin{dunefigure}
[Separation of photons and electrons by $dE/dx$
in the pre-shower region]
{fig:emdedx}
{Separation of photons and electrons by $dE/dx$
in the pre-shower region.  Alternative wire angles and wire
pitches are also shown.}
  \includegraphics[width=0.7\linewidth]{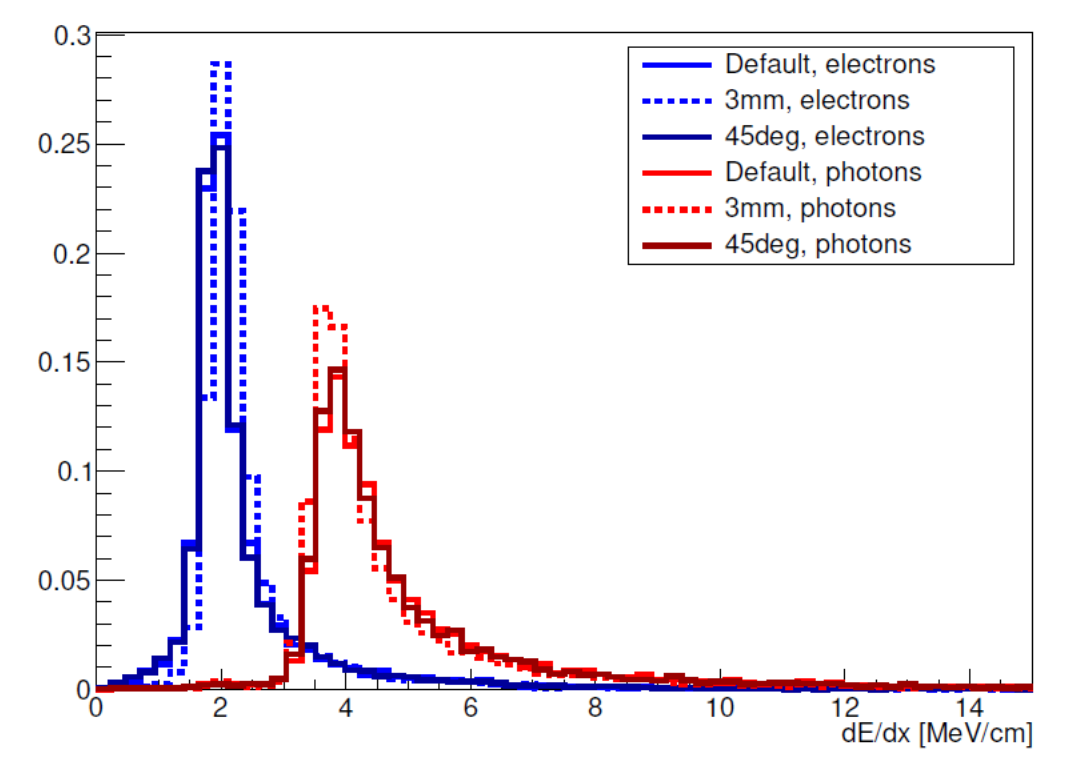}
\end{dunefigure}

In the analysis presented in Chapter~\ref{ch:osc}, and in
the preliminary \dword{cpv} sensitivity in
Figure~\ref{fig:cpv_staging_execsum}, neutrino flavor classification
is accomplished using a modern convolutional neural network
technique that takes directly as input the TPC wire hits
in the three detector views.

Event timing requirements for beam events flow from the
need to establish the fiducial volume.  This is discussed
generally in the section on light yield below.

\noindent {\bf Supernova burst neutrinos.}  A core-collapse
supernova at 10~kiloparsecs will provide $\sim$1000 neutrino
interactions in the \dword{fd} over the course of
$\sim$10 seconds with typical energies between 5 and 30~MeV.
Charged current \nue{} events make up the majority of these.
Much of the desired astrophysical information comes via the
time-dependent energy spectrum of these neutrinos.  As shown
earlier in Sec.~\ref{sec:exec-sensitiv-results} and later
in more detail in Chapter~\ref{ch:snb-lowe},
DUNE capabilities are quantified through sensitivities
both to generic pinched-thermal spectral parameters 
and to
specific phenomena within the star.

Figure~\ref{fig:specpars} shows the precision with which DUNE
can measure two of the spectral parameters, $\varepsilon$, related to
the binding energy of the neutron star remnant, and $\langle
E_{\nu_e}\rangle$, the average energy of the $\nu_e$ component, for the
time-integrated spectrum.
The assumed measured spectrum takes into account some degradation
from the neutrino interaction process itself 
({\em e.g.}, energy lost to neutrons), via the \dword{marley} event generator.
The colored contours show increasing levels of energy smearing.
A 10\% resolution is noticeable but insignificant, and the
overall precision on the spectral parameters up to 30\%
resolution does not change dramatically
As shown later,
the additional smearing introduced by the detector's response
falls within the resolution envelope suggested here, and according to
detector simulation is closest to the 20\% level.
In an eventual detailed analysis, the spectral
fits will be done in time slices to study the evolution of
the supernova, so the minimum contour size in each time slice
will be larger due to reduced event counts in each slice.

\begin{dunefigure}
[90\%~C.L.\ contours for spectral parameters for a supernova
at 5~kpc.]
{fig:specpars}
{90\%~C.L.\ contours for the luminosity and average $\nu_e$ energy
spectral parameters for a supernova
at 5~kpc.  The contours are obtained using the time-integrated
spectrum.  As discussed in the text, the allowed regions
change noticeably but not drastically as one moves from
no detector smearing (pink) to various realistic resolutions (wider regions).}
  \includegraphics[width=0.7\linewidth]{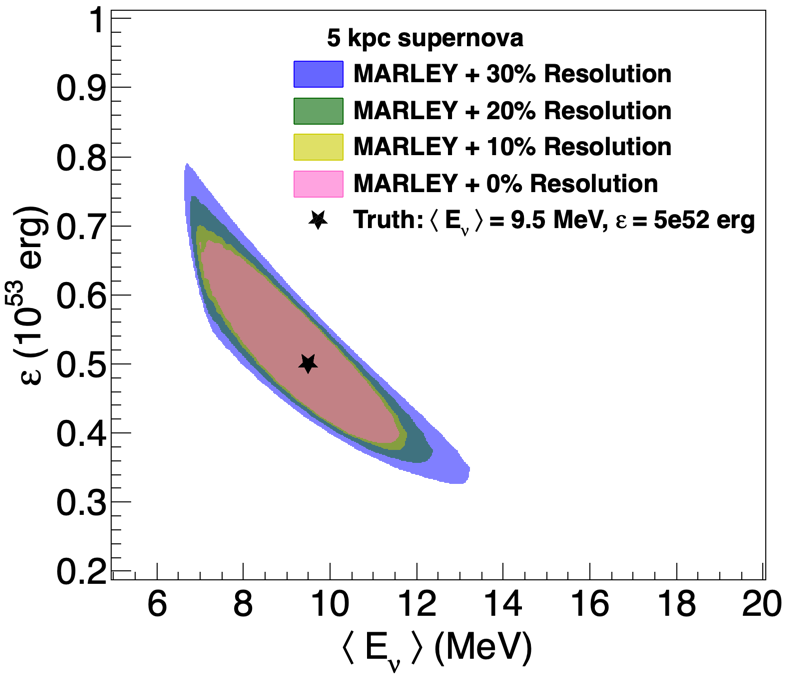}
\end{dunefigure}

Given the dominance of \nue{} charged current events in the
supernova neutrino sample, particle identification is not a
requirement for the primary physics measurements.  However,
additional capability may be possible by identifying separately
neutral current and elastic scattering interactions.  Studies
are on-going, and these possibilities are discussed in 
Chapter~\ref{ch:snb-lowe}.

Timing for \dword{snb} events is provided by both the
TPC and the photon detector system.  Basic timing requirements
flow from event vertexing and fiducialization needs.  These are
discussed generally for DUNE in the light yield section, but
here we note a few supernova-specific design considerations.
During the first 50~ms of a 10-kpc-distant supernova, the
mean interval between successive neutrino interactions is
$0.5 - 1.7~\rm{ms}$ depending on the model.  The TPC alone
provides a time resolution of 0.6~ms (at 500~V/cm), commensurate
with the fundamental statistical limitations at this distance.
However nearly half of galactic supernova candidates lie closer
to Earth than this, so the rate can be tens or (less likely)
hundreds of times higher.  A resolution of $\mathord{<}10~\mu\rm{s}$,
as already provided by the photon detector system, ensures that
DUNE's measurement of the neutrino burst time profile is always
limited by rate and not detector resolution.  The hypothesized
oscillations of the neutrino flux due to standing accretion shock
instabilities would lead to features with a characteristic time
of $\sim$10~ms, comfortably greater than the time resolution.
The possible neutrino trapping notch at the start of the burst
has a width of $1 - 2~\rm{ms}$.  Observing the trapping notch
could be possible for the closest progenitors.

\subsubsection{Key High-level Detector Design Specifications}

With the discussion above and in later chapters of this 
document, it is possible to identify several high-level 
detector design parameters that together characterize the  
overall function of DUNE single-phase \lartpc modules.  These 
parameters and specified operating points are given in 
Table~\ref{tab:exec-fd-specs}.
\begin{dunetable}[High-level DUNE single-phase far detector design specifications]
{lll}
{tab:exec-fd-specs}
{High-level DUNE single-phase far detector design parameters 
and specifications}
Parameter & Specification & Goal\\ \toprowrule
Drift field       & $>\SI{250}{\volt/\cm}$     & 500~V/cm\\
Electron lifetime & $>\SI{3}{\milli\second}$   & 10~ms\\
System noise      & $<1000$~enc & --- \\
Light yield (at cathode)  & $>\SI{0.5}{pe/\MeV}$ & $>\SI{5}{pe/\MeV}$\\
Time resolution   & $<\SI{1}{\micro\second}$    & 100~ns\\
\end{dunetable}

The column headings in the table are defined as follows:
\paragraph{Specification:} This is the intended value for the parameter or, more often, the 
upper or lower limit for the parameter.  Fixed values are given for parameters that are not intrinsically dynamic ({\em e.g.}, wire pitch).  Limits are set by the more stringent driver, either
 the physics or engineering needs.

\paragraph{Goal:} This is an improved value that offers some benefit, and the collaboration
aims to achieve this value where it is cost effective to do so.  While in some cases the goal offers potential physics benefit directly, more often the goal provides risk mitigation, since improving the performance on one parameter can mean relaxing the requirements on other correlated parameters, thus protecting against unforeseen performance issues.

The first three parameters (drift field, electron lifetime, 
and TPC system noise) in Table~\ref{tab:exec-fd-specs} 
enter directly into the ability to discriminate between 
ionization signals due to physics events and noise.  Physics 
capability degrades if readout noise is not small compared to 
the ionization signal expected for minimum-ionizing particles
located anywhere within the active volume of the detector.
The remaining parameters (light yield for events at the cathode, 
and timing resolution) pertain to the ability 
of the scintillation photon detection system to enable 
localization of events within the TPC, needed for the 
non-accelerator based far detector physics program, both 
for fiducialization and for corrections to TPC charge 
attenuation.  The general 
arguments for the specifications listed for each parameter 
are given below.

\paragraph{Drift field}
The basic operating principle of the TPC involves the transport 
of ionization electrons out of the argon volume and to the 
detection plane.   
A higher drift field reduces electron transport time 
and thus electron loss due to impurities; 
reduces ion-electron recombination (increasing ionization signal at the expense of reduced scintillation photon yield); increases induction 
signals due to increased electron velocity; and reduces 
electron diffusion.

The argon volume in the \dword{fd} single phase design is 
divided into four separate drift regions, each with a maximum 
drift distance of 3.5~m.  The design goal of 500~V/cm field 
implies a voltage across the drift region of approximately 
180~kV.  At this field, the electron drift velocity is 
1.6~mm/$\mu$s , implying a maximum drift time $t=2.2~\rm{ms}$.  
This drift time can be compared with the electron lifetime 
$\tau$ set by the argon purity.  At $\tau=3~\rm{ms}$, signals
originating near the cathode will be attenuated to 
$e^{-t/\tau} = 48\%$ of their original strength.  
For the minimum field of 250~V/cm, this transmission becomes 
23\%.  Additionally, electron/ion recombination happens more 
readily at lower field.  From 500~V/cm to 250~V/cm, 
an additional signal loss of 11\% (taking 23\% to 20\%) is 
introduced due to recombination.  The lowered field also 
reduces the drift velocity and, in proportion, signal pick-up 
on the induction wires.  Moving from 500~V/cm to 250~V/cm drops 
the induction signal by an additional 34\% to an effective
transmission for low-field depositions near the cathode 
(relative to ``500~V/cm near the anode'') of 14\%.

These signal attenuations are acceptable as long as the readout 
maintains good \dword{s/n} and charge resolution.  High \dword{s/n} for 
\dword{mip} signals has been demonstrated at ProtoDUNE-SP -- \dword{s/n}=30 
(collection), 15 (induction) -- and the minimum transmissions 
above would not significantly damage the ability to identify 
wire hits.  The charge resolution on individual wires, while not 
a driver of overall event resolution, feeds into the $dE/dx$ 
estimation for short segments of tracks and thus into 
particle identification.  Studies of selection efficiencies at 
varied signal levels continue, but notably the \nue 
selection efficiency exhibits no dependence on drift 
distance in the default simulation, which is based on a 
3~ms electron lifetime.  As mentioned below, ProtoDUNE readily 
achieved higher lifetimes.

Electrons drifting across the full 3.5~m will experience 
transverse diffusion of 1.7~mm (2.0~mm) at 500~V/cm (250~V/cm).  
The change in diffusion with field strength is insignificant in comparison to the wire pitch of 5~mm.

The reduced recombination at higher field results in 
smaller scintillation photon yields.  At 500 V/cm, the 
yield is 60\% of that at zero field.  Thus any reduction 
in field strength will improve this detection channel.  
However, the incremental nature of this improvement and 
the more critical dependence of successful execution of 
the science program on the TPC performance together make
optimization with respect to scintillation a secondary consideration.

ProtoDUNE-SP is currently operating at 500~V/cm.

\paragraph{Electron lifetime}
Electronegative impurities ({\em e.g.}, $\rm{H}_{2}\rm{O}$, $\rm{O}_{2}$) within the liquid argon must be kept at 
low levels to prevent the capture of drifting electrons after ionization.  Electron lifetime is inversely proportional to the level of these impurities.

The values in Table~\ref{tab:exec-fd-specs} correspond to 
contamination levels of of 100 ppt $\rm{O}_2$-equivalent 
for 3~ms and 30~ppt for 10~ms.  The influence of electron 
lifetime on physics capabilities has been discussed in the 
section on drift field above.  
Indeed, one can largely trade off purity for field. 
Note that the lower lifetime of 3~ms was assumed 
versus the goal of 10~ms.  ProtoDUNE-SP has achieved electron lifetimes exceeding 5~ms.

\paragraph{Electronics system noise}
Noise in the electronics system can 
limit the ability to identify and correctly associate wire hits 
and can worsen charge resolution.  From engineering
considerations, the noise level in the front-end electronics 
drives the specification.  All other pieces of the electronics 
chain are to be kept well below this level.  
The specification is given in units of $e^-$ equivalent 
noise charge (enc).

At current gain settings, 1000~enc corresponds 
to 6.5 ADC counts.  Initial ProtoDUNE analyses are 
showing 3.5 (4.5) ADC counts on collection (induction) channels. 
The current \dword{fd} simulation assumes a noise level similar to 
ProtoDUNE performance, but higher noise levels are being 
explored.  It is not expected that these relatively small
adjustments (factor of $\sim$2) will impact physics analysis 
in any significant way.  Noise assumptions (level and 
correlations) do influence DAQ design choices.

\paragraph{Light yield and photon-based timing}
The photon detector system provides an event time 
based on the scintillation light produced in the liquid argon.  
In conjunction with the TPC ionization signal, this allows 
one to determine where the event occurred along 
the drift direction for event vertexing, fiducialization, 
and electron attenuation corrections.  The 
specifications here are given for the worst-case event 
location in the fiducial volume, typically near the cathode 
and thus far from any photon detector on the anode planes.

A photon-based time resolution of 1~$\mu$s corresponds to the 
time resolution for single TPC wire hits, allowing for useful 
event matching between the TPC and photon detector systems.  
Given the drift velocity, 1~$\mu$s also corresponds to an 
effective spatial granularity in the drift direction 
($\sim$2~mm) that is similar to the wire pitch.  The resulting 
three-dimensional event vertex provided by combining TPC and 
photon detector information has essential uses in DUNE physics 
analyses.  A fiducial volume must be defined at the 
$\mathord{<}1\%$ level for the accelerator-based neutrino 
oscillation measurements and for nearby supernovas, and at less 
stringent levels for other measurements.  Additionally, most 
cosmogenic and environmental backgrounds for non-accelerator 
measurements ({\em e.g.}, neutral particles produced by cosmic 
rays in the surrounding rock) have tell-tale distributions in the
active volume and can thus be mitigated or eliminated through 
event localization.

The precise event time, and thus event location, also allows a 
correction for electron attenuation, which otherwise could have 
a large effect on energy resolutions due to the non-uniformity
of response across the drift volume.  
The minimum TPC performance 
considered (with $E=250~\rm{V/cm}$, $\tau=3~\rm{ms}$) would 
correspond to an energy smearing of 22\% due to electron loss.  
This effect is made negligible at 1~$\mu$s time resolution.

This attenuation correction is only possible when photon signals can be successfully associated with a TPC-recorded event.  For low energy supernova neutrino events, this association is not 
100\% efficient.  Figure~\ref{fig:tpcres} shows the smearing on visible energy in the TPC for supernova neutrino events with and without a drift correction based on different photon detector system performance.  The difference between no correction and any correction is dramatic.  The small differences between different light levels (cast as effective photodetector area in the figure) stem not from improved spatial resolution but from a higher efficiency at reconstructing and associating light signals with the TPC signals.  An effective area of 23~$\rm{cm}^2$ corresponds roughly to a light yield of 0.5~p.e./MeV at the cathode, {\em i.e.}\ the minimum specification.
\begin{dunefigure}
[Dependence of reconstructed \dshort{snb} event energy on timing-based drift correction]
{fig:tpcres}
{Energy residuals for supernova neutrino events without (black) 
and with (color) a timing-based drift correction to the 
reconstructed energy.  The red histogram assumes the event 
vertex is known perfectly, and the realistic cases approach that
ideal quickly.  The $23~\rm{cm}^2$ histogram roughly corresponds 
to the specification of 0.5~p.e./MeV.}
  \includegraphics[width=0.7\linewidth]{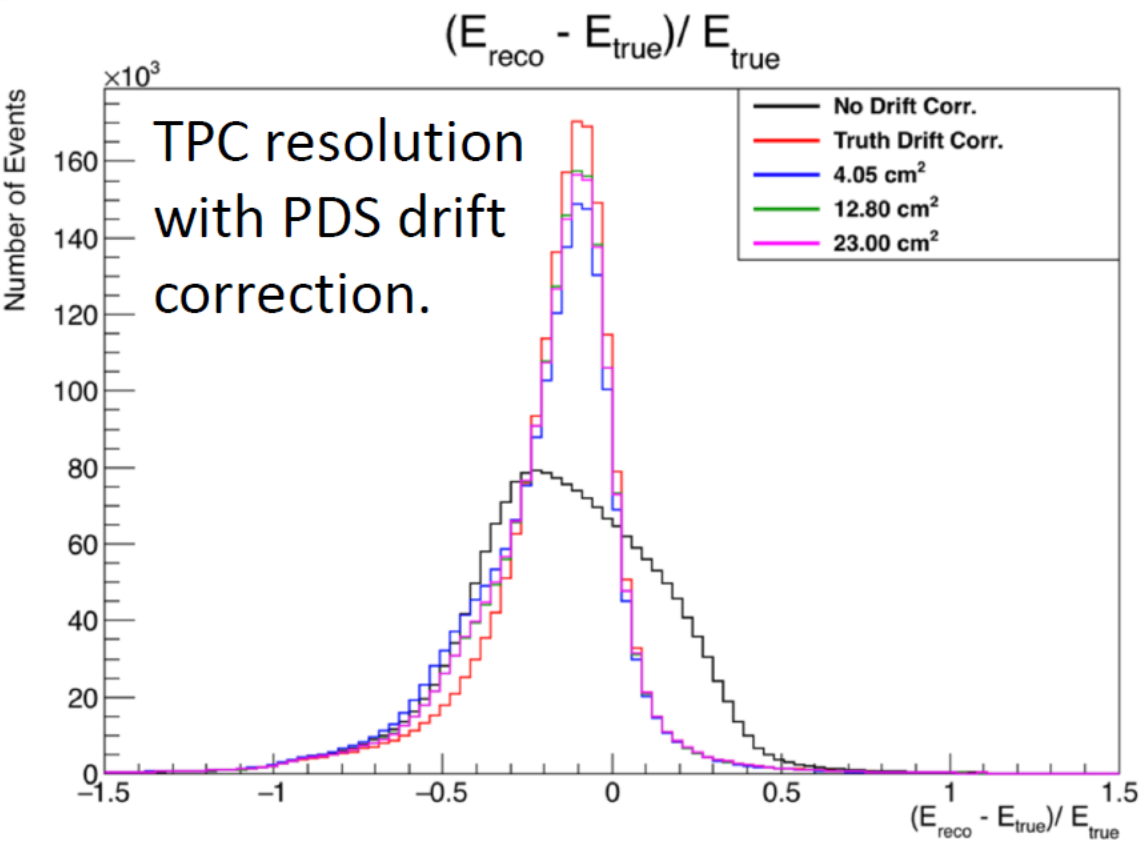}
\end{dunefigure}

The use of photon signals for direct event calorimetry in 
supernova neutrino events is under study. Initial tests suggest 
resolutions around 25\% are possible at 0.5 p.e./MeV, which is
competitive with the TPC resolution at these energies.

Two light-collection bar designs and one segmented design 
(ARAPUCA) are operating in ProtoDUNE-SP.  Initial performance 
evaluation is excellent, and a full quantitative assessment, 
described in Volume~\volnumbersp{} 
is in progress.

\subsubsection{Detector Design Driver Summary}

The above discussion provides the basic guidelines for key far 
detector performance specifications in the context of the 
single-phase module design.  Further elaboration is given in the 
chapters devoted to science capabilities in this document.  
Discussion of other significant detector specifications 
and their impact on 
physics sensitivity is given in Volumes~\volnumbersp{} and 
\volnumberdp{}.  While it is not practical to carry out 
comprehensive physics sensitivity studies comprehensively, in 
which every major detector parameter is varied individually or 
in conjunction with others, such studies have been done 
for a few significant parameters (such as anode wire pitch for 
the single-phase \lartpc design).  These are reported in the 
corresponding detector volume.






\section{Scope and Organization of this Document}
\label{sec:exec-scope}

The scope and organization of this document follow from
both programmatic and practical considerations.

First, while this volume is strongly interconnected with
the other TDR volumes, it is written so as to stand on its
own 
to be of best use to the
community outside DUNE.  To accomplish this, some duplication
of material presented in other volumes is unavoidable.
At the same time, the full utility of this volume
is as just one element within an integrated set of TDR volumes.
Thus, explicit and implicit reference
to material presented in other volumes is made freely.

Second, while two volumes describe the technical designs
for far detector modules based on the single-phase and
dual-phase liquid argon TPC technologies, the exact
configuration of all four modules is not yet established.  As of this writing,
it is understood that the first two modules will likely
be one of each technology.  Practical considerations,
including the current state of development of event
reconstruction and other software tools, have
led the DUNE science collaboration to undertake a rigorous
evaluation of capabilities for a DUNE program consisting
solely of single-phase far detector modules.  Based
on the considerable progress already made toward the
realization of effective reconstruction software for the
dual-phase far detector implementation, current
understanding is that its capabilities are at least
as well optimized for the key physics goals as those of
the single-phase implementation.

Thus it should be understood that the studies and results
reported in this document were undertaken with the
specification of single-phase detector modules.  Where possible,
comments on how performance and/or capabilities of
dual-phase modules might differ are provided.


\cleardoublepage

\chapter{Introduction to LBNF and DUNE}
\label{ch:physics-intro}


The \dword{dune} will be a world-class neutrino observatory and nucleon decay detector designed to answer fundamental questions about elementary particles and their role in the universe. The international \dword{dune} experiment, hosted by the U.S. Department of Energy's \dword{fnal}{}, will consist of a \dword{fd} located about \SI{1.5}{km} underground at the \dword{surf} in South Dakota, USA, \SI{1300}{\km} from \dword{fnal}{}, and a \dword{nd} located on site at \dword{fnal} in Illinois. The far detector will be a very large, modular \dword{lartpc} with a total mass of nearly \SI{70}{kt} of \dword{lar}, at least \fdfiducialmass (\SI{40}{\giga\gram}) of which is fiducial. The \dword{lar} technology 
has the unique capability to reconstruct neutrino interactions with image-like precision and unprecedented resolution. 

The \dword{dune} detectors will be exposed to the world's most intense neutrino beam originating at \dword{fnal}{}. A high-precision near detector, \SI{574}{m} from the neutrino source on the \dword{fnal} site, will be used to characterize the intensity and energy spectrum of this wide-band beam. The ability to compare the energy spectrum of the neutrino beam between the \dword{nd} and \dword{fd}
is crucial for discovering new phenomena in neutrino oscillations. The \dword{lbnf}, also hosted by \dword{fnal}, provides the infrastructure for this complex system of detectors at the Illinois and South Dakota sites. \dword{lbnf} is responsible for the neutrino beam, the deep-underground site, and the infrastructure for the \dword{dune} detectors. 

\section{The LBNF Facility}
\label{sec:physics-intro-lbnf}

The \dword{lbnf} project is building the facility that will house and provide infrastructure for the first two \dword{dune} \dword{fd} modules  in South Dakota  and the \dword{nd} in Illinois.  Figure~\ref{fig:lbnf} shows
a schematic of the facilities at the two sites, and Figure~\ref{fig:caverns} shows a diagram of the cavern layout for the \dword{fd}.  
The organization and management of \dword{lbnf} is separate from the \dword{dune} collaboration. \dword{lbnf} is also hosted by \dword{fnal} and its design and construction are organized as a \dword{doe}/\dword{fnal} project incorporating international partners. 

\begin{dunefigure}[ 	
LBNF/DUNE project: beam from Illinois to South Dakota]{fig:lbnf}{ 	
LBNF/DUNE project: beam from Illinois to South Dakota.}
\includegraphics[width=0.9\textwidth]{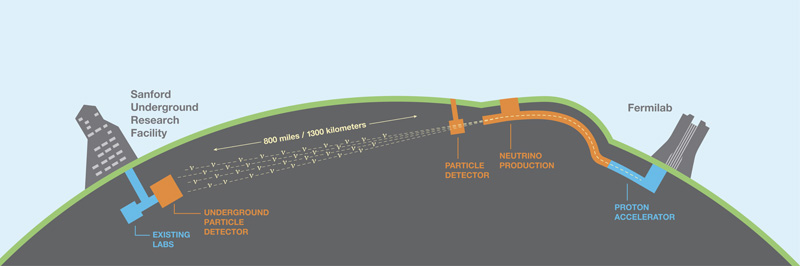}
\end{dunefigure}

\begin{dunefigure}[ 	
Underground caverns for DUNE in South Dakota]{fig:caverns}{Underground caverns for DUNE FD and cryogenics systems at \dword{surf}, in South Dakota. The drawing, which looks towards the northeast, shows the first two far detector modules in place.}
\includegraphics[width=0.99\textwidth]{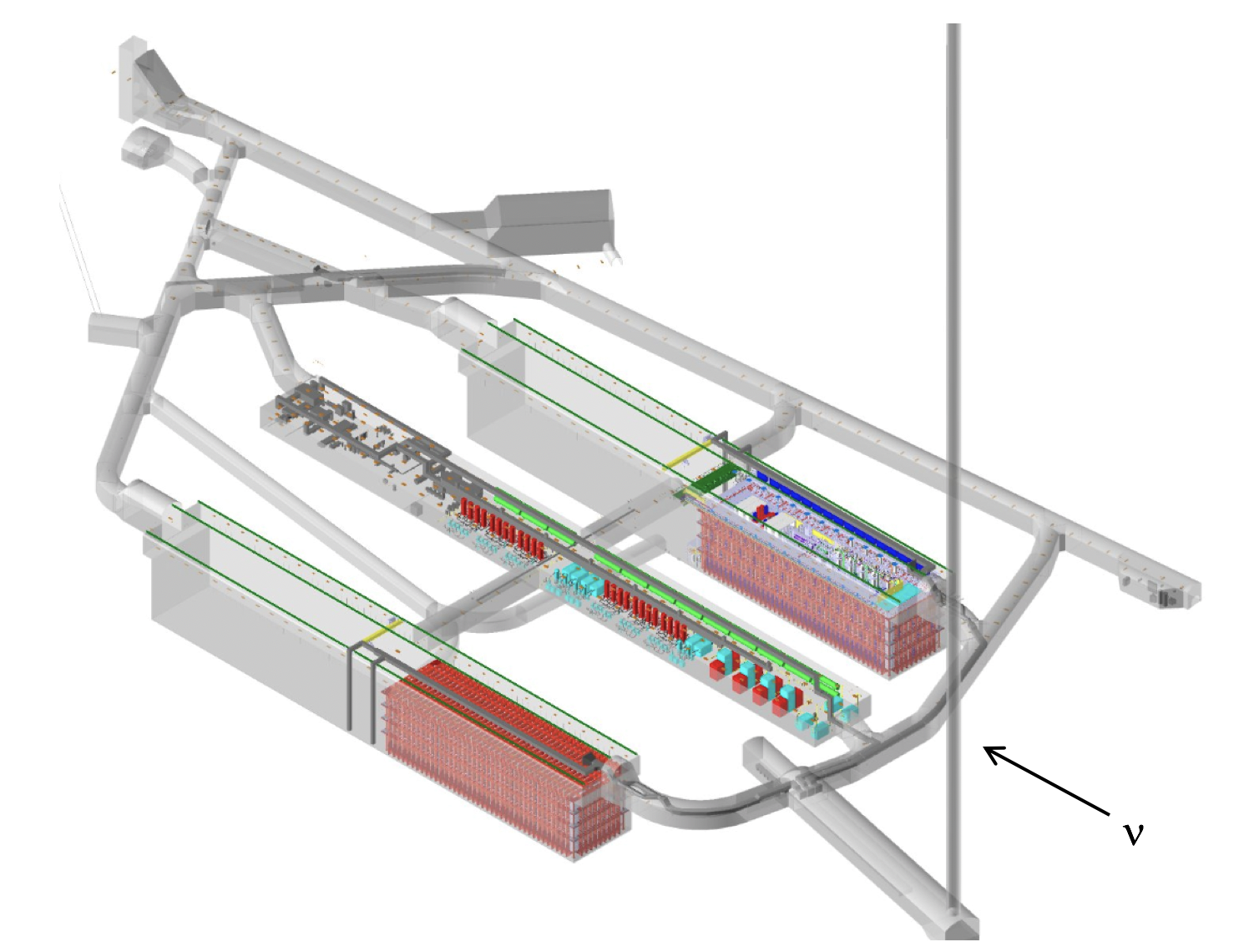}
\end{dunefigure}

The \dword{lbnf} project provides to \dword{dune}
\begin{itemize}
\item  the  technical and conventional facilities for a powerful neutrino beam utilizing the \dword{pip2} upgrade~\cite{pip2-2013} of the \dword{fnal} accelerator 
complex. The \dword{pip2} project will deliver between \SIrange{1.0}{1.2}{MW} of proton beam power from \dword{fnal}'s Main Injector in the energy range  \SIrange{60}{120}{GeV} at the start of \dword{dune} operations and provide a platform for extending beam power to \dword{dune} to 
$>\,$\SI{2}{MW}. 
A further planned upgrade 
of the accelerator complex will enable it to provide up to \SI{2.4}{\MW} of beam power by 2030. 

\item  the civil construction, or \dword{cf}, for the \dword{nd} systems at \dword{fnal}; (see Figure~\ref{fig:beamline});

\item the excavation of three underground caverns at \dword{surf} to house the \dword{dune} \dword{fd}. The north and south caverns will each house two cryostats with 
a minimum \nominalmodsize fiducial mass of liquid argon, while the \dword{cuc} will house cryogenics and data acquisition facilities for all four detector modules;

\item surface, shaft, and underground infrastructure to support 
the outfitting of the caverns with four free-standing, steel-supported cryostats 
and the required cryogenics systems to enable rapid deployment of the first two \nominalmodsize \dword{fd} modules. 
The intention is to install the third and fourth cryostats as rapidly as funding will 
allow.

\end{itemize}

\begin{dunefigure}[Neutrino beamline and DUNE near detector hall in Illinois
]{fig:beamline}{Neutrino beamline and DUNE near detector hall at Fermilab in Illinois}
\includegraphics[width=0.95\textwidth]{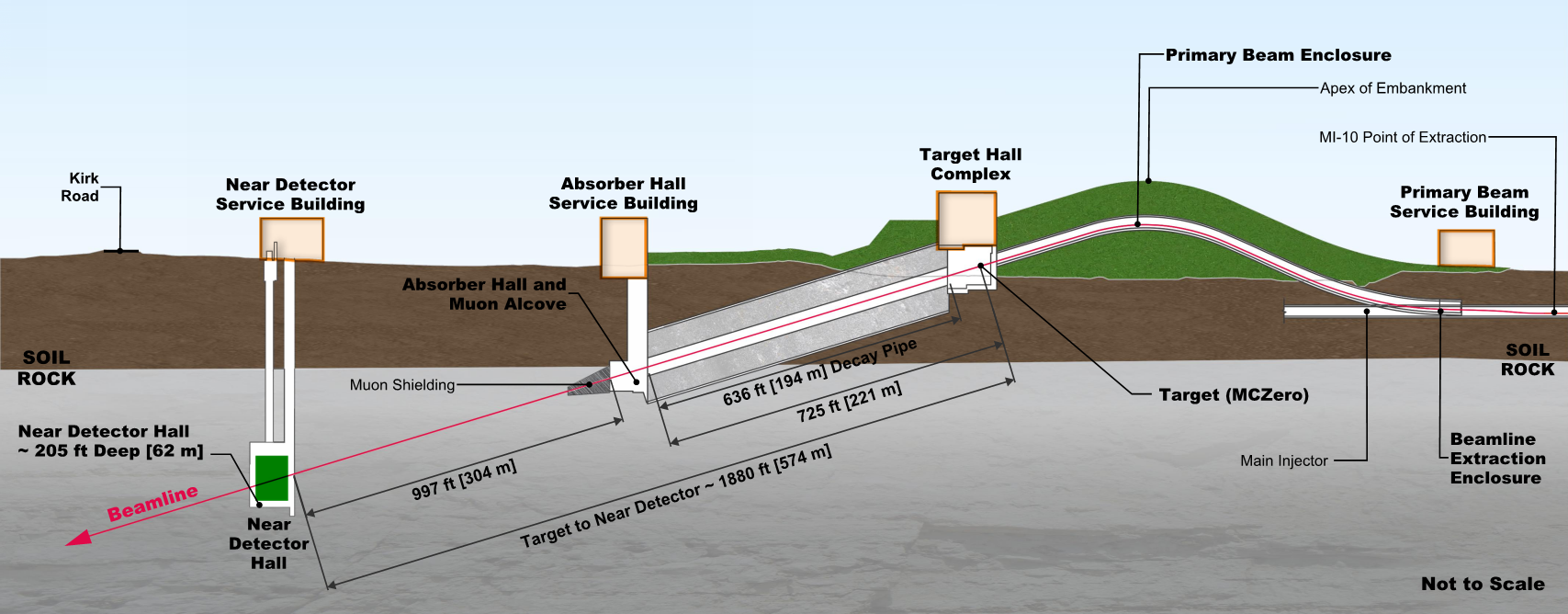}
\end{dunefigure}

\section{\dshort{dune}: Far Detector Modules}
\label{sec:physics-intro-dunefd}

The 
\dword{dune} \dword{fd} consists of four \dword{lartpc} \dwords{detmodule}, each contained in a cryostat that holds \larmass of \dword{lar}. Each module, installed approximately \SI{1.5}{km} underground, has a fiducial mass of at least \nominalmodsize. The \dword{lartpc} technology provides
excellent tracking and calorimetry performance, making it an ideal
choice for the \dword{dune} \dword{fd}. Each of the \dwords{lartpc} fits inside a cryostat of internal dimensions
\cryostatwdth (W) $\times$ \cryostatht (H) $\times$ \cryostatlen~(L) that contains a total \dword{lar} mass of about \larmass{}.
 The four identically sized modules provide flexibility for staging construction and for evolution of \dword{lartpc} technology.

\dword{dune} is planning for and is prototyping two \dword{lartpc} technologies:
\begin{itemize}
\item \Dword{sp}: In the \dword{sp} technology, ionization charges are drifted horizontally in \dword{lar} and read out on wires in the liquid.  The maximum drift length in the first \dword{dune} \dword{spmod} is \spmaxdrift, and the nominal drift field is \spmaxfield, corresponding to a cathode \dword{hv} of \sptargetdriftvoltpos. This design requires very low-noise electronics to achieve readout with good \dword{s/n} because no signal amplification occurs in the liquid. This technology was pioneered by the \dword{icarus} project, and after several decades of worldwide R\&D, is now a mature technology. It is the technology used for \dword{fnal}'s currently operating \dword{microboone} detector, as well as the \dword{sbnd} detector, which is under construction. 

\item \Dword{dp}: This technology was pioneered at a large scale by the \dword{wa105} collaboration at \dword{cern}. It is less established than the \dword{sp} technology but offers a number of potential advantages. Here, ionization charges drift vertically in \dword{lar} and are transferred into a layer of gas above the liquid. Devices called \dwords{lem} amplify the signal charges  in the gas phase. The gain achieved in the gas reduces stringent requirements on the electronics, and increases the possible drift length, which, in turn, requires a correspondingly higher voltage. The nominal drift field is \dpnominaldriftfield, as for the \dword{sp} detector, but in this case corresponds to a cathode \dword{hv} of \dptargetdriftvoltpos.
The maximum drift length in the \dword{dpmod} is \dpmaxdrift{}.  
\end{itemize}
In both technologies, the drift volumes are surrounded by a \dword{fc} that defines the volume(s) and ensures uniformity of the \efield to 1\% within the volume.

\dword{lar} is an excellent scintillator at a wavelength of \SI{126.8}{\nano\meter}. This fast scintillation light, once shifted into the visible spectrum, is collected by \dwords{pd} in both designs. The \dwords{pd} provide a time $t_{0}$ for every event, indicating when the ionization electrons begin to drift. Comparing the time at which the ionization signal reaches the anode relative to the $t_{0}$ allows reconstructing event topology in the drift coordinate; the precision of the measured $t_{0}$, therefore, directly corresponds to the precision of the spatial reconstruction in this direction. 

Two key factors affect the performance of the \dword{dune} \dwords{lartpc}.  First, the \dword{lar} purity must be high enough to achieve minimum charge attenuation over the longest drift lengths in a given \dword{detmodule}.  Thus, the levels of electro-negative contaminants (e.g., oxygen and water) must be maintained at ppt levels.  The \dword{sp} and \dword{dp} designs have slightly different purity requirements (expressed in minimum electron lifetimes of \SI{3}{ms} versus \SI{5}{ms}, respectively) due to the different drift lengths.

Second, the electronic readout of the \dword{lartpc} requires very low noise levels to allow the signal from the drifting electrons to be clearly discernible over the baseline of the electronics.  This requires using low-noise cryogenic electronics, especially in the case of the \dword{sp} design. 

The plans for the \dword{sp} and \dword{dp} \dwords{tpc} are described briefly in the following sections. 
The \dword{dune} collaboration is committed to deploying both technologies.
For planning purposes, we assume that the first \dword{detmodule} will be
\dword{sp} and the second will be \dword{dp}.
%
Studies are also under way toward a more advanced \dword{detmodule} design that could be realized as 
the fourth module, for example. 
%
The actual sequence of \dword{detmodule} installation will depend on results from the prototype detectors, described below, and on available resources.


\subsection{Single-phase Technology}
\label{sec:physics-intro-dunefd-splar}

Figure~\ref{fig:LArTPC1ch1} shows the general operating principle of the \dword{sp} \dword{lartpc}, as has been previously demonstrated by ICARUS~\cite{Icarus-T600}, \dword{microboone}~\cite{microboone}, \dword{argoneut}~\cite{Anderson:2012vc}, \dword{lariat}~\cite{Cavanna:2014iqa}, and \dword{protodune}~\cite{Abi:2017aow}. Figure~\ref{fig:DUNESchematic1ch1} shows the configuration of a \dword{dune} \dword{spmod}. Each of the four drift volumes of \dword{lar} is subjected to a strong electric field (\efield{}) of \spmaxfield. Charged particles passing through the \dword{tpc} ionize the argon, and the ionization electrons drift in the \efield to the anode planes.

\begin{dunefigure}[The SP LArTPC operating principle]{fig:LArTPC1ch1}
{The general operating principle of the \dword{sp} \dword{lartpc}.}
\includegraphics[width=0.75\textwidth]{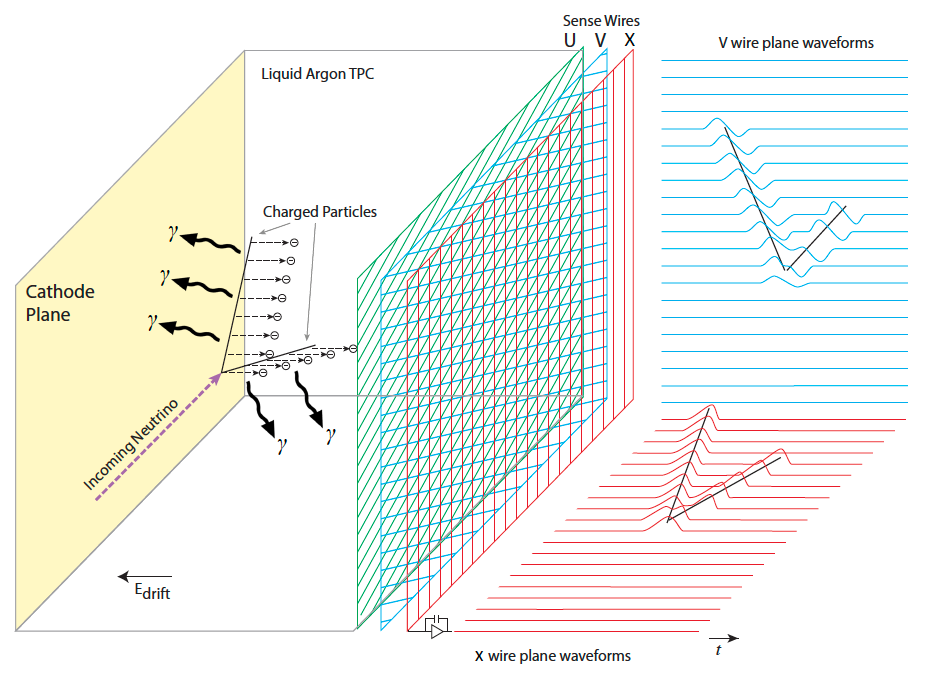} 
\end{dunefigure}

\begin{dunefigure}[A \nominalmodsize DUNE far detector SP module]{fig:DUNESchematic1ch1}
{Schematic of a \nominalmodsize \dword{dune} \dword{fd} \dword{spmod}, showing the alternating anode (A) and cathode (C) planes that divide the \dword{lartpc} into four separate drift volumes. The red arrows point to one top and one bottom \dword{fc} module and to the rear endwall field cage.}
\includegraphics[width=0.65\textwidth]{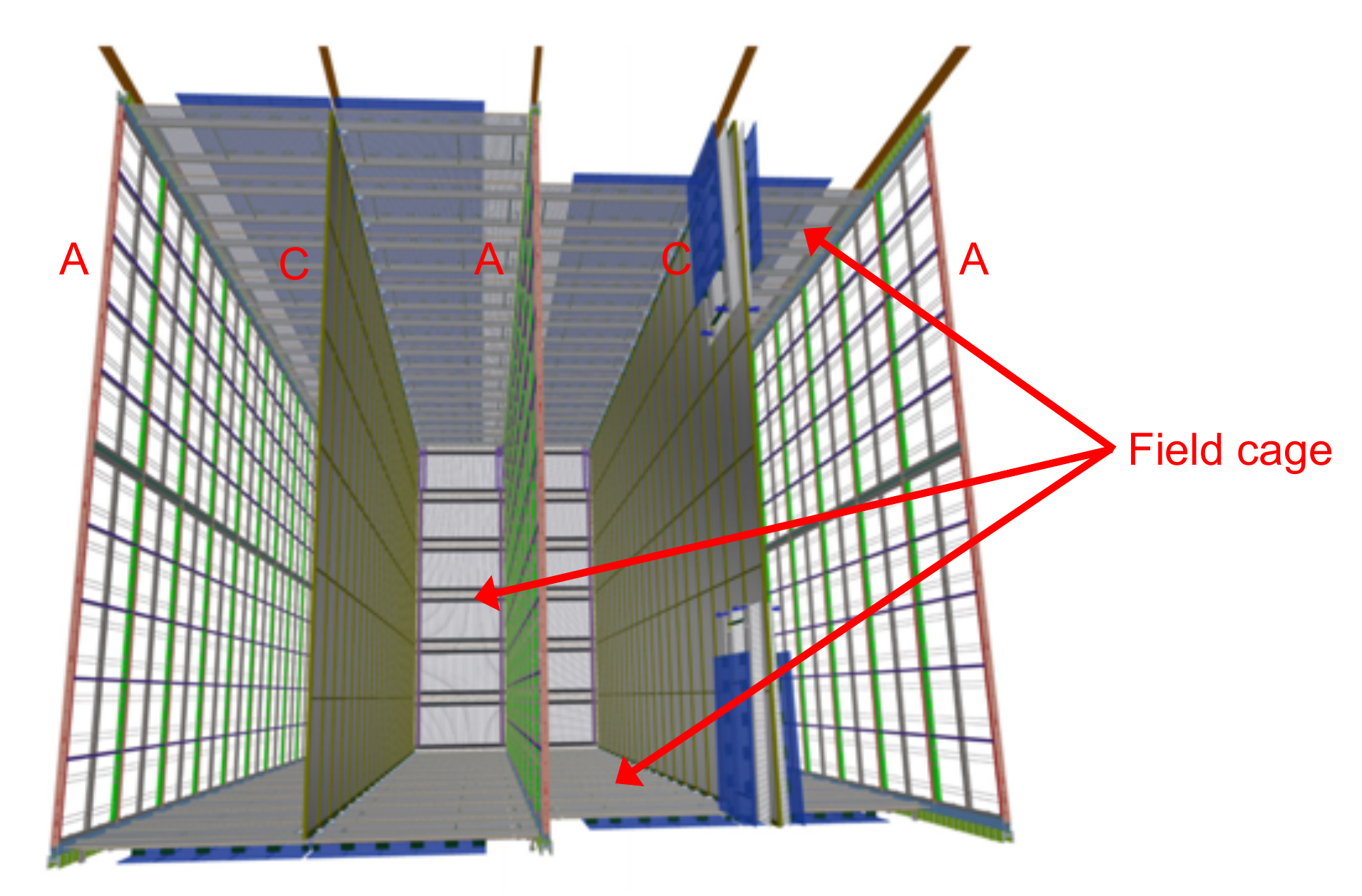}
\end{dunefigure}

A \dword{spmod} is instrumented with three module-length anode planes constructed from \SI{6}{m} high by \SI{2.3}{m} wide \dword{apa}s, stacked two \dword{apa}s high and 25 wide, for 50 \dword{apa}s per plane, or 150 total. 
Each \dword{apa} is two-sided with three layers of active wires forming a grid on each side of the \dword{apa}.
 The relative voltage between the layers is chosen to ensure the transparency to the drifting electrons of the first two layers ($U$ and $V$). These layers produce bipolar induction signals as the electrons pass through them. The final layer ($X$) collects the drifting electrons, resulting in a unipolar signal. The pattern of ionization collected on the grid of anode wires provides the reconstruction in the remaining two coordinates perpendicular to the drift direction.

Scintillation photons are detected in 
novel \dword{pd} modules, based on  
a light-trap concept known as 
ARAPUCA~\cite{arapuca_jinst,arpkLNLS,Segreto:2018jdx} that   
utilizes dichroic filters, wavelength-shifting plates and 
\dword{sipm} read-out. The variant of this technology in 
the DUNE baseline design (X-ARAPUCA) is described in Volume~\volnumbersp{}.  
The \dword{pd} modules 
are placed in the inactive space between the 
innermost wire planes of the \dword{apa}s, installed through 
slots in a pre-wound \dword{apa} frame. 
There are ten \dword{pd} modules per \dword{apa} for a total of 
\num{1500} per \dword{spmod}.  Of these, \num{500} are mounted in 
central \dword{apa} frames and must collect light from both 
directions, 
and \num{1000} are mounted in frames  near the 
cryostat walls and collect light from only one direction. 

\FloatBarrier
\subsection{Dual-phase Technology}
\label{sec:fddp-exec-splar}

The \dword{dp} operating principle, illustrated in Figure~\ref{fig:DPprinciplech1}, is very similar to that of the \dword{sp}. 
 Charged particles that traverse the active volume of the \dword{lartpc} ionize the medium while also producing scintillation light.  The ionization electrons drift along an \efield towards a segmented anode where they deposit their charge. Scintillation light is measured in \dwords{pd} that view the volume from below. 

\begin{dunefigure}[The DP LArTPC operating principle]{fig:DPprinciplech1}{The general operating principle of the \dword{dp} \dword{lartpc}.}
\includegraphics[width=0.5\textwidth]{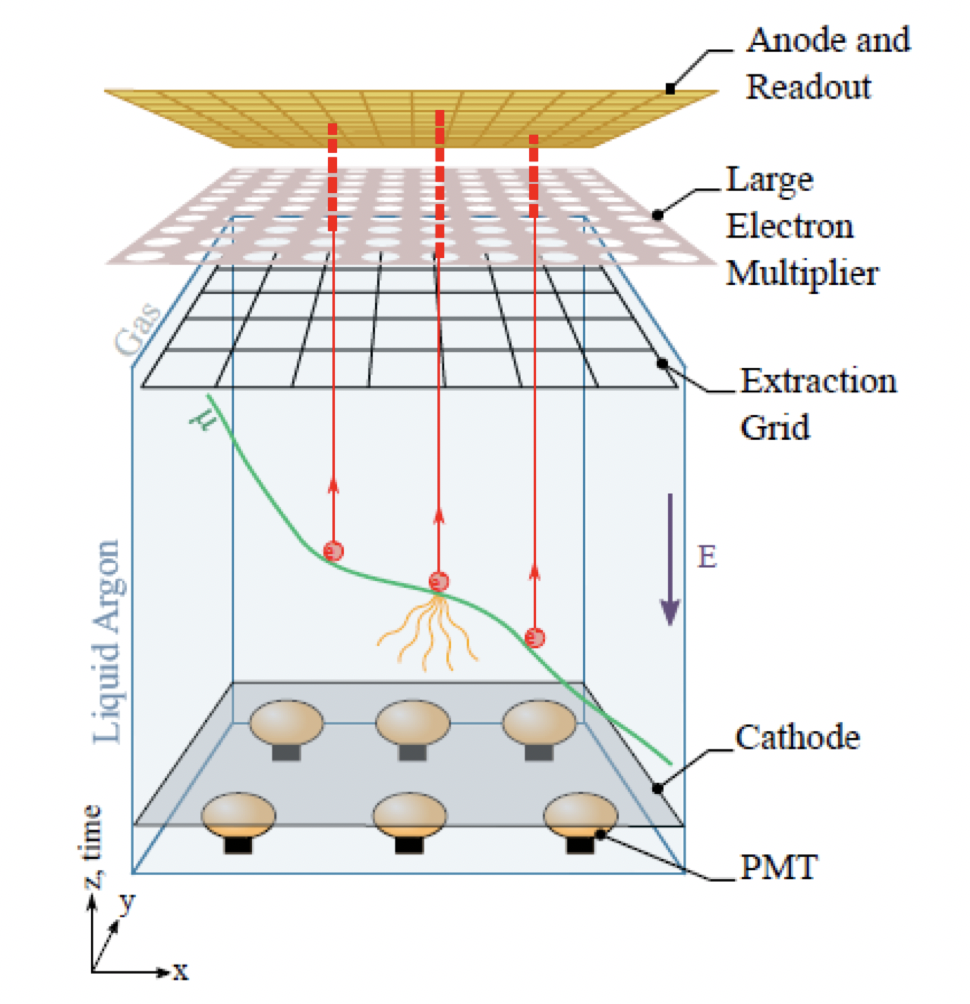}
\end{dunefigure}

In this design, shown in Figure~\ref{fig:DPdet1ch1}, electrons drift upward toward an extraction grid just below the liquid-vapor interface. 
After reaching the grid, an \efield stronger than the \dpnominaldriftfield{} drift field extracts the electrons from the liquid up into the gas phase. Once in the gas, electrons encounter micro-pattern gas detectors, called \dwords{lem}, with high-field regions. The \dwords{lem} amplify the electrons in avalanches that occur in these high-field regions. The amplified charge is then collected and recorded on a \twod anode
consisting of two sets of 
gold-plated copper strips that provide the $x$ and $y$ coordinates (and thus two views) of an event. 

\begin{dunefigure}[A \nominalmodsize DUNE far detector DP module]{fig:DPdet1ch1}
  {Schematic of a \nominalmodsize \dword{dune} \dword{fd} \dword{dp} \dword{detmodule} with cathode, \dwords{pmt}, \dword{fc}, and anode plane with \dwords{sftchimney}. The drift direction is vertical in the case of a \dword{dpmod}.}
  \includegraphics[width=0.9\textwidth]{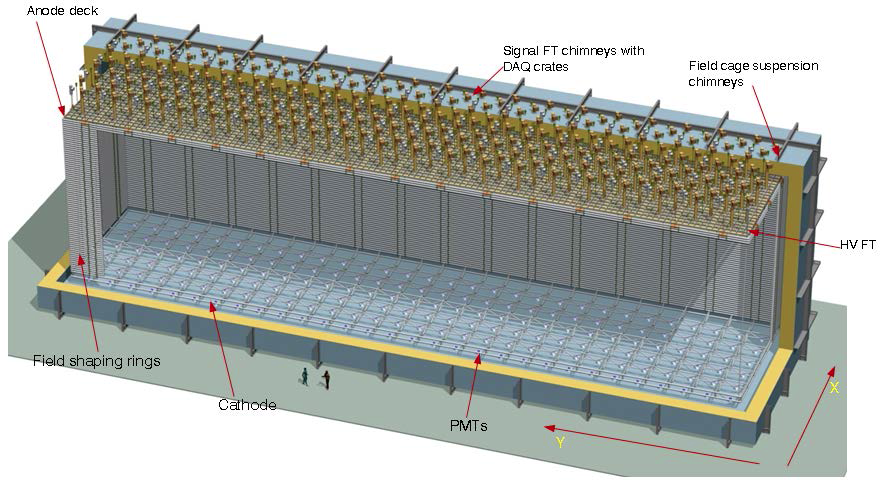}
\end{dunefigure}

 The extraction grid, \dword{lem}, and anode are assembled into three-layered sandwiches with precisely defined inter-stage distances and inter-alignment,  which are then connected horizontally into \num{9}~m$^2$ modular detection units. These detection units are called \dwords{crp}.

The precision tracking and calorimetry offered by the \dword{dp} technology provides excellent capabilities for identifying interactions of interest while mitigating sources of background.  Whereas the \dword{sp} design has multiple drift volumes, the \dword{dpmod} design allows a single, fully homogeneous \dword{lar} volume with a much longer drift length.

A simple array of \dwords{pmt} coated with a wavelength-shifting material is located below the cathode. The \dwords{pmt} record  the time and pulse characteristics of the incident light.

\FloatBarrier

\subsection{ProtoDUNEs: Far Detector Prototypes}

The \dword{dune} collaboration has constructed 
two large prototype detectors (\dwords{protodune}), \dword{pdsp} and \dword{pddp}, located at CERN. 
 Each is approximately one-twentieth the size of a \dword{dune} \dword{detmodule} and uses components identical in size to those of the full-scale module. \dword{pdsp} has the same \spmaxdrift maximum drift length as the full \dword{spmod}. \dword{pddp} has a \SI{6}{m} maximum drift length, half that planned for the \dword{dpmod}. See the photos in Figures~\ref{fig:protodunes_northarea} and~\ref{fig:protodunes_interior}.

\begin{dunefigure}[ProtoDUNE cryostats at the CERN Neutrino Platform]
{fig:protodunes_northarea}
{ProtoDUNE-SP and ProtoDUNE-DP cryostats in the CERN Neutrino Platform in CERN's North Area.  The view is from the downstream end of the hall with respect to the beam lines.  At front and  center is the top of the \dword{pdsp} cryostat.  The \dword{pddp} cryostat with its painted red steel support frame visible is located at the rear of the photo on the right side of the hall.} 
\includegraphics[width=0.9\linewidth]{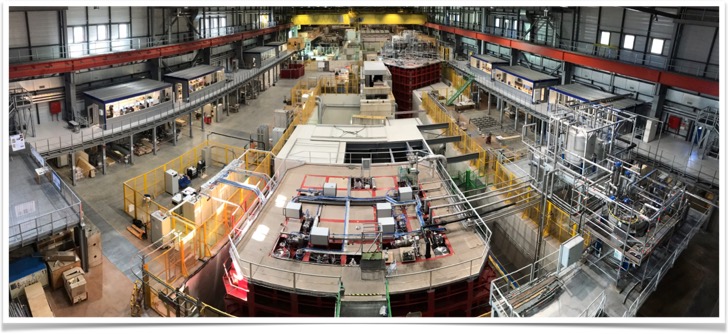}
\end{dunefigure}

\begin{dunefigure}[Interior views of the ProtoDUNEs]
{fig:protodunes_interior}
{Interior views of ProtoDUNE-SP (left) and ProtoDUNE-DP (right). For ProtoDUNE-SP, one of two identical drift volumes is shown.}
\includegraphics[width=0.46\linewidth]{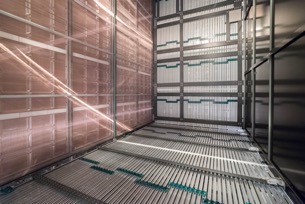}\hspace{0.05\linewidth}
\includegraphics[width=0.44\linewidth]{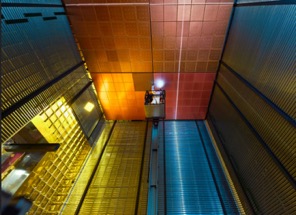}
\end{dunefigure}

These large-scale prototypes allow us to validate key aspects of 
the \dword{tpc} designs, test engineering procedures, and collect 
valuable calibration data using a hadron test beam. 

The construction phase of \dword{pdsp} was finished in July 2018, 
and the detector was filled with \dword{lar} in August 2018. The 
detector collected hadron beam data and cosmic rays during the 
fall of 2018 and continues to collect cosmic-ray data.
The construction of the \dword{pddp} detector was completed in 
June of 2019, and started operations in September 2019.  

Data taken with the \dword{pdsp} detector demonstrates its 
excellent performance and has already provided valuable 
information on the design, calibration, and simulation of the 
\dword{dune} \dword{fd}. In all, $99.7\%$ of the 15360 \dword{tpc} 
electronics channels are responsive in the \dword{lar}. The 
equivalent noise charge
amounts to $\approx 550$ $e^{-}$ on the collection 
wires and $\approx 650$ $e^{-}$ on the induction wires. An average 
\dword{s/n} of 38 for the collection plane is measured using 
cosmic-ray muons, while for the two induction planes, the 
\dword{s/n} is 14 (U) and 17 (V), exceeding the requirement  
for the \dword{dune} \dword{fd}. 
The \dword{pdsp} photon detection system has also operated stably, 
demonstrating the principle of effective collection of 
scintillation light in a large-volume \dword{lartpc} with 
detectors embedded within the \dwords{apa}.

\section{Near Detector Complex}
\label{sec:physics-nd-overview}

The \dword{dune} \dword{nd} is crucial for the success of the \dword{dune} physics program. It is used to precisely measure the neutrino beam flux and flavor composition. Comparing the measured neutrino energy spectra at the near and far site allows us to disentangle the different energy-dependent effects that modulate the beam spectrum and to reduce the systematic uncertainties to the level required for discovering \dword{cp} violation. In addition, the \dword{nd} will measure neutrino-argon interactions with high precision using both gaseous and liquid argon, which will further reduce the systematic uncertainties associated with the modeling of these interactions.

The \dword{nd} hall will be located \SI{574}{m} downstream from the target and will include three primary detector components, shown in Figure~\ref{fig:neardetectors}  and listed in Table~\ref{tab:NDsumm}. Two of them can move off beam axis, providing access to different neutrino energy spectra. The movement off axis, called \dword{duneprism}, provides a crucial extra degree of freedom for the \dword{nd} measurement and is an integral part of the \dword{dune} \dword{nd} concept.

\begin{dunefigure}[DUNE near detector]
{fig:neardetectors}
{DUNE Near Detector. The beam enters from the right and encounters
the \dword{lartpc}, the MPD, and the SAND on-axis beam monitor.}
\includegraphics[width=0.9\textwidth]{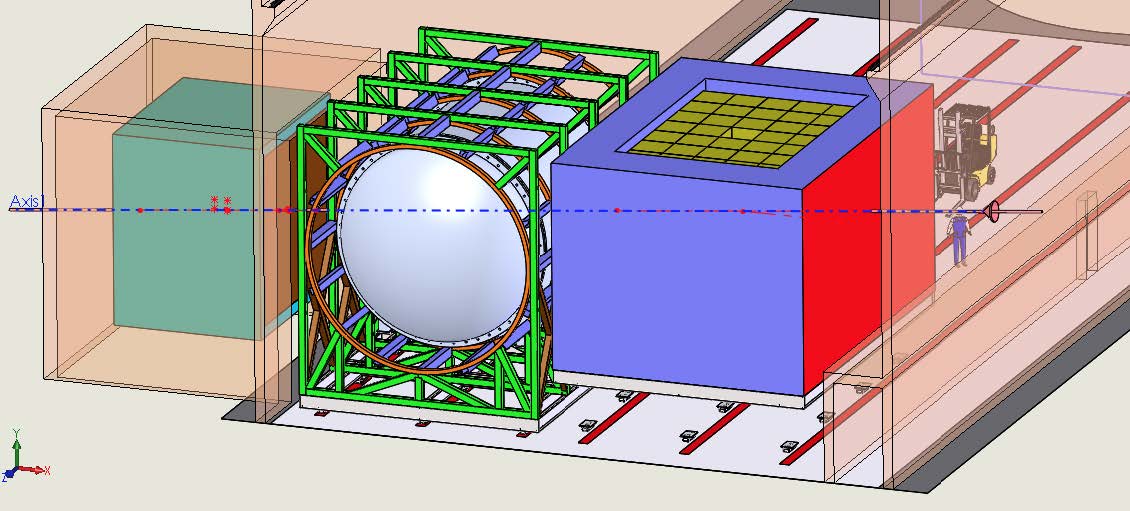}
\end{dunefigure}

The three detector components -- a \dword{lartpc} called \dword{arcube}; a \dword{hpgtpc} within a magnet surrounded by an \dword{ecal}, together called \dword{mpd}; and an on-axis beam monitor called \dword{sand} -- serve important individual and overlapping functions in the mission of the \dword{nd}. 
The \dword{dune} \dword{nd} is shown schematically in the \dword{dune} \dword{nd} hall in Figure~\ref{fig:NDHallconfigs}.  
Table~\ref{tab:NDsumm} provides a high-level overview of the three components of the \dword{dune} \dword{nd} along with the off-axis capability.  

\begin{dunetable}[Components of the DUNE ND]
{p{.22\textwidth}p{.22\textwidth}p{.22\textwidth}p{.22\textwidth}}
{tab:NDsumm}{This table gives a high-level breakdown of the three major detector components and the capability of movement for the DUNE ND along with function and primary physics goals.}
Component & Essential Features  & Primary function & Select physics aims \\
 \toprowrule
LArTPC (ArgonCube) & Mass  & Experimental control for the Far Detector & $\numu$($\overline{\nu}_{\mu}$) CC \\
          & Target nucleus Ar &  Measure unoscillated $E_\nu$ spectra   & $\nu$-e$^{-}$ scattering   \\
          &  Technology FD-like    &  Flux determination  &  $\nue +$$\overline{\nu}_{e}$ CC  \\
          &  &  &  Interaction model \\ \colhline
Multipurpose detector (MPD) & Magnetic field & Experimental control for the LArTPCs & $\numu$($\overline{\nu}_{\mu}$) CC \\
  &  Target nucleus Ar & Momentum analyze liquid Ar $\mu$ & $\nue$ CC, $\overline{\nu}_{e}$ \\
  & Low density & Measure exclusive final states with low momentum threshold & Interaction model \\  \colhline
%
On-axis beam monitor \hfill (\dword{sand})
 & On-axis & Beam flux monitor &  On-axis flux stability \\ 
  & Mass & Neutrons & Interaction model \\ 
& Magnetic field &  & A dependence \\
    & CH target & & $\nu$-e$^{-}$ scattering \\ \colhline \colhline
    
    DUNE-PRISM (capability) & LArTPC$+$MPD move off-axis & Change flux spectrum &  Deconvolve xsec*flux \\ 
 & & & Energy response \\
 & & & Provide FD-like energy spectrum at ND\\ 
 & & & ID mismodeling \\ \colhline
\end{dunetable}

The \dword{arcube} detector contains the same target nucleus and shares some aspects of form and functionality with the \dword{fd}. The differences are necessitated by the expected high intensity of the neutrino beam at the \dword{nd}.  This similarity in target nucleus and, to some extent, technology, reduces sensitivity to nuclear effects and detector-driven systematic uncertainties in extracting the oscillation signal at the  \dword{fd}. 
The \dword{arcube} \dword{lartpc} is large enough to provide high statistics ($\num{1e8} {\numu \text{ charged current events/year on axis}}$), and its volume is sufficiently large to provide good hadron containment.  The tracking and energy resolution, combined with the mass of the \dword{lartpc}, will allow measurement of the flux in the beam using several techniques, including the rare process of $\nu$-e$^{-}$ scattering.

\begin{dunefigure}[DUNE ND Hall with component detectors]
{fig:NDHallconfigs}
{\dword{dune} \dword{nd} hall shown with component detectors all in the on-axis configuration (left) and with the \dword{lartpc} and \dword{mpd} in an off-axis configuration (right). The on-axis monitor \dword{sand} is shown in position on the beam axis. 
}
\includegraphics[width=0.49\textwidth]{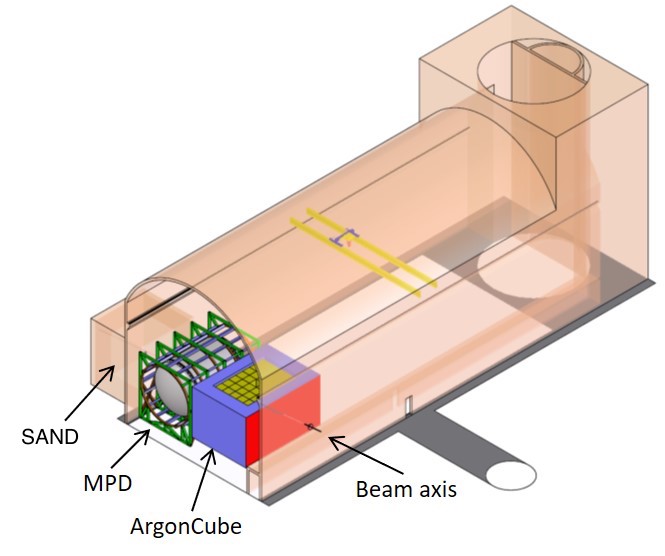}
\includegraphics[width=0.49\textwidth]{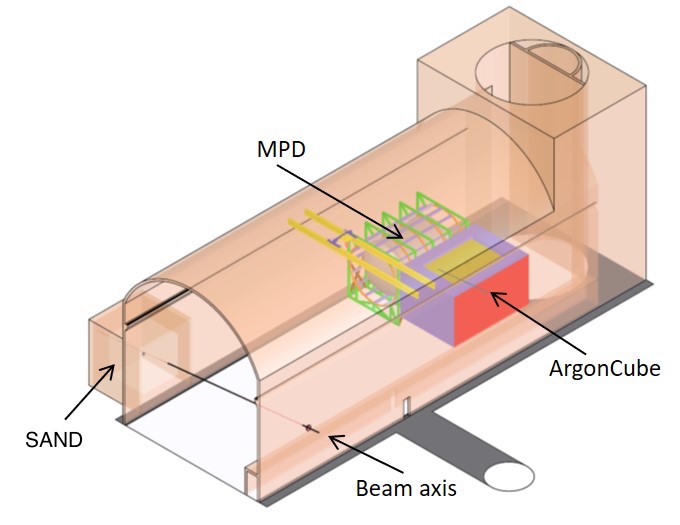}
\end{dunefigure}

The \dword{lartpc} begins to lose acceptance for muons with a measured momentum higher than 
$\approx$0.7~GeV/c because the muons will not be contained in the \dword{lartpc} volume.  Because the muon momentum is a critical component of determining the neutrino energy, a magnetic spectrometer is needed downstream of the \dword{lartpc} to measure the charge sign and momentum of the muons.  In the \dword{dune} \dword{nd} concept, this function is accomplished by the \dword{mpd}, which consists of a \dword{hpgtpc} surrounded by an \dword{ecal} in a \SI{0.5}{T} magnetic field. The \dword{hpgtpc} provides a lower density medium with excellent tracking resolution for the muons from the \dword{lartpc}.  

In addition, neutrinos interacting with the argon in the gas \dword{tpc} constitute a sample of $\nu$-argon events that can be studied with a very low charged-particle tracking threshold and excellent resolution superior to \dword{lar}. The high pressure yields a sample of $\num{2e6}{\numu \text{-CC events/year}}$ for these studies. These events will be valuable for studying the charged particle activity near the interaction vertex because this detector can access lower momenta protons than the \dword{lar} detector and can better identify charged pions.  The lack of secondary interactions in these samples will be helpful for identifying the particles produced in the primary interaction and modeling secondary interactions in denser detectors.

The \dword{ecal} adds neutral particle (mainly $\gamma$'s and 
neutrons) detection capability otherwise lacking in the \dword{mpd}.
NC-$\pi^0$ backgrounds to $\nu_e$ CC interactions can be studied, 
for example.  
Additionally, neutron production in neutrino-nucleus interactions 
is poorly understood: the presence of the \dword{ecal} 
opens the possibility of identifying neutrons via time-of-flight.

The \dword{lartpc} and \dword{mpd} can be moved sideways up to 33 m to take data in positions off the beam axis.  This capability is referred to as \dword{duneprism}. As the detectors move off-axis, the incident neutrino flux spectrum changes, with the mean energy dropping and the spectrum becoming more monochromatic.  Though the neutrino interaction rate drops off-axis, the intensity of the beam and the size of the \dword{lartpc}  combine to yield ample statistics even in the off-axis positions.
The \dword{dune} concept is based on reconstructing the energy-dependent neutrino spectrum and
comparing the far and near sites. The ability to modify the energy spectrum at the near site by measuring at the off-axis locations will allow disentangling otherwise degenerate effects due to systematic biases of the energy reconstruction.

The final component of the \dword{dune} \dword{nd} suite (\dword{sand}) is the on-axis beam monitor that remains in fixed position at all times and serves as a dedicated neutrino spectrum monitor. 
It can also provide an excellent on-axis neutrino flux determination that can be used as an important point of comparison and a systematic crosscheck for the flux as determined by \dword{arcube}.


\cleardoublepage

\chapter{Scientific Landscape}
\label{ch:physics-landscape}

The aim of this chapter is to set the stage for the discussions of 
DUNE's scientific capabilities that are presented in the chapters 
that follow.  This is implemented as a series of brief 
descriptions of the theoretical and experimental contexts relevant 
for key areas of the DUNE physics program. 

It is important to state at the outset that a fully comprehensive 
review is not possible here.  Rather, the descriptions presented 
in this chapter are intended to be illustrative, so as to 
convey broadly the array of scientific opportunities for which 
DUNE is designed to realize.
Furthermore, the supporting literature is vast, and it is 
not within the scope or purpose of this chapter to provide an 
exhaustive list of references. 
More details, including concrete references to the literature, 
are provided in subsequent chapters.

\section{Neutrino Oscillation Physics}
\label{sec:landscape-osc}

The first positive hint for neutrino flavor-change was uncovered in the 1960's with the first measurement of the flux of neutrinos from the sun. The hint compounded in the late 1980's, with high-statistics measurements of the differential flux of muon-type neutrinos produced by the collisions of cosmic rays with the earth's atmosphere. Both hints were ultimately confirmed in the late 1990's and early 2000's by the Super-Kamiokande and SNO experiments. Concurrently, neutrino oscillations were confirmed as the dominant physics behind neutrino flavor change. 

Neutrino oscillations imply nonzero neutrino masses and flavor-mixing in the leptonic charged-current interactions.  That the neutrino masses are not zero is among the most important discoveries in fundamental particle physics of the twenty-first century. Understanding the mechanism behind nonzero neutrino masses is among the unresolved mysteries that drive particle physics today; they remain one of the few unambiguous facts that point to the existence of new particles and interactions, beyond those that make up the remarkable standard model of particle physics. Learning more about the properties of neutrinos is a very high priority for particle physics, and neutrino oscillations remain, as of today, the only phenomenon capable of observing the neutrino masses and lepton mixing in action. Precision measurements of neutrino oscillations have the potential to play a leading role in shaping particle physics in the next few decades. 

Almost all neutrino data can be understood within the three-flavor paradigm with massive neutrinos, the simplest extension of the standard model capable of reconciling theory with observations. A handful of intriguing results, including those from the LSND, MiniBooNE, and short-baseline reactor experiments, remain unexplained and are currently the subject of intense experimental and theoretical scrutiny. If confirmed as the manifestation of new physics involving neutrinos -- e.g., new neutrino states --  these will open the door to more neutrino-related questions, many of which can be further explored with DUNE. We will return to those later but assume, conservatively, that the resolutions to these so-called short-baseline anomalies lie outside of neutrino-related particle physics. 

\subsection{Oscillation Physics with Three Neutrino Flavors}

The three-flavor paradigm with massive neutrinos consists of introducing distinct, nonzero, masses for at least two neutrinos, while maintaining the remainder of the standard model of particle physics. Hence, neutrinos interact only via the standard model charged-current and neutral-current weak interactions. The neutrino mass eigenstates -- defined as $\nu_1,\nu_2, \nu_3$ with masses, $m_1, m_2, m_3$, respectively -- are distinct from the neutrino charged-current interaction eigenstates, also referred to as the flavor eigenstates -- $\nu_e$, $\nu_{\mu}$, $\nu_{\tau}$, labeled according to the respective charged-lepton $e,\mu,\tau$ to which they couple in the charged-current weak interaction. The flavor eigenstates can be expressed as linear combinations of the mass eigenstates (and vice-versa). The coefficients of the respective linear combinations define a unitary $3\times 3$ mixing matrix, referred to as the neutrino mixing matrix, the leptonic mixing matrix, or the Pontecorvo-Maki-Nakagawa-Sakata (PMNS) matrix, as follows:
\begin{equation}
\left(\begin{array}{c} \nu_e \\ \nu_{\mu} \\ \nu_{\tau} \end{array}\right) = \left(\begin{array}{ccc} U_{e1} & U_{e2} & U_{e3} \\  U_{\mu1} & U_{\mu2} & U_{\mu3}  \\  U_{\tau1} & U_{\tau2} & U_{\tau3}  \end{array}\right) \left(\begin{array}{c} \nu_1 \\ \nu_2 \\ \nu_3 \end{array}\right).
\end{equation}
The PMNS matrix is the leptonic-equivalent of the Cabibbo-Kobayashi-Maskawa (CKM) matrix that describes the charged-current interactions of quark mass eigenstates. If the neutrinos are Dirac fermions, taking advantage of the unitary nature of the matrix and the ambiguity in defining the relative phases among the standard model lepton fields, the neutrino mixing matrix, like the CKM matrix, can be unambiguously parameterized with three mixing angles and one complex phase. If, however, the neutrinos are Majorana fermions, there are fewer field-redefinitions available and one ends up with at most two other physical complex phases.\footnote{Majorana phases can also be interpreted as complex phases of the neutrino mass eigenvalues and need not be considered as part of the neutrino mixing matrix.}  Strictly speaking, these so-called Majorana phases can manifest themselves in ``neutrino--antineutrino'' oscillations \cite{deGouvea:2002gf} and could be observed in neutrino oscillation experiments. These effects, however, are expected to be unobservably small and will be henceforth ignored, along with the Majorana phases. For all practical purposes, neutrino oscillation experiments cannot distinguish Majorana from Dirac neutrinos. Majorana phases are expected to play a significant role in experiments that are sensitive to the Majorana versus Dirac nature of the neutrinos, including searches for neutrinoless double-beta decay. 

The PDG-parameterization \cite{Tanabashi:2018oca}, used throughout this report, makes use of three mixing angles $\theta_{12}$, $\theta_{13}$, and $\theta_{23}$, defined as
\begin{eqnarray}
\sin^2\theta_{12} &\equiv&       \frac{|U_{e2}|^2}{1-|U_{e3}|^2}, \\
\sin^2\theta_{23} &\equiv& \frac{|U_{\mu3}|^2}{1-|U_{e3}|^2}, \\
\sin^2\theta_{13} &\equiv& |U_{e3}|^2,
\end{eqnarray} 
and one phase $\deltacp$, which in the conventions of~\cite{Tanabashi:2018oca}, is given by
\begin{equation}
\deltacp \equiv -{\rm arg}(U_{e3}).
\end{equation}
For values of $\deltacp\neq 0,\pi$, and assuming none of the $U_{\alpha i}$ vanish ($\alpha=e,\mu,\tau$, $i=1,2,3$), the neutrino mixing matrix is complex and CP-invariance is violated in the lepton sector. This, in turn, manifests itself as different oscillation probabilities, in vacuum, for neutrinos and antineutrinos: $P(\nu_{\alpha}\to\nu_{\beta})\neq P(\bar{\nu}_{\alpha}\to\bar{\nu}_{\beta})$, $\alpha,\beta=e,\mu,\tau$, $\alpha\neq\beta$.\footnote{For neutrino disappearance, in vacuum, the relation $P(\nu_{\alpha}\to\nu_{\alpha}) =P(\bar{\nu}_{\alpha}\to\bar{\nu}_{\alpha})$ is a consequence of the CPT-theorem.} 

Information on the values of the neutrino masses comes from measurements of the neutrino oscillation frequencies, which are proportional to the differences of the squares of the neutrino masses, $\Delta m^2_{ij}\equiv m_i^2-m_j^2$. Since all positive evidence for nonzero neutrino masses comes from measurements of neutrino oscillations, there is no direct information concerning the values of the masses themselves, only the mass-squared differences. As far as neutrino oscillation data are concerned, the hypothesis that the lightest neutrino mass is exactly zero is just as valid as the hypothesis that all neutrino masses are nonzero and almost degenerate. Three neutrino masses allow for two independent mass-squared differences and the existing neutrino data point to two hierarchically different $\Delta m^2$, one whose magnitude is of order $10^{-4}$~eV$^2$, the other with magnitude of order $10^{-3}$~eV$^2$. 

With this information, it is possible to unambiguously define the neutrino masses in a convenient way, as follows.\footnote{Equivalently, one can use the neutrino mixing matrix to define the neutrino mass eigenstates. $\nu_1$ could be defined as the state associated to the largest $|U_{ei}|^2$ ($i=1,2,3$), $\nu_2$ to the second largest $|U_{ei}|^2$, and $\nu_3$ to the smallest $|U_{ei}|^2$: $|U_{e1}|^2>|U_{e2}|^2>|U_{e3}|^2$.} The mass-squared difference with the smallest magnitude is defined to be $\Delta m^2_{21}$, positive-definite so $m_2^2>m_1^2$. The third mass eigenvalue is such that $|\Delta m^2_{31}|\sim|\Delta m^2_{32}|$, of order $10^{-3}$~eV$^2$ while the sign of  $\Delta m^2_{31}, \Delta m^2_{32}$ defines the neutrino mass ordering, or the neutrino mass hierarchy. If $\Delta m^2_{31}, \Delta m^2_{32}>0$, the neutrino mass ordering is defined to be `normal' and $m_1^2<m_2^2<m_3^2$. If $\Delta m^2_{31}, \Delta m^2_{32}<0$, the neutrino mass ordering is defined to be `inverted' and $m_3^2<m_1^2<m_2^2$. This definition allows one to change from a normal to an inverted ordering without having to change the relationship between the neutrino mixing matrix and the various experimental results. The distinct neutrino mass orderings are illustrated in Figure~\ref{massordering}.
\begin{dunefigure}[Mass Ordering Illustration]{massordering}{
   Fractional flavor content, $|U_{\alpha i}|^2$ ($\alpha = e, \mu, \tau$) of the three mass eigenstates $\nu_i$, based on the current best-fit values of the mixing angles. $\deltacp$ is varied from 0 (bottom of each colored band) to $180^\circ$ (top of colored band), for normal and inverted mass ordering on the left and right, respectively. The different colors correspond to the $\nu_e$ fraction (red), $\nu_\mu$ (green) and $\nu_\tau$ (blue). 
}
  \includegraphics[width=0.6\textwidth]{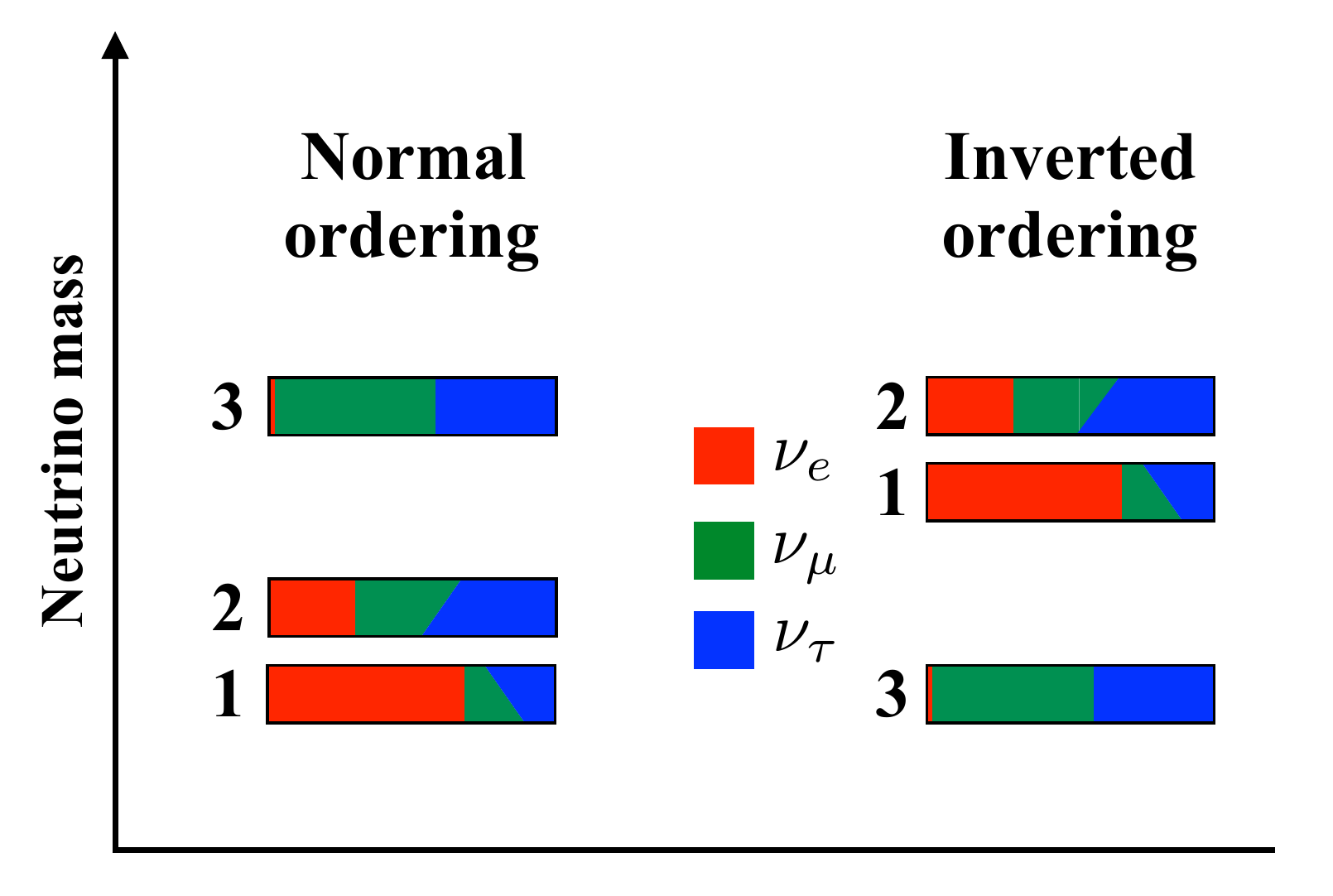}
\end{dunefigure}

\subsubsection{Synthesis of Experimental Inputs} 

The world's neutrino data significantly constrain all of the oscillation parameters in the three-flavor paradigm. The results of a recent global fit \cite{Esteban:2018azc} to all neutrino data, except those associated to the short-baseline anomalies, are depicted in Fig.~\ref{nufit40}. The magnitudes of both mass-squared differences are known at better than 3\%, while, at the one sigma level, $\sin^2\theta_{12}$, $\sin^2\theta_{13}$, $\sin^2\theta_{23}$ are known at better than the 5\% level. Note, however, that the error bars are rather non-Gaussian, especially for $\sin^2\theta_{23}$. At the three sigma level,  according to \cite{Esteban:2018azc}, $\sin^2\theta_{23}$ is constrained to lie between 0.43 and 0.62 so values of $\sin^2\theta_{23}>0.5$ and $\sin^2\theta_{23}<0.5$ are allowed.
\begin{dunefigure}[Nufit 4.0 global fit]{nufit40}{Global three-neutrinos-oscillation analysis from \cite{Esteban:2018azc}. Each panel depicts the two-dimensional projection of the allowed six-dimensional region after marginalization with respect to the undisplayed parameters. The different contours correspond to the two-dimensional allowed regions at 1$\sigma$, 90\%, 2$\sigma$, 99\%, 3$\sigma$ CL (2~dof). Note that the top panel refers to $\Delta m^2_{31}$ in the case of the normal mass-ordering and $\Delta m^2_{32}$ in the case of the inverted one. The regions in the lower four panels are defined using $\Delta \chi^2$ relative to minimum value of $\chi^2$ obtained for a fixed choice of the mass ordering, normal ordering on the left-hand-side, inverted ordering on the right-hand-side.
}
  \includegraphics[width=0.7\textwidth]{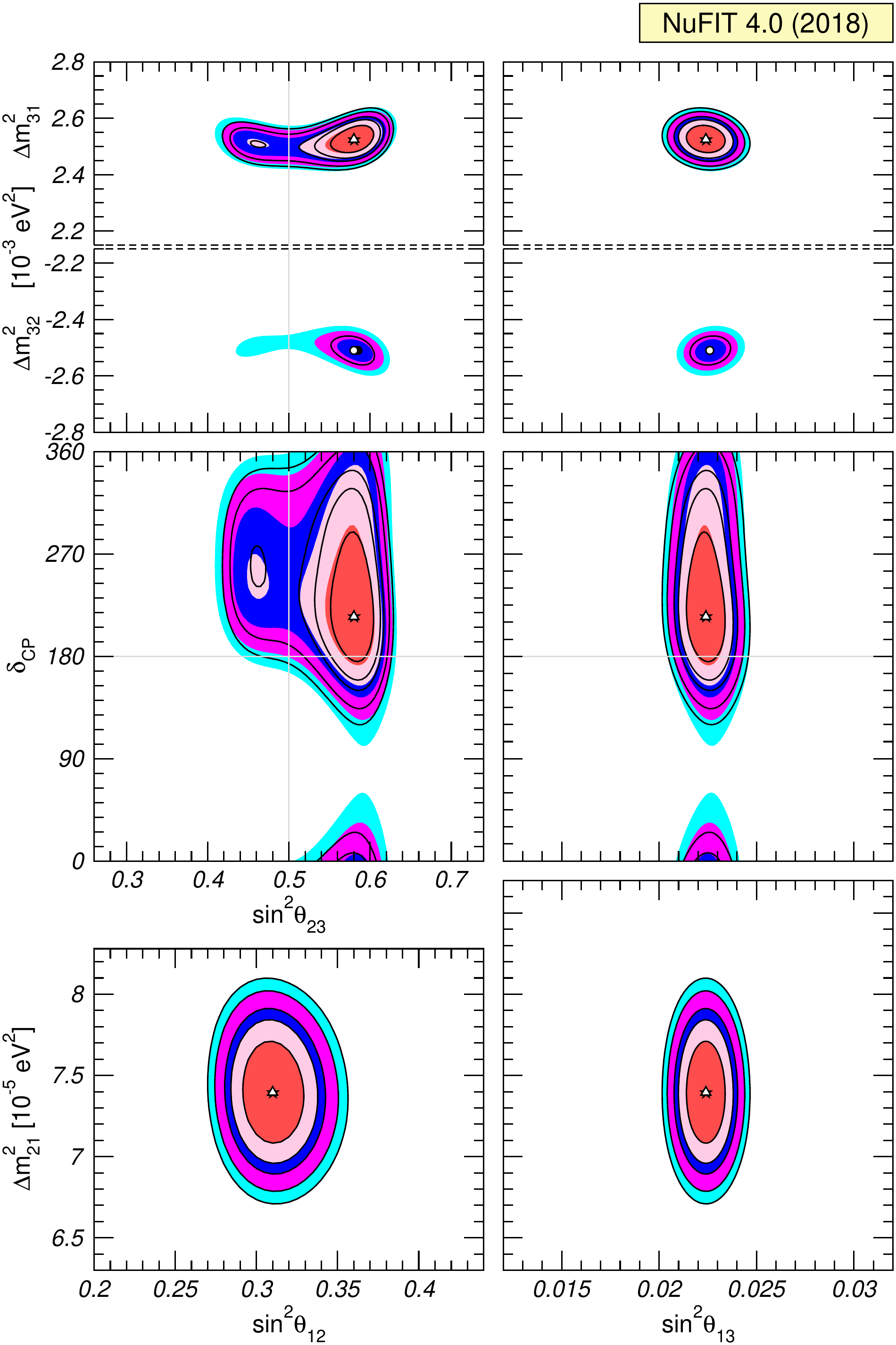}
\end{dunefigure}

Critical questions remain open. The neutrino mass ordering is unknown. Current data prefer the normal ordering but the inverted one still provides a decent fit to the data. The octant of $\theta_{23}$ (whether $\sin^2\theta_{23}<0.5$ [$\theta_{23}<\pi/4$] or $\sin^2\theta_{23}>0.5$ [$\theta_{23}>\pi/4$]) remains unknown. The value of $\deltacp$ is only poorly constrained. 
While positive values of $\sin\deltacp$ are disfavored, all $\deltacp$ values between $\pi$ and $2\pi$, including the CP-conserving values $\deltacp=0,\pi$, are consistent with the world's neutrino data.\footnote{It should be noted that recent results from the T2K experiment~\cite{Abe:2019vii} show only marginal consistency with CP-conserving values of $\deltacp$.} 
That the best fit to the world's data 
favors large \dword{cpv} is intriguing, providing further impetus 
for experimental input to resolve this particular question.
It is central to the \dword{dune} mission that all of the questions 
posed here can be addressed 
by neutrino oscillation experiments.

Other fundamental questions, including the nature of the neutrino -- Majorana versus Dirac -- and the determination of the values of the neutrino masses -- oscillation experiments only measure mass-squared differences -- are not accessible to oscillation experiments and must be addressed using different experimental tools.  

At a more fundamental level, the three-flavor paradigm is yet to be significantly challenged by precision experiments. The overall picture described briefly above, while minimalistic and appealing, may turn out to be incomplete. While we don't know what new neutrino physics, if any, lies beyond the three-flavor paradigm, many possibilities have been identified and are currently the subject of intense phenomenological and theoretical scrutiny. We list a few here; these and a few others are discussed in more detail in this report. There may be more neutrino-like states and hence new oscillation frequencies and mixing parameters. This is true regardless of the solution to the short-baseline anomalies. New neutrino-like states are often a ``side-effect'' of the physics responsible for nonzero neutrino masses and serve as a natural connection between the standard model and would-be dark sectors that may contain the elusive dark matter particle. Indeed, new neutrino states may, themselves, be a component of the dark matter. Neutrinos may also participate in new, currently, unknown interactions. These can be mediated by new heavy gauge bosons or new, weakly coupled, light particles. The heavier of the known neutrinos may also be much more short-lived than what is expected of the standard model interactions. The neutrino lifetimes are only poorly constrained, and some are best constrained by existing neutrino oscillation data. The quantum interferometric nature of the neutrino oscillation phenomenon also allows searches for new phenomena that manifest themselves as violations of $CPT$-invariance or violations of the law that governs the time-evolution of quantum states. 

Currently, the information that goes into determining the parameters of the three-flavor paradigm comes from a large variety of experiments that make use of different neutrino sources, neutrino flavors, and neutrino energies. Different parameters are determined by different experiments in such a way that there is only limited information on whether the formalism is complete. For example, $\sin^2\theta_{12}$ is best constrained by measurements of the differential flux of solar neutrinos -- mostly electron neutrinos with energies between 100~keV and 10~MeV  -- while $\Delta m^2_{21}$ is best constrained by the KamLAND experiment -- electron antineutrinos from nuclear reactors and baselines around 100~km. While solar experiments are also sensitive to $\Delta m^2_{21}$ and KamLAND to $\sin^2\theta_{12}$, the respective uncertainties are not relatively competitive. 
Another example, the mixing parameters $\sin^2\theta_{13}$ is best constrained by reactor experiments with baselines around 1~km. Long-baseline experiments sensitive to $\nu_{\mu}\to\nu_e$ oscillations -- baselines between 100~km and 1000~km, neutrino energies between a few 100~MeV and a few GeV -- are also sensitive to $\sin^2\theta_{13}$, but the associated uncertainties cannot compete with those from the reactor experiments. It is, therefore, not possible to compare, in any effective way, the reactor measurement of $\theta_{13}$ with the long-baseline measurement of $\theta_{13}$ and perform a simple, non-trivial check of the three-flavor paradigm, which predicts those two numbers to be the same. 

\subsubsection{Opportunities for DUNE}

The DUNE experiment is well positioned to over-constrain  the three-flavor paradigm and reveal what may potentially lie beyond. The high-statistics of DUNE is, for example, capable of extracting $\sin^2\theta_{13}$ via the electron neutrino appearance channel, $\nu_{\mu}\to\nu_e$, with precision that approaches that of the reactor electron antineutrino disappearance measurements, $\bar{\nu}_e\to\bar{\nu}_e$. If the three-flavor paradigm is incomplete, these two independent values for $\sin^2\theta_{13}$ need not agree. The high-statistics of DUNE also allow one to directly determine whether CP-invariance is violated by comparing how neutrinos and antineutrinos -- after matter effects are taken into account -- oscillate. The neutrino energies and the baseline of LBNF-DUNE imply that the oscillation probabilities will be significantly impacted by matter effects. These, in turn, allow DUNE to establish the neutrino mass ordering independent from the results of other neutrino oscillation experiments.\footnote{The current hint for the normal ordering relies on the reactor measurement of $\sin^2\theta_{13}$, the atmospheric neutrino sample from Super-Kamiokande, and the results from the beam experiments T2K and NO$\nu$A.} The presence of significant matter effects make DUNE sensitive to new neutrino interactions, which can modify neutrinos oscillation probabilities in a way that cannot be constrained by other experiments. The broadband character of the LBNF-DUNE neutrino beam allow one to ``see'' the oscillations and hence ultimately measure the $L/E$ ($L$ is the baseline and $E$ is the neutrino energy) behavior of the oscillation probabilities in a way that is outside the capabilities of off-axis experimental setups and with better control of systematics than what can be expected of high-statistics measurements of atmospheric neutrinos. Measurements of the oscillation probabilities as a function of $L/E$ performed within the same experimental setup are, for example, sensitive to new oscillation frequencies -- and hence new neutrino mass eigenstates -- and provide excellent tests of Lorentz-invariance in the neutrino sector.\footnote{$L/E$ is proportional to the neutrino proper time. Lorentz-invariance dictates that oscillation probabilities, once matter effects are accounted for, only depend on $L/E$, not on $L$ or $E$ independently. This is true for a large class of phenomena, including allowing for the possibility that the neutrinos decay.} Finally, DUNE energies are high enough that one can start to more seriously explore the dominant $\nu_{\mu}\to\nu_{\tau}$ oscillations via the charged-current production and subsequent detection of $\tau$-leptons. 

DUNE may also reveal that the three-flavor paradigm provides a complete description of the neutrino oscillation phenomenon. In this case, the impact of DUNE, as far as neutrino oscillation physics is concerned, can be quantified mostly via (i) precision measurements of the neutrino oscillation parameters and (ii) information on CP-invariance in the lepton sector. We comment on those in turn in the section below.


\subsection{Fermion Flavor Physics: Masses, Mixing Angles 
   and CP-odd Phases}

The patterns defined by the fermion masses and mixing parameters have been the subject of intense theoretical activity for the last several decades. The values of masses and mixing parameters, and potential relations among them, may contain invaluable information for physics beyond the standard model and may reveal more fundamental structures and symmetries. The discovery of neutrino masses and lepton mixing provided more and different information that is still being deciphered. Progress depends on how well masses and mixing parameters are known, and one can define, in a mostly model-independent way, useful goals and guidelines. 

Grand unified theories posit that quarks and leptons are different manifestations of the same fundamental entities so their masses and mixing parameters are related. While it is very clear that the CKM and PMNS matrices are very different, they may come from the same seed processed in different ways. Different models make different predictions but, in order to compare different possibilities, it is important that lepton mixing parameters be known as precisely as quark mixing parameters. Currently, the precision with which quark mixing parameters are known \cite{Tanabashi:2018oca} varies from 0.2\% (for $V_{us}$) to 5\% (for $V_{ub}$). The unitarity-triangle phase $\gamma$ (or $\phi_3$) is known at the 10\% level. Future Belle II data are expected to reduce this uncertainty to one or two percent \cite{Kou:2018nap}. These naively indicate that equal-footing comparisons between quark and lepton mixing require that the mixing angles be determined at the few percent level while $\deltacp$ should be measured at the 10\% level or better.

There are other well-motivated scenarios that relate the values of the different lepton mixing parameters in such a way that knowledge of a subset of parameters is enough to determine the entire set. These relations can often be expressed as mathematical constrains of the form:
\begin{equation}
f(\theta_{12},\theta_{13}, \theta_{23}, \deltacp)=0~,
\end{equation}
where $f$ is some model-dependent function. The ability to test these relations is limited by how well the different mixing parameters -- sometimes all of them -- are constrained. Optimal power requires all mixing parameters to be known equally well. Right now, $\theta_{23}$ is the least well measured mixing parameter other than the CP-odd phase $\deltacp$, which is virtually unconstrained. Improving, very significantly, the uncertainty on both of these is among the neutrino-oscillation goals of DUNE. Note that sometimes these relations among mixing parameters are guided by the physics responsible for nonzero neutrino masses and may include the mass-squared differences (or even the masses themselves).

A concrete example was discussed in Ref.~\cite{Antusch:2007rk} (for many other examples and details, see, for example Ref.~\cite{Ballett:2013wya}). For a large subclass of phenomenological models aimed at explaining the structure of the neutrino mixing matrix, one can derive the following relation (in the limit $\theta_{13}\ll 1$):
$$
\sin\theta_{12}-\sin\theta_{13}\tan\theta_{23}\cos\deltacp = A,
$$
where $A$ is a parameter that characterizes the model (e.g., $A=1/\sqrt{2}, 1/\sqrt{3}, 0.22$, etc), i.e., different models make different quantitative predictions for $A$.  While $\sin\theta_{12}$ is rather well constrained experimentally, the uncertainty in $\tan\theta_{23}$ and $\deltacp$ -- we currently only suspect that $\cos\deltacp\le0$ and, at the three sigma level, $\tan\theta_{23}\sim 1.1\pm0.2$ --  practically prevents one from testing whether the sum rule is obeyed for most values of $A$. Indeed, it is challenging to use the sum rule to, for example, predict the value of $\cos\deltacp$ because of the large current error on $\tan\theta_{23}$. 

The neutrino mass ordering also contains invaluable clues concerning the pattern of fermion masses and mixing matrices. If the neutrino mass ordering is ``normal,'' the pattern of neutrino masses may mirror that of the charged-fermions: $m_{\rm lightest}\ll m_{\rm middle}\ll m_{\rm largest}$, barring the possibility, which cannot be tested in oscillation experiments, that $m_1\sim m_2$. If, however, the mass ordering were inverted, we would learn that at least the two heavier neutrinos are almost degenerate in mass. No other matter particles with nonzero masses are quasi-degenerate; quasi-degenerate neutrino masses would inevitably be interpreted as evidence of an internal symmetry that lurks deep inside the neutrino sector and would invite vigorous new research efforts to tease out the nature of this new symmetry. 

Even within the three-flavor paradigm, the CP-odd Dirac phase $\deltacp$ is a new source of CP-invariance violation. Indeed, if the neutrinos are Majorana fermions, the standard model accommodates at most five independent CP-odd parameters. Three of these -- the majority -- ``live'' in the neutrino sector and one of them can only be probed, at least for the foreseeable future, in neutrino oscillations. If we are to ever understand how and why nature chooses to distinguish matter from antimatter, we will need to explore, in as much detail as possible, CP-violation in the neutrino sector.

\subsection{Impacts of DUNE for other Experimental Programs}



The information on neutrino properties obtained with DUNE data will also serve as invaluable input for other experiments in fundamental physics, including those beyond the realm of neutrino properties. We highlight some of these here. 

Long-baseline neutrino oscillation experiments are sensitive to the neutrino mass ordering via matter effects. Information on the mass ordering will also be obtained in atmospheric neutrino experiments and by looking for the $\Delta m^2_{21}$--$\Delta m^2_{31}$ interference in reactor neutrino oscillations in vacuum. Given the importance of this measurement, it is critical to have multiple techniques to corroborate the findings. As DUNE will be able to achieve a 5$\sigma$ determination of the ordering in a very controlled environment, this input will allow the study of subdominant effects in atmospheric neutrino oscillations, which depend on the Earth matter profile, and in supernova neutrinos.
%

The predictions for the decay rate of neutrinoless double beta decay critically depend on the neutrino mass ordering, via the effective Majorana mass parameter $m_{\beta \beta}$. If DUNE determines that the neutrino mass ordering is inverted, $m_{\beta \beta}$ is predicted to be bigger than $15~\mathrm{meV}$, within reach of the next generation of neutrinoless double beta decay experiments. Further conclusions could be obtained depending on future experimental results. For concreteness, let us first assume that the ordering is established to be inverted and consider a few relevant possibilities. (i) If  $|m_{\beta \beta}| \geq 15~\mathrm{meV}$ is measured, one would conclude that neutrinos are Majorana particles and that Majorana neutrino exchange is, most likely, the dominant mechanism behind neutrinoless double-beta decay. In principle, if a very precise measurement of the masses is derived from, for example, cosmic surveys and neutrino oscillation experiments, these data combined with a very accurate determination of $m_{\beta \beta}$ might allow one to search for CP-violating effects due to the Majorana phases. (ii) If, on the other hand, $m_{\beta \beta}$ is experimentally constrained to be smaller than  $15~\mathrm{meV}$, the simplest conclusion would be that neutrinos are Dirac particle unless a cancellation with other sources of lepton-number violation suppresses the decay rate of neutrinoless double beta decay. It would be critical to test this second hypothesis by looking for new particles and interactions which could provide sizable contributions to neutrinoless double beta decay. Second, let us consider the scenario in which DUNE establishes that the ordering is normal, as first hints from current neutrino data seem to indicate. In this case, expectations for $m_{\beta \beta}$ range from the current upper bounds to exactly zero. Information from cosmic surveys on the sum of neutrino masses, combined with data from DUNE, would help evaluate whether $m_{\beta \beta}$ is just around the corner or whether it might be severely suppressed. In the latter case, vigorous research towards multi-ton-scale ultralow-background neutrinoless double beta decay experiments will be required.  

%

Neutrinos have a strong impact on the evolution of the universe as their presence suppresses the growth of cosmological structures such as galaxies and clusters of galaxies at the small scales. This is due to the fact that, being light, they free-streamed from high-density to low-density regions, weakening the effects of the gravitational pull of high-density regions. The effect is greater the larger the neutrino mass. As it is a gravitational effect, it does not depend on the flavor and the relevant parameter is, given current and future expected sensitivities, the sum $\Sigma_i m_i$.  If DUNE establishes that the ordering is inverted, this implies that $\Sigma_i m_i \geq 0.1$~eV, while for normal ordering the sum can be as low as 0.06~eV. Future cosmological observations claim to be able to distinguish these two possibilities, under the assumption of the standard cosmological model. A precise measurement from cosmology would allow an accurate determination of the values of neutrino masses, with implications for neutrinoless double beta decay as discussed above.
There is also the possibility that incompatibilities are observed. For instance, if DUNE finds that the ordering is inverted and cosmological observations constrain $ \Sigma_i m_i < 0.1$~eV, one would have to conclude that there are new cosmological or particle physics effects which reduce the impact of neutrino masses in the formation of large scale structures or which counter them.

\subsection{Neutrino Masses, CP-violation and Leptogenesis}


The information which can be obtained in neutrino experiments, in particular DUNE, is essential to understand the origin of neutrino masses and possibly of the baryon asymmetry of the universe. The latter can be explained in the context of neutrino mass models, invoking the leptogenesis mechanism~\cite{Fukugita:1986hr}. The simplest extension of the Standard Model for neutrino masses requires right-handed (RH) neutrinos, which are singlets with respect to the Standard Model gauge group. They can couple to the Higgs doublet and the leptonic doublet via Yukawa couplings. Dirac masses arise for neutrinos as they do for all the other known fermions. This mechanism, although minimal, requires the promotion of the lepton-number symmetry from an accidental to a fundamental one and does not provide any insight on the smallness of neutrino masses or a rationale for the very different leptonic and quark mixing matrices.


If lepton-number is not imposed as a fundamental symmetry, Majorana masses for the RH neutrinos are also allowed and their magnitudes are unrelated to the scale of electroweak symmetry breaking. Once the Higgs gets a vacuum expectation value, both the Majorana and Dirac mass terms need to be included. If the RH-neutrino Majorana masses are much larger than the Dirac masses, this leads to small Majorana masses for the mostly-active neutrinos (those in the lepton-doublets) that manifest themselves via the Weinberg operator. This is the so-called seesaw mechanism and a strong suppression, without requiring very small Yukawa couplings, can be obtained if the RH neutrino masses are much heavier than the weak scale. 

Models for nonzero neutrino masses, including the seesaw models, offer an explanation of the baryon asymmetry of the universe via the leptogenesis mechanism. 
This problem is one of the most compelling questions in cosmology. The baryon asymmetry of the universe has been measured precisely by Planck~\cite{Aghanim:2018eyx}
\begin{equation}
Y_B^{\mathrm{CMB}}\simeq (8.67\pm 0.09) \times 10^{-10}\,,
\end{equation}
where $Y_B$ is the baryon to photon ratio at recombination. These results are in good agreement with data on big bang nucleosynthesis. Assuming that the universe initially had the same amount of baryons and antibaryons,\footnote{A period of inflation in the early universe implies that this assumption is effectively unavoidable.} the baryon asymmetry can be generated dynamically if the Sakharov conditions~\cite{Sakharov:1967dj} are satisfied: lepton or baryon number violation, for instance in presence of RH neutrino Majorana masses, C and CP violation, and out-of-equilibrium dynamics, satisfied by the expansion of the universe. 

We restrict the discussion here to high-energy, type-I seesaw models in which RH neutrinos are introduced with very heavy Majorana masses. 
These models can satisfy all of the Sakharov conditions because of the Majorana nature of the RH neutrinos and of the presence of complex Yukawa couplings. The basic picture is the following. In the early universe, RH neutrinos were in thermal equilibrium for large temperatures. Once the temperature dropped below their mass, the bath does not have sufficient energy to keep them in equilibrium and they decouple, decaying into leptons and Higgs bosons. If there is CP violation, the decays of this channel and of the conjugated one can proceed with different rates, controlled by the CP-violating phases in the Yukawa couplings. 
This asymmetry is partially washed out by inverse processes and the remaining lepton asymmetry is converted into a baryon asymmetry later on by non-perturbative \dword{sm} effects. 

The question of whether and how the CP-violation involved in leptogenesis and that observable
 in DUNE and other long-baseline experiments are related has been debated extensively in the
 literature. Restricting the discussion to high-energy seesaw models only, for simplicity, the link is
 provided by the complex Yukawa couplings which control on one side the baryon asymmetry and
 on the other neutrino masses and consequently the PMNS matrix which diagonalizes them. In
 general, relationships are rather complex and very indirect because the high-energy theory contains
  more parameters -- including more CP-odd phases -- than are measurable at low-energy experiments.
  In a completely model-independent way, it is not possible to draw a direct link between the two.
However, in many models that have a reduced number of parameters, for instance because of flavor symmetries,
 experimentally accessible CP-odd phases can be directly connected to the baryon asymmetry generated via leptogenesis.

 Even without resorting to a restriction of the number of parameters, rather general models present such connection
if in the Early universe the thermal bath distinguished between charged lepton flavors in the so-called flavored leptogenesis.
 It is possible to show that, in these circumstances, the PMNS mixing matrix and specifically the CP-violating phase $\deltacp$ does explicitly contribute to the CP asymmetry, and consequently the baryon asymmetry, and can even generate enough CP-violation to reproduce the observed baryon
 asymmetry. This is a highly non-trivial statement since its CP-violating effects are suppressed by $\theta_{13}$ and hence enough early-universe CP-violation relies crucially on the relatively large observed value of $\theta_{13}$.

The consensus in the community is that one should be able to conclude that, generically, the observation of lepton-number violation (e.g., neutrinoless double beta decay) combined with that of  CP-violation in long-baseline neutrino oscillation experiments (or, possibly, neutrinoless double beta decay) constitutes strong circumstantial evidence -- albeit not a proof -- of the leptogenesis mechanism as the origin of the baryon asymmetry of the universe.

\section{Nucleon Decay and $\Delta$B=2 Physics}
\label{sec:landscape-ndk}
Are protons stable?  Few questions within elementary 
particle physics can be posed as simply and at the same time 
have implications as immediate.  In more general terms, the 
apparent stability of protons suggests that baryon number 
is conserved in nature, although no known symmetry 
requires it to be so.  Indeed, baryon number conservation is 
implicit in the formulation of the \dword{sm} Lagrangian, and 
thus observation of \dword{bnv} processes such 
as nucleon decay or neutron-antineutron oscillation 
would be evidence for physics beyond the \dword{sm}.\footnote{Non-perturbative 
effects that involve tunneling between vacua with differing baryon number 
do allow for \dword{bnv} processes within the \dword{sm}, but at rates many orders of 
magnitude below directly observable levels (see, e.g., Ref.~\cite{Nath:2006ut}).}
On the other hand, continued non-observation of \dword{bnv} processes will 
demand an answer to what new symmetry is at play that forbids 
them.
 
Especially compelling is that the observation of \dword{bnv} processes 
could be the harbinger for \dwords{gut}, in which strong, weak and 
electromagnetic forces are unified.  Numerous \dword{gut} models 
have been proposed, each with distinct features.  Yet, \dword{bnv} processes 
are expected on general grounds, and it is a feature of many models 
that nucleon decay channels can proceed at experimentally 
accessible rates (see, e.g., Refs.~\cite{Nath:2006ut,Babu:2013jba} 
and references therein).

The theoretical literature on nucleon decay, and \dword{bnv} processes in general,  
is vast, and has been well summarized in recent 
reviews~\cite{Nath:2006ut,Babu:2013jba}.  
It may be sufficient here to simply note that the 
theoretical motivations for baryon number non-conservation give strong 
arguments for the discovery potential of experimental searches, 
and that the existing array of null results from highly sensitive experiments  
provides hard constraints that models of new physics must abide by.
Some additional theoretical context is provided in Chapter~\ref{ch:nonaccel}.
The remainder of the discussion in this section focuses on the experimental 
landscape so as to illustrate the scientific opportunities for DUNE in \dword{bnv} physics.

\subsection{Experimental Considerations for Nucleon Decay Searches}
\label{subsec:landscape-ndk-expt}

The articulation of early \dword{gut} ideas led to the development of large-scale 
detectors located deep underground dedicated toward the search for proton and \dword{bnv}
bound-neutron decay.  Illustrating the present context, the limits on a subset 
of possible nucleon decay modes, from a succession of sensitive experimental searches, 
are plotted in Fig.~\ref{fig:nucleondecay_exptsummary}.
\begin{dunefigure}
[Summary of nucleon decay experimental limits]
{fig:nucleondecay_exptsummary}
{Summary of nucleon decay experimental lifetime limits from past and currently 
running experiments for decays to anti-lepton plus meson final states.  Recently 
reported improvements in limits~\cite{TheSuper-Kamiokande:2017tit} are highlighted, 
indicating the ongoing nature of experimental effort in this area.
The limits shown are 90\% \dword{cl} lower limits on the partial lifetimes, 
$\tau/B$, where $\tau$ is the total mean life and $B$ is the branching fraction. 
Updated from~\cite{Babu:2013jba}.}
\includegraphics[width=0.8\textwidth]{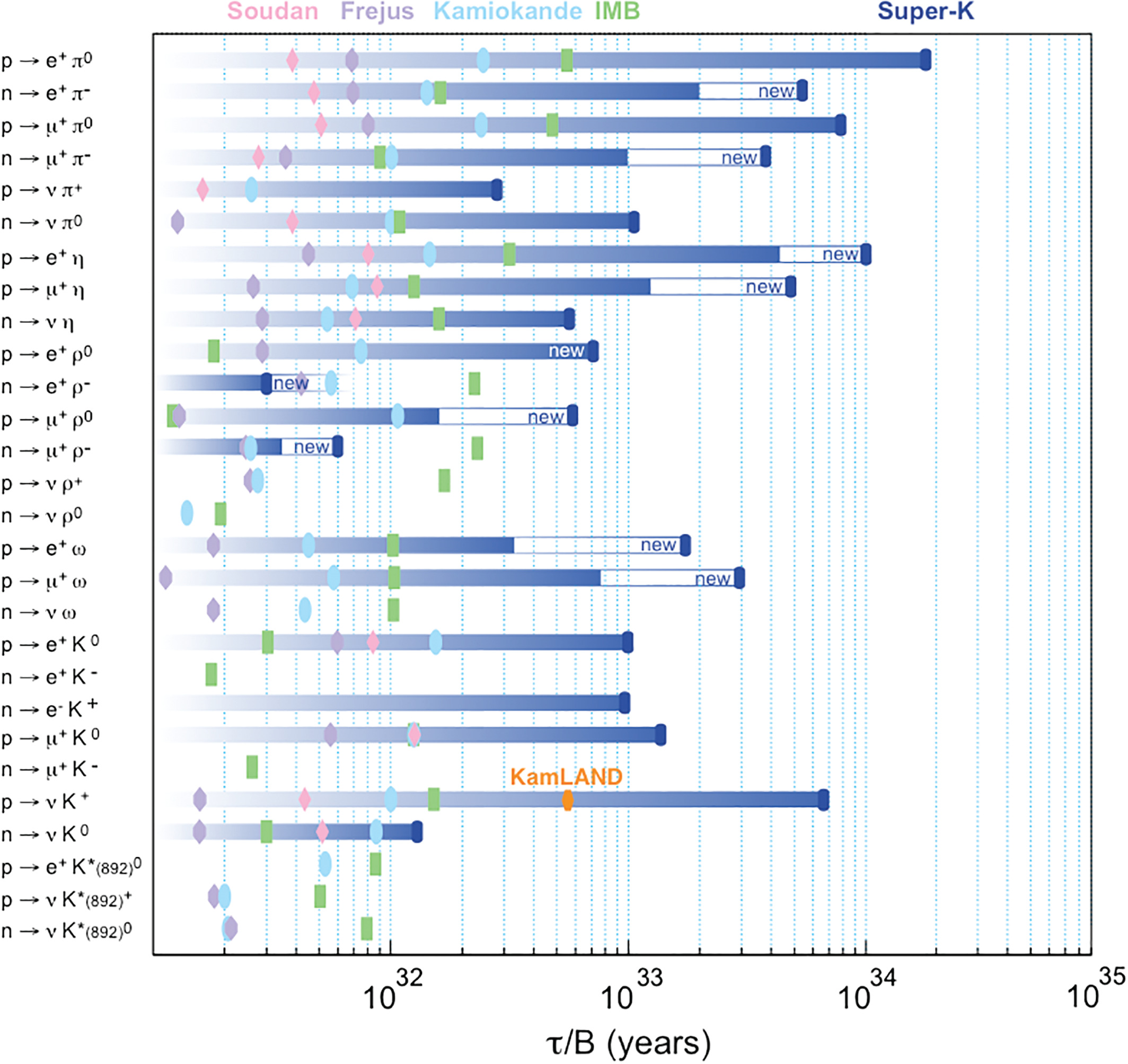}
\end{dunefigure}

Particularly sensitive limits have been obtained with water-based Cherenkov ring 
imaging detectors, most notably \superk.  The strengths of this approach include 
the cost-effectiveness of utilizing large volumes of water (\SI{22.5}{\kt} fiducial 
mass in the case of \superk) as a source of nucleons and 
capabilities for particle identification, timing, energy and direction resolution.
The technology is scalable to even larger masses, as in the proposed 
\hyperk~\cite{Abe:2018uyc} experiment, with a \SI{187}{\kt} fiducial mass in its  
single-tank configuration.  The combination of deep underground location with active 
shielding enables rejection of backgrounds from atmospheric muons.  As a result, the 
dominant backgrounds are due to interactions of atmospheric neutrinos, which are suppressed 
by event selection on the distinctive kinematic and signal timing features of
the various nucleon decay channels. 

With published results~(see, e.g., 
Refs.~\cite{Abe:2014mwa,Miura:2016krn,TheSuper-Kamiokande:2017tit})
based on exposures up to \SI{0.32}{\Mtyr}, \superk 
nucleon decay branching ratio sensitivity continues to increase linearly 
with exposure for many channels where background estimates are at the 
one-per-\si{\Mtyr} level.  However, as exposure increases further, the 
rate of improvement will be diminished as backgrounds enter.    
Candidate events are starting to appear~\cite{TheSuper-Kamiokande:2017tit} 
in channels where the estimated background rate exceeds this level.  

With a fiducial mass of \fdfiducialmass{}, DUNE can capitalize on the 
potential for discovery of nucleon decay in channels where backgrounds 
can be reduced below the one-per-\si{\Mtyr} level thanks to the 
excellent imaging, calorimetric and particle identification capabilities 
of the LArTPC for events with 200 to \SI{1000}{\MeV} of deposited energy.  
In a background-free analysis, sensitivity to channels with partial lifetimes 
in the range of $10^{33}$ to a few times $10^{34}$ \si{years} may be achievable, 
depending on event selection efficiency.  The limiting factor for DUNE 
is likely to be the combined impact of nucleon Fermi motion and final state 
interactions of decay hadrons as they escape the argon nucleus.  Detailed 
analyses carried out for several prominent nucleon decay channels are 
described in Chapter~\ref{ch:nonaccel}.

Should nucleon decays occur at rates not far beyond current best limits, 
as predicted in numerous \dword{gut} models, a handful of candidate 
events could be observed by DUNE in a given decay mode.  
Even just one or two candidate events may be 
sufficient on their own to indicate evidence for nucleon decay, 
or provide confirmation 
for an excess above background observed in one of the contemporaneous 
large water or liquid scintillator experiments, e.g., \dword{hyperk}~\cite{Abe:2018uyc}
and JUNO~\cite{Djurcic:2015vqa,An:2015jdp} respectively.

%
\section{Low-Energy Neutrinos from Supernovae and Other Sources}
\label{sec:landscape-snb}


The burst of neutrinos from the celebrated core-collapse supernova 1987A in the Large Magellanic Cloud, about
50~kpc from Earth, heralded the era of extragalactic neutrino
astronomy.  The few dozen recorded $\bar{\nu}_e$ events
have confirmed the basic physical
picture of core collapse and yielded constraints on a wide range of new
physics~\cite{Schramm:1990pf, Vissani:2014doa}.   This sample has nourished physicists and
astrophysicists for many years, but has
by now been thoroughly picked over.  The community anticipates a
much more sumptuous feast of data when the next nearby star collapses.

Core-collapse supernovae within a few hundred kiloparsecs of Earth --
within our own galaxy and nearby -- are quite rare on a human
timescale.  They are expected once every few decades in the Milky Way
(within about 20~kpc), and with a similar rate in Andromeda, about
700~kpc away.  However core collapses should be common enough to have
a reasonable chance of occurring during the few-decade long lifetime
of a typical large-scale neutrino detector.  The rarity of these
spectacular events makes it all the more critical for the community to
be prepared to capture every last bit of information from them.

The information in a supernova neutrino burst available in principle
to be gathered by experimentalists is the \textit{flavor, energy and
  time structure} of several-tens-of-second-long, all-flavor,
few-tens-of-MeV neutrino burst~\cite{Mirizzi:2015eza, Horiuchi:2017sku}.  Imprinted on
the neutrino spectrum as a function of time is information about the
progenitor, the collapse, the explosion, and the remnant, as well as
information about neutrino parameters and potentially exotic new
physics.  Neutrino energies and flavor content of the burst can be
measured only imperfectly, due to intrinsic nature of the weak
interactions of neutrinos with matter, as well as due to imperfect
detection resolution in any real detector.  For example, supernova
burst energies are below charged-current threshold for $\nu_\mu$,
$\nu_\tau$, $\bar{\nu}_\mu$ and $\bar{\nu}_{\tau}$ (collectively
$\nu_x$), which represent two-thirds of the flux; so these flavors are
accessible only via neutral-current interactions, which tend to have
low cross sections and indistinct detector signatures. These issues make a
comprehensive unfolding of neutrino flavor, time and energy structure
from the observed interactions a challenging problem.

Much has occurred since 1987, both for experimental and theoretical
aspects of supernova neutrino detection.
There has been huge progress in the modeling of supernova explosions,
and there have been many new theoretical insights about
neutrino oscillation and exotic collective effects that may occur in
the supernova environment.    Experimentally,
worldwide detection capabilities have increased enormously, such that
we now expect several thousands of events from a core collapse at the center
of the Galaxy.

\subsection{Current Experimental Landscape}
At the time of this writing, Super-Kamiokande is the leading supernova
neutrino detector; it expects $\sim$8000 events at 10~kpc.  As for
the 1987A sample, these will be primarily $\bar{\nu}_e$ flavor via
inverse beta decay (IBD) on free protons.  Super-K will soon be
enhanced with the addition of gadolinium, which will aid in IBD
tagging.  IceCube is another water detector, with a different kind of
supernova neutrino sensitivity -- it cannot reconstruct individual
neutrino events, given that any given interaction in the ice rarely
leads to more than one photoelectron detected.  However it can measure
the overall supernova neutrino ``light curve'' as a glow of photons
over background counts.  Scintillator detectors, made of hydrocarbon,
also have high IBD rates.  There are several kton-scale scintillator
detectors online currently: these are KamLAND, LVD, and Borexino.
There is one small lead-based detector, HALO.  Some surface or
near-surface detectors will also usefully record counts even in the
presence of significant cosmogenic background: these include NOvA,
Daya Bay, and MicroBooNE.

In the world's current supernova neutrino flavor sensitivity
portfolio~\cite{Scholberg:2012id, Mirizzi:2015eza}, the sensitivity is primarily to electron antineutrino
flavor, via IBD. There is only minor sensitivity to the $\nu_e$
component of the flux, which carries with it particularly interesting
information content of the burst (e.g., neutronization burst neutrinos
are created primarily as $\nu_e$).  While there is some $\nu_e$
sensitivity in other detectors via elastic scattering on electrons and
via subdominant channels on nuclei, statistics are relatively small,
and it can be difficult to disentangle the flavor content.
Neutral-current channels are also of particular interest, given their
sensitivity to the entire supernova flux; the only way to access the
$\nu_x$ component is via NC.  NC channels are subdominant in large
neutrino detectors, and typically difficult to tag, although
scintillator has some sensitivity via NC excitation of $^{12}$C as
well as elastic scattering on protons.  Dark matter detectors have
access to the entire supernova flux via NC coherent elastic
neutrino-nucleus scattering on nuclei, with statistics at the level of
 of $\sim$10 events per ton at 10~kpc.

\subsection{Projected Landscape in the DUNE Era}
The next generation of supernova neutrino detectors, in the era of
DUNE, will be dominated by Hyper-Kamiokande, JUNO and DUNE.  Hyper-K
and JUNO are sensitive primarily to $\bar{\nu}_e$, and will have
potentially enormous statistics.  The next-generation long-string
water detectors, IceCube and KM3Net, will bring their timing
strengths.   New tens-of-ton scale
noble liquid detectors such as DARWIN will bring new full-flux NC
sensitivity. 
DUNE will bring unique $\nu_e$ sensitivity: it will offer
a new opportunity to measure the $\nu_e$ content of the burst with
high statistics and good event reconstruction.

The past decade has also brought rapid evolution of
\textit{multi-messenger astronomy}.  With the advent of gravitational
waves detection, and high-energy extragalactic neutrino detection in
IceCube, a broad community of physicists and astronomers are now
collaborating to extract maximum information from observation in a
huge range of electromagnetic wavelengths, neutrinos, charged particles
and gravitational waves.  This collaboration resulted in the
spectacular multimessenger observation of a kilonova~\cite{kilonova}.  The
next core-collapse supernova will be a similar multimessenger
extravaganza.  Worldwide neutrino detectors are currently participants
in SNEWS, the SuperNova Early Warning System~\cite{snews}, which will be
upgraded to have enhanced capabilities over the next few
years.  Information from DUNE will enhance the SNEWS
network's reach.

Neutrino pointing information is vital for prompt multi-messenger
capabilities.  Only some supernova neutrino detectors have the ability
to point back to the source of neutrinos.  Imaging water Cherenkov
detectors like Super-K can do well at this, via directional
reconstruction of neutrino-electron elastic scattering events. However other detectors
lack pointing ability, due to intrinsic quasi-isotropy of the neutrino
interactions, combined with lack of detector sensitivity to
final-state directionality.  Like Super-K, DUNE is capable of pointing
to the supernova via its good tracking ability.

\subsection{The Role of DUNE}
Supernova neutrino detection is more of a collaborative than a
competitive game.  The more information gathered by detectors
worldwide, the more extensive the knowledge to be gained; the whole is
more than the sum of the parts.  The flavor sensitivity of DUNE is
highly complementary to that of the other detectors, and will bring
critical information for reconstruction of the entire burst's flavor and
spectral content as a function of time~\cite{Ankowski:2016lab}.

\subsection{Beyond Core Collapse}
While a core-collapse burst is a known source of a
low-energy ($<$100 MeV) neutrinos, there are other potential
interesting sources of neutrinos in this energy range.  Nearby
thermonuclear or pair instability supernova events may create bursts
as well, although they are expected to be fainter in neutrinos than
core-collapse supernovae.  Mergers of neutron stars and black holes
will be low-energy neutrino sources, although the rate of these nearby
enough to detect will be small.  There are also interesting
steady-state sources of low-energy neutrinos -- in particular, there
may still be useful oscillation and solar physics information to
extract via measurement of the solar neutrino flux. DUNE will have the
unique capability of measuring solar neutrino energies event by event
with the $\nu_e$CC interactions with large statistics, in contrast to
other detectors primarily make use of recoil spectra.  The technical
challenge for solar neutrinos is overcoming radiological and
cosmogenic backgrounds, although preliminary studies are promising.
The diffuse supernova neutrino background neutrinos are another target
which have a bit higher energy, but which are much more challenging due to very low
event rate.  There may also be surprises in store for us, both from burst
and steady-state signals, enabled by unique DUNE liquid argon tracking
technology.


\section{Beyond-SM Searches}
\label{sec:landscape-bsm}

With the advent of a new generation of neutrino experiments which leverage high-intensity neutrino beams for precision measurements, 
the opportunity arises to explore in depth physics topics Beyond the Standard neutrino-related physics. 
Given that the realm of \dword{bsm} physics  has been mostly sought at high-energy regimes at colliders, 
such as the LHC at CERN, the exploration of \dword{bsm} physics in neutrino experiments will enable complementary 
measurements at the energy regimes that balance those  of the LHC. 
This, furthermore, is  in concert with new ideas for high-intensity beams for fixed target and beam-dump experiments 
world-wide, e.g., those proposed at CERN~\cite{Beacham:2019nyx}.

The combination of the high intensity proton beam facilities and massive detectors for precision neutrino oscillation parameter measurements and for CP violation phase measurements will help make \dword{bsm} physics reachable even in low energy regimes in the accelerator based experiments.
Large mass detectors with highly precise tracking and energy measurements, excellent timing resolution, and low energy thresholds will enable the searches for \dword{bsm} phenomena from cosmogenic origin, as well.
Therefore, it can be anticipated that \dword{bsm} physics topics studied with the next-generation neutrino 
experiments may have a large impact in the foreseeable future, 
as the precision of the neutrino oscillation parameter and CPV measurements continues to improve.
A recent review of the current landscape of \dword{bsm} theory in neutrino experiments in two selected areas of the \dword{bsm} topics -- dark matter and neutrino related \dword{bsm} -- has been recently reported in~\cite{Arguelles:2019xgp}.

The DUNE experiment has two important assets that will play a significant role in 
future searches for \dword{bsm} physics.
The unique combination of the high-intensity LBNF proton beams with a highly-capable precision
 DUNE Near Detector (ND), and massive liquid argon time-projection chamber (LArTPC) far detector modules at a \SI{1300}{\km} baseline (FD), enables a variety of opportunities for \dword{bsm} physics, either novel or with unprecedented sensitivity.
The planned Near Detector can basically act as a stand alone experiment,
to catch long lived particles produced in the proton target beam dump. On the other hand the Far Detector 
will allow for precision measurements on oscillation parameters, and for measurements cosmogenic and
 non-accelerator related phenomena, e.g. the detection of dark matter particles in certain scenarios.

In this section we give a few examples of particle searches in New Physics scenarios than can be conducted with the DUNE experiment, for which the sensitivities are discussed in the next chapters of this volume. For those searches
for new particles in the `beam-dump' mode, i.e. for  
searches for long-lived particles that  pass through, or decay in, the Near Detector, a few scenarios have been
studied in detail, but it will be important in the near future to connect with the
Physics Beyond Collider study~\cite{Beacham:2019nyx} and compare the potential sensitivity of DUNE for these
benchmark scenarios,
especially for so called ``feebly interacting  particle'' sensitivity projections as made for 
potential new beam dump experiments for the next 10-15 years. DUNE is an already planned facility, which has the potential to cover interesting regions in the coupling/mass phase space
for dark photons, dark scalars and axion-like particles, for which the sensitivity has not been studied yet.
In addition the precision measurements of the oscillation phenomena will allow also to search for e.g. non-standard interactions, CPT violating effects as discussed before.

 
 \subsection{Search for low-mass dark matter}
 Various cosmological and astrophysical observations strongly support the existence of dark matter
(DM) representing 27\% of the mass-energy of the universe, but its nature and potential non
gravitational interactions with regular matter remain undetermined. The lack of evidence for
 weakly interacting massive particles (WIMP) at direct detection and the LHC experiments has
 resulted in a reconsideration of the WIMP paradigm. For instance, if dark matter has a mass
 which is much lighter than the electroweak scale (e.g., below GeV level), it motivates theories for
 dark matter candidates that interact with ordinary matter through a new vector portal mediator.
 High flux (neutrino) beam experiments,  have been shown to provide coverage of
DM+mediator parameter space which cannot be covered by either direct detection or collider
 experiments. In LBNF, low-mass dark matter may be produced through proton interactions in
the target, and can be detected in the ND through neutral current (NC)-like interactions either
 with electrons or nucleons in the detector material via elastic scattering. Since these experimental
 signatures are virtually identical to those of neutrinos, neutrinos are a significant background that
 can be suppressed using timing and kinematics of the final-state electron or nucleons in the ND.
Therefore, it is essential for the ND to be able to differentiate arrival time differences of the order
 a few ns or smaller, which determines the reachable range of the dark matter, and to measure
 precisely the kinematic parameters of the recoil electrons, such as the scattering angle and the
energy. These capabilities will enable DUNE's search for light dark matter to be competitive and
 complementary to other experiments at mass range below 1-2 GeV. 

The capability has recently been demonstrated in a dedicated search by MiniBooNE~\cite{Aguilar-Arevalo:2018wea,Aguilar-Arevalo:2017mqx}, which placed new limits on the well-motivated vector portal dark matter model~\cite{Pospelov:2007mp}, as shown in Figure~\ref{fig:MB-plot}.

\begin{dunefigure}[Results from the MiniBooNE-DM light dark matter search]{fig:MB-plot}
{Results from the MiniBooNE-DM search for light dark matter from Ref.~\cite{Aguilar-Arevalo:2018wea}
}
\includegraphics[width=0.7\textwidth,height=3.5in]{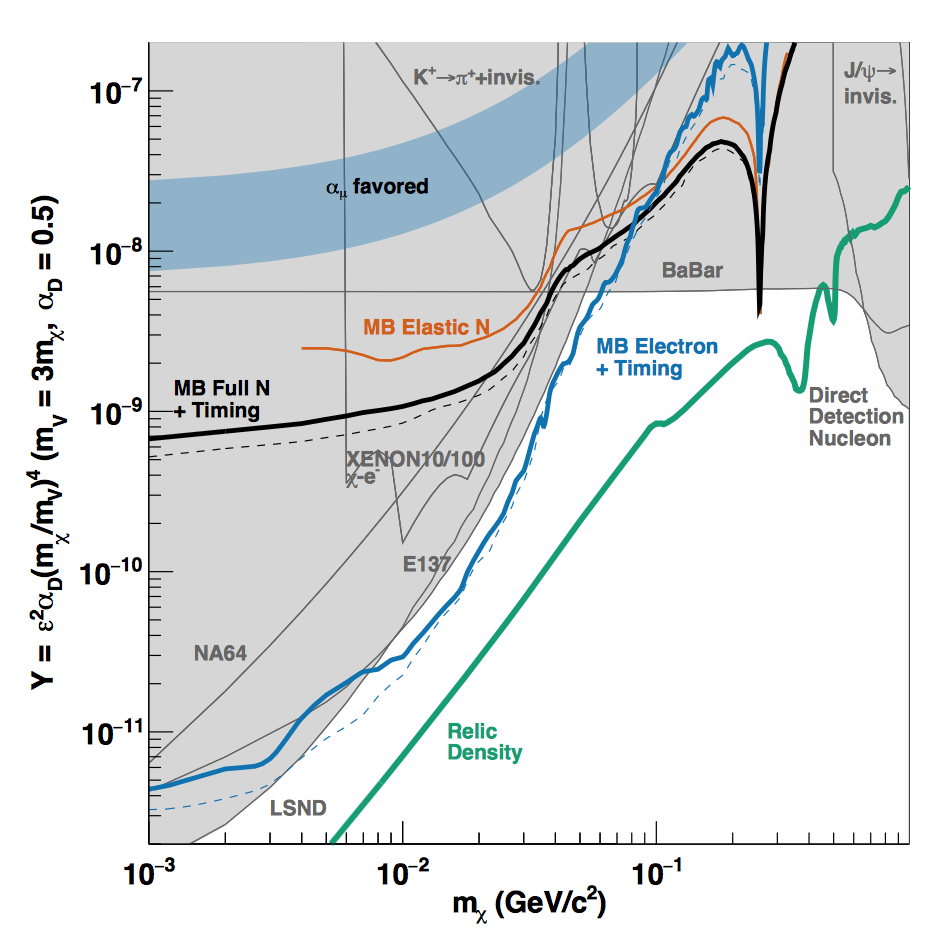}
\end{dunefigure}

More scenarios for \dword{dm} detection will become accessible for DUNE, due to its improved sensitivity
using LarTPC technology, and the large FD volume, These scenarios include boosted dark matter, 
produced in models with 
a multi-particle dark sector. Sensitivities of such scenarios will be 
examined further in Chapter~\ref{ch:bsm} of this volume.

\subsection{Sterile neutrino search}
 Experimental results in tension with the three-neutrino-flavor paradigm, which may be interpreted
 as mixing between the known active neutrinos and one or more sterile states, have led to a rich
 and diverse program of searches for oscillations into sterile neutrinos. DUNE will be sensitive over a
 broad range of values of the sterile neutrino mass splitting by looking for disappearance of charged
 current (CC) and NC interactions over the long distance separating the near and far detectors,
 as well as over the short baseline of the ND. 

The present lead in the search for sterile neutrinos, those which couple to standard neutrinos but not to the weak interaction, comes from disappearance experiments such as muon-neutrino accelerators and reactor anti-neutrino experiments, where unitarity is a necessary assumption. All the most precise measurements of the standard oscillation parameters have been made by disappearance experiments as shown in the left panel of Figure~\ref{fig:disappearance}. The \dword{lsnd} and \miniboone anomalies are expected to be elucidated by \microboone due to its unprecedented event reconstruction capabilities. After the recent measurement from MINOS+ and IceCube are combined with unitarity constraints (see e.g.~\cite{Parke:2015goa}), most of the favored parameter space to explain \dword{lsnd} and \miniboone, with a sterile neutrino, is now disfavored as shown in the right panel of Figure~\ref{fig:disappearance}. Addressing the apparent excess of electron events appearing in the muon-neutrino beam at \miniboone and \dword{lsnd} is also the main goal for the future SBN program at Fermilab, using for the first time near and far detectors with the same technology for this study. Furthermore  in the next years conclusive results will become available from very short baseline 
reactor experiments, which measure the rate of inverse beta decay 
as function of length to the reactor core. These aim to see
small modulations as function of distance, which could be caused
by sterile neutrinos. However given the ensemble of 
all present data so far, if the anomalies survive it seems to 
indicate that a (or a few) 'standard' sterile neutrino(s) 
hypothesis does not fit the data and the explanation may turn
out to be much more complex, in which case certainly the capabilities of the DUNE experiment will play an important role 
in unraveling the exact nature of the new phenomenon.

\begin{dunefigure}[Exclusion limits for muon neutrino disappearance to sterile species in a 3+1 model]{fig:disappearance}
{Left panel: Comparison of present exclusion limits from various experiments obtained through searches for disappearance of muon neutrinos into sterile species assuming a 3+1 model.  The Gariazzo et al. region represents a global fit to neutrino oscillation data~\cite{Gariazzo:2015rra}. 
    Right panel: The combined results of the disappearance measurements from \dword{minosplus}, \dword{dayabay}, and \dword{bugey}, compared to the appearance measurements from \dword{lsnd} and \miniboone.}
    \includegraphics[height=3.5in,width=0.43\textwidth]{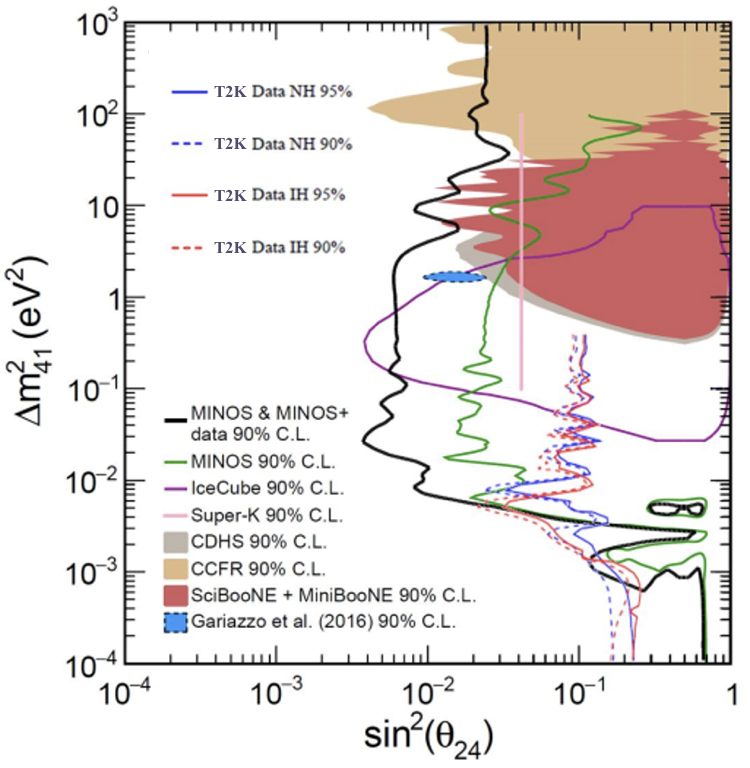}
    \includegraphics[height=3.7in,width=0.47\textwidth]{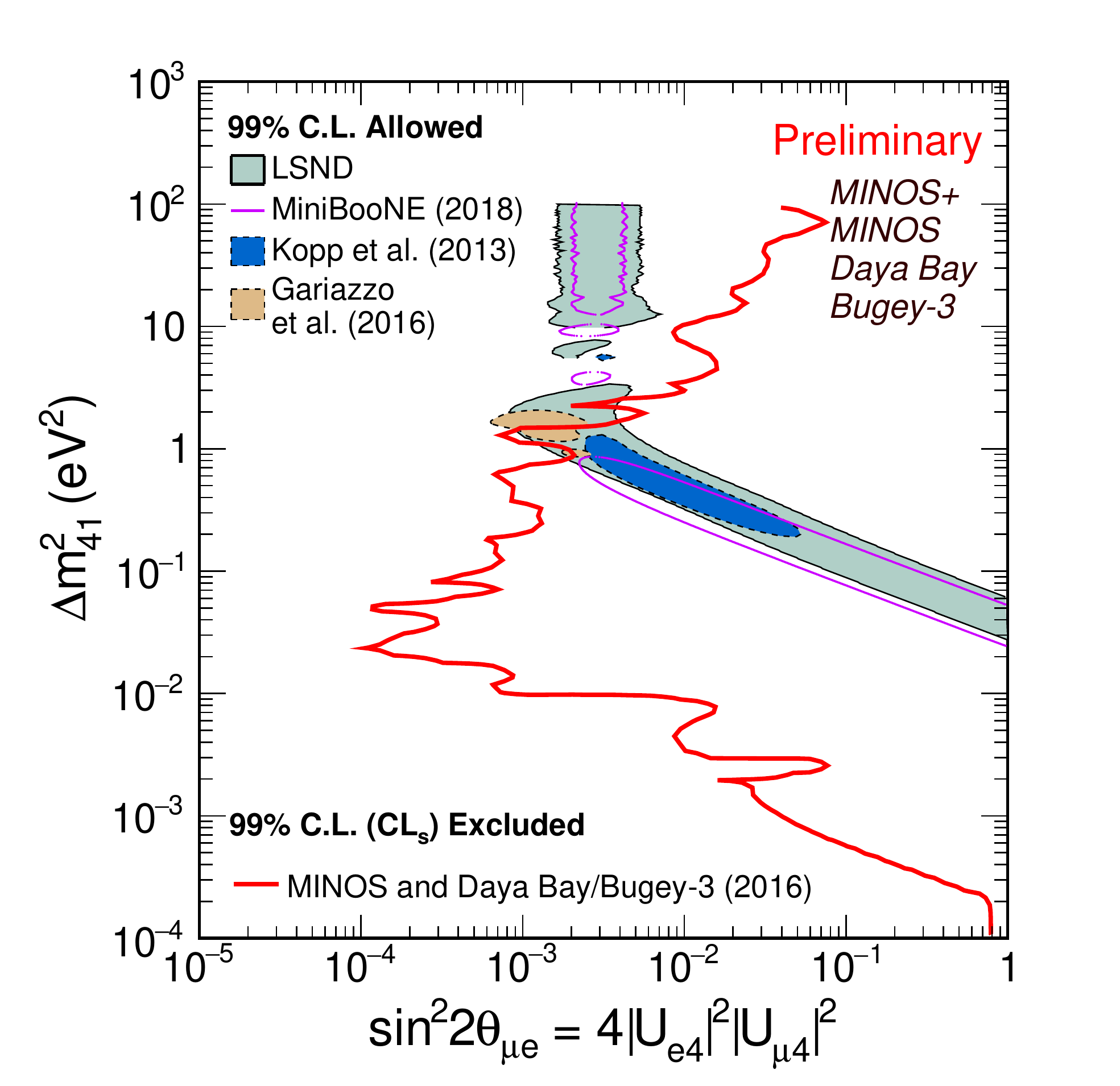}
\end{dunefigure}

\subsection{Neutrino tridents}
 Neutrino trident production is a rare weak process in which a neutrino, scattering off the Coulomb
 field of a heavy nucleus, generates a pair of charged leptons. The typical final state of a neutrino
 trident interaction contains two leptons of opposite charge. Measurements of muonic neutrino
 tridents were carried out at the CHARM-II, CCFR, and NuTeV experiments, and yielded results
 consistent with \dword{sm} predictions, but those measurements leave ample room for 
 potential searches for New Physics. As an example, a class of models that modify the trident cross
 section are those that contain an additional neutral gauge boson, $Z^\prime$, 
 that couples to neutrinos and charged leptons. This $Z^\prime$
 boson can be introduced by gauging an anomaly-free global symmetry
of the \dword{sm}, with a particular interesting case realized by gauging L$_{\mu}$--L$_{\tau}$~\cite{He:1990pn,He:1991qd}. Such a $Z^\prime$
 is not very tightly constrained and could address~\cite{Baek:2001kca,Harigaya:2013twa} the observed discrepancy between the Standard
Model prediction and measurements of the anomalous magnetic moment of the muon, (g--2)$_{\mu}$
 The DUNE ND offers an excellent environment to generate a sizable number of trident events,
  offering very promising prospects to both improve the above measurements, and to
 look for an excess of events above the \dword{sm} prediction, which would be an 
  indication of new physics.

  Another category of \dword{bsm} Physics
 models that can be probed through neutrino trident measurements are dark neutrino sectors. In
 these scenarios, SM neutrinos mix with heavier SM singlet fermions (dark neutrinos) with their
 own new interactions. Due to this mixing, neutrinos inherit some of this new interaction and may
 up-scatter to dark neutrinos. These heavy states in turn decay back to SM fermions, giving rise
 to trident signatures. These scenarios can explain the smallness of neutrino masses and possibly
 the MiniBooNE low energy excess of events, discussed above.

\subsection{Heavy neutral leptons}
 The DUNE ND can be used to search topologies of rare event interactions and decays that originate
 from very weakly-interacting long-lived particles, including heavy neutral leptons -- right-handed
 partners of the active neutrinos, vector, scalar, or axion portals to the hidden sector, and light
 supersymmetric particles. The high intensity of the NuMI source and the capability of production
 of charm mesons in the beam allow accessing a wide variety of lightweight long-lived, exotic,
 particles. Competitive sensitivity is expected for the case of searches for decay-in-flight of sub-GeV
 particles that are also candidates for dark matter, and may provide an explanation for leptogenesis
 in the case of charge-parity symmetry violation (CPV) indications. DUNE would probe the lighter
 particles of their hidden sector, which can only decay in SM particles in the form of pairs like $e^+e^-$ ,
 $\mu^+\mu^-$ , qq. The parameter space explored by the DUNE ND extends to the cosmologically relevant
region that is complementary to the LHC dark-matter searches through missing energy
 and mono-jets. 

A recent study on the present limits  and capabilities with future experiments for covering the coupling-mass phase space 
is shown in Figure \ref{fig:bc7_pbc_2}, taken from \cite{Beacham:2019nyx}. These future prospects include proposed experiments, such as SHiP, 
which would operate over the period of the next 10 to 15 years, hence the same period as for DUNE.  While a dedicated analysis of DUNE's sensitivity has not yet been carried out, the sensitivity from a previous study with the formerly-proposed LBNE Near Detector (shown as a dark-green dashed curve at low values of $m_N$) may give a representative indication.
The FCC curve corresponds to a study aimed much further in the future.

\begin{dunefigure}[Sensitivities to heavy neutral leptons]{fig:bc7_pbc_2}
    {Sensitivity to Heavy Neutral Leptons with coupling to the second lepton generation only.
    Current bounds (filled areas) and 10-15 years prospects for PBC projects (SHiP, MATHUSLA200, CODEX-b and FASER2) (dotted and solid lines).
    Projections for the formerly-proposed LBNE near detector with $5 \times 10^{21}$ protons on target (dark green dashed line starting 
    from lower left region of the plot) and FCC-ee with $10^{12}$ $Z^0$
    decays (light green dashed line at higher $m_N$ values) also shown .}
  \includegraphics[width=0.8\linewidth]{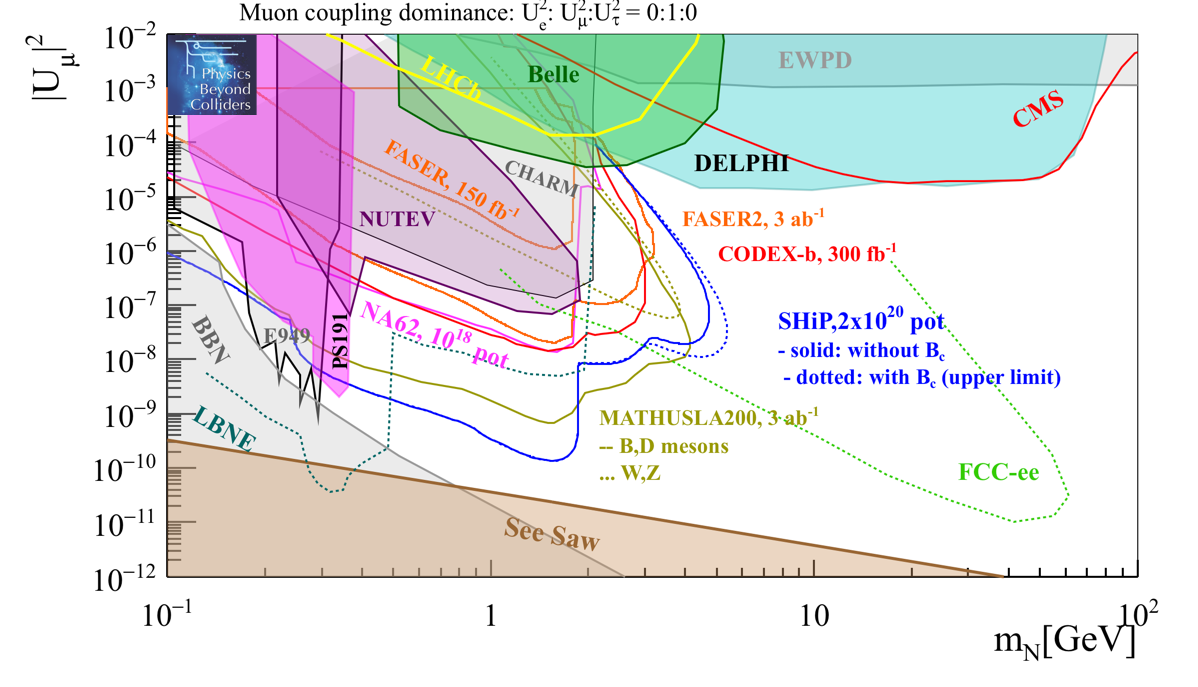}
\end{dunefigure}



\section{Other Scientific Opportunities }
\label{sec:landscape-othertop}
The high rate of charged-current muon-neutrino argon interactions occurring in 
the near detector will provide important data samples to understand better neutrino-argon 
interactions in the relevant energy range for the DUNE far detector. The next chapters
will give examples of scenarios where detailed understand of such interactions with 
precision measurements will have a significant impact on the physics reach for some
topics. Effects of final state interactions, event topology and kinematics, neutron production and more can be studied in detail with such large statistics 
data samples.

The collection of the expected statistics and the determination of the neutrino and
antineutrino fluxes to unprecedented precision would solve  two main limitations of past neutrino experiments. At the same time, we can then exploit the unique
properties of the neutrino probe for the study of fundamental interactions with a broad program of precision \dword{sm} measurements. These potential measurements
have not yet been studied in detail in this \dword{tdr}, as the capabilities
depend critically on the final design choice of the near detector, and this is still under discussion.

Neutrinos and anti-neutrinos are effective probes for investigating Electroweak 
physics. A precise determination of the weak mixing angle (sin$^2\theta_W$) in neutrino
scattering at the DUNE energies is twofold: (a) it provides a direct measurement of neutrino
couplings to the Z boson and (b) it probes a different scale of momentum transfer than LEP
did by virtue of not being at the Z boson mass peak. The unprecedented large statistics
of deep inelastic scattering events will allow for significant measurements of the mixing 
angle. Other \dword{sm} measurements include those of nucleon structure functions, the strange content of nucleons, and a precise verification of a number of sum rules. 
Some of these measurements would need cross section measurements 
on hydrogen targets.  These expected sensitivity of these measurements will be addressed
in future studies.


\cleardoublepage

\chapter{Tools and Methods}
\label{ch:tools}

Evaluation of the capabilities of DUNE/LBNF to realize the scientific program envisioned requires a detailed understanding of the experimental signatures of the relevant physical processes, the response of detection elements, and the performance of calibration systems and event reconstruction and other tools that enable analysis of data from the DUNE detectors.  It is the aim of this chapter to introduce the network of calibration, simulation, and reconstruction tools that form the basis for the demonstration of science capabilities presented in the chapters that follow.  The presentation here covers general components, namely those that are commonly utilized across the science program, although many of these are geared toward application to the long-baseline oscillation physics at the heart of this program.  Other tools and methods developed for specific physics applications are described in the corresponding chapters that follow.

Where appropriate, the performance of reconstruction tools and algorithms is quantified.  Some of these characterizations form the basis for parameterized-response simulations used by physics sensitivity studies that have not yet advanced to the level of analysis of fully reconstructed simulated data.  They also serve as metrics that allow linkages to be drawn between detector configuration specifications and physics sensitivity.

Another critical role for the simulation and reconstruction tools described in this chapter, implicit above, is to enable detailed study of sources of systematic error that can affect physics capability, which can also lead to the development of mitigation strategies.  Thus, where possible, assessments of systematic uncertainties in the modeling of LBNF/DUNE conditions and performance are presented.

\section{Monte Carlo Simulations}
\label{sec:tools-mc}

Many physics processes are simulated in the DUNE \dword{fd}; these include the interactions of beam neutrinos, atmospheric neutrinos, \dword{snb} neutrinos, proton decays and cosmogenic events. Figure~\ref{fig:dune_tpc} shows a portion of the DUNE \single TPC consisting of \dword{apa}s and \dword{cpa}s.

\begin{dunefigure}[Schematic view of a DUNE \single TPC module]{fig:dune_tpc}{A portion of DUNE \single TPC is shown. Four separate drift regions are separated by \dword{apa}s and \dword{cpa}s.}
\includegraphics[width=0.7\textwidth]{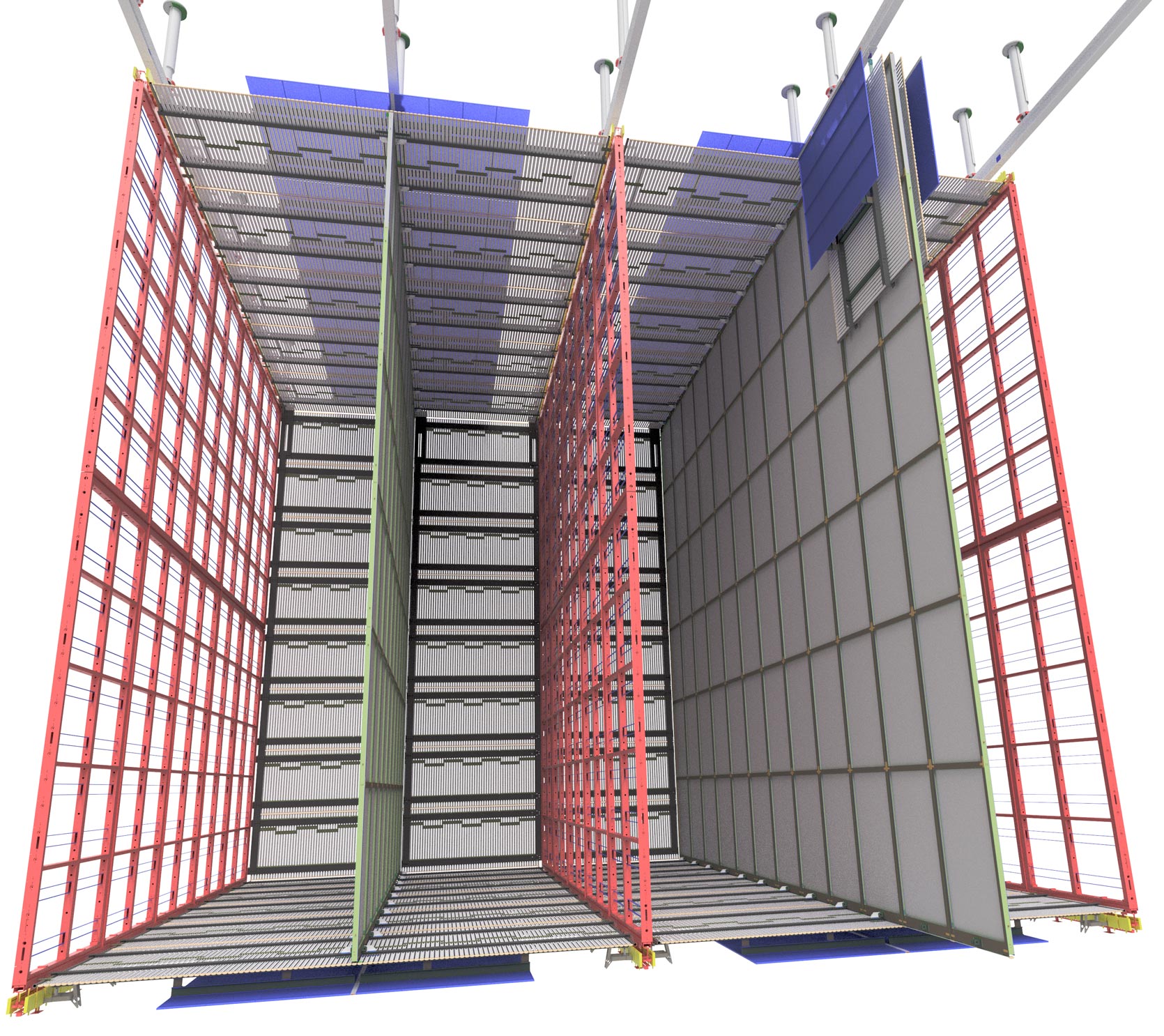}
\end{dunefigure}

To save processing time, all the \dword{fd} samples except the cosmogenics sample were simulated using a smaller version of the full \nominalmodsize far \dword{detmodule} geometry. This geometry is \SI{13.9}{m} long, \SI{12}{m} high and \SI{13.3}{m} wide, which consists of 12 \dword{apa}s and 24 \dword{cpa}s. 
Figure~\ref{fig:dune_apa} shows the detailed structure of an \dword{apa}. 
\begin{dunefigure}
[Detailed structure of the APA]
{fig:dune_apa}
{The detailed structure of the \dword{apa} is shown. Each \dword{apa} consists of four wrapped induction wire planes and two collection wire planes.
The \dword{pd} is sandwiched between the two collection wire planes.}
\includegraphics[width=0.7\textwidth]{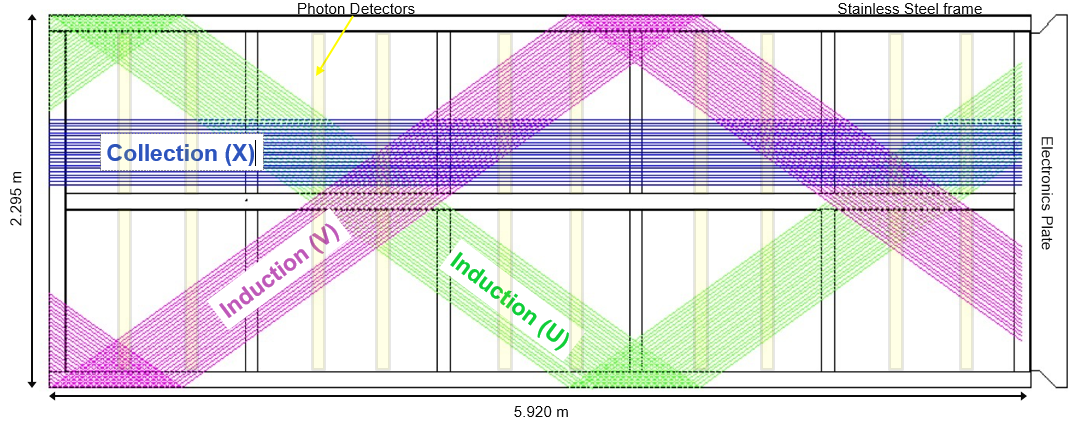}
\end{dunefigure}

For the simulation chain, each sample is simulated in three steps: generation (gen), {\sc geant4} tracking (g4), TPC signal simulation, and digitization (detsim). The first step is unique for each sample while the second and the third steps are mostly identical for all samples. 

\subsection{Neutrino Flux Modeling}
\label{sec:tools-mc-flux}

Neutrino fluxes were generated using G4LBNF, a \textsc{Geant}4\xspace-based simulation of the LBNF neutrino beam.  The simulation was configured to use a detailed description of the LBNF optimized beam design~\cite{optimizedbeamcdr}.  That design starts with a \SI{1.2}{MW}, \SI{120}{\GeV} primary proton beam that impinges on a \SI{2.2}{m} long, \SI{16}{mm} diameter cylindrical graphite target.  Hadrons produced in the target are focused by three magnetic horns operated with \SI{300}{kA} currents.  The target chase is followed by a \SI{194}{m} helium-filled decay pipe and a hadron absorber.  The focusing horns can be operated in forward or reverse current configurations, creating neutrino and antineutrino beams, respectively.   

\begin{dunefigure}[Visualization of the focusing system as simulated in g4lbnf]{fig:beam_vis}
{Visualization of the focusing system as simulated in g4lbnf.}
    \includegraphics[width=0.7\textwidth]{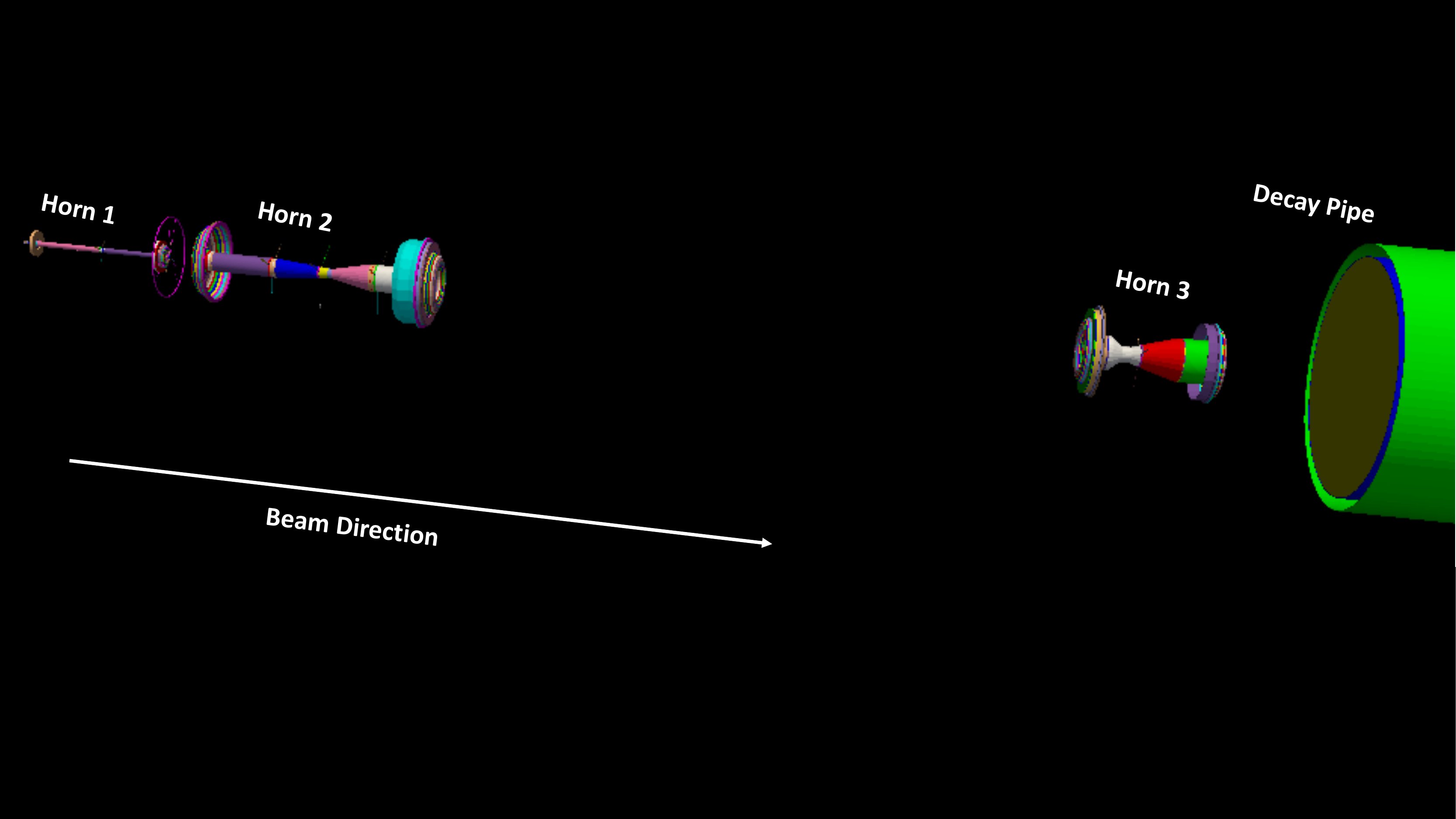}\end{dunefigure}

The optimized LBNF neutrino beam design is the result of several years of effort by LBNF and DUNE to identify a focusing system optimized to DUNE's long-baseline physics goals.  The optimization process requires scanning many parameters describing the hadron production target, focusing horns, and the decay pipe. Genetic algorithms have been used successfully in the past to scan the large parameter space to find the optimal beam design~\cite{Calviani:2014cxa}. The  LBNF beam optimization process began with a genetic algorithm that scanned simulations of many different horn and target geometries to identify those that produced the optimal sensitivity to \dword{cpv}.  The specific metric used was estimated sensitivity to 75\% of \dword{cp} phase space after \ktmwyr{300}  
of exposure, taking into account the number and neutrino spectra of all neutrino flavors. The resulting beam effectively optimized flux at the first and second oscillation maxima, which also benefits measurements of other oscillation parameters.  The output of the genetic algorithm was a simple design including horn conductor and target shapes.  This design was transformed into a detailed conceptual design by LBNF engineers, and iterated with DUNE physicists to ensure that engineering changes had minimal impact on physics performance.  Relative to the previous NuMI-like design, the optimized design reduces the time to three-sigma coverage of 75\% of \dword{cp} phase space by 42\%, which is equivalent to increasing the mass of the far detector by 70\%.  It also substantially increases sensitivity to the mass hierarchy and improves projected resolution to quantities such as $\sin^22\theta_{13}$ and $\sin^2\theta_{23}$~\cite{fields_doc_2901}.        

\subsubsection{On-axis Neutrino Flux and Uncertainties}

\begin{dunefigure}[Neutrino fluxes at the near detector]{fig:flux_flavor_parent}
{Predicted neutrino fluxes at the near detector for neutrino mode (left) and antineutrino mode (right). From top to bottom shown are muon neutrino, muon antineutrino, electron neutrino, and electron antineutrino fluxes.}
\includegraphics[width=0.9\textwidth]{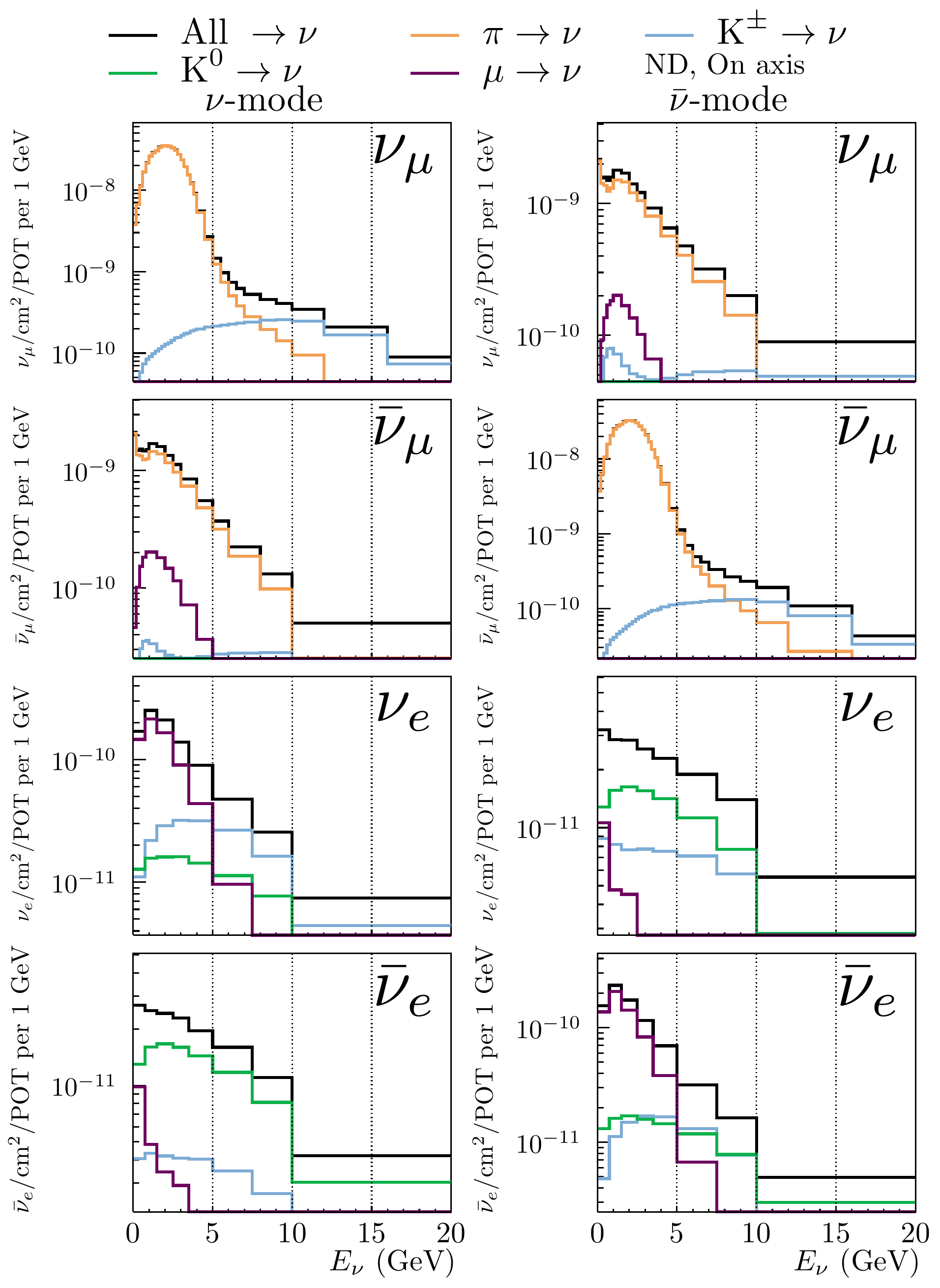}

\end{dunefigure}

The predicted neutrino fluxes for neutrino and antineutrino mode configurations of LBNF are shown in Figure~\ref{fig:flux_flavor_parent}.  In neutrino (antineutrino) mode, the beams are 92\% (90.4\%) muon neutrinos (antineutrinos), with wrong-sign contamination making up 7\% (8.6\%) and electron neutrino and antineutrino backgrounds 1\% (1\%).  Although 
we expect a small nonzero intrinsic tau neutrino flux, this is not simulated by G4LBNF.  Nor are neutrinos arising from particle decay at rest. 

\begin{dunefigure}[Flux uncertainties at the \dword{fd} as a function of neutrino energy]{fig:flux_uncertainties_flavor}
{Flux uncertainties at the far detector as a function of neutrino energy in neutrino mode (left) and antineutrino mode (right) for, from top to bottom, muon neutrinos, muon antineutrinos, electron neutrinos and electron antineutrinos. }
    \includegraphics[width=0.85\textwidth]{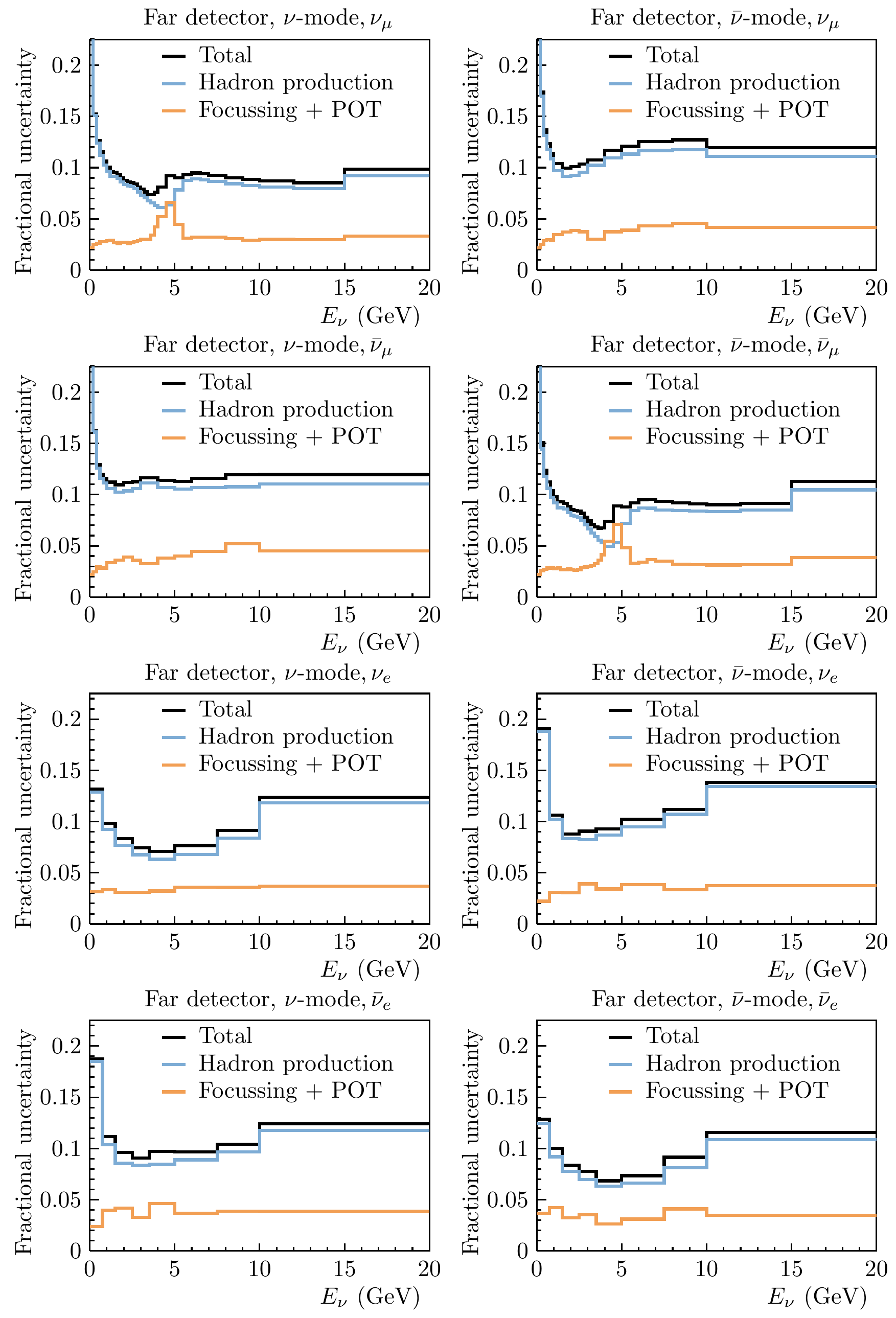}

    \end{dunefigure}

\begin{dunefigure}[Focusing and hadron production uncertainties on the $\nu$ mode $\nu_{\mu}$ flux]{fig:flux_uncertainty_breakdown}
{Focusing (left) and hadron production (right) uncertainties on the neutrino mode muon neutrino flux at the \dword{fd}.}
  \includegraphics[width=0.45\textwidth]{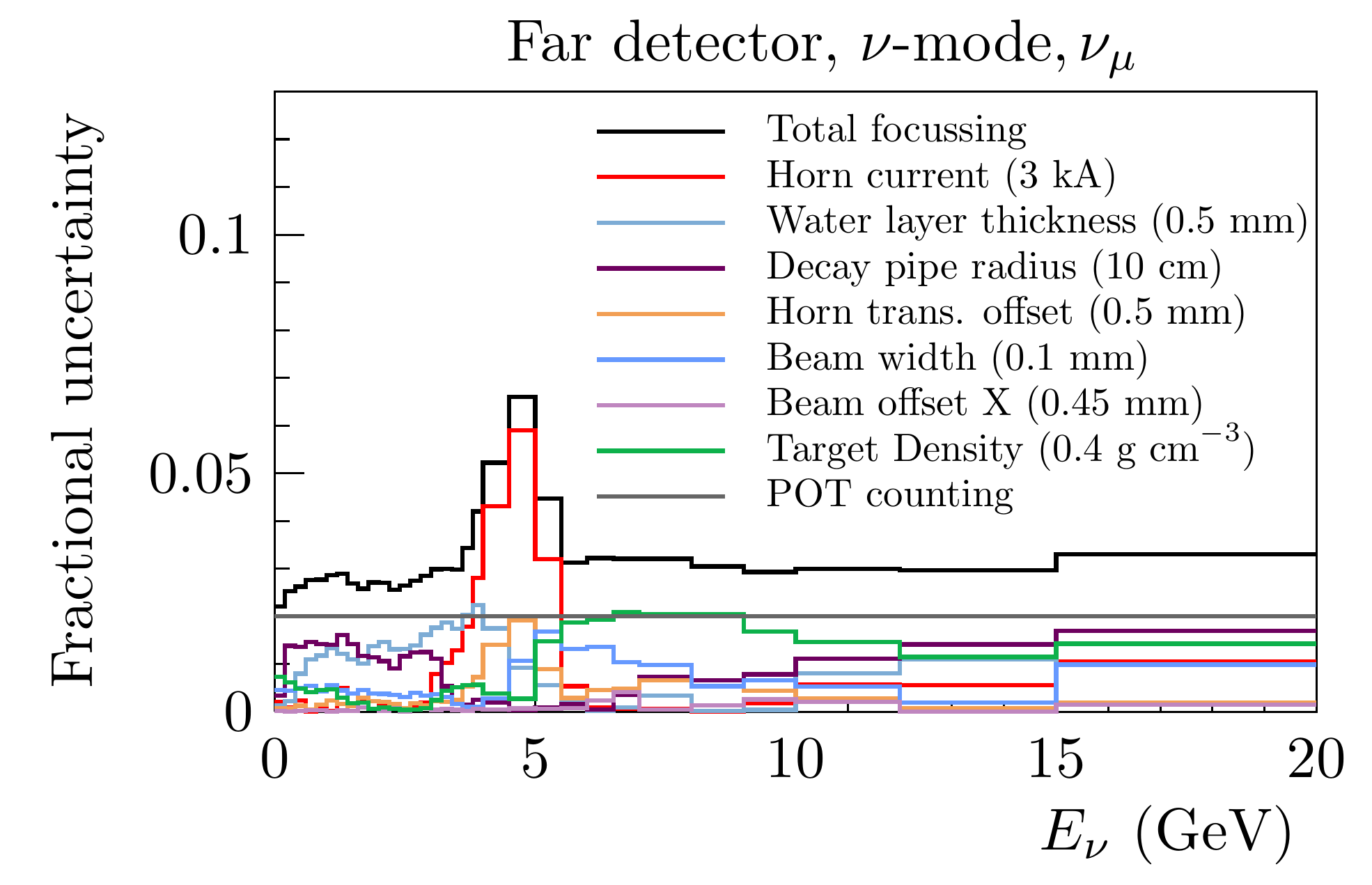}
  \includegraphics[width=0.45\textwidth]{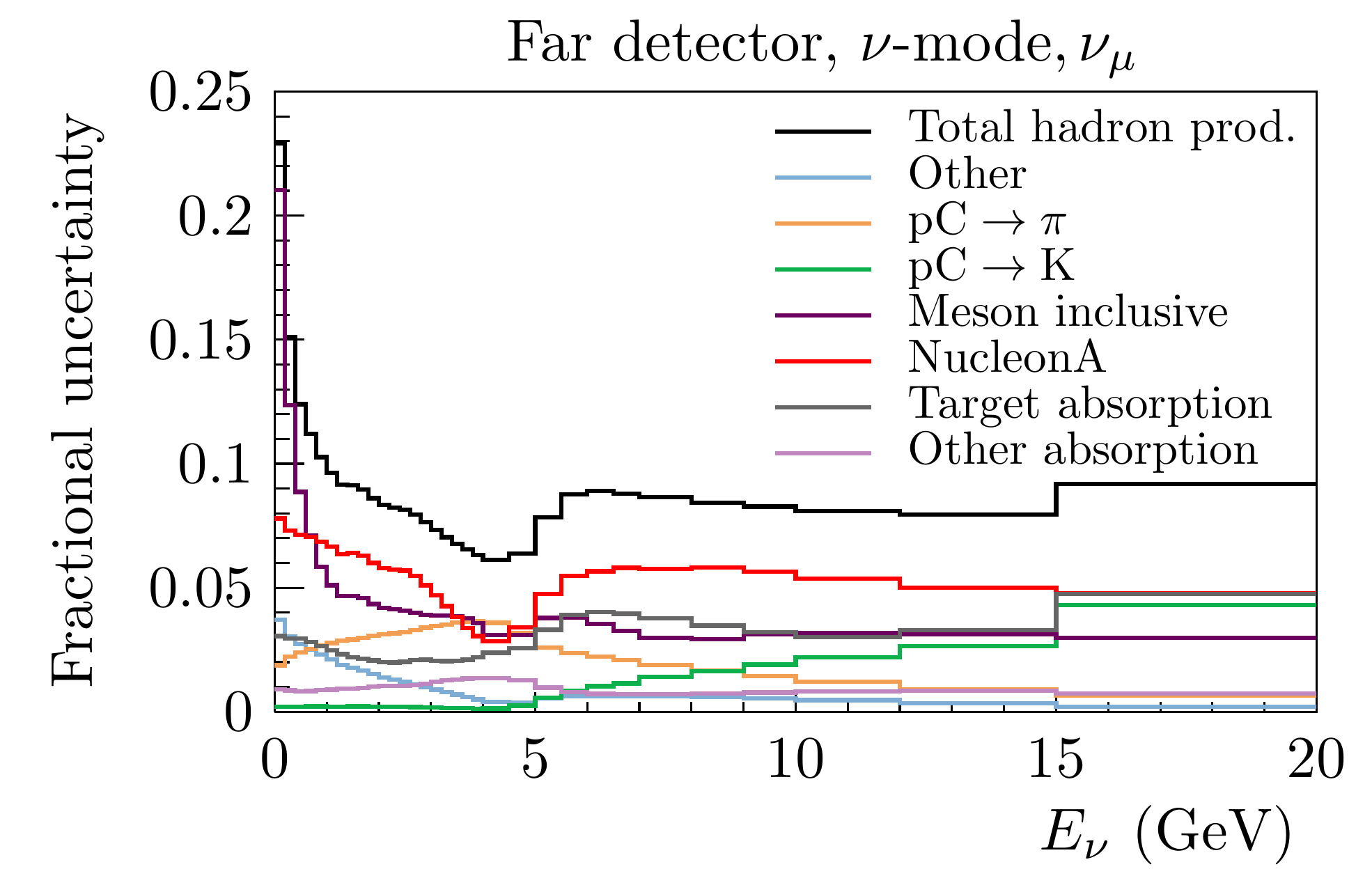}

\end{dunefigure}

Uncertainties on the neutrino fluxes arise primarily from uncertainties in hadrons produced off the target and uncertainties in parameters of the beam such as horn currents and horn and target positioning (commonly called ``focusing uncertainties'').  Uncertainties on the neutrino fluxes arising from both of these categories of sources are shown in Figure~\ref{fig:flux_uncertainties_flavor}.  Hadron production uncertainties are estimated using the Package to Predict the FluX (PPFX) framework developed by the \minerva collaboration~\cite{Aliaga:2016oaz, AliagaSoplin:2016shs}, which assigns uncertainties for each hadronic interaction leading to a neutrino in the beam simulation, with uncertainties taken from thin target data (from e.g., the NA49~\cite{NA49} experiment) where available, and large uncertainties assigned to interactions not covered by data.  Focusing uncertainties are assessed by altering beamline parameters in the simulation within their tolerances and observing the resulting change in predicted flux.  A breakdown of the hadron production and focusing uncertainties into various components are shown in Figure~\ref{fig:flux_uncertainty_breakdown} for the neutrino mode muon neutrino flux at the \dword{fd}.    

At most energies, hadron production uncertainties are dominated by the ``NucleonA'' category, which includes proton and neutron interactions that are not covered by external data.  At low energies, uncertainties due to pion reinteractions (denoted ``Meson Inc'') dominate.   The largest source of focusing uncertainty arises from a 1\% uncertainty in the horn current, followed by a 2\% 
uncertainty in the number of protons impinging on the target.   For all neutrino flavors and all neutrino energies, hadron production uncertainties are larger than focusing uncertainties.  However, hadron production uncertainties are expected to decrease in the next decade, as more thin target data becomes available.  Hadron production measurements taken with a replica target are also being considered and would substantially reduce the uncertainties.  

\begin{dunefigure}[Correlation of flux uncertainties]{fig:flux_uncertainty_correlation}
{Correlation of flux uncertainties.  Each block of neutrino flavor corresponds to bins of energy with bin boundaries of
[0.0, 0.2, 0.4, 0.6, 0.8, 1.0, 1.2, 1.4, 1.6, 1.8, 2.0, 2.2, 2.4, 2.6, 2.8, 3.0, 3.2, 3.4, 3.6, 3.8, 4.0, 4.5, 5.0, 5.5, 6.0, 6.5, 7.0, 8.0, 9.0, 10.0, 12.0, 15.0, 20.0] GeV for right sign muon neutrinos, [0.0, 0.2, 0.4, 0.6, 0.8, 1.0, 1.5, 2.0, 2.5, 3.0, 4.0, 5.0, 6.0, 8.0, 10.0, 20.0] GeV for wrong sign muon neutrinos, and [0.0, 0.75, 1.5, 2.5, 3.5, 5.0, 7.5, 10.0, 20.0] for electron neutrinos and antineutrinos. }
    \includegraphics[width=0.9\textwidth]{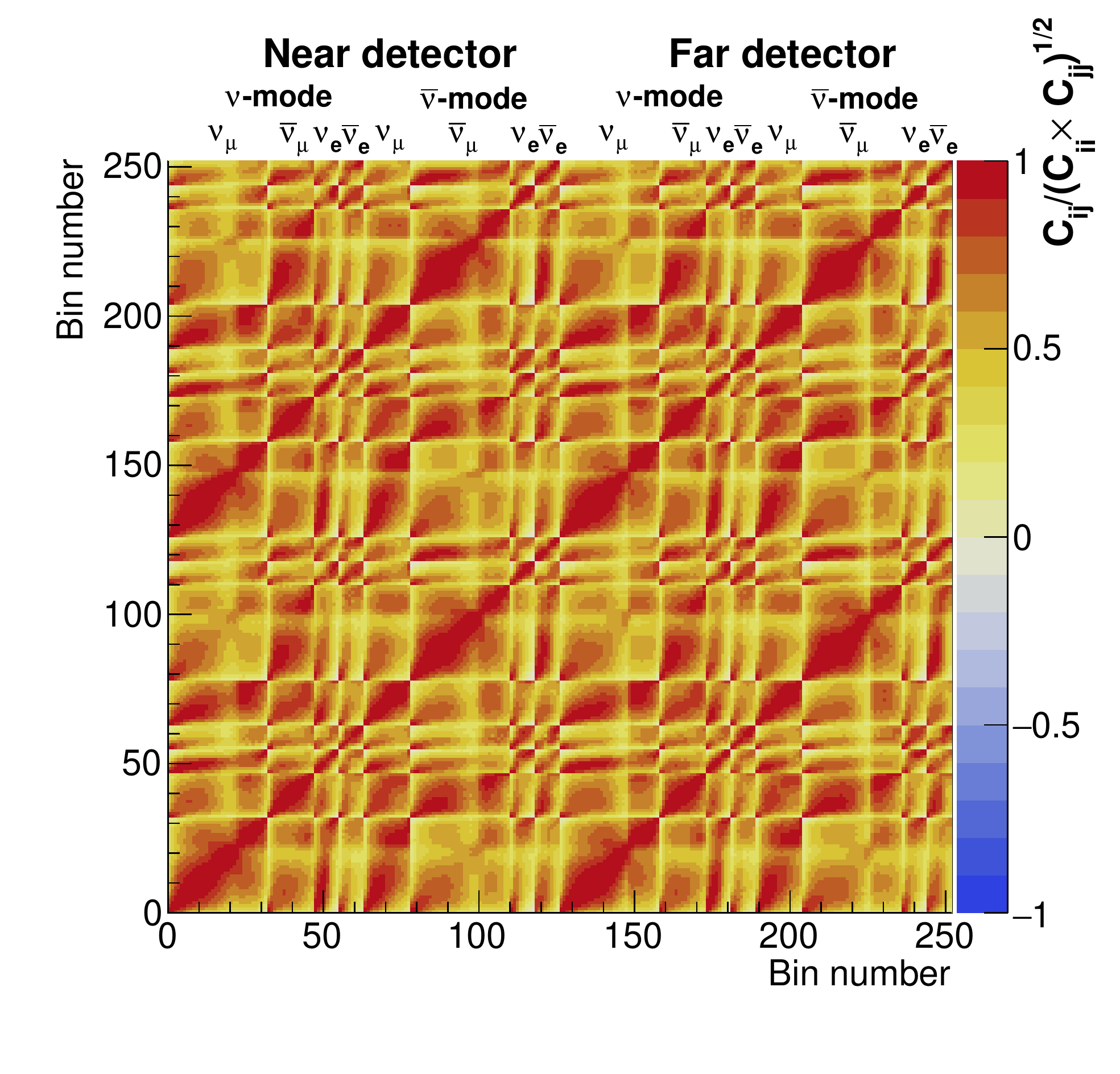}
\end{dunefigure}

Figure~\ref{fig:flux_uncertainty_correlation} shows correlations of the total flux uncertainties.  In general, the uncertainties are highly correlated across energy bins. However, the flux in the very high energy, coming predominantly from kaons, tends to be uncorrelated with flux at the peak, 
arising predominantly from pion decays.  Flux uncertainties are also highly correlated between the near and far detectors and between neutrino-mode and antineutrino-mode running.  The focusing uncertainties do not affect wrong-sign backgrounds, which reduces correlations between e.g., muon neutrinos and muon antineutrinos in the same running configuration in the energy bins where focusing uncertainties are significant.    

The unoscillated fluxes at the \dword{nd} and \dword{fd} are similar but not identical. Figure~\ref{fig:flux_nearfar} shows the ratio of the near and far neutrino-mode muon neutrino unoscillated fluxes 
and the uncertainties on the ratio.  The uncertainties are approximately 1\% or smaller except at the falling edge of the focusing peak, where they rise to 2\%, but are still much smaller than the uncertainty on the absolute fluxes.    And unlike the case for absolute fluxes, the uncertainty on the near-to-far flux ratio is dominated by focusing rather than hadron production uncertainties.  This ratio and its uncertainty are for the fluxes at the center of the near and far detectors, 
and do not take into account small variations in flux across the face of the \dword{nd}.     

\begin{dunefigure}[Ratio of neutrino-mode muon neutrino fluxes at the near and far detectors]{fig:flux_nearfar}
{Ratio of neutrino-mode muon neutrino fluxes at the near and far detectors (left) and uncertainties on the ratio (right). }
    \includegraphics[width=0.45\textwidth]{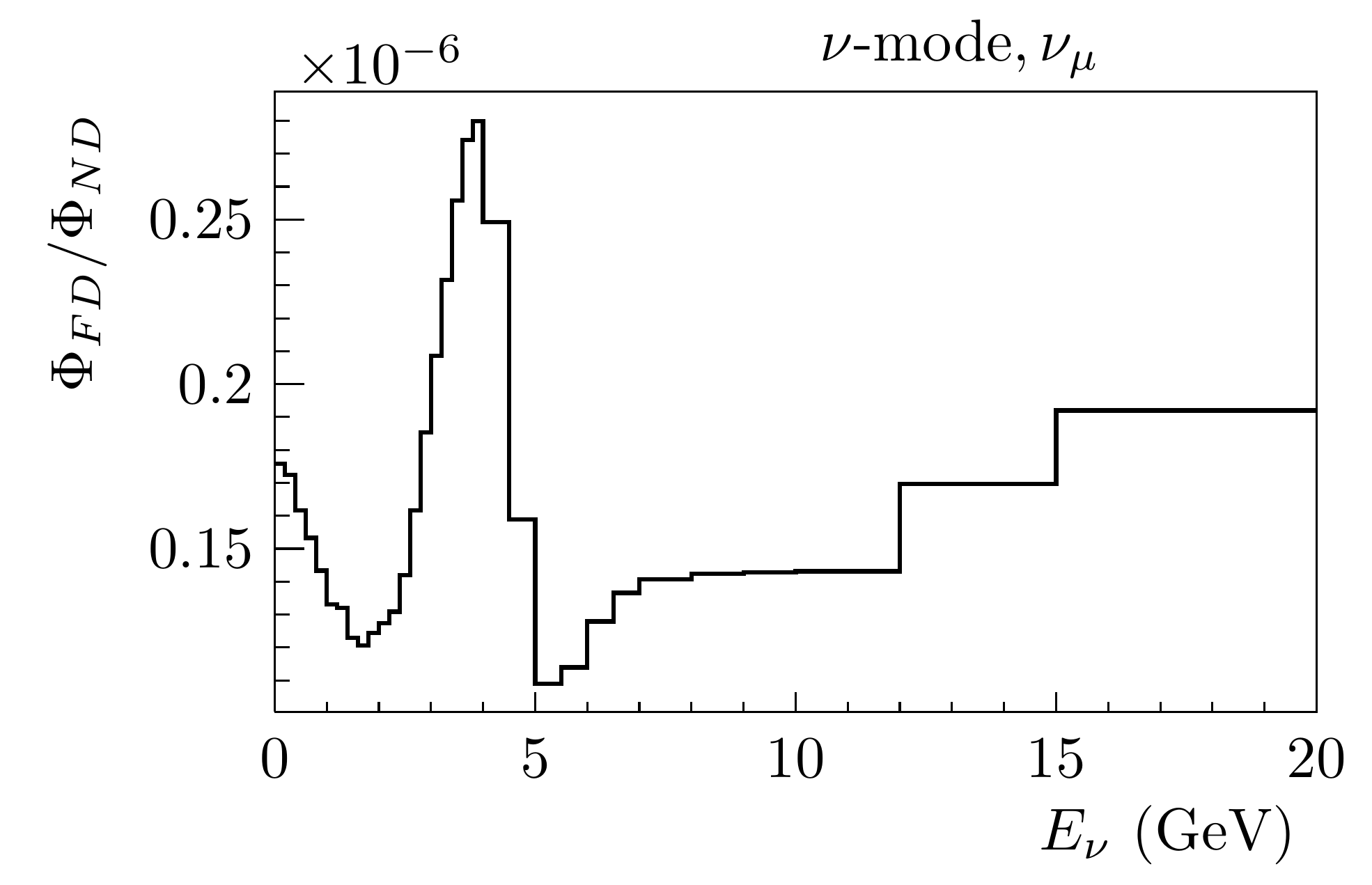}
     \includegraphics[width=0.45\textwidth]{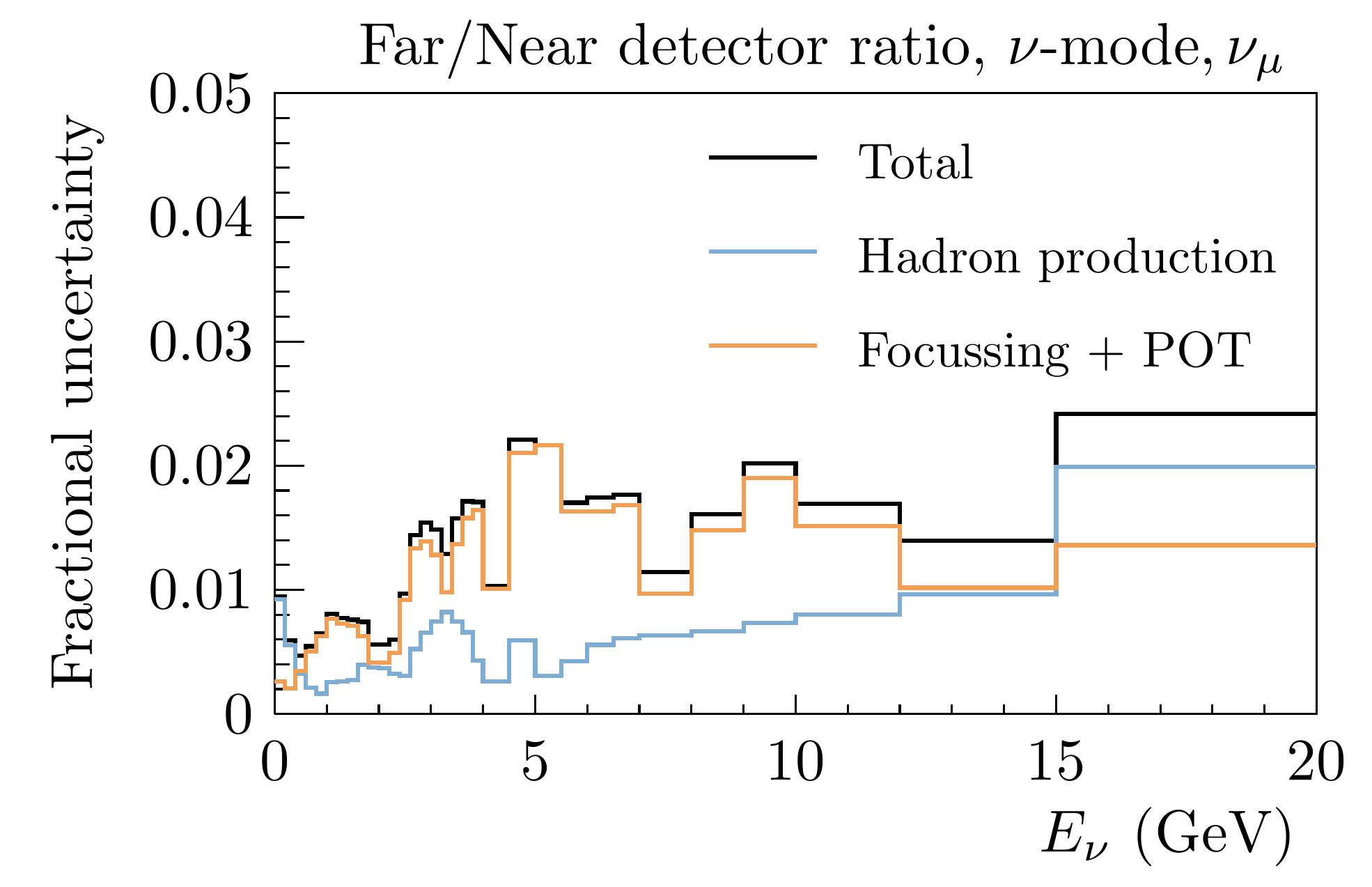}

\end{dunefigure}

\subsubsection{Off-axis Neutrino Flux and Uncertainties}

The neutrino flux has a broad angular distribution and extends outward at the \dword{nd} hall. At an ``off-axis'' angle relative to the initial beam direction, the subsequent neutrino energy spectrum is narrower and peaked at a lower energy than the on-axis spectrum.
The relationship between the parent pion energy and neutrino energy is shown in Figure~\ref{fig:OAAFluxFigs}.  At \SI{575}{m}, the location of the \dword{nd} hall, a lateral shift of \SI{1}{m} corresponds to approximately a \ang{0.1} change in off-axis angle. 
The DUNE-PRISM concept, in which the near detector \lartpc can be moved to enable  off-axis measurements, relies on this feature to help constrain systematic errors for the \dword{lbl} oscillation program as described in Section~\ref{sec:ch-nu-osc-06-ndconcept-offaxis}.

\begin{dunefigure}[$\nu$ energy as  function of parent pion energy for off-axis angles] 
{fig:OAAFluxFigs}
{(left) The neutrino energy as a function of parent pion energy for different angles away from the pion momentum vector. Figure from Ref.~\cite{Duffy:2016owt}. (right) The DUNE near detector flux predictions over a range of off-axis positions for a near detector at \SI{575}{m} downstream of the target station. }
    \includegraphics[width=0.4\textwidth]{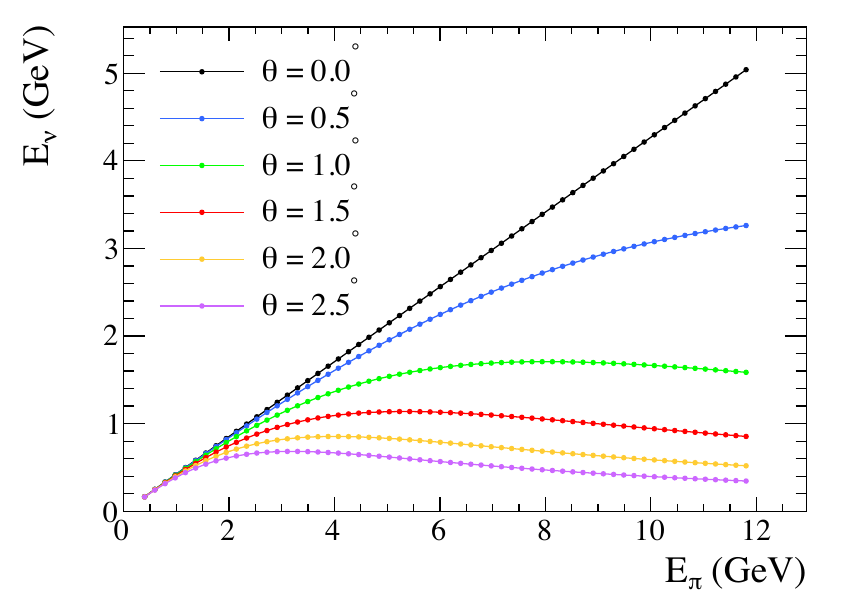}
    \raisebox{0.5em}{\includegraphics[width=0.5\textwidth]{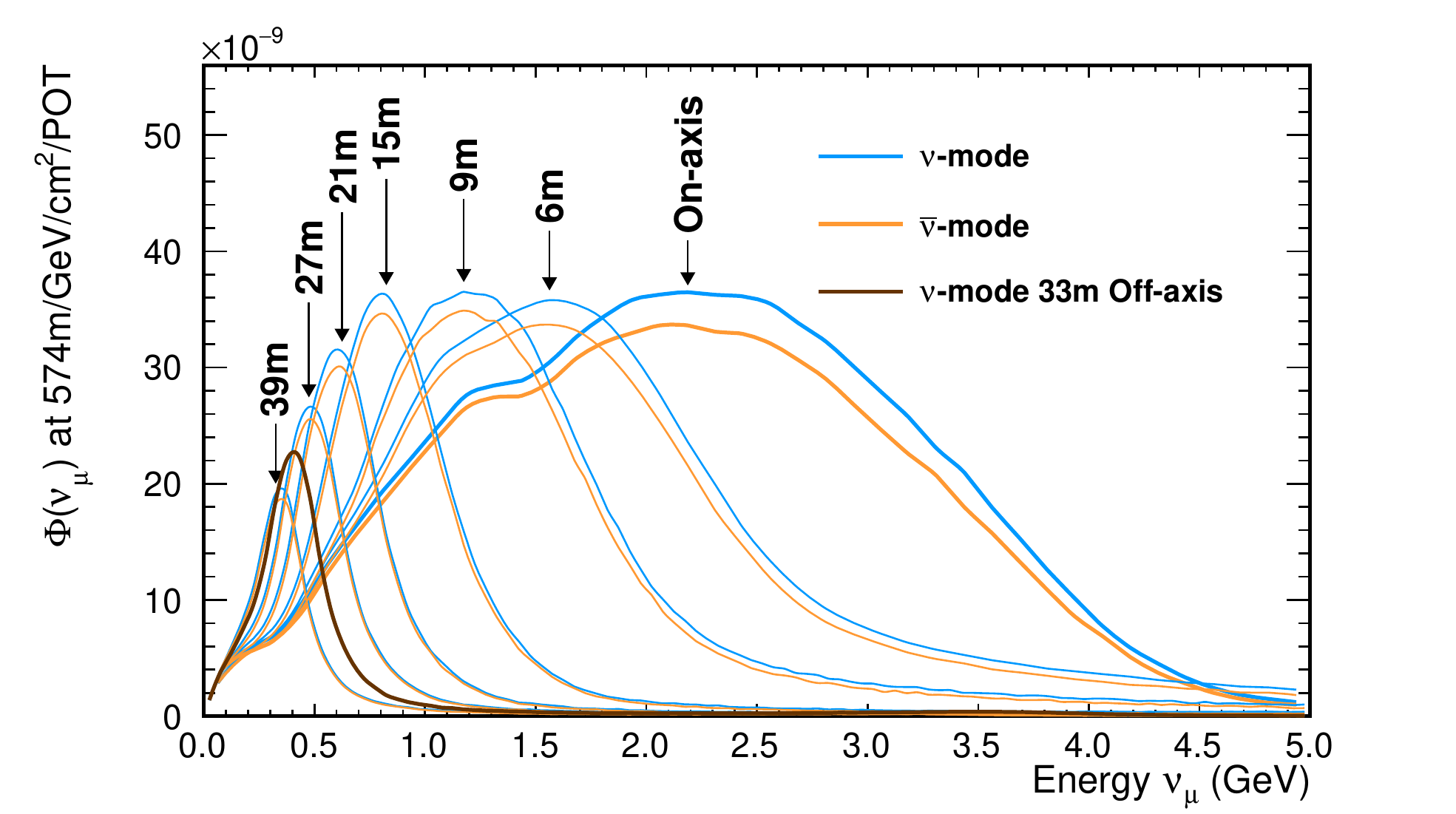}}
\end{dunefigure}

The intrinsic neutrino flavor content of the beam varies with off-axis angle. Figure~\ref{fig:OffAxisFluxes_1D_AllSpec} shows the neutrino-mode and anti-neutrino-mode predictions for the four neutrino flavors at the on-axis position, and a moderately off-axis position. At the \SI{30}{m} position, a second, smaller energy peak at approximately \SI{4}{\GeV} is due to the charged kaon neutrino parents. 

\begin{dunefigure}[The predicted ND \numu energy spectra, on axis and \SI{30}{m} off axis]{fig:OffAxisFluxes_1D_AllSpec}
{The predicted muon neutrino energy spectra at two \dword{nd} positions, on axis and \SI{30}{m} off axis. (a) The predicted neutrino flavor-content of the neutrino-mode (FHC) and anti-neutrino-mode (RHC) beam. (b) The neutrino-mode, muon-flavor predicted flux, separated by the particle that decayed to produce the neutrino. The off-axis spectrum displays a double peak structure due to charged kaon parent decay kinematics. The on-axis kaon-peak occurs at higher neutrino energy and will have a significantly broader energy spread. Top: Beam neutrino flavor content, middle: Beam neutrino flavor content; bottom: Beam neutrino decay-parent species}
    \includegraphics[width=0.8\textwidth]{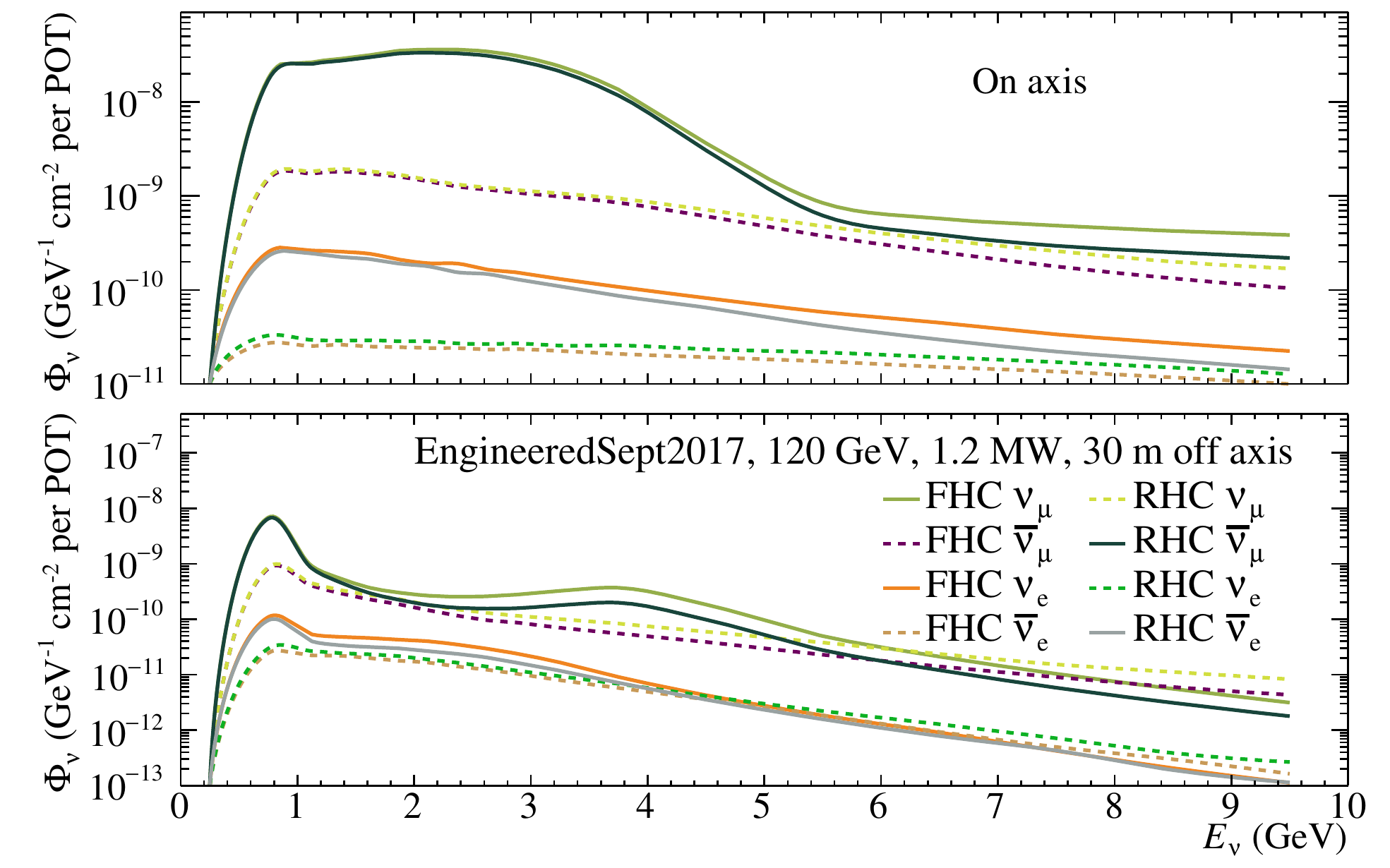}
  \includegraphics[width=0.8\textwidth]{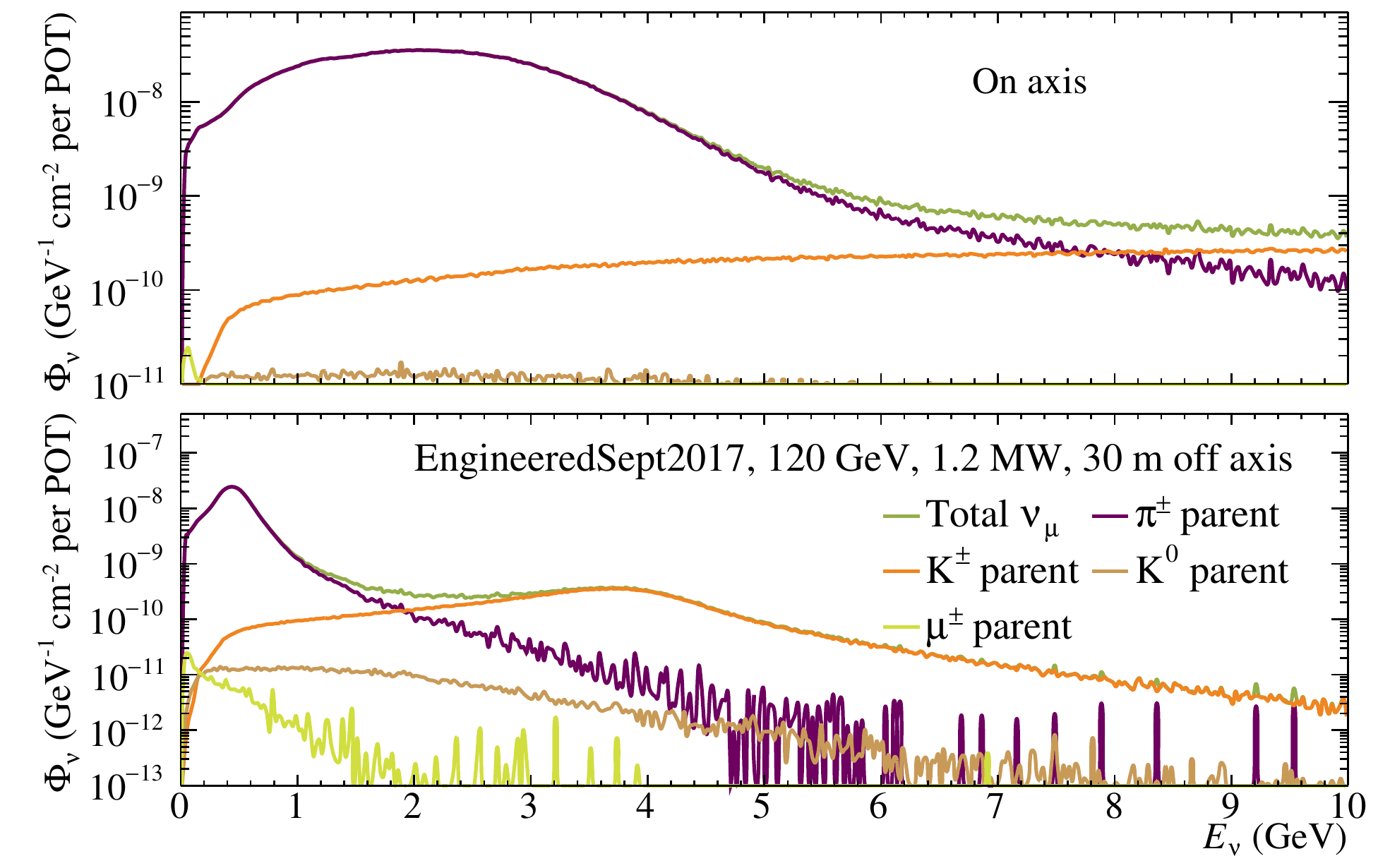}    
    \end{dunefigure}

The same sources of systematic uncertainty that affect the on-axis spectra also modify the off-axis spectra. 
Figure \ref{fig:onvsoff_flux_uncertainty} shows the on-axis and off-axis hadron production and focusing uncertainties. 
Generally, the size of the off-axis uncertainties is comparable to the on-axis uncertainties and the uncertainties are highly correlated across off-axis and on-axis positions. While the hadron production uncertainties are similar in size, the focusing uncertainties are smaller for the off-axis flux. The systematic effects have different shapes as a function of neutrino energy at different off-axis locations, making off-axis flux measurements useful to diagnose beamline physics. Measuring on-axis and off-axis flux breaks degeneracy between various systematics and allows better flux constraint.

\begin{dunefigure}[On-axis and off-axis flux uncertainties]{fig:onvsoff_flux_uncertainty}
{The flux uncertainty for the on-axis flux, and several off-axis positions. Shown is the total hadron production uncertainty and several major focusing uncertainties.}
    \includegraphics[width=0.8\textwidth]{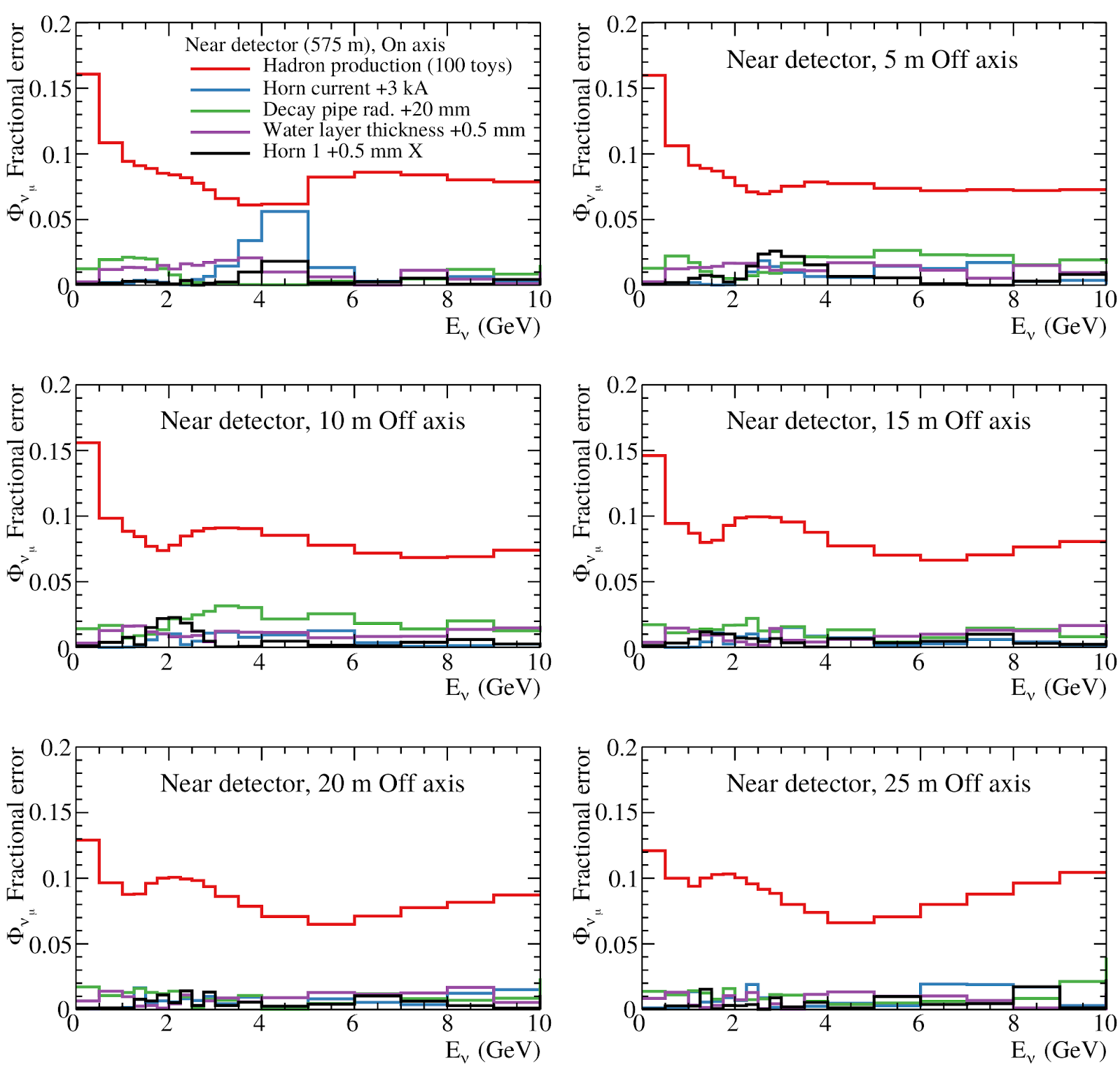}
\end{dunefigure}

\subsubsection{Alternate Beamline Configurations}

Although the LBNF beamline is expected to run for many years in a \dword{cp}-optimized configuration, it could potentially be modified in the future for other physics goals.  For example, it could be altered to produce a higher-energy spectrum 
to measure tau neutrino appearance.  In the standard \dword{cp}-optimized configuration, we expect about 130 tau neutrino \dword{cc} interactions per year 
at the \dword{fd}, before detector efficiency and assuming \SI{1.2}{MW} beam power.  However, replacing the three \dword{cp}-optimized horns with two NuMI-like parabolic horns can raise this number to approximately \num{1000} tau neutrinos per year.  Figure~\ref{fig:tau-optimized} shows the muon neutrino flux for one such configuration.  Although the flux in the \SIrange{0}{5}{\GeV} region critical to $\delta_{CP}$ measurements is much smaller, the flux above \SI{5}{\GeV}, where the tau neutrino interaction cross section becomes significant, is much larger.  Many other energy distributions are possible by modifying the position of the targets and horns.  Even altering parameters of the \dword{cp}-optimized horns offers some variablity in energy spectrum, but the parabolic NuMI horns offer more configurability.  Because the LBNF horns are not expected to be remotely movable, such reconfigurations of the beamline would require lengthy downtimes to reconfigure target chase shielding and horn modules.   

\begin{dunefigure}[Comparison of standard and tau-optimized neutrino fluxes]{fig:tau-optimized}
{Comparison of standard and tau-optimized neutrino fluxes.  The tau optimized flux was simulated with a \SI{120}{\GeV} proton beam and two NuMI parabolic horns, with the second horn starting \SI{17.5}{m} 
downstream from the start of the first horn, and a \SI{1.5}{m} long, \SI{10}{mm} wide carbon fin target starting \SI{2}{m} from the upstream face of the first horn.  
}
\includegraphics[width=0.5\textwidth]{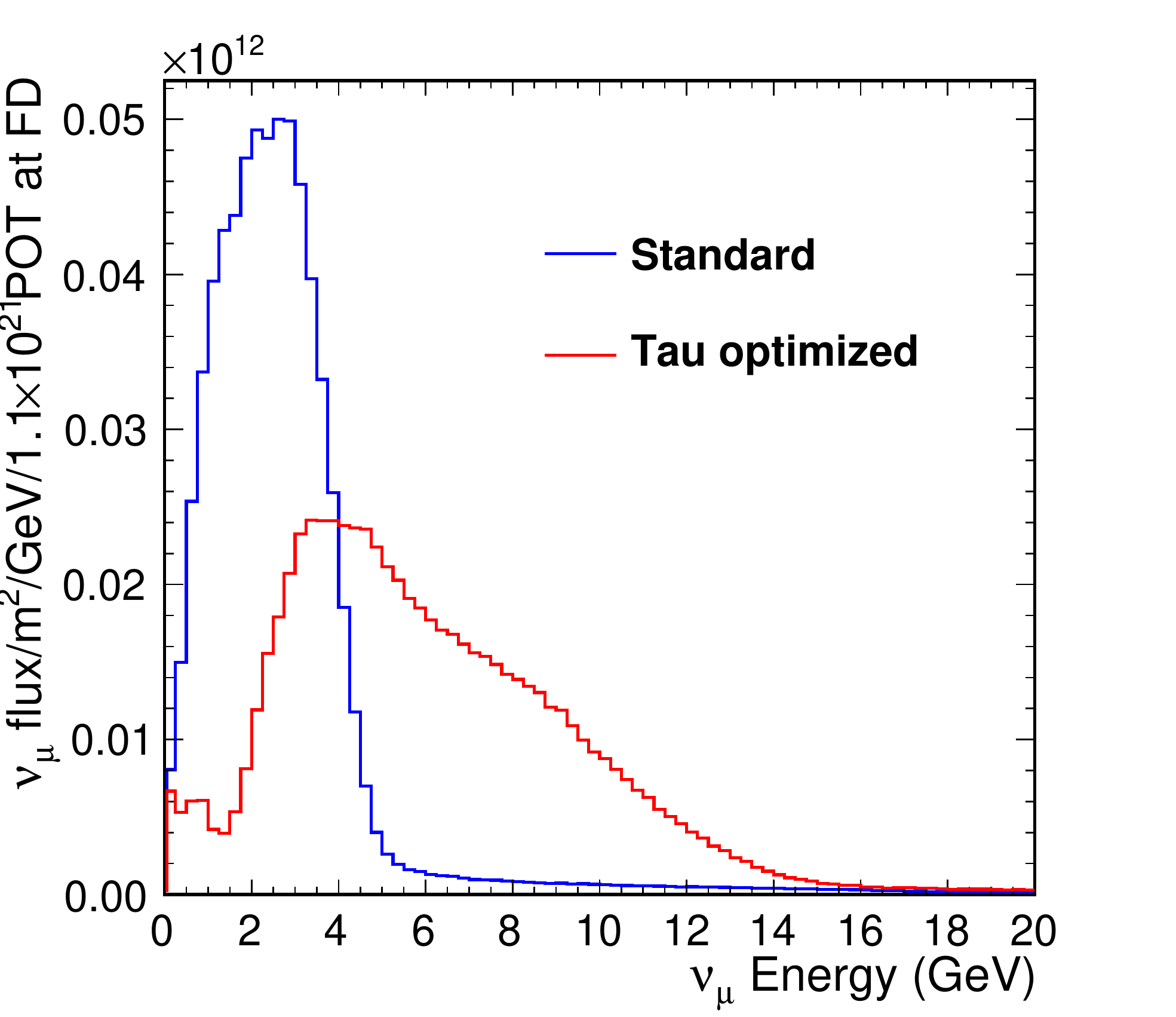}
\end{dunefigure}

\subsection{Neutrino Interaction Generators} 
\label{sec:tools-mc-gen}

\subsubsection{Supernova Neutrinos}

The \dword{snb} neutrino events were generated using custom code wrapped in a \dword{larsoft} module.
This code simulates \dword{cc} $\nu_e$-$^{40}$Ar interactions.
For each electron neutrino it calculates probabilities to produce a $^{40}$K nucleus
in different excited states (using a model from~\cite{Bhattacharya:1998hc}),
randomly selects one, and (with energy levels from~\cite{Cameron:2004myb}) 
produces several de-excitation $\gamma$s and an electron carrying the remaining energy.
All particles are produced isotropically, and 
there is no delay between the electron and corresponding de-excitation $\gamma$s
(in this model the $^{40}$K nucleus de-excites instantaneously) and they share a vertex,
which is simulated with equal probability anywhere in the active volume.
The primary neutrino energy distribution used in these samples is the cross-section-weighted 
energy spectrum obtained from SNOwGLoBES~\cite{snowglobes} (using the ``GKVM'' flux~\cite{GKVM}).
The \dword{snb} neutrino generator also allows to simulate a Poisson-distributed random number 
of neutrino interactions per event. These samples were simulated with, on average, 2 or 20 neutrinos.
In addition, one of the samples was generated with $1.01$~Bq/kg of $^{39}$Ar background.

\subsubsection{GENIE}

The DUNE \dword{mc} simulation chain is interfaced to the \dword{genie} event generator \cite{Andreopoulos:2009rq}. This is an open-source product of the \dword{genie} collaboration\footnote{www.genie-mc.org}  that provides state-of-the-art modeling of neutrino-nucleus interactions, as well as simulation of several other non-neutrino processes (nucleon decay, neutron-antineutron oscillation, boosted dark matter interactions, hadron and charged lepton scattering off nuclei). The generator product also includes off-the-shelf components (flux drivers and interfaces to outputs of detailed neutrino beamline simulations, detector geometry drivers, and several specialized event generation applications) for the simulation of realistic experimental setups. The \dword{genie} collaboration performs an advanced global analysis of neutrino scattering data, and is leading the development and characterization of comprehensive interaction models. The \dword{genie} comprehensive models and physics tunes which are developed using its proprietary Comparisons and Tuning products, are fully integrated in the \dword{genie} Generator product. Finally, the open-source \dword{genie} Reweight product provides means for propagating modeling uncertainties. 

At the time of the \dword{tdr} writing, the DUNE simulation uses a version in the v2 series of the \dword{genie} generator, which includes empirical comprehensive models, based on home-grown hadronic simulations (AGKY model \cite{Yang:2009zx} for neutrino-induced hadronization and INTRANUKE/hA model \cite{Dytman:2015taa} for hadronic re-interactions) and nuclear neutrino cross sections calculated within the framework of the simple relativistic Fermi gas model \cite{Bodek:1981wr}. Several processes are simulated within that framework with the most important ones, in terms of the size of the corresponding cross section at a few GeV, being: (1) quasi-elastic scattering, simulated using an implementation of the Llewellyn Smith model \cite{LlewellynSmith:1971uhs}, (2) multi-nucleon interactions, simulated with an empirical model motivated by the Lightbody model \cite{Lightbody:1988gcu} and using a nucleon cluster model for the simulation of the hadronic system, (3) baryon resonance neutrino-production simulated using an implementation of the Rein-Sehgal model \cite{Rein:1980wg}, and (4) deep-inelastic scattering, simulated using the model of Bodek and Yang \cite{Bodek:2002ps}.  These comprehensive models, as well as the \dword{genie} procedure for tuning the cross section model in the transition region, have been used for several years and are well understood and documented \cite{Andreopoulos:2009rq}. The actual tune used is the one produced for the analysis of data from the MINOS experiment and, as was already known at that time, it has several caveats as it emphasizes inclusive data and does not address tensions with exclusive data. The future DUNE simulation will be done using the v3 \dword{genie} Generator where improved models and tunes are available. 

Besides simulation of neutrino-nucleus interactions, \dword{genie} provides simulation of several \dword{bsm} physics channels:

\textit{\dword{bsm}}: The implementation of a \dword{bsm} \dword{mc} simulation has been motivated by several theory studies \cite{Agashe:2014yua, 
Berger:2014sqa, Kong:2014mia, Cherry:2015oca, Kopp:2015bfa, Necib:2016aez, Alhazmi:2016qcs, Kim:2016zjx}. The current implementation focuses on two models presented in  \cite{Berger:2014sqa}. The first has a fermionic \dword{dm} candidate, a $Z^\prime$ mediator, and velocity independence of the spin-dependent cross section in the non-relativistic limit. The second model has a scalar \dword{dm} candidate, a $Z^\prime$ mediator, and a $u^2$ velocity dependence of the spin-dependent cross section in the non-relativistic limit.

\textit{Nucleon decay}: \dword{genie} simulates several nucleon decay topologies. For the initial nuclear state environment and intranuclear hadron transport, it uses the same modeling as it does for neutrino event simulation. In the nucleon decay simulation, the nucleon binding and momentum distribution is simulated using one of the nuclear models implemented in \dword{genie} (typically a Fermi gas model), and it is decayed to one of many topologies using a phase space decay. The decay products are produced within the nucleus and further re-interactions of hadrons are simulated by the \dword{genie} hadron transport models. The simulated nucleon decay topologies are given in Table~\ref{tab:genie_ndk}, presented in 
Sec.~\ref{sec:ndksim}.

\textit{Neutron-antineutron oscillation}: \dword{genie} simulates several event topologies that may emerge following the annihilation of the antineutron produced from a bound neutron to antineutron transition. For the initial nuclear state environment and intranuclear hadron transport, the simulation, as in the case of nucleon decay, uses the same modeling as it does for the neutrino event simulation. The simulated reactions are listed in Table~\ref{tab:nnbar-br} in 
Sec.~\ref{subsec:nonaccel-nnbar-dunesensitivity}.

\subsection{Detector Simulation}
\label{sec:tools-mc-detsim}


The detector simulation consists of particle propagation in the liquid argon using {\sc geant} and the TPC and photon detector response simulation. This step is done in the common framework \dword{larsoft} and is validated by other \dword{lartpc} experiments such as ArgoNeuT, MicroBooNE, LArIAT and ProtoDUNE.

\subsubsection{LArG4}\label{sec:larg4}

The truth particles generated in the event generator step are passed to a {\sc geant4} v4\_10\_1\_p03-based detector simulation. In this step, each primary particle from the generator and its decay or interaction daughter particles are tracked when they traverse \lar. The energy deposition is converted to ionization electrons and scintillation photons. Some electrons are recombined with the positive ions~\cite{Acciarri:2013met,Amoruso:2004dy} while the rest of the electrons are drifted towards the wire planes. The number of electrons is further reduced due to the existence of impurities in the \lar, which is commonly parameterized as the electron lifetime. 
Unless otherwise specified, an electron
lifetime of \SI{3}{ms} is assumed in the simulations.
The longitudinal diffusion smears the arrival time of the electrons at the wires and the transverse diffusion smears the electron location among neighboring wires. More details
regarding the recent measurements of diffusion coefficients can be found
in~\cite{Li:2015rqa,larpropertiesbnl}.

\subsubsection{Photon Simulation}

When ionization is calculated, the amount of scintillation light is also calculated. The response of the \dwords{pd} is simulated using a ``photon library,'' a pre-generated table giving the likelihood that photons produced within a voxel in the detector volume  will reach any of the \dwords{pd}. The photon library is generated using \dword{geant4}'s photon transport simulation, including \SI{66}{cm} Rayleigh scattering length, \SI{20}{m} attenuation length, and reflections off of the interior surface detectors. The library also incorporates the response versus location of the \dwords{pd}, capturing the attenuation between the initial conversion location of the photon and the \dwords{sipm}.

\subsubsection{TPC Detector Signal Simulation}\label{sec:tpc_sim}

When ionization electrons drift through the induction wire planes toward the collection wire plane, current is
induced on nearby wires. The principle of current induction is described by the Ramo theorem~\cite{Shockley1938,Ramo:1939vr}. 
For an element of ionization charge, the instantaneous induced current $i$ is proportional to the amount of drifted charge $q$: 
\begin{equation}\label{eq:shockley_ramo}
  i = - q \cdot \vec{E}_w \cdot \vec{v}_q.
\end{equation}
The proportionality factor is a product of the weighting field $\vec{E}_w$ at the location of the charge and 
the charge's drifting velocity $\vec{v}_q$. The weighting field $\vec{E}_w$ depends on the geometry of 
the electrodes. The charge's drifting velocity $\vec{v}_q$ is a function of the external \efield, which 
also depends on the geometry of the electrodes as well as the applied drifting and bias voltages. The current induced at a given electrode and electron drift path ($x$)
  sampled over a period of time ($t$) is called a ``field response function'' $R(x,t)$.
\begin{dunefigure}
[Garfield configuration for simulating the field response functions]
{field_resp_geometry}
{Garfield configuration for simulating the field response functions.}
\includegraphics[width=0.85\textwidth]{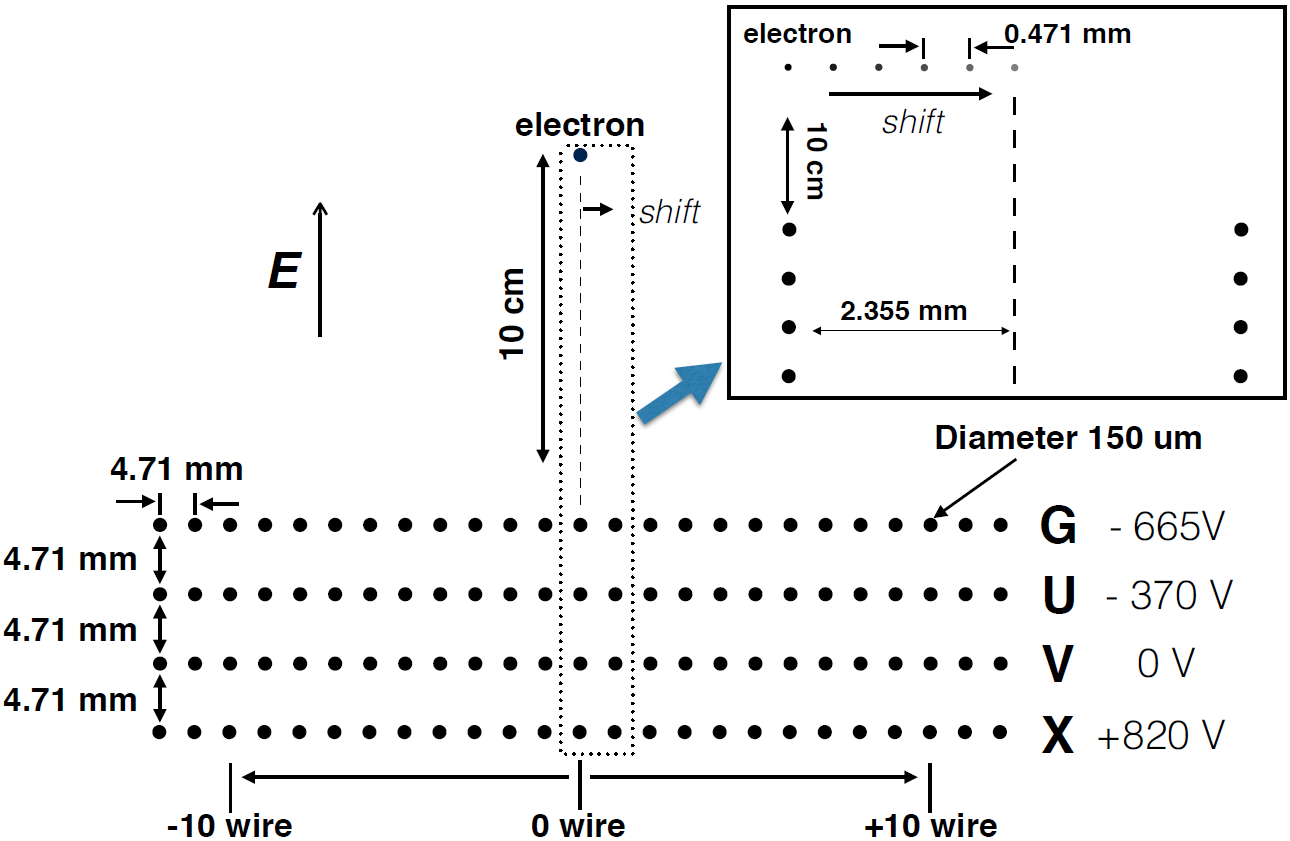}
\end{dunefigure}

The field response functions for a single ionization electron are simulated with Garfield~\cite{garfield}. 
In the Garfield simulation, a \SI{22}{cm} (along the \efield or drift direction) $\times$ 
\SI{30}{cm} (perpendicular to the field direction and wire orientation) region is configured.  
Figure~\ref{field_resp_geometry} shows a part of the region close to the anode wire planes. There are five wire planes with \SI{4.71}{mm} spacing, referred to as G, U, V, X, and 
M with operating bias voltages of \SI{-665}{V}, \SI{-370}{V}, \SI{0}{V}, \SI{820}{V}, \SI{0}{V}, respectively.  
These bias voltages ensure 100\% transmission of electrons through the grid plane (G) and the 
first two induction planes (U and V) and complete collection by the collection plane X with the main drift 
field at \SI{500}{V/cm}. In the simulation, each wire plane contains 101 wires with \SI{150}{\micro\meter} diameter
  separated at $\sim\,$\SI{4.71}{\mm} wire pitch. The electron drift velocity as a function of electric
  field is taken from recent measurements~\cite{Li:2015rqa,larpropertiesbnl}. In this simulation, the motion of the positive ions is not included as their drift velocity is about five orders of magnitude slower than that of ionization electrons. In the underground condition, the distortion in E-field caused by the space charge (accumulated positive ions) is expected to be a factor of 100 smaller than that in the surface-operating ProtoDUNE detectors. This leads the maximal position distortion to be less than 3 mm. 
\begin{dunefigure}
[Position-dependent (long-range) field response simulated with the Garfield program]
{field_resp}
{Position-dependent (long-range) field response simulated with the Garfield program 
for two induction and one collection planes.
The $z$-axis scale is logarithmic ($\propto{\rm sgn}(i)\log(|i|))$.
The wire of interest 
is assumed to locate at position zero. When a cloud of ionization electrons are drifting through a particular transverse position, the waveform on the wire of interest is shown in z-axis along the x-axis (drift time). Obviously, as the magnitude of transverse positions are large, the induced signal becomes small.}
\includegraphics[width=0.7\textwidth]{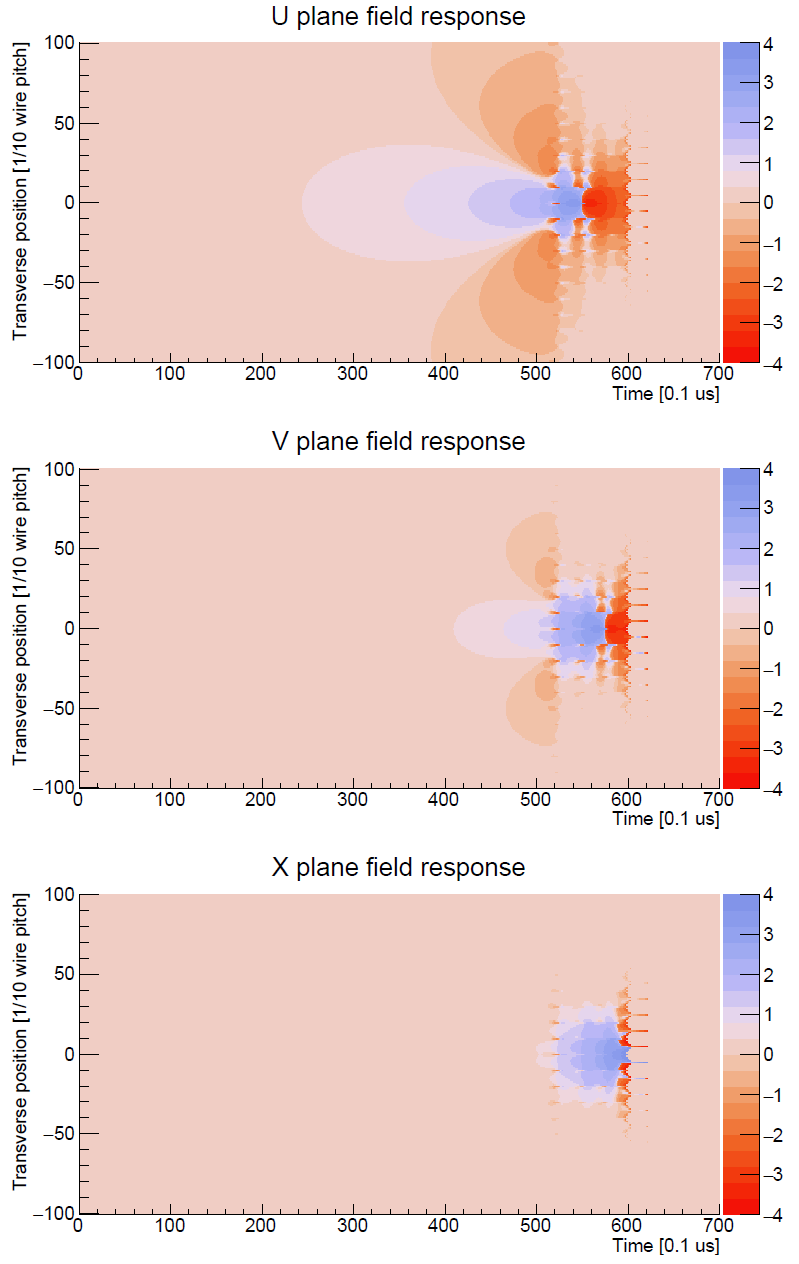}
\end{dunefigure}

 Given the above configuration, the field response function can then be calculated in Garfield
  for each individual wire  for an
  electron starting from any position within the region of simulation. 
  The field response functions for a range $\pm\,$10 wires on both sides of the central wire (covering 21 wires in total) are 
  recorded and stored for later
  application in the TPC detector signal simulation.
  Figure~\ref{field_resp} shows the simulated field response.

  Following the earlier work in MicroBooNE~\cite{Adams:2018dra}, the TPC detector signal simulation
  is implemented in the software package Wire-Cell Toolkit~\cite{ref:wire_cell_toolkit,ref:full_simulation}, which is 
  further interfaced with \dword{larsoft}. This simulation procedure has been validated in the MicroBooNE experiment~\cite{Adams:2018gbi}. In
  the following, we summarize the major features. The TPC signal simulation takes input from the \dword{geant4}-simulated energy deposition when particles traverse the detector, and outputs digitized waveforms on the \dword{fe} electronics.
   A data-driven, analytical simulation of the inherent electronics noise is also performed. 
Figure~\ref{fig:simevent} shows the example waveform for minimum ionizing particles traveling parallel to 
the wire plane, but perpendicular to the wire orientation.  

\begin{dunefigure}
[Waveform for minimum ionizing particles traveling parallel to the wire plane]
{fig:simevent}
{Waveform for minimum ionizing particles traveling parallel to the wire plane. For different
        wire plane, the corresponding track is assumed to travel perpendicular to the wire orientation.}
\includegraphics[width=0.6\textwidth]{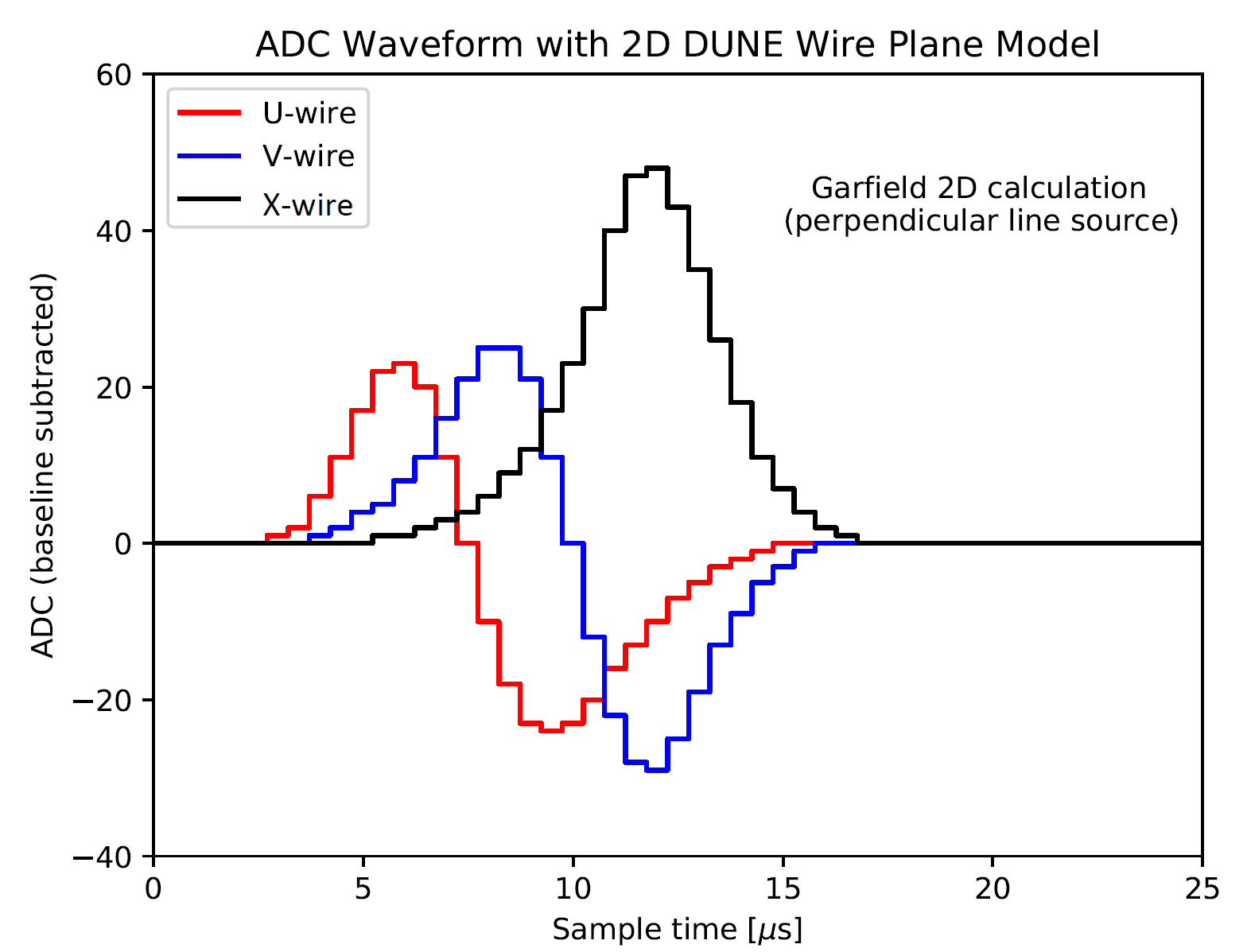}
\end{dunefigure}

The signal simulation, i.e., the \dword{adc} waveform on a given channel, 
\begin{equation}
  \label{eq:sim-convolution}
    M = (Depo \otimes Drift \otimes Duct + Noise) \otimes Digit, 
\end{equation}
is conceptually a convolution of five functions:
\begin{description}
\item[$Depo$] represents the initial distribution of the ionization electrons created by energy depositions in space and time as
discussed in Section~\ref{sec:larg4}.
\item[$Drift$] represents a function that transforms an initial charge cloud to a distribution of electrons arriving at the wires. Physical processes related to drifting, including attenuation due to impurities, diffusion and possible distortions to the nominal applied \efield, are applied in this function.
\item[$Duct$] is a family of functions, each is a convolution $F \otimes E$ of the field response functions $F$ associated with the sense wire and the element of drifting charge and the electronics response function $E$ corresponding to the shaping and amplification of the \dword{fe} electronics. More details can be found in Section~\ref{sec:tools-mc-daq}.
\item[$Digit$] models the digitization electronics according to a given sampling rate, resolution, and dynamic voltage range and baseline offset resulting in an \dword{adc} waveform. 
\item[$Noise$] simulates the inherent electronics noise by producing a voltage level waveform from a random sampling of a Rayleigh distributed amplitude spectrum and uniformly distributed phase spectrum.  The noise spectra used are from measurements with the \dword{pdsp} detector after software noise filters, which have excess (non-inherent) noise effects removed.
\end{description}
These functions are defined over broad ranges and with fine-grained resolution. The resolutions are set by the variability (sub millimeter) and extent (several centimeters) of the field response functions and the sampling rate of the digitizer (\SI{0.5}{\micro\second}). Their ranges are set by the size of the detector (several meters) and the length of time over which one exposure is digitized (several milliseconds).

\subsection{Data Acquisition Simulations and Assumptions}
\label{sec:tools-mc-daq}


  The electrons ($\sim$5300 electrons per mm for MIP signals) on each wire are converted into raw wire signal (\dword{adc} vs time) by convolution with the field response and electronics response, which is implemented in the Wire-Cell Toolkit software package~\cite{ref:wire_cell_toolkit}.
  The \dword{asic} electronics response was simulated with the BNL SPICE~\cite{spice} simulation.  For most samples, the \dword{asic} gain was set to \SI{14}{mV/fC} and the shaping time was set to \SI{2}{\micro\second}. There are several considerations in choosing the \SI{2}{\micro\second} shaping time setting (out of 0.5, 1.0, 2.0, 3.0 \SI{}{\micro\second}):
  \begin{itemize}
\item Since the digitization frequency is at 2 MHz, an anti-aliasing filter to ensure the satisfaction of the Nyquist theorem is required. This essentially excludes the \SI{0.5}{\micro\second} shaping time, which is not enough  to ensure complete anti-aliasing. 
\item A smaller shaping time in principle leads to a slightly better two-peak separation. However, since the drifting time of ionization electrons through one wire plane is about \SI{3}{\micro\second}, the difference between \SI{1}{\micro\second} and \SI{2}{\micro\second} shaping time is limited. 
\item The electronics noise, as parameterized by the standard deviation of the ADC values on each sample, is slightly lower for the \SI{2}{\micro\second} and \SI{3}{\micro\second} shaping-time settings than that of the \SI{1}{\micro\second} (by about 10\% or so). 
\end{itemize}
Te digitization is performed by a 12-bit ADC, which covers a range of about \SI{1.6}{V}. The number of bits is chosen so that the intrinsic noise
introduced by the digitization is negligible.  The intrinsic noise level was set to around 2.5 \dword{adc} RMS, based on extrapolation 
  from the MicroBooNE experiment~\cite{Acciarri:2017sde}. This value was further validated in \dword{pdsp}. 
  Figure~\ref{elec_resp} shows the expected electronics shaping functions.

\begin{dunefigure}
[ASIC's electronics shaping functions]
{elec_resp}
{The shaping functions of the Front-End ASIC, shown for the four shaping
time settings at \SI{14}{mV/fC} gain.}
\includegraphics[width=0.7\textwidth]{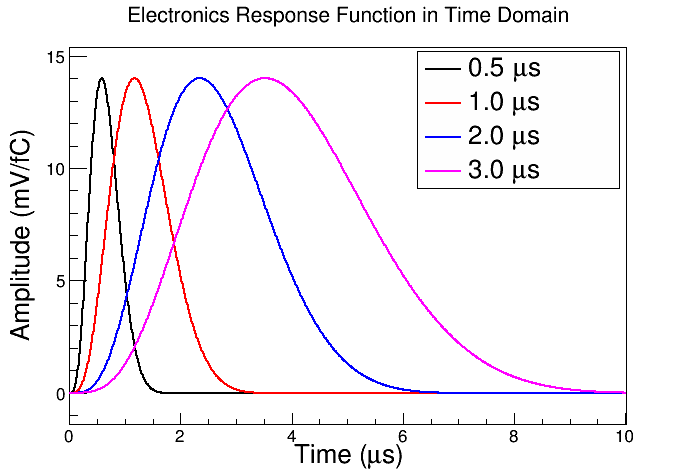}
\end{dunefigure}

The \dword{pd} electronics simulation separately generates waveform 
for each channel (\dword{sipm}) of a \dword{pd} that has been hit by photons.
Every detected photon 
appears as a single \phel{} pulse
(with the shape taken from~\cite{http://lss.fnal.gov/archive/2015/pub/fermilab-pub-15-488-nd-ppd.pdf:2015gov}) on a randomly selected channel
(belonging to the \dword{pd} in which the photon was registered).
Then dark noise (with the rate of \SI{10}{Hz}) and 
line noise (Gaussian noise with the RMS of $2.6$ \dword{adc} counts) are added.
Each photon (or a dark-noise pulse) has a probability of appearing
as $2$ \phel{}s on a waveform (the cross-talk probability is $16.5~\%$).
The final step of the digitization process is recording only fragments
of the full simulated waveform that have a signal in them.
This is accomplished by passing the waveform through a hit finder
described in Sec.~\ref{sec:gaushit} 
and storing parts of the waveform corresponding to the hits found.

\section{Event Reconstruction in the \dword{fd}}
\label{sec:tools-fdreco}

This section describes various reconstruction algorithms used to reconstruct events in the \dword{fd} TPC. A successful \lartpc reconstruction needs to deliver reconstructed tracks and showers, particle and event identification, particle momentum and event energy. The reconstruction starts with finding signals on each wire above a threshold and building ``hits'' out of each pulse. All the LArTPC \threed reconstruction algorithms share the same principle. The $x$ coordinate is determined by the drift time and the $y$ and $z$ coordinates are determined by the intersection of two wires on different planes with coincident hits. There are currently three different reconstruction approaches in the DUNE reconstruction package. The \twod$\rightarrow$\threed reconstruction approach starts with clustering together nearby hits on each plane, 
followed by the use of time information to match \twod clusters between different planes to form \threed tracks and showers. Examples of this approach include TrajCluster and \dword{pandora}.  The direct \threed approach reconstructs \threed points directly from hits and then proceeds to perform pattern recognition using those \threed points. Examples of this approach include SpacePointSolver and \dword{wirecell}. The third approach uses a deep-learning technique, known as a \dword{cnn}. There are several tools needed to complete the task of LArTPC reconstruction. These tools include track fitter (\dword{pma} or KalmanFilter), calorimetry, \dword{pid} and track momentum reconstruction using range for contained tracks or multiple Coulomb scattering for exiting tracks. In addition to the TPC reconstruction, the \dword{pd} reconstruction provides trigger and $t_0$ information for non-beam physics.

\subsection{TPC Signal Processing}\label{sec:tpc_sp}

The raw data are in the format of \dword{adc} counts as a function of TPC ticks (\SI{0.5}{\micro\second} on each channel. The signal has a 
unipolar shape for a collection wire and a bipolar shape for an induction wire. The first step 
in the reconstruction is to reconstruct the distribution of ionization electrons arriving at the anode plane. 
This is achieved by passing the raw data through a deconvolution algorithm. In real detectors, excess 
noise may exist and require removal 
through a dedicated noise filter~\cite{Acciarri:2017sde}. 

The deconvolution technique was introduced to LArTPC signal processing in the context ArgoNeuT 
data analysis~\cite{Baller:2017ugz}. The goal of the deconvolution is to ``remove'' the impact of field and 
electronics responses from the measured signal along the time dimension in order to reconstruct the number of ionized
electrons. This technique has the advantages of being robust and fast, and is an essential step in the overall drifted-charge profiling process. This 1D deconvolution procedure was improved to a \twod deconvolution procedure 
by the MicroBooNE collaboration~\cite{Adams:2018dra,Adams:2018gbi}, which further took into account the long-range 
induction effects in the spatial dimension. Two-dimensional software filters (channel and time) are implemented to 
suppress high-frequency noise after the deconvolution procedure. For induction plane signals, regions of interest 
(ROIs) are selected to minimize the impact of electronics noise. More details of this algorithm can be found in~\cite{Adams:2018dra}.

This procedure, implemented in the Wire-Cell toolkit software package~\cite{ref:wire_cell_toolkit},  has been used in the TPC signal processing in \dword{pdsp}. Figure~\ref{pDUNE_sp_wf}
shows an example induction U-plane waveform before and after the signal processing procedure. The bipolar 
shape is converted into a unipolar shape after the \twod deconvolution. Figure~\ref{pDUNE_sp_example}
shows the full \twod image of induction U-plane signal from a \dword{pdsp} 
event~\cite{ref:pdune_signal_processing}. The measured signal (left) has a bipolar shape with red (blue) color 
representing positive (negative) signals. The deconvolved signal after 
the \twod deconvolution procedure (right) represents the reconstructed 
distribution of ionization electrons arriving at the anode wire plane. The 
deconvolved signal becomes unipolar, and the long-range 
induction effect embedded in the field response is largely removed. 

\begin{dunefigure}
[Measured and deconvolved waveform from an induction U-plane channel of ProtoDUNE-SP]
{pDUNE_sp_wf}
{An example of measured (black) and deconvolved waveform from an induction U-plane channel of \dword{pdsp}
before and after the signal processing procedure. For the measured waveform, the unit is \dword{adc}. For the deconvolved waveform, the unit is number of electrons after scaling down by a factor of 125.
Bipolar signal shapes are converted into unipolar signal shapes after \twod deconvolution.}
\includegraphics[width=0.95\textwidth]{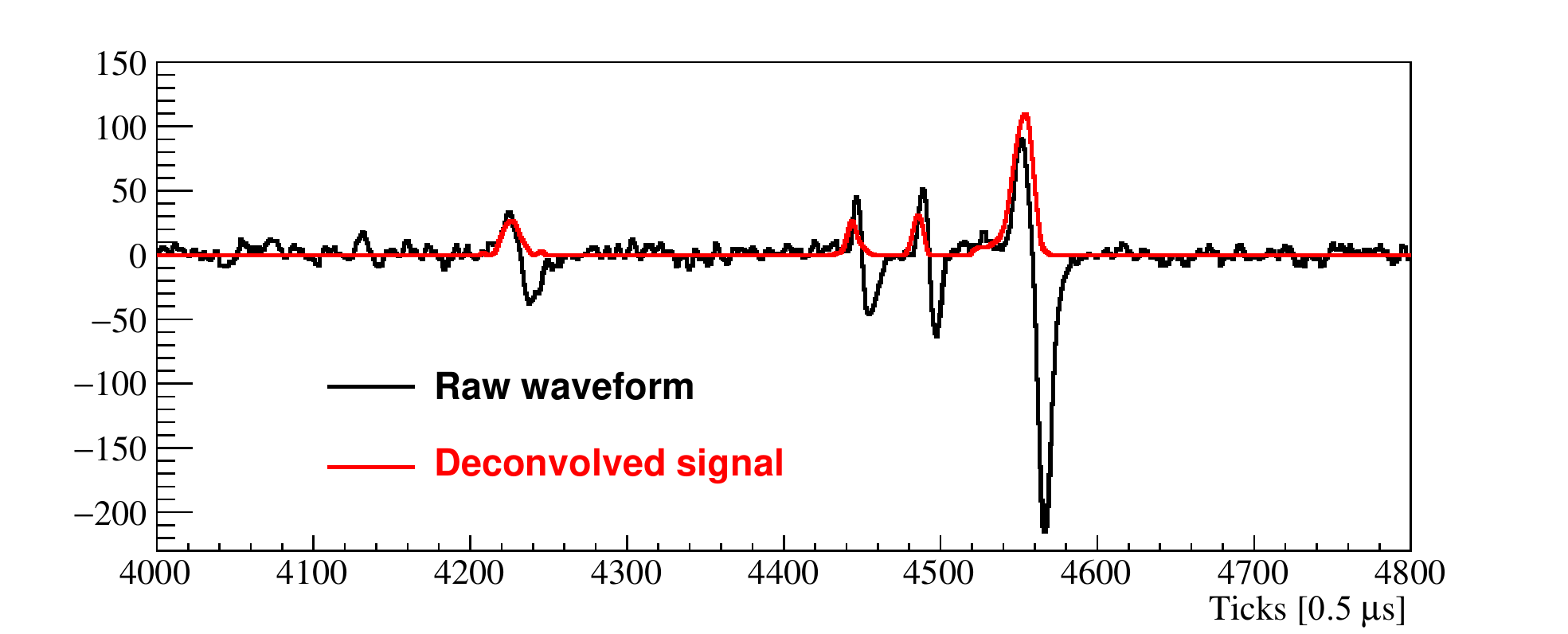}
\end{dunefigure}

\begin{dunefigure}
[Raw  and deconvolved induction U-plane signals from a ProtoDUNE-SP event]
{pDUNE_sp_example}
{Comparison of raw (left) and deconvolved induction U-plane signals (right) before and after 
the signal processing procedure from a \dword{pdsp} event. The bipolar shape with red (blue) color representing
positive (negative) signals is converted to the unipolar shape after the \twod deconvolution.}
\includegraphics[width=0.49\textwidth]{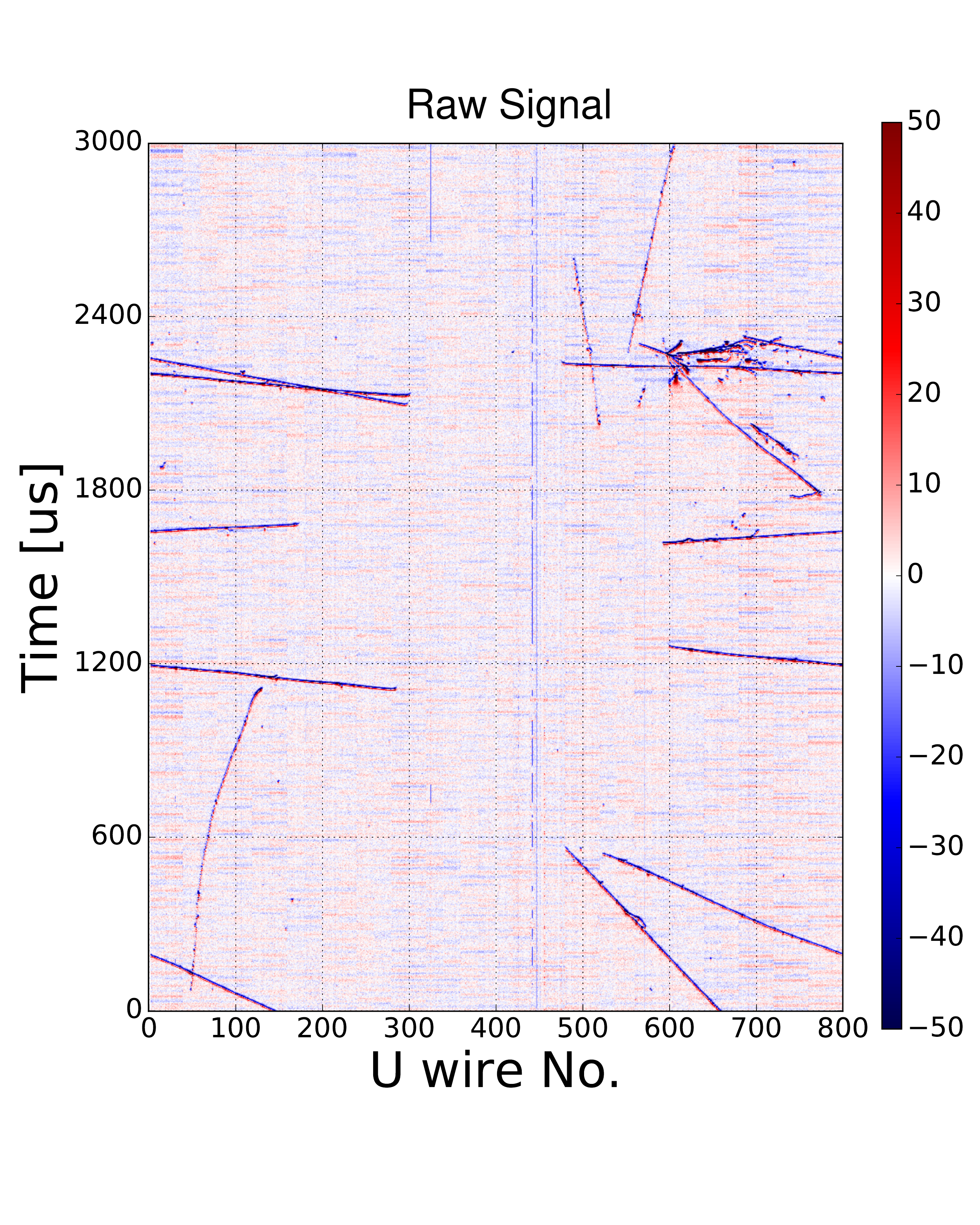}
\includegraphics[width=0.49\textwidth]{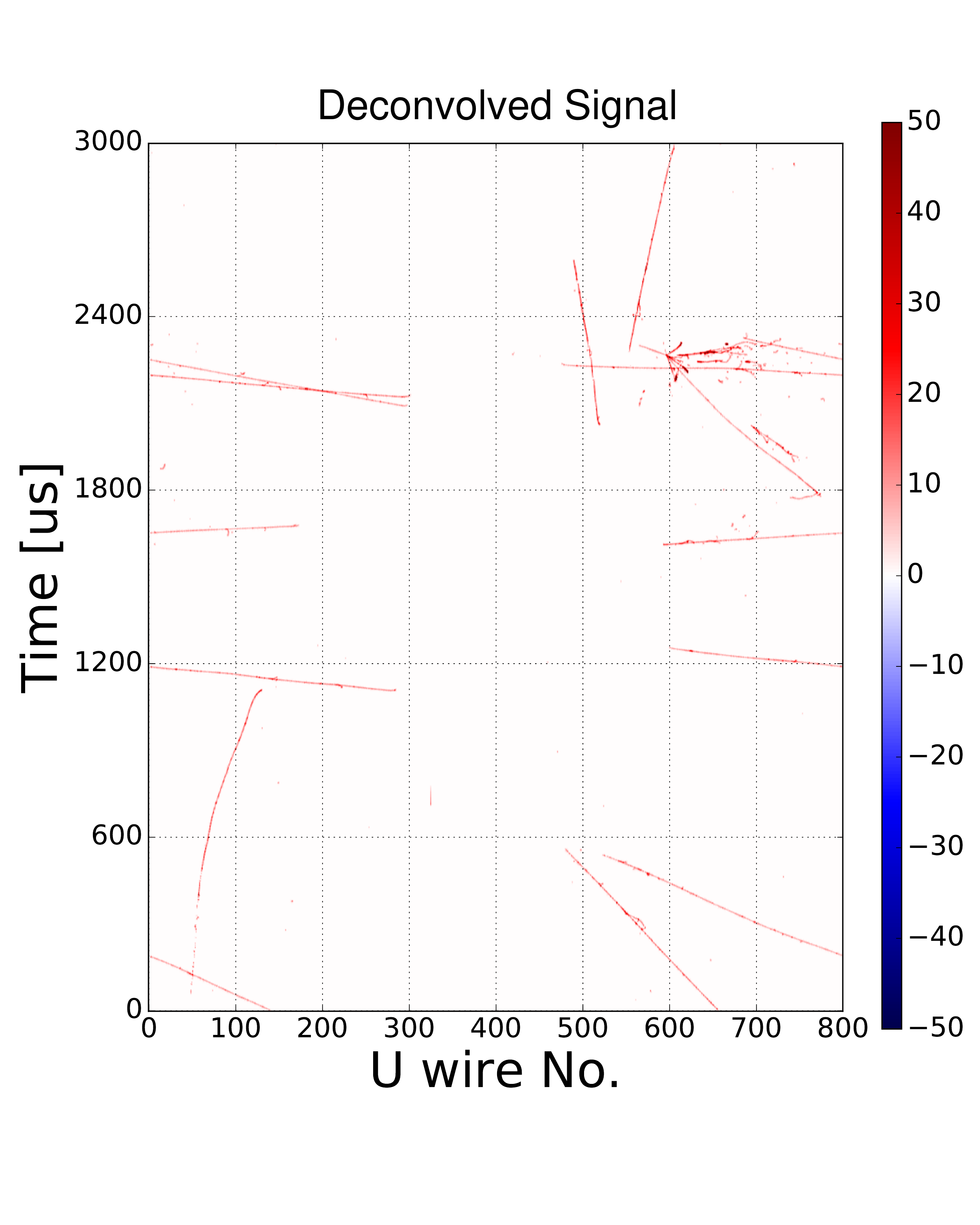}\end{dunefigure}

\subsection{Hit and Space-Point Identification}

\subsubsection{Gaussian Hit Finder}\label{sec:gaushit}

The reconstruction algorithms currently employed by \larsoft are based on finding hits on the deconvolved waveforms for each plane. A key assumption is that the process of deconvolution will primarily result in Gaussian-shaped charge deposits on the waveforms and this drives the design of the Gaussian hit finder module. Generally, the module loops over the input deconvolved waveforms and handles each in three main steps: first it searches the waveforms 
for candidate pulses, it then fits these candidates 
to a Gaussian shape and, finally, it places the resulting hit 
in the output hit collection. Not all charge deposits will be strictly Gaussian shaped, for example a track can emit a delta ray and it can take several wire spacings before the two charge depositions are 
completely separated. Alternatively, a track can have a trajectory at large angles to the sense wire plane creating a charge deposition over a large number of waveform ticks. The candidate peak-finding stage of the hit finder attempts to resolve the individual hits in both of these cases, still under an assumption that the shape of each individual charge deposition is 
Gaussian. If this results in candidate peak trains that are ``too long'' then special handling  breaks these into a number of evenly-spaced hits and bypasses the hit-fitting stage. 

Figure~\ref{pDUNE_sp_hits} displays the results of the Gaussian hit finder for the case of two or three hits only barely separated in \dword{pdsp} data. In this figure the deconvolved waveform is shown in blue, the red line represents the fit of the candidate peak to two or three Gaussian shapes, the crosses represent the centers of the fit peaks, the pulse heights above the waveform baseline and their fit widths. 

\begin{dunefigure}
[An example of reconstructed hits in ProtoDUNE-SP data]
{pDUNE_sp_hits}
{An example of reconstructed hits in \dword{pdsp} data. The deconvolved waveform is shown in blue, the red line represents the fit of the candidate peak to two or three Gaussian shapes, the crosses represent the centers of the fit peaks, the pulse heights above the waveform baseline and their fit widths.}
\includegraphics[width=\textwidth]{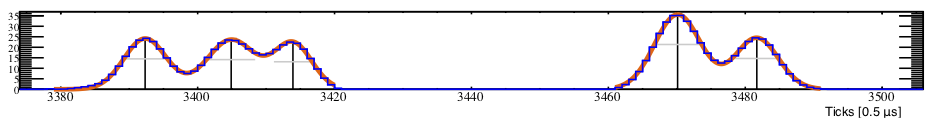}
\end{dunefigure}

\subsubsection{Space Point Solver}

\begin{figure}
\includegraphics[width=.5\linewidth]{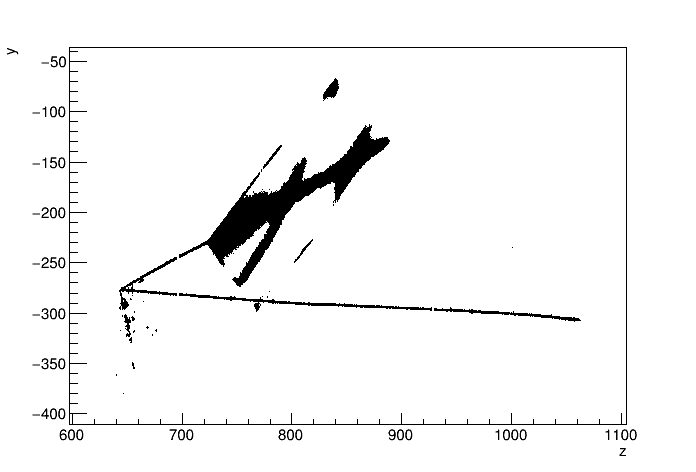}
\includegraphics[width=.5\linewidth]{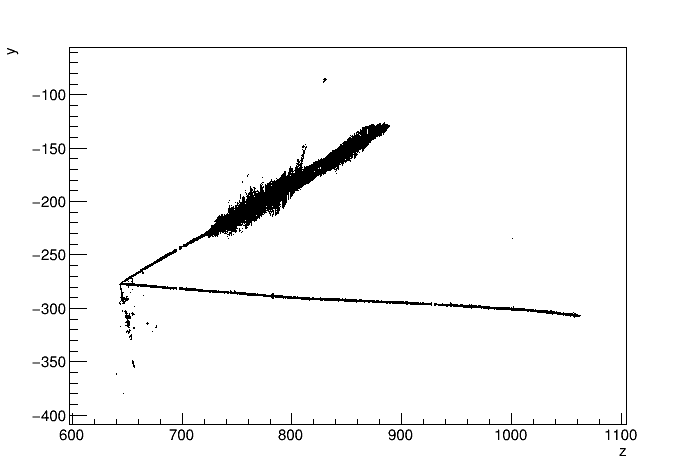}\\
\makebox[.5\linewidth][c]{(a) All coincidences}
\makebox[.5\linewidth][c]{(b) Without regularization}\\
\includegraphics[width=.5\linewidth]{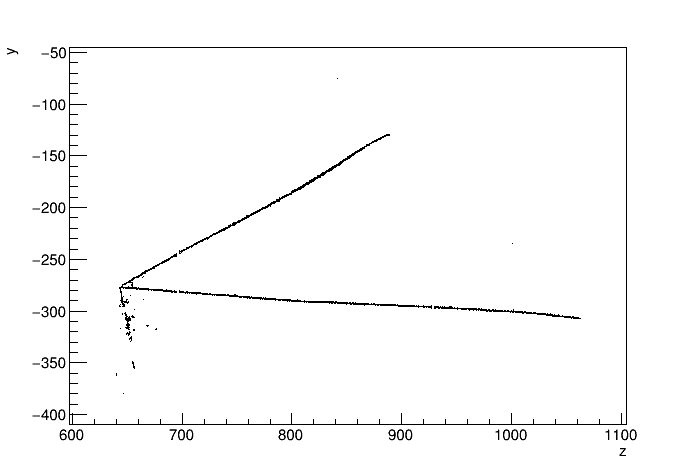}
\includegraphics[width=.5\linewidth]{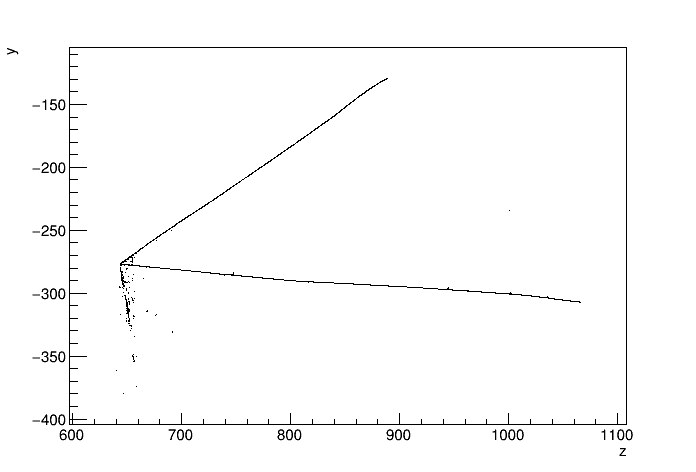}
\makebox[.5\linewidth][c]{(c) With regularization}
\makebox[.5\linewidth][c]{(d) True charge distribution}\\

\caption[Event displays of SpacePointSolver performance]{Performance of SpacePointSolver on a simulated \dword{fd} neutrino interaction. The first panel shows the position of all triplet coincidences in the $zy$ view (looking from the side of the detector), displaying multiple ambiguous regions. The second and third panels show the solution with and without regularization, the regularization disfavoring various erroneous scattered hits. The final panel shows the true charge distribution, demonstrating 
the fidelity of the regularized reconstruction.}

\label{fig:spacepoint}
\end{figure} 

The SpacePointSolver algorithm aims to transform the three \twod views provided by the wire planes into a single collection of \threed ``space points.''

First, triplets of wires are found with hits that are coincident in time within a small window (corresponding to \SI{2}{mm} in the drift direction) and where the crossing positions of the wires are consistent within \SI{3.55}{mm}. In some cases a collection wire hit may have only a single candidate pair of induction hits and the space point can be formed immediately. Often though, there are multiple candidate triplets, for example when two tracks are overlapped as seen in one view.

SpacePointSolver resolves these ambiguities by distributing the charge from each collection wire hit between the candidate space points so as to minimize the deviations between the expected and observed charges of the induction wire hits

\begin{equation}
\chi^2 = \sum_i^{\rm wires} \left(q_i - \sum_j^{\rm points}T_{ij}p_j\right)^2
\end{equation}

where $q_i$ is the charge observed in the $i^{\rm th}$ induction hit, $p_j$ is the solved charge at space point $j$, and $T_{ij}\in\{0,1\}$ encodes whether space point $j$ is coincident in space and time with wire hit $i$.

The minimization is subject to the condition that each predicted charge $p_j\ge0$, and that the total predicted charge for each collection wire hit exactly matches observations:

\begin{equation}
\sum_j^{\rm points}U_{jk}p_j=Q_k
\end{equation}

where $Q_k$ is the charge observed on the $k^{\rm th}$ collection wire, and $U_{jk}$ encodes the coincides of space point charges with the collection wires.

The problem as formulated is convex and can thus be solved exactly in a deterministic fashion. A single extra term can be added to the expression while retaining this property:

\begin{equation}
\chi^2 \to \chi^2 - \sum_{ij}^{\rm points}V_{ij}p_ip_j .
\end{equation}

By setting $V_{ij}$ larger for neighboring points this term acts as a regularization such that solutions with a denser collection of space points are preferred. The $V$ function is chosen empirically to have an exponential fall-off with constant \SI{2}{cm}.

Figure \ref{fig:spacepoint} shows the performance of this algorithm on a sample \dword{fd} \dword{mc} event, demonstrating good performance at eliminating spurious coincidences, and the importance of the regularization term.

SpacePointSolver was developed with the intention of acting as the first stage of a fully \threed reconstruction for \dword{fd} neutrinos, but it has been successfully put to use in a more restricted role to solve the disambiguation problem in \dword{protodune}. The full problem is solved, but for this application the information retained is restricted to the drift volume to which the corresponding space points for each induction hit are assigned. This technique correctly resolves more than 99\% of hits while requiring less CPU time than the standard disambiguation algorithm.

The outcome of the SpacePointSolver reconstruction, which associates a \threed point with three hits on three wire planes, is used in the process of disambiguation for \dword{protodune} and has been tested and used as well for \dword{fd}. This process of disambiguation determines which wire segment corresponds to the energy deposited by the particle in the TPC, since the induction wires are wrapped in the \dword{fd} TPC design in order to save cost on electronics and minimize dead regions between \dword{apa}s, which as a consequence produces that multiple induction wire segments will be read out by the same electronic channel.

\subsection{Hit Clustering, Pattern Recognition and Particle Reconstruction}

There are different approaches for hit clustering, pattern recognition and particle reconstruction that are being explored in the context of DUNE \dword{fd} interactions. The main ones are described in this section.  

\subsubsection{Line Cluster}\label{sec:LineCluster}
The intent of the Line Cluster algorithm is to construct \twod line-like clusters using local information. The algorithm was originally known as Cluster Crawler. The ``Crawler'' name is derived from the similarity of this technique to ``gliders'' in \twod cellular automata. The concept is to construct a short line-like ``seed'' cluster of proximate hits in an area of low hit density where hit proximity is a good indication that the hits are indeed associated with each other. Additional nearby hits are attached to the leading edge of the cluster if they are similar to the hits already attached to it. The conditions are that the impact parameter between a prospective hit and the cluster projection is similar to those previously added and the hit charge is similar as well. These conditions are moderated to include high charge hits that are produced by large $dE/dx$ fluctuations and the rapid increase in $dE/dx$ at the end of stopping tracks while rejecting large charge hits from $\delta$-rays.
Seed clusters are formed at one end of the hit collection so that crawling in only one direction is sufficient. LineCluster uses disambiguated 
hits as input and produces a new set of refined hits. More details on the Line Cluster algorithm can be found in~\cite{ref:linecluster}.

\subsubsection{TrajCluster}\label{sec:TrajCluster}
TrajCluster reconstructs \twod trajectories in each plane. It incorporates elements of pattern recognition and Kalman Filter fitting. The concept is to construct a short ``seed'' trajectory of nearby hits. Additional nearby hits are attached to the leading edge of the trajectory if they are similar to the hits already attached to it. The similarity requirements use the impact parameter between the projected trajectory position and the prospective hit, the hit width and the hit charge. This process continues until a stopping condition is met such as lack of hits, an abnormally high or low charge environment, or encountering a \twod vertex or a Bragg peak.

\twod vertices are found between trajectories in each plane. The set of \twod vertices is matched between planes to create \threed vertices. A search is made of the ``incomplete'' \threed vertices, those that are only matched in two planes, to identify trajectories in the third plane that were poorly reconstructed.

Two recent additions to TrajCluster are matching trajectories in \threed and tagging of shower-like trajectories. More details on the TrajCluster algorithm can be found in~\cite{ref:trajcluster}.

\subsubsection{Pandora}\label{sec:Pandora}

The \dword{pandora} software development kit~\cite{Marshall:2015rfa} was created to address the problem of identifying energy deposits from individual particles in fine-granularity detectors, using a multi-algorithm approach to solving pattern-recognition problems. Complex and varied topologies in particle interactions, especially with the level of detail provided by \lartpc{}s, are unlikely to be solved successfully by a single clustering algorithm. Instead, the \dword{pandora} approach is to break the pattern recognition into a large number of decoupled algorithms, where each algorithm addresses a specific task or targets a particular topology. The overall event is then built up carefully using a chain of many tens of algorithms. The \dword{pandora} multi-algorithm approach has already been applied to \lartpc{} detectors, and has been successfully used in different analyses for the automated reconstruction of cosmic-ray muons and neutrino interactions in the MicroBooNE experiment~\cite{Acciarri:2017hat} as well as test beam interactions in the \dword{pdsp} detector (see Section~\ref{sec:Pandora:ProtoDUNE}).

The input to the \dword{pandora} pattern recognition is a list of reconstructed and disambiguated \twod hits, alongside detector information (such as dimensions, unresponsive or dead material regions). The specified chain of pattern-recognition algorithms is applied to these input hits (once translated into native \dword{pandora} \twod hits). The results of the pattern recognition are persisted in the \dword{art}/\dword{larsoft} framework, with the major output being a list of reconstructed \threed particles (termed \dwords{pfparticle}). A \dword{pfparticle} corresponds to a distinct track or shower in the event, and has associated objects such as collections of \twod hits for each view (Clusters), \threed positions (SpacePoints) and a reconstructed Vertex position that defines its interaction point or first energy deposit. Navigation along \dword{pfparticle} hierarchies is achieved using the \dword{pfparticle} interface, which connects parent and daughter \dword{pfparticle}s, providing a particle flow description of the interaction. The identity of each particle is currently not reconstructed by \dword{pandora}, but \dword{pfparticle}s are instead characterized as track-like or shower-like based on their topological features. 

The main stages of the \dword{pandora} pattern recognition chain are outlined below, and are illustrated in Figure~\ref{reco_steps}. Note that both the individual pattern recognition algorithms and the overall reconstruction strategy are under continual development and will evolve over time, with a current emphasis 
on the inclusion of machine-learning approaches to drive decisions in some key algorithms. The current chain of pattern-recognition algorithms has largely been tuned for neutrino interactions from the \fnal Booster Neutrino Beam; however, the algorithms are designed to be generic and easily reusable, and they are in the process of being adapted for neutrino interactions in the energy regime of DUNE. A more detailed description of the algorithms can be found in~\cite{Acciarri:2017hat}.

\begin{enumerate}
\item{\bf Input hits:} The input list of reconstructed and disambiguated \twod hits are translated into native \dword{pandora} \twod hits and separated into the different views and into ``drift volumes'', defined as the regions of the detector with a common drift readout.
\item{\bf \twod track-like clusters:}  The first phase of the \dword{pandora} pattern recognition is track-oriented \twod clustering, creating ``proto-clusters'' that represent continuous, unambiguous lines of \twod hits. This early clustering phase is careful to ensure that the proto-clusters have high purity (i.e., represent energy deposits from exactly one true particle) even if this means they are initially of low completeness (i.e., only contain a small fraction of the total hits within a single true particle). A series of cluster-merging and cluster-splitting algorithms then examine the \twod proto-clusters and try to extend them, making decisions based on topological information, aiming to improve completeness without compromising purity.
\item{\bf \threed vertex reconstruction:} The neutrino interaction vertex is an important feature point. Once identified, any \twod clusters can be split at the projected vertex position, reducing chances of merging particles in any view. Cluster-merging operations also take proximity to the vertex into account, in order to protect primary particles emerging from the vertex region, and ensure good reconstruction performance for interactions with many final-state particles. Pairs of \twod clusters from different views are first used to produce lists of possible \threed vertex positions. These candidate vertices are examined and scored, and the best vertex is selected. \dword{pandora} has developed different algorithms for the selection of the neutrino vertex, including the use of machine-learning approaches in MicroBooNE. Similar approaches can be harnessed in the future for interactions in 
the \dword{fd}, where a score-based approach is currently used.
\item{\bf \threed track reconstruction:} The aim of the \threed track reconstruction is to identify the combinations of \twod clusters (from the different views) that represent the same true, track-like particle. These \twod clusters are formally associated by the construction of a \threed track particle. During this process, \threed information can also be used to improve the quality of the \twod clustering. A \dword{pandora} algorithm considers all possible combinations of \twod clusters, one from each view, and builds (what is loosely termed) a rank-three tensor to store a comprehensive set of cluster-consistency information. This tensor can be queried to identify and understand any cluster-matching ambiguities. \threed track particles are first built for any unambiguous combinations of \twod clusters. Cases of cluster-matching ambiguities are then addressed, with iterative corrections to the \twod clustering being made to resolve the ambiguities and so enable \threed particle creation.
\item{\bf \twod and \threed shower reconstruction:} A series of topological metrics (additional use of some calorimetric information would be desirable in the future) are used to characterize each \twod cluster as track-like or shower-like. This information is analyzed to identify the longest shower-like clusters, which form the ``seeds'' or ``spines'' for \twod and \threed shower reconstruction. A recursive algorithm is used to add shower branches onto each top-level shower seed, then branches onto branches, etc. The \twod showers are then matched between views to form \threed showers, reusing ideas from the \threed track-matching procedure.
\item{\bf \twod and \threed particle refinement and event building:} Following the \threed track and shower reconstruction, a series of algorithms is used to improve the completeness of the reconstructed particles by merging together any nearby particles that are just fragments of the same true particle. Both \twod and \threed approaches are used, 
where a typical approach uses combinations of \twod clusters (from different views) to identify features in \threed, or projects \threed features into each of the \twod views. This is a powerful demonstration of the \dword{pandora} rotational coordinate transformation system, which allows seamless use of \twod and \threed information to drive pattern-recognition decisions. Finally, \threed space points are created for each \twod input hit, and the \threed particle trajectories are used to organize the reconstructed particles into a hierarchy. Final-state particles can be navigated via parent-daughter links, thus reconstructing their subsequent interactions or decays. For neutrino interactions, a top-level reconstructed neutrino particle is created; it represents the primary particle in the hierarchy linking together the daughter final-state particles and provides the information about the neutrino interaction vertex.

\end{enumerate}

\begin{dunefigure}
[Main stages of the PANDORA pattern recognition chain]
{reco_steps}
{Illustration of the main stages of the \dword{pandora} pattern recognition chain: (1) Input Hits; (2) \twod track-like cluster creation and association; (3) \threed vertex reconstruction; (4) \threed track reconstruction; (5) Track/Shower separation; (6) \twod and \threed particle refinement and event building.}
\includegraphics[width=3.7cm, height=5.5cm]{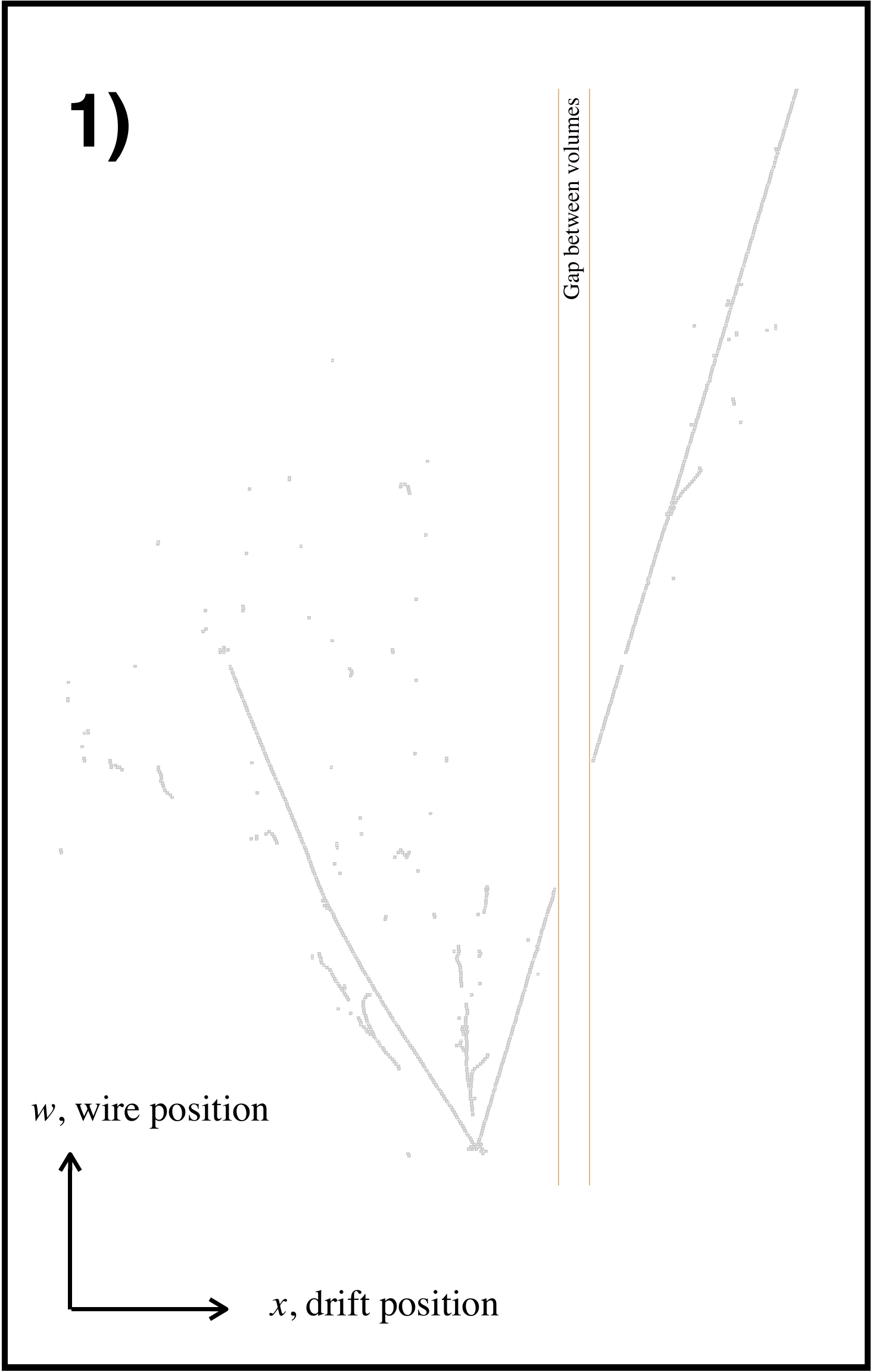}
\includegraphics[width=9.5cm, height=5.5cm]{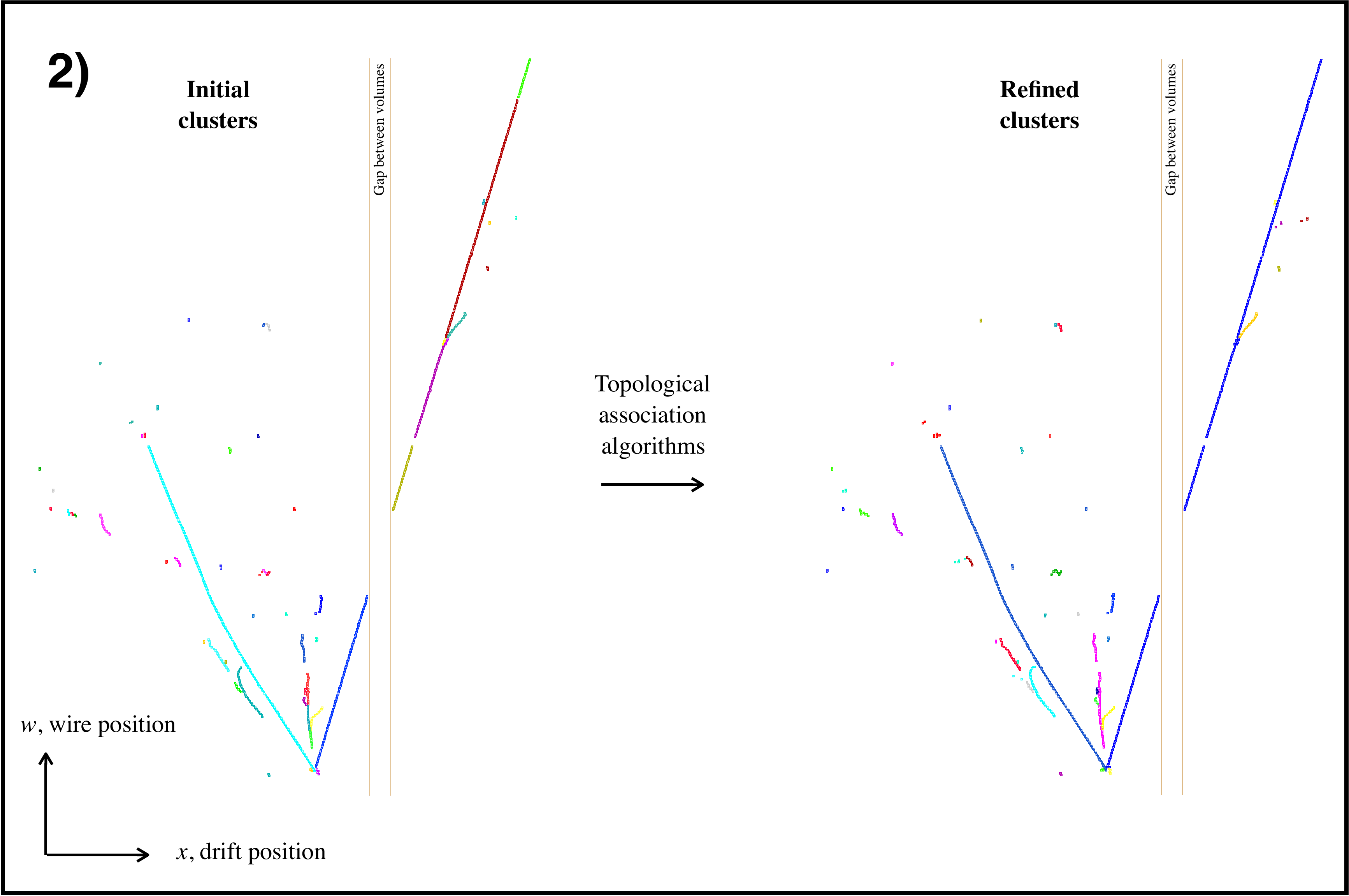}
\includegraphics[width=3.7cm, height=5.5cm]{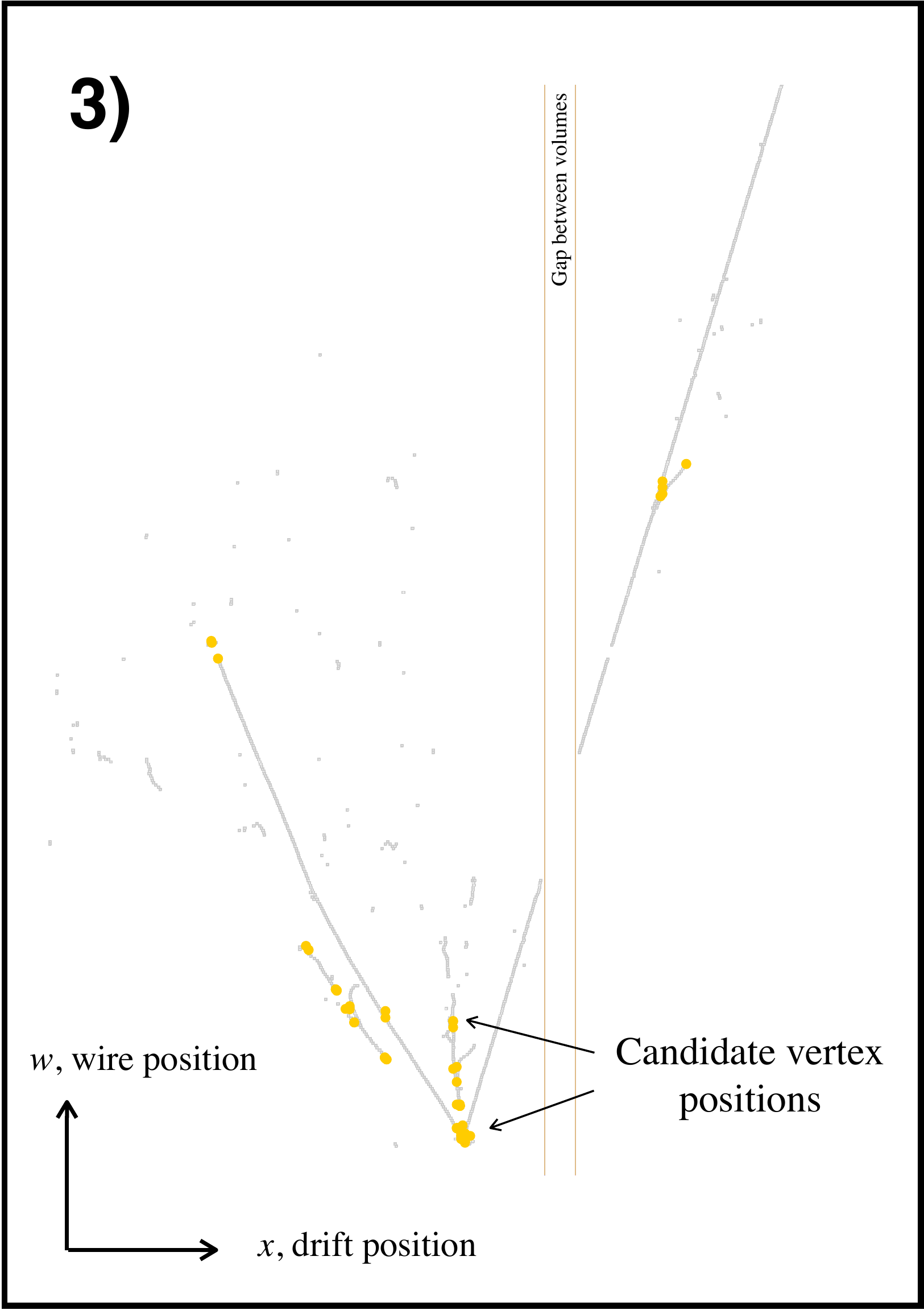}
\includegraphics[width=6.0cm, height=5.5cm]{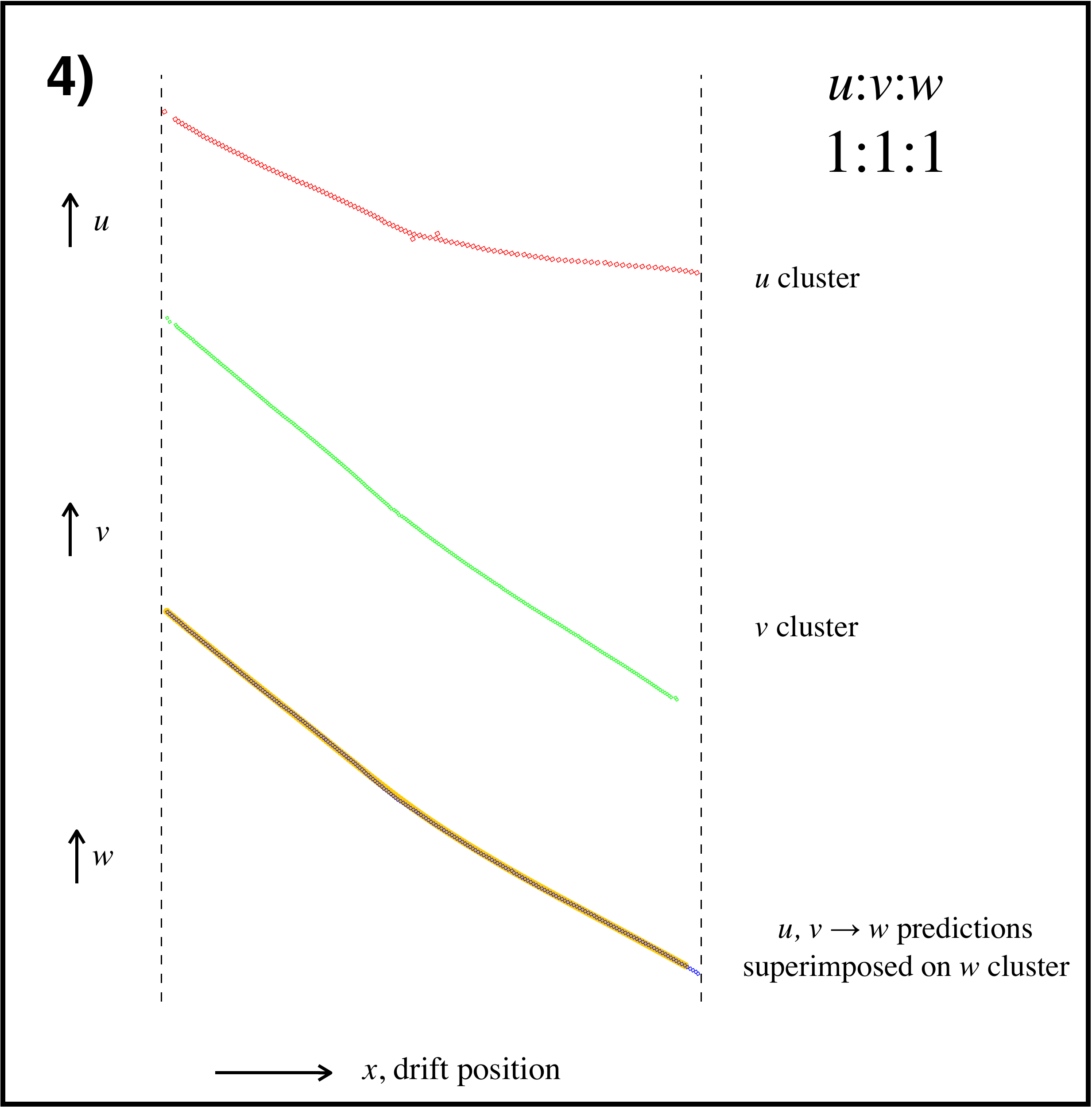}
\includegraphics[width=4.0cm, height=5.5cm]{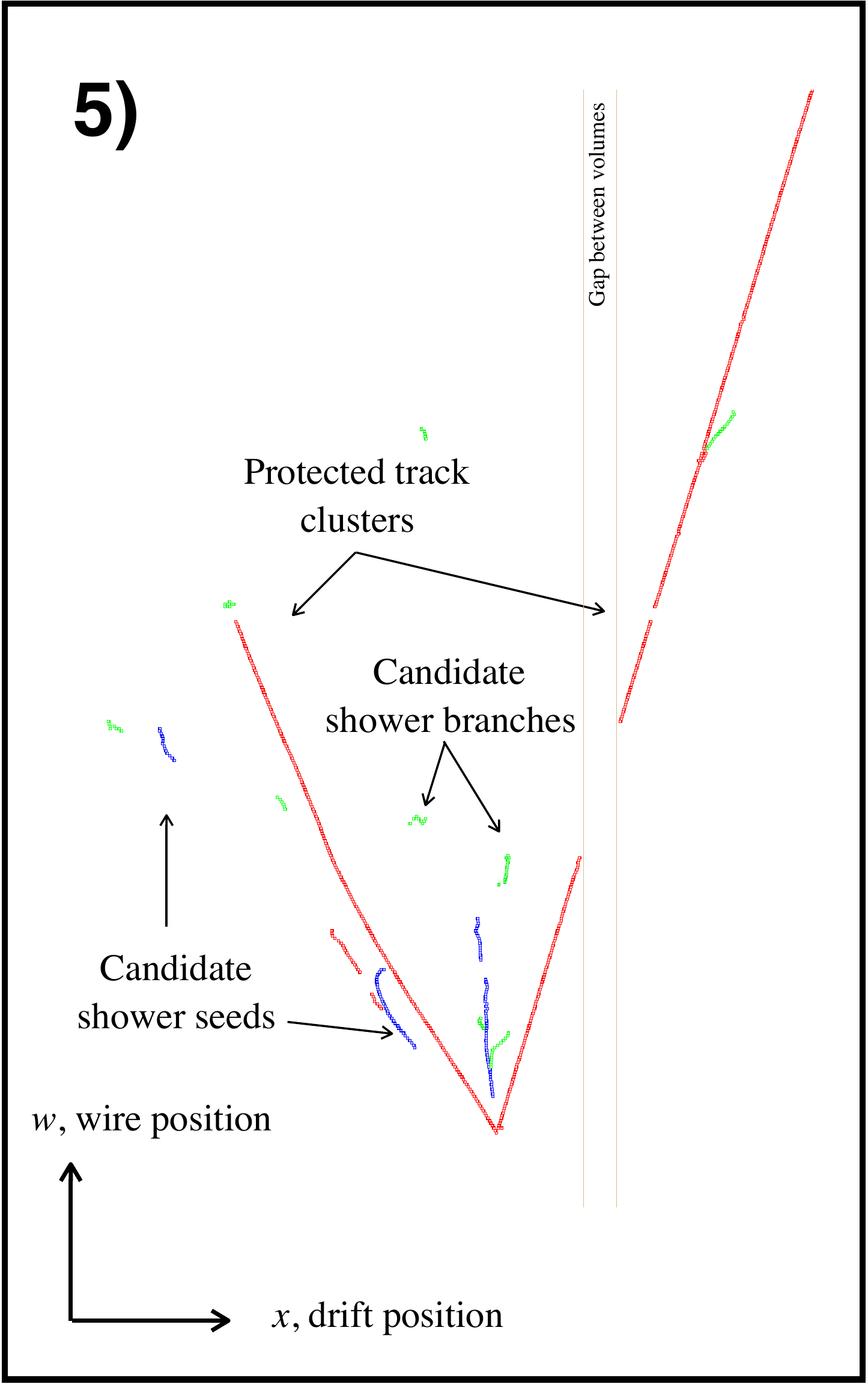}
\includegraphics[width=8.8cm, height=5.5cm]{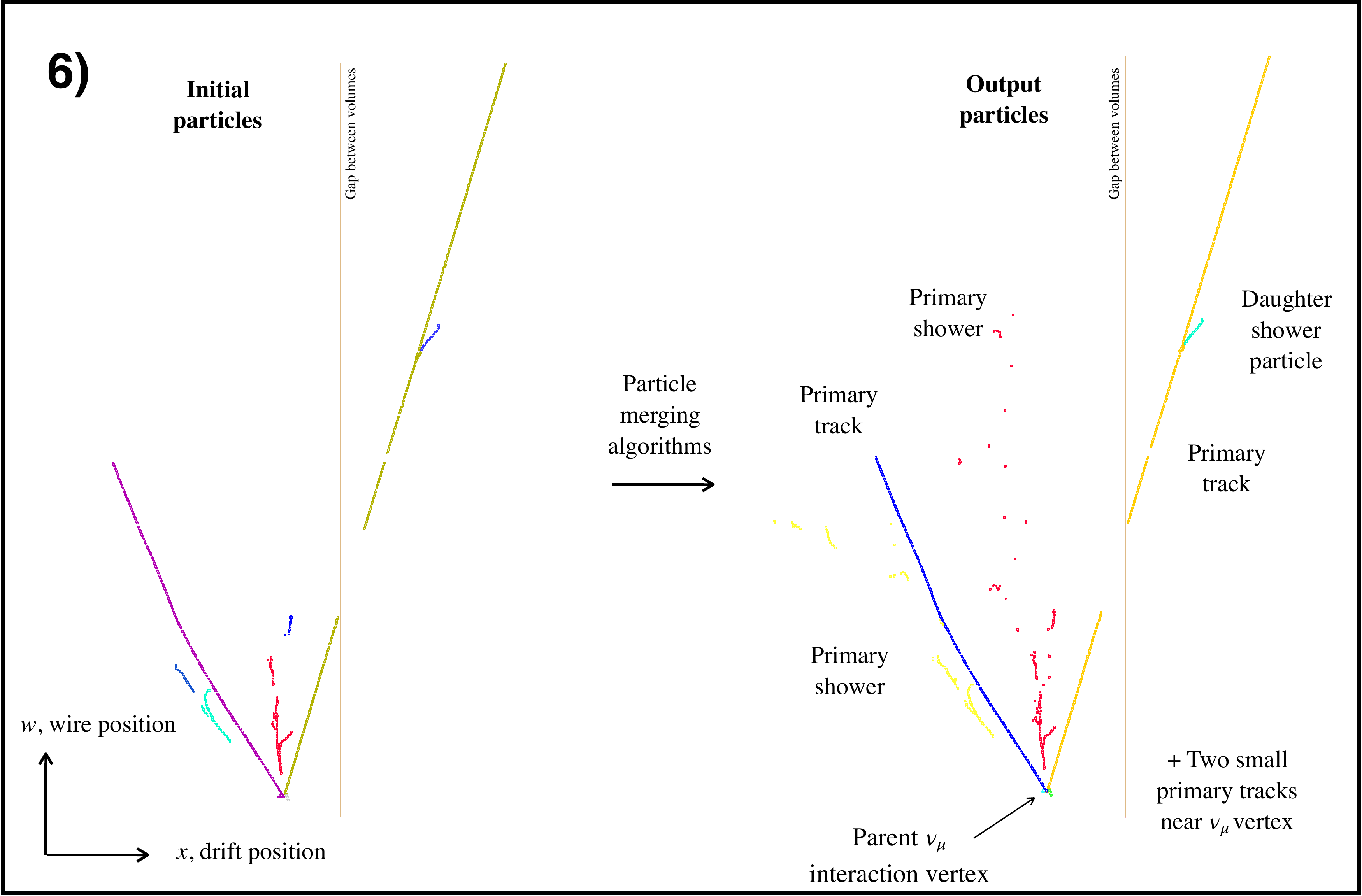}
\end{dunefigure}

The algorithms forming the stages described above can be used in different ways, thanks to the multi-algorithm approach. Currently, two \dword{pandora} reconstruction paths ({\it Pandora Cosmic} and {\it Pandora Neutrino}) have been created, using chains of tens of algorithms each (note that over 130 algorithms and tools are used in total). Although many algorithms are shared between the two paths, the overall algorithm selection results in different key features:
\begin{enumerate}
\item Pandora Cosmic: Strongly track-oriented, optimized for the reconstruction of cosmic-ray muons and their daughter (shower-like) delta rays. 
\item Pandora Neutrino: Optimized for the reconstruction of neutrino or test beam particle interactions, carefully building the event using the reconstructed interaction vertex (protecting particles emerging from it) and including a careful treatment of tracks versus showers. 
\end{enumerate}

These two chains of algorithms are harnessed together to provide a consolidated output in the case of surface detectors exposed to cosmic rays, such as MicroBooNE and \dword{pdsp} (without significant cosmic-ray background, only the Pandora Neutrino algorithm chain is necessary for the \dword{fd}). The overall reconstruction strategy in such detectors is illustrated in Figure~\ref{consolidated_reco}. It starts by running the Pandora Cosmic reconstruction on the entire collection of input hits, then identifies ``clear'' cosmic rays. This identification uses a geometrical approach to tag through-going cosmic rays and examines the consistency of the cosmic rays with the $t_{0}$ appropriate to the neutrino beam spill. Clear cosmic rays are output at this stage. For the remaining ambiguous hits, however, additional stages are required. A \textit{slicing} process is applied to the remaining hits, dividing them into smaller regions (slices) that represent separate, distinct interactions. Each slice is reconstructed using both the Pandora Neutrino and Pandora Cosmic reconstruction chains and the results are compared directly to identify whether the slice corresponds to a cosmic ray or a neutrino interaction (in the case of MicroBooNE) or test beam interaction (in the case of \dword{pdsp}). The consolidated event output is formed of three classes of reconstructed particles: (1) clear cosmic rays, (2) cosmic rays that are spatially and temporally consistent with being a neutrino interaction in the detector (remaining cosmic-rays) and (3) candidate neutrino or test beam interactions.

\begin{dunefigure}
[Schema of PANDORA consolidated output and reconstruction strategy for surface LArTPCs]
{consolidated_reco}
{Schema of the \dword{pandora} consolidated output and overall reconstruction strategy for surface \dwords{lartpc} such as \dword{microboone} and \dword{pdsp}. See text for more details.}
\includegraphics[width=0.65\textwidth]{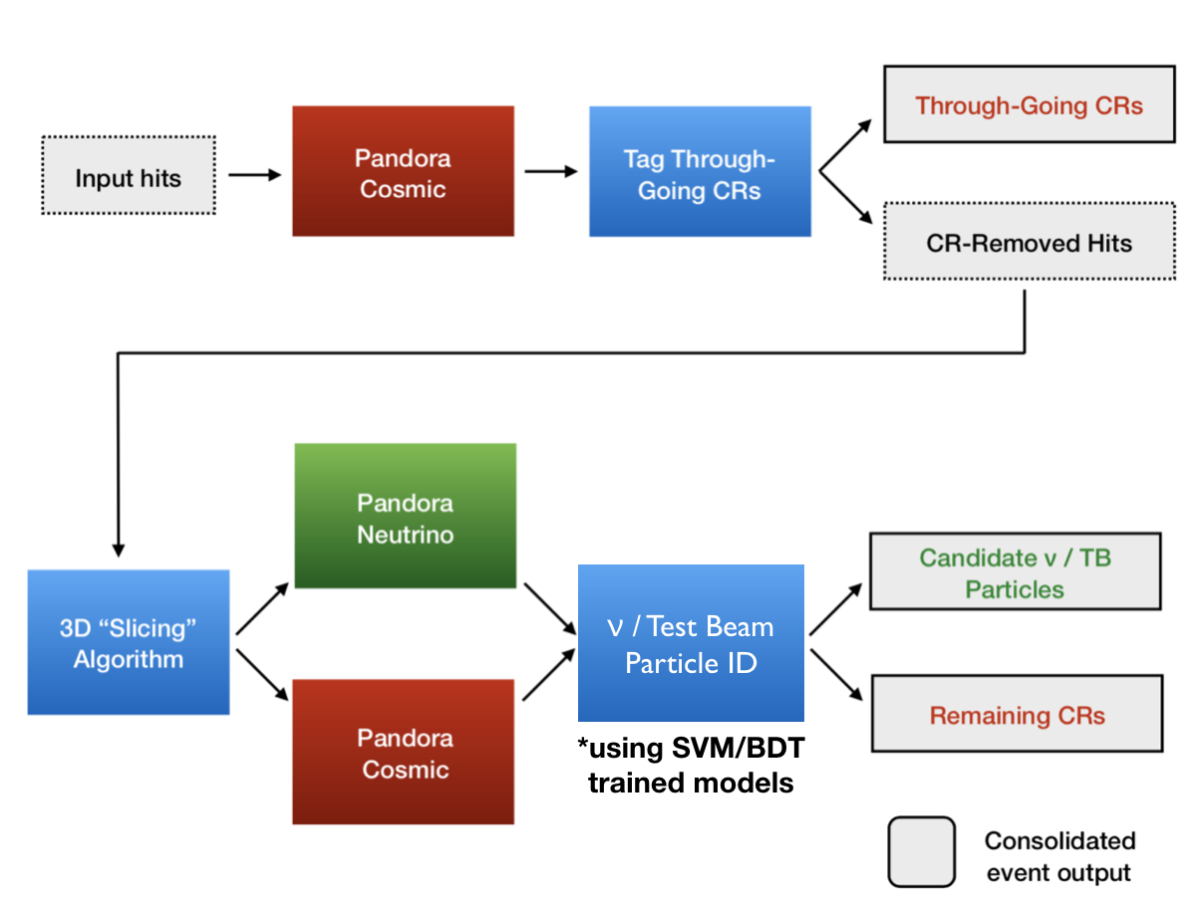}
\end{dunefigure}

Of particular importance in this overall reconstruction strategy is the neutrino (MicroBooNE) or test beam particle (\dword{pdsp}) identification tool. This tool is responsible for deciding whether to output the cosmic ray or neutrino (or test beam) reconstruction outcomes for a given slice. For \dword{pdsp}, this decision is based on the output from adaptive \dwords{bdt}, trained to distinguish between cosmic-ray and test beam particles, which has proved to be highly efficient across the momentum range of \dword{pdsp} data (see Section ~\ref{sec:Pandora:ProtoDUNE}). 

The performance obtained with the current algorithms are shown in Section~\ref{sec:performance}, both for the \dword{fd} and \dword{pdsp}.  As previously mentioned, both the individual pattern recognition algorithms and the overall reconstruction strategy are under continual development. Many algorithms 
still require explicit tuning for the DUNE energy ranges, and new algorithms, designed specifically for DUNE, will be added to the multi-algorithm pattern recognition. The performance presented in this document therefore represents a current snapshot and is expected to improve with future dedicated work.

\subsubsection{Projection Matching Algorithm}\label{sec:PMA}
\dword{pma}  was primarily developed as a technique of \threed reconstruction of individual particle trajectories (trajectory fit) Ref~\cite{Antonello:2012hu}. \dword{pma} was designed to address a challenging issue of transformation from a set of independently reconstructed \twod projections of objects into a \threed representation. Reconstructed \threed objects are also providing  basic physics quantities like particle directions and $dE/dx$ evolution along the trajectories. \dword{pma} uses as its input the output from \twod pattern recognition: clusters of hits. For the purposes of the DUNE reconstruction chain the Line Cluster algorithm (Section~\ref{sec:LineCluster}) is used as input to \dword{pma}, however the use of hit clusters prepared with other algorithms may be configured as well. As a result of \twod pattern recognition, particles may be broken into several clusters of \twod projections, fractions of particles may be missing in individual projections, and clusters obtained from complementary projections 
may not cover corresponding sections of trajectories. Such behavior is expected since ambiguous configurations of trajectories can be resolved only if the information from multiple \twod projections is used. Searching for the best matching combinations of clusters from all \twod projections was introduced to the \dword{pma} implementation in the \dword{larsoft} framework. The algorithm also attempts to correct hit-to-cluster assignments using properties of \threed reconstructed objects. In this sense \dword{pma} is also a pattern-recognition algorithm.
The underlying idea of \dword{pma} is to build and optimize objects in \threed space (formed as polygonal lines with 
the number of segments iteratively increased) by minimizing the cost function calculated simultaneously in all available \twod projections. 
Several features were developed in \dword{larsoft}'s \dword{pma} implementation to address detector-specific issues like stitching the particle fragments found in different TPCs or 
performing disambiguation at the \threed reconstruction stage. Since algorithms existing within or interfaced to the \dword{larsoft} framework (see Section \ref{sec:Pandora}) can provide pattern reconstruction results that include the particle hierarchy description, the mode for applying \dword{pma} to calculate 
trajectory fits alone was developed. In this mode the collections of clusters forming particles are taken from the ``upstream'' algorithm and  hit-to-cluster associations remain unchanged. 

\subsubsection{Wire Cell}

\begin{dunefigure}
[Overview of wire-cell reconstruction paradigm]
{wire-cell-overview}
{Overview of the \dword{wirecell} reconstruction paradigm, taken from~\cite{ref:wc_talk}. 
See text for more details.}
\includegraphics[width=0.98\textwidth]{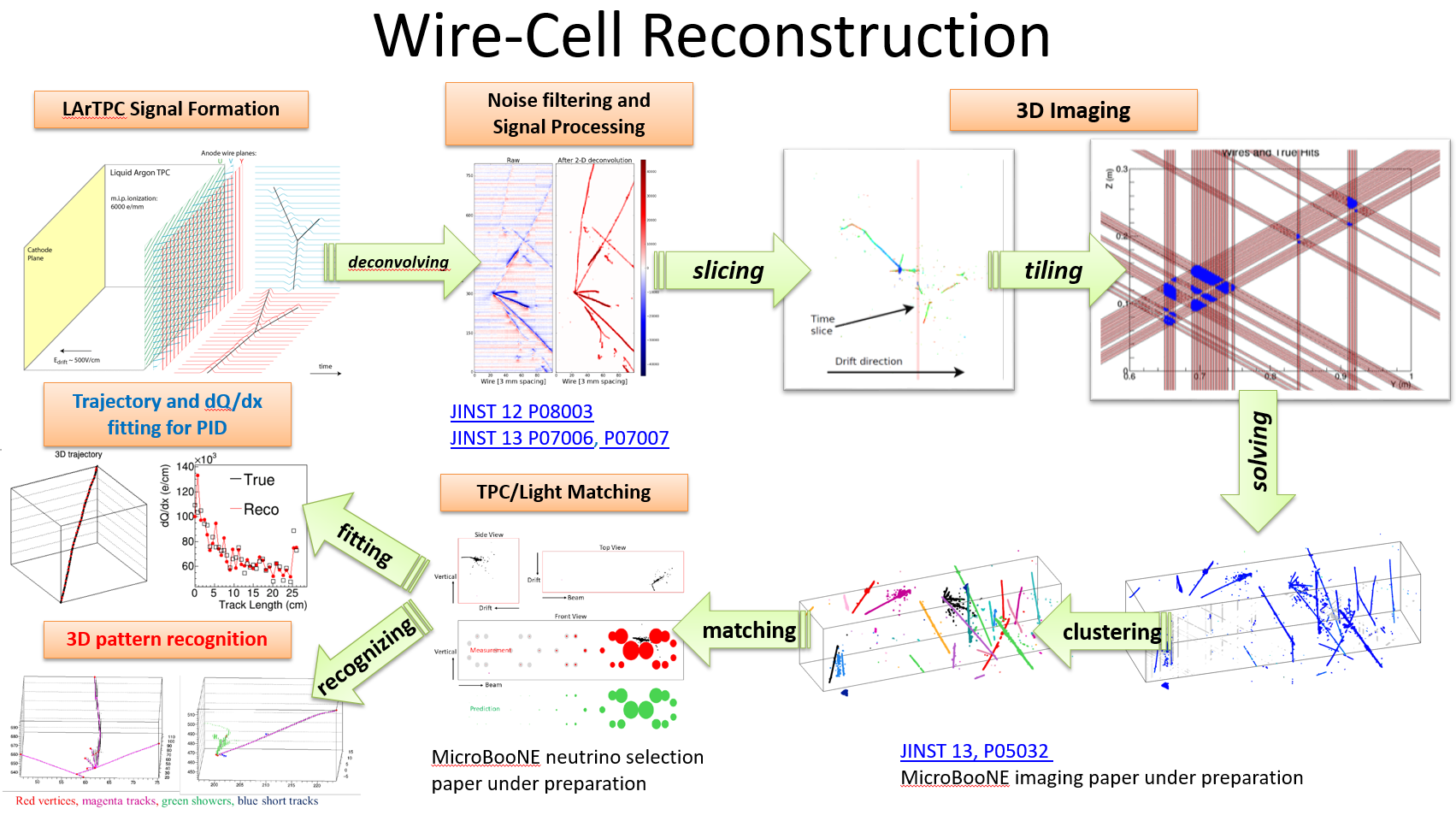}\end{dunefigure}

\Dword{wirecell} is a new reconstruction package under development. The current 
status of this reconstruction paradigm is shown in Figure~\ref{wire-cell-overview}. The 
simulation of the induction signal in a LArTPC and the overall signal processing process,
which are general to all reconstruction methods, are described in Sections~\ref{sec:tpc_sim} 
and~\ref{sec:tpc_sp}, respectively. The subsequent reconstruction in \dword{wirecell} adopts 
a different approach from the aforementioned algorithms. Instead of directly performing pattern 
recognition on each of the \twod views (drift time versus wire number), \threed imaging of events 
is obtained with time, geometry, and charge information. This step is independent from 
the event topologies, and the usage of the charge information takes advantage of a unique 
feature of the projection views, as each of the wire plane detects the same 
amount of the ionization electrons under transparency condition. The strong requirement of the time, geometry, and charge 
information provides a natural way to suppress electronic noise
 while combining with successful signal processing maintains high hit efficiency. Details of this step is described in~\cite{Qian:2018qbv}. The subsequent reconstruction involves the object clustering and
TPC and light matching, which has been crucial for selecting neutrino interactions in the 
MicroBooNE experiment~\cite{uboone_wc_note}. The current focus of the \dword{wirecell} algorithm 
development is on the trajectory and $dQ/dx$ fitting, which aims at enabling precision particle
identification in a \lartpc. 
Development of \threed pattern recognition also needs
to be revisited before reaching a complete reconstruction chain. 

\begin{dunefigure}
[\threed display of interaction in ProtoDUNE-SP]
{wire-cell-bee}
{This \threed display shows the full size of the \dword{pdsp} detector (gray box) and 
the direction of the particle beam (yellow arrow). Particles from other sources (such as cosmic rays) 
can be seen throughout the white box, while the red box highlights the region of interest: 
in this case, an interaction resulting from the 7 GeV beam particle through the detector. 
The \threed points are obtained using the Space~Point~Solver reconstruction algorithm. This event
can be accessed through interactive web-based event display Bee at \url{https://www.phy.bnl.gov/twister/bee/set/protodune-gallery/event/0/}.}
\includegraphics[width=0.85\textwidth]{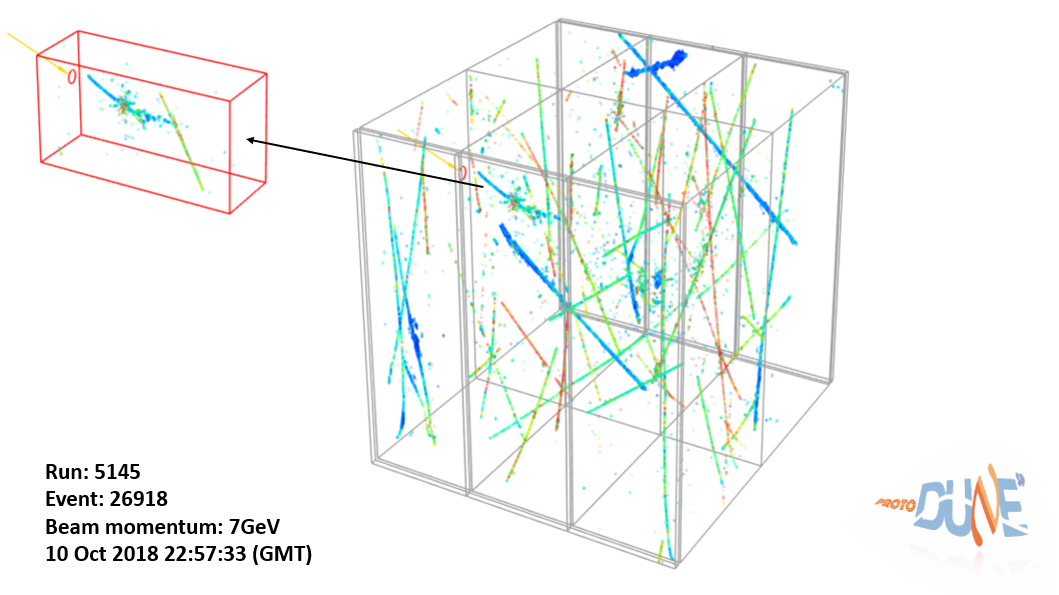}
\end{dunefigure}

The \dword{wirecell} team also created an advanced web-based \threed event display, ``Bee''~\cite{wire-cell-bee}, to aid the reconstruction development and provide interactive visualizations to end users.  Bee, together with \twod Magnify event 
display tools, have played important roles in 
the development of various reconstruction algorithms, including signal processing, \threed event 
imaging, object clustering, TPC and light matching, and trajectory and $dQ/dx$ fitting. The Bee event display 
was also used during the ProtoDUNE data-taking period to stream real-time reconstructed events to the users.
Figure~\ref{wire-cell-bee} shows an example of a data event from the \dword{pdsp} detector~\cite{ref:wc_bee}. 
The full video of this event can be found in~\cite{ref:bee_video}.

\subsubsection{Deep Learning}\label{sec:deeplearning}

Deep learning methods are used 
in two main areas of the DUNE event reconstruction. 
Both of these algorithms are based on \dwords{cnn}. 
In recent years \dwords{cnn} have become the method of choice for many image recognition tasks in commerce and industry, and lately have been applied to high energy physics. The \dwords{cnn} contain a series of filters that are applied to the input detector data images in order to extract the features required to classify the images.

\subsubsection{\dword{cnn} for track and shower separation}
The hit-level \dword{cnn} aims to classify each reconstructed hit as either track-like or shower-like by looking at the local region surrounding the hit in (charge, time) coordinates.  The \dword{cnn} is trained using a large number of simulated images with the known true origin of the energy deposits. Once trained, the \dword{cnn} provides the track-like or shower-like classification for each hit object in the event. This algorithm is applied to each readout view in each TPC separately. 

\subsubsection{\dword{cnn} for event selection}
The algorithm used for the classification of neutrino interaction types is called the \dword{cvn} and is  based on a \dword{cnn}. The primary goal of the \dword{cvn} is to provide a probability for each neutrino interaction to be \dword{cc}$\,\nu_\mu$, \dword{cc}$\,\nu_e$, \dword{cc}$\,\nu_\tau$ or \dword{nc}. The \dword{cvn} takes three $500\,\times\,500$ pixel images of the neutrino interactions as input, one from each view. The images contain the charge and the peak time of the reconstructed hits and does not use any information beyond the hit reconstruction. The \dword{cvn} is discussed in more detail in Chapter~\ref{ch:osc}.

\subsection{Calorimetric Energy Reconstruction and Particle Identification}

As charged particles traverse a \lar{} volume, they deposit energy through ionization and scintillation. It is important to measure the energy deposition, as it provides information on particle energy and species. The algorithm for reconstructing the ionization energy in \dword{larsoft} is optimized for line-like tracks and is being extended to more complicated event topology such as showers. The algorithm takes all the hits associated with a reconstructed track and for 
each hit, it converts the hit area or amplitude, in \dword{adc} counts, 
to the charge $Q_{\rm det}$, in units of \si{\femto\coulomb}, on the wire using an \dword{adc}-to-\si{\femto\coulomb} conversion factor that was determined by muons or test stand measurements. To account for the charge loss along the drift due to impurities, a first correction is applied to $Q_{\rm det}$ to get the free charge after recombination $Q_{\rm free} = Q_{det}/e^{-t/\tau_{e}}$, where $t$ is the electron drift time for the hit and $\tau_{e}$ is the electron lifetime measured by the muons or purity monitors. The charge $Q_{\rm{free}}$ is divided by the track pitch $dx$, which is defined as wire spacing divided by the cosine of the angle between the track direction and the direction normal to the wire direction in the wire plane, to get the $dQ_{\rm{free}}/dx$ for the hit. Finally, to account for charge loss due to recombination, also known as ``charge quenching,'' a second correction is applied to convert $dQ_{\rm{free}}/dx$ to $dE/dx$ based on the modified Box's model~ \cite{Acciarri:2013met} or the Birks's model~\cite{Amoruso:2004dy}. The total energy deposition from the track is obtained by summing the $dE/dx$ from each hit: $\sum\limits_{i}^{\rm all\ hits}(dE/dx)_{i}\cdot dx_{i}$.

If the incident particle stops in the LArTPC active volume, the energy loss $dE/dx$ as a function of the residual range ($R$), the path length to the endpoint of the track, is used as a powerful method for particle identification. There are two methods in \dword{larsoft} to determine particle species using calorimetric information. The first method calculates four $\chi^{2}$ values for each track by comparing measured $dE/dx$ 
versus $R$ to hypotheses for the proton, charged kaon, charged pion and muon, and identifies the track as the particle that gives the smallest $\chi^{2}$ value. The second method calculates the quantity $PIDA = \langle A_{i}\rangle = \left\langle(dE/dx)_{i}R_{i}^{0.42}\right\rangle$ \cite{Acciarri:2013met}, which is defined to be the average of $A_{i} = (dE/dx)_{i}R_{i}^{0.42}$ over all track points where the residual range $R_{i}$ is less than \SI{30}{cm}. The particle species can be determined by making a selection on the $PIDA$ value.

\subsection{Optical Reconstruction}

\subsubsection{Optical Hit Finder}
\label{sec:OpticalHitFinder}
The first step of the DUNE optical reconstruction is reading
individual waveforms from the simulated \dword{pd} electronics
and finding optical hits -- regions of the waveforms containing pulses.
The optical hit contains the optical channel (\dword{sipm}) that the hit
was found on, time corresponding to the hit, its width,
area, amplitude, and number of \phel{}s.

The current DUNE optical-hit-finder algorithm then searches for regions of the waveform
exceeding a certain threshold ($13$ \dword{adc} counts), checking whether that region
is wider than $10$ optical time ticks\footnote{The current simulation assumes a 
\SI{150}{MHz} digitizer like that used in ProtoDUNE, though the final far detector electronics
will use an \SI{80}{MHz} digitizer.}, and, if it is, calculating the aforementioned
optical-hit parameters for the region (including parts of the waveform around it
that have \dword{adc} values greater than $1$) and recording it as an optical hit.
The number of \phel{}s is calculated by dividing the full area of the hit
by the area of a single-\phel{} pulse.
The pedestal is assumed to be constant and is specified in the hit finder as $1500$ \dword{adc} counts (always correct for the MC).

\subsubsection{Optical Flash Finder}
After optical hits are reconstructed, they are grouped into higher-level objects called optical flashes.
The optical flash contains the time and time width of the flash,
its approximate $y$ and $z$ coordinates (and spatial widths along those axes),
its location and size in the wire planes,
the distribution of \phel{}s across all \dwords{pd},
and the total number of \phel{}s in the flash, among other parameters.

The flash-finding algorithm searches for an increase in \dword{pd} activity
(the number of \phel{}s) in time using information from optical hits
on all photon detectors.
When a collection of hits with the total number of \phel{}s  
greater than or equal to $2$ is found, the algorithm begins creating an optical flash.
It starts with the largest hit and adds hits from the found hit collection 
that lie closer than half the combined widths of the flash under construction
a nd the hit being added to it.
The flash is stored after no more hits can be added to it
and if it has more than two \phel{}s.

The algorithm also estimates spatial parameters of the optical flash
by calculating the number-of-photoelectron-weighted mean and 
root mean square of locations of the optical hits
(defined as centers of \dwords{pd} where those hits were detected)
contained in the flash.

\section{Reconstruction Performance}
\label{sec:performance}

An automated reconstruction of the neutrino interaction events in DUNE, often complex topologies with multiple final state particles, is a significant challenge. The current chain of \dword{pandora} pattern recognition algorithms 
has been tuned for neutrino interactions from the \fnal Booster Neutrino Beam, and is in the process of being adapted for the wide range of energies of the DUNE \dword{fd}. Despite this, and thanks to the reusability of \dword{pandora} algorithms for different single phase \lartpc detectors,  good performance is already achieved with this first-pass pattern recognition, and output from Pandora is used in the computation of the energy reconstruction in the oscillation analysis. Significant improvements are expected  in the upcoming years with a more dedicated tune of the current algorithms, and the development of new ones, as needed. 

The current reconstruction performance, evaluated using metrics introduced in \ref{sec:Pandora:assessment}, is presented for simulated neutrino interactions in a \single \nominalmodsize \dword{fd} module in \ref{sec:Pandora:DUNEFD}, and for simulated and real data test beam events in \dword{pdsp} in \ref{sec:Pandora:ProtoDUNE}. These results outline the baseline performance on which improvements will continue to be made in the next years. In addition, examples of current high-level reconstruction performance are presented in~\ref{sec:Pandora:High}.

\subsection{Pandora Performance Assessment}
\label{sec:Pandora:assessment}
The performance of the \dword{pandora} pattern recognition is assessed by matching reconstructed \dword{pfparticle}s to the simulated \dword{mcparticle}s. These matches are used to evaluate the efficiency with which \dword{mcparticle}s are reconstructed as \dword{pfparticle}s, and to calculate the completeness and purity of each reconstructed \dword{pfparticle}.

The following procedure is used to match reconstructed \dword{pfparticle}s with simulated \dword{mcparticle}s:

\begin{itemize}
\item \textit{Selection of \dword{mcparticle}s:} The full hierarchy of true particles is extracted from the simulated neutrino interaction. A list of ``target'' particles is then compiled by navigating through this hierarchy and selecting the final-state ``visible'' particles producing a minimum number of reconstructed hits (allowed to be: $e^{\pm}$, $\mu^{\pm}$, $\gamma$, $\pi^{\pm}$, $\kappa^{\pm}$, $p$)\footnote{A minimum number of 15 reconstructed hits, with at least two views with 5 or more hits, is required in the definition of ``target'' \dword{mcparticle}. This  corresponds to true momentum thresholds of approximately 60 MeV for muons and 250 MeV for protons in the MicroBooNE simulation~\cite{Acciarri:2017hat}. Note that this selection is purely for performance assessment purposes, and that particles with fewer hits might still be created by \dword{pandora}.}. Any downstream daughter particles are folded in these target particles.
\item \textit{Matching of reconstructed \twod hits to \dword{mcparticle}s:} Each reconstructed \twod hit is matched to the target \dword{mcparticle} responsible for depositing the most energy within the region of space covered by the hit. The collection of \twod hits matched to each target \dword{mcparticle} is known as its ``true hits''.
\item \textit{Matching of \dword{mcparticle}s to reconstructed \dword{pfparticle}s:} The reconstructed \dword{pfparticle}s are matched to target \dword{mcparticle}s by analyzing their shared \twod hits. A \dword{pfparticle} and \dword{mcparticle} will be matched if the \dword{mcparticle} contributes the most hits to the \dword{pfparticle}, and if the \dword{pfparticle} contains the largest collection of hits from the \dword{mcparticle}. The matching procedure is iterative, such that once each set of matched particles has been identified, these \dword{pfparticle}s and \dword{mcparticle}s are removed from consideration when making the next set of matches. 
\end{itemize}

Using the output of this matching scheme, the following performance metrics can be calculated:

\begin{itemize}
\item \textit{Efficiency:} Fraction of \dword{mcparticle}s with a matched \dword{pfparticle},
\item \textit{Completeness:} The fraction of \twod hits in a \dword{mcparticle} that are shared with its matched reconstructed \dword{pfparticle}, and 
\item \textit{Purity:} The fraction of \twod hits in a \dword{pfparticle} that are shared with its matched \dword{mcparticle}.
\end{itemize}

\subsection{Reconstruction Performance in the DUNE \dword{fd}}
\label{sec:Pandora:DUNEFD}

The performance of the \dword{pandora} pattern recognition has been evaluated using a sample of accelerator neutrino and antineutrino interactions simulated using the reference DUNE neutrino energy spectrum and the \nominalmodsize \dword{detmodule} geometry. The breakdown of the different interaction channels as a function of the true neutrino energy in the samples used is presented in Fig. \ref{breakdown_nuenergy}, for the events in the neutrino mode in which at least one ``target'' reconstructable \dword{mcparticle} is created and therefore evaluated. The following plots show that a good efficiency has already been achieved, and indicate particular regions and channels in which improvements can be made. 

\begin{dunefigure}
[Interaction channels vs true $\nu$ energy; simulated events used to assess performance]
{breakdown_nuenergy}
{Breakdown of the different interaction channels as a function of the true neutrino energy in the samples used in the assessment of reconstruction performance, for the simulated events in the neutrino mode in which at least one ``target'' reconstructable \dword{mcparticle} particle is created and therefore evaluated. Percentages indicate the fraction of each channel in the total number of events.}
\includegraphics[width=0.8\textwidth]{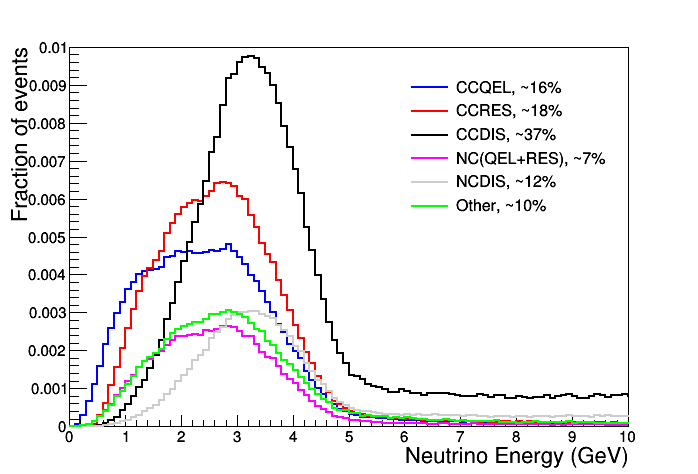}
\end{dunefigure}

\begin{dunefigure}
[Reconstruction efficiency of PANDORA pattern recognition for a range of final-state particles at the FD]
{pandora_particle_efficiency}
{The reconstruction efficiency of the \dword{pandora} pattern recognition obtained for a range of final-state particles produced in all types of accelerator neutrino interactions except deep inelastic ones at DUNE FD. The efficiency is plotted as a function of the total number of \twod hits associated with the final-state \dwords{mcparticle} (summed across all views) on the top row, and as a function of the true momentum of the particle on the bottom row. Plots are shown for track-like particles (left) and shower-like particles (right) of each type leading in the event.}
\includegraphics[width=0.49\textwidth]{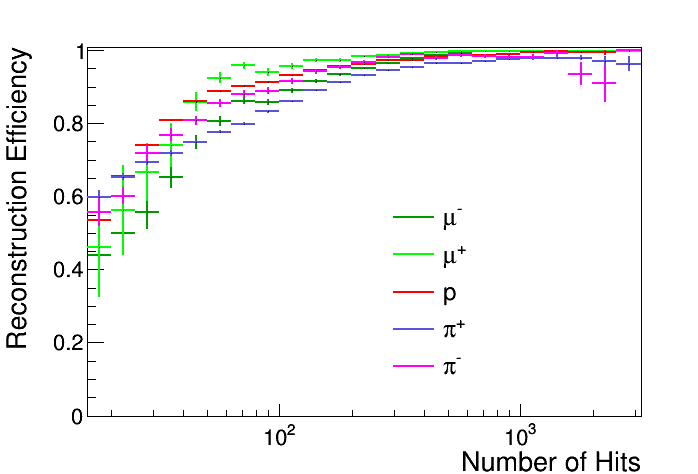}
\includegraphics[width=0.49\textwidth]{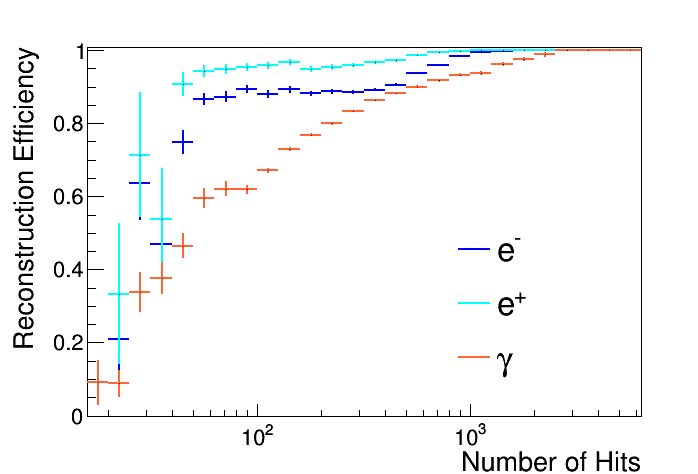}
\includegraphics[width=0.49\textwidth]{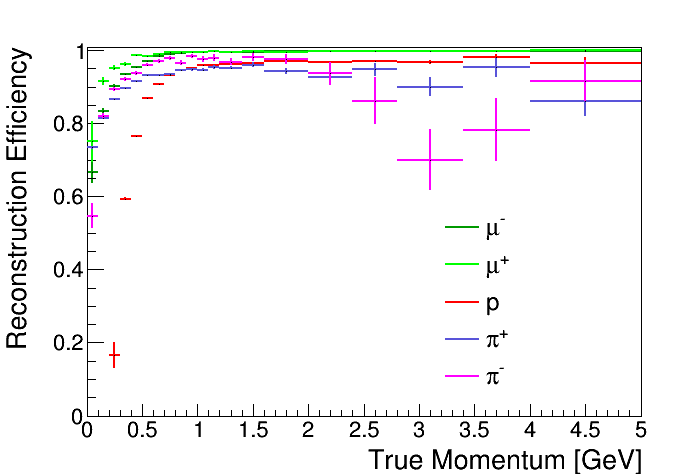}
\includegraphics[width=0.49\textwidth]{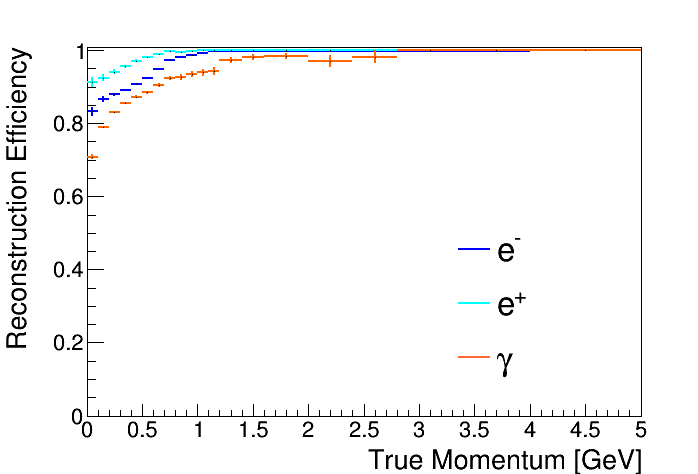}
\end{dunefigure}

Figure~\ref{pandora_particle_efficiency} shows the reconstruction efficiency as a function of the number of total true \twod hits and as a function of the true momentum for a range of final-state particles. The typical reconstruction efficiencies obtained for track-like \dword{mcparticle}s ($\mu^{\pm}$, $\pi^{\pm}$, $p$) rise from \SIrange{65}{85}{\%} for simulated particles depositing 100\,hits to \SIrange{85}{100}{\%} for particles with 1000\,hits. It should be emphasized that inefficiencies almost always result from accidental merging of multiple nearby true particles, rather than an inability to cluster hits from a true particle. The reconstruction efficiency for shower-like \dword{mcparticle}s ($e^{-}$,$\gamma$) is a bit lower than the equivalent for track-like particles at lower number of hits, but comparable with $>$100 hits.

Figure~\ref{pandora_completeness_purity} shows distributions of completenesses and purities for a range of final-state particles. In the case of final-state track-like particles, good completeness and purity are 
achieved, indicating that the track-based pattern recognition algorithms currently provide a high-quality reconstruction. It can be seen that final-state shower-like particles are typically reconstructed with high purity, but somewhat lower completeness, indicating that, although the shower reconstruction is fairly good already, there is room for addition of new algorithms specifically targeting an increase in shower completeness at DUNE.

For deep inelastic interactions, in which tens of final-state particles may be produced, a breakdown such as in Figures~\ref{pandora_particle_efficiency} and~\ref{pandora_completeness_purity} is less representative and informative (however, no significant impact has been observed when adding DIS events in the calculation of such quantities). Instead, Figure~\ref{pandora_dis} presents an assessment of the reconstruction of such events 
by comparing the number of reconstructed particles as a function of the number of true final-state particles in the event for \dword{nc} (left) and \dword{cc} (middle) deep inelastic interactions. These distributions are more populated in the diagonal, as they should be for perfect 1:1 reconstruction, indicating a good level of reconstruction of such events up to >5 final-state particles. In addition, the number of reconstructed particles matching the leading lepton in \dword{cc} deep inelastic interactions is also presented (right), which shows a consistently predominant single match for the leading lepton. 

Figure~\ref{pandora_vertex_resolution} shows distributions of the displacements $\Delta x$, $\Delta y$, $\Delta z$ and $\Delta R^{2} = (\Delta x)^2 + (\Delta y)^2 + (\Delta z)^2$ between the reconstructed and simulated neutrino interaction positions for all types of accelerator neutrino events. It can be seen that, for the vast majority of events, the reconstructed neutrino interaction vertex lies within $2$\,cm of the \dword{mc} truth in $x$, $y$ and $z$. While the $\Delta x$ and $\Delta y$ distributions are both symmetrical and sharply peaked around the origin, a small forward bias can be seen in the $\Delta z$ distribution. The reason for this bias comes from the fact that the neutrino interaction will be boosted in the forward $z$ direction, so vertex candidates are more likely created at $\Delta z>0$ than $\Delta z\,<\,0$.  

\begin{dunefigure}
[Completeness and purities for a range of final-state track-like  and shower-like particles]
{pandora_completeness_purity}
{Distributions of completenesses (top) and purities (bottom) for a range of final-state particles divided into track-like (left) and shower-like (right), produced in all types of accelerator neutrino interactions except deep inelastic ones at DUNE FD.}
\includegraphics[width=0.49\textwidth]{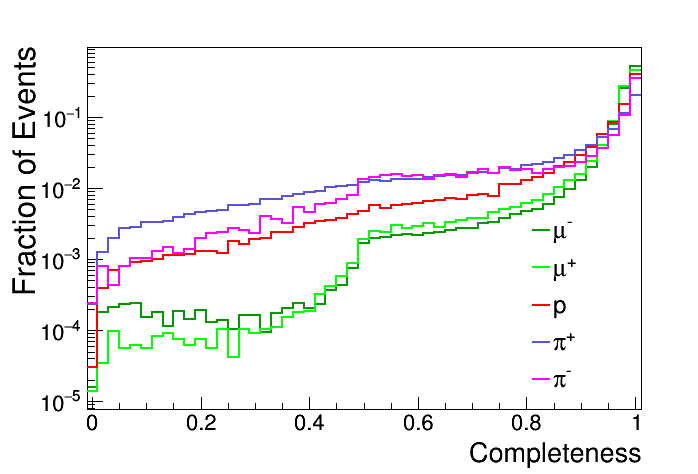}
\includegraphics[width=0.49\textwidth]{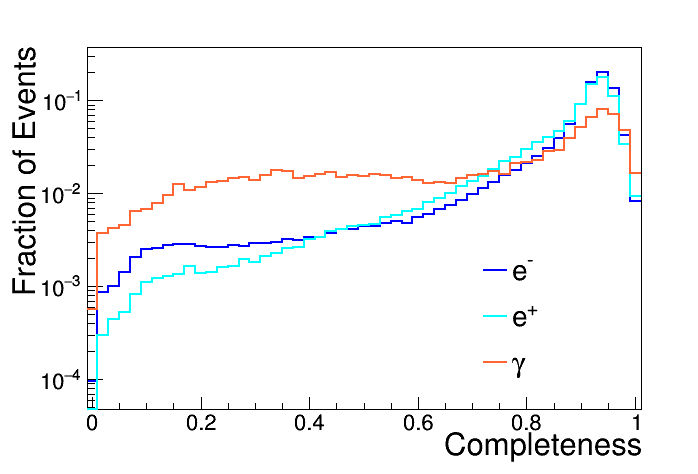}
\includegraphics[width=0.49\textwidth]{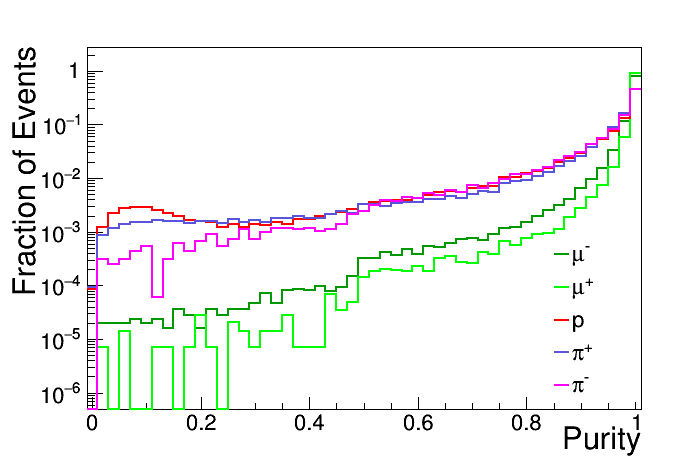}
\includegraphics[width=0.49\textwidth]{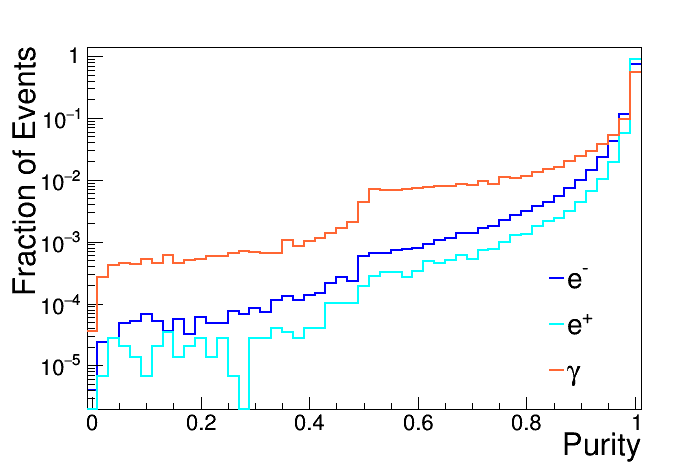}\end{dunefigure}

\begin{dunefigure}
[Number of reconstructed particles vs number of true final-state particles, CC and NC DIS events]
{pandora_dis}
{Distributions of number of reconstructed particles as a function of number of true final-state particles in deep inelastic events for neutral-current (left) and charged-current (right) 
interactions. In addition, the number of reconstructed particles matching the leading lepton in charged-current deep inelastic interactions is also presented (bottom).} 

\includegraphics[width=0.49\textwidth]{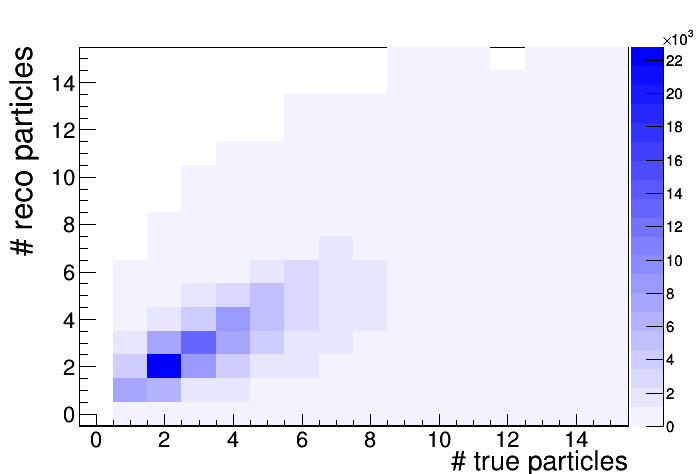}
\includegraphics[width=0.49\textwidth]{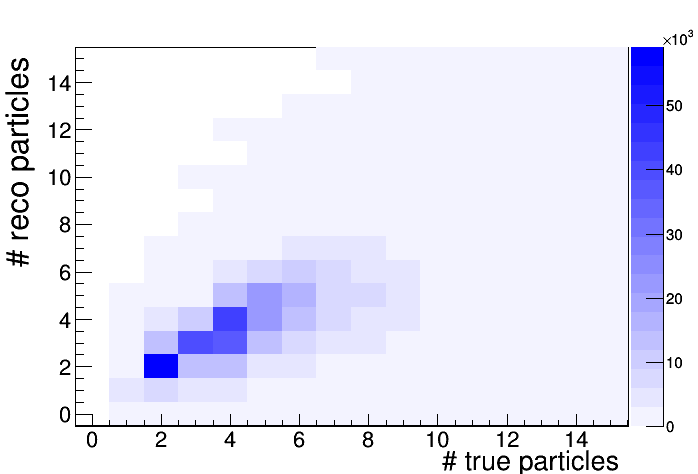}
\includegraphics[width=0.49\textwidth]{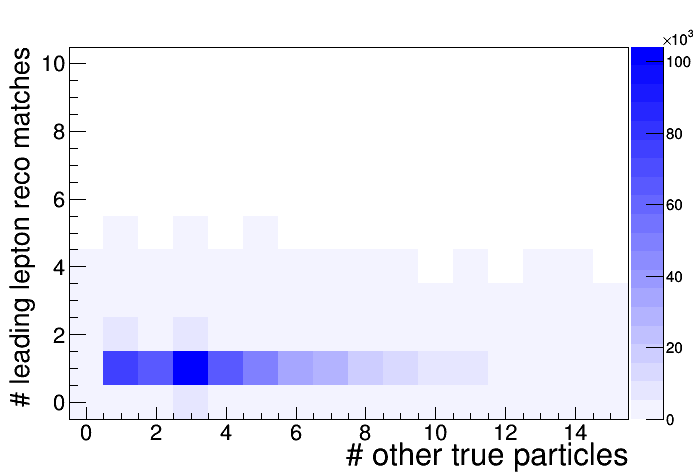}
\end{dunefigure}

\begin{dunefigure}
[Displacements between reconstructed and simulated $\nu$ interaction vertices]
{pandora_vertex_resolution}
{The displacements between the reconstructed and simulated neutrino interaction vertices. The distributions are plotted for $x$ (top left), $y$ (top right), $z$ (bottom left) and $R^2$ (bottom right) and include all types of accelerator neutrino interaction (also deep inelastic events).}
\includegraphics[width=0.49\textwidth]{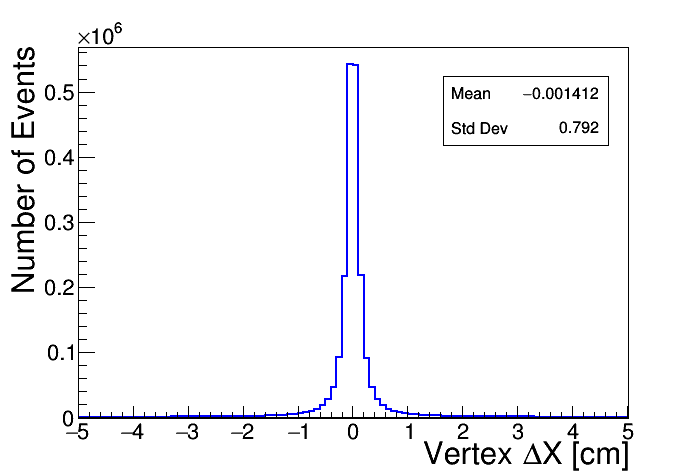}
\includegraphics[width=0.49\textwidth]{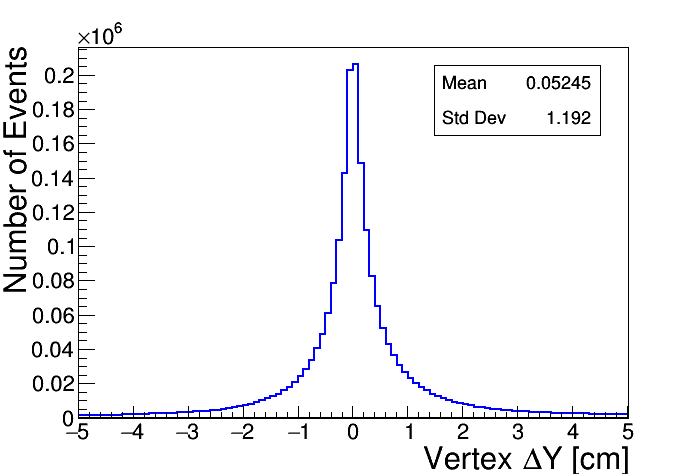}
\includegraphics[width=0.49\textwidth]{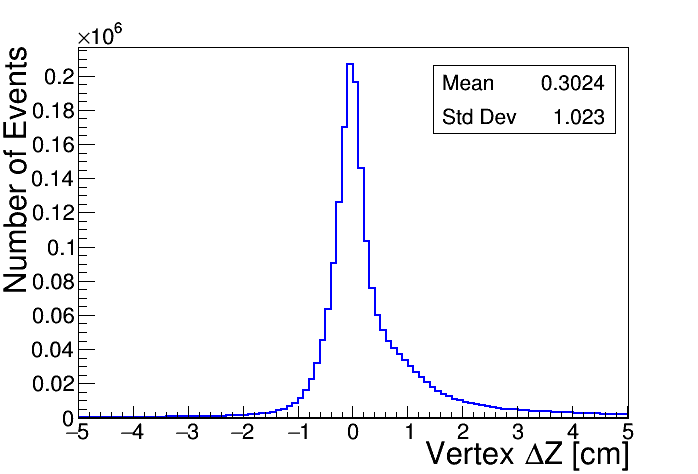}
\includegraphics[width=0.49\textwidth]{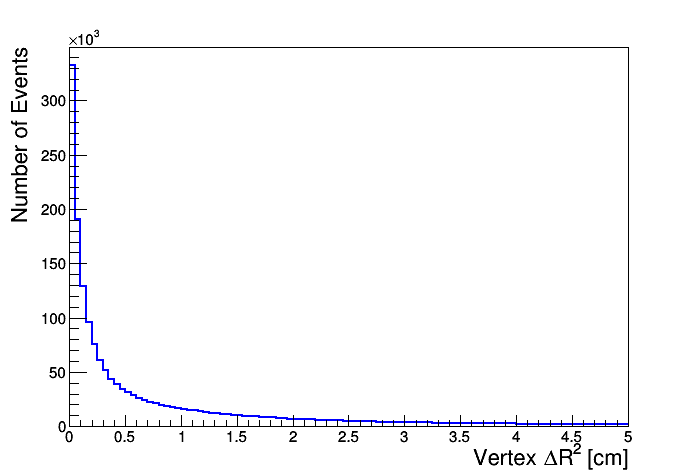}
\end{dunefigure}

\subsection{Reconstruction Performance in \dword{pdsp}}
\label{sec:Pandora:ProtoDUNE}

\begin{figure}[!ht]
\centering
\subfloat[]{\label{fig:reco3d}\includegraphics[width=0.75\textwidth]{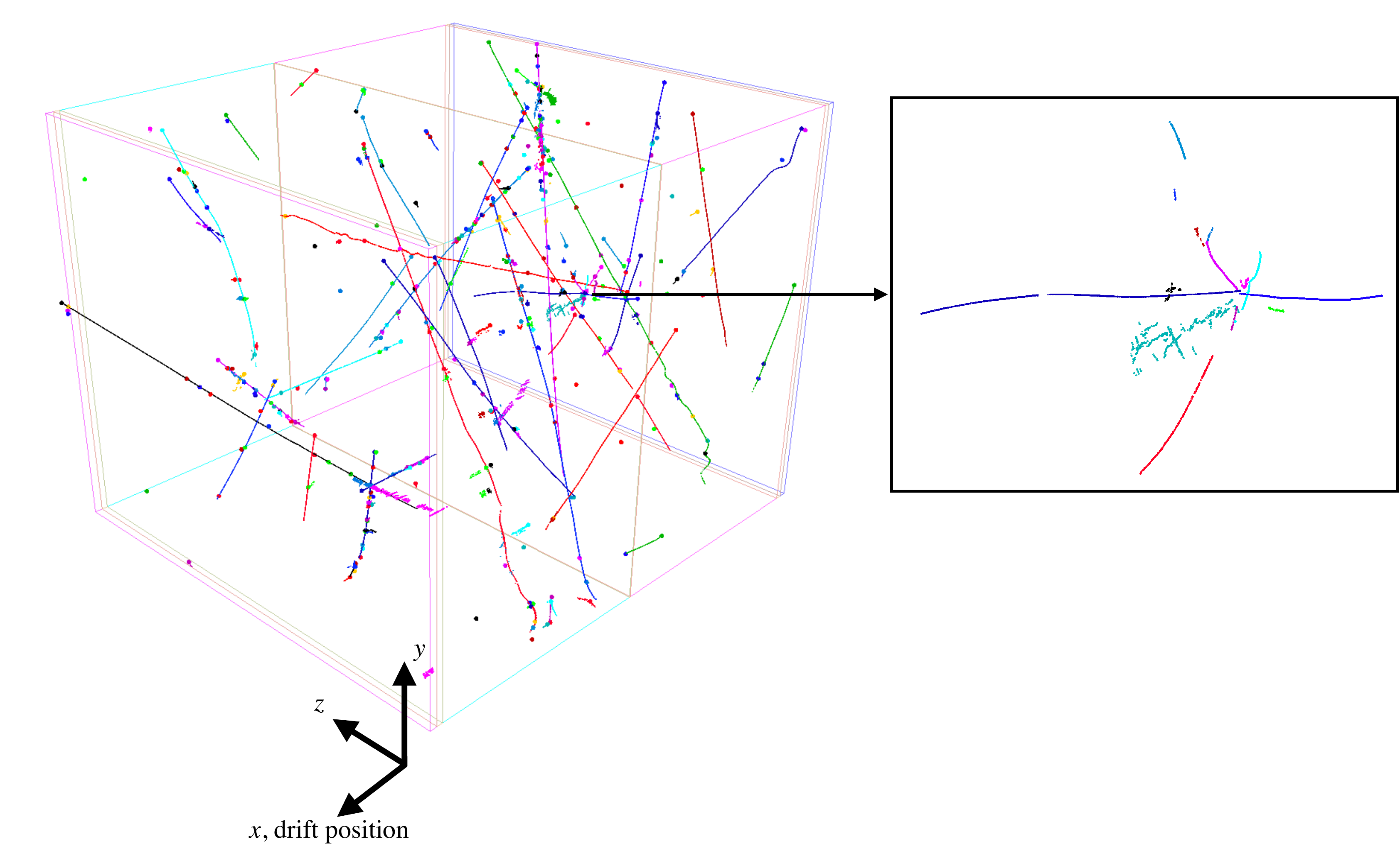}} \\ 
\subfloat[]{\label{fig:recou}\includegraphics[width=0.33\textwidth]{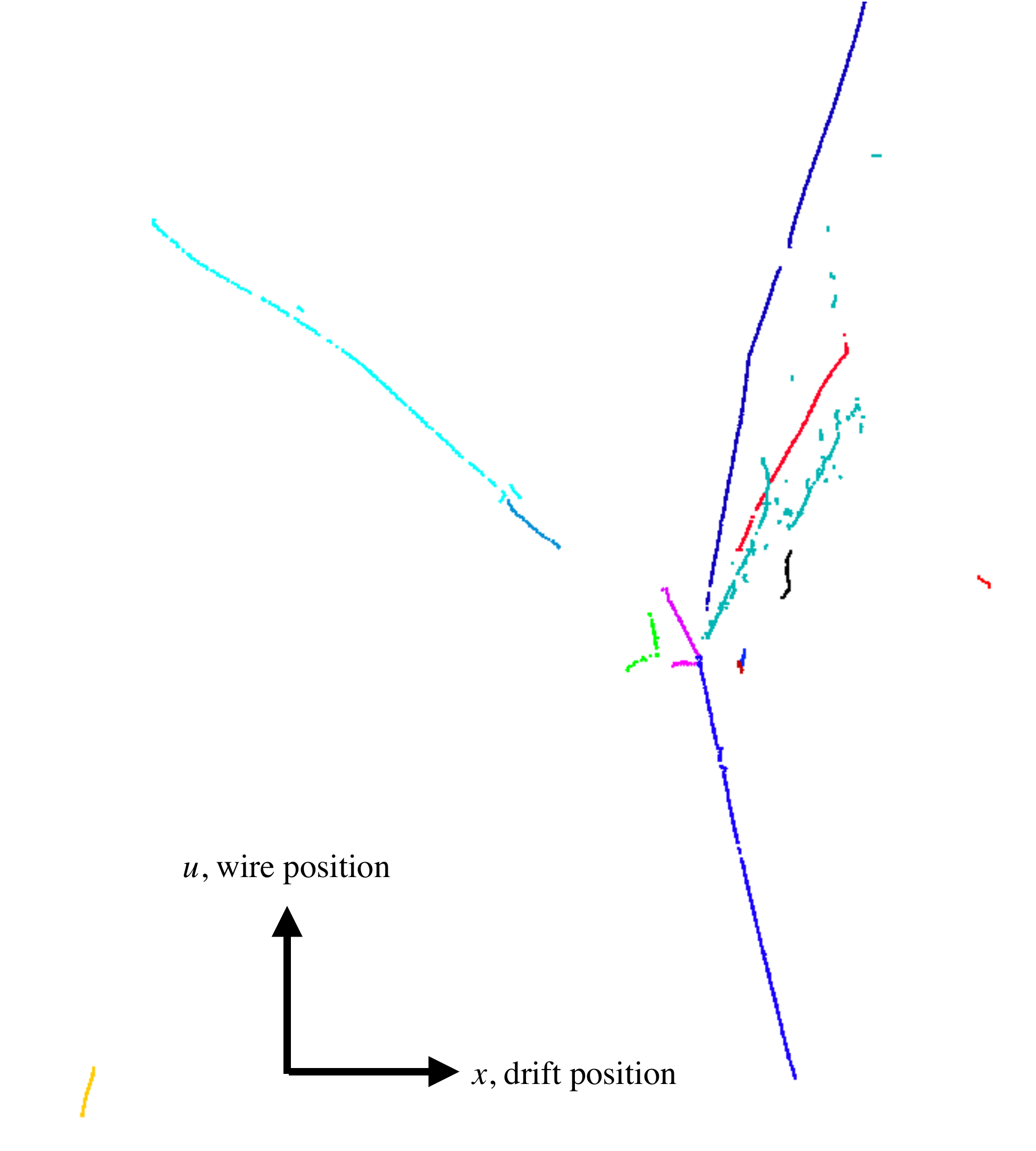}}
\subfloat[]{\label{fig:recov}\includegraphics[width=0.33\textwidth]{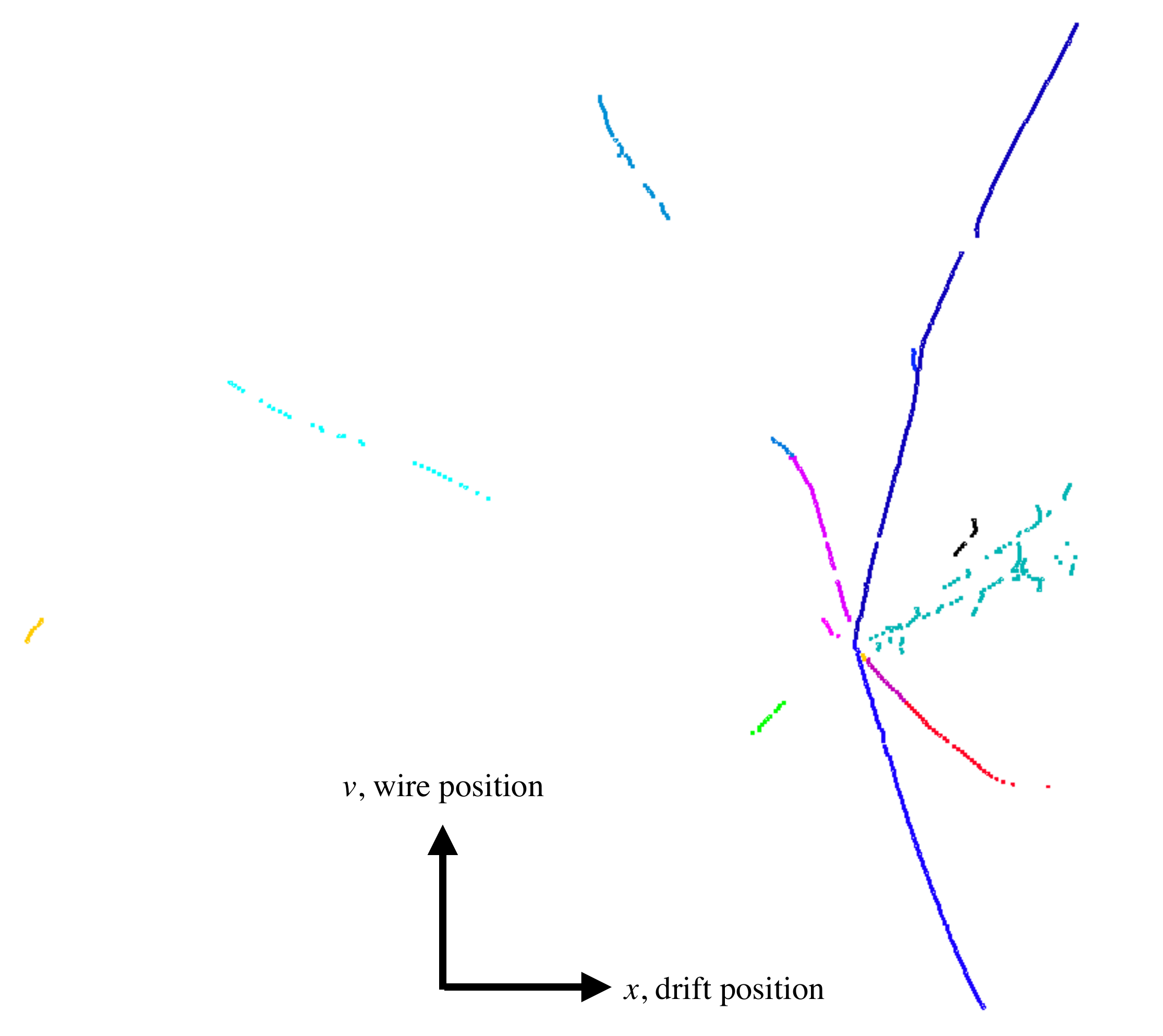}}
\subfloat[]{\label{fig:recow}\includegraphics[width=0.33\textwidth]{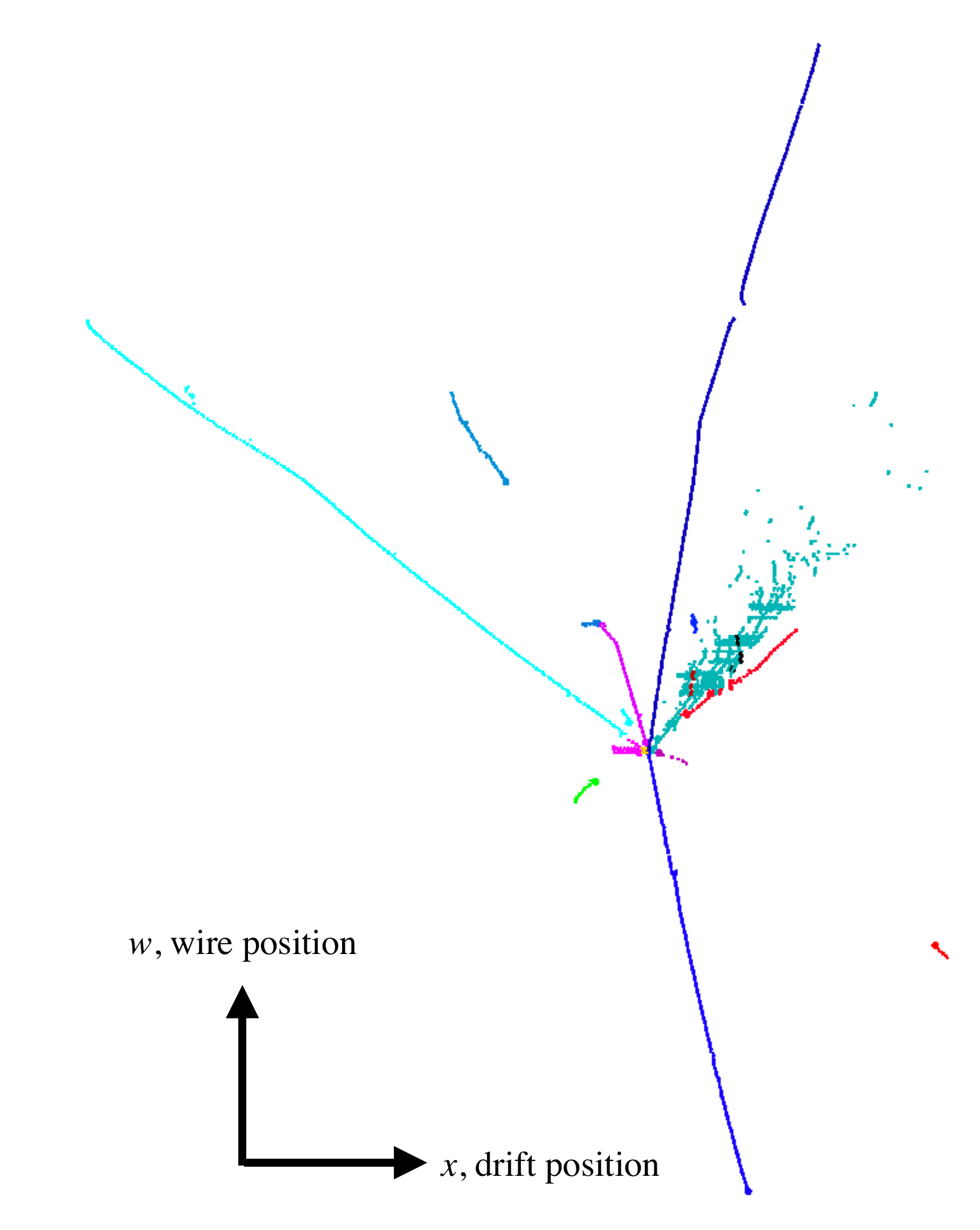}}
\caption[Pandora reconstruction output for \SI{7}{GeV} MC test beam event]{An example of the \dword{pandora} reconstruction output for a 7 GeV Monte Carlo test beam event.  Figure \protect\subref{fig:reco3d} shows the \threed reconstruction output for this event where the correctly reconstructed and tagged triggered test beam particle has been highlighted.  Figures \protect\subref{fig:recou}, \protect\subref{fig:recov} and \protect\subref{fig:recow} show the \twod hits for the reconstructed test beam particle where each colored cluster of hits represents a different particle in the reconstructed particle hierarchy.}
\label{pandora_protodune_reco} 
\end{figure}

Further examination of the performance of the \dword{pandora} pattern recognition is provided through studies of the test-beam data taken by \dword{pdsp}.  Figure \ref{pandora_protodune_tbrecoeff} shows the reconstruction efficiency for triggered test-beam particles as a function of the momentum recorded by the trigger.  The reconstruction efficiency metric folds in many effects, including reconstruction, removal of cosmic-ray background and identification of the reconstructed particle as originating from the test beam.  An example of the \dword{pandora} reconstruction output for ProtoDUNE \dword{mc} simulations is shown in Figure~\ref{pandora_protodune_reco}.  For high-momenta test-beam particle interactions, a close agreement between the reconstruction efficiency for \dword{mc} simulations and data is observed in Figure~\ref{pandora_protodune_tbrecoeff}.  At high-momenta, the effect of beam-halo particles in the simulation appears to be overestimated, which results in the marginally lower reconstruction efficiency observed in simulation when comparison to data.  For low-momenta test-beam particle interactions, the reconstruction efficiency for data is significantly lower than that see in \dword{mc} simulations.  This is due to particles interacting between the trigger and the LArTPC before reaching its active volume in data.

\begin{dunefigure}
[Reconstruction efficiency for triggered test-beam particles as a function of particle momentum]
{pandora_protodune_tbrecoeff}
{The efficiency of reconstruction for the triggered test-beam particle as a function of particle momentum in data (red) and simulation (black).}
\includegraphics[width=0.75\textwidth]{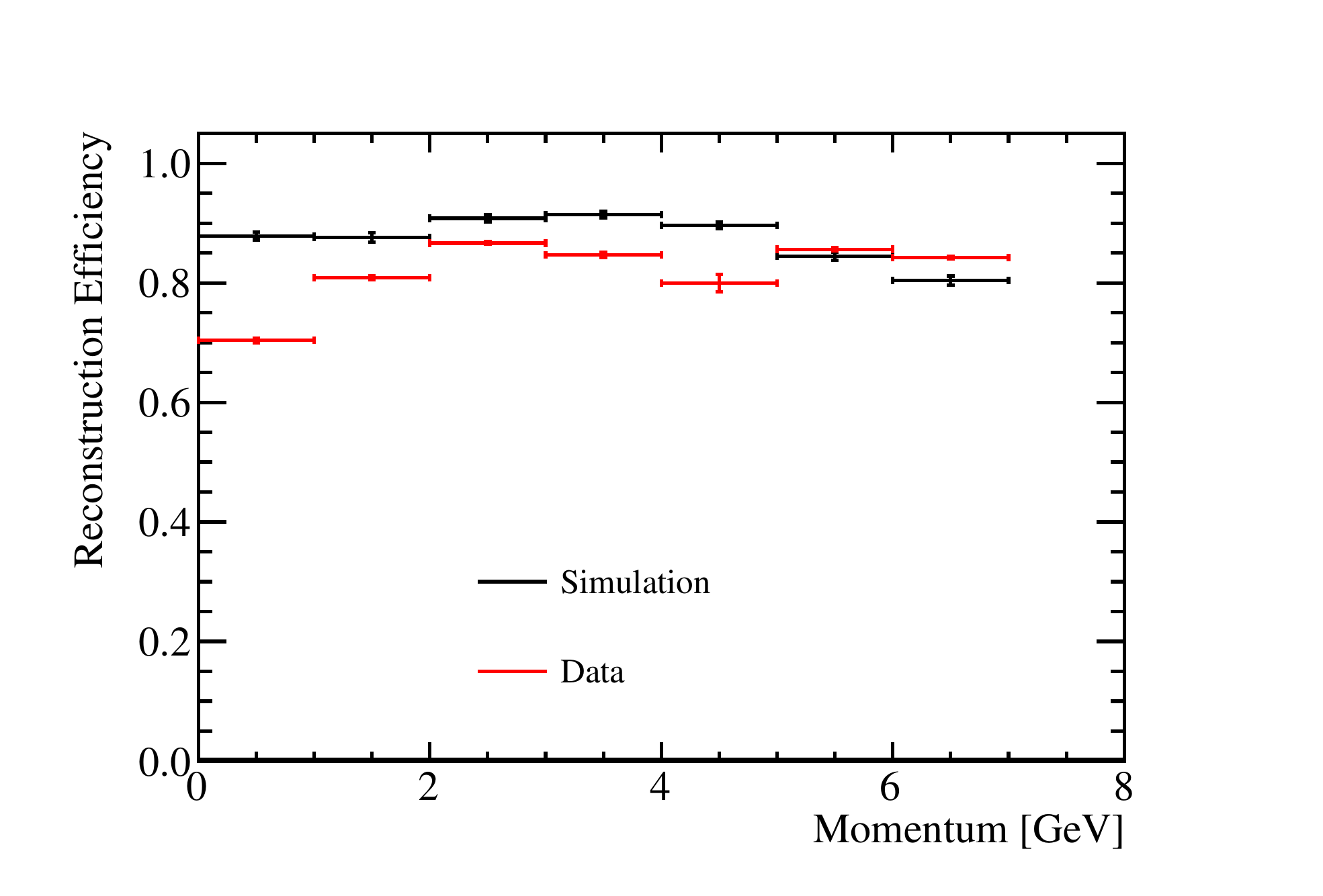}
\end{dunefigure}

\begin{figure}[!ht] 
\centering
\subfloat[]{\label{fig:pandora_protodune_tbrecoeffbrkdwn_p}\includegraphics[width=0.5\textwidth]{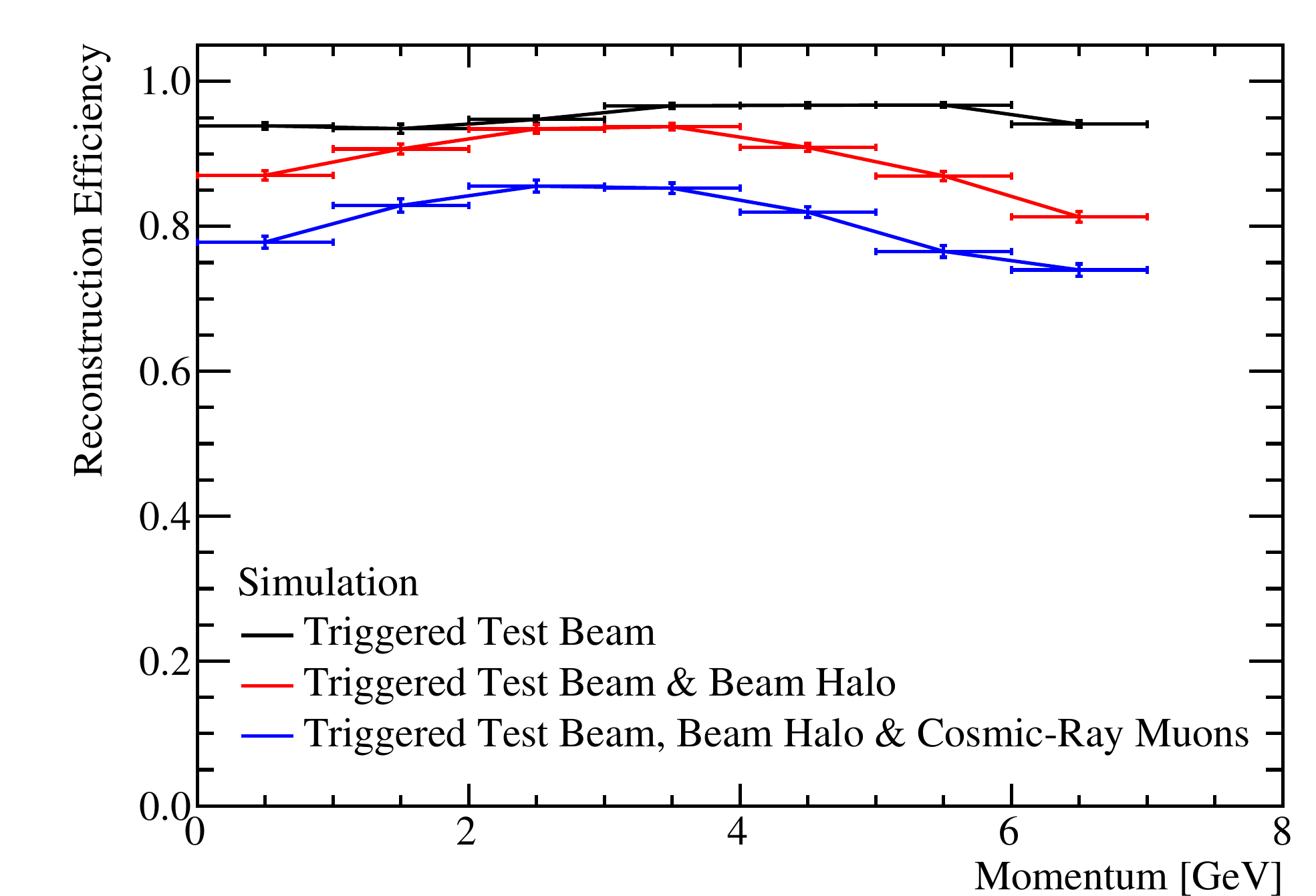}}
\subfloat[]{\label{fig:pandora_protodune_tbrecoeffbrkdwn_nhits}\includegraphics[width=0.5\textwidth]{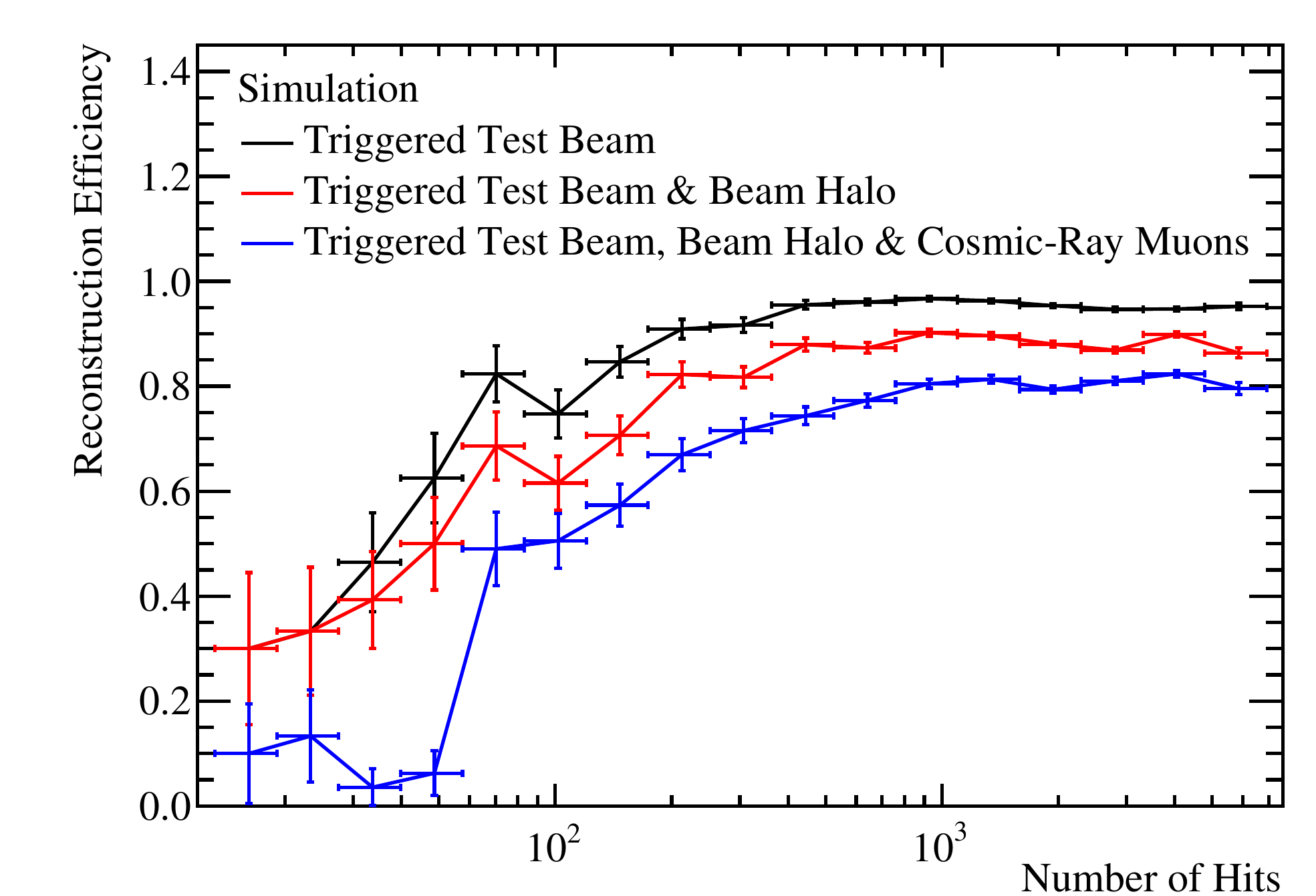}} \\
\subfloat[]{\label{fig:pandora_protodune_tbrecoeffbrkdwn_evt}\includegraphics[width=0.6\textwidth]{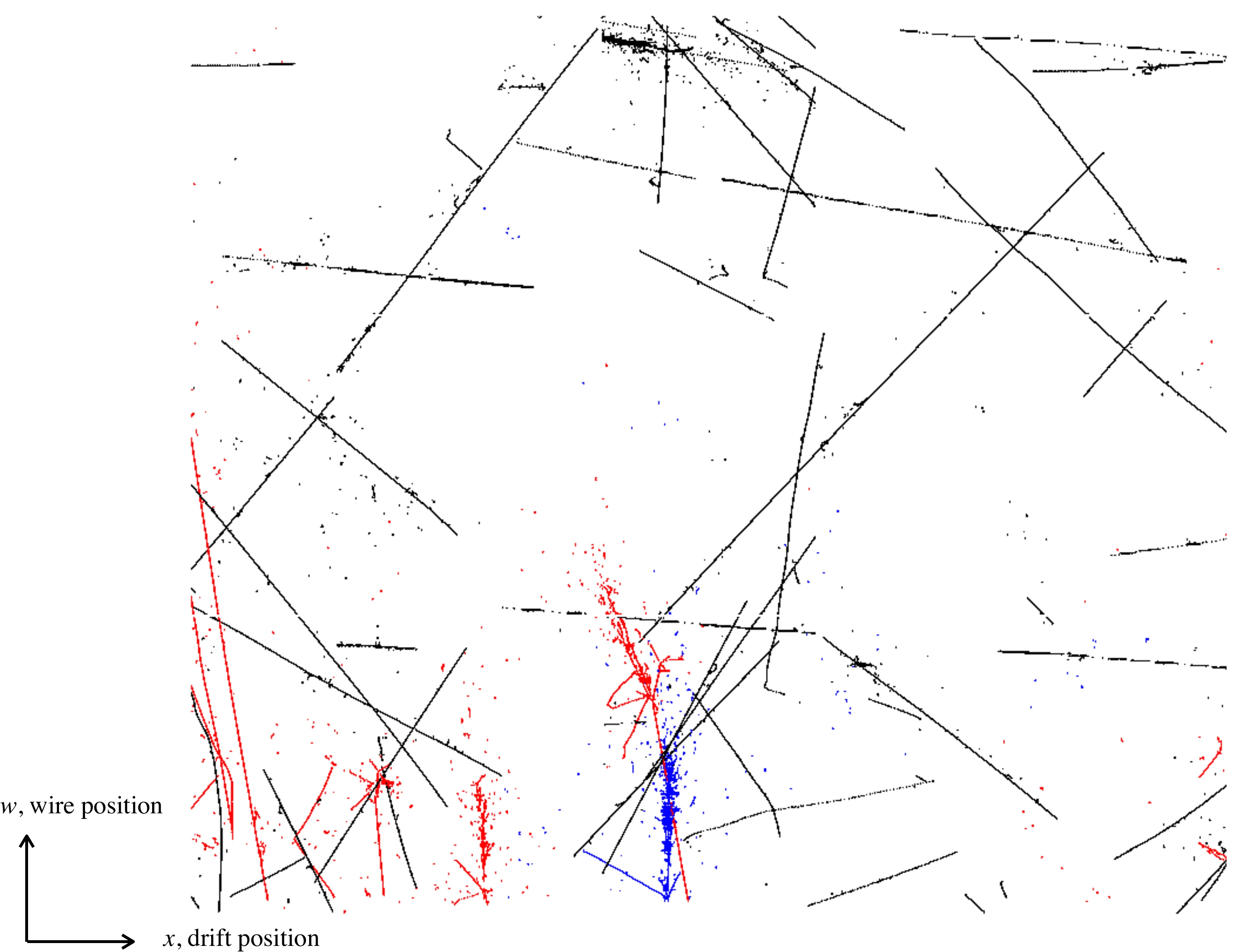}}
\caption[Reconstruction efficiency for test beam particle in MC per momentum and hits]{The efficiency of reconstruction for the triggered test beam particle in Monte-Carlo as a function of \protect\subref{fig:pandora_protodune_tbrecoeffbrkdwn_p} the triggered beam momenta and \protect\subref{fig:pandora_protodune_tbrecoeffbrkdwn_nhits} the number of hits made by the triggered particle.  The three curves show the reconstruction efficiency of the triggered test beam particle in isolation (black), with beam particle halo overlaid (red) and with both beam particle halo and cosmic-ray backgrounds overlaid (blue).  Figure \protect\subref{fig:pandora_protodune_tbrecoeffbrkdwn_evt} shows the W plane view for a Monte-Carlo event where the triggered beam particle is shown in blue, the beam halo in red and the cosmic-ray backgrounds in black.}
\label{fig:pandora_protodune_tbrecoeffbrkdwn}
\end{figure}

The effect of cosmic-ray backgrounds and the test beam particle halo on the reconstructed test beam particle efficiency is illustrated in Figure~\ref{fig:pandora_protodune_tbrecoeffbrkdwn}, where the efficiency is shown as a function of the momentum of the triggered particle (\ref{fig:pandora_protodune_tbrecoeffbrkdwn_p}) and the number of hits produced by the triggered particle (\ref{fig:pandora_protodune_tbrecoeffbrkdwn_nhits}).  These figures indicate that the primary loss mechanisms in the test beam particle reconstruction, accounting for $\approx 70\%$ of all inefficiencies, are due to irreducible cosmic-ray and  beam halo backgrounds.

Alongside the test beam particle reconstruction metrics, the \dword{pandora} cosmic ray reconstruction has been studied using \dword{pdsp} data. 

\begin{figure}[!ht] 
\centering
\subfloat[]{\label{fig:pandora_protodune_cr_n}\includegraphics[width=0.5\textwidth]{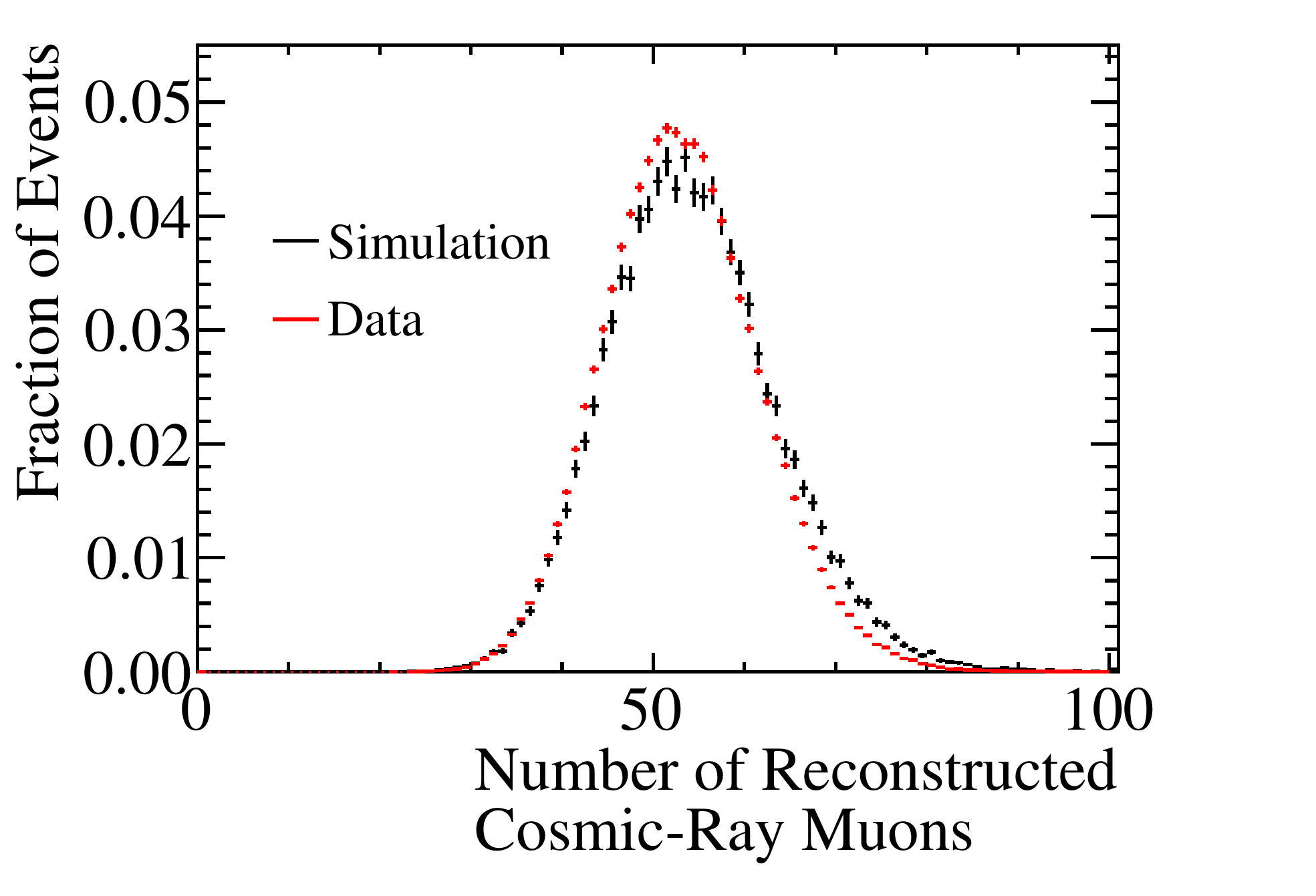}}
\subfloat[]{\label{fig:pandora_protodune_cr_recovsmc}\includegraphics[width=0.357\textwidth]{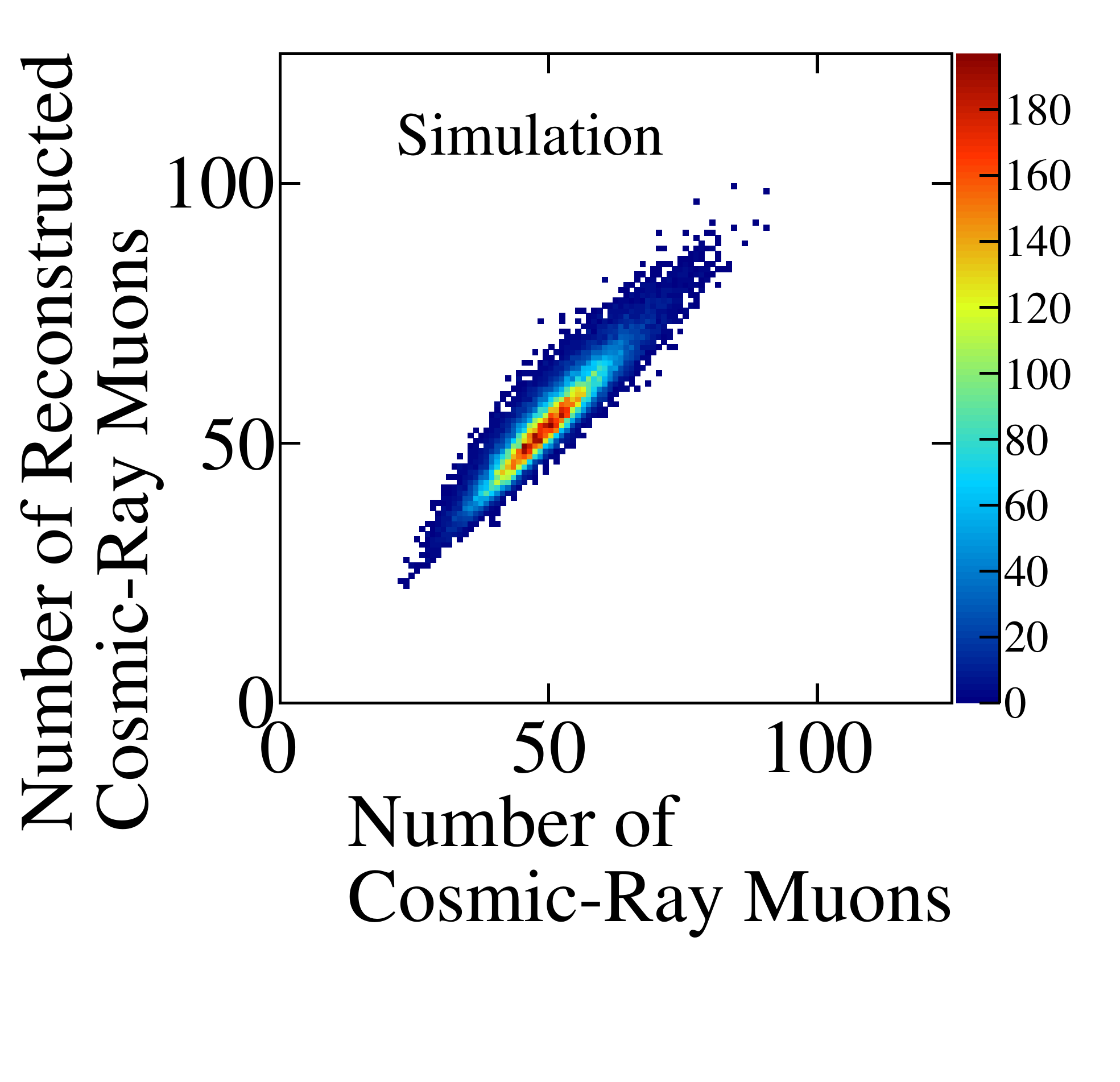}} \\
\caption[Reconstructed cosmic rays per event for data and MC]{The number of distinct, i.e., containing at least 100 hits, reconstructed cosmic rays per event is shown in figure \protect\subref{fig:pandora_protodune_cr_n} for data and \dword{mc}.  Figure \protect\subref{fig:pandora_protodune_cr_recovsmc} shows the number of matched reconstructed cosmic rays per event as a function of the true number of reconstructable distinct cosmic rays passing through the detector per event for \dword{mc}.}
\label{fig:pandora_protodune_cr_number}
\end{figure}

\begin{figure}[!ht]  
\centering
\subfloat[]{\label{fig:pandora_protodune_cr_t}\includegraphics[width=0.5\textwidth]{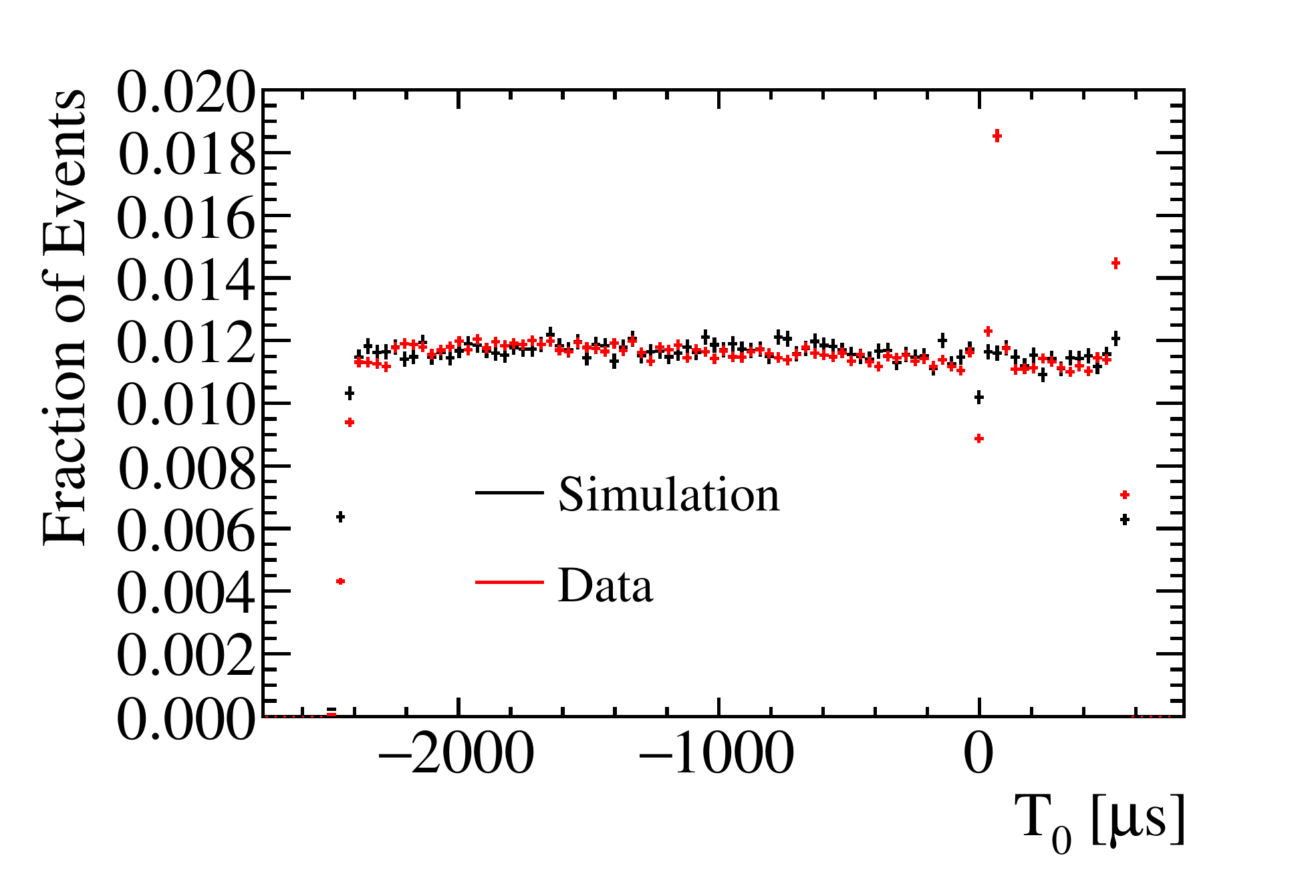}}
\subfloat[]{\label{fig:pandora_protodune_cr_tres}\includegraphics[width=0.5\textwidth]{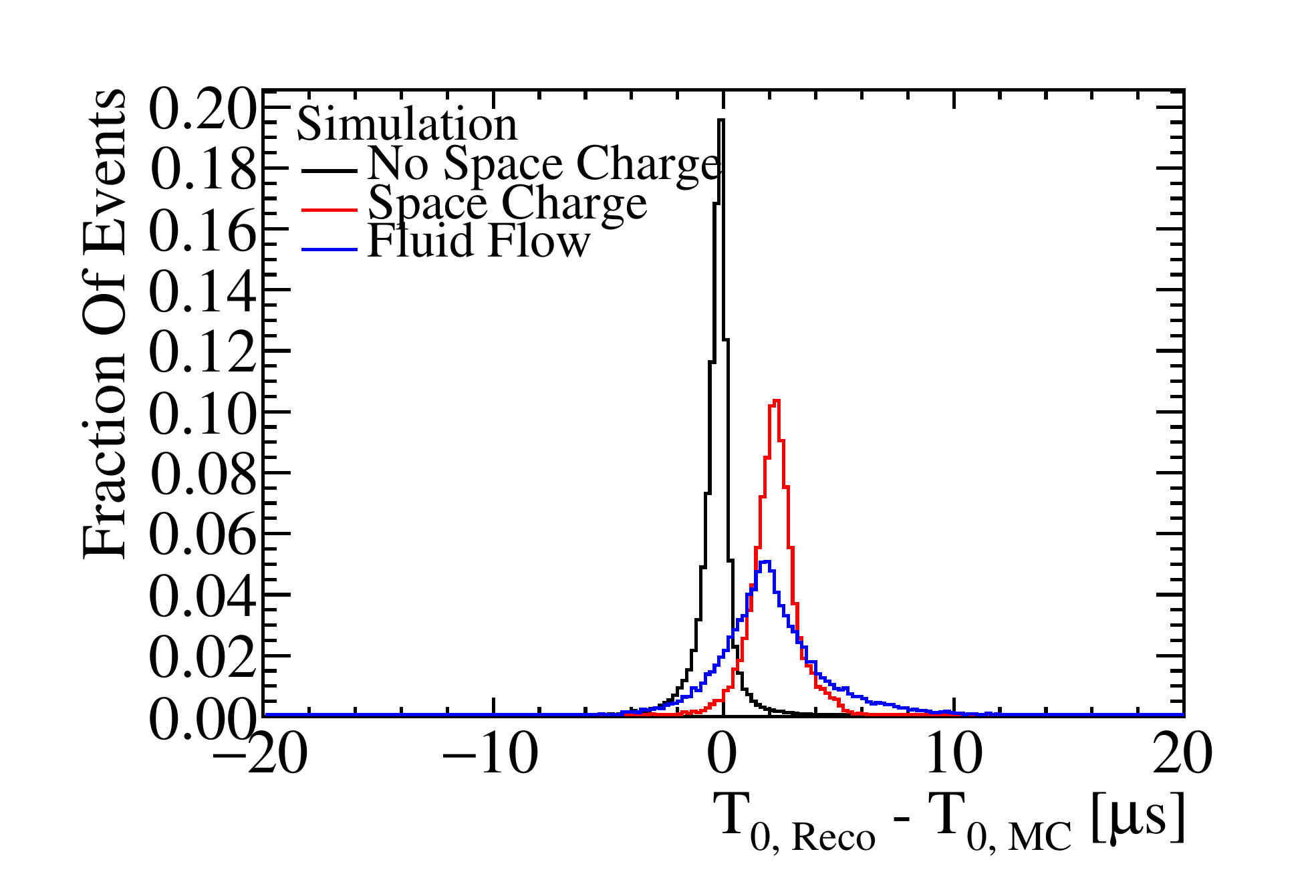}}
\caption[Distribution of reconstructed $t_{0}$ for cosmic rays]{\protect\subref{fig:pandora_protodune_cr_t} The distribution of the reconstructed $t_{0}$ for cosmic rays crossing the \dword{cpa} in both data and \dword{mc}.  \protect\subref{fig:pandora_protodune_cr_tres} the resolution on the reconstructed $t_{0}$ in \dword{mc} with different space charge effects applied; No space charge (black), space charge (red) and fluid flow (blue).}
\label{fig:pandora_protodune_cr_t0}
\end{figure}

Figure \ref{fig:pandora_protodune_cr_n} shows the number of distinct, i.e. that contain at least 100 hits, reconstructed cosmic rays per event.  Both data and \dword{mc} have a similar average number of cosmic rays per event; $53.17\pm0.02$ for data and $54.34\pm0.06$ for simulation.  However, the \dword{mc} distribution has a larger tail suggesting differences between the cosmic-ray profile in data and that used in simulation. Figure~\ref{fig:pandora_protodune_cr_recovsmc} shows the number of matched reconstructed cosmic rays per event as a function of the number of ``target'' reconstructable (as explained in Section~\ref{sec:Pandora:assessment}) distinct cosmic rays per event for \dword{mc} simulation, illustrating that the \dword{pandora} cosmic ray reconstruction is highly efficient.  The \dword{pandora} reconstruction is also able to tag the true time that a cosmic ray passes through the detector, $t_{0}$, should it cross a drift volume boundary, either \dword{cpa} or \dword{apa}.  This allows us to compare the $t_{0}$ distribution for tagged cosmic rays in data and \dword{mc}, shown in Figure \ref{fig:pandora_protodune_cr_t}.  There is excellent agreement between data and \dword{mc} in this instance.  The peak in the data distribution at $\approx 75$~ns appears due to channels affected by a known issue with the cold electronics that is now mitigated in the latest reconstruction.

Figure~\ref{fig:pandora_protodune_cr_tres} shows the resolution on the reconstructed $t_{0}$ for \dword{mc}, which indicates that the \dword{pandora} $t_{0}$ tagging is precise to the order of microseconds.  The shift in the mean of the distribution when applying the space charge effects is due to the effect of bowing of the tracks when space charge is applied.  Furthermore, the broadening of the distribution when applying the fluid flow model, in comparison to space charge, is due to the fact that the bowing effect is no longer correlated between tracks in the asymmetric fluid flow model. The size of the space charge effect is computed from simulations and efforts are ongoing to produce a data-driven space charge simulation.

\subsection{High-Level Reconstruction}
\label{sec:Pandora:High}

This section presents a series of studies to illustrate the results of current efforts on high-level reconstruction, analyzing different reconstructed quantities for tracks and showers. After the pattern recognition stage provided by \dword{pandora}, further fits to the reconstructed \threed particles can be made in order to characterize their properties. For the moment, the results presented here use only the output provided by \dword{pandora}, which includes a first pass of high-level reconstruction to build these objects. For tracks, \dword{pandora} sliding linear fits are used to calculate the trajectory of the particle, whereas for showers a principal component analysis (\dword{pca}) is used to estimate directions and opening angles. 

\begin{figure}[!ht]   
\centering
\subfloat[]{\label{fig:pandora_dunefd_highlevel_reco_length1}}\includegraphics[width=0.5\textwidth]{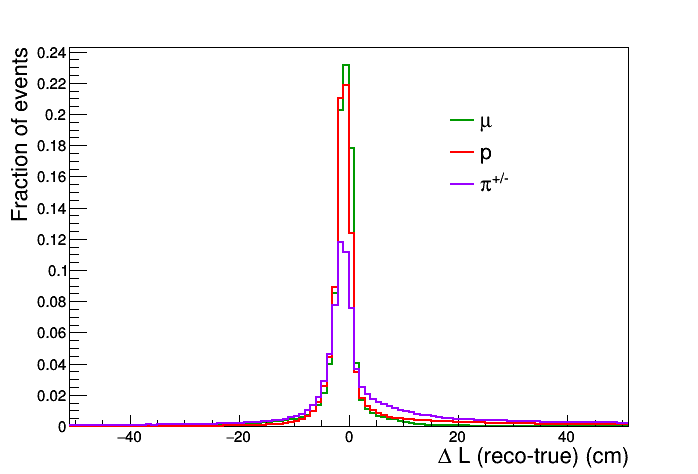}\subfloat[]{\label{fig:pandora_dunefd_highlevel_reco_length2}}\includegraphics[width=0.5\textwidth, height=6cm]{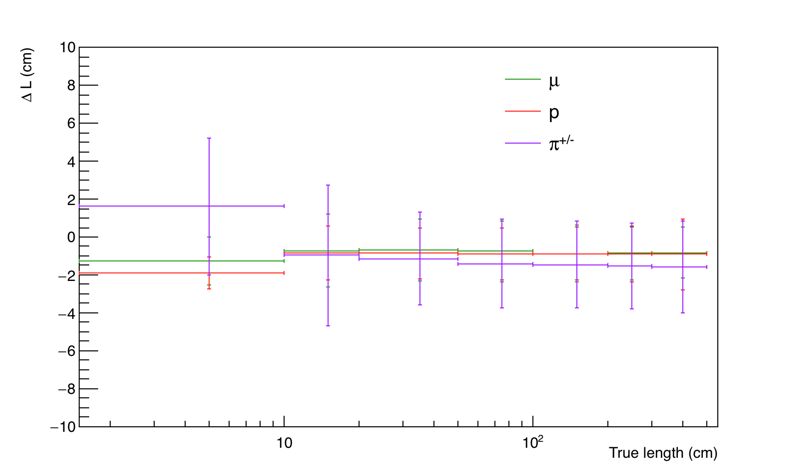}
\caption[Distribution of reconstructed-true length for different track-like particles]{Distribution of reconstructed - true length (\threed distance between start and end positions) for different track-like particles \protect\subref{fig:pandora_dunefd_highlevel_reco_length1}, and mean and sigma (as marker and error bar respectively) of the $\Delta L$ distribution in different ranges of true length\protect\subref{fig:pandora_dunefd_highlevel_reco_length2},}
\label{fig:pandora_dunefd_highlevel_reco_length}
\end{figure}

The opening angle between the reconstructed and the true \threed direction of tracks and showers is presented in Fig. \ref{fig:pandora_dunefd_highlevel_reco_angle} in simulated \dword{fd} neutrino events\footnote{The distributions in Figs. \ref{fig:pandora_dunefd_highlevel_reco_angle} and \ref{fig:pandora_dunefd_highlevel_reco_length} include only good reco-true matches, requiring a minimum of 10\% completeness and 50\% purity for the match. In addition, Fig.~\ref{fig:pandora_dunefd_highlevel_reco_length} is made using only contained tracks, by requiring that both true start and end point are within the fiducial volume. }. The reconstructed direction of tracks is obtained as the initial momentum of the track, after a \dword{pandora} sliding linear fit is performed to its reconstructed \threed points. For showers, the reconstructed direction corresponds to the primary eigenvector result of the \dword{pca} fit to its reconstructed \threed points. In both cases, the opening angle is very small, indicating a good agreement between the reconstructed and true direction of the particles. The few cases in which an opening angle of $~\pi$ is obtained are explained by a good reconstruction of the particle (hit clustering) but the particle vertex placed at its wrong end. 

\begin{dunefigure}
[Opening angle between reconstructed and true direction for track- and shower-like particles]
{fig:pandora_dunefd_highlevel_reco_angle}
{Distribution of opening angle (in radians) between reconstructed and true direction for track-like (left) and shower-like (right) particles in simulated \dword{fd} neutrino events.}
\includegraphics[width=0.49\textwidth]{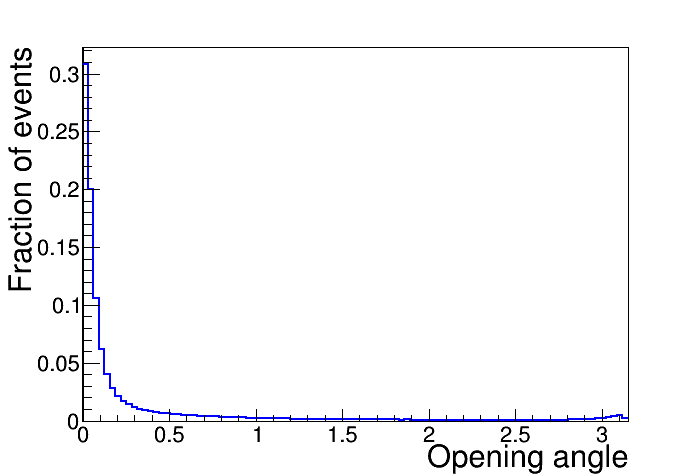}
\includegraphics[width=0.49\textwidth]{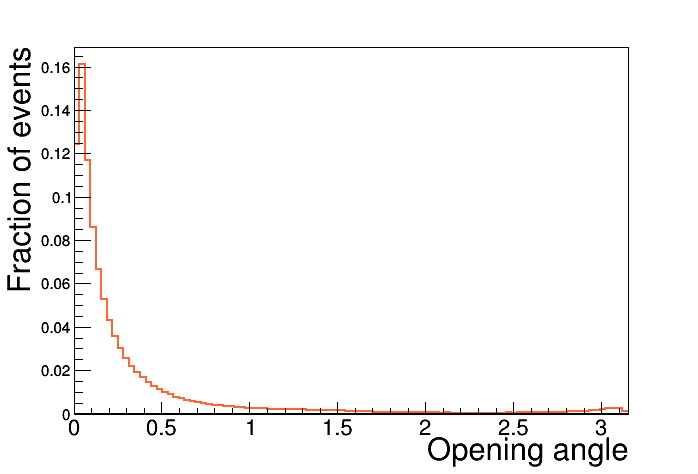}
\end{dunefigure}

For track-like particles, another quantity that can be explored in the high-level reconstruction is the length. Figure~\ref{fig:pandora_dunefd_highlevel_reco_length1} shows the difference between reconstructed and true particle length ($\Delta L$), computed as the \threed distance between start and end positions, for simulated track-like particles of various types in the \dword{fd}. The difference in length $\Delta L$ clearly depends on the particle type: for example, $\sim$90\% ($\sim$83\%) of muons have a $\Delta L$ smaller than 10 cm (5cm), whereas for protons (pions) the fraction within 5 cm is $\sim$78\% ($\sim$54\%). $\Delta L$ depends on the true length of the particle, as shown in Fig.~\ref{fig:pandora_dunefd_highlevel_reco_length2}, which presents the mean and sigma (as marker and error bar respectively) of the $\Delta L$ distribution in different ranges of true length for different particle types\footnote{A Gaussian fit is performed to the $\Delta L$ distributions in each range of true length, except the first one (true length < 10cm) which presents a larger tail and its behavior is better represented by a Landau distribution, of which the most probable value is given instead}. In general, small values of $\Delta L$ can be understood in terms of the efficiency of 3D points creation, and resolution of the vertex reconstruction. Particles presenting kinks due to scattering, such as pions and to a lesser extent protons, have the additional risk of merging parent and daughter particles when the scattering angle is small, increasing the value of $\Delta L$. Short pions are in particular subject to this effect, in addition to merges with other close or overlapping particles in complex topologies, which might translate into larger values of $\Delta L$. 

A number of these variables can be also explored in experimental data taken by the \dword{pdsp} detector. 
For example, figure~\ref{fig:pandora_protodune_vertex} presents a measurement of the test beam particle interaction vertex by comparing the end point of the primary test beam particle track and the fitted interaction vertex for \dword{pdsp} data and \dword{mc} events. 


\begin{dunefigure}
[Resolution on test beam interaction vertex on ProtoDUNE data and MC events]
{fig:pandora_protodune_vertex}
{Resolution on the test beam particle interaction vertex on \dword{protodune} data and \dword{mc} events, calculated comparing the end point of the primary test beam particle track and the fitted interaction vertex.}
\includegraphics[width=0.75\textwidth]{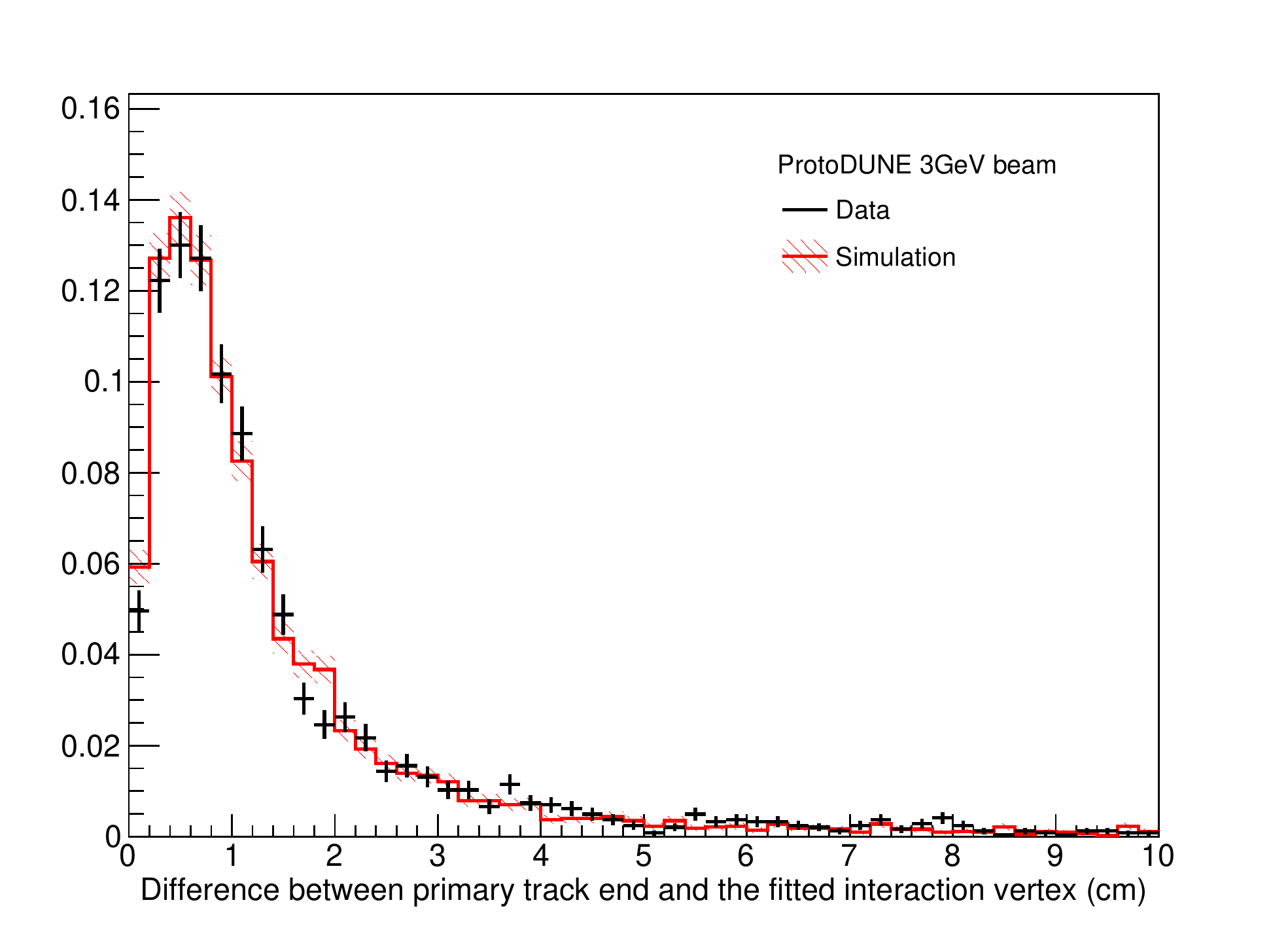}
\end{dunefigure}

\begin{figure}[!ht]   
\centering
\subfloat[]{\label{fig:muon_dedx_resrange_run5387}\includegraphics[width=0.33\textwidth]{muon_run5387_dedx.pdf}}
\subfloat[]{\label{fig:muon_dedx_resrange_sce}\includegraphics[width=0.33\textwidth]{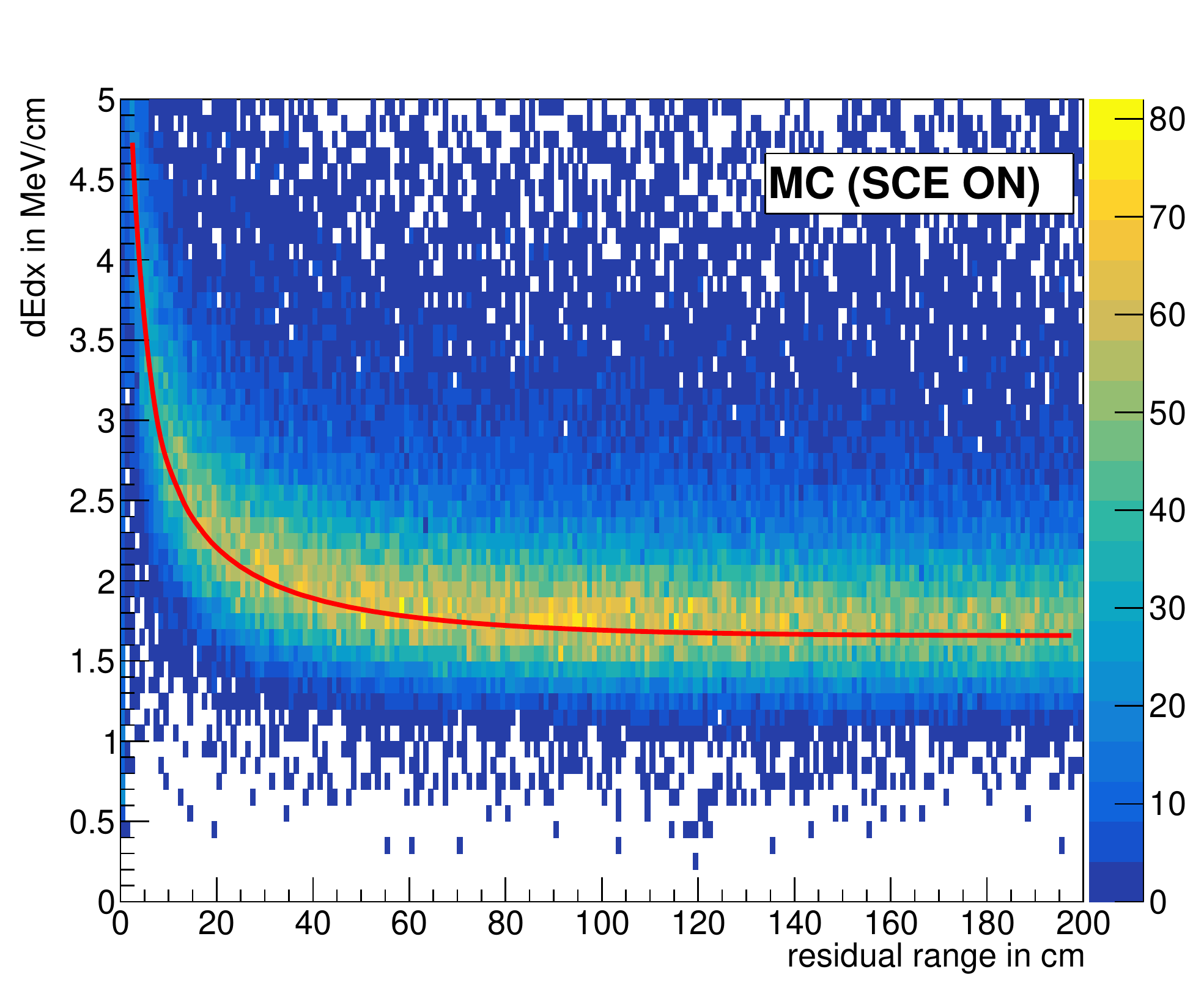}}
\subfloat[]{\label{fig:muon_comparison_dedx_sce_data}\includegraphics[width=0.33\textwidth]{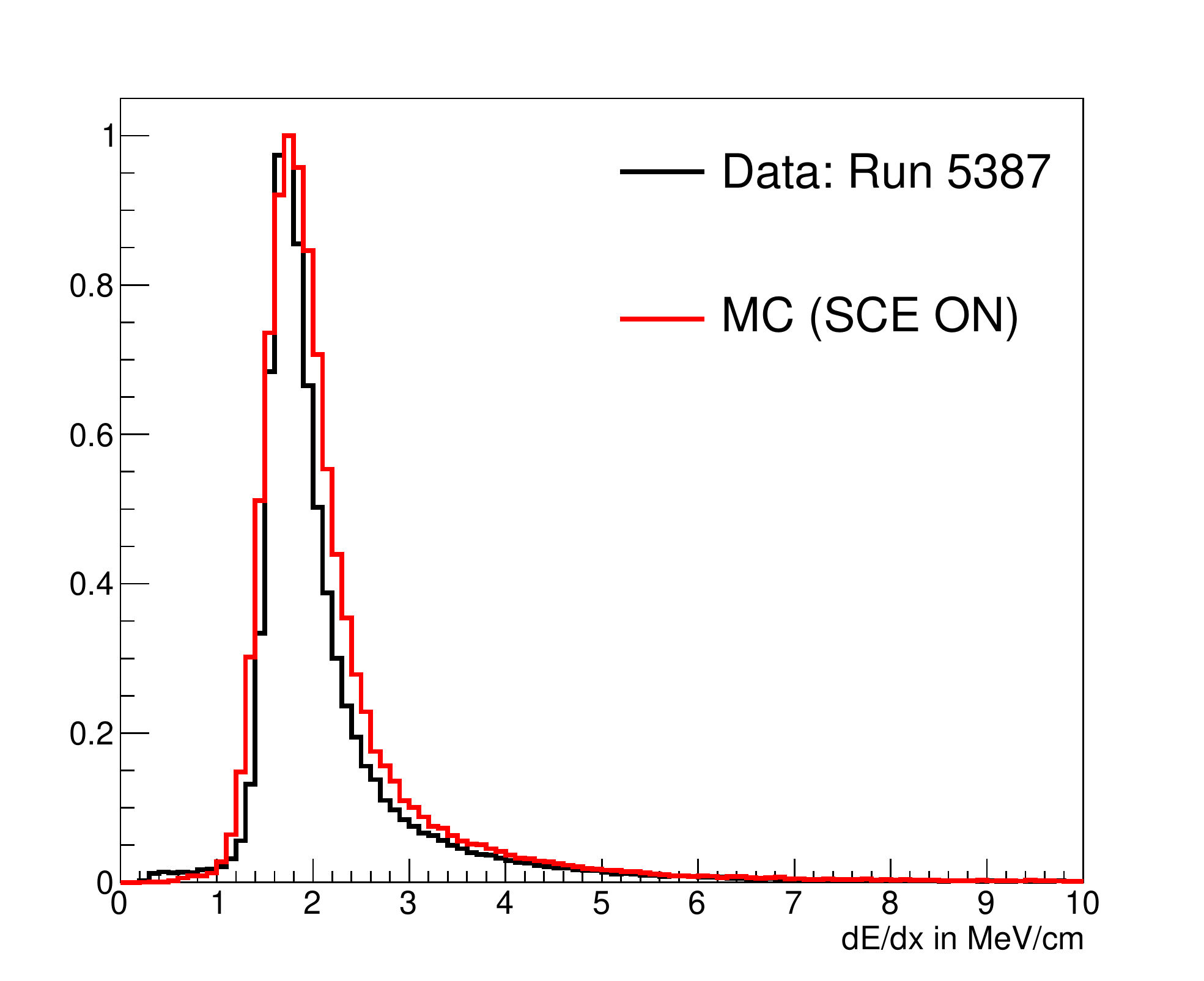}}
\caption[Stopping muon $dE/dx$ distributions for the ProtoDUNE-SP cosmic data and MC]{Stopping muon $dE/dx$ distributions for the \dword{pdsp} cosmic data and \dword{mc}. The red curves in \protect\subref{fig:muon_dedx_resrange_run5387} and \protect\subref{fig:muon_dedx_resrange_sce} are the expected most probable value of $dE/dx$ versus residual range.}
\label{fig:pandora_protodune_mu}
\end{figure}

Cosmic-ray muons in the \dword{pdsp} detector are also used to calibrate the detector nonuniformity and determine the absolute energy scale. Cathode crossing cosmic-ray muons with $t_{0}$ information are used to correct for the attenuation effect caused by impurities in the \lar. Stopping cosmic-ray muons are used to determine the calorimetry constants that convert the calibrated \dword{adc} counts to the number of electrons so that the $dE/dx$ versus residual range distributions match the expectation, as shown in Figures~\ref{fig:muon_dedx_resrange_run5387} and~\ref{fig:muon_dedx_resrange_sce} for \dword{pdsp} data and \dword{mc} simulation with space charge effects after calibration. The data $dE/dx$ distribution has better resolution because the purity in data is better than in the simulation. 

\begin{figure}[!ht]  
\centering
\subfloat[]{\label{fig:proton_dedx_resrange_run5387}\includegraphics[width=0.33\textwidth]{proton_dedx_resrange_run5387.pdf}}
\subfloat[]{\label{fig:proton_dedx_resrange_sce}\includegraphics[width=0.33\textwidth]{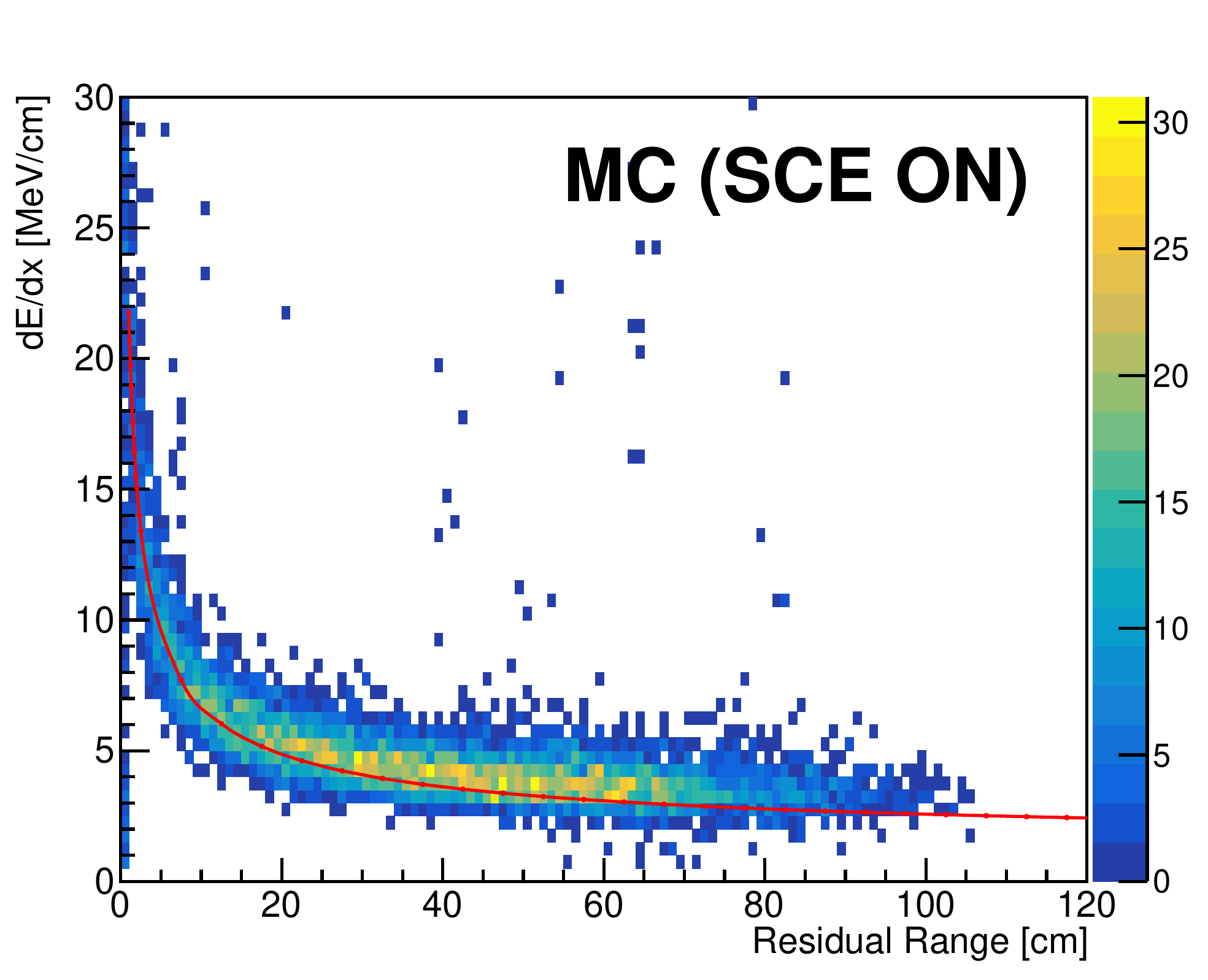}}
\subfloat[]{\label{fig:proton_comparison_dedx_sce_data}\includegraphics[width=0.33\textwidth]{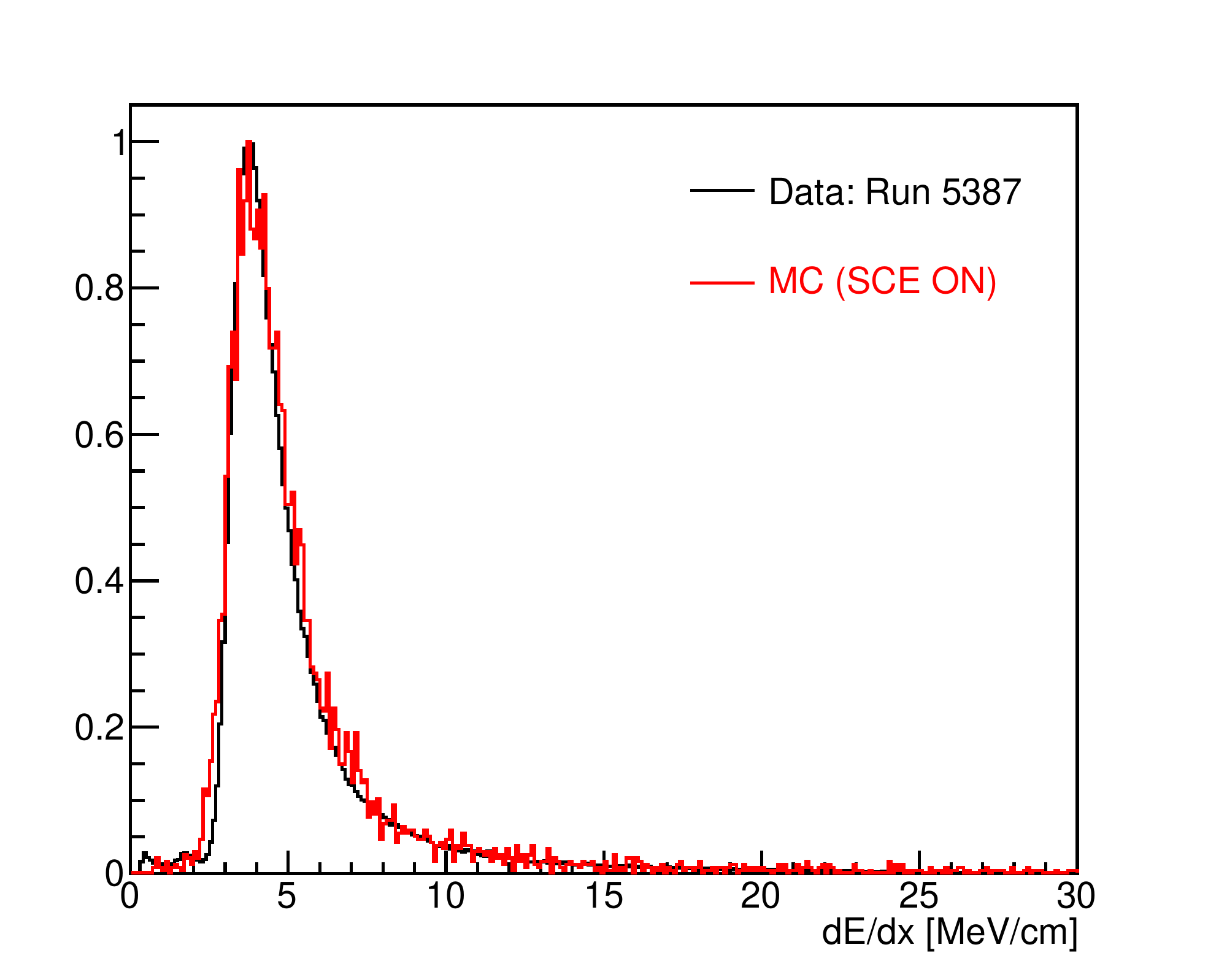}}
\caption[Proton $dE/dx$ distributions for the ProtoDUNE-SP 1 GeV beam data and MC]{Proton $dE/dx$ distributions for the \dword{pdsp} 1 GeV beam data and \dword{mc}. The red curves in \protect\subref{fig:proton_dedx_resrange_run5387} and \protect\subref{fig:proton_dedx_resrange_sce} are the expected most probable value of $dE/dx$ vs residual range.}
\label{fig:pandora_protodune_proton}
\end{figure}

The same attenuation correction and calorimetry constants are applied to the beam proton data and \dword{mc} and the resulting $dE/dx$ distributions are shown in Figure~\ref{fig:pandora_protodune_proton}. The data and \dword{mc} $dE/dx$ distributions agree well. Discrepancy with expectation is observed in the large residual range region, which corresponds to the beam entering point on the TPC front face where space charge effects are large. Good progress is being made on the space charge effects calibration, which will lead to more accurate $dE/dx$ measurements.

\clearpage

\section{DUNE Calibration Strategy}
\label{sec:phys-calib-strat}

The DUNE \dword{fd} presents a unique challenge for calibration in many ways. It differs from existing \dword{lbl} neutrino detectors and existing \dwords{lartpc} because of its size -- the largest \dword{lartpc} ever constructed -- but also because of its deep underground location. 
The DUNE \dword{nd}, which we expect to include a \dword{lartpc}, will also differ from previous experiments (e.g., MINOS and \nova). In particular, while the ND will be highly capable, pile-up and readout will be different, and this may complicate extrapolation of all relevant detector characteristics.

As for any \lartpc, full exploitation of DUNE's capability for precision tracking and calorimetry requires a detailed understanding of the detector response. The inherently highly convolved detector response model and the strong correlations that exist between various calibration quantities make this challenging. 
For example, the determination of energy associated with an event of interest will depend on the simulation model, associated calibration parameters, non-trivial correlations between the parameters, and spatial and temporal dependence of those parameters caused by the non-static nature of the \dword{fd}. 
Changes can be abrupt (e.g., noise, a broken resistor in the \dword{fc}), or ongoing (e.g., exchange of fluid through volume, ion accumulation).

Convincing physics measurements will require a demonstration that the overall detector response is well understood. The systematic uncertainties for the \dword{lbl} and low-energy (\dword{snb}) program will determine the required precision on dedicated calibration systems.
The calibration program must provide measurements at the few-percent-or-better level stably across an enormous volume and over a long period of time, and provide sufficient redundancy.

This section describes the current calibration strategy for DUNE that uses existing sources of particles, external measurements, and dedicated external calibration hardware systems. Existing calibration sources for DUNE include beam or atmospheric neutrino-induced samples, cosmic rays, argon isotopes, and instrumentation devices such as \lar purity and temperature monitors. Dedicated calibration hardware systems currently include laser and pulsed neutron system (PNS).
The responsibility of these hardware systems and assessment of alternative calibration system designs fall under the joint \dword{sp} and \dword{dp} calibration consortium. External measurements by \dword{protodune} and SBN will validate techniques, tools and the design of systems applicable to the DUNE calibration program;  \dword{protodune} will also perform essential measurements of charged particle interactions in \dword{lar}.

Under current assumptions, the calibration strategy described in this document is applicable to both \dwords{spmod} and \dwords{dpmod}. 
Section~\ref{sec:phys-calib-req} briefly describes the physics-driven calibration requirements. 
The nominal \dword{dune} \dword{fd} calibration design is described in Section~\ref{sec:phys-calib-sources}. Finally,
Section \ref{sec:phys-calib-approach} describes a staging plan for calibration
from after the \dword{tdr} through to the operation of the experiment including design validation at \dword{protodune}.


\subsection{Physics-driven Calibration Requirements}
\label{sec:phys-calib-req}

To perform adequate calibrations the physics processes that lead to the formation of the signals required for DUNE's broad physics program,
expected (and unexpected) detector effects must be carefully understood, as they ultimately affect the detector's energy, position and particle identification response. 
Other categories of effects, such as the neutrino interaction model or reconstruction pathologies, can impact measurements of physical quantities. These other effects are beyond the scope of the \dword{fd} calibration effort and would only lead to a higher overall error budget.

\subsubsection{\dword{lbl} physics}
\label{sec:phys-calib-lbl}

Calibration information needs to provide an approximately 1-2\% understanding of normalization 
and position resolution within the detector to support \dword{dune} \dword{lbl} physics. 
A bias on the lepton energy has a significant impact on the sensitivity to \dword{cpv}. 
A \num{3}\% bias in the hadronic state (excluding neutrons) is important, as the inelasticity  distribution for neutrinos and antineutrinos is quite different.  Different fractions of their energies go into the hadronic state. Finally, while studies largely consider a single, absolute energy scale, DUNE will need to monitor and correct relative spatial differences across the enormous DUNE \dword{fd} volume; this is also true for time-dependent changes~\cite{ebias}.

A number of in situ calibration sources will be required to address these broad range of requirements. 
Michel electrons, neutral pions and radioactive sources (both intrinsic and external) are needed for calibrating detector response to electromagnetic activity in the tens-to-hundreds of MeV energy range. Stopping protons and muons from cosmic rays or beam interactions form an important calibration source for calorimetric reconstruction and particle identification. 
\Dword{protodune}, as a dedicated test beam experiment, provides important measurements to characterize and validate particle identification strategies in a \SI{1}{kt}-scale detector and is an essential input to the overall program. Dedicated calibration systems, like lasers, will be useful to provide in situ full-volume measurements of \efield distortions. 
Measuring the strength and uniformity of the \efield is a key aspect of calibration, as  estimates of calorimetric response and \dword{pid} depend on the \efield through recombination. The stringent physics requirements on energy scale and fiducial volume also put similarly stringent requirements on detector physics quantities such as \efield, drift velocity, electron lifetime, and the time dependencies of these quantities; this is discussed in more detail in 
the dedicated laser system discussion under Section~\ref{sec:phys-calib-hardware}.

\subsubsection{\dword{snb} and low-energy neutrino physics}
A combination of 6~MeV (direct neutron capture response), 9~MeV (peak visible $\gamma$-energy of interest to \dword{snb} and $^{8}B$/hep solar neutrinos), 15~MeV (upper visible energy of $^{8}B$/hep solar neutrinos) and $\sim$30~MeV (decay electrons) is needed to map the low energy response. Supernova signal events present specific reconstruction and calibration challenges, and observable energy is shared between different charge clusters and types of energy depositions. In particular, the supernova signal will have a low-energy electron, gamma and neutron capture component, and each needs to be characterized. As discussed further in Section~\ref{ch:snb-lowe}, primary requirements for this physics include 
(1) calibration of absolute energy scale and energy resolution, which is important for resolving spectral features of \dword{snb} events;
(2) calibration of time and light yield response of optical photon detectors;  
(3) absolute timing of events;  
(4) measurement of trigger efficiency at low energies;  and 
(5) understanding of detector response to radiological backgrounds. Further details on the necessary energy scale, energy resolution and trigger efficiency targets needed can be in Ref~\cite{bib:docdb14068}.
Potential calibration sources in this energy range include Michel electrons from muon decays (successfully utilized by ICARUS and \dword{microboone}~\cite{Acciarri:2017sjy}), which have a well known spectrum up to $\sim\,$\SI{50}{\MeV}.
Photons from neutral pion decay (from atmospheric and beam induced $\pi^0$) will provide an overall energy scale between \SI{50}{\MeV} and \SI{100}{\MeV}, in addition to cosmic ray muon energy loss. 
However, the limited statistical power of those samples (see Table~\ref{tab:cosmic-ray-calib-rates}) mean that it is not possible for these samples to provide the energy scale or resolution at the spatial and temporal granularity needed.
The pulsed neutron system can provides a source of direct neutron capture across the entire DUNE volume, providing a timing and energy calibration.
The proposed radioactive source system provides an in situ source of the electrons and de-excitation $\gamma$ rays, which are directly relevant for physics signals from \dword{snb} or $^{8}B$ solar neutrinos. These two systems (discussed in more detail in Section~\ref{sec:phys-calib-hardware}) can provide calibrations of photon, electron, and neutron response for energies below \SI{10}{\MeV}, where photons and electrons may have very different characteristics in \dword{lar}.

\subsubsection{Nucleon decay and other exotic physics}
The calibration needs for nucleon decay and other exotic physics are comparable to those for the \dword{lbl} program, as listed in Section~\ref{sec:phys-calib-lbl}. Signal channels for light \dword{dm} and sterile neutrino searches will be \dword{nc} interactions that are background to the \dword{lbl} physics program. 
Based on the widths of $dE/dx$-based metrics of \dword{pid}, qualitatively, we need to calibrate $dE/dx$ across all drift and track orientations at the few-percent level, similar to the \dword{lbl} effort.

\subsection{Calibration Sources, Systems and External Measurements}
\label{sec:phys-calib-sources}
Calibration sources and systems provide measurements of the detector response model parameters, or provide tests of the response model itself. Calibration measurements can also provide corrections to data, data-driven efficiencies, systematics and particle responses. 
Figure~\ref{fig:calibneeds} shows the broad range of categories of measurements that calibrations can provide, and lists 
important calibration parameters for DUNE's detector response model applicable to both \dword{sp} or \dword{dp}. Due to the significant interdependencies of many parameters (e.g., recombination, \efield, and \lar purity), a calibration strategy will either need to  measure parameters iteratively, or find sources that break these correlations.

Table~\ref{tab:calibsystem} provides a list of the calibration sources and dedicated calibration systems, along with their primary usage, that will comprise the current
nominal DUNE \dword{fd} calibration design. 
The next sections provide more details on each of them. \Dword{protodune} and previous measurements provide independent tests of the response model, indicating that the choice of parameterization and values correctly reproduces real detector data. 
Not all of the ex situ measurements can be directly extrapolated to DUNE, however, due to other detector effects and conditions -- only those considered to be universal (e.g., argon ionization energy). 

Each of the many existing calibration sources comes with its own challenges. For example, while electrons from muon decay (Michel electrons) are very useful for studying the detector response to low-energy electrons (\SI{50}{\MeV}), these low-energy electrons present reconstruction challenges due to the loss of charge from radiative photons, as demonstrated in \dword{microboone}~\cite{Acciarri:2017sjy}.  
Michel electrons are therefore considered an important, independent, and necessary test of the TPC energy response model, but they will not provide a measurement of a particular response parameter.

\begin{dunefigure}[Categories of measurements provided by calibration]{fig:calibneeds}{Categories of measurements provided by calibration.}
\includegraphics[width=.9\textwidth]{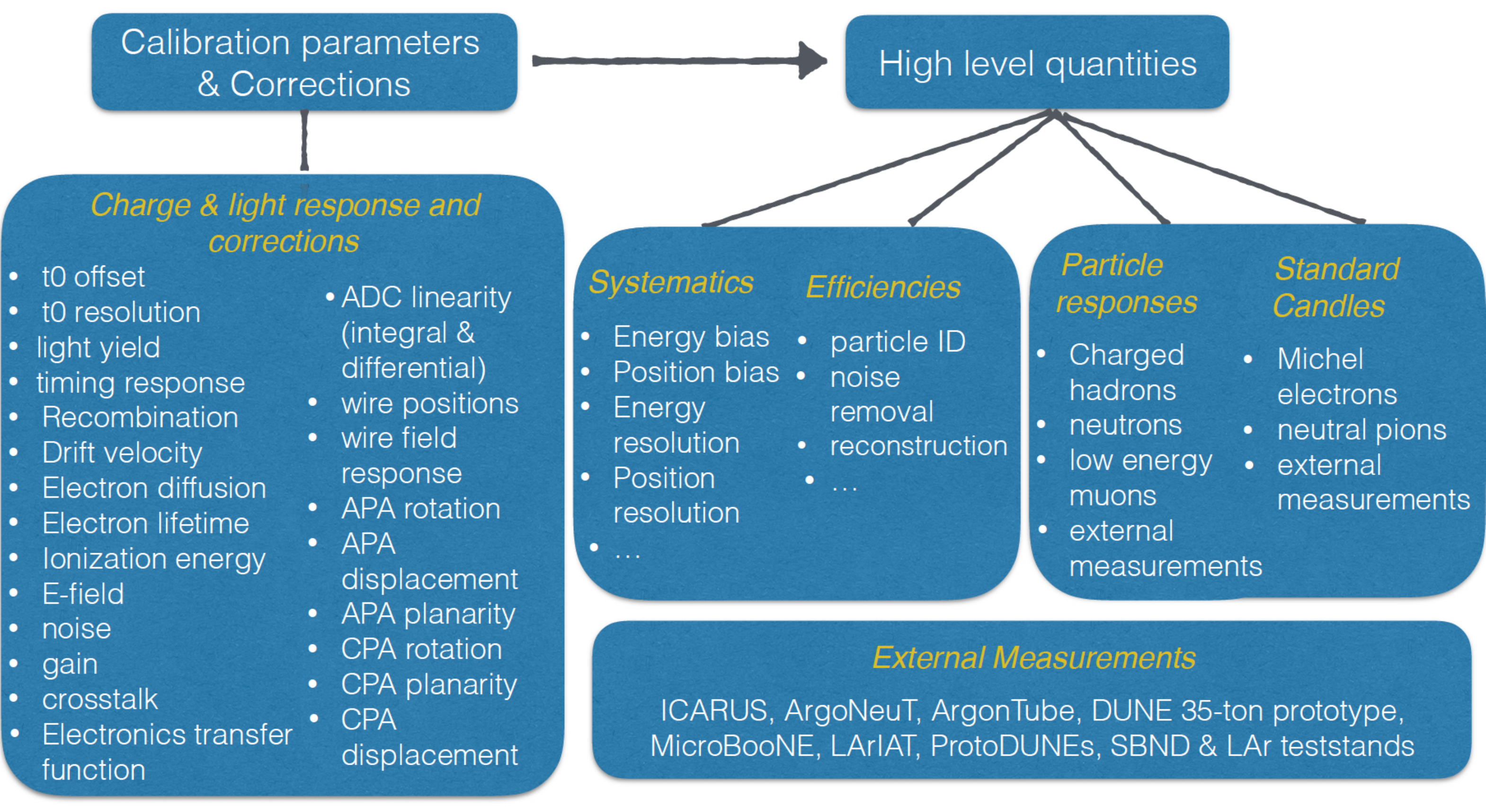}
\end{dunefigure}

\begin{dunetable}[Calibration systems and sources of the nominal DUNE FD calibration design]
{p{.4\textwidth}p{.55\textwidth}}{tab:calibsystem}{Primary calibration systems and sources that comprise the nominal DUNE FD calibration design along with their primary usage.}
System & \textbf{Primary Usage}  \\ \toprowrule 
& \\
\textbf{Existing Sources} & \textbf{Broad range of measurements} \\ \toprowrule
$\mu$, predominantly from cosmic ray & Position (partial), angle (partial), 
electron lifetime, wire response, $dE/dx$ calibration etc.\\ \colhline 
Decay electrons, $\pi^0$ from beam, cosmic, atm $\nu$ & Test of electromagnetic response model \\ \colhline
$^{39}$Ar beta decays &  electron lifetime (x,y,z,t), diffusion, wire response \\  \colhline
& \\ 
\textbf{External Measurements} & \textbf{Tests of detector model, techniques and systems} \\ \toprowrule
ArgoNeuT~\cite{Acciarri:2013met}, ICARUS~\cite{Amoruso:2004dy, Antonello:2014eha, Cennini:1994ha}, MicroBooNE & Model parameters (e.g., recombination, diffusion) \\ \colhline 
DUNE \dword{35t}~\cite{Warburton:2017ixr} & Alignment and \textit{t0} techniques\\ \colhline 
ArgonTUBE~\cite{Ereditato:2014tya}, MicroBooNE~\cite{Acciarri:2016smi}, SBND, ICARUS~\cite{Auger:2016tjc},  \Dword{protodune}~\cite{Abi:2017aow} & Test of systems (e.g., Laser) \\ \colhline
ArgoNeuT~\cite{Acciarri:2015ncl}, MicroBooNE~\cite{bib:uBlifetime, MICROBOONE-NOTE-1018-PUB, MICROBOONE-NOTE-1028-PUB, Acciarri:2017sjy, Abratenko:2017nki, Acciarri:2013met}, ICARUS~\cite{Ankowski:2008aa,  Ankowski:2006ts,Antonello:2016niy},  \Dword{protodune} & Test of calibration techniques and detector model (e.g., electron lifetime, Michel electrons, $^{39}$Ar beta decays) \\ \colhline
\Dword{protodune}, LArIAT~\cite{Cavanna:2014iqa}, CAPTAIN~\cite{Bhandari:2019rat} & Test of particle response models and fluid flow models \\  \colhline
\dword{lartpc} test stands~\cite{Cancelo:2018dnf, Moss:2016yhb, Moss:2014ota, Li:2015rqa} & Light and LAr properties; signal processing techniques \\ \colhline 
& \\
\textbf{Monitoring Systems} & \textbf{Operation, Commissioning and Monitoring} \\ \toprowrule
Purity monitors & Electron lifetime \\ \colhline
Photon detection monitoring System & \dword{pds} response \\ \colhline
Thermometers & Temperature, velocity; test of fluid flow model \\ \colhline
Charge injection & Electronics response \\ \colhline
& \\
\textbf{Dedicated Calibration Systems} & \textbf{Targeted (near) independent, precision calibration}\\ \toprowrule
Direct ionization via laser & Position, angle, electric field (x,y,z,t) \\ \colhline
Photoelectric ejection via laser & Position, electric field (partial) \\ \colhline
Neutron injection & Test of \dword{snb} signal, neutron capture model \\ \colhline
Proposed Radioactive source deployment & Test of \dword{snb} signal model \\ \colhline
\end{dunetable}


\subsubsection{Existing sources} 
\label{sec:phys-calib-exis}
Cosmic rays and neutrino-induced interactions provide commonly used ``standard candles,'' e.g., electrons from muon decays, and photons from neutral pions, which have characteristic energy spectra. Cosmic ray muons are also used to determine detector element locations (alignment), timing offsets or drift velocity, electron lifetime, and channel-by-channel response, and to help constrain \efield distortions. 
Table~\ref{tab:cosmic-ray-calib-rates} summarizes the rates for cosmic ray events. Certain measurements (e.g., channel-to-channel gain uniformity and cathode panel alignment) are estimated to take several months of data. Table~\ref{tab:atmos_rates} gives the atmospheric $\nu$ interaction rates, which  
 are comparable to beam-induced events -- neither occurs at sufficient rates to provide meaningful spatial or temporal calibration; they will likely provide supplemental measurements only. (The beam will not yet be operational for calibration of the first \dword{detmodule} during early data taking.) Instead, we can use the reconstructed energy spectrum of ${}^{39}$Ar beta decays to make a precise measurement of electron lifetime with spatial and temporal variations. 
 This can also provide other necessary calibrations, such as measurements of wire-to-wire response variations and diffusion measurements using the signal shapes associated with the beta decays. The ${}^{39}$Ar beta decay rate in commercially-provided argon is about \SI{1}{\becquerel\per\kilo\gram}, so $O(\mathrm{50k})$ ${}^{39}$Ar beta decays are expected in a single \SI{5}{\milli\s} event readout in an entire \nominalmodsize \detmodule. 
 The ${}^{39}$Ar beta decay cutoff energy is \SI{565}{\keV},  which is close to the energy deposited on a single wire by a \dword{mip}. However, several factors can impact the observed charge spectrum from ${}^{39}$Ar beta decays, such as electronics noise, electron lifetime and recombination fluctuations; more details can be found in the Appendix~\ref{app:ar39}. MicroBooNE~\cite{MICROBOONE-NOTE-1050-PUB} and ProtoDUNE are actively pursuing this technique, thus providing valuable inputs for DUNE.

\begin{dunetable}
[Annual rates for classes of cosmic-ray events useful for calibration]
{lrl}
{tab:cosmic-ray-calib-rates}
{Annual rates for classes of cosmic-ray events described in this section assuming 100\% reconstruction efficiency.  Energy, angle, and fiducial requirements
have been applied. Rates and geometrical features apply to the single-phase far detector design. }
Sample & Annual Rate & Detector Unit \\ \colhline
Inclusive & $1.3\times 10^6$ & Per \nominalmodsize module \\ \colhline
Vertical-Gap crossing & 3300 & Per gap \\ \colhline
Horizontal-Gap crossing & 3600 & Per gap \\ \colhline
\dword{apa}-piercing & 2200 & Per \dword{apa} \\ \colhline
\dword{apa}-\dword{cpa} piercing & 1800 & Per active \dword{apa} side \\ \colhline
\dword{apa}-\dword{cpa} piercing, \dword{cpa} opposite to \dword{apa} & 360 & Per active \dword{apa} side \\ \colhline
Collection-plane wire hits & 3300 & Per wire \\ \colhline
Stopping Muons & 28600 & Per \nominalmodsize module \\ \colhline
$\pi^0$ Production & 1300 & \nominalmodsize module \\ \colhline
\end{dunetable}

\subsubsection{Monitors} 
 
Chapter~8 of \voltitlesp{} and \voltitledp{} discuss several instrumentation and detector monitoring devices in detail. These devices, including liquid argon temperature monitors, \lar purity monitors, gaseous argon analyzers, cryogenic (cold) and inspection (warm) cameras, and liquid level monitors, will provide valuable information for early calibrations and for tracking the space-time dependence of the \dwords{detmodule}. 
The \dword{cfd} simulations play a key role for calibrations initially in the design of the cryogenics recirculation system, and later for physics studies when the cryogenics instrumentation data can be used to validate the simulations. Chapters~4 and~5 of the \dword{detmodule} volumes discuss other instrumentation devices essential for calibration, such as drift \dword{hv} current monitors and external charge injection systems. 

\subsubsection{External measurements} 

DUNE will use external measurements from past experimental runs (e.g., ArgoNeuT, the DUNE \dword{35t}, ICARUS, and \lariat), from ongoing and future experiments (e.g., \dword{microboone}, \dword{protodune}, and SBND), and from small scale \dword{lartpc} test stands. External measurements provide a test bed for dedicated calibration hardware systems and techniques for the \dword{fd}. In particular, \dword{protodune} will provide validation of the fluid flow model using cryogenic instrumentation data. 
Early calibration for physics in DUNE will utilize \lar physical properties from \Dword{protodune} or SBN  for tuning detector response models in simulation. Table~\ref{tab:calibsystem} provides  references for specific external measurements. The usability of ${}^{39}$Ar has been demonstrated with \microboone data~\cite{MICROBOONE-NOTE-1050-PUB}. 
Use of  ${}^{39}$Ar  and other radiological sources and, in particular, the \dword{daq} readout challenges associated with their use, will be tested on the \dword{protodune} detectors. Dedicated systems for DUNE, 
including the laser system, have been used by previous experiments (ARGONTUBE~\cite{Zeller:2013sva,Ereditato:2014lra}, CAPTAIN, and MicroBooNE experiments) and at SBND in the future, and will provide more information on use of the system and optimization of the design.  The small-scale \lar test stand planned at Brookhaven National Lab, USA, will provide important information on simulation and calibration of field response for DUNE.

External measurements of particle response (e.g. pion interactions in \dword{lar}) are also important inputs to the detector model. These include dedicated measurements made with ProtoDUNE, LArIAT, and CAPTAIN~\cite{Bhandari:2019rat}; the DUNE ND, with both a \dword{lar} and low density gas detector, will also make measurements which characterize the relevant cross sections and outgoing final state particles.


\subsubsection{Dedicated Calibration Hardware Systems}
\label{sec:phys-calib-hardware}

This section briefly describes the physics motivation and measurement goals for the calibration hardware systems and the designs currently envisioned. The calibration chapters in \voltitlesp{} and \voltitledp{} of the \dword{tdr} provide further details on the design and development plan for these systems. We plan to deploy prototype designs of these systems in 
the phase 2 of \dword{protodune} to demonstrate proof-of-principle.

\textbf{Laser systems} 

The primary purpose of a laser system is to provide an independent, fine-grained estimate of the \efield in space or time, which is a critical parameter for physics signals as it ultimately impacts the spatial resolution and energy response of the detector. External measurements, e.g.,  MicroBooNE's, use both a laser system and cosmic rays to estimate the \efield, however the expected cosmic rate at the deep underground installation of the \dword{fd} will not provide sufficient spatial or temporal granularity to study local distortions.

\efield distortions can arise from multiple sources. Current simulation studies indicate that positive ion accumulation and drift (space charge) due to ionization sources such as cosmic rays or ${}^{39}$Ar are small in the \dword{fd}; however, the fluid flow pattern in the \dword{fd} is not yet sufficiently understood to exclude the possibility of stable eddies that may amplify the effect for both \single and \dual modules. The \dword{dpmod} risks significant further amplification due to  accumulation in the liquid of ions created by the electron multiplication process in the gas phase.
Detector imperfections can also cause localized \efield distortions. Examples include \dword{fc} resistor failures, non-uniform resistivity in the voltage dividers, \dword{cpa} misalignment, \dword{cpa} structural deformations, and \dword{apa} and \dword{cpa} offsets and  deviations from flatness. Individual \efield distortions may add in quadrature with other effects, and can reach 4-5\% under certain conditions, which corresponds to a 1-2\% impact on charge, 
and a $\sim 2$ cm impact on position (and fiducial volume). Both charge and position distortions affect energy scale. Understanding all these effects requires an in situ calibration of the E field with a precision of about 1\% with a coverage of at least 75\% of the detector volume.

The laser calibration system offers secondary uses, e.g., alignment (especially modes that are weakly constrained by cosmic rays, see Figure~\ref{fig:apacurtainalign}), stability monitoring, and diagnosing detector failures in systems such as \dword{hv}.  

\begin{dunefigure}[Sample distortion that may be difficult to detect with cosmic rays]{fig:apacurtainalign}
{An example of a distortion that may be difficult to detect with cosmic rays.  The \dword{apa} frames are shown as rotated rectangles, as viewed from the top.}
\includegraphics[width=0.8\textwidth]{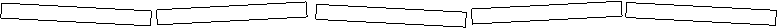}
\end{dunefigure}

Two systems are under consideration to extract the \efield map: \phel{}s from the \lartpc cathode and direct ionization of the \dword{lar}, both driven by a \SI{266}{\nano\m} laser.  The reference design from \dword{microboone}~\cite{bib:uBlaser2019} and SBND uses direct ionization laser light with multiple laser paths. This can provide field map information in $(x, y, z, t)$; a \phel laser only provides an integrated measurement of the \efield along the drift direction.
The ionization-based system can characterize the \efield with fewer dependencies compared to other systems. If two laser tracks enter the same spatial voxel in a \dword{detmodule}, the relative position of the tracks provides an estimate of the local \threed \efield. The deviation from straightness of single ``laser tracks'' can also be used to constrain local \efield{}s. Comparison of the known laser track path against the path reconstructed from cosmic or beam data, assuming uniform \efield, can also be used to estimate local \efield distortions. A schematic of the ionization laser setup and a laser track from \dword{microboone} is shown in Figure~\ref{fig:uB_laser_schematic}.

\begin{dunefigure}[\microboone laser calibration system schematics]{fig:uB_laser_schematic}
{Left: Schematics of the ionization laser system in \dword{microboone}~\cite{Antonello:2015lea}. Right: A UV laser event in the MicroBooNE detector~\cite{bib:uBlaser2019}. The laser track can be identified by the endpoint on the cathode (larger charge visible at the top of the image) and the absence of charge fluctuations along the track. The charge released at the cathode comes from photoelectric effect. Other tracks seen in the display are from cosmic muons.}
\includegraphics[width=0.45\linewidth]{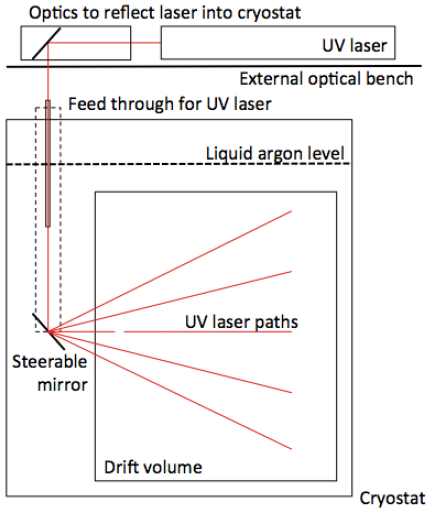}
\includegraphics[width=0.45\linewidth]{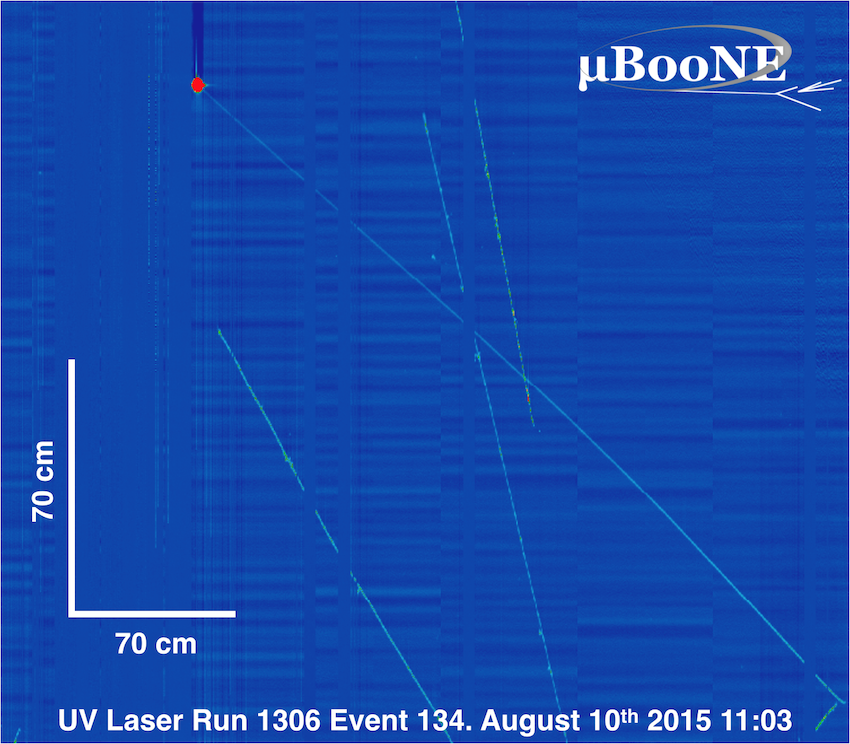}
\end{dunefigure}

A \phel{}-based calibration system was used in the T2K gaseous (predominantly Ar), TPCs~\cite{Abgrall:2010hi}. 
Thin metal surfaces placed at surveyed positions on the cathode provided point-like and line sources of \phel{}s when illuminated by a laser. The T2K \phel system provided measurements 
of adjacent electronics modules' relative timing response, drift velocity with a few \si{\nano\s} resolution over their \SI{870}{\milli\m} drift distance, electronics gain, 
transverse diffusion, and an integrated measurement of the \efield along the drift direction. DUNE would use the system similarly to diagnose electronics or TPC response issues on demand, and to provide an integral field measurement across drift as well as measure relative distortions of $y$, $z$ positions with time, $x$ and/or drift velocity. \microboone has also observed ejection of \phel{}s from the cathode using the direct ionization laser system. 

\textbf{Pulsed neutron source} 

An external neutron generator system would provide a triggered, well defined energy deposition from neutron capture in $^{40}$Ar detectable throughout the \dword{detmodule} volume. Neutron capture is a critical component of signal processes for \dword{snb} and \dword{lbl} physics; this system would enable direct testing of the detector response  spatially and temporally for the low-energy program.  This is important to measure energy scale, energy resolution and detection threshold spatially and temporally across the enormous DUNE volume.

\begin{dunefigure}[Cross sections enabling the PNS concept]{fig:pns_xsec}
{Illustration of interference anti-resonance dip in the cross section of  \isotope{Ar}{40}. Elastic scattering cross section data is obtained from ENDF VIII.0}
\includegraphics[width=8cm]{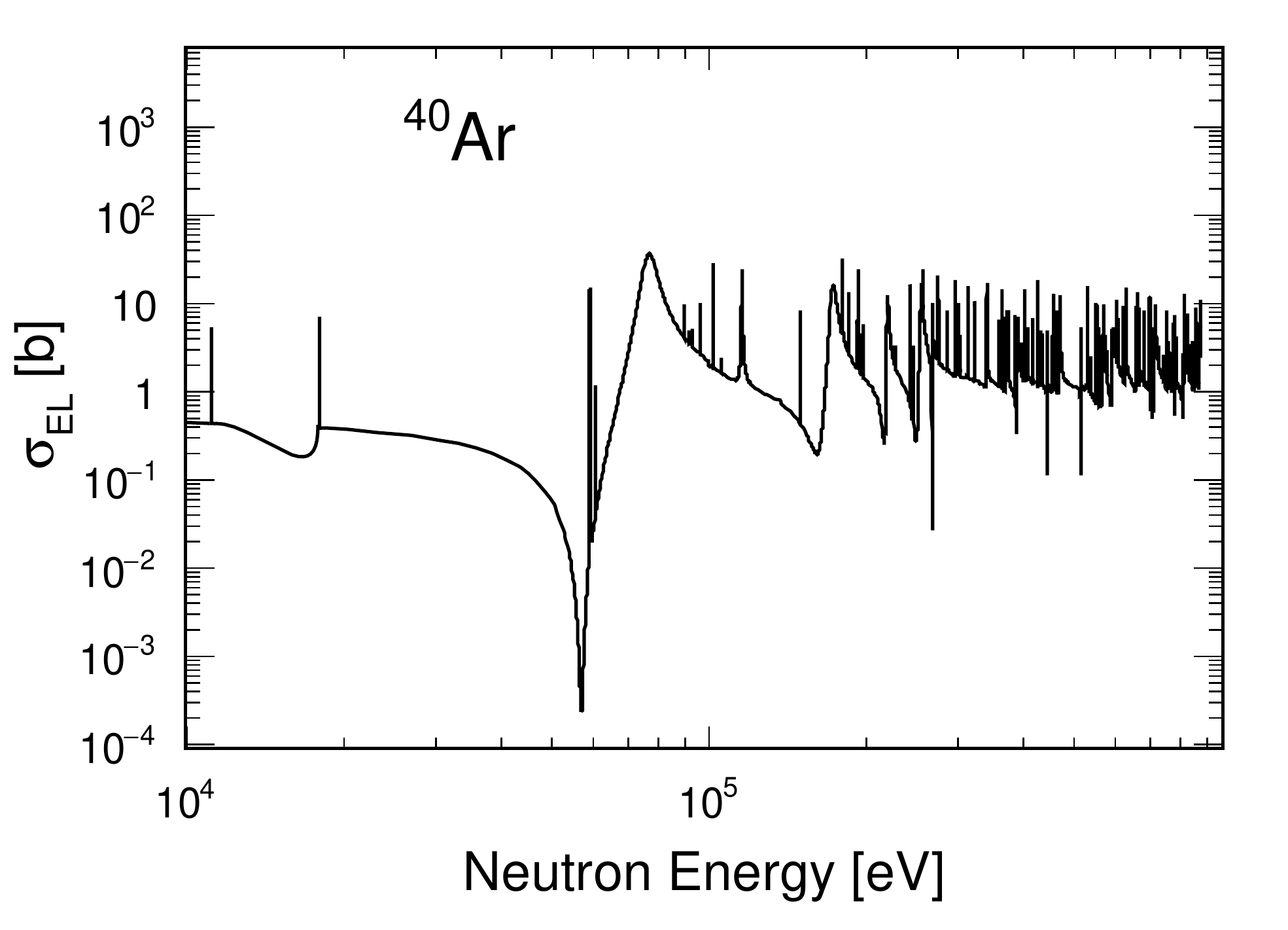}
\end{dunefigure}

A triggered pulse of neutrons can be generated outside the TPC and injected into the \dword{lar}, where it spreads through the entire volume to produce a mono-energetic cascade of photons via the $^{40}$Ar(n,$\gamma$)$^{41}$Ar capture process. The uniform population of neutrons throughout the \dword{detmodule} volume exploits a remarkable property of argon -- the near transparency to neutrons of energy near \SI{57}{\keV}. 
This is due to a deep minimum in the cross section caused by the destructive interference between two high-level states of the \isotope{Ar}{40} nucleus (see Fig.~\ref{fig:pns_xsec}). This cross section ``anti-resonance'' is approximately  \SI{10}{\keV} wide, and \SI{57}{keV} neutrons consequently have a scattering length of \SI{859}{m}; the scattering length averaged over the isotopic abundance in natural Ar is approximately \SI{30}{m}. 
For neutrons moderated to this energy the DUNE \dword{lartpc} is essentially transparent. The \SI{57}{keV} neutrons that do scatter quickly leave the anti-resonance and thermalize, at which time they capture. Each neutron capture releases exactly the binding energy difference between \isotope{Ar}{40} and \isotope{Ar}{41}, about \SI{6.1}{\MeV}, in the form of gamma rays. 
The neutron capture cross-section and the $\gamma$ spectrum have been measured and characterized. Recently, the ACED Collaboration performed a neutron capture experiment using  the Detector  for Advanced  Neutron  Capture  Experiments  at DANCE (ACED)  at the  Los  Alamos  Neutron  Science  Center  (LANSCE). The result of neutron capture cross-section was published~\cite{Fischer:2019qfr} and will be used to prepare a database for the neutron capture studies. The data analysis of the energy spectrum of correlated gamma cascades from neutron captures is underway.

DUNE plans
to place a fixed, shielded deuterium-deuterium ($DD$) neutron generator  above a penetration in the hydrogenous insulation of the \dword{detmodule} cryostat. Between the generator and the cryostat, layers of water or plastic and intermediate fillers would provide sufficient degradation of the neutron energy. 

\textbf{Additional Systems}

There are 
additional systems under consideration for DUNE calibration. Radioactive source deployment provides an in situ source of low energy electrons and de-excitation gamma rays at a known location and with a known activity, which are directly relevant for detection of \dword{snb} or $^{8}B$ solar neutrinos. As shown in Section~\ref{ch:snb-lowe}, the electron and photon response in the TPC is quite different (electrons leave worm-like tracks, photons leave `blips'). The PNS source will provide a 6.1 MeV multi-photon signal; radioactive sources can provide a single photon signal to measure detection threshold and demonstrate sufficient uncertainty on energy resolution at the peak of the \dword{snb} photon signal.  The radioactive source system is under study, and feasibility and safety of deployment would be established with a dedicated run using a prototype system in ProtoDUNE.

The utility of internal source injection (e.g., ${}^{222}$Rn or ${}^{220}$Rn injection) for mapping electron lifetime and fluid flow in the \dword{tpc}, used in dark matter experiments, will also be considered in the future. The major challenge for this system is if the 
coverage of the \dword{pds} is sufficient, and whether or not it will be able to identify a signal and trigger over the massive amount of ${}^{39}$Ar present. Recognizing that the presence of radioactive impurities can also impact such a system, the newly formed DUNE \dword{fd} Background Task Force will address this concern. This system would not require any cryostat penetrations or affect major \dword{daq} requirements.


\subsection{Calibration Staging Plan}
\label{sec:phys-calib-approach}


The calibration strategy for DUNE will need to address the evolving operational and physics needs at every stage of the experiment in a timely manner using the primary sources and systems listed in Table~\ref{tab:calibsystem}. 
Here we describe the validation plan for calibration systems at ProtoDUNE and a staging plan to deploy calibration systems during different phases of the experiment: commissioning, early data taking, and stable operations. 

This \dword{tdr} presents the baseline calibration systems and strategy. Post-\dword{tdr}, once the calibration strategy is set, the calibration consortium will need to develop the necessary designs for calibration hardware along with tools and methods to be used with various calibration sources. To allow for flexibility in this process, the physical interfaces for calibration such as flanges or ports on the cryostat will be designed to accommodate the calibration hardware. As described in the calibration \dword{sp} detector volume, the calibration task force has provided the necessary feedthrough penetration design 
for the \dword{spmod} and will soon finalize the design for the \dword{dpmod}.  As DUNE physics turns on at different rates and times, a calibration strategy at each stage for physics and data taking is required. The strategy described in this section assumes that all systems are commissioned and deployed according to the nominal DUNE run plan.

\textit{Design Validation:} A second run of ProtoDUNE will be used to validate the designs of dedicated calibration systems, including the laser, PNS, and possibly the proposed radioactive source. In addition, ProtoDUNE data (and the SBN program) will provide data analysis techniques, tools, and detector model simulation improvements in advance of DUNE operation.

\textit{Commissioning:} When a \dword{detmodule} is filled, data from various instrumentation devices validate the argon fluid flow model and purification system. Once filled and at the desired high voltage, the \dword{detmodule} immediately becomes live for \dword{snb} and proton decay signals (beam and atmospheric neutrino physics will require a few years of data accumulation)
at which point it is critical that early calibration track the space-time dependence of the detector. Noise data 
and pulser data (taken with signal calibration pulses injected into the electronics) are needed to understand the TPC electronics response. Essential systems at this stage include temperature monitors, purity monitors, \dword{hv} monitors, robust \dword{fe} charge injection system for cold electronics, and a \dword{pds} monitoring system. 
In addition, as the $^{39}$Ar data is available immediately, DUNE must be ready (in terms of reconstruction tools and methods) to utilize $^{39}$Ar decays for understanding both low-energy response and space-time uniformity. 
Dedicated calibration systems as listed in Table~\ref{tab:calibsystem} are deployed and commissioned at this stage. Commissioning data from these systems must verify the expected configuration for each system and identify any needed adjustments to tune for data taking.

\textit{Early data taking:} Since DUNE will not yet have all in situ measurements of \lar physical properties at this stage, early calibration of the detector will use \lar measurements from \dword{protodune} or SBN, and \efield{}s from calculations tuned to measured \dword{hv} values.
This early data will most likely need to be recalibrated at a later stage with dedicated calibration runs when in situ measurements are available and as data taking progresses. 
The early physics will also require analysis of cosmic ray muon data to develop methods and tools for muon reconstruction from MeV to TeV and a well validated cosmic ray event generator with data. 
Dedicated early calibration runs using calibration hardware systems will develop and tune calibration tools to beam data taking and correct for any space-time irregularities observed in the TPC. Given the expected low rate of cosmic ray events at the underground location (see Section~\ref{sec:phys-calib-sources}), calibration with cosmic rays is not possible over short time scales and will proceed from coarse-grained to fine-grained over the course of years, as statistics accumulate. 
The experiment will rely on calibration hardware systems, such as a laser system, for calibrations that require an independent probe with reduced or removed interdependencies, fine-grained measurements (both in space and time), and detector stability monitoring on the time scales required by physics. Some measurements are simply not possible with cosmic rays (e.g., \dword{apa} flatness, global alignment of all \dword{apa}s). 

\textit{Stable operations:} Once the detector is running stably, dedicated calibration runs, ideally before, during and after each run period, will ensure that detector conditions have not significantly changed.
As statistics accumulate, DUNE can use standard-candle data samples (e.g., Michel electrons and neutral pions) from cosmic rays and beam-induced and atmospheric neutrinos to validate and improve the detector response models needed for precision physics. 
As DUNE becomes systematics-limited, dedicated precision-calibration campaigns using the calibration hardware systems will become crucial for meeting the stringent physics requirements on energy scale reconstruction and detector resolution. For example, understanding electromagnetic (EM) response in the FD will require both cosmic rays
and external systems. The very high energy muons from cosmic rays at that depth that initiate
EM showers (which would be rare at ProtoDUNE or SBND), will provide information to study EM response at high energies. External systems such as the pulsed neutron source system or the proposed radioactive source system will provide low energy EM response at the precision required for low energy supernovae
physics. Dedicated measurements of charged hadron interactions, initially in \Dword{protodune} and later with DUNE \dword{nd} will also be important in this phase.


\subsection{$^{39}$Ar beta decays}
\label{app:ar39}

Assuming the $^{39}$Ar beta decays are uniformly distributed in the drift direction, one is able to precisely determine the expected reconstructed energy spectrum 
provided a given set of well measured detector response parameters.  This can be done independently of using timing information (e.g.~from prompt scintillation light). 

A number of factors can impact accurately measuring the end point energy, including noise, wire response, electron lifetime, recombination (and electric field), cosmogenic activity, and other radiological backgrounds.
Many of the detector effects may be determined in-situ.  For instance, measuring the electronics response can be done in situ with pulser data (charge injection on the front-end ASICs); measuring the wire field response can be done with cosmic tracks and other dedicated measurements ex-situ. There are also plans to measure recombination parameters ex-situ (e.g. ProtoDUNE, MicroBooNE). Figure~\ref{fig:ar39} illustrates the different possible reconstructed $^{39}$Ar beta decay electron energy spectra one might see in the SP DUNE far detector after correcting for all other detector effects except for electron lifetime.
Also shown in Figure~\ref{fig:ar39} is the impact of varying the 
recombination model.
The impact on the reconstructed energy spectrum is very different for the two detector effects, allowing for simultaneous determination of both quantities.

This method is one foreseeable way to obtain a fine-grained (spatially and temporally) electron lifetime measurement in the DUNE FD.  It can also provide other necessary calibrations, such as measurements of wire-to-wire response variations and diffusion measurements, 
and could serve as an online monitor of 
\efield distortions in the detector by looking at the relative number of decays 
near the edges of the 
detector.  

One important consideration is whether or not the DUNE 
\dword{daq} can provide the necessary rate and type of data 
to successfully carry out this calibration at the desired frequency and level of spatial precision.  Knowing that the $^{39}$Ar beta decay rate is about 1~Bq/kg in natural (atmospheric) argon, one finds that $O(\mathrm{50k})$ $^{39}$Ar beta decays are expected in a single 5~ms event readout in an entire 10~kt module.  
From studies at MicroBooNE, $O(\mathrm{250k})$ will be needed 
for percent-level calibration of electron lifetime which means that for DUNE one would only need roughly five readout events in order to make a single measurement. 
However, to allow for the electron lifetime to spatially vary throughout the entire 10~kt module, it may be necessary to collect much more data in order to obtain a precise electron lifetime measurement throughout the detector.  
Studies of data rates and alternative methods for recording special $^{39}$Ar calibration data are currently in progress.

\begin{dunefigure}[Impact of different detector effects on the reconstructed \Ar39 $\beta$ decay energy spectrum]{fig:ar39}{Illustration of the impact of different detector effects on the reconstructed \Ar39 beta decay electron energy spectrum for decays observed in the SP DUNE far detector.  On the left are examples of the reconstructed energy spectrum for various different electron lifetimes, as well as the nominal 
\Ar39 beta decay spectrum (corresponding to an infinite electron lifetime).  On the right are examples of the reconstructed energy spectrum when the true recombination model is different from the one assumed in energy reconstruction (varying the $\alpha$ parameter of the modified Box model, $\mathcal{R} = \ln(\alpha + \xi)/\xi$, where $\xi = \beta\frac{dE}{dx}/{\rho}E_{\mathrm{drift}}$ and with fixed $\beta = 0.212$) and the electron lifetime is infinite.  All curves have been normalized to have the same maximal value.}
\includegraphics[width=.49\textwidth]{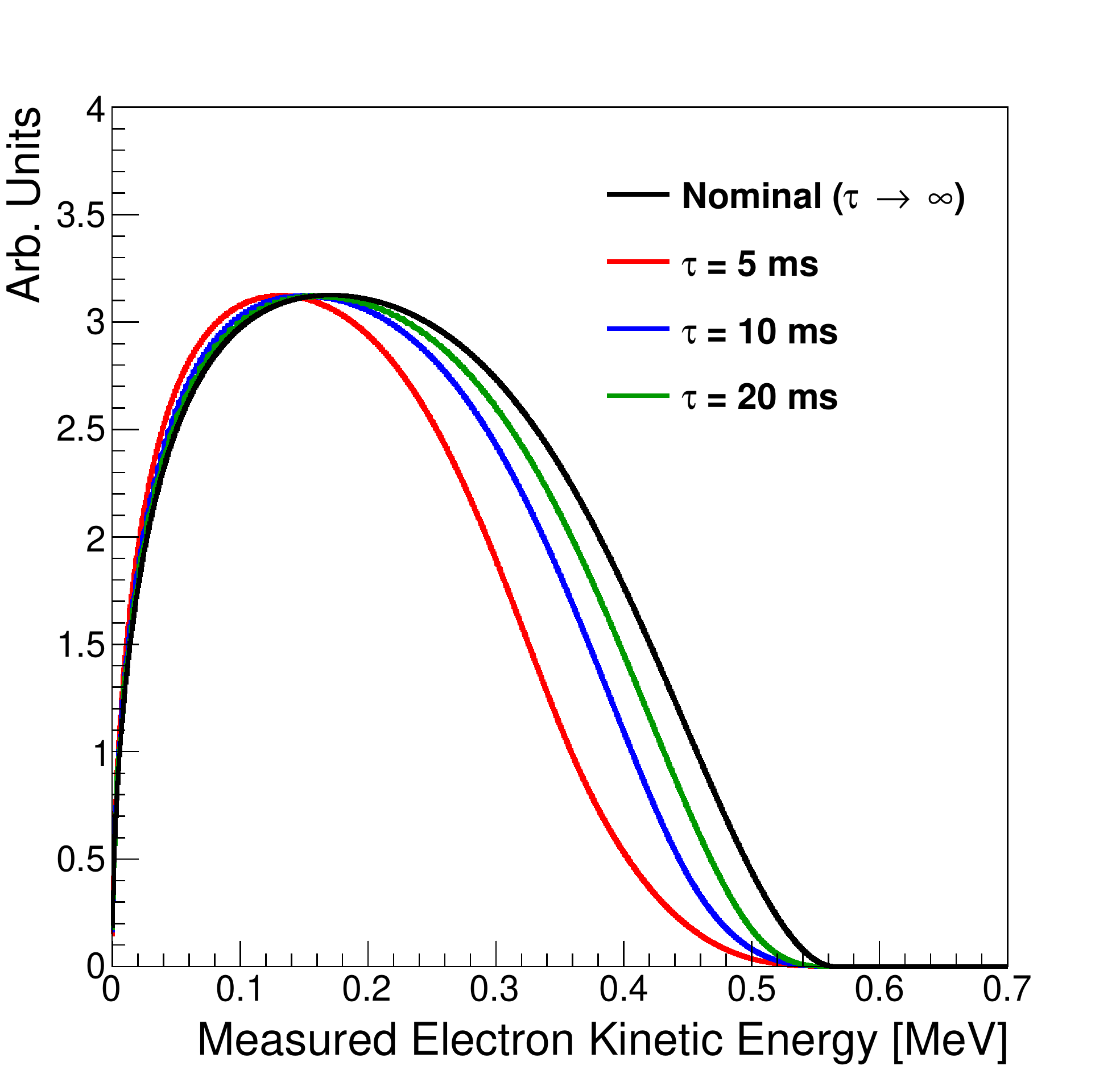}
\includegraphics[width=.49\textwidth]{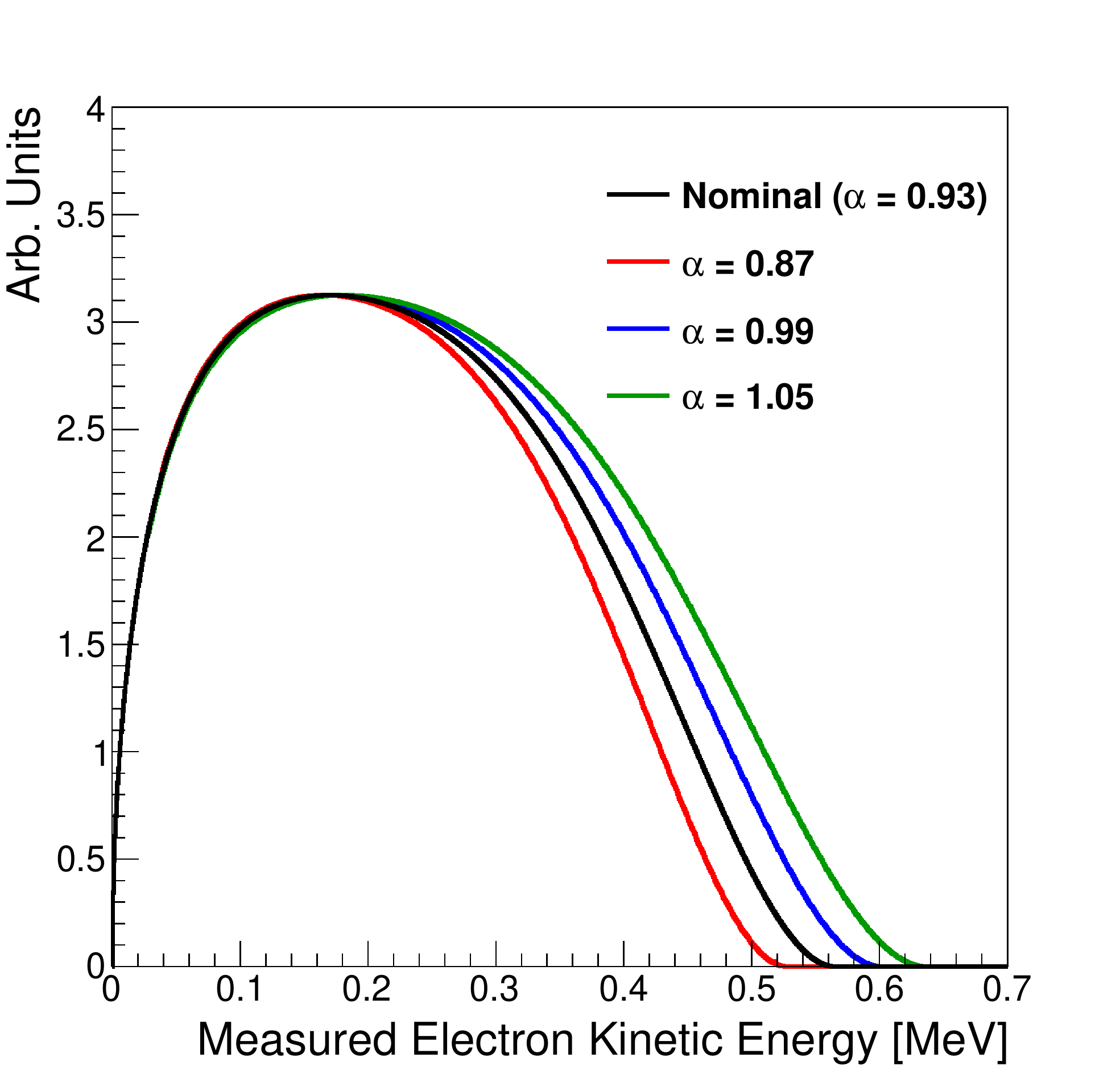}
\end{dunefigure}

\cleardoublepage

\chapter{Standard neutrino oscillation physics program}
\label{ch:osc}

\section{Overview and Theoretical Context}

\label{sec:physics-lbnosc-context}

The \dword{sm} of particle physics presents a remarkably accurate
description of the elementary particles and their
interactions. However, its limitations pose deeper questions about
Nature. With the discovery of the Higgs boson at the \dword{cern}, the Standard
Model would be ``complete'' except for the discovery of neutrino
mixing, which indicates neutrinos have a very small but nonzero
mass. In the \dword{sm}, 
the simple Higgs mechanism is responsible
for both quark and charged lepton masses, quark mixing and
\dword{cpv}. 
However, the small size of neutrino
masses and their relatively large mixing bears little resemblance to
quark masses and mixing, suggesting that different physics -- and
possibly different mass scales -- in the two sectors may be present,
thus motivating precision study of mixing and \dword{cpv} 
in the
lepton sector of the \dword{sm}. 


The \dword{dune} plans to pursue a detailed study of neutrino mixing, resolve the
neutrino mass ordering, and search for \dword{cpv} 
in the lepton
sector by studying the oscillation patterns of
high-intensity \numu and \anumu 
beams measured over a long baseline.  Neutrino oscillation arises from
mixing between the flavor 
(\nue, \numu, \nutau) and mass $(\nu_1,\, \nu_2,\, \nu_3)$ eigenstates
of neutrinos.  
In direct correspondence with mixing in the quark sector, the transformation
between basis states is expressed in the form of a complex unitary
matrix, known as the \dword{pmns} matrix: 

\begin{equation}
\left(\begin{array}{ccc} \nu_e \\ \nu_\mu \\ \nu_\tau \end{array} \right)= 
\underbrace{
  \left(\begin{array}{ccc}
      U_{e 1} &  U_{e 2} & U_{e 3} \\ 
      U_{\mu1} &  U_{\mu2} & U_{\mu 3} \\ 
      U_{\tau 1} &  U_{\tau 2} & U_{\tau 3} 
    \end{array} \right)
}_{U_{\rm PMNS}} \left(\begin{array}{ccc} \nu_1 \\ \nu_2 \\ \nu_3 \end{array} \right).
\label{eqn:pmns0}
\end{equation}
The \dword{pmns} matrix in full generality depends on just three mixing angles
and a \dword{cp}-violating phase\footnote{In the case of Majorana neutrinos, there are two additional \dword{cp} phases, but they are unobservable in the oscillation processes.}.  The mixing angles and phase are designated
as $(\theta_{12},\, \theta_{23},\, \theta_{13})$ and
\deltacp.
This matrix can be expressed as the product of three
two-flavor mixing matrices as follows~\cite{Schechter:1980gr}, where $c_{\alpha \beta}=\cos \theta_{\alpha \beta}$ and $s_{\alpha
 \beta}=\sin \theta_{\alpha \beta}$:

\begin{equation}
U_{\rm PMNS} = 
  \underbrace{
    \left( \begin{array}{ccc}
        1 & 0 & 0 \\ 
        0 & c_{23} & s_{23} \\ 
        0 & -s_{23} & c_{23}
    \end{array} \right)
  }_{\rm I}
\underbrace{
  \left( \begin{array}{ccc}
        c_{13} & 0  & e^{-i\mdeltacp} s_{13} \\ 
         0 & 1 & 0 \\ 
        -e^{i\mdeltacp} s_{13} & 0 & c_{13}
  \end{array} \right)   
  }_{\rm II}
\underbrace{
 \left( \begin{array}{ccc}
      c_{12} & s_{12} & 0 \\ 
      -s_{12} & c_{12} & 0 \\ 
      0 & 0 & 1
  \end{array} \right)
}_{\rm III}.
\label{eqn:pmns}
\end{equation}

The parameters of the \dword{pmns}
matrix determine the probability amplitudes of the neutrino
oscillation phenomena that arise from mixing.  The frequency of neutrino oscillation 
depends on the difference in the squares of the neutrino
masses, $\Delta m^{2}_{ij} \equiv m^{2}_{i} - m^{2}_{j}$; a set of three
neutrino mass states implies two independent mass-squared differences
(the ``solar'' mass splitting, $\Delta m^{2}_{21}$, and the ``atmospheric'' mass splitting, 
$\Delta m^{2}_{31}$), where $\Delta m^{2}_{31} = \Delta m^{2}_{32} + \Delta m^{2}_{21}$. 
The use of numbers to label the neutrino mass states is arbitrary; by convention, the numbering is defined such that the solar mass splitting is positive, in accordance to the ordering determined from solar matter effects.
This leaves two possibilities for
the ordering of the
mass states, known as the \emph{neutrino mass ordering} or \emph{neutrino mass hierarchy}. An ordering of
$m_1 < m_2 < m_3$ is known as the \emph{normal ordering} since it matches
the mass ordering of the charged leptons in the \dword{sm}, whereas an ordering of $m_3 < m_1 < m_2$
is referred to as the \emph{inverted ordering}.

The entire complement of neutrino experiments to date has measured
five of the mixing parameters~\cite{Esteban:2018azc,deSalas:2017kay,Capozzi:2017yic}: the three angles $\theta_{12}$,
$\theta_{23}$, and $\theta_{13}$, and the two mass differences
$\Delta m^{2}_{21}$ and $\Delta m^{2}_{31}$. 
The neutrino mass ordering (i.e., the sign of $\Delta m^{2}_{31}$) is unknown.
The values of $\theta_{12}$ and $\theta_{23}$ are large, while 
$\theta_{13}$ is smaller. The value of \deltacp is not well known, though neutrino oscillation data are beginning to provide some information on its value.
The absolute values of the entries of the \dword{pmns} matrix, which
contains information on the strength of flavor-changing weak decays in
the lepton sector, can be expressed in approximate form as
\begin{equation}
|U_{\rm PMNS}|\sim \left(\begin{array}{ccc} 0.8 & 0.5 & 0.1 \\ 0.5 & 0.6 & 0.7 \\ 0.3 & 0.6 & 0.7\end{array} \right),
\label{eq:pmnsmatrix}
\end{equation}
using values for the mixing angles given in Table~\ref{tab:oscpar_nufit}. 
While the three-flavor-mixing scenario for neutrinos is now well
established, the mixing parameters are not known to the same precision 
as are those in the
corresponding quark sector, and several important quantities, including
the value of \deltacp and the sign of the large mass splitting, are
still undetermined.

The oscillation probability of \numu $\rightarrow$ \nue through matter in a constant density
approximation is,  
to first order~\cite{Nunokawa:2007qh}:
\begin{eqnarray}
P(\nu_\mu \rightarrow \nu_e) & \simeq & \sin^2 \theta_{23} \sin^2 2 \theta_{13} 
\frac{ \sin^2(\Delta_{31} - aL)}{(\Delta_{31}-aL)^2} \Delta_{31}^2\\ \nonumber
& & + \sin 2 \theta_{23} \sin 2 \theta_{13} \sin 2 \theta_{12} \frac{ \sin(\Delta_{31} - aL)}{(\Delta_{31}-aL)} \Delta_{31} \frac{\sin(aL)}{(aL)} \Delta_{21} \cos (\Delta_{31} + \mdeltacp)\\ \nonumber
& & + \cos^2 \theta_{23} \sin^2 2 \theta_{12} \frac {\sin^2(aL)}{(aL)^2} \Delta_{21}^2, \\ \nonumber
\label{eqn:appprob}
\end{eqnarray}
where $\Delta_{ij} = \Delta m^2_{ij} L/4E_\nu$, $a = G_FN_e/\sqrt{2}$, $G_F$ is the Fermi constant, $N_e$ is the number density of electrons in the Earth, $L$ is the baseline in km, and $E_\nu$ is the neutrino energy in GeV. 
In the equation above, both \deltacp and $a$ 
switch signs in going from the
$\nu_\mu \to \nu_e$ to the $\bar{\nu}_\mu \to \bar{\nu}_e$ channel; i.e.,
a neutrino-antineutrino asymmetry is introduced both by \dword{cpv} (\deltacp)
and the matter effect ($a$). The origin of the matter effect asymmetry 
is simply the presence of electrons and absence of positrons in the Earth.  
In the few-GeV energy range, the asymmetry from the matter effect increases with baseline as the neutrinos
pass through more matter; therefore an experiment with a longer baseline will be
more sensitive to the neutrino mass ordering. For baselines longer than 
$\sim$\SI{1200}\km, the degeneracy between the asymmetries from matter
and \dword{cpv} effects can be resolved~\cite{Bass:2013vcg}. \dword{dune}, with a baseline of  \SI{1300}{\km}, 
will be able to unambiguously
determine the neutrino mass ordering \textit{and} measure the value of \deltacp~\cite{Diwan:2004bt}. 

\begin{figure}
  \centering
\includegraphics[width=0.45\linewidth]{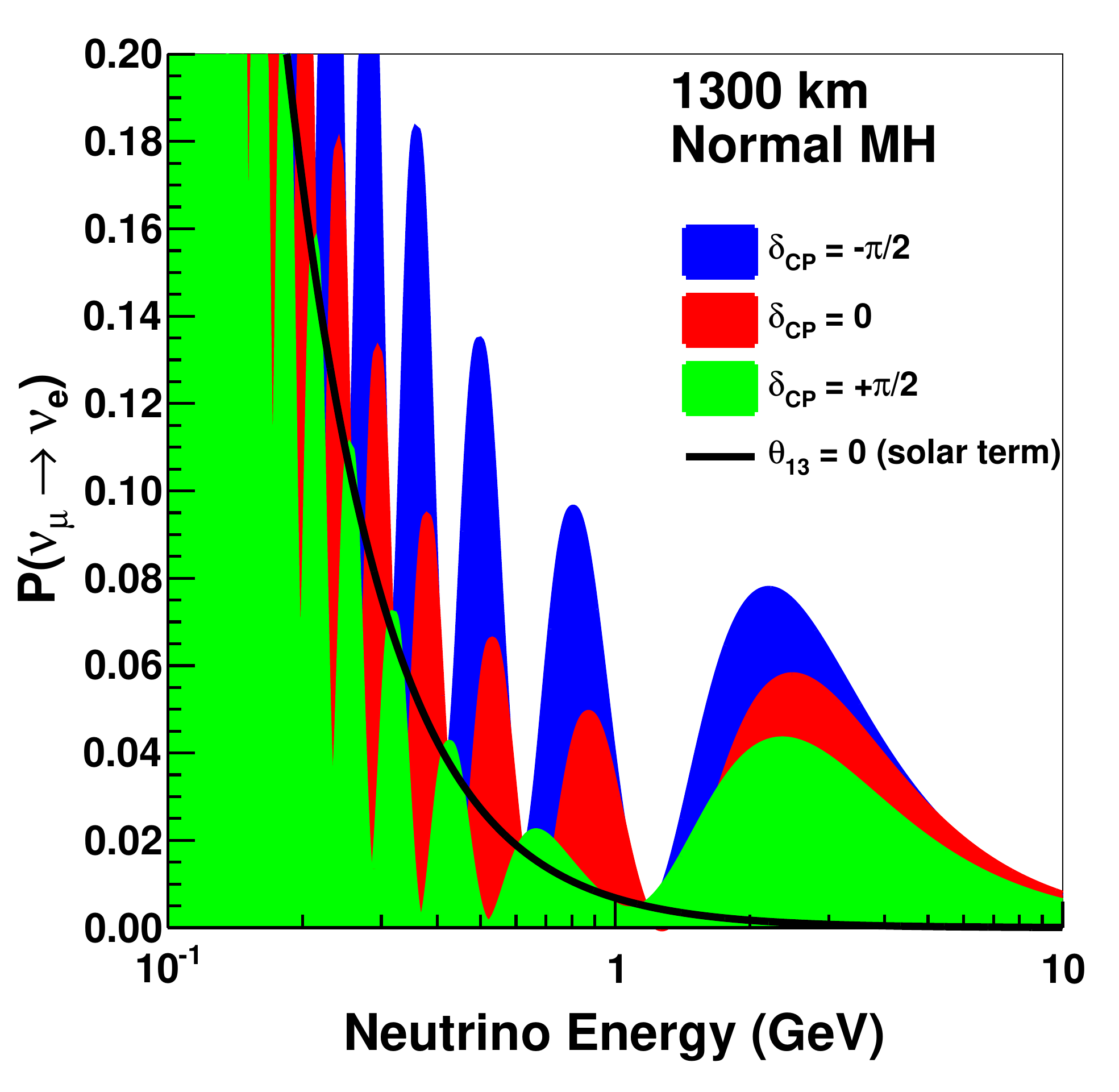}
\includegraphics[width=0.45\linewidth]{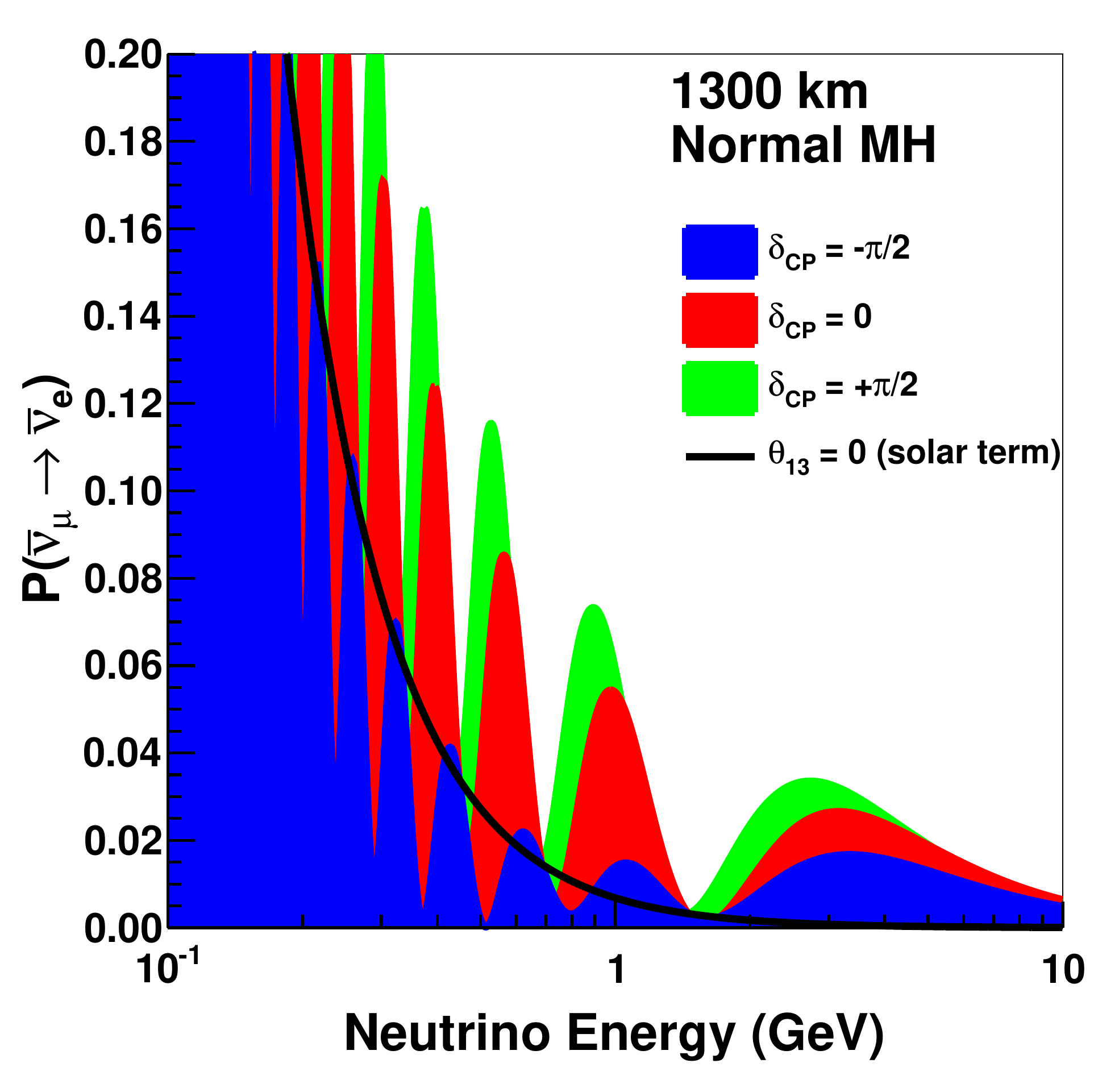}
  \caption[Appearance probabilities for \nue and \anue at \SI{1300}{\km}]{The appearance probability at a baseline of \SI{1300}\km{},
  as a function of neutrino energy, for \deltacp = $-\pi/2$ (blue), 
  0 (red), and $\pi/2$ (green), for neutrinos (left) and antineutrinos
  (right), for normal ordering. The black line indicates the oscillation
  probability if $\theta_{13}$ were equal to zero. Note that DUNE will be built at a baseline of \SI{1300}{\km}}
  \label{fig:oscprob}
\end{figure}

The electron neutrino appearance probability, $P(\nu_\mu \rightarrow \nu_e)$, 
is shown in 
Figure~\ref{fig:oscprob} at a baseline of \SI{1300}\km{} as a function of neutrino 
energy for several values of \deltacp. As this figure illustrates, the value 
of \deltacp affects both the amplitude and phase of
the oscillation. The difference in probability amplitude
for different values of \deltacp is larger at higher oscillation nodes, which 
correspond to energies less than 1.5~GeV. Therefore, a broadband experiment, 
capable of measuring not only the rate of \nue appearance but of mapping out the 
spectrum of observed oscillations down to energies of at least 500~MeV, is desirable. 

In the particular expression of the PMNS matrix shown in
Equation~\ref{eqn:pmns}, the middle factor labeled ``II'' describes
the mixing between the $\nu_1$ and $\nu_3$ mass states, and depends on
the CP-violating phase \deltacp. The variation in the $\nu_\mu \rightarrow
\nu_e$ oscillation probability with the value of \deltacp
indicates that it is experimentally possible to measure the value of
\deltacp at a fixed baseline using only the observed shape of the
$\nu_\mu \rightarrow \nu_e$ {\em or} the 
$\bar{\nu}_\mu \rightarrow \bar{\nu}_e$
appearance signal measured over an energy range that encompasses at
least one full oscillation interval. A measurement of the value of
$\mdeltacp \neq 0 \ {\rm or} \ \pi$, assuming that neutrino mixing follows the three-flavor model, would imply \dword{cpv}. In the approximation for the electron neutrino appearance
probability given in Equation~\ref{eqn:appprob}, expanding the middle
term results in the presence of CP-odd terms (dependent on $\sin
\mdeltacp$) that have opposite signs in $\nu_{\mu} \rightarrow \nu_e$
and $\bar{\nu}_{\mu} \rightarrow \bar{\nu}_e$ oscillations.
For $\mdeltacp \neq 0$ or $\pi$, these terms introduce an asymmetry in
neutrino versus antineutrino oscillations. 
Regardless of the measured value obtained for \deltacp, the explicit
observation of the asymmetry in $\nu_{\mu}
\rightarrow \nu_e$ and $\bar{\nu}_{\mu} \rightarrow
\bar{\nu}_e$ oscillations is sought to directly demonstrate the
leptonic \dword{cpv} effect.  
Furthermore, for long-baseline
experiments such as \dword{dune} where the neutrino beam propagates through
the Earth's mantle, the leptonic \dword{cpv} effects must be
disentangled from the matter effects.

The \SI{1300}{\km} baseline establishes one of \dword{dune}'s key strengths: sensitivity to the matter effect. This effect leads to a large asymmetry in the $\nu_\mu\to \nu_e$ versus $\bar{\nu}_\mu \to \bar{\nu}_e$ oscillation probabilities, the sign of which depends on the neutrino mass ordering.  At \SI{1300}{\km} this asymmetry is approximately $\pm 40\%$ in the region of the peak flux; this is larger than the maximal possible \dword{cp}-violating asymmetry associated with \deltacp, meaning that both the \dword{mh} and \deltacp can be determined
unambiguously with high confidence within the same experiment using the beam neutrinos. Concurrent analysis of the corresponding atmospheric-neutrino samples may provide an independent measurement of the neutrino mass ordering.

The rich oscillation structure that can be observed by \dword{dune} will enable precision measurement  in a single experiment of all the mixing parameters governing $\nu_1$-$\nu_3$ and $\nu_2$-$\nu_3$ mixing. Higher-precision measurements of the known oscillation parameters improves sensitivity to physics beyond the three-flavor oscillation model, particularly when compared to independent measurements by other experiments, including reactor measurements of $\theta_{13}$ and
measurements with atmospheric neutrinos. \dword{dune} will seek not only to demonstrate explicit \dword{cpv} by observing a difference in the neutrino and antineutrino oscillation probabilities, but also to precisely measure the value of \deltacp. 

 The mixing angle $\theta_{13}$ has been measured accurately in reactor experiments. While the constraint on $\theta_{13}$ from the reactor experiments will be important in the
early stages of \dword{dune}, 
\dword{dune} itself will eventually be able to measure
$\theta_{13}$ independently with a similar precision to reactor experiments. 
Whereas the reactor experiments measure $\theta_{13}$ using $\bar{\nu}_e$ disappearance, \dword{dune} will measure it through $\nu_e$ and $\bar{\nu}_e$ appearance, thus providing an independent constraint on
the three-flavor mixing matrix.   

Current world measurements of \sinst{23} leave an ambiguity as to whether the value of $\theta_{23}$ is in the lower octant (less than 45\mbox{$^{\circ}$}), the upper octant (greater than 45\mbox{$^{\circ}$}), or exactly 45\mbox{$^{\circ}$}.  The value of $\sin^2 \theta_{23}$ from \dword{nufit}~\cite{Esteban:2018azc,nufitweb} is in the upper octant, but the distribution of the $\chi^{2}$ has another local minimum in the lower octant. A \emph{maximal} mixing value of $\sin^2 \theta_{23} =0.5$ is therefore still allowed by the data and the octant is still largely undetermined.  A value of
$\theta_{23}$ exactly equal to 45\mbox{$^{\circ}$} would indicate that $\nu_{\mu}$ and $\nu_{\tau}$ have equal contributions from $\nu_3$, which could be evidence for a previously unknown symmetry.  It is therefore important to experimentally determine the value of $\sin ^2
\theta_{23}$ with sufficient precision to determine the octant of $\theta_{23}$.

The magnitude of the
CP-violating terms in the oscillation depends most directly on the
size of the Jarlskog invariant~\cite{Jarlskog:1985cw}, a function that
was introduced to provide a measure of CP violation independent of the
mixing-matrix parameterization. In terms of the parameterization
presented in Equation~\ref{eqn:pmns}, the Jarlskog invariant is:
\begin{equation}
J_{CP}^{\rm PMNS} \equiv \frac{1}{8} \sin 2 \theta_{12} \sin 2 \theta_{13}
\sin 2 \theta_{23} \cos \theta_{13} \sin \mdeltacp.
\end{equation}
The relatively large values of the mixing angles in the lepton sector imply that
leptonic \dword{cpv} effects may be quite large, though this depends on
the value of \deltacp, which is currently unknown. Given the current best-fit values of the mixing angles~\cite{Esteban:2018azc,nufitweb} and assuming normal ordering,
\begin{equation}
J_{CP}^{\rm PMNS} \approx 0.03 \sin \mdeltacp.
\end{equation}
This is in sharp contrast to the very small mixing in the quark sector,  
which leads to a very small value of the corresponding quark-sector
Jarlskog invariant~\cite{Tanabashi:2018oca}, 
\begin{equation}
J_{CP}^{\rm CKM} \approx 3 \times 10^{-5},
\end{equation}
despite the large value of $\delta^{\rm CKM}_{CP}\approx70^{\circ}$.

A comparison among the values of the parameters in the neutrino
and quark sectors suggest that mixing in the two sectors may be
qualitatively different. Illustrating this difference, the value of
the entries of the \dword{ckm} 
quark-mixing matrix (analogous to the \dword{pmns} matrix for
neutrinos, and thus indicative of the strength of flavor-changing weak
decays in the quark sector) can be expressed in approximate form as
\begin{equation}
|V_{\rm CKM}|\sim \left(\begin{array}{ccc} 1 & 0.2 & 0.004\\ 0.2 & 1 & 0.04 \\ 0.008 & 0.04 & 1\end{array} \right),
\label{eq:ckmmatrix}
\end{equation}
for comparison to the entries of the \dword{pmns} matrix given in Equation~\ref{eq:pmnsmatrix}.
As discussed in \cite{King:2014nza}, the question of why the quark mixing angles are
smaller than the lepton mixing angles is an important part of the 
flavor pattern question. Data on the patterns of neutrino mixing are already contributing to the quest to understand whether there is a relationship between quarks and leptons and their seemingly arbitrary generation structure.

DUNE is designed to make significant contributions to completion of the standard three-flavor 
mixing picture. Scientific goals are definitive determination of the neutrino mass ordering, definitive observation of CP violation for more than 50\% of possible true \deltacp values,  
and precise measurement of oscillation parameters, particularly \deltacp, \sinstt{13}, and the octant of \sinst{23}. There is 
great value in obtaining this set of measurements in a single experiment using a broadband beam, so that the oscillation pattern may be clearly observed and a detailed test of the three-flavor neutrino model may be performed. 

\section{Expected Event Rate and Oscillation Parameters}
\label{sec:physics-lbnosc-osc}

The signal for \nue (\anue) appearance is an excess of \dword{cc} 
\nue and \anue interactions over the expected background in the far detector.  The background to \nue appearance is composed of: (1) \dword{cc} interactions of \nue and \anue intrinsic to the beam; (2) misidentified \dword{nc} 
interactions;  (3) misidentified \numu and \anumu \dword{cc} interactions; and (4) $\nu_\tau$ and $\bar{\nu}_\tau$ \dword{cc} interactions in which the $\tau$s decay leptonically into electrons/positrons. \dword{nc} and $\nu_\tau$ backgrounds emanate from interactions of higher-energy neutrinos that feed down to lower reconstructed neutrino energies due to missing energy in unreconstructed final-state neutrinos. The selected NC and \dword{cc} \numu generally include an asymmetric decay of a relatively high energy $\pi^0$ coupled with a prompt photon conversion.

A full simulation chain that includes the beam flux, the \dword{genie} 
neutrino interaction
generator~\cite{Andreopoulos:2009rq}, and \dword{geant4}-based 
detector models has been implemented. Section~\ref{sec:physics-lbnosc-flux} describes the beam design, simulated flux, and associated uncertainties.
Event rates are based on a 1.2~MW neutrino beam and corresponding protons-on-target per year assumed to be 1.1 $\times 10^{21}$ POT.  These numbers assume a combined uptime and efficiency of the \dword{fnal} accelerator complex and the \dword{lbnf} beamline of 56\%.
An upgrade to 2.4 MW is assumed after six years of data collection. The neutrino interaction model has been generated using \dword{genie} 2.12 and the choices of models and tunes as well as associated uncertainties are described in detail in Section~\ref{sec:nu-osc-05}. The performance parameters for the near and far detectors are described in detail in Sections~\ref{sec:physics-lbnosc-ND} and~\ref{sec:physics-lbnosc-FD}. 
 Near Detector Monte Carlo has been generated using \dword{geant4} and a parameterized reconstruction based on true energy deposits in the active detector volumes has been used as described in Section~\ref{sec:physics-lbnosc-ND}.
 Far detector Monte Carlo has been generated using LArSoft and the reconstruction and event selection in the Far Detector has been fully implemented, as described in Section~\ref{sec:physics-lbnosc-FD}. Uncertainties associated with detector effects in the near and far detectors are described in Section~\ref{sec:physics-lbnosc-syst}. The methods used in calculating the DUNE sensitivity results are described in Section~\ref{sec:physics-lbnosc-sens} and these results based on the full framework are shown in Section~\ref{sec:physics-lbnosc-results}.

The neutrino oscillation parameters and the uncertainty on those parameters are taken from the \dword{nufit}~\cite{Esteban:2018azc,nufitweb} global fit to neutrino data; the
values are given in Table~\ref{tab:oscpar_nufit}.  (See also
\cite{deSalas:2017kay} and \cite{Capozzi:2017yic} for other recent global fits.) The sensitivities in this chapter are shown assuming normal ordering; this is an arbitrary choice for simplicity of presentation.

Event rates are presented as a function of calendar years and are calculated with the following assumed deployment plan, which is based on a technically limited schedule.
\begin{itemize}
    \item Start of beam run: Two \dword{fd} module 
    volumes for total fiducial mass of 20 kt, 1.2 MW beam
    \item After one year: Add one \dword{fd} module  volume for total fiducial mass of 30 kt
    \item After three years: Add one \dword{fd} module  volume for total fiducial mass of \fdfiducialmass
    \item After six years: Upgrade to 2.4 MW beam
\end{itemize}

\begin{table}[]
    \centering
    \begin{tabular}{lcc}
 Parameter &    Central Value & Relative Uncertainty \\
\toprowrule
$\theta_{12}$ & 0.5903 & 2.3\% \\ \colhline
$\theta_{23}$ (NO) & 0.866  & 4.1\% \\ 
$\theta_{23}$ (IO) & 0.869  & 4.0\% \\ \colhline
$\theta_{13}$ (NO) & 0.150  & 1.5\% \\ 
$\theta_{13}$ (IO) & 0.151  & 1.5\% \\ \colhline
$\Delta m^2_{21}$ & 7.39$\times10^{-5}$~eV$^2$ & 2.8\% \\ \colhline
$\Delta m^2_{32}$ (NO) & 2.451$\times10^{-3}$~eV$^2$ &  1.3\% \\
$\Delta m^2_{32}$ (IO) & -2.512$\times10^{-3}$~eV$^2$ &  1.3\% \\
    \end{tabular}
    \caption[Parameter values and uncertainties from a global fit to neutrino oscillation data]{Central value and relative uncertainty of neutrino oscillation parameters from a global fit~\cite{Esteban:2018azc,nufitweb} to neutrino oscillation data. Because the probability distributions are somewhat non-Gaussian (particularly for $\theta_{23}$), the relative uncertainty is computed using 1/6 of the 3$\sigma$ allowed range from the fit, rather than the 1$\sigma$ range.   For $\theta_{23}$, $\theta_{13}$, and $\Delta m^2_{31}$, the best-fit values and uncertainties depend on whether normal mass ordering (NO) or inverted mass ordering (IO) is assumed.}
    \label{tab:oscpar_nufit}
\end{table}

Figures~\ref{fig:appspectra} and~\ref{fig:disspectra} show the expected rate of selected events for \nue appearance and \numu disappearance, respectively, including expected flux, cross section, and oscillation probabilities, as a function of reconstructed neutrino energy at a baseline of
\SI{1300}{\km}. The spectra are shown for a \num{3.5}~year (staged) exposure each for neutrino and antineutrino beam mode, for a total run time of seven 
years. Tables~\ref{tab:apprates} and~\ref{tab:disrates} give the integrated rate for the \nue 
appearance and \numu 
disappearance spectra, respectively.  

\begin{dunefigure}[\nue and \anue appearance spectra]{fig:appspectra}
{\nue and \anue appearance spectra: Reconstructed energy distribution of selected \nue \dword{cc}-like events assuming 3.5 years (staged) running in the neutrino-beam mode (left) and antineutrino-beam mode (right), for a total of seven years (staged) exposure.  The plots assume normal mass ordering and include curves for $\mdeltacp = -\pi/2, 0$, and $\pi/2$.}
 \includegraphics[width=0.49\textwidth]{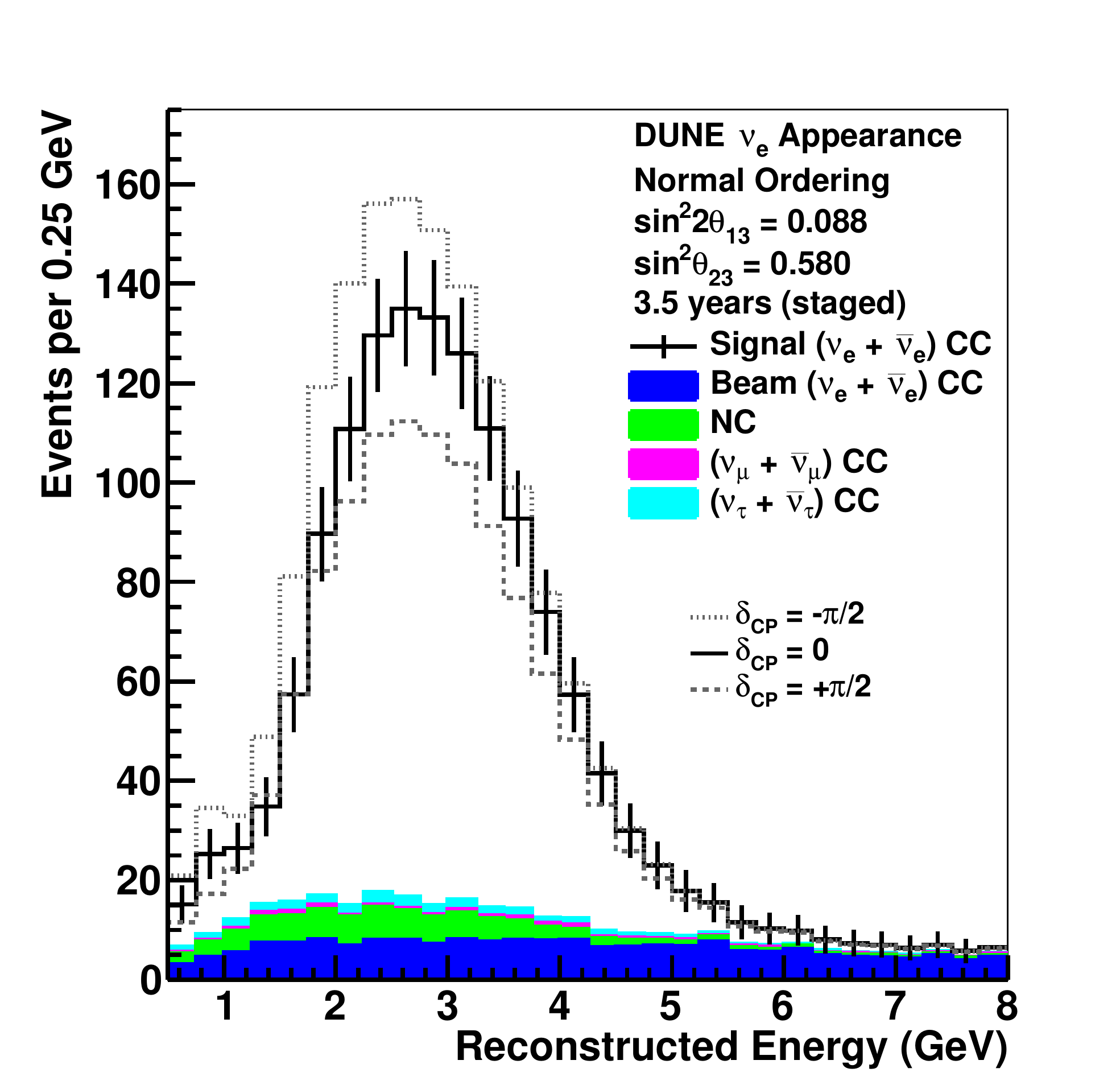}
 \includegraphics[width=0.49\textwidth]{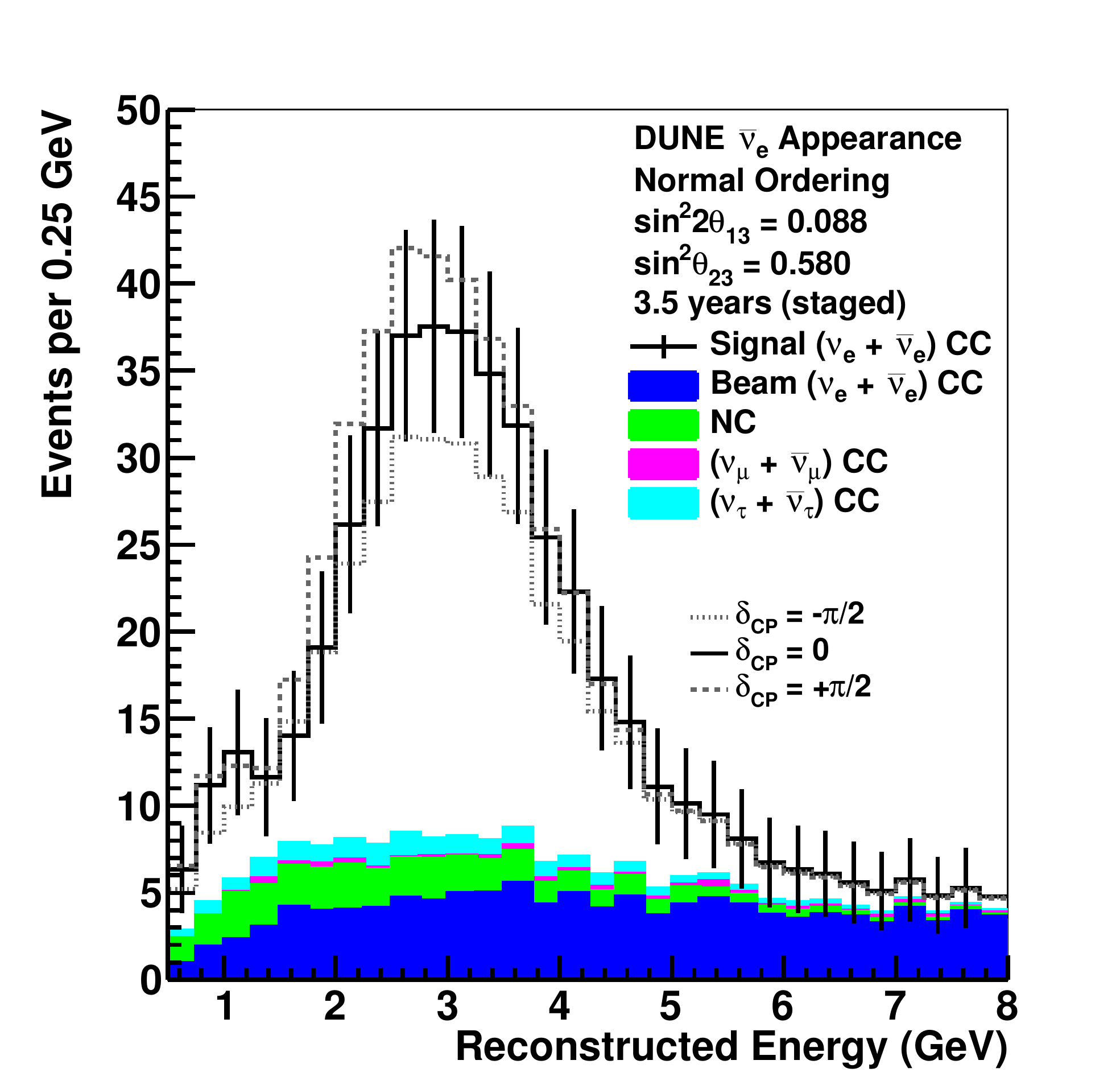}
\end{dunefigure}

\begin{dunefigure}[\numu and \anumu disappearance spectra]{fig:disspectra}
{\numu and \anumu disappearance spectra: Reconstructed energy distribution of selected $\nu_{\mu}$ \dword{cc}-like events assuming 3.5 years (staged) running in the neutrino-beam mode (left) and antineutrino-beam mode (right), for a total of seven years (staged) exposure. The plots assume normal mass ordering.}
\includegraphics[width=0.49\textwidth]{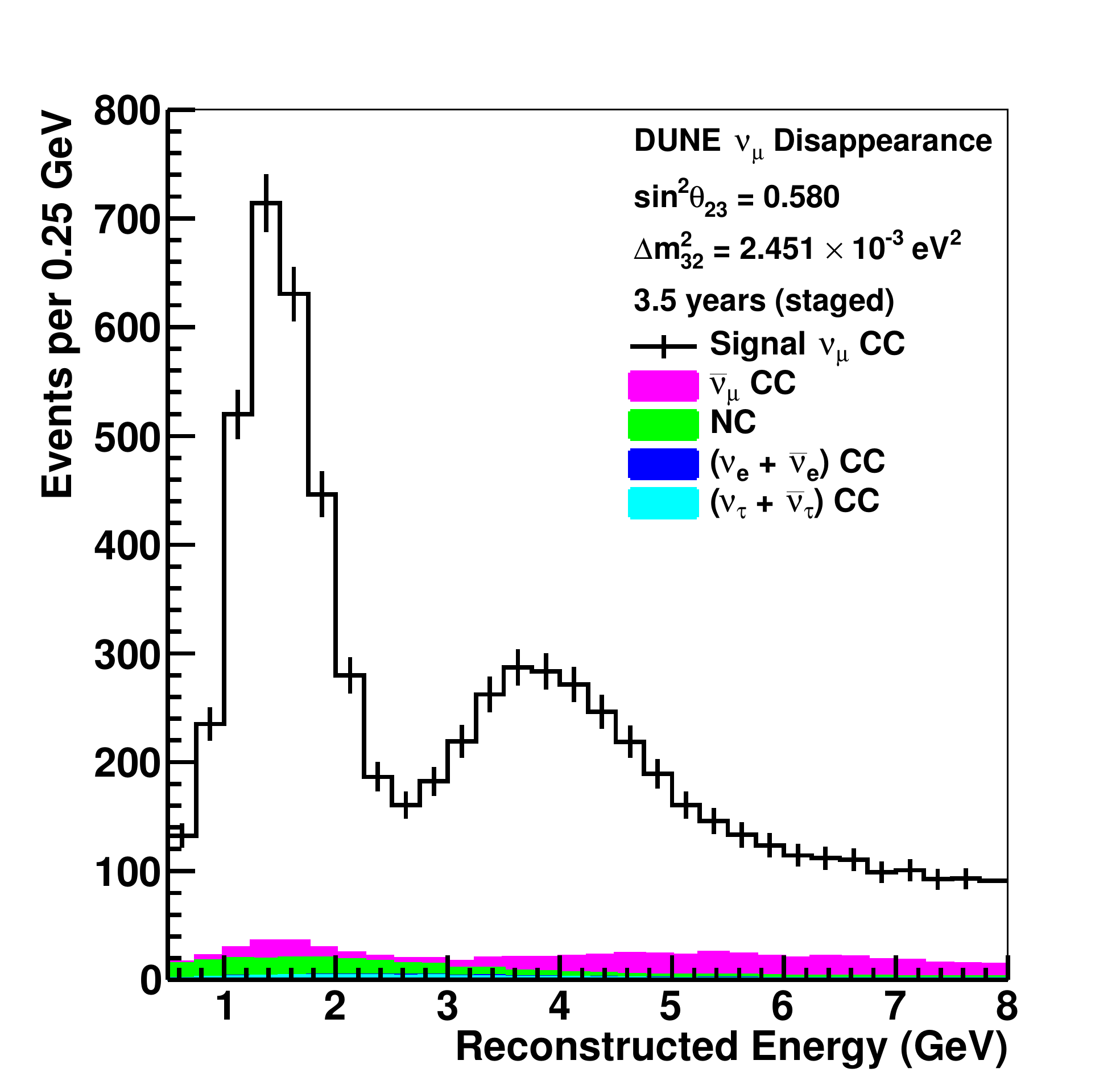}
\includegraphics[width=0.49\textwidth]{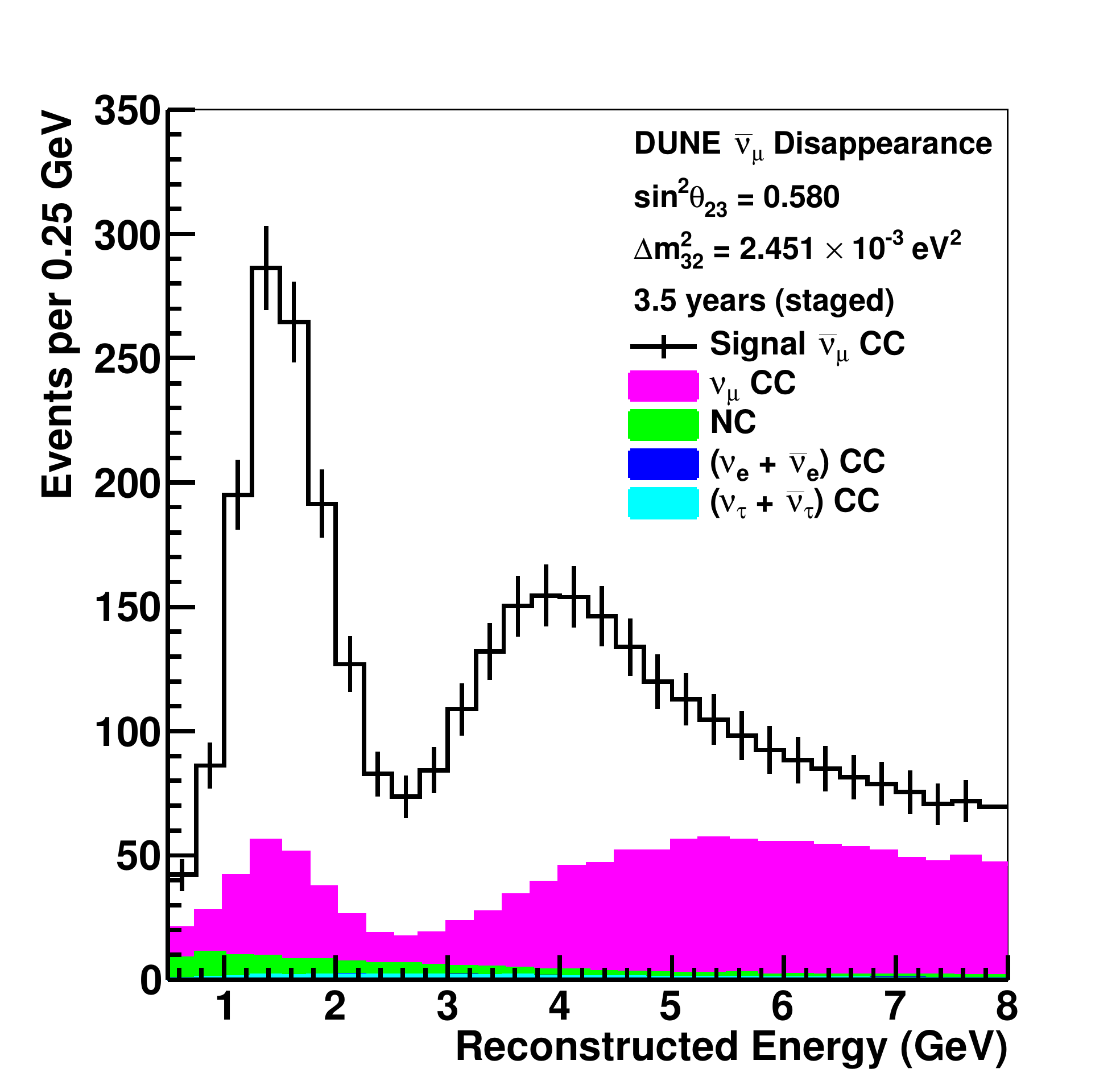}
\end{dunefigure}

\begin{dunetable}
[\nue and \anue appearance rates]
{lrr}
{tab:apprates}
{\nue and \anue appearance rates: Integrated rate of selected $\nu_e$ \dword{cc}-like events between 0.5 and 8.0~GeV assuming a \num{3.5}-year (staged) exposure in the neutrino-beam mode and antineutrino-beam mode.  The signal rates are shown for both normal mass ordering (NO) and inverted mass ordering (IO), and all the background rates assume normal mass ordering.  All the rates assume $\mdeltacp = 0$.}
& Expected Events (3.5 years staged) \\ \toprowrule
 $\nu$ mode & \\
 \colhline 
 \nue Signal NO (IO) & 1092 (497) \\
 \anue Signal NO (IO) & 18 (31) \\
  \colhline
 Total Signal NO (IO) & 1110 (528) \\
  \colhline 
 Beam $\nu_{e}+\bar{\nu}_{e}$ \dword{cc} background & 190 \\
 \dword{nc} background & 81 \\
 $\nu_{\tau}+\bar{\nu}_{\tau}$ \dword{cc} background & 32 \\
 $\nu_{\mu}+\bar{\nu}_{\mu}$ \dword{cc} background & 14 \\
  \colhline
 Total background & 317 \\
 \toprowrule
 $\bar{\nu}$ mode & \\
 \colhline 
 \nue Signal NO (IO) & 76 (36) \\
 \anue Signal NO (IO) & 224 (470) \\
  \colhline
 Total Signal NO (IO) & 300 (506) \\
  \colhline 
 Beam $\nu_{e}+\bar{\nu}_{e}$ \dword{cc} background & 117 \\
 \dword{nc} background & 38 \\
 $\nu_{\tau}+\bar{\nu}_{\tau}$ \dword{cc} background & 20 \\
 $\nu_{\mu}+\bar{\nu}_{\mu}$ \dword{cc} background & 5 \\
  \colhline 
 Total background & 180 \\
\end{dunetable}

\begin{dunetable}
[\numu and \anumu disappearance rates]
{lrr}
{tab:disrates}
{\numu and \anumu disappearance rates: Integrated rate of selected $\nu_{\mu}$ \dword{cc}-like events between 0.5 and 8.0~GeV assuming a \num{3.5}-year (staged) exposure in the neutrino-beam mode and antineutrino-beam mode.  The rates are shown for normal mass ordering and $\mdeltacp = 0$.}
& Expected Events (3.5 years staged)\\ \toprowrule
  $\nu$ mode & \\
 \colhline 
 \numu Signal & 6200 \\
 \colhline 
  \anumu \dword{cc} background & 389 \\
 \dword{nc} background & 200 \\
 $\nu_{\tau}+\bar{\nu}_{\tau}$ \dword{cc} background & 46 \\
 $\nu_e+\bar{\nu}_e$ \dword{cc} background & 8 \\
 \toprowrule
 $\bar{\nu}$ mode  & \\
\colhline 
 \anumu Signal & 2303 \\
\colhline 
  \numu \dword{cc} background & 1129 \\
 \dword{nc} background & 101 \\
 $\nu_{\tau}+\bar{\nu}_{\tau}$ \dword{cc} background & 27 \\
 $\nu_e+\bar{\nu}_e$ \dword{cc} background & 2 \\
\end{dunetable}

\section{Neutrino Beam Flux and Uncertainties}\label{sec:nu-osc-04}
\label{sec:physics-lbnosc-flux}

The neutrino fluxes are described in detail in Section~\ref{sec:tools-mc-flux}.  They were generated using G4LBNF, a \textsc{Geant}4\xspace-based simulation of the LBNF neutrino beam.  The simulation is configured to use a detailed description of the \dword{lbnf} optimized beam design~\cite{optimizedbeamcdr}, which includes horns and target designed to maximize sensitivity to \dword{cpv} given the physical constraints on the beamline design.   

\begin{dunefigure}[Neutrino fluxes at the far detector]{fig:flux_flavor}
{Neutrino fluxes at the \dword{fd} for neutrino mode (left) and
antineutrino mode (right). }
    \includegraphics[width=0.45\textwidth]{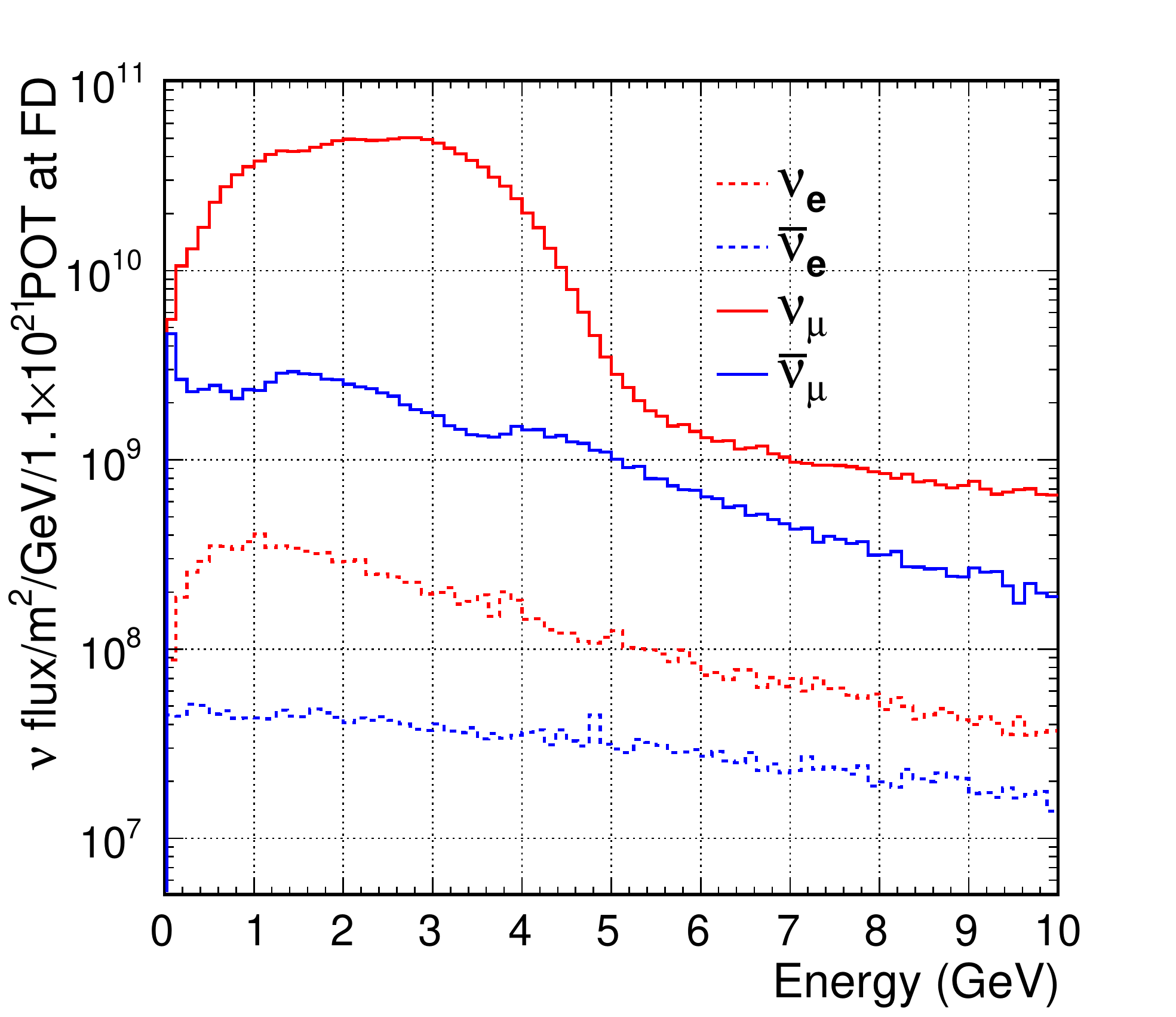}
     \includegraphics[width=0.45\textwidth]{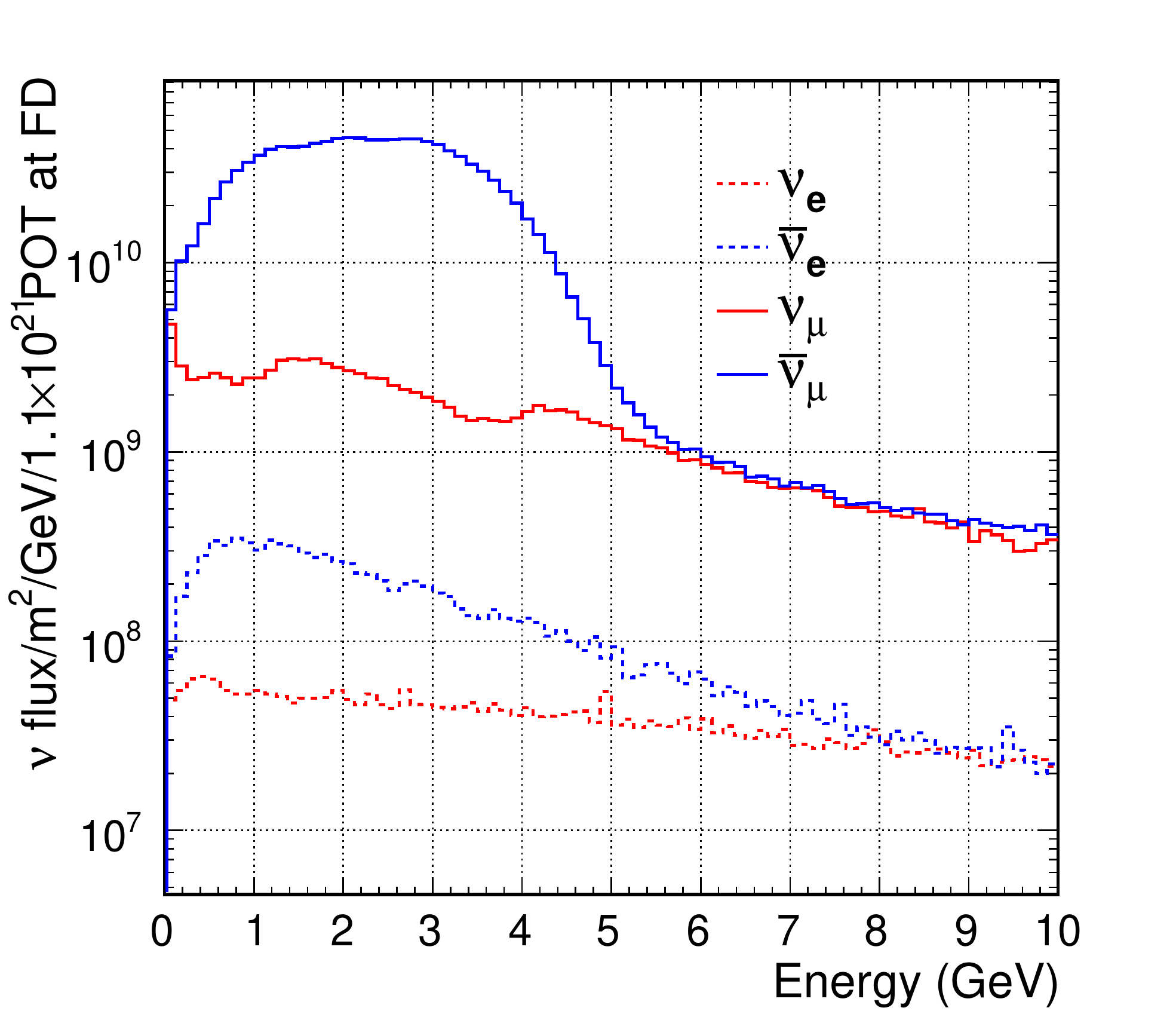}
\end{dunefigure}

Neutrino fluxes for neutrino and antineutrino mode configurations of \dword{lbnf} are shown in Figure~\ref{fig:flux_flavor}.   Uncertainties on the neutrino fluxes arise primarily from uncertainties in hadrons produced off the target and uncertainties in the design parameters of the beamline, such as horn currents and horn and target positioning (commonly called ``focusing uncertainties''). Given current measurements of hadron production and LBNF estimates of alignment tolerances, flux uncertainties are approximately 8\% at the first oscillation maximum and 12\% at the second.  These uncertainties are highly correlated across energy bins and neutrino flavors

 Future hadron production measurements are expected to improve the quality of and the resulting constraints on these flux uncertainty estimates.  Approximately 40\% of the interactions that produce neutrinos in the LBNF beam simulation have no data constraints whatsoever.  Large uncertainties are assumed for these interactions. The largest unconstrained sources of uncertainty are proton quasielastic interactions and meson incident interactions.  The proposed EMPHATIC experiment~\cite{Akaishi:2019dej} at Fermilab will be able to constrain quasielastics and low energy interactions that dominate the lowest neutrino energy bins.  The NA61 experiment at CERN has taken data that will constrain many higher energy interactions, including pion reinteractions. It  also plans to measure hadrons produced off of a replica LBNF target, which would provide tight constraints on all interactions occurring in the target.  A similar program at NA61 has reduced flux uncertainties for T2K from ~10\% to ~5\%~\cite{Vladisavljevic:2018prd}, and NOvA is currently analyzing NA61 replica target data~\cite{Aduszkiewicz:2222876}.  Another proposed experiment, the LBNF spectrometer, would measure hadrons after both production and focusing in the horns, effectively constraining nearly all hadron production uncertainties, and could also enable measurement of the impact on focused hadrons of shifted alignment parameters (which is currently taken from simulations).  
The neutrino flux uncertainties, as well as their bin-to-bin and  flavor-to-flavor correlations, are very sensitive to correlations in hadron production measurements.  None of the currently available measurements have provided correlations, so the uncertainty estimates make basic assumptions that statistical uncertainties are not correlated between bins but systematic uncertainties are completely correlated.  New hadron production measurements that cover phase space similar to past measurements but that provide bin-to-bin correlations would also improve the quality of the estimated neutrino flux uncertainties at DUNE.     

The unoscillated fluxes at the \dword{nd} and \dword{fd} are similar, but not identical (since the ND sees a line source, while the FD sees a point source. The relationship is well understood, and flux uncertainties mostly cancel for the ratio of fluxes between the two detectors.  Uncertainties on the ratio are around 1\% or smaller except at the falling edge of the focusing peak, where they rise to 2\%. The far to near flux ratio and uncertainties on this ratio are shown in Fig.~\ref{fig:flux_nearfar}. 

The peak energy of neutrino flux falls off and the width of the peak narrows as the distance from the beams central axis increases. The flux at these ``off-axis'' positions can be understood through the relationship between the parent pion energy and neutrino energy, as shown in Figure~\ref{fig:OAAFluxFigs}. For an off-axis angle relative to the initial beam direction, the subsequent neutrino energy spectra is narrower and peaked at a lower energy than the on-axis spectra. At $575\,\textrm{m}$, the location of the \dword{nd} hall, a lateral shift of $1\,\textrm{m}$ corresponds to approximately a $0.1^\circ$ change in off-axis angle.


The same sources of systematic uncertainty which affect the on-axis spectra also modify the off-axis spectra.  Generally, the size of the off-axis uncertainties is comparable to the on-axis uncertainties and the uncertainties are highly correlated across off-axis and on-axis positions. 
The impact of focusing and alignment uncertainties varies depending on off-axis angle.
Therefore, off-axis flux measurements are useful to diagnose beamline aberrations, and to further constrain flux uncertainties.

\section{Neutrino Interactions and Uncertainties}\label{sec:nu-osc-05} 

\subsection{Interaction Model Summary}

The goal of parameterizing the neutrino interaction model uncertainties is to provide a framework for considering how these uncertainties affect the oscillation analysis at the \dword{fd}, and for considering how constraints at the \dword{nd} can limit those uncertainties.
 
The model developed for this purpose generally factorizes the neutrino interaction on nuclei into an incoherent sum of hard scattering neutrino interactions with the single nucleons in the nucleus. The effect of the nucleus is implemented as initial and final state interaction effects, with some (albeit few) nucleus-dependent hard scattering calculations. Schematically, we express this concept as $\text{Scattering Process} = \text{Initial State} \otimes \text{Nucleon Interaction} \otimes \text{Final State Propagation}$.

The initial state effects relate to the description of the momentum and position distributions of the nucleons in the nucleus, kinematic modifications to the final state (such as separation energy, or sometimes described as a binding energy), and Coulomb effects.   The concept of binding energy reflects the idea that the struck nucleon may be off the mass-shell inside the nucleus.
Final state interactions refer to the propagation and interaction of hadrons produced in the nucleon interaction through the nucleus. The \dword{fsi} alter both the momentum and energy of the recoiling particles produced in the final state, and may also alter their identity and multiplicity in the case of inelastic reinteractions (e.g., in a nucleus a hadron may be absorbed, rescattered, or create a secondary hadron).  The \dword{fsi} model implemented in the \dword{genie}, \dword{nuwro}, and \dword{neut} neutrino interaction generators is a semi-classical cascade model. In particular, \dword{genie}'s $hA$ model is a single step scaled model, based on hadron-nucleus and hadron-nucleon scattering data and theoretical corrections. 

Generators vary in their attempts to accurately model the largely undetected final state ``spectator'' nuclear system.  The nuclear system can carry away significant undetected momentum---hundreds of MeV is not unusual---in the form of one or more heavy, non-relativistic particles.  These particles typically carry off very little kinetic energy; however they can absorb on the order of tens of MeV of energy from the initial state from breakup or excitation of the target nucleus.  This energy and momentum will typically be invisible to the detector.

The factorization outlined above is not present in all parts of the model.  Most modern generators include ``2p2h'' (two particle, two hole) interactions that model meson exchange processes and scattering on highly correlated pairs of nucleons in the nucleus.  These interactions are often implemented as another process that incorporates both hard scattering and initial state effects in processes that create multiple final state nucleons, with a different prescription for different nuclei.
Neutrino scattering on atomic electrons and the coherent production of pions (which scatters off the entire nucleus) also
do not follow this factorization.

The interaction model and its variations are implemented in the  \dword{genie} generator.  The fixed version of \dword{genie} used for this report, v2.12.10\footnote{At the time of the development of this model for interactions and their uncertainties, initial pieces of \dword{genie} 3 had just recently been released (October 2018) and reweighting and documentation followed after this. The timing made it impractical to use \dword{genie} 3 for this work.}, will not contain all of the possible cross section variations which need to be modeled.  Therefore, the variations in the cross sections to be considered are implemented as some combination of: \dword{genie} weighting parameters (sometimes referred to as ``\dword{genie} knobs''), {\em ad hoc} weights of events that are designed to parameterize uncertainties or cross section corrections currently not implemented within \dword{genie}, and discrete alternative model comparisons, achieved through alternative generators, alternative \dword{genie} configurations, or custom weightings. For the studies presented in this chapter we have  identified classes of uncertainties that are intended to span a representative range of alternative models such as those found in other generators. 

In this work, two example alternative models are used directly to evaluate additional uncertainties in the case where the assumptions about the near detector are relaxed. These studies are described in Section~\ref{sec:ndimpact}. The first is based on the \dword{nuwro} generator and the second is designed to produce the same on-axis visible energy distributions as the nominal model, but with a different relationship between true neutrino energy and visible energy.

\subsection{Interaction Model Uncertainties}

The interaction uncertainties are divided into seven roughly exclusive groups: (1) initial state uncertainties, (2) hard scattering uncertainties and nuclear modifications to the quasielastic process, (3) uncertainties in multinucleon (2p2h) hard scattering processes, (4) hard scattering uncertainties in pion production processes, (5) uncertainties governing other, higher $W$ and neutral current processes, (6) final state interaction uncertainties, (7) neutrino flavor dependent uncertainties. Uncertainties are intended to reflect current theoretical freedom, deficiencies in implementation, and/or current experimental knowledge.  
There are constraints on nuclear effects because of measurements on lighter targets, however for the argon nuclear target some additional sources of uncertainty are identified.  We also discuss cases where the parameterization is limited or simplified.

\subsubsection{Initial State Uncertainties}
The default nuclear model in \dword{genie} is a modified global Fermi gas model of the nucleons in the nucleus.  There are significant deficiencies that are known in global Fermi gas models. These include a lack of consistent incorporation of the tails that result from correlations among nucleons, the lack of correlation between location within the nucleus and momentum of the nucleon, and an incorrect relationship between momentum and energy of the off-shell, bound nucleon within the nucleus. \dword{genie} modifies the nucleon momentum distribution empirically to account for short-range correlation effects, which populates tails above the Fermi cutoff, but the other deficiencies persist. Alternative initial state models, such as spectral functions~\cite{Benhar:1994hw,Nieves:2004wx}, the mean field model of GiBUU~\cite{Gallmeister:2016dnq}, or continuum random phase approximation (CRPA) calculations~\cite{Pandey:2014tza} may provide better descriptions of the nuclear initial state~\cite{Sobczyk:2017mts}. 

\subsubsection{Quasielastic uncertainties}
The primary uncertainties considered in quasielastic interactions are the axial form factor of the nucleon and nuclear screening---from the so-called  \dword{rpa} calculations---of low momentum transfer reactions.

The axial form factor uncertainty has been historically described with a single parameter uncertainty with the dipole form by varying $M_A$, and we will continue this for these studies.  Unfortunately, this framework overconstrains the form factor at high $Q^2$, and an alternative parameterization based on the $z$-expansion has been proposed as a replacement~\cite{Meyer:2016oeg}.  However, this parameterization is multi-dimensional and poses problems for the analysis framework of this study which factorizes all $N$-dimensional variations out into $N\times{}1$-dimensional analysis bin response functions. For some multi-dimensional parameterizations, this simplification is an adequate approximation, e.g., the BeRPA described below. 

One part of the Nieves et al.\cite{Nieves:2011pp,Gran:2013kda} description of the $0\pi$ interaction on nuclei includes RPA, used to sum the $W^\pm$ self-energy terms. In practice, this modifies the 1p1h/Quasi-Elastic cross-section in a non-trivial way. The calculations from Nieves et al. have associated uncertainties presented in~\cite{nieves_uncert}, 
\fixme{could it be \cite{Valverde:2006zn}, Theoretical uncertainties on quasielastic charged-current neutrino-nucleus cross sections?} which were evaluated as a function of $Q^2$~\cite{sanchez-private}. In 2018, \minerva and \nova parameterized the central value and uncertainty in $(q_0, q_3)$ using RPA uncertainties as parameterized in~\cite{RikRPA}, whereas T2K used central values and uncertainties in $Q^2$ only. Here we use T2K's 2017/8 parameterization of the RPA effect~\cite{Abe:2018wpn}
due to its simplicity. The shape of the correction and error is parameterized with a Bernstein polynomial up to $Q^2=1.2\text{ GeV}^2$ which switches to a decaying exponential. The BeRPA (Bernstein RPA) function has three parameters controlling the polynomial ($A, B, C$), where the parameters control the behavior at increasing $Q^2$ and a fourth parameter $E$ controls the high $Q^2$ tail.

The axial form factor parameterization we use is known to be inadequate.  However, the convolution of BeRPA uncertainties with the limited axial form factor uncertainties do provide more freedom as a function of $Q^2$, and the two effects likely provide adequate freedom for the $Q^2$ shape in quasielastic events.

\subsubsection{$\boldsymbol{2p2h}$ uncertainties}
We start with the Nieves et al.\ or ``Valencia'' model~\cite{Nieves:2011pp,Gran:2013kda} 
for multinucleon ($2p2h$) contributions to the cross section.  However, \minerva has shown directly~\cite{Rodrigues:2015hik}, and \nova indirectly, that this description is missing observed strength on carbon.  As a primary approach to the model, we add that missing strength to a number of possible reactions.  We then add uncertainties for energy dependence of this missing strength and uncertainties in scaling the $2p2h$ prediction from carbon to argon.

The extra strength from the ``\minerva tune'' to $2p2h$ is applied in $(q_0,q_3)$ space (where $q_0$ is energy transfer from the leptonic system, and $q_3$ is the magnitude of the three momentum transfer) to fit reconstructed \minerva CC-inclusive data~\cite{Rodrigues:2015hik} in $E_\text{avail}$\footnote{$E_\text{avail}$ is calorimetrically visible energy in the detector, roughly speaking total recoil hadronic energy, less the masses of $\pi^\pm$ and the kinetic energies of neutrons} and $q_3$.  Reasonable fits to \minerva's data are found by attributing the missing strength to any of $2p2h$ from $np$ initial state pairs, $2p2h$ from $nn$ initial state pairs, or $1p1h$ or quasielastic processes.  The default tune uses an enhancement of the $np$ and $nn$ initial strengths in the ratio predicted by the Nieves model, and alternative systematic variation tunes (``MnvTune'' 1-3) attribute the missing strength to the individual hypotheses above. Implementation of the ``MnvTune'' is based on weighting in true $(q_0,q_3)$. The weighting requires \dword{genie}'s Llewelyn-Smith $1p1h$ and Valencia $2p2h$ are used as the base model. To ensure consistency in using these different tunes as freedom in the model, a single systematic parameter is introduced that varies smoothly between applying the $1p1h$ tune at one extreme value to applying the $nn$ tune at the other extreme via the default tune which is used as the central value. The $np$ tune is neglected in this prescription as being the most redundant, in terms of missing energy content of the final state, of the four discrete hypotheses.

The rates for $1p1h$ and $2p2h$ processes could be different on argon and carbon targets.  There is little neutrino scattering data to inform this, but there are measurements of short-ranged correlated pairs from electron scattering on different nuclei~\cite{Colle:2015ena}.  These measurements directly constrain $2p2h$ from short range correlations, although the link to dynamical sources like meson exchange current processes (MEC) is less direct. Interpolation of that data in $A$ (Nucleon number) suggests that scaling from carbon relative to the naive $\propto A$ prediction for $2p2h$ processes would give an additional factor of $1.33\pm 0.13$ for $np$ pairs, and $0.9\pm 0.4$ for $pp$ pairs.
\dword{genie}'s prediction for the ratio of $2p2h$ cross-sections in $\text{Ar}^{40}/\text{C}^{12}$ for neutrinos varies slowly with neutrino energy in the DUNE energy range: from $3.76$ at $1$~GeV to $3.64$ at $5$~GeV. The ratio for antineutrino cross sections is consistent with $3.20$ at all DUNE energies. Since the ratio of $A$ for $\text{Ar}^{40}/\text{C}^{12}$ is $3.33$, this is consistent with the ranges suggested above by the measured $pp$ and $np$ pair scaling.  A dedicated study by the SuSA group using their own theoretical model for the relevant MEC process also concludes that the transverse nuclear response (which drives the $\nu-A$ MEC cross section) ratio between $\text{Ca}^{40}$ (the isoscalar nucleus with the same $A$ as $\text{Ar}^{40}$) and $\text{C}^{12}$ is $3.72$ \cite{Amaro:2017eah}. We vary \dword{genie}'s Valencia model based prediction, including the \minerva tune, for $2p2h$ by $\sim 20\%$ to be consistent with the correlated pair scaling values above. This is done independently for neutrino and antineutrino scattering.

The \minerva tune may be $E_\nu$ dependent. \minerva separated its data into an $E_\nu<6$~GeV and an $E_\nu>$6~GeV piece, and sees no dependence with a precision of better than $10\%$~\cite{Rodrigues:2015hik}.  The mean energy of the $E_\nu<6$~GeV piece is roughly $\left< E_\nu\right>\approx 3$~GeV.  In general, an exclusive cross-section will have an energy dependence $\propto \frac{A}{E_\nu^2}+\frac{B}{E_\nu}+C$~\cite{llewelyn-smith}; therefore, unknown energy dependence may be parameterize by an {\em ad hoc} factor of the form $1/\left(1+ \frac{A^{'}}{E_\nu^2 }+\frac{B^{'}}{E_\nu}\right)$.  The \minerva constraints suggest $A^{'}<0.9$~GeV$^2$ and $B^{'}<0.3$~GeV.  The variations for neutrinos and antineutrinos could be different since this is an effective modification. Ideally this energy dependent factor would only affect the \minerva tune, but practically, because of analysis framework limitations already discussed, this is not possible. As a result, this energy dependent factor is applied to all true $2p2h$ events.

\subsubsection{Single pion production uncertainties}
\dword{genie} uses the Rein-Sehgal model for pion production. Tunes to $D_2$ data have been performed, both by the \dword{genie} collaboration itself and in subsequent re-evaluations~\cite{Rodrigues:2016xjj}; we use the latter tune as our base model. For simplicity of implementation, the `v2.8.2 (no norm.)' results are used here. 

\minerva single pion production data~\cite{Altinok:2017xua,McGivern:2016bwh,Eberly:2014mra} indicates disagreement at low $Q^2$ which may correspond to an incomplete nuclear model for single pion production in the generators. A similar effect was observed at MINOS~\cite{Adamson:2014pgc} 
and \nova implements a similar correction in analyses~\cite{nova_2018}.
A fit to \minerva data~\cite{StowellThesis} measured a suppression parameterized by
\begin{align}
R(Q^2<x_3) & = \frac{R_{2} (Q^2-x_1)(Q^2-x_3)}{(x_2-x_1)(x_2-x_3)} \notag\\
             & + \frac{(Q^2-x_1)(Q^2-x_2)}{(x_3-x_1)(x_3-x_2)}  \\
W(Q^{2}) &= 1-(1-R_{1})(1-R(Q^{2}))^{2}
\end{align}
where $R_{1}$ defines the magnitude of the correction function at the intercept, $x_{1}=0.0$. $x_{2}$ is chosen to be $Q^2=0.35\text{ GeV}^2$ so that $R_{2}$ describes the curvature at the center point of the correction. The fit found $R_1\approx0.3$ and $R_2\approx0.6$. The correction is applied to events with a resonance decay inside the nucleus giving rise to a pion, based on \dword{genie} event information.

An improved Rein-Sehgal-like resonance model has recently been developed~\cite{minoo} which includes a non-resonant background in both $I=\frac{1}{2}$ and $I=\frac{3}{2}$ channels and interference between resonant-resonant and resonant-non-resonant states. 
It also improves on the Rein-Sehgal model in describing the outgoing pion and nucleon kinematics using all its resonances.
A template weighting in $(W, Q^2, E_\nu)$ is implemented to cover the differences between the two models as a systematic uncertainty. The weighting also suppresses \dword{genie} non-resonant pion production events (deep inelastic scattering events with $W<1.7\text{ GeV}$) as the new model already includes the non-resonant contribution coherently. 
The weighting is only applied to true muon-neutrino charged-current resonant pion production interactions.

Coherent inelastic pion production measurements on carbon are in reasonable agreement with the \dword{genie} implementation of the Berger-Sehgal model~\cite{Mislivec:2017qfz}.  The process has not been measured at high statistics in argon. While coherent interactions provide a very interesting sample for oscillation analyses, they are a very small component of the event rate and selections will depend on the near detector configuration. Therefore we do not provide any evaluation of a systematic uncertainty for this extrapolation or any disagreements between the Berger-Sehgal model and carbon data.

\subsubsection{Other hard scattering uncertainties}
\nova oscillation analyses~\cite{nova_2018} have found the need for excursions beyond the default \dword{genie} uncertainties to describe their single pion to deep inelastic scattering (DIS) transition region data.  Following suit, we drop \dword{genie}'s default ``Rv[n,p][1,2]pi'' knobs and instead implement separate, uncorrelated uncertainties for all perturbations of 1, 2, and $\geq 3$ pion final states, CC/NC, neutrinos/anti-neutrinos, and interactions on protons/neutrons, with the exception of CC neutrino 1-pion production, where interactions on protons and neutrons are merged, following \cite{Rodrigues:2016xjj}. This leads to 23 distinct uncertainty channels ([3 pion states] x [n,p] x [nu/anti-nu] x [CC/NC] - 1), all with a value of 50\% for $W \leq 3$ GeV.  For each channel, the uncertainty drops linearly above $W = 3$ GeV until it reaches a flat value of 5\% at $W = 5$ GeV, where external measurements better constrain this process.

\subsubsection{Final state interaction uncertainties}\label{sec:fsi}
\dword{genie} includes a large number of final state uncertainties to its $hA$ final state cascade model which are summarized in Table~\ref{table:HadTranspKnobs}.  These uncertainties have been validated in neutrino interactions primarily on light targets such as carbon, but there is very little data available on argon targets.
The lack of tests against argon targets is difficult to address directly because there are many possible \dword{fsi} processes that could be varied.

\subsubsection{Neutrino flavor dependent uncertainties}
The cross sections include terms proportional to lepton mass, which are significant contributors at low energies where quasielastic processes dominate.  Some of the form factors in these terms have significant uncertainties in the nuclear environment.  Ref.~\cite{Day-McFarland:2012} ascribes the largest possible effect to the presence of poorly constrained second-class current vector form factors in the nuclear environment, and proposes a variation in the cross section ratio of $\sigma_\mu/\sigma_e$ of $\pm 0.01/{\rm\textstyle Max}(0.2~{\rm\textstyle GeV},E_\nu)$ for neutrinos and $\mp 0.018/{\rm\textstyle Max}(0.2~ {\rm\textstyle GeV},E_\nu)$ for anti-neutrinos.  Note the anticorrelation of the effect in neutrinos and antineutrinos.

In addition, radiative Coulomb effects may also contribute, which for T2K is of order $\pm5$ MeV shifts in reconstructed lepton momentum.  Like the second class current effect in the cross section, it flips sign between neutrinos and antineutrinos and is significant only at low energies.  This effect is not implemented herein.

Finally, some electron neutrino interactions occur at four momentum transfers where a corresponding muon neutrino interaction is kinematically forbidden, therefore the nuclear response has not been constrained by muon neutrino cross section measurements.  This region at lower neutrino energies has a significant overlap with the Bodek-Ritchie tail of the Fermi gas model. There are significant uncertainties in this region, both from the form of the tail itself, and from the lack of knowledge about the effect of RPA and $2p2h$ in this region. The allowed phase space in the presence of nonzero lepton mass is $E_\nu-\sqrt{\left( E_\nu-q_0\right) ^2-m_l^2}\leq q_3\leq E_\nu+\sqrt{\left( E_\nu-q_0\right) ^2-m_l^2}$. Here, a 100\% variation is allowed in the phase space present for $\nu_e$ but absent for $\nu_\mu$.

A similar prescription cannot applied for differences between interactions of $\nu_\mu$ and $\nu_\tau$ because the $\tau$ mass scale is of the same order of magnitude as the neutrino energies, and is thus a leading effect. No specific uncertainties were developed for $\nu_\tau$ interactions as there is little theoretical guidance.

\subsection{Listing of Interaction Model Uncertainties}

The complete set of interaction model uncertainties includes \dword{genie} implemented uncertainties 
(Tables~\ref{table:NuXSecKnobs}, and \ref{table:HadTranspKnobs}), 
and new uncertainties developed for this effort (Table~\ref{tab:nuintsystlist}) which represent uncertainties beyond those implemented in the \dword{genie} generator.  

\begin{table}[ptb]
\center
\global\long\def\arraystretch{1.75}
\scalebox{0.9}{
\begin{tabular}{llr|l}
\hline
$x_{P}$  & Description of $P$  & $P_\textsc{cv}$ & $\delta{P}/{P}$  \tabularnewline
\hline
&\textbf{Quasielastic}&&\tabularnewline
$x_{M_{A}}^{CCQE}$ & Axial mass for CCQE & & ${}^{+0.25}_{-0.15}$~GeV \tabularnewline
$x_{VecFF}^{CCQE}$  & Choice of CCQE vector form factors (BBA05 $\leftrightarrow$ Dipole)  &  & N/A \tabularnewline
$x_{kF}^{CCQE}$  & Fermi surface momentum for Pauli blocking &   & $\pm$30\% \tabularnewline
&\textbf{Low $\mathbf{W}$}&&\tabularnewline
$x_{M_{A}}^{CCRES}$ & Axial mass for CC resonance & 0.94 & $\pm$0.05~GeV \tabularnewline
$x_{M_{V}}^{CCRES}$  & Vector mass for CC resonance &  & $\pm$10\% \tabularnewline
$x_{\eta\ BR}^{\Delta Decay}$  & Branching ratio for $\Delta\rightarrow\eta$ decay &   & $\pm$50\% \tabularnewline
$x_{\gamma\ BR}^{\Delta Decay}$  & Branching ratio for $\Delta\rightarrow\gamma$ decay &   & $\pm$50\% \tabularnewline
$x_{\theta_{\pi}^{\Delta Decay}}$  & $\theta_{\pi}$ distribution in decaying $\Delta$ rest frame (isotropic $\rightarrow$ RS) &   & N/A \tabularnewline
&\textbf{High $\mathbf{W}$}&\tabularnewline
$x_{A_{HT}^{BY}}^{DIS}$  & $A_{HT}$ higher-twist param in BY model scaling variable $\xi_{w}$  &   & $\pm$25\% \tabularnewline
$x_{B_{HT}^{BY}}^{DIS}$  & $B_{HT}$ higher-twist param in BY model scaling variable $\xi_{w}$  &  & $\pm$25\% \tabularnewline
$x_{C_{V1u}^{BY}}^{DIS}$  & $C_{V1u}$ valence GRV98 PDF correction param in BY model  &  & $\pm$30\% \tabularnewline
$x_{C_{V2u}^{BY}}^{DIS}$  & $C_{V2u}$ valence GRV98 PDF correction param in BY model  &  & $\pm$40\% \tabularnewline
&\textbf{Other neutral current}&&\tabularnewline
$x_{M_{A}}^{NCEL}$  & Axial mass for NC elastic  &  & $\pm$25\% \tabularnewline
$x_{\eta}^{NCEL}$  & Strange axial form factor $\eta$ for NC elastic  &  & $\pm$30\%  \tabularnewline
$x_{M_{A}}^{NCRES}$  & Axial mass for NC resonance & &$\pm$10\% \tabularnewline
$x_{M_{V}}^{NCRES}$  & Vector mass for NC resonance & &$\pm$5\% \tabularnewline
&\textbf{Misc.}&&\tabularnewline
$x_{FZ}$  & Vary effective formation zone length &  & $\pm$50\% \tabularnewline
\hline
\end{tabular}
}
\\[2pt]
\caption[Neutrino interaction cross-section systematic parameters considered in GENIE]
{Neutrino interaction cross-section systematic parameters considered in \dword{genie}. \dword{genie} default central values and uncertainties are used for all parameters except $x_{M_{A}}^{CCRES}$. Missing \dword{genie} parameters were omitted where uncertainties developed for this analysis significantly overlap with the supplied \dword{genie} freedom, the response calculation was too slow, or the variations were deemed unphysical.
}
\label{table:NuXSecKnobs}
\end{table}

\begin{table}[ptb]
\center

\global\long\def\arraystretch{1.75}
\begin{tabular}{llll}
\hline
$x_{P}$  & Description of $P$  & $\delta{P}/{P}$  & \tabularnewline
\hline
$x_{cex}^{N}$  & Nucleon charge exchange probability  & $\pm$50\%  & \tabularnewline
$x_{el}^{N}$  & Nucleon elastic reaction probability  & $\pm$30\%  & \tabularnewline
$x_{inel}^{N}$  & Nucleon inelastic reaction probability  & $\pm$40\%  & \tabularnewline
$x_{abs}^{N}$  & Nucleon absorption probability  & $\pm$20\%  & \tabularnewline
$x_{\pi}^{N}$  & Nucleon $\pi$-production probability  & $\pm$20\%  & \tabularnewline
$x_{cex}^{\pi}$  & $\pi$ charge exchange probability  & $\pm$50\%  & \tabularnewline
$x_{el}^{\pi}$  & $\pi$ elastic reaction probability  & $\pm$10\%  & \tabularnewline
$x_{inel}^{\pi}$  & $\pi$ inelastic reaction probability  & $\pm$40\%  & \tabularnewline
$x_{abs}^{\pi}$  & $\pi$ absorption probability  & $\pm$20\%  & \tabularnewline
$x_{\pi}^{\pi}$  & $\pi$ $\pi$-production probability  & $\pm$20\%  & \tabularnewline
\hline
\end{tabular}\\[2pt] \caption[Intra-nuclear hadron transport systematic parameters implemented in GENIE]
{The intra-nuclear hadron transport systematic parameters implemented in \dword{genie} with associated uncertainties considered in
this work. Note that the 'mean free path' parameters are omitted for both N-N and $\pi$-N interactions as they produced unphysical variations in observable analysis variables. Table adapted from Ref~\cite{Andreopoulos:2015wxa}.
}

\label{table:HadTranspKnobs}
\end{table}

Table~\ref{tab:nuintsystlist} separates the interaction model parameters into three categories based on their treatment in the analysis:
\begin{itemize}
\item Category 1:  On-axis near detector data is expected to constrain these parameters; the uncertainty is implemented in the same way in near and far detectors.  All \dword{genie} uncertainties (original or modified) are all treated as Category 1.
\item Category 2: These uncertainties are implemented in the same way in near and far detectors, but on-axis data alone is not sufficient to constrain these parameters. We use two sub-categories. The first category (2A) corresponds to interaction effects which may be difficult to disentangle from detector effects. A good example of this is the $E_b$ parameter, which may be degenerate with the energy scale of the near detector.  This may be constrained with electron scattering and dedicated studies carefully selected samples of near detector data, but would be difficult to constrain with inclusive near detector samples. The second category (2B) corresponds to parameters that can be constrained by off-axis samples, described in Section~\ref{sec:nu-osc-06}.
\item Category 3: These uncertainties are implemented only in the far detector.  Examples are $\nu_e$ and $\overline{\nu}_e$ rates which are small and difficult to precisely isolated from background at the near detector.  Therefore, near detector data is not expected to constrain such parameters.
\end{itemize}

\begin{table}[ptb]
\scalebox{0.9}{
\begin{tabular}{lccc}\hline
Uncertainty & Mode & Description & Category  \\  \hline \hline
BeRPA & 1p1h/QE & RPA/nuclear model suppression &  1  \\  \hline
MnvTune1 & 2p2h & Strength into (nn)pp only &  1 \\  \hline
MnvTuneCV & 2p2h & Strength into 2p2h  &  1 \\  \hline
MnvTune2 & 1p1h/QE & Strength into 1p1h  &  1 \\  \hline
ArC2p2h & 2p2h Ar/C scaling & Electron scattering SRC pairs & 1 \\ \hline
$E_{2p2h}$ & 2p2h & 2p2h Energy dependence  &  2B \\  \hline
Low $Q^2$ $1\pi$ & RES & Low $Q^2$ (empirical) suppression &   1 \\  \hline
MK model & $\nu_\mu$ CC-RES & alternative strength in W &   1 \\  \hline
CC Non-resonant $\nu\rightarrow\ell+1\pi$ & $\nu$ DIS & Norm. for $\nu+n/p\rightarrow\ell+1\pi$ (\it{c.f.}\cite{Rodrigues:2016xjj}) & 1 \\ \hline
Other Non-resonant $\pi$ & $N\pi$ DIS & Per-topology norm. for $1<W<5$ GeV. & 1 \\ \hline
$E_{avail}/q_0$ & all & Extreme \dword{fsi}-like variations  &  2B \\
\hline
Modified proton energy & all & 20\% change to proton E &  2B \\
\hline
$\nu_\mu\to\nu_e$ & $\nu_e$/$\overline{\nu}_e$ & 100\% uncertainty in $\nu_e$ unique phase space &  3 \\ \hline
$\nu_e$/$\overline{\nu}_e$ norm & $\nu_e$,$\overline{\nu}_e$ & Ref.~\cite{Day-McFarland:2012}  &  3 \\
\hline
\hline
\end{tabular}
} 
\caption[List of extra interaction model uncertainties in addition to those provided by GENIE]{List of extra interaction model uncertainties in addition to those provided by GENIE.}
\label{tab:nuintsystlist}
\end{table}

Finally, there are a number of tunes applied to the default model, to represent known deficiencies in \dword{genie}'s description of neutrino data, and these are listed in Table~\ref{table:NuXSecKnobs_Central}.

\begin{table}[ptb]
\center
\global\long\def\arraystretch{1.75}
\scalebox{0.9}{
\begin{tabular}{ll|l}
\hline
$x_{P}$  & Description of $P$  & $P_\textsc{cv}$  \tabularnewline
\hline
&\textbf{Quasielastic} & \tabularnewline
BeRPA & Random Phase Approximation tune & $A: 0.59$ \tabularnewline
& $A$ controls low $Q^2$, $B$ controls low-mid $Q^2$ & $B: 1.05$ \tabularnewline
& $D$ controls mid $Q^2$, $E$ controls high $Q^2$ fall-off & $D: 1.13$ \tabularnewline
& $U$ controls transition from polynomial to exponential & $E: 0.88$ \tabularnewline
& & $U: 1.20$ \tabularnewline
&\textbf{2p2h}&\tabularnewline
MINERvA 2p2h tune & $q0,q3$ dependent correction to 2p2h events&\\
&\textbf{Low $\mathbf{W}$ single pion production} & \tabularnewline
$x_{M_{A}}^{CCRES}$ & Axial mass for CC resonance in \dword{genie}& $0.94$ \tabularnewline
Non-res CC1$\pi$ norm. & Normalization of CC1$\pi$ non-resonant interaction & $0.43$  \tabularnewline
\hline
\end{tabular}
}
\\[2pt]
\caption[Neutrino interaction cross-section systematic parameters that receive a central-value tune]{Neutrino interaction cross-section systematic parameters that receive a central-value tune}
\label{table:NuXSecKnobs_Central}
\end{table}

\section{The Near Detector Simulation and Reconstruction}\label{sec:nu-osc-06}\label{sec:physics-lbnosc-ND}

Oscillation parameters are determined by comparing observed charged-current event spectra at the \dword{fd} to predictions that are, {\em a priori}, subject to uncertainties on the neutrino flux and cross sections at the level of tens of percent as described in the preceding sections. To achieve the required few percent precision of \dword{dune}, it is necessary to constrain these uncertainties with a highly capable \dword{nd} suite. The \dword{nd} is described in more detail in Volume~\volnumberexec{}, \voltitleexec{}.

The broad \dword{nd} concept is described briefly in Section~\ref{sec:ndconcept}, along with an outline of the \dword{nd}'s role in the oscillation analysis. The parameterized reconstruction and event selection is described in Section~\ref{sec:ndsimreco}. 
\dword{nd} and \dword{fd} uncertainties, including those due to energy estimation and selection efficiencies, are discussed in Section~\ref{sec:physics-lbnosc-syst}.  

\subsection{The Near Detector Concept}
\label{sec:ndconcept}

The \dword{dune} \dword{nd} system consists of a \dword{lartpc} functionally coupled to a magnetized \dword{mpd}, and a 
\dword{sand}. 
The \dword{nd} hall is located at \dword{fnal} 574 m from the neutrino beam source and 60 m underground. The long dimension of the hall is oriented at 90 degrees with respect to the beam axis to facilitate measurements at both on-axis and off-axis locations with a movable detector system. 

The \dword{lartpc} is modular, with fully-\threed  pixelated readout and optical segmentation. These features greatly reduce reconstruction ambiguities that hamper monolithic, projective-readout \dwords{tpc}, and enable the \dword{nd} to function in the high-intensity environment of the \dword{dune} \dword{nd} site. Each module is itself a \dword{lar} \dword{tpc} with two anode planes and a central cathode. The active dimensions of each module are $1 \times 3 \times 1$~m ($x \times y \times z$), where the $z$ direction is $6^{\circ}$ upward from the neutrino beam, and the $y$ direction points upward. Charge drifts in the $\pm x$ direction, with a maximum drift distance of 50 cm for ionization electrons produced in the center of a module. The module design is described in detail in Ref.~\cite{Asaadi:2018xfh}. The full \dword{lar} detector consists of an array of modules in a single cryostat. The minimum active size for full containment of hadronic showers is $3 \times 4 \times 5$~m. High-angle muons can also be contained by extending the width to 7 m. For this analysis, 35 modules are arranged in an array 5 modules deep in the $z$ direction and 7 modules across in $x$ so that the total active dimensions are $7 \times 3 \times 5$~m. The total active \dword{lar} volume is $105$~m$^{3}$, corresponding to a mass of 147 tons.

The anode planes are tiled with readout pads, such that the $yz$ coordinate is given by the pad location and the $x$ coordinate is given by the drift time, and the three-dimensional position of an energy deposit is uniquely determined. A dedicated, low-power readout ASIC is being developed, which will enable single-pad readout without analog multiplexing~\cite{Dwyer:2018phu}. The module walls orthogonal to the anode and cathode are lined with a photon detector that is sensitive to scintillation light produced inside the module, called ArCLight~\cite{Auger:2017flc}. The detector is optically segmented, and tiled so that the vertical position of the optical flash can be determined with $\sim$30 cm resolution. It is therefore possible to isolate flashes to a volume of roughly 0.3 m$^{3}$, and associate them to a specific neutrino interaction even in the presence of pile-up. The neutrino interaction time, $t_{0}$, is determined from the prompt component of the scintillation light.

The \dword{mpd} consists of a high-pressure \dword{gartpc} in a cylindrical pressure vessel at 10 bar, surrounded by a granular, high-performance electromagnetic calorimeter. The \dword{mpd} sits immediately downstream of the \dword{lar} cryostat so that the beam center crosses the exact center of both the \dword{lar} and gaseous argon active volumes. The pressure vessel is 5 m in diameter and 5 m long. The \dword{tpc} is divided into two drift regions by a central cathode, and filled with a 90/10 Ar/CH$_{4}$ gas mixture, such that 97\% of neutrino interactions will occur on the Ar target. The gas TPC is described in detail in Ref.~\cite{bib:docdb12388}.

The electromagnetic calorimeter is composed of a series of absorber layers followed by arrays of scintillator read out by SiPMs mounted directly onto boards. The inner-most layers will be tiled, giving \threed position information for each hit, and sufficient granularity to enable reconstruction of the angle of incoming photons. The \dword{ecal} design is described in Ref. ~\cite{Emberger:2018pgr}. The entire \dword{mpd} sits inside a magnetic field with a strength of at least 0.4 T. A superconducting magnet is preferred, to reduce the total amount of mass near the detectors.

The optimization of the detector design is still underway at the time of preparing this document, and the eventual parameters may be somewhat different from what is simulated. For the oscillation analysis presented herein, only the \dword{lar} event sample is explicitly used. The \dword{ecal}, pressure vessel, and magnet design have a small impact on the acceptance of muons originating in the \dword{lar}. The \dword{ecal} is assumed to be 30 layers of alternating planes of 5mm CH and 2mm Cu. The pressure vessel is assumed to be 3 cm thick titanium. The magnet is a solenoid with an inner radius of 320 cm, with a yoke cut out of the upstream barrel to minimize the passive material between the two \dword{tpc} detectors. 
%
The on-axis neutrino beam monitor \dword{sand}
is not functionally coupled to the \dword{lar} detector and thus is not included in these simulations. Small changes in these parameters are not expected to significantly impact the acceptance. A profile view visualization of the ND as implemented in this analysis is shown in Figure~\ref{fig:NDvis}.

\begin{dunefigure}[ND visualization]{fig:NDvis}
{The near detector shown from the side. The neutrinos are incident from the left along the axis shown, which intersects the center of the LAr and GAr detectors.}
 \includegraphics[width=0.6\textwidth]{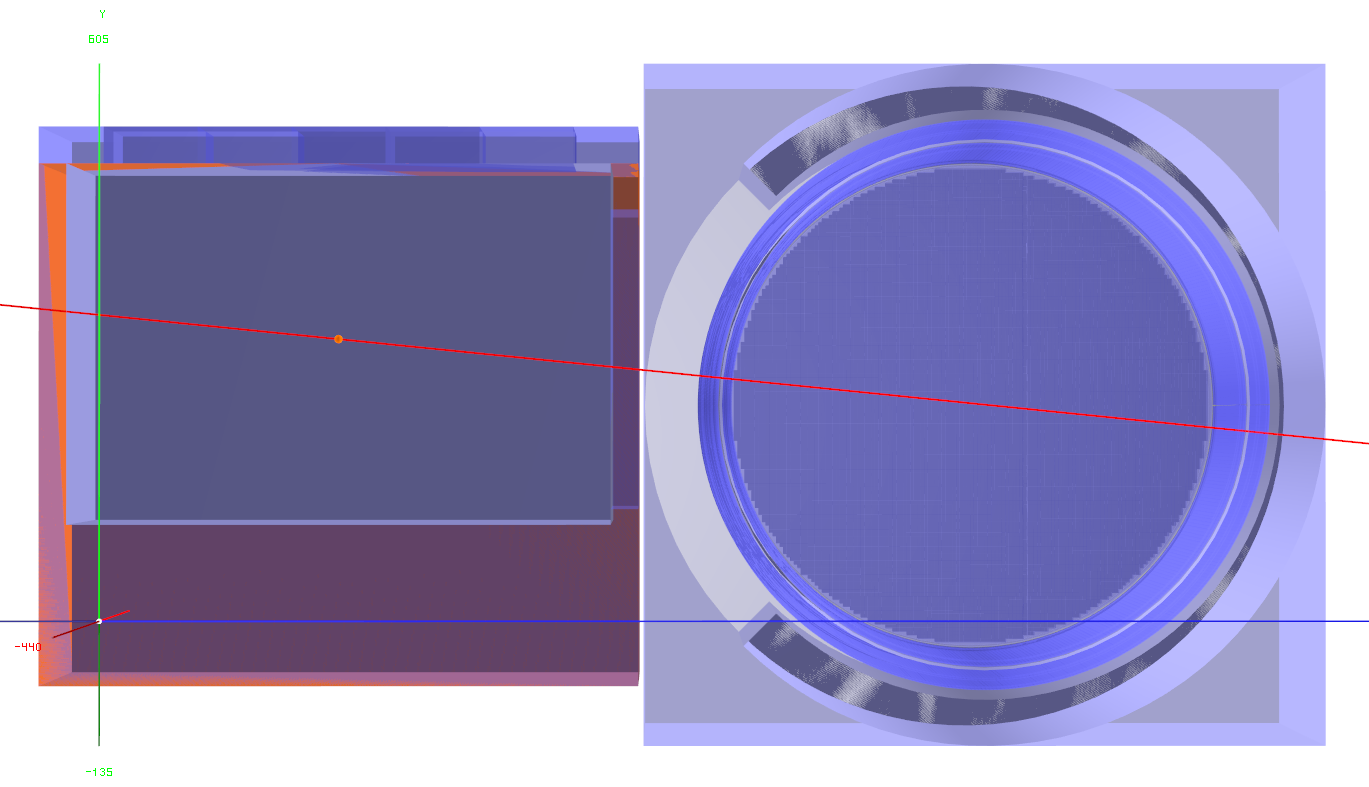}
\end{dunefigure}

The oscillation analysis includes samples of $\nu_{\mu}$ and $\bar{\nu}_{\mu}$ charged-current interactions originating in the \dword{lar}. The samples are binned in \twod as a function of neutrino energy and inelasticity, $y = 1 - E_{\mu}/E_{\nu}$, where $E_{\mu}$ and $E_{\nu}$ are the muon and neutrino energies, respectively. 

\subsection{Event Simulation and Parameterized Reconstruction}
\label{sec:ndsimreco}

Neutrino interactions are simulated in the active volumes of the \dword{lar} and \dword{hpg} \dwords{tpc}. The neutrino flux prediction is described in Section~\ref{sec:physics-lbnosc-flux}. Interactions are simulated with the \dword{genie} event generator using the model configuration described in Section~\ref{sec:nu-osc-05}. The propagation of neutrino interaction products through the detector volumes is simulated using a \dword{geant4}-based model. Pattern recognition and reconstruction software has not yet been developed for the \dword{nd}. Instead, we perform a parameterized reconstruction based on true energy deposits in active detector volumes as simulated by \dword{geant4}.

\subsubsection{Liquid Argon charged-current interactions}

Liquid argon events are required to originate in a fiducial volume that excludes 50 cm from the sides and upstream edge, and 150 cm from the downstream edge of the active region, for a total of $6 \times 2 \times 3$~m$^{2}$. A hadronic veto region is defined as the outer 30 cm of the active volume on all sides. Events with more than 30 MeV total energy deposit in the veto region are excluded from analysis, as this energy near the detector edge suggests leakage, resulting in poor energy reconstruction. Even with the containment requirement, events with large shower fluctuations to neutral particles can still be very poorly reconstructed. Neutrons, in particular, are largely unreconstructed energy.

Electrons are reconstructed calorimetrically in the liquid argon. The radiation length is 14 cm in \dword{lar}, so for fiducial interactions and forward-going electrons there are between 10 and 30 radiation lengths between the vertex and the edge of the \dword{tpc}. As there is no magnetic field in the \dword{lar} \dword{tpc} region, electrons and positrons cannot be distinguished and the selected $\nu_{e}$ sample contains both neutrino- and antineutrino-induced events.

Muons with kinetic energy greater than $\sim$1 GeV typically exit the \dword{lar}. An energetic forward-going muon will pass through the \dword{ecal} and into the gaseous \dword{tpc}, where its momentum and charge are reconstructed by curvature. For these events, it is possible to differentiate between $\mu^{+}$ and $\mu^{-}$ event by event. Muons that stop in the \dword{lar} or \dword{ecal} are reconstructed by range. Exiting muons that do not match to the \dword{hpg} \dword{tpc} are not reconstructed, and events with these tracks are rejected from analysis.
These are predominantly muon CC, where the muon momentum cannot be determined. Forward exiting muons will enter the magnetized \dword{mpd}, where their momenta and charge sign are reconstructed by curvature. The asymmetric transverse dimensions of the \dword{lar} volume make it possible to reconstruct wide-angle muons with some efficiency. High-angle tracks are typically lost when the $\nu-\mu$ plane is nearly parallel to the $y$ axis, but are often contained when it is nearly parallel to the $x$ axis. 

The charge of stopping muons in the \dword{lar} volume cannot be determined. However, the wrong-sign flux is predominantly concentrated in the high-energy tail, where leptons are likelier to be forward and energetic. In FHC mode, the wrong-sign background in the focusing peak is negligibly small, and $\mu^{-}$ is assumed for all stopping muon tracks. In RHC mode, the wrong-sign background is larger in the peak region. Furthermore, high-angle leptons are generally at higher inelasticity, $y$, which enhances the wrong-sign contamination in the contained muon subsample. To mitigate this, a Michel electron is required. The wrong-sign $\mu^{-}$ captures on Ar with 75\% probability, effectively suppressing the relative $\mu^{-}$ component by a factor of four.

Events are classified as either $\nu_{\mu}$ CC, $\bar{\nu}_{\mu}$ CC, $\nu_{e}$+$\bar{\nu}_{e}$ CC, or NC. True muons and charged pions are evaluated as potential muon candidates. The track length is determined by following the true particle trajectory until it hard scatters or ranges out. The particle is classified as a muon if its track length is at least 1 m, and the mean energy deposit per centimeter of track length is less than 3 MeV. The mean energy cut rejects tracks with detectable hadronic interactions. The minimum length requirement imposes an effective threshold on true muons of about 200 MeV kinetic energy, but greatly suppresses potential NC backgrounds with short, non-interacting charged pions.

True electrons are reconstructed with an ad-hoc efficiency that is zero below 300 MeV, and rises linearly to unity between 300 and 700 MeV. Neutral-current backgrounds arise from photon and $\pi^{0}$ production. Photons are misreconstructed as electrons when the energy deposit per centimeter in the first few cm after conversion is less than 4 MeV. This is typically for Compton scatters, and can also occur due to a random downward fluctuation in the $e^{+}e^{-}$ dE/dx. The conversion distance must also be small so that no visible gap can be identified. We consider a photon gap to be clear when the conversion distance is greater than 2 cm, which corresponds to at least four pad widths. For $\pi^{0}$ events, the second photon must also be either less than 50 MeV, or have an opening angle to the first photon less than 10 mrad. Electrons are generally contained in the \dword{lar} and are reconstructed calorimetrically. It is possible for CC $\nu_{\mu}$ events to be reconstructed as CC $\nu_{e}$ when the muon is too soft and a $\pi^{0}$ fakes the electron.

\dword{lar} events are classified as $\nu_{\mu}$ CC, $\bar{\nu}_{\mu}$ CC, $\nu_{e}$ + $\bar{\nu}_{e}$ CC, or NC. Charged-current events are required to have exactly one reconstructed lepton of the appropriate flavor. The muon-flavor samples are separated by reconstructed charge, but the electron-flavor sample is combined because the charge cannot be determined. The neutral-current sample includes all events with zero reconstructed leptons. Spectra for selected \numu \dword{cc} events in \dword{fhc} are shown in Figure~\ref{fig:recoSamples} as a function of both neutrino energy and inelasticity.

\begin{dunefigure}[ND selected samples]{fig:recoSamples}
{Reconstructed neutrino energy and $y$ for events classified as $\nu_{\mu}$ \dword{cc} in \dword{fhc} mode. Background events are predominantly neutral currents and are shown in red.}
 \includegraphics[width=0.48\textwidth]{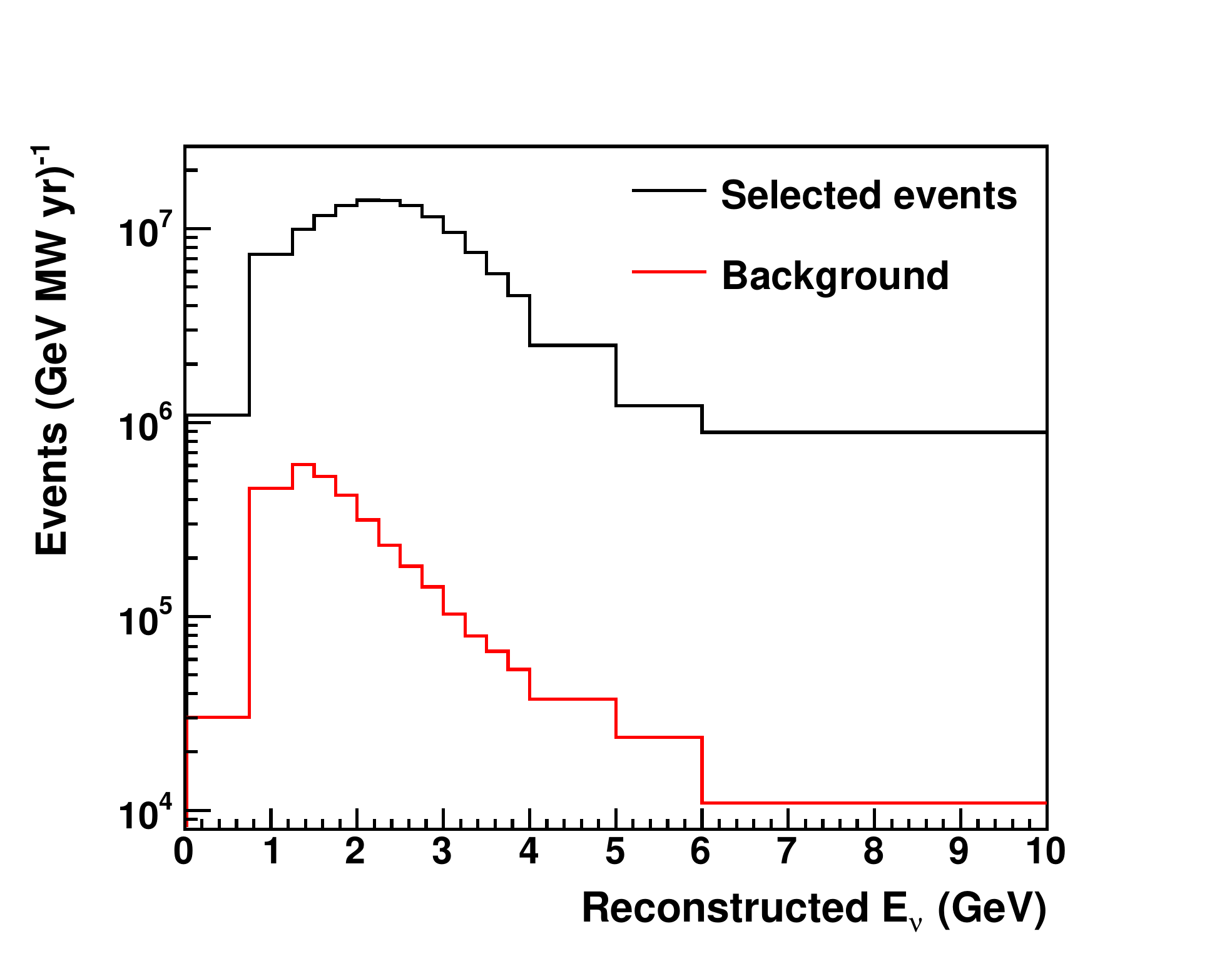}
 \includegraphics[width=0.48\textwidth]{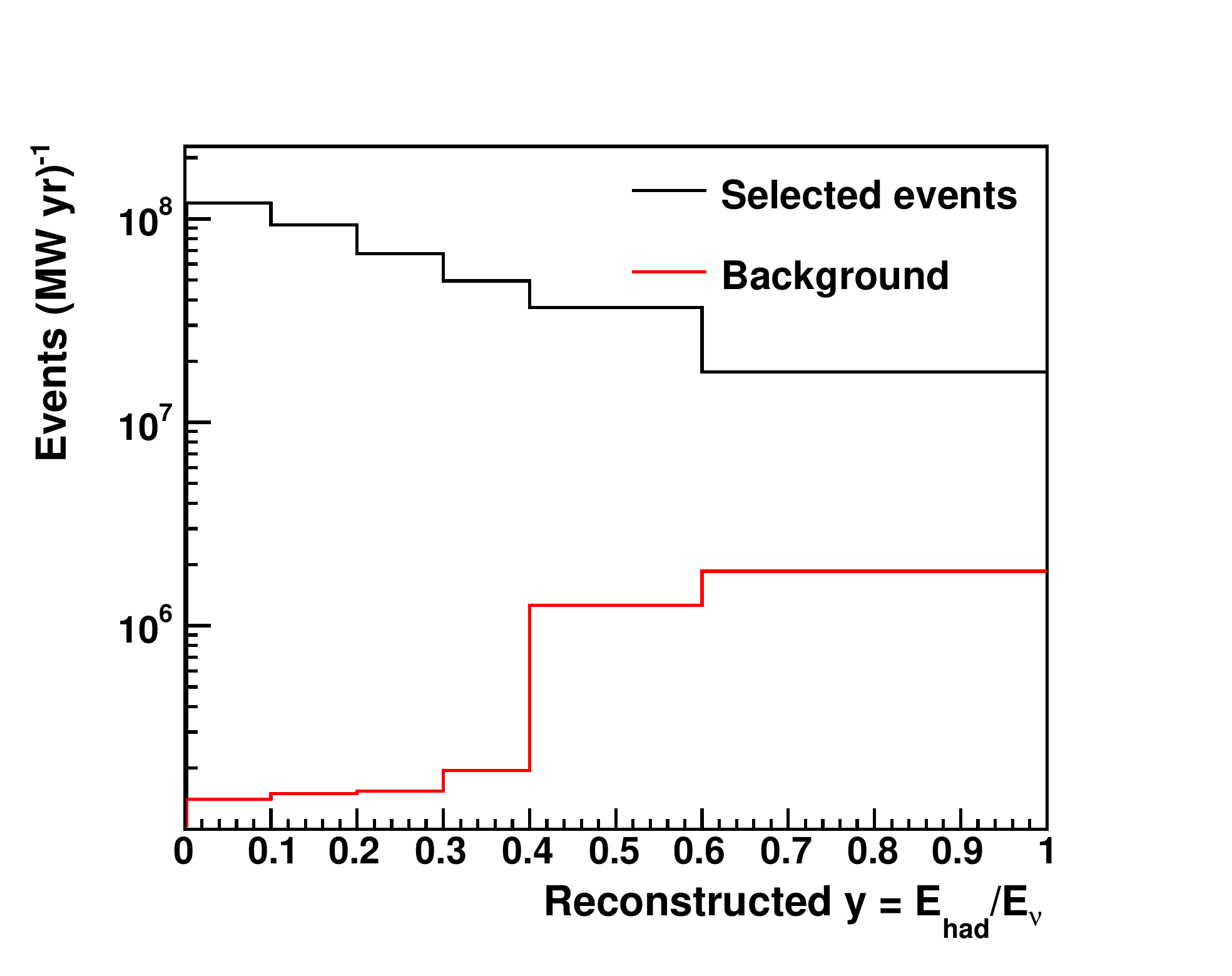}
\end{dunefigure}

Hadronic energy is estimated by summing visible energy deposits in the active \dword{lar} volume. Events are rejected when energy is observed in the outer 30 cm of the detector, which is evidence of poor hadronic containment.  Events with more than 30 MeV of visible hadronic energy in the veto region are also excluded.  This leads to an acceptance that decreases with hadronic energy, as shown in the right panel of Figure~\ref{fig:NDacceptance}.


Events are classified as either $\nu_{\mu}$ \dword{cc}, $\bar{\nu}_{\mu}$ \dword{cc}, $\nu_{e}$+$\bar{\nu}_{e}$ \dword{cc}, or \dword{nc} based on the presence of charged leptons. Backgrounds to $\nu_{\mu}$ \dword{cc} arise from \dword{nc} $\pi^{\pm}$ production where the pion leaves a long track and does not shower. Muons below about 400 MeV kinetic energy have a significant background from charged pions, so these \dword{cc} events are excluded from the selected sample. Backgrounds to $\nu_{e}$ \dword{cc} arise from photons that convert very near the interaction vertex. The largest contribution is from $\pi^{0}$ production with highly asymmetric decay.

\begin{dunefigure}[ND acceptance]{fig:NDacceptance}
{Left: Detector acceptance for $\nu_{\mu}$ \dword{cc} events as a function of muon transverse and longitudinal momentum. Right: Acceptance as a function of hadronic energy; the black line is for the full fiducial volume while the red line is for a $1 \times 1 \times 1$~m$^{3}$ volume in the center, and the blue curve is the expected distribution of hadronic energy given the \dword{dune} flux.}
 \includegraphics[width=0.48\textwidth]{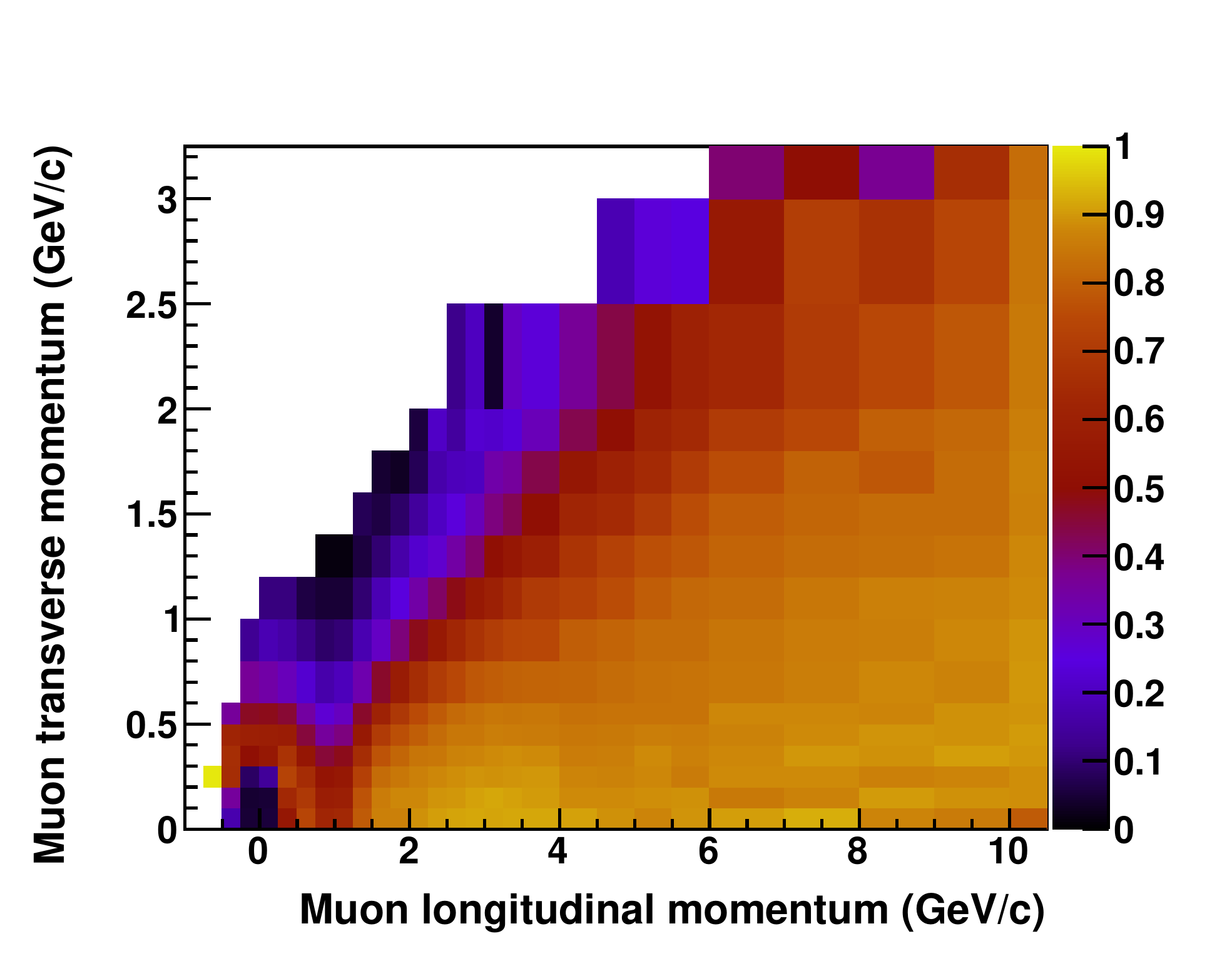}
 \includegraphics[width=0.48\textwidth]{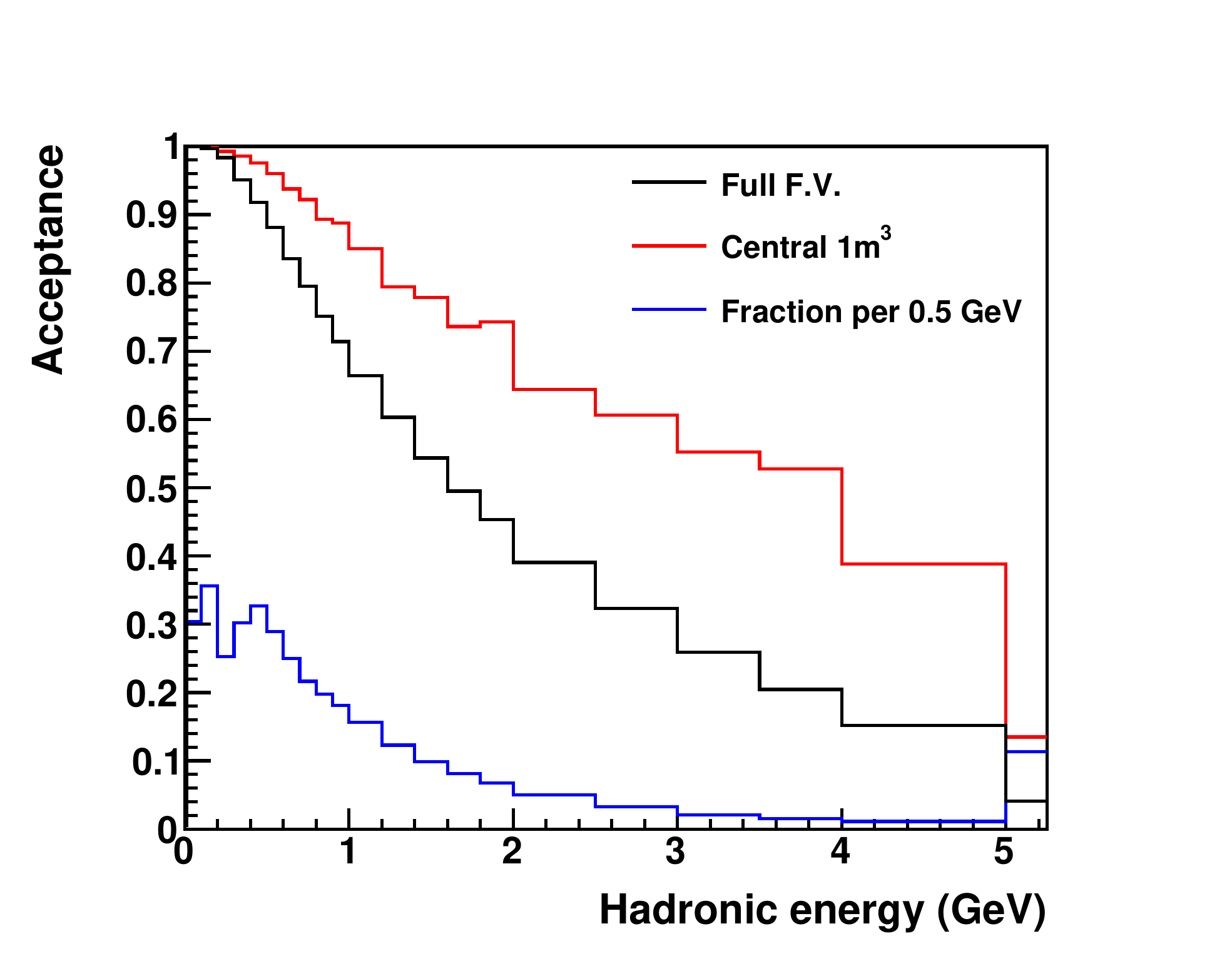}
\end{dunefigure}

\subsubsection{Neutrino-electron elastic scattering}



In addition to the CC event selections, neutrino-electron elastic scattering is also selected. Measurements of neutrino-nucleus scattering are sensitive to the product of the flux and cross section, both of which are uncertain. This can lead to a degeneracy between flux and cross section nuisance parameters in the oscillation fit, and results in significant anti-correlations, even when the uncertainty on the diagonal component is small. One way to break this degeneracy is by including a sample for which the a priori cross section uncertainties are very small. 

Neutrino-electron scattering is a pure-electroweak process with calculable cross section. It is therefore possible to directly constrain the flux by measuring the event rate of $\nu+ e \rightarrow \nu +e$ and dividing by the known cross section. The final state consists of a single electron, subject to the kinematic limit 

\begin{equation}
1 - \cos \theta = \frac{m_{e}(1-y)}{E_{e}},
\end{equation}

where $\theta$ is the angle between the electron and incoming neutrino, $E_{e}$ and $m_{e}$ are the electron mass and total energy, respectively, and $y = T_{e}/E_{\nu}$ is the fraction of the neutrino energy transferred to the electron. For DUNE energies, $E_{e} \gg m_{e}$, and the angle $\theta$ is very small, such that $E_{e}\theta^{2} < 2m_{e}$.

The overall flux normalization can be determined by counting $\nu e \rightarrow \nu e$ events. Events can be identified by searching for a single electromagnetic shower with no other visible particles. Backgrounds from $\nu_{e}$ charged-current scattering can be rejected by looking for large energy deposits near the interaction vertex, which are evidence of nuclear breakup. Photon-induced showers from neutral-current $\pi^{0}$ events can be distinguished from electrons by the energy profile at the start of the track. The dominant background is expected to be $\nu_{e}$ charged-current scattering at very low $Q^{2}$, where final-state hadrons are below threshold, and $E_{e}\theta^{2}$ happens to be small. The background rate can be constrained with a control sample at higher $E_{e}\theta^{2}$, but the shape extrapolation to $E_{e}\theta^{2} \rightarrow 0$ is uncertain at the 10-20\% level.

For the DUNE flux, approximately 100 events per year per ton of fiducial mass are expected with electron energy above 0.5 GeV. For a LAr TPC mass of 25 tons, this corresponds to 2500 events per year, or 12500 events in the full 5-year FHC run, assuming the ND stays on axis. Given the very forward signal, it may be possible to expand the fiducial volume to enhance the rate. The statistical uncertainty on the flux normalization from this technique is expected to be $\sim$1\%.

To evaluate the impact of neutrino-electron scattering, a dedicated high-statistics signal-only sample is generated. Due to the simple nature of the signal, it is possible to estimate backgrounds without a full detector simulation. A single electromagnetic shower (electron, positron or photon) is required. To reject $\pi^{0}$ events with clearly-identifiable second photons, no additional showers over 50 MeV are allowed.

Charged-current $\nu_{e}$ interactions can be rejected when there is evidence of nuclear breakup in the form of final-state charged hadrons. A conservative cut of 40 MeV total charged hadron kinetic energy is applied. For a single proton, this corresponds to $\sim$ 1 cm of track length, which will leave energy on two or three readout pads and be easily identified. Finally, a cut requiring low $E_{e}\theta_{e}^{2}$ isolates the $\nu+e$ signal. Alternatively, templates in $(E_{e}, \theta_{e})$ can be formed, and the unique shape of the signal can be used in a fit to extract the flux normalization.

\subsubsection{Off-axis \dword{nd} measurements}
\label{sec:ch-nu-osc-06-ndconcept-offaxis}

Neutrino energy reconstruction is one of the biggest challenges in a precision long-baseline oscillation experiment like \dword{dune}. Even with a highly capable \dword{fd}, a fraction of the final-state hadronic energy is typically not observed. For example, neutrons may travel meters without interacting, and can exit the detector with significant kinetic energy. This missing energy is typically corrected with a neutrino interaction generator, which is used to relate the true neutrino energy to the observed energy. These models have many tens of uncertain parameters, which can be constrained by \dword{nd} measurements. However, there may be many different parameter combinations that adequately describe the \dword{nd} data. These degenerate solutions can extrapolate differently to the \dword{fd}, where the flux is significantly different due to oscillations. This can lead to biases in the fitted oscillation parameters, including $\delta_{CP}$, despite an apparently good quality of fit.

While these biases can be partially mitigated by an on-axis \dword{nd} capable of making numerous exclusive measurements, the energy dependence of the interaction cross section and the bias in reconstructed neutrino energy cannot be measured in a single beam. To gain sensitivity to these, the \dword{lartpc} and \dword{mpd} combination is movable, and the \dword{nd} hall is oriented to facilitate both on-axis measurements and measurements at positions up to 33 m off axis. The flux spectrum varies as a function of off-axis angle, peaking lower in energy as the angle is increased, from $\sim$2.5 GeV in the on-axis position down to $\sim$0.5 GeV at 33 m off axis. As uncertainties in the flux prediction are strongly correlated across off-axis angles, off-axis measurements of reconstructed neutrino energy constrain cross section uncertainties and provide further handles on possible degeneracies in the fit. 

By taking linear combinations of such measurements, it is also possible to reproduce the predicted \dword{fd} oscillated flux for some set of oscillation parameters, and directly compare visible energy between \dword{nd} and \dword{fd} over essentially the same ``oscillated'' flux, and with greatly reduced model dependence. Figure~\ref{fig:nd_prism} shows the result of such a linear combination, overlaid with the \dword{fd} flux. The oscillated flux is well reproduced between $\sim$0.5 GeV and $\sim$3.5 GeV. The lower $E_{\nu}$ bound is determined by the range of accessible off-axis angles; to cover down to 0.5 GeV, measurements out to 33 m off axis are required. The off-axis technique cannot reproduce the high-energy flux tail seen in the \dword{fd} spectrum. This is because the off-axis spectra all provide lower peak energies; it is not possible to produce a peak energy higher than that of the \dword{fd} because the \dword{fd} is on axis.

\begin{dunefigure}[DUNE-PRISM fluxes]{fig:nd_prism}
{The predicted \dword{fd} flux (black), and a prediction made up of linear combinations of \dword{nd} fluxes (green).}
 \includegraphics[width=0.49\textwidth]{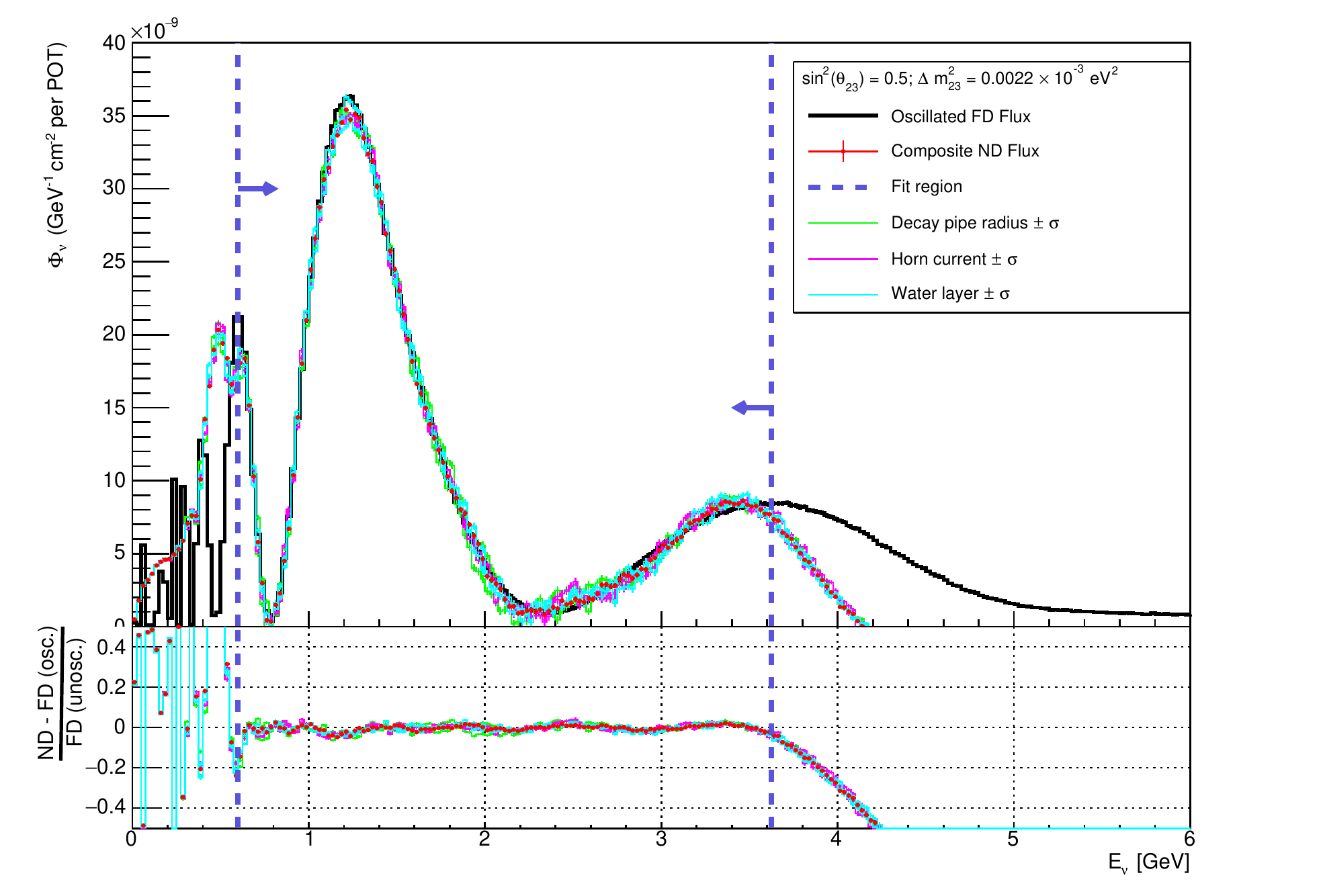}
\end{dunefigure}

A potential run plan is to take on-axis data approximately 50\% of the time, with the other 50\% split among enough off-axis positions so that the fiducial volumes of adjacent ``stops'' overlap, giving a continuous range of angles. Event selection at an off-axis location of the \dword{lar} detector is identical to the on-axis case. The selection efficiency varies as a function of muon energy due to containment and matching to the downstream magnetized tracker, and as a function of hadronic energy due to containment. As muon and hadronic energy are correlated to neutrino energy, the efficiency varies with off-axis position. The efficiency also varies as a function of vertex position; interactions occurring near the edges of the detector are more likely to fail containment requirements. These effects are corrected with simulation; as they are largely geometric, the uncertainties that arise from the corrections are small compared to the uncertainties on neutrino cross sections and energy reconstruction. 

\subsubsection{Gaseous argon charged-current interactions}

With over 30 million charged-current events per year, the \dword{lartpc} event sample can be analyzed in many different exclusive channels and provide powerful constraints. However, its relatively high density makes certain hadronic topologies challenging to reconstruct. The gaseous \dword{tpc} complements the \dword{lar} detector by providing low reconstruction thresholds, excellent pion/proton separation, and charge-sign reconstruction. In particular, measurements of proton and charged pion multiplicities as a function of neutrino energy constrain cross section uncertainties not accessible to the \dword{lar} alone.

In addition, the gas \dword{tpc} provides a useful check on the reconstruction efficiency of the \dword{lar} selection. Due to the combining of contained muons with gas \dword{tpc}-matched events, there are kinematic regions where the acceptance of the \dword{lartpc} is uncertain. Also, without a magnetic field, the wrong sign contamination cannot be directly measured, especially at high angle and low energy. The gas \dword{tpc}, however, has uniform acceptance over the full 4$\pi$, as well as charge measurement capability except when the muon is nearly parallel to the magnetic field lines.

Unlike the \dword{lartpc}, where the hadronic energy is determined by a calorimetric sum of energy deposits, the gas \dword{tpc} hadronic energy is reconstructed particle-by-particle, including pion masses. For this analysis, samples of $\nu_{\mu}$ \dword{cc} events are selected in slices of charged pion multiplicity, and fit as a function of reconstructed neutrino energy. The threshold for charged pion selection is 5 MeV, and $\pi^{+}$ can be reliably separated from protons up to momenta of 1.3 GeV/c.


\section{The Far Detector Simulation and Reconstruction}\label{sec:nu-osc-07}\label{sec:physics-lbnosc-FD}
\label{sec:physics-lbnosc-simreco}

The calculation of \dword{dune} sensitivities to oscillation parameter measurements requires predictions for the number of events to be observed in the \dword{fd} fiducial volume, the reconstructed neutrino energy for each of these events, and the probability that they will be correctly identified as signal for each analysis samples. To build these analysis samples a  \dword{geant4} simulation of the \dword{fd} has been developed. The output of that simulation has been used to build neutrino energy estimators, and an event selection discriminant that can separate $\nu_{e}$ \dword{cc}, $\nu_{\mu}$ \dword{cc}, and \dword{nc} events. Each of these components is described in detail in this section. The uncertainties associated with each step in the simulation and reconstruction chain, including the \dword{fd} simulation, reconstructed energy estimators, and selection efficiencies are discussed in Section~\ref{sec:nu-osc-09}.

\subsection{Simulation}
The neutrino samples were simulated using a smaller version of the full \nominalmodsize far \dword{detmodule} geometry. This geometry is 13.9 m long, \tpcheight high and 13.3 m wide, which consists of 12 \dwords{apa} and 24 \dwords{cpa}. The reference flux was used (Section~\ref{sec:physics-lbnosc-flux}) and samples were produced with both the forward-horn-current (neutrino enhanced) and inverted-horn-current (antineutrino enhanced) beam configurations. Three samples were generated. The first sample keeps the original neutrino flavor composition of  the neutrino beam. The second sample converts all the muon neutrinos to electron neutrinos. The third sample converts all the muon neutrinos to tau neutrinos. Oscillation probabilities are used to weight \dword{cc} events to build oscillated \dword{fd} predictions from the three event samples. \dword{genie} 2.12.10 was used to model the neutrino-argon interactions in the volume of cryostat. The produced final-state (after \dword{fsi}) particles were propagated in the detector through \dword{geant4}. 
The ionization electrons and scintillation light were digitized to produce signals in the wire planes and \dwords{pd}. More details on the simulation can be found in Section~\ref{sec:tools-mc-detsim}.

\subsection{Event Reconstruction and Kinematic Variables}
The first step in the reconstruction is to convert the raw signal from each wire to a standard (e.g., Gaussian) shape. This is achieved by passing the raw data through a calibrated deconvolution algorithm to remove the impact of the \dword{lartpc} \efield and the electronic response from the measured signal. The resulting wire waveform possesses calibrated charge information. 

The hit-finding algorithm scans the processed wire waveform looking for local minima. If a minimum is found, the algorithm follows the waveform after this point until it finds a local maximum. If the maximum is above a specified threshold, the program scans to the next local minimum and identifies this region as a hit. Hits are fit with a Gaussian function whose features identify the correct position (time coordinate), width, height, and area (deposited charge) of the hit. A single Gaussian function is used to describe hits produced by isolated single particles. In regions where there are overlapping particles (e.g., around the neutrino interaction vertex) single Gaussian fits may fail, and fits to multiple Gaussian functions may be used. The reconstructed hits are used by reconstruction and event selection pattern recognition algorithms. In particular the \dword{cvn} event selection algorithm is described later in this section.

The reconstruction algorithms (\dword{pandora} and \dword{pma}) define clusters as hits that may be grouped together due to proximity in time and space to one another. Clusters from different wire planes are matched to form high-level objects such as tracks and showers. These high level objects are used as inputs to the neutrino energy reconstruction algorithm. More details on the reconstruction can be found in section~\ref{sec:tools-fdreco}.


The energy of the incoming neutrino in \dword{cc} events is estimated by adding the reconstructed lepton and hadronic energies. 
If the event is selected as $\nu_{\mu}$ \dword{cc}, the neutrino energy is estimated as the sum of the energy of the longest reconstructed track and the hadronic energy. The energy of the longest reconstructed track is estimated from its range if the track is contained in the detector, and this is calibrated using simulated $\nu_{\mu}$ \dword{cc} events with true muon energies from 0.2-1.7 GeV. If the longest track exits the detector, its energy is estimated from multi-Coulomb scattering, and corrected using simulated events with true muon energies from 0.5-3 GeV. The hadronic energy is estimated from the charge of reconstructed hits that are not in the longest track, and corrections are applied to each hit charge for recombination and the electron lifetime. An additional correction is then made to the hadronic energy to account for missing energy due to neutral particles and final-state interactions, and this is done using simulated events with true hadronic energies from 0.1-1.6 GeV. The same hadronic shower energy calibration is used for both $\nu$ and $\bar{\nu}$ based on a sample of $\nu$ and $\bar{\nu}$ events.

If the event is selected as $\nu_{e}$ \dword{cc}, the energy of the 
neutrino is estimated as the sum of the energy of the reconstructed shower with the highest energy and the hadronic energy. The former is estimated from the charges of the reconstructed hits in the shower, and the latter from the hits not in the shower; the recombination and electron lifetime corrections are applied to the charge of each hit. Subsequently the shower energy is corrected using simulated events with true electron energies from 0.5-3 GeV, and the missing energy correction is applied to the hadronic energy.

\begin{figure}
    \centering
    \begin{minipage}[t]{0.3\textwidth}
        \centering
        \hspace*{-0.5in}
        \includegraphics[width=1.37\textwidth]{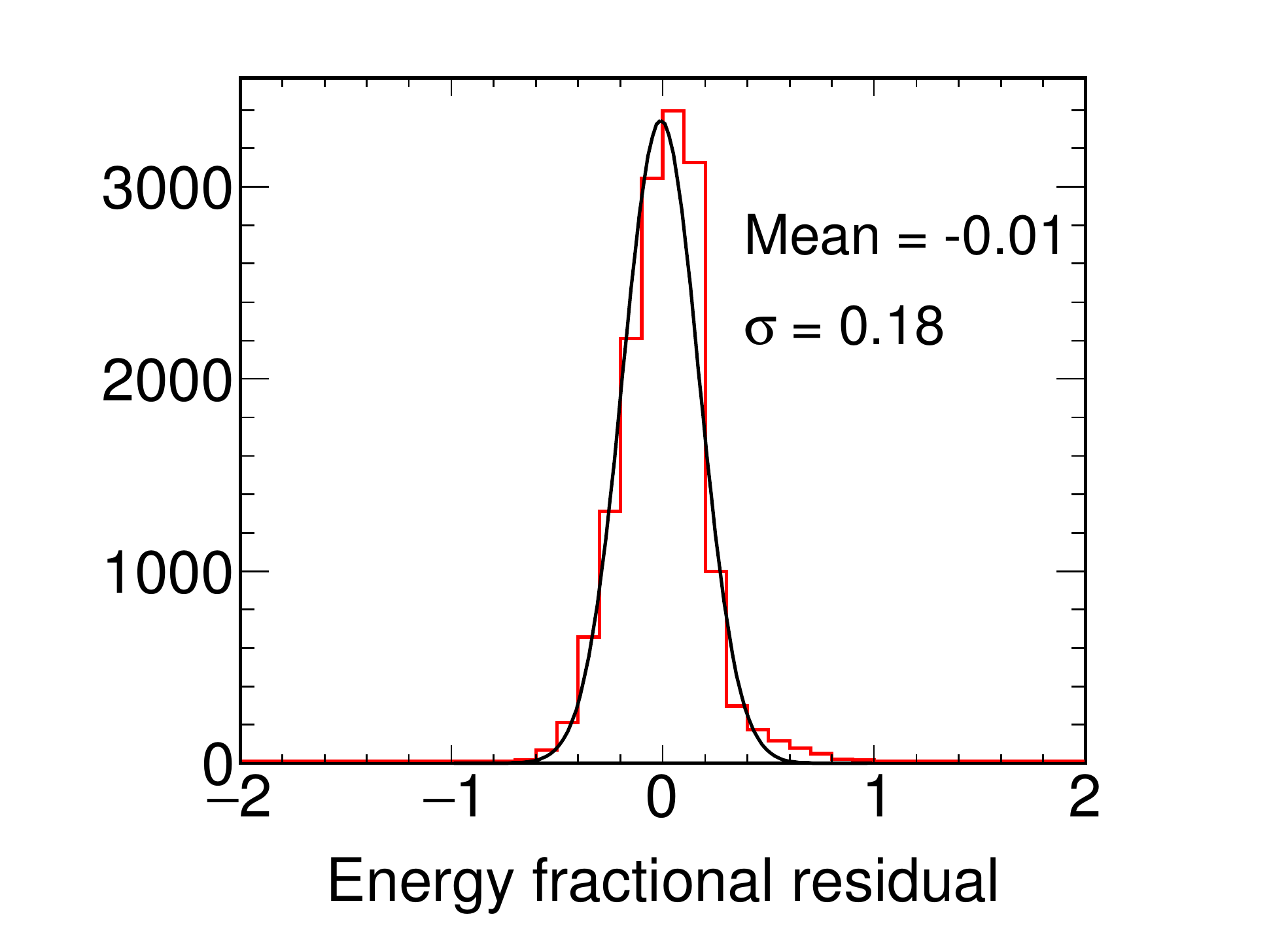}
        \caption[Fractional residuals of reconstructed $\nu_{\mu}$ energy in CC events with contained tracks]{Fractional residuals of reconstructed $\nu_{\mu}$ energy in $\nu_{\mu}$ \dword{cc} events with contained tracks} 
        \label{fig:enresnumucont}
    \end{minipage}\hfill
    \begin{minipage}[t]{0.3\textwidth}
        \centering
        \hspace*{-0.5in}
        \includegraphics[width=1.37\textwidth]{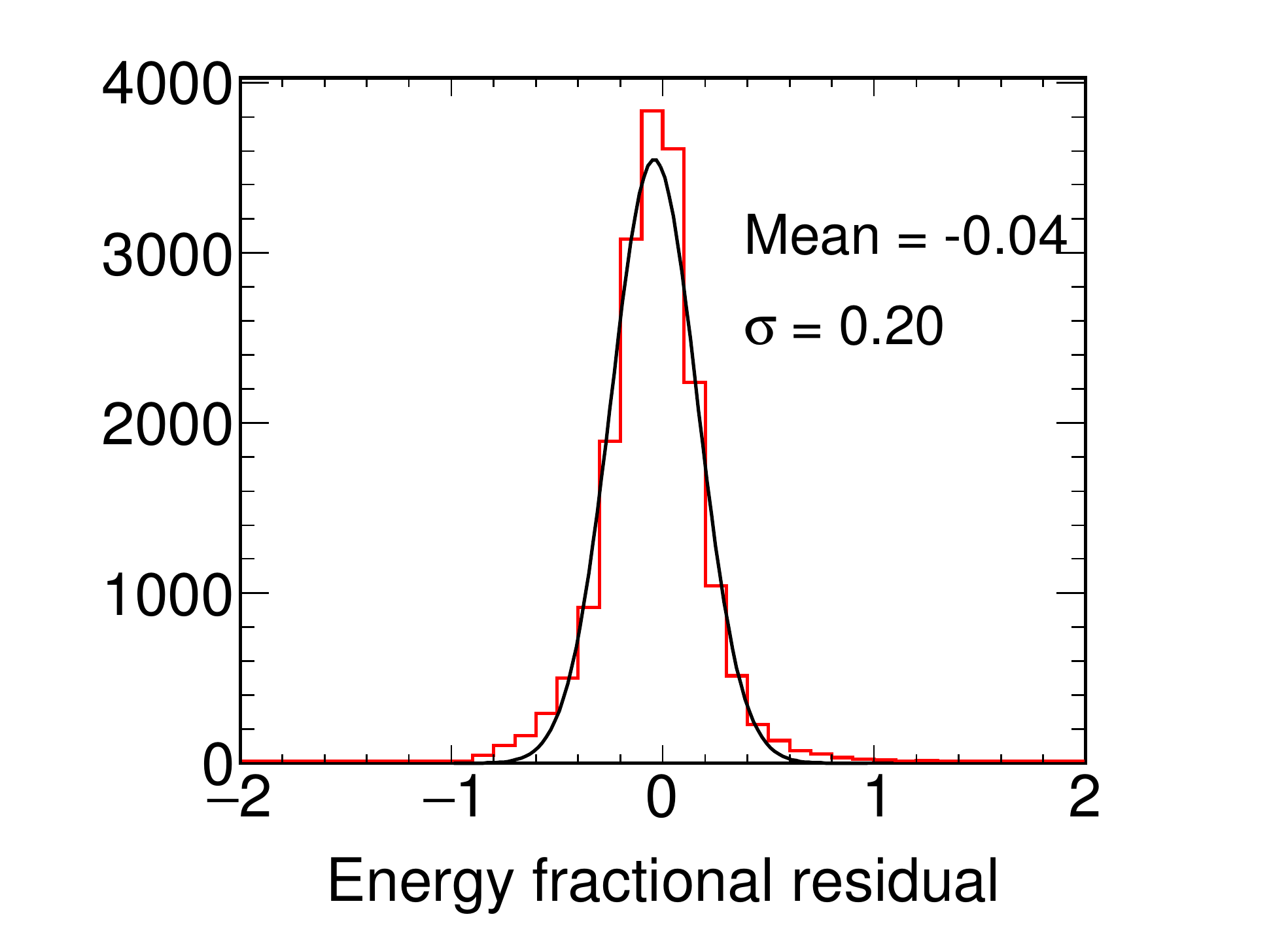}
        \caption[Fractional residuals of reconstructed $\nu_{\mu}$ energy in CC events with exiting tracks]{Fractional residuals of reconstructed $\nu_{\mu}$ energy in $\nu_{\mu}$ \dword{cc} events with exiting tracks}
        \label{fig:enresnumuexit}
    \end{minipage}\hfill
    \begin{minipage}[t]{0.3\textwidth}
        \centering
        \hspace*{-0.5in}
        \includegraphics[width=1.37\textwidth]{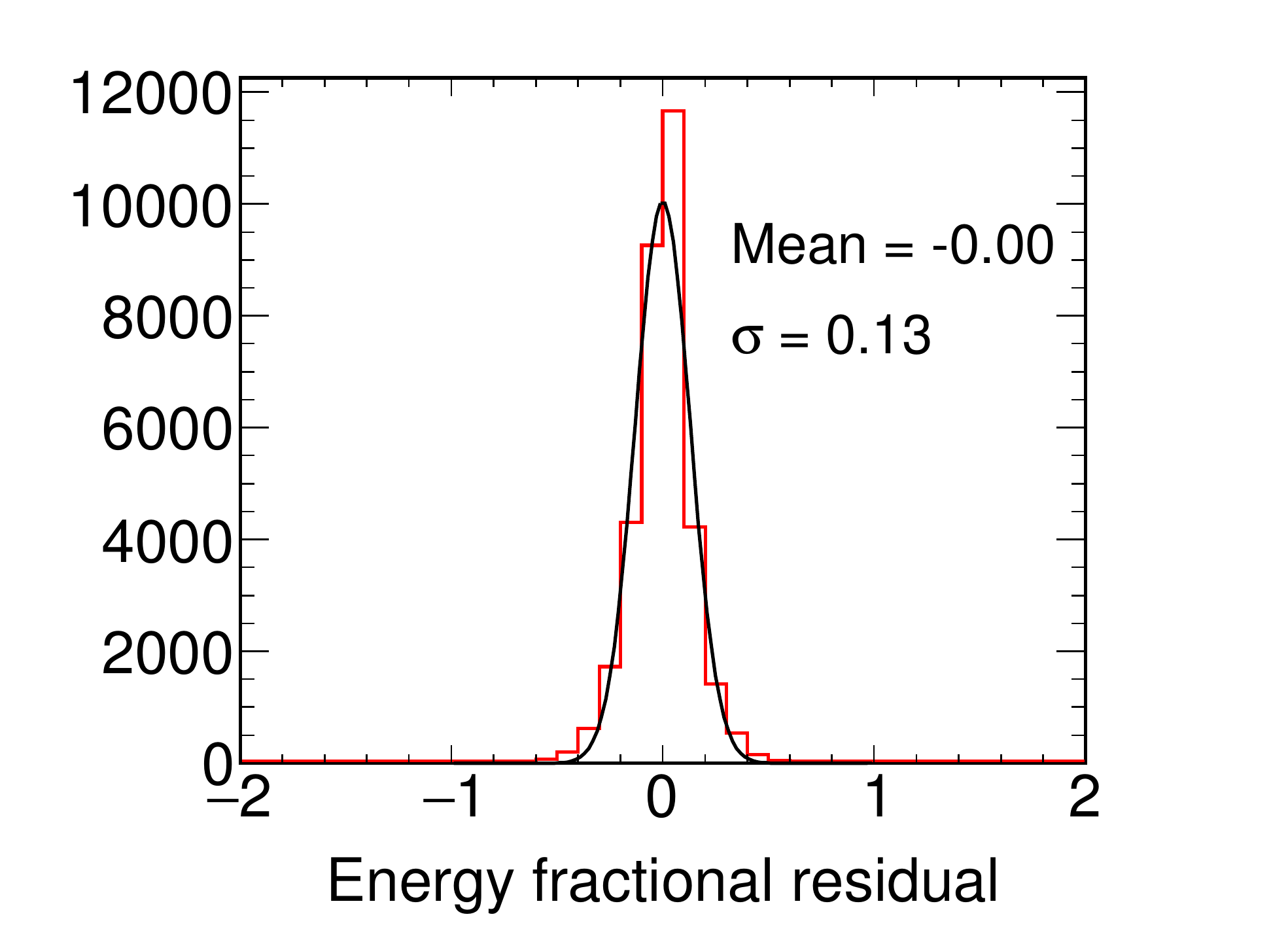}
        \caption[Fractional residuals of reconstructed $\nu_{e}$ energy in CC events]{Fractional residuals of reconstructed $\nu_{e}$ energy in $\nu_{e}$ \dword{cc} events }
        \label{fig:enresnue}
    \end{minipage}
\end{figure}

The fractional residuals of reconstructed neutrino energy are shown for $\nu_{\mu}$ \dword{cc} events with contained tracks in figure \ref{fig:enresnumucont}, for $\nu_{\mu}$ \dword{cc} events with exiting tracks in figure \ref{fig:enresnumuexit} and for $\nu_{e}$ \dword{cc} events in figure \ref{fig:enresnue}. The biases and resolutions of reconstructed neutrino energy are summarized in Table~\ref{tab:ressummary}.

\begin{table}[h!]
\begin{center}
\begin{tabular}{|c|c|c|}
\hline  
 Event selection  &   Bias (\%) & Resolution (\%) \\ \hline
\hline
 $\nu_{\mu}$ \dword{cc} with contained track  &   -1  &  18   \\ \hline
 $\nu_{\mu}$ \dword{cc} with exiting track  &  -4   &  20 \\ \hline
 $\nu_{e}$ \dword{cc}    &  0 & 13    \\ \hline
\end{tabular}
\caption[Summary of biases and resolutions of reconstructed neutrino energy]{Summary of biases and resolutions of reconstructed neutrino energy}
\label{tab:ressummary}
\end{center}
\end{table}


\subsection{Neutrino Event Selection using \dword{cvn}}
The \dword{dune} \dword{cvn} classifies neutrino interactions in the \dword{dune} \dword{fd} through image recognition techniques. In general terms it is a \dword{cnn}. Similar techniques have been demonstrated to outperform traditional methods in many aspects of high energy physics~\cite{Radovic:2018dip}.

The primary goal of the \dword{cvn} is to efficiently and accurately produce event selections of the following interactions: $\nu_{\mu}$ \dword{cc} and $\nu_{e}$ \dword{cc} in the FHC beam mode, and $\bar{\nu}_\mu$ \dword{cc} and $\bar{\nu}_e$ \dword{cc} in the \dword{rhc} beam mode. Future goals will include studies of exclusive neutrino interaction final states since separating the event selections by interaction type can improve the sensitivity as interaction types have different energy resolutions and systematic uncertainties. Detailed descriptions of the \dword{cvn} architecture can be found in~\cite{Aurisano:2016jvx}.

An important feature for the \dword{dune} \dword{cvn} is the fine-grained detail of a \dword{lartpc} encoded in the input images to be propagated further into the \dword{cvn}. This detail is more than what would be possible using a traditional \dword{cnn}, such as the GoogLeNet-inspired network (also called Inception v1)~\cite{GoogLeNet} used by \dword{nova}~\cite{Aurisano:2016jvx}. To accomplish this, the \dword{cvn} design is based on the SE-ResNet architecture, which consists of a standard ResNet (residual neural network) architecture~\cite{He-et-al-2015-deep} along with Squeeze-and-Excitation blocks~\cite{Hu-et-al-2017-squeeze}. Residual neural networks allow the $n^{th}$ layer access to the output of both the $(n-1)^{th}$ layer and the $(n-k)^{th}$ layer via a residual connection, where $k$ is a positive integer ($\ge 2$).

\begin{figure}[htb] 
\centering
	\begin{tabular}{ccc}
		\includegraphics[trim={1cm 4cm 5cm 3cm}, clip, scale=2.0]{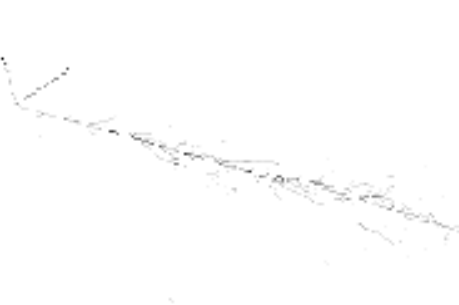}
    \end{tabular}

\caption[A simulated \SI{2.2}{GeV} \nue CC interaction viewed by collection wires in the SP \lartpc]{A simulated 2.2\,GeV $\nu_{e}$ \dword{cc} interaction shown in the collection view of the \dword{dune} \dwords{lartpc}. The horizontal axis shows the wire number of the readout plane and the vertical axis shows time. The greyscale shows the charge of the energy deposits on the wires. The interaction looks similar in the other two views.}
	\label{fig:views}
\end{figure}

In order to build the training input to the \dword{dune} \dword{cvn} three images of the neutrino interactions are produced, one for each of the three readout views, using the reconstructed hits on the individual wire planes. The images are not dependent on any further downstream reconstruction algorithms. The images contain 500 $\times$ 500 pixels, each in the (wire, time) parameter space, where the wire is the wire channel number and the time is the peak time of the reconstructed hit. 
The value of each pixel represents the integrated charge of the reconstructed hit. An example simulated 2.2\,GeV $\nu_{e}$ \dword{cc} interaction is shown in all three views in Figure~\ref{fig:views} demonstrating the fine-grained detail available from the \dword{lartpc} technology.

The \dword{cvn} is trained using approximately three million neutrino interactions from the \dword{mc} simulation. An independent sample is used to generate the physics measurement sensitivities. The training sample is chosen to ensure similar numbers of training examples from the different neutrino flavors. Validation is performed to ensure that similar classification performance is obtained for the training and test samples, i.e., the \dword{cvn} is not overtrained.

For the analysis presented here, we have used the primary output of the \dword{cvn}, namely the neutrino flavor which returns probabilities that each interaction is one of the following classes: $\nu_{\mu}$ \dword{cc}, $\nu_{e}$ \dword{cc}, $\nu_{\tau}$ \dword{cc} and \dword{nc}.

\subsubsection{Neutrino Flavor Identification Efficiency}
\begin{figure}
    \centering
    \begin{tabular}{cc}
		\includegraphics[width=0.45\linewidth]{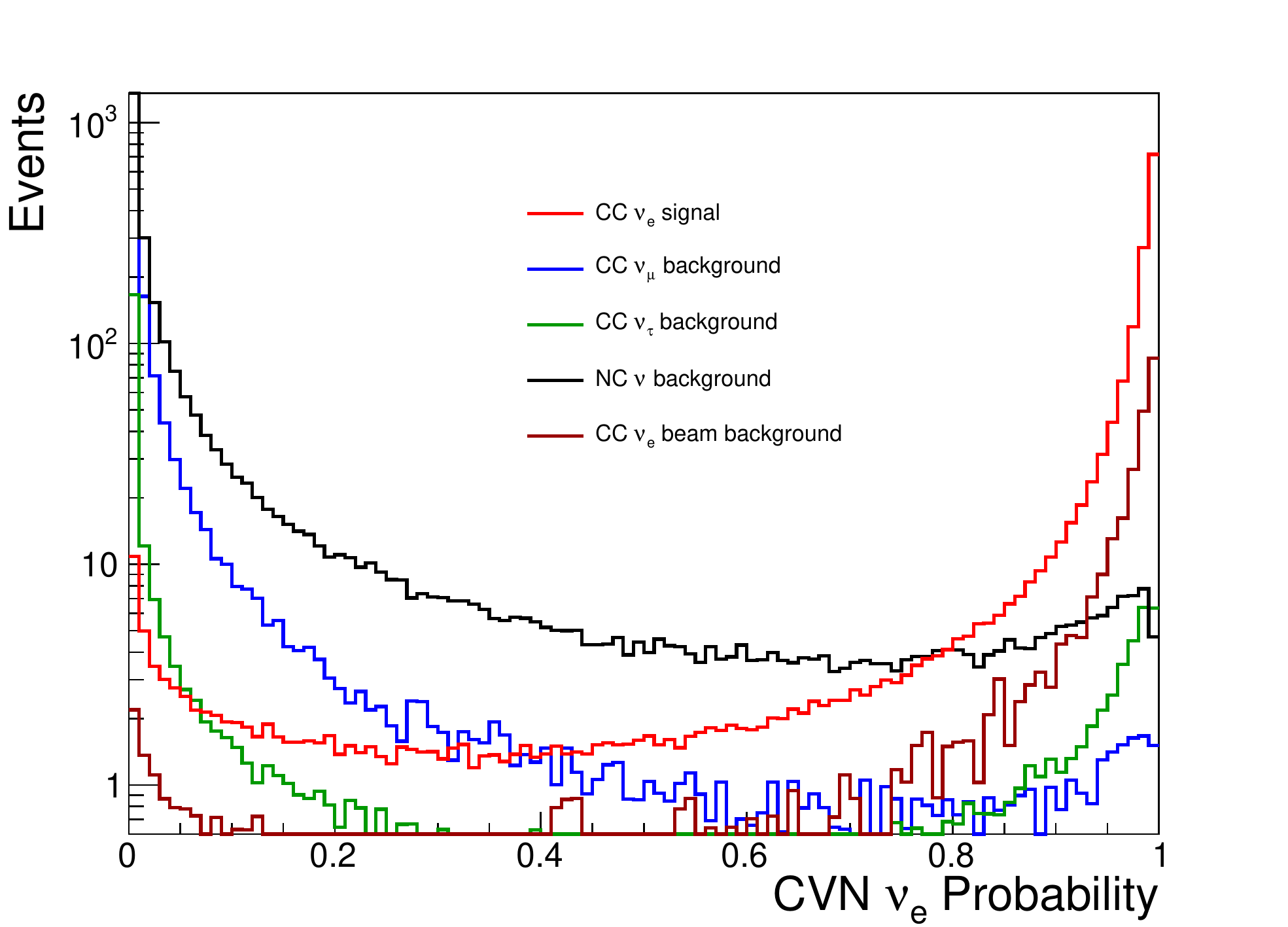} &
		\includegraphics[width=0.45\linewidth]{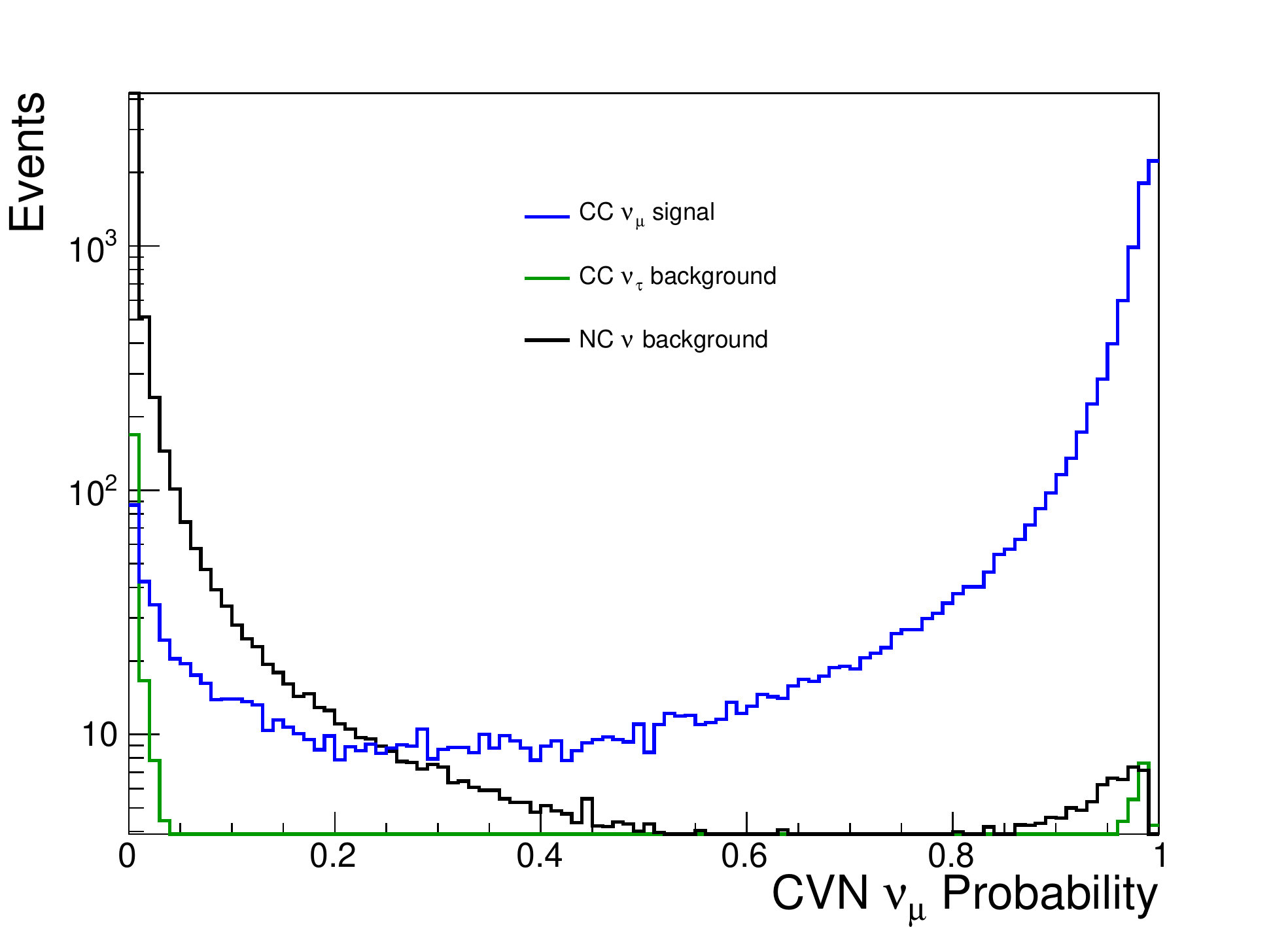} 
	\end{tabular}
	\caption[The CVN \nue CC probability and \numu CC probability for the FHC beam mode]{The \dword{cvn} $\nu_e$ \dword{cc} probability (left) and $\nu_\mu$ \dword{cc} probability (right) for the \dword{fhc} beam mode shown with a log scale.}
    \label{fig:cvnprob}
\end{figure}

The primary goal of the \dword{cvn} algorithm is to accurately identify $\nu_{e}$ \dword{cc} interactions and $\nu_{\mu}$ \dword{cc} interactions to allow for the selection of the samples required for the neutrino oscillation analysis. The $\nu_{e}$ \dword{cc} probability distribution, $P(\nu_e \textrm{ \dword{cc}})$, and the $\nu_\mu$ \dword{cc} probability distribution, $P(\nu_\mu \textrm{ \dword{cc}})$, are shown on the left and right of Figure~\ref{fig:cvnprob}, respectively. Excellent separation between the signal and background interactions is seen in both cases.

The $\nu_e$ \dword{cc} event selection uses events where $P(\nu_e \textrm{ \dword{cc}}) > 0.85$ for an interaction to be considered a candidate event of this type. Similarly, interactions are selected as $\nu_\mu$ \dword{cc} candidates if $P(\nu_\mu \textrm{ \dword{cc}}) > 0.5$. Note that since all of the flavor classification probabilities must sum to one, the interactions selected in the two event selections are completely independent. The same selection criteria are used for both \dword{fhc} and \dword{rhc} beam modes. The values used in the selection criteria were optimized to produce the best $\delta_{CP}$ sensitivity.

\begin{figure}
    \centering
		\includegraphics[width=0.9\linewidth]{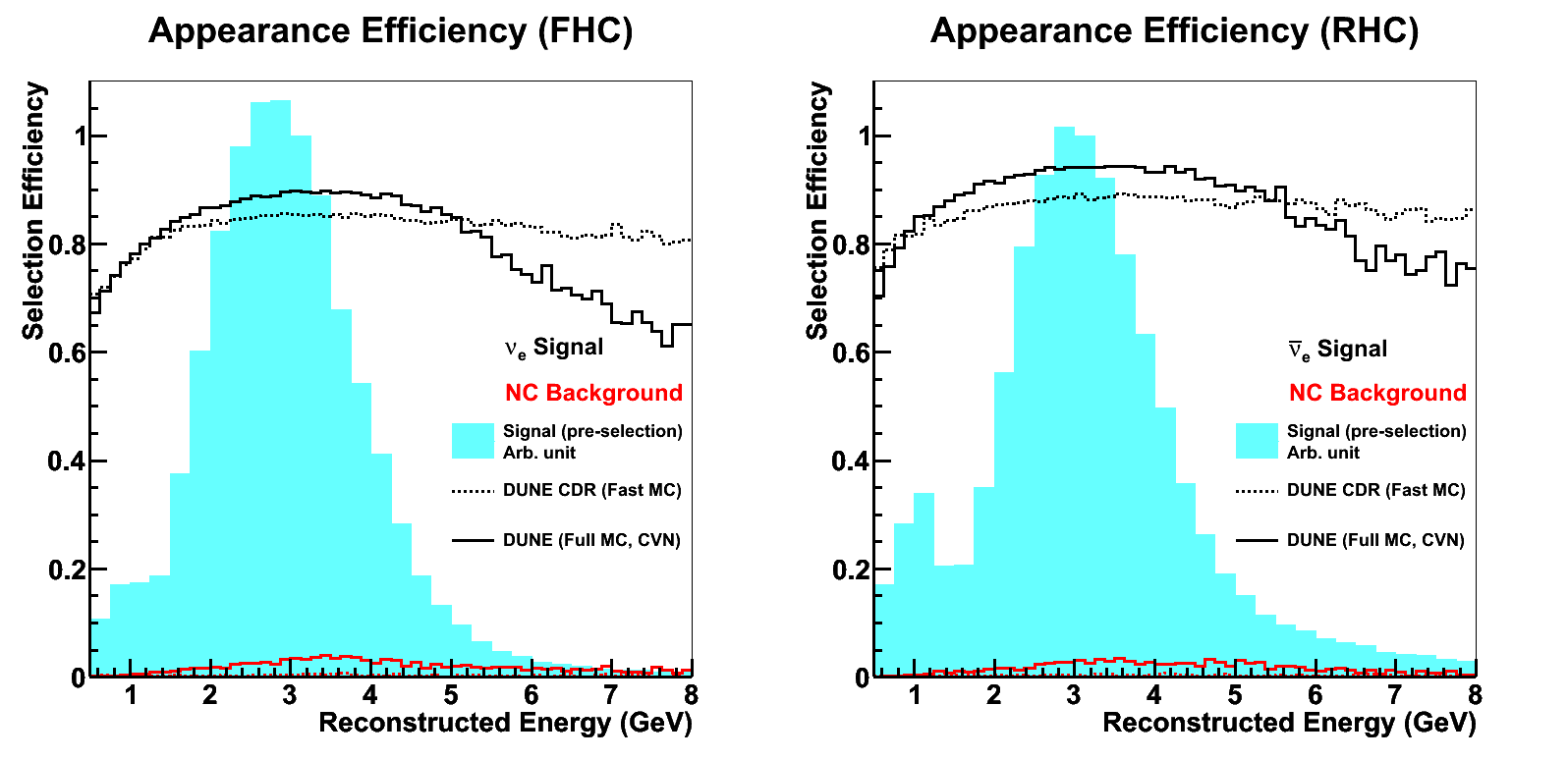} 
	\caption[The $\nu_e$ CC selection efficiency for $P(\nu_e \textrm{CC}) > 0.85$]{The $\nu_e$ \dword{cc} selection efficiency for FHC-mode (left) and RHC-mode (right) simulation with the criterion $P(\nu_e \textrm{ \dword{cc}}) > 0.85$. The solid (dashed) lines show results from the \dword{cvn} (\dword{cdr}) for signal $\nu_e$ \dword{cc} and $\bar{\nu}_e$ \dword{cc} events in black and \dword{nc} background interaction in red. The blue region shows the oscillated flux (A.U.) to illustrate the most important regions of the energy distribution.}
    \label{fig:nueeff}
\end{figure}

\begin{figure}
    \centering
		\includegraphics[width=0.9\linewidth]{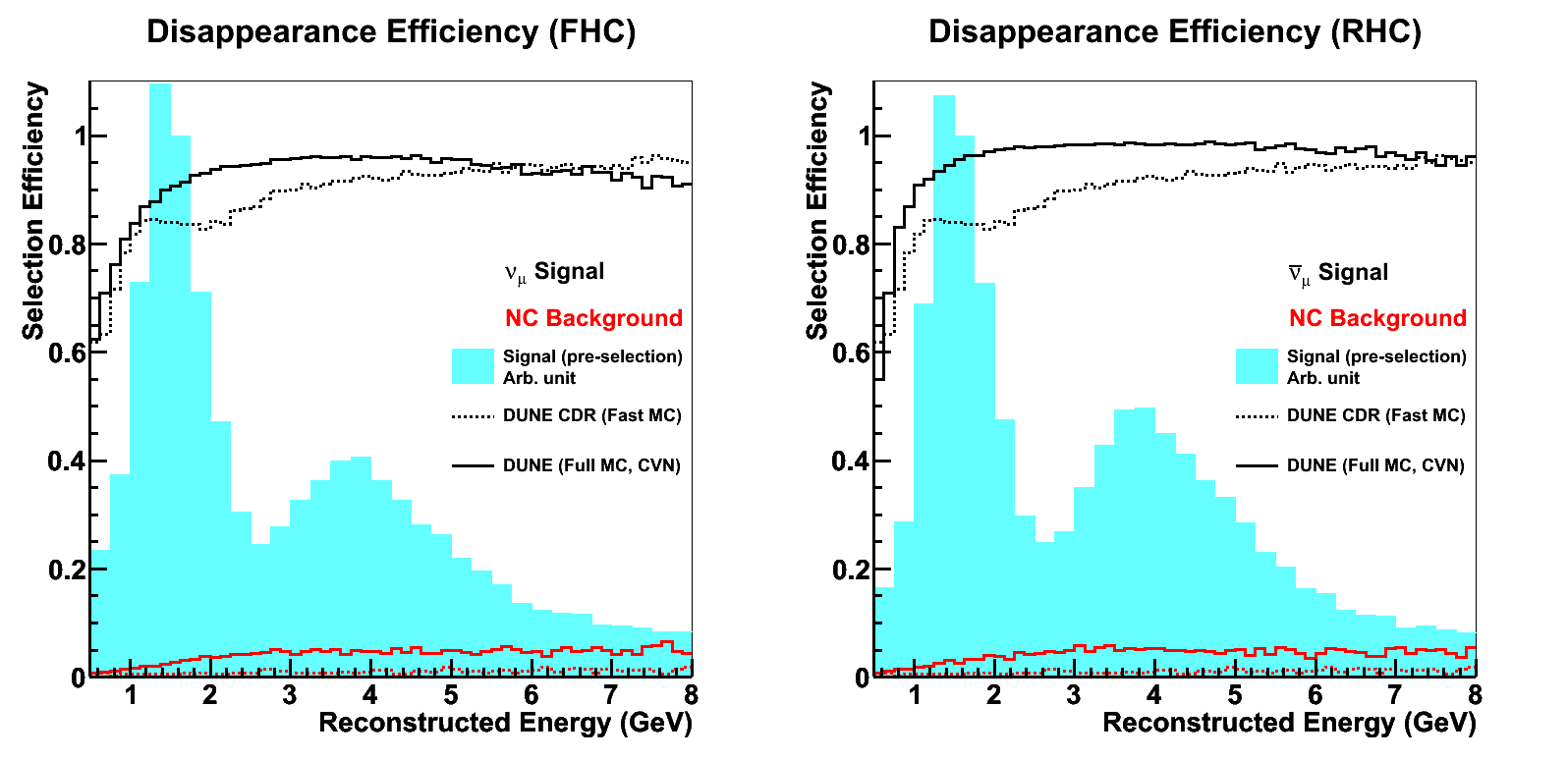} 
	\caption[The $\numu$ CC selection efficiency for $P(\numu \textrm{CC}) > 0.5$]{The $\numu$ \dword{cc} selection efficiency for FHC-mode (left) and RHC-mode (right) simulation with the criterion $P(\numu \textrm{ \dword{cc}}) > 0.5$. The solid (dashed) lines show results from the \dword{cvn} (\dword{cdr}) for signal $\numu$ \dword{cc} and $\bar{\nu}_\mu$ \dword{cc} events in black and \dword{nc} background interaction in red. The blue region shows the oscillated flux (A.U.) to illustrate the most important regions of the energy distribution.}
    \label{fig:numueff}
\end{figure}

Figure~\ref{fig:nueeff} shows the efficiency as a function of reconstructed energy (under the electron neutrino hypothesis) for the $\nu_e$ event selection. The efficiency in both the FHC and RHC beam modes exceeds 90\% in the neutrino flux peak.  
Figure~\ref{fig:numueff} shows the corresponding excellent selection efficiency for the $\nu_\mu$ event selection.

\subsubsection{Neutrino Flavor Identification Robustness}
A common concern on the applications of Deep Learning in high energy physics is the potential for differences in performance between data and simulation. Work is in progress to evaluate the \dword{dune} \dword{cvn} using data from a large \dword{dune} prototype, \dword{pdsp}~\cite{Abi:2017aow}.
While the data-based validation is underway a thorough investigation of the selection efficiency as a function of various event kinematics was carried out. The results of the investigation is that the \dword{cvn} selection does not suffer from model dependence at a level that would undermine the conclusions of the oscillation analysis studies. All efficiency curves are consistent with a few key observations. 

The ability of the \dword{cvn} to identify neutrino flavor is dependent on its ability to resolve and identify the charged lepton. 
Backgrounds are induced by mis-identification of charged pions for $\nu_{\mu}$ disappearance, and photons for $\nu_{e}$ appearance samples. Efficiency for these backgrounds tracks directly with the momentum and isolation of the energy depositions from the pions and photons. Efficiency was also observed to drop as a function of track/shower angle when energy depositions aligned with wire planes. The shapes of the efficiency functions in lepton momentum, lepton angle, and hadronic energy fraction (inelasticity) were all observed to be consistent with results from previous studies, including hand scans of \dword{lartpc} simulations. It is still conceivable that the efficacy is increased, especially at low charged lepton momentum, by the CVN identifying fine details of model dependent event kinematics. However, these effects are small enough to be covered by the assigned uncertainties.



Experience in ensuring robustness of deep learning image recognition techniques already exists within the community; similar techniques will be applied to future DUNE analyses. For example, the \dword{nova} experiment uses a technique that takes clear $\nu_{\mu}$ \dword{cc} interactions identified in data and simulation and removes all of the reconstructed hits associated with the reconstructed muon track. The reconstructed muon is replaced by a simulated electron with the same kinematic variables~\cite{Sachdev:2015hpa,Gandrajula:2018ytr}. This procedure was originally developed by \dword{minos}~\cite{Boehm:2009zz}, and allows a large sample of data-like electron neutrino interactions to be studied and excellent agreement was seen between the performance of the event selection for data and simulation. This approach will prove critical once \dword{dune} begins data taking to ensure the performance of the \dword{cvn} is the same for data and simulation.

\subsection{FD Neutrino Interaction Samples}

A complete neutrino interaction event simulation has been implemented, including realistic neutrino energy reconstruction and event selection algorithms which yields an appropriately accurate representation of the \dword{fd} samples to be used in the long-baseline oscillation analysis. The samples used in the sensitivity studies presented in this document require event by event simulations that effectively produce the convolution of the neutrino flux model, neutrino-argon scattering models, and models of the detector response. This last step must include estimates of energy smearing and bias, as well as the impact of a realistic event selection on signal acceptance and background rejection rates. This section has outlined the methods used to implement these algorithms. The final product is the  selected \dword{fd} event samples shown as a function of reconstructed neutrino energy in Section~\ref{sec:physics-lbnosc-osc}. Figure~\ref{fig:appspectra} shows the \numutonue and $\bar{\nu}_\mu \to \bar{\nu}_e$ appearance spectra and Figure~\ref{fig:disspectra} shows the \numutonumu and  $\bar{\nu}_\mu \to \bar{\nu}_\mu$ disappearance spectra. Tables~\ref{tab:apprates} and \ref{tab:disrates} provide the signal and background event rates for the appearance and disappearance analyses, respectively. Based on these predictions we observe the largest background to the \nue \dword{cc} appearance signal to be the intrinsic beam \nue interactions. There is also a contribution from misidentified neutral current interactions as well as small contributions from misidentified \numu and \nutau interactions. The \numu disappearance signal has negligible background, though there is a significant ``wrong-sign''  \numu component in the $\bar{\numu}$ sample. 

\section{Detector Model and Uncertainties}
\label{sec:physics-lbnosc-syst}\label{sec:nu-osc-09}


Detector effects impact the event selection efficiency as well as the reconstruction of quantities used in the oscillation fit, such as neutrino energy. The main sources of detector systematic uncertainties are limitations of calibration and modeling of particles in the detector. While neutrino interaction uncertainties can also affect reconstruction, this section is focused on effects that arise from the detectors.

The near \dword{lartpc} detector uses a similar technology as the far detector, namely they are both \dwords{lartpc}. However, important differences lead to uncertainties that do not fully correlate between the two detectors. First, the readout technology is different, as the near \dword{lartpc} uses pixels as well as a different, modular photon detector. Therefore, the charge response to particle types (e.g., muons and protons) will be different between near and far due to differences in electronics readout, noise, and local effects like alignment.  Second, the high-intensity environment of the \dword{nd} complicates associating detached energy deposits to events, a problem which does not exist in the \dword{fd}. Third, the calibration programs will be different. For example, the \dword{nd} has a high-statistics calibration sample of through-going, momentum-analyzed muons from neutrino interactions in the upstream rock, which does not exist for the \dword{fd}. 
Finally, the reconstruction efficiency will be inherently different due to the relatively small size of the \dword{nd}. Containment of charged hadrons will be significantly worse at the \dword{nd}, especially for events with energetic hadronic showers or with vertices near the edges of the fiducial volume. Detector systematic uncertainties in the \dword{gartpc} at the near site will be entirely uncorrelated to the \dword{fd}.

\subsection{Energy Scale Uncertainties}
\label{sec:EnergyScaleSysts}

An uncertainty on the overall energy scale is included in the analysis presented here, as well as particle response uncertainties that are separate and uncorrelated between four species: muons, charged hadrons, neutrons, and electromagnetic showers. In the \dword{nd}, muons reconstructed by range in \dword{lar} and by curvature in \dword{mpd} are treated separately. The energy scale and particle response uncertainties are allowed to vary with energy; each term is described by three free parameters:

\begin{equation}
\label{eq:escale_unc}    
E^{\prime}_{rec} = E_{rec} \times (p_{0} + p_{1}\sqrt{E_{rec}} + \frac{p_{2}}{\sqrt{E_{rec}}})
\end{equation}

\noindent
where $E_{rec}$ is the nominal reconstructed energy, $E^{\prime}_{rec}$ is the shifted energy, and $p_{0}$, $p_{1}$, and $p_{2}$ are free fit parameters that are allowed to vary within \textit{a priori} constraints. The energy scale and resolution parameters are conservatively treated as uncorrelated between the \dword{nd} and \dword{fd}. With a better understanding of the relationship between \dword{nd} and \dword{fd} calibration and reconstruction techniques, it may be possible to correlate some portion of the energy response. The full list of energy scale uncertainties is given as Table~\ref{tab:EscaleSysts}. Uncertainties on energy resolutions are also included and are taken to be 2\% for muons, charged hadrons, and EM showers and 40\% for neutrons.

\begin{dunetable}[Energy scale systematics]{c|ccc}{tab:EscaleSysts}
{Uncertainties applied to the energy response of various particles. $p_{0}$, $p_{1}$, and $p_{2}$ correspond to the constant, square root, and inverse square root terms in the energy response parameterization given in Equation~\ref{eq:escale_unc}. All are treated as uncorrelated between the \dword{nd} and \dword{fd}.}
    Particle           & $p_{0}$ & $p_{1}$ & $p_{2}$ \\ \toprowrule
    all (except muons) & 2\%   & 1\%   & 2\%   \\
    $\mu$ (range)      & 2\%   & 2\%   & 2\%   \\
    $\mu$ (curvature)  & 1\%   & 1\%   & 1\%   \\
    p, $\pi^{\pm}$     & 5\%   & 5\%   & 5\%   \\
    e, $\gamma$, $\pi^{0}$ & 2.5\%   & 2.5\%   & 2.5\%   \\
    n                  & 20\%  & 30\%  & 30\%  \\
    \hline
\end{dunetable} 

The scale of these uncertainties is derived from recent experiments, including calorimetric based approaches (\dword{nova}, \dword{minerva}) and \dwords{lartpc} (\dword{lariat}, \dword{microboone}, \dword{argoneut}). On \dword{nova}~\cite{NOvA:2018gge}, the muon (proton) energy scale achieved is $<1$\% (5\%). Uncertainties associated to the pion and proton re-interactions in the detector medium are expected to be controlled from \dword{protodune} and \dword{lariat} data, as well as the combined analysis of low density (gaseous) and high density (\dword{lar}) \dwords{nd}. Uncertainties in the \efield also contribute to the energy scale uncertainty, and calibration is needed (with cosmics at \dword{nd}, laser system at \dword{fd}) to constrain the overall energy scale. The recombination model will continue to be validated by the suite of \dword{lar} experiments and is not expected to be an issue for nominal field provided minimal \efield distortions. Uncertainties in the electronics response are controlled with dedicated charge injection system and validated with intrinsic sources, Michel electrons and \Ar39.

The response of the detector to neutrons is a source of active study and will couple strongly to detector technology. The validation of neutron interactions in \dword{lar} will continue to be characterized by dedicated measurements (e.g., CAPTAIN~\cite{Berns:2013usa,Bhandari:2019rat}) and the \dword{lar} program (e.g., \dword{argoneut}~\cite{Acciarri:2018myr}).  However, the association of the identification of a neutron scatter or capture to the neutron's true energy has not been demonstrated, and significant reconstruction issues exist, so a large uncertainty (20\%) is assigned comparable to the observations made by \dword{minerva}~\cite{Elkins:2019vmy} assuming they are attributed entirely to the detector model. Selection of photon candidates from $\pi^0$ is also a significant reconstruction challenge, but a recent measurement from \dword{microboone} indicates this is possible and the $\pi^0$ invariant mass has an uncertainty of 5\%, although with some bias~\cite{Adams:2018sgn}.

\subsection{Acceptance and Reconstruction Efficiency Uncertainties}

The \dword{nd} and \dword{fd} have different acceptance to \dword{cc} events due to the very different detector sizes. The \dword{fd} is sufficiently large that acceptance is not expected to vary significantly as a function of event kinematics. However, the \dword{nd} selection requires that hadronic showers be well contained in \dword{lar} to ensure a good energy resolution, resulting in a loss of acceptance for events with energetic hadronic showers. The \dword{nd} also has regions of muon phase space with lower acceptance due to tracks exiting the side of the \dword{tpc} but failing to match to the \dword{mpd}.

Uncertainties are evaluated on the muon and hadron acceptance of the \dword{nd}. The detector acceptance for muons and hadrons is shown in Figure~\ref{fig:NDacceptance}. Inefficiency at very low lepton energy is due to events being misreconstructed as neutral current, which can also be seen in Figure~\ref{fig:NDacceptance}. For high energy, forward muons, the inefficiency is only due to events near the edge of the fiducial volume where the muon happens to miss the \dword{mpd}. At high transverse momentum, muons begin to exit the side of the \dword{lar} active volume, except when they happen to go along the 7 m axis. The acceptance is sensitive to the modeling of muons in the detector. An uncertainty is estimated based on the change in the acceptance as a function of muon kinematics. This uncertainty can be constrained with the \dword{mpd} by comparing the muon spectrum in \dword{cc} interactions between the liquid and gaseous argon targets. The acceptance in the \dword{mpd} is expected to be nearly 4$\pi$ due to the excellent tracking and lack of scattering in the detector. Since the target nucleus is the same, and the two detectors are exposed to the same flux, the ratio between the two detectors is dominated by the \dword{lar} acceptance. Given the rate in the \dword{mpd}, the expected constraint is at the level of $\sim$0.5\% in the peak and $\sim$3\% in the tail.

Inefficiency at high hadronic energy is due to the veto on more than 30 MeV deposited in the outer 30 cm collar of the active volume. Rejected events are typically poorly reconstructed due to low containment, and the acceptance is expected to decrease at high hadronic energy. Similar to the muon reconstruction, this acceptance is sensitive to detector modeling, and an uncertainty is evaluated based on the change in the acceptance as a function of true hadronic energy. This is more difficult to constrain with the \dword{mpd} because of the uncertain mapping between true and visible hadronic energy in the \dword{lar}.

\section{Sensitivity Methods}
\label{sec:physics-lbnosc-sens}

Sensitivities to the neutrino mass ordering, CP violation, and $\theta_{23}$ octant, as well as expected resolution for neutrino oscillation parameter measurements, are obtained by simultaneously fitting the \numutonumu, $\bar{\nu}_\mu \rightarrow \bar{\nu}_\mu$, \numutonue, and $\bar{\nu}_\mu \rightarrow \bar{\nu}_e$ far detector spectra along with selected samples from the near detector.  It is assumed that 50\% of the total exposure is in neutrino beam mode and 50\% in antineutrino beam mode.  A 50\%/50\% ratio of neutrino to antineutrino data has been shown to produce a nearly optimal \deltacp and mass ordering sensitivity, and small deviations from this (e.g., 40\%/60\%, 60\%/40\%) produce negligible changes in these sensitivities. 

In the sensitivity calculations, neutrino oscillation parameters governing long-baseline neutrino oscillation are allowed to vary. In all sensitivities presented here (unless otherwise noted) \sinstt{13} is constrained by a Gaussian prior with 1$\sigma$ width as given by the relative uncertainty shown in Table~\ref{tab:oscpar_nufit}, while \sinst{23}, $\Delta m^{2}_{32}$, and \deltacp are allowed to vary freely. The oscillation parameters $\theta_{12}$ and \dm{12} are allowed to vary constrained by the uncertainty in Table~\ref{tab:oscpar_nufit}. The matter density of the earth is allowed to vary constrained by a 2\% uncertainty on its nominal value. Systematic uncertainty constraints from the near detector are included either by explicit inclusion of \dword{nd} samples within the fit or by applying constraints expected from the \dword{nd} data to \dword{fd}-only fits.

The experimental sensitivity is quantified using a test statistic, $\Delta\chi^2$, which is calculated by comparing the predicted spectra for alternate hypotheses.  The details of the sensitivity calculations are described in Section~\ref{sect:methods-dunefits}. 
A ``typical experiment'' is defined as one with the most probable data given a set of input parameters, i.e., in which no statistical fluctuations have been applied. In this case, the predicted spectra and the true spectra are identical; for the example of \dword{cpv}, $\chi^2_{\mdeltacp^{true}}$ is identically zero and the $\Delta\chi^2_{CP}$ value for a typical experiment is given by $\chi^2_{\mdeltacp^{test}}$. The interpretation of $\sqrt{\Delta\chi^2}$ has been discussed in \cite{Qian:2012zn,Blennow:2013oma}; it may be interpreted as approximately equivalent to significance in $\sigma$ for $\Delta\chi^2>1$.

\dword{dune} sensitivity has been studied using several different fitting frameworks. \dword{globes}~\cite{Huber:2004ka,Huber:2007ji} 
-based fits have been used extensively in the past, in particular for sensitivity studies presented in the \dword{dune} \dword{cdr}; details are available in \cite{Acciarri:2015uup,Alion:2016uaj,Bass:2014vta}. \dword{globes} is now used primarily for studies in support of algorithm development and optimization. \dword{valor}\cite{valorweb} has also been used for internal studies.  The sensitivities presented in this document are calculated using the CAFAna analysis framework described below.

\subsection{The \dword{dune} Analysis Framework}
\label{sect:methods-cafana}

To demonstrate the sensitivity reach of \dword{dune}, we have adopted the analysis framework known as \dword{cafana}~\cite{CAFAna}. This framework was developed for the \dword{nova} experiment and has been used for $\nu_\mu$-disappearance, $\nu_e$-appearance, and joint fits, plus sterile neutrino searches and cross-section analyses.  Unless otherwise noted, sensitivity results presented in this document are performed within \dword{cafana}. 

In the sensitivity studies, the compatibility of a particular oscillation hypothesis with the data is evaluated using the likelihood appropriate for Poisson-distributed data \cite{Tanabashi:2018oca}:

\begin{equation}
    \chi^2 = -2\log\mathcal{L} = 2\sum_i^{N_{\rm bins}}\left[ M_i-D_i+D_i\ln\left({D_i\over M_i}\right) \right]
\end{equation}

where $M_i$ is the \dword{mc} 
expectation in bin $i$ and $D_i$ is the observed count. Most often the bins here represent reconstructed neutrino energy, but other observables, such as reconstructed kinematic variables or event classification likelihoods may also be used. Multiple samples with different selections can be fit simultaneously, as can multi-dimensional distributions of reconstructed variables.

Event records representing the reconstructed properties of neutrino interactions and, in the case of  \dword{mc}, the true neutrino properties are processed to fill the required histograms. Oscillated \dword{fd} predictions are created by populating \twod histograms, with the second axis being the true neutrino energy, for each oscillation channel ($\nu_\alpha\to\nu_\beta$). These are then reweighted as a function of the true energy axis according to an exact calculation of the oscillation weight at the bin center 
and summed to yield the total oscillated prediction:

\begin{equation}
    M_i = \sum_\alpha^{e,\mu}\sum_\beta^{e,\mu,\tau}\sum_j P_{\alpha\beta}(E_j)M_{ij}^{\alpha\beta}
    \label{eqn:cafana_ll}
\end{equation}

where $P_{\alpha\beta}(E)$ is the probability for a neutrino created in flavor state $\alpha$ to be found in flavor state $\beta$ at the \dword{fd}. $M_{ij}^{\alpha\beta}$ represents the number of selected events in bin $i$ of the reconstructed variable with true energy $E_j$, taken from a simulation where neutrinos of flavor $\alpha$ from the beam have been replaced by equivalent neutrinos in flavor $\beta$. Oscillation parameters that are not displayed in a given figure are profiled over using {\sc minuit}~\cite{James:1994vla}. That is, their values are set to those that produce the best match with the simulated data at each point in displayed parameter space.

Systematic uncertainties are included to account for the expected uncertainties in the beam flux, neutrino interaction, and detector response models used in the simulation at the time of the analysis. The neutrino interaction systematic uncertainties expand upon the existing GENIE systematic uncertainties to include recently exposed data/MC differences that are not expected to be resolved by the time DUNE starts running. The impact of systematic uncertainties is included by adding additional nuisance parameters into the fit. Each of these parameters can have arbitrary effects on the \dword{mc} prediction, and can affect the various samples and channels within each sample in different ways. These parameters are profiled over in the production of the result. The range of these parameters is controlled by the use of Gaussian penalty terms to reflect our prior knowledge of reasonable variations.

For each systematic parameter under consideration, the matrices $M_{ij}^{\alpha\beta}$ are evaluated for a range of values of the parameter, by default $\pm1,2,3\sigma$. The predicted spectrum at any combination of systematic parameters can then be found by interpolation. Cubic interpolation is used, which guarantees continuous and twice-differentiable results, advantageous for gradient-based fitters such as {\sc minuit}. 

For many systematic variations, a weight can simply be applied to each event record as it is filled into the appropriate histograms. For others, the event record itself is modified, and for a few systematic uncertainties it is necessary to use an entirely separate sample that has been simulated with some alteration made to the simulation parameters. 

\subsection{\dword{dune} Sensitivity Studies}
\label{sect:methods-dunefits}

\dword{dune} sensitivity studies are performed using the \dword{cafana} framework, which works as described in the previous section. Sensitivity calculations for \dword{cpv}, neutrino mass ordering, and octant are performed, in addition to studies of oscillation parameter resolution in one and two dimensions.
The experimental sensitivity and resolution functions are quantified using a test statistic, $\Delta\chi^2$, which is calculated by comparing the predicted spectra for alternate hypotheses. These quantities are defined for neutrino mass ordering, $\theta_{23}$ octant, and \dword{cpv} sensitivity as follows:

\begin{eqnarray}
\Delta\chi^2_{\textrm{ordering}} & = & \chi^2_{\textrm{opposite}} - \chi^2_{\textrm{true}} \label{eq:dx2_MH}\\
\Delta\chi^2_{\textrm{octant}} & = & \chi^2_{\textrm{opposite}} - \chi^2_{\textrm{either}} \\
\Delta\chi^2_{\textrm{CPV}} & = & \textrm{Min}[\Delta\chi^2_{CP}(\mdeltacp^{\textrm{test}}=0),\Delta\chi^2_{CP}(\mdeltacp^{\textrm{test}}=\pi)],
\end{eqnarray}
where $\Delta\chi^2_{CP} = \chi^{2}_{\mdeltacp^{test}} - \chi^2_{\mdeltacp^{true}}$, and $\chi^2$ is defined in Equation~\ref{eqn:cafana_ll}. Where appropriate, a scan is performed over all possible values of $\mdeltacp^{true}$, and the neutrino mass ordering and the $\theta_{23}$ octant are also assumed to be unknown and are free parameters. The lowest value of $\Delta\chi^2$ is obtained by finding the combination of fit parameters that best describe the simulated data. The size of $\Delta\chi^2$ is a measure of how well those data can exclude this alternate hypothesis given the uncertainty in the model.

The expected resolution for oscillation parameters is determined from the spread in best-fit values obtained from an ensemble of data sets that vary both statistically and systematically.  For each data set, the true value of each nuisance parameter is chosen randomly from a distribution determined by the a priori uncertainty on the parameter. For some studies, oscillation parameters are also randomly chosen as described in Table~\ref{table:OA_throw}. Poisson fluctuations are then applied to all analysis bins, based on the mean event count for each bin after the systematic adjustments have been applied.  For each simulated data set in the ensemble, the test statistic is minimized, and the best-fit value of all parameters is determined. When calculating $\Delta \chi^{2}$ values from Equation~\ref{eq:dx2_MH}, both of the individual $\chi^{2}$ values used are calculated with the same data set. The one-sigma resolution is defined as the width of the interval around the true value containing 68\% of simulated data sets.
An alternative method of determining parameter resolutions, namely by identifying the range of parameters satisfying $\Delta\chi^2<1$, is also used for some studies.

\begin{dunetable}
[Oscillation parameter throws]
{ccc}
{table:OA_throw}
{Treatment of the oscillation parameters for the simulated data set studies. The width of the $\theta_{13}$ range is determined from the \dword{nufit} result.}
Parameter & Prior & Range\\ \toprowrule
$\sin^{2}\theta_{23}$ & Uniform & [0.4; 0.6] \\
$|\Delta m^{2}_{32}|$ ($\times 10^{-3}$ eV$^{2}$) & Uniform & |[2.3;2.7]| \\
\deltacp ($\pi$) & Uniform & [-1;1] \\
$\theta_{13}$ & Gaussian & \dword{nufit} \\
\end{dunetable}

The \dword{dune} oscillation sensitivities presented here include four \dword{fd} \dword{cc} samples binned as a function of reconstructed neutrino energy: \numutonumu, \numubartonumubar, \numutonue, and \numubartonuebar. Systematic parameters are constrained by unoscillated \dword{nd} \numu and \anumu \dword{cc} samples selected from the \dword{lar} \dword{tpc} and binned in two dimensions as a function of reconstructed neutrino energy ($E_{\nu}$) and reconstructed Bjorken $y$ (i.e. inelasticity).

For some systematic uncertainties, such as uncertainties on the neutrino flux (Section \ref{sec:nu-osc-05}), the natural treatment leads to a large number of parameters that have strongly-correlated effects on the predicted spectrum. In this case, \dword{pca} is used to create a greatly reduced set of systematic parameters which cover the vast majority of the allowed variation, and remove degenerate parameters. The flux \dword{pca}  is described in Section~\ref{sec:fluxPCA}.

Information from the \dword{nd}, which is used to constrain systematic uncertainties, is included via additional $\chi^2$ contributions (Equation \ref{eqn:cafana_ll}) without oscillations. Specific \dword{nd} samples such as neutrino-electron elastic scattering and off-axis samples may be included separately. 
External constraints, for example from solar neutrino experiments, can be included as an arbitrary term in the $\chi^2$ depending on the oscillation parameters. In practice, a quadratic term, corresponding to a Gaussian likelihood, is used.

\subsubsection{Covariance matrix for \dword{nd} uncertainties}

Far detector energy scale and resolution uncertainties are treated as nuisance parameters in the oscillation fits. These parameters are allowed to vary, and in practice become very weakly constrained in Asimov fits due to the limited statistics of the \dword{fd}. Detector uncertainties in the \dword{nd}, in contrast, are included by adding a covariance matrix to the $\chi^{2}$ calculation.  This choice protects against overconstraining that could occur given the limitations of the parameterized \dword{nd} reconstruction described in Section~\ref{sec:ndsimreco} taken together with the high statistical power at the \dword{nd}.  This covariance matrix is constructed with a many-universes technique. In each universe, all \dword{nd} energy scale, resolution, and acceptance parameters are simultaneously thrown according to their respective uncertainties. The resulting spectra, in the same binning as is used in the oscillation sensitivity analysis, are compared with the nominal prediction to determine the bin-to-bin covariance.

\subsubsection{Implementation of flux uncertainties}
\label{sec:fluxPCA}

Uncertainties on the flux prediction are described by a covariance matrix, where each bin corresponds to an energy range of a particular beam mode, neutrino species, and detector location. The covariance matrix includes all beam focusing uncertainties evaluated by reproducing the simulation many times, each with simultaneous random variations in the underlying hadron production model. Each random model variation is referred to as a universe. The  matrix used is $208 \times 208$ bins, despite having only $\sim$30 input uncertainties (and thus $\sim$30 significant eigenvalues). To evaluate the impact of these uncertainties on the long-baseline oscillation sensitivity, it is possible to include each focusing parameter, and each hadron production universe, as separate nuisance parameters. It is also possible to treat each bin of the prediction as a separate nuisance parameter, and include the covariance matrix in the log-likelihood calculation. However, both of these options are computationally expensive, and would include many nuisance parameters with essentially no impact on any distributions.

Instead, a principal component analysis is used, primarily to improve the computational performance of the analysis by reducing the number of parameters while still capturing the same physical effects. The covariance matrix is diagonalized, and each principal component is treated as an uncorrelated nuisance parameter. The 208 principal components are ordered by the magnitude of their corresponding eigenvalues, and only the first $\sim$30 are large enough that they need to be included. By the 10th principal component, the eigenvalue is 1\% of the 0th eigenvalue. Since the time required to perform a fit scales $\sim$linearly with the number of nuisance parameters, including only 30 principal components reduces the computing time by an order of magnitude.

This is purely a mathematical transformation; the same effects are described by the \dword{pca} as by a full analysis, including correlations between energy bins. As expected, the largest uncertainties correspond to the largest principal components. This can be seen in Figure~\ref{fig:fluxPCA}. The largest principal component matches the hadron production uncertainty on nucleon-nucleus interactions in a phase space region not covered by data (N+A unconstrained). Components 3 and 7 correspond to the data-constrained uncertainty on proton interactions in the target producing pions and kaons, respectively. Components 5 and 11 correspond to two of the largest focusing uncertainties, the density of the target and the horn current, respectively. Other components not shown either do not fit a single uncertain parameter and may represent two or more degenerate systematics or ones that produce anticorrelations in neighboring energy bins.

\begin{dunefigure}[Flux principal components]{fig:fluxPCA}
{Select flux principal components are compared to specific underlying uncertainties from the hadron production and beam focusing models. See text. }
    \includegraphics[width=0.9\textwidth]{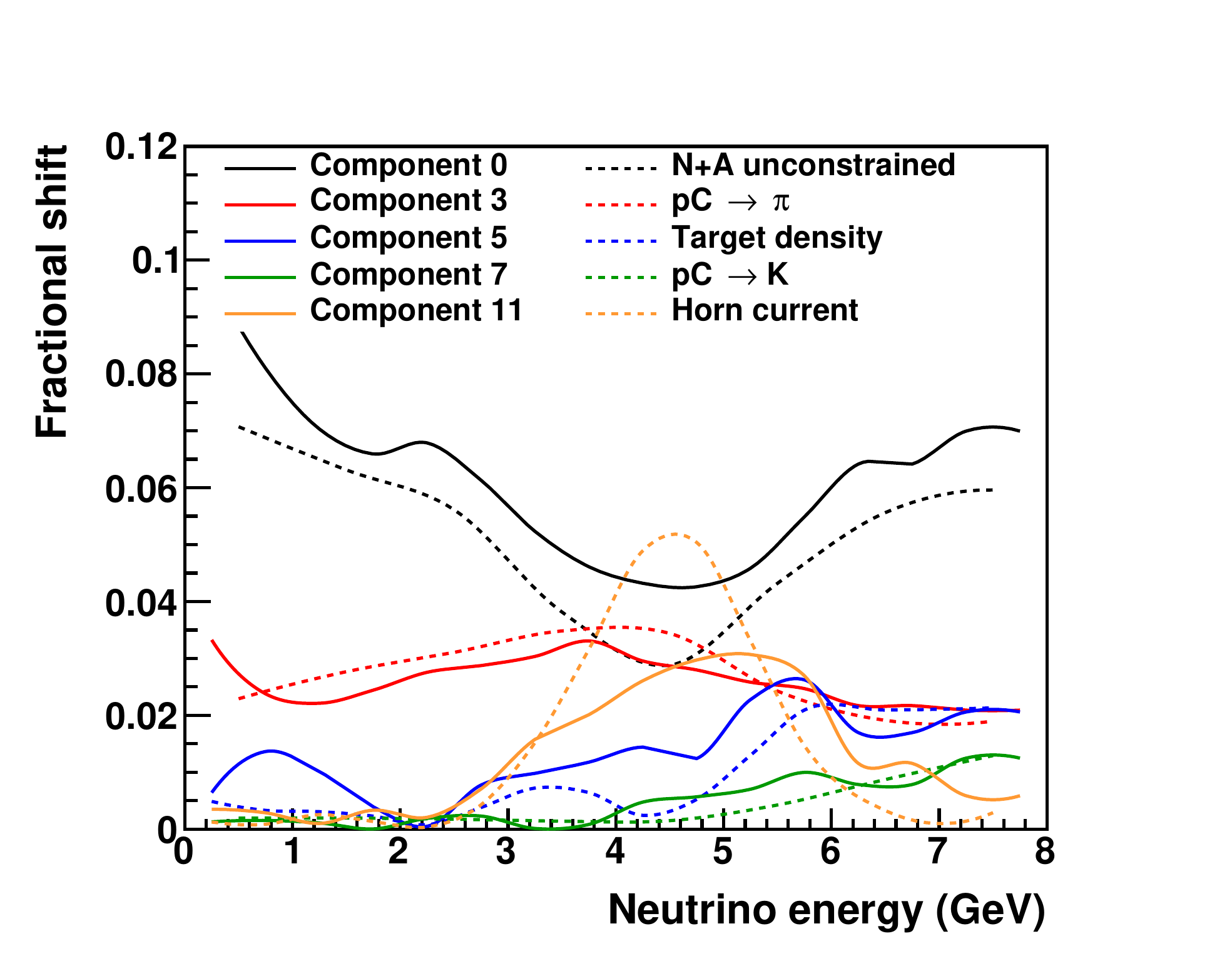}
\end{dunefigure}

\section{Sensitivities}
\label{sec:physics-lbnosc-results}

Using the analysis framework described in the preceding sections, the simulated data samples for the far and near detectors are input to fits for CP violation sensitivity, mass ordering sensitivity, parameter measurement resolutions, and octant sensitivity. The results of these fits are presented in the following sections. Unless otherwise noted, all results include samples from both the near and far detectors and all systematic uncertainties are applied. Nominal exposures of seven, ten, and fifteen years are considered, where the staging plan described in Section~\ref{sec:physics-lbnosc-osc}, including a beam upgrade to 2.4 MW after six years, has been assumed. Results are shown as a function of the true values of oscillation parameters and/or as a function of exposure in staged years and/or kt-MW-years. In all cases, equal running in neutrino and antineutrino mode is assumed; no attempt is made to anticipate a realistic schedule of switching between neutrino and antineutrino mode. For the sake of simplicity, only true normal ordering is shown.

Possible variations of sensitivity are presented in several ways. For results at the nominal exposures, the sensitivity is calculated by performing fits in which the systematic parameters, oscillation parameters, and event rates are chosen at random, constrained in some cases by pre-fit uncertainties, as described in Section~\ref{sect:methods-dunefits}. A fit is performed for each of these simulated data sets or ``throws;'' the nominal result is the median of these fit results and the uncertainty band is calculated to be the interval containing 68\% of the fit results. For these results, the uncertainty band is drawn as as transparent filled area. In other cases, ranges of possible sensitivity results are explored by considering different true values of oscillation parameters or different analysis assumptions, such as removal of external constraints or variation in systematic uncertainties assumptions. For these results, a solid band indicates the range of possible results; this band is not intended to be interpreted as an uncertainty.

The exposures required to reach selected sensitivity milestones for the nominal analysis are summarized in Table~\ref{tab:milestones}. CP violation sensitivity is discussed in Section~\ref{sec:physics-lbnosc-cpv}, neutrino mass ordering sensitivity is discussed in Section~\ref{sec:physics-lbnosc-mh}, and precision measurements of oscillation parameters are discussed in Section~\ref{sec:physics-lbnosc-prec}. The impact of the true values of oscillation parameters, systematic uncertainties, and near detector measurements are explored in Sections~\ref{sec:physics-lbnosc-oscvar}, \ref{sec:physics-lbnosc-systresults}, and \ref{sec:ndimpact}, respectively.

\begin{table}[]
    \centering
    \begin{tabular}{lcc}
 Physics Milestone & Exposure (staged years, \sinst{23} = 0.580) \\
\toprowrule
 5$\sigma$ Mass Ordering & 1 \\
 \deltacp = -$\pi/2$ & \\ \hline
 5$\sigma$ Mass Ordering & 2 \\
 100\% of \deltacp values & \\ \hline
 3$\sigma$ CP Violation & 3 \\
 \deltacp = -$\pi/2$ & \\ \hline
 3$\sigma$ CP Violation & 5 \\
 50\% of \deltacp values & \\ \hline
 5$\sigma$ CP Violation & 7 \\
 \deltacp = -$\pi/2$ & \\ \hline
 5$\sigma$ CP Violation & 10 \\
 50\% of \deltacp values & \\ \hline
 3$\sigma$ CP Violation & 13 \\
 75\% of \deltacp values & \\ \hline
 \deltacp Resolution of 10 degrees & 8 \\
 \deltacp = 0 & \\ \hline
 \deltacp Resolution of 20 degrees & 12 \\
 \deltacp = -$\pi/2$ & \\ \hline
 \sinstt{13} Resolution of 0.004 & 15 \\ \hline
    \end{tabular}
    \caption[Projected DUNE oscillation physics milestones]{Exposure in years, assuming true normal ordering and equal running in neutrino and antineutrino mode, required to reach selected physics milestones in the nominal analysis, using the \dword{nufit} best-fit values for the oscillation parameters. As discussed in Section~\ref{sec:physics-lbnosc-oscvar}, there are significant variations in sensitivity with the value of \sinst{23}, so the exact values quoted here are strongly dependent on that choice. The staging scenario described in Section~\ref{sec:physics-lbnosc-osc} is assumed. Exposures are rounded to the nearest year.
For reference, 30, 100, 200, 336, 624, and \SI{1104}{\ktMWyr} correspond to 1.2, 3.1, 5.2, 7, 10, and 15 staged years, respectively.
}
    \label{tab:milestones}
\end{table}

\subsection{CP-Symmetry Violation}
\label{sec:physics-lbnosc-cpv}

\begin{figure}[h!]
    \centering
		\includegraphics[width=0.95\linewidth]{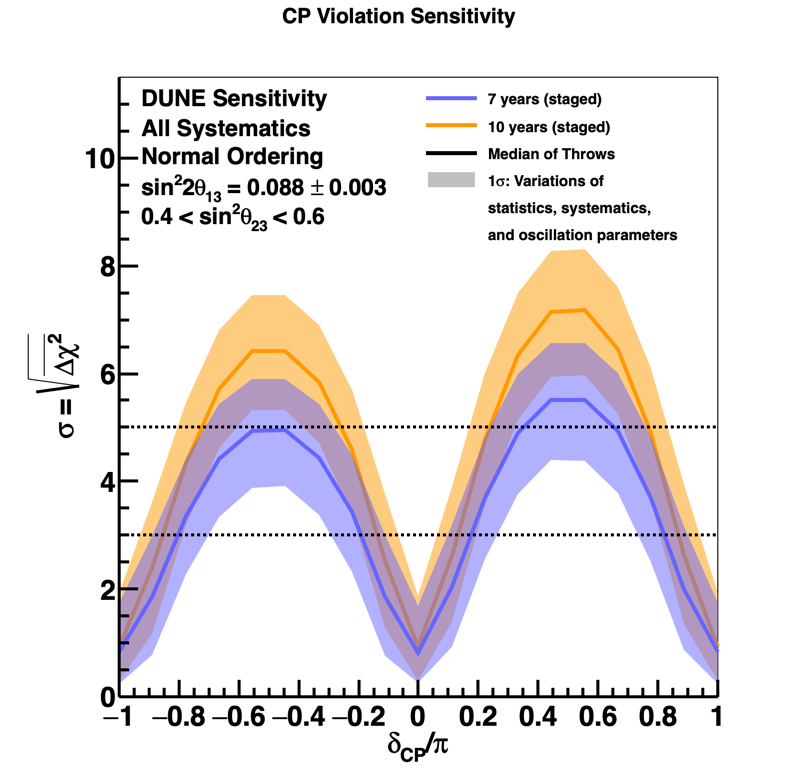}
	\caption[Significance of the DUNE determination of CP-violation as a function of \deltacp]{Significance of the DUNE determination of CP-violation (i.e.: \deltacp $\neq 0$ or $\pi$) as a function of the true value of \deltacp, for seven (blue) and ten (orange) years of exposure. True normal ordering is assumed. The width of the transparent bands cover 68\% of fits in which random throws are used to simulate statistical variations and select true values of the oscillation and systematic uncertainty parameters, constrained by pre-fit uncertainties. The solid lines show the median sensitivity.}
    \label{fig:cpv_nominal}
\end{figure}

Figure~\ref{fig:cpv_nominal} shows the significance with which CP
violation ($\mdeltacp \neq 0 \ {\rm or} \ \pi$) can be observed as a
function of the true value of \deltacp for exposures corresponding to seven and ten years of data, with equal running in neutrino and antineutrino mode, using the staging scenario described in Section~\ref{sec:physics-lbnosc-osc}.
This sensitivity has a characteristic double peak
structure because the significance of a \dword{cpv} measurement
necessarily drops to zero where there is no \dword{cpv}: at the
CP-conserving values of $-\pi,~0,~{\rm and}~\pi$. The width of the transparent band represents 68\% of fits when random throws are used to simulate statistical variations and select true values of the oscillation and systematic uncertainty parameters, constrained by pre-fit uncertainties. The solid curve is the median sensitivity. As illustrated in Section~\ref{sec:physics-lbnosc-oscvar}, variation in the true value of \sinst{23} is responsible for a significant portion of this variation.

\begin{figure}[h!]
    \centering
		\includegraphics[width=0.95\linewidth]{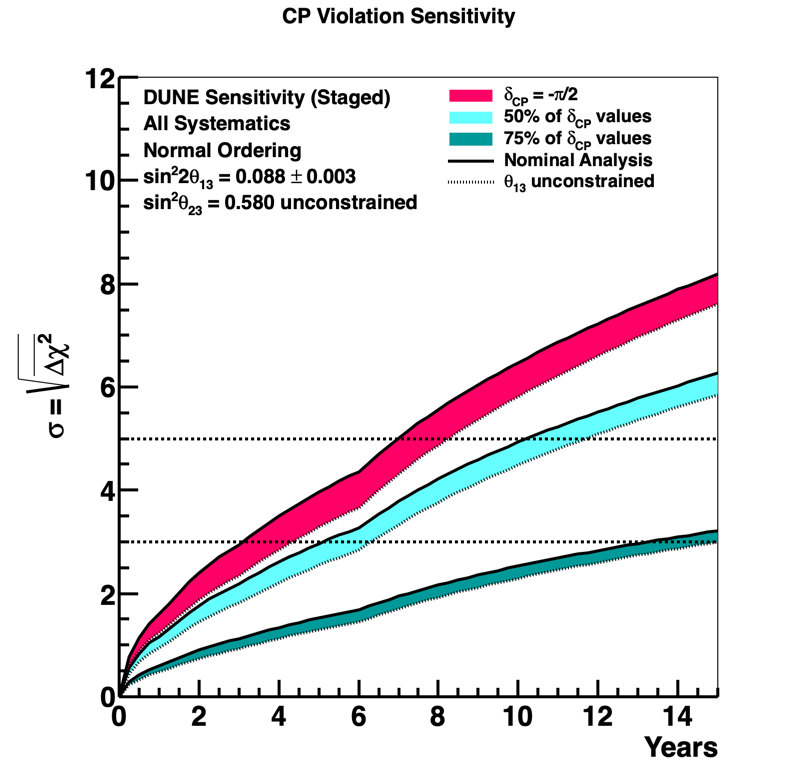}
	\caption[Significance of the DUNE determination of CP-violation as a function of time]{Significance of the DUNE determination of CP-violation (i.e.: \deltacp $\neq 0$ or $\pi$) for the case when \deltacp=$-\pi/2$, and for 50\% and 75\% of possible true \deltacp values, as a function of time in calendar years. True normal ordering is assumed. The width of the band shows the impact of applying an external constraint on \sinstt{13}.}
    \label{fig:cpv_staging}
\end{figure}

\begin{figure}[h!]
    \centering
    \begin{tabular}{cc}
		\includegraphics[width=0.475\linewidth]{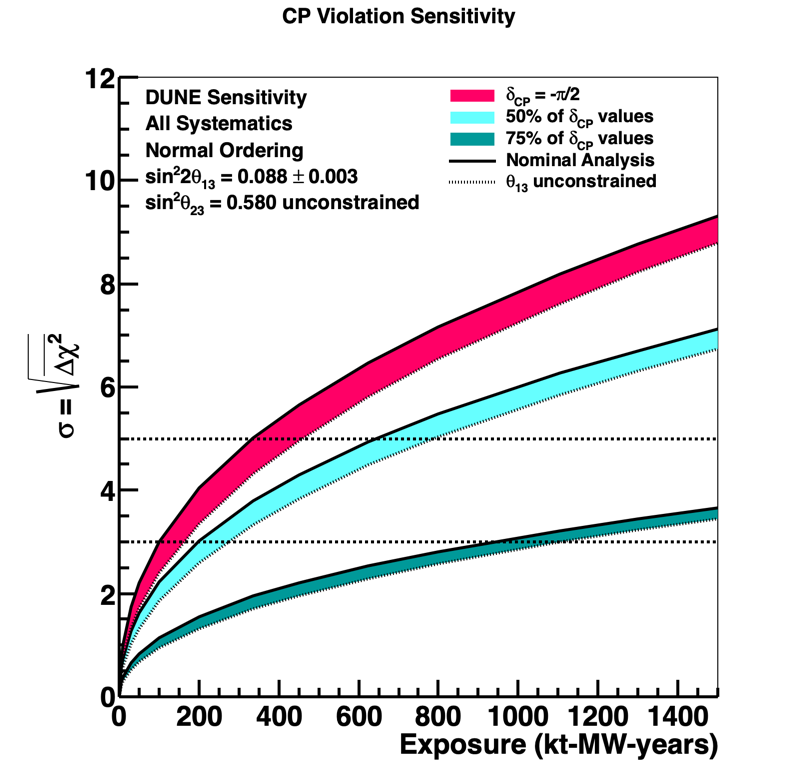} &
		\includegraphics[width=0.475\linewidth]{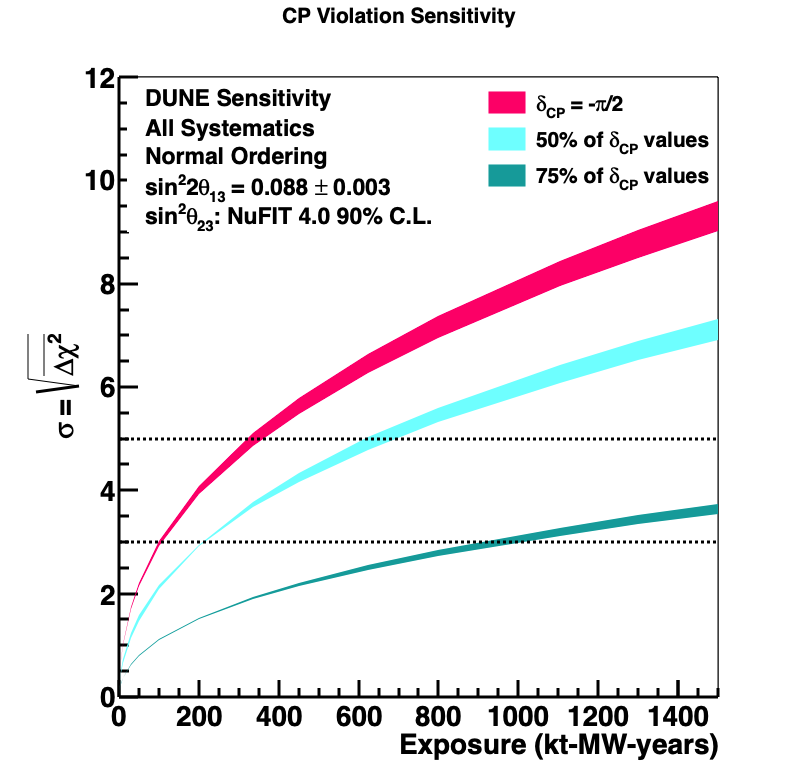}
    \end{tabular}
	\caption[Significance of the DUNE determination of CP-violation as a function of exposure]{Significance of the DUNE determination of CP-violation (i.e.: \deltacp $\neq 0$ or $\pi$) for the case when \deltacp=$-\pi/2$, and for 50\% and 75\% of possible true \deltacp values, as a function of exposure in kt-MW-years. True normal ordering is assumed. Left: The width of the band shows the impact of applying an external constraint on \sinstt{13}. Right: The width of the band shows the impact of varying the true value of \sinst{23} within the \dword{nufit} 90\% C.L. region.
For reference, 30, 100, 200, 336, 624, and \SI{1104}{\ktMWyr} correspond to 1.2, 3.1, 5.2, 7, 10, and 15 staged years, respectively.
}
    \label{fig:cpv_exposure}
\end{figure}

Figure~\ref{fig:cpv_staging} shows the significance
with which CP violation can be determined for 75\% and 50\% of \deltacp values, and when $\deltacp=-\pi/2$, as a function of exposure in years, using the staging scenario described in Section~\ref{sec:physics-lbnosc-osc}. It is not possible for any experiment to provide 100\% coverage in \deltacp for a \dword{cpv} measurement because \dword{cpv} effects vanish at certain values of \deltacp. The changes in trajectory of the curves in the first three years results from the staging of far detector module installation; the change at 6 years is due to the upgrade from 1.2- to 2.4-MW beam power. The width of the bands show the impact of applying an external constraint on \sinstt{13}. As seen in Table~\ref{tab:milestones}, CP violation can be observed with 5$\sigma$ significance after about 7 years if \deltacp = $-\pi/2$ and after about 10 years for 50\% of \deltacp values. CP violation can be observed with 3$\sigma$ significance for 75\% of \deltacp values after about 13 years of running. Figure~\ref{fig:cpv_exposure} shows the same CP violation sensitivity as a function of exposure in kt-MW-years. In the left plot, the width of the bands shows the impact of applying an external constraint on \sinstt{13}, while in the right plot, the width of the bands is the result of varying the true value of \sinst{23} within the \dword{nufit} 90\% C.L. allowed region.

\subsection{Mass Hierarchy}
\label{sec:physics-lbnosc-mh}

\begin{figure}[h!]
    \centering
		\includegraphics[width=0.95\linewidth]{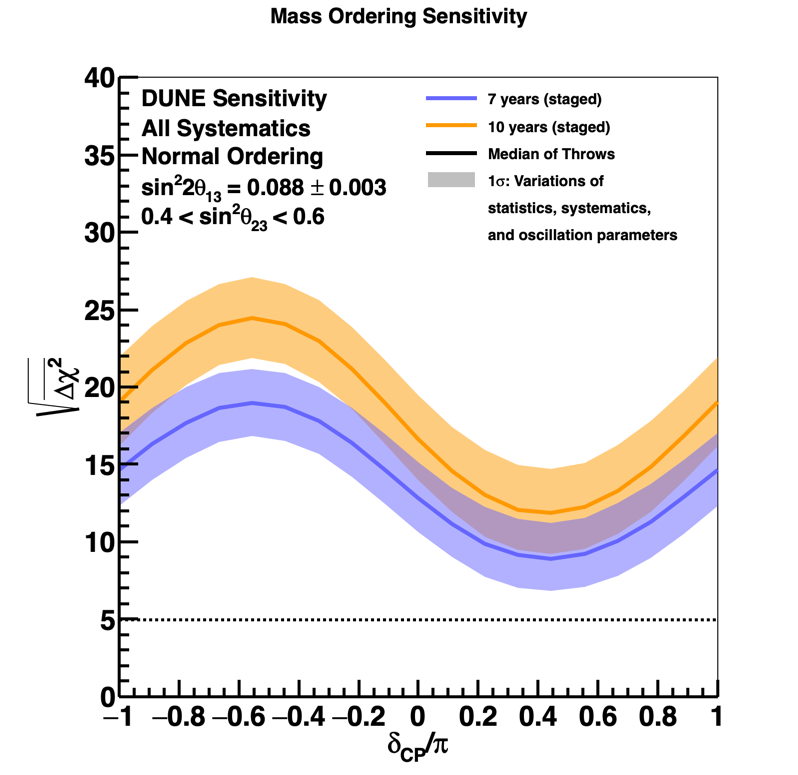}
	\caption[Significance of the DUNE neutrino mass ordering determination, as a function of \deltacp]{Significance of the DUNE determination of the neutrino mass ordering, as a function of the true value of \deltacp, for seven (blue) and ten (orange) years of exposure. True normal ordering is assumed. The width of the transparent bands cover 68\% of fits in which random throws are used to simulate statistical variations and select true values of the oscillation and systematic uncertainty parameters, constrained by pre-fit uncertainties. The solid lines show the median sensitivity. }
    \label{fig:mh_nominal}
\end{figure}

Figure~\ref{fig:mh_nominal} shows the significance with which the neutrino mass ordering can be determined as a function of the true value of \deltacp, using the same exposures and staging assumptions described in the previous section. The characteristic shape results from near degeneracy between matter and CP-violating effects that occurs near $\deltacp=\pi/2$ for true normal ordering.
As in the CP violation sensitivity, the solid curve represents the median sensitivity, the width of the transparent band represents 68\% of fits when random throws are used to simulate statistical variations and select true values of the oscillation and systematic uncertainty parameters, constrained by pre-fit uncertainties, and variation in the true value of \sinst{23} is responsible for a significant portion of this variation.

\begin{figure}[h!]
    \centering
	\includegraphics[width=0.95\linewidth]{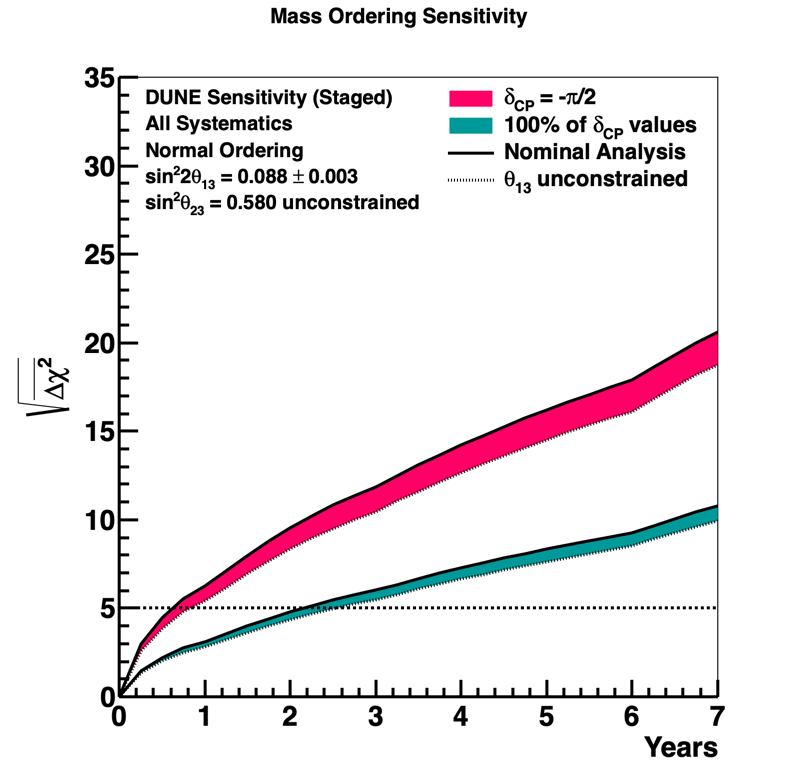}	
	\caption[Significance of the DUNE neutrino mass ordering determination, as a function of time]{Significance of the DUNE determination of the neutrino mass ordering for the case when \deltacp=$-\pi/2$, and for 100\% of possible true \deltacp values, as a function of time in calendar years. True normal ordering is assumed. The width of the band shows the impact of applying an external constraint on \sinstt{13}.}
    \label{fig:mh_staging}
\end{figure}

\begin{figure}[h!]
    \centering
    \begin{tabular}{cc}
    \includegraphics[width=0.475\linewidth]{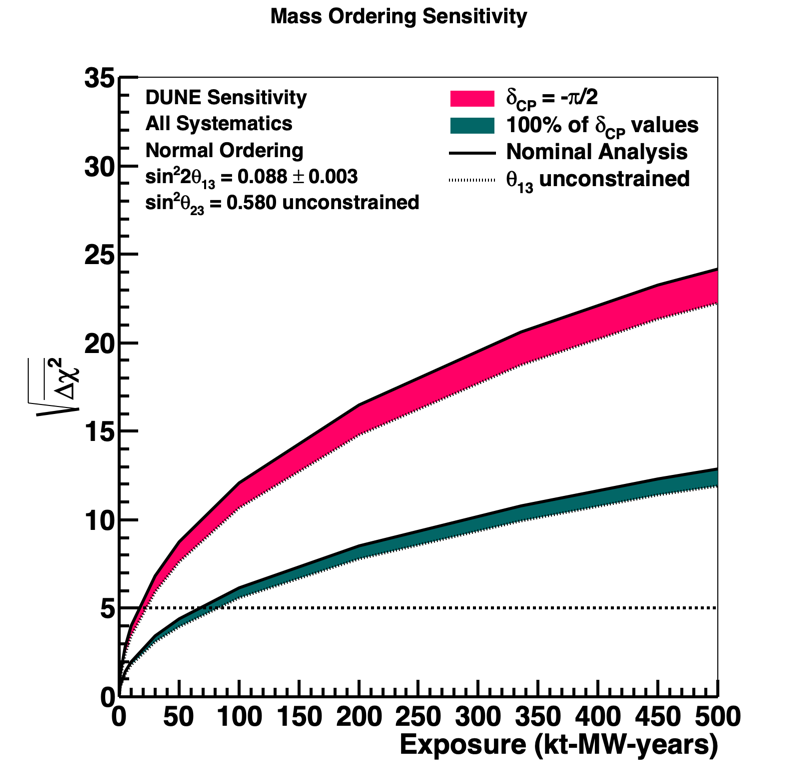} &
	\includegraphics[width=0.475\linewidth]{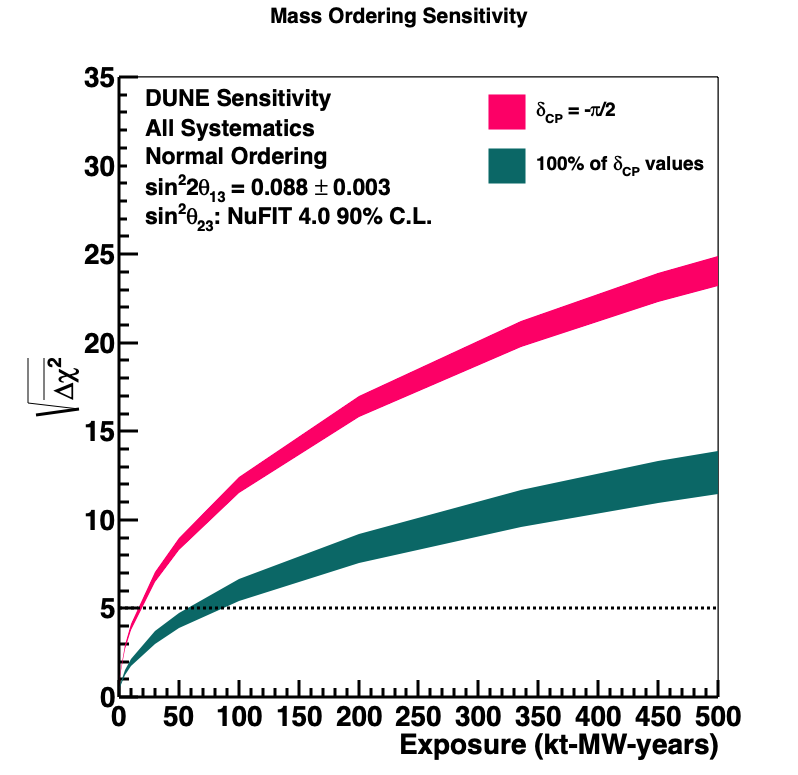} 
	\end{tabular}
	\caption[Significance of the DUNE neutrino mass ordering determination as a function of exposure]{Significance of the DUNE determination of the neutrino mass ordering for the case when \deltacp=$-\pi/2$, and for 100\% of possible true \deltacp values, as a function of exposure in kt-MW-years. True normal ordering is assumed. Left: The width of the band shows the impact of applying an external constraint on \sinstt{13}. Right: The width of the band shows the impact of varying the true value of \sinst{23} within the \dword{nufit} 90\% C.L. region.
For reference, 30, 100, 200, and \SI{336}{\ktMWyr} correspond to 1.2, 3.1, 5.2, and 7  staged years, respectively.
}
    \label{fig:mh_exposure}
\end{figure}

Figure~\ref{fig:mh_staging} shows the significance
with which the neutrino mass ordering can be determined for 100\% of \deltacp values, and when $\deltacp=-\pi/2$, as a function of exposure in years. The width of the bands show the impact of applying an external constraint on \sinstt{13}. Figure~\ref{fig:mh_exposure} shows the same sensitivity as a function of exposure in kt-MW-years. As DUNE will be able to establish the neutrino mass ordering at the 5-$\sigma$ level for 100\% of \deltacp values after between two and three years, these plots extend only to seven years and 500 kt-MW-years, respectively.

\begin{figure}[h!]
    \centering
		\includegraphics[width=0.95\linewidth]{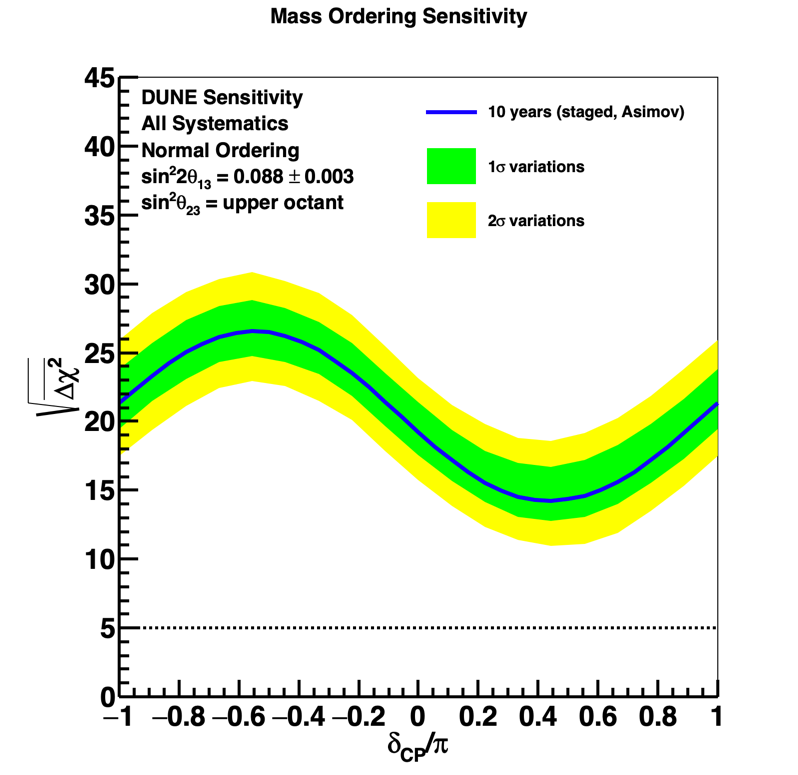}
	\caption[Significance of the DUNE determination of the neutrino mass ordering: statistical and systematic variations]{Significance of the DUNE determination of the neutrino mass ordering, as a function of the true value of \deltacp, for ten years of exposure. True normal ordering is assumed. The width of the bands are 1- and 2-$\sigma$ statistical and systematic variations. The blue curve shows sensitivity for the Asimov set.}
    \label{fig:mh_stats}
\end{figure}

Studies have indicated that special attention must be paid to the statistical interpretation of neutrino mass ordering sensitivities~\cite{Qian:2012zn,Blennow:2013oma} because the $\Delta\chi^2$ metric does not follow the expected chi-squared function for one degree of freedom, so the interpretation of the sensitivity given by the Asimov data set is less straightforward. The error band on the mass ordering sensitivity shown in Figure~\ref{fig:mh_nominal} includes this effect using the technique of statistical throws described in Section~\ref{sect:methods-dunefits}. The effect of statistical fluctuation and systematic uncertainties in the neutrino mass ordering sensitivity for values of \sinst{23} in the range 0.56 to 0.60 is explored using random throws to determine the 1- and 2-$\sigma$ ranges of possible sensitivity. The resulting range of sensitivities is shown in Figure~\ref{fig:mh_stats}, for 10 years of exposure.

\subsection{Precision Oscillation Parameter Measurements}
\label{sec:physics-lbnosc-prec}

In addition to the discovery potential for neutrino mass hierarchy and \dword{cpv}, 
\dword{dune} will improve the precision on key parameters that govern neutrino oscillations, including: \deltacp, $\sin^22\theta_{13}$, \dm{31}, $\sin^2\theta_{23}$ and the octant of $\theta_{23}$. 

\begin{figure}[h!]
    \centering
		\includegraphics[width=0.95\linewidth]{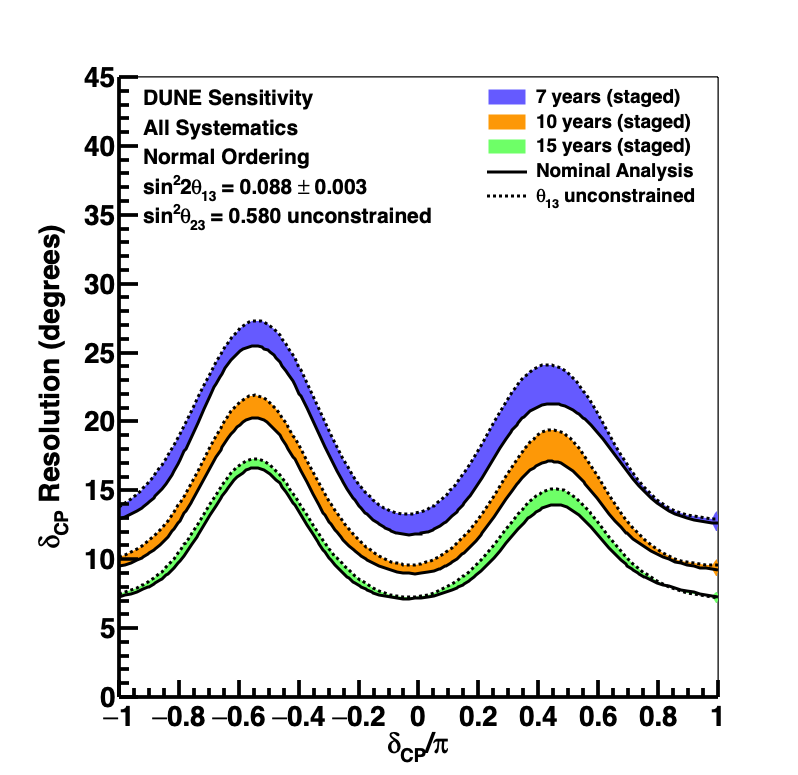}
	\caption[Resolution for the DUNE measurement of \deltacp as a function of \deltacp]
	{Resolution in degrees for the DUNE measurement of \deltacp, as a function of the true value of \deltacp, for seven (blue), ten (orange), and fifteen (green) years of exposure. True normal ordering is assumed. The width of the band shows the impact of applying an external constraint on \sinstt{13}.}
    \label{fig:dcpresvdcp}
\end{figure}

Figure~\ref{fig:dcpresvdcp} shows the resolution, in degrees, of DUNE's measurement of \deltacp, as a function of the true value of \deltacp. The resolution of this measurement is significantly better near CP-conserving values of \deltacp, compared to maximally CP-violating values. For fifteen years of exposure, resolutions between five and fifteen degrees are possible, depending on the true value of \deltacp. A smoothing algorithm has been applied to interpolate between values of \deltacp at which the full analysis has been performed.

\begin{figure}[h!]
    \centering
    \begin{tabular}{cc}
		\includegraphics[width=0.475\linewidth]{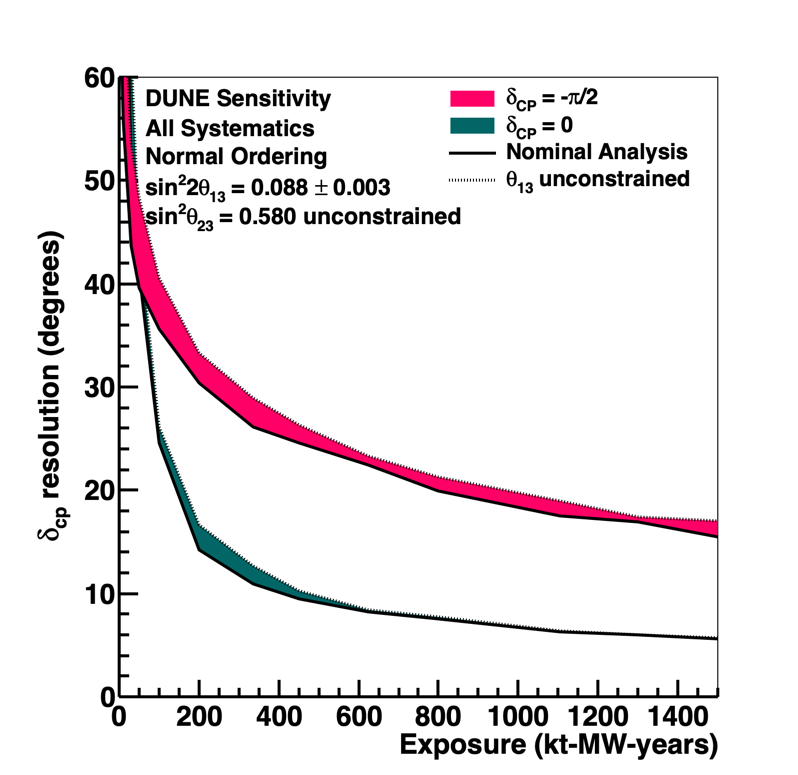} &
		\includegraphics[width=0.475\linewidth]{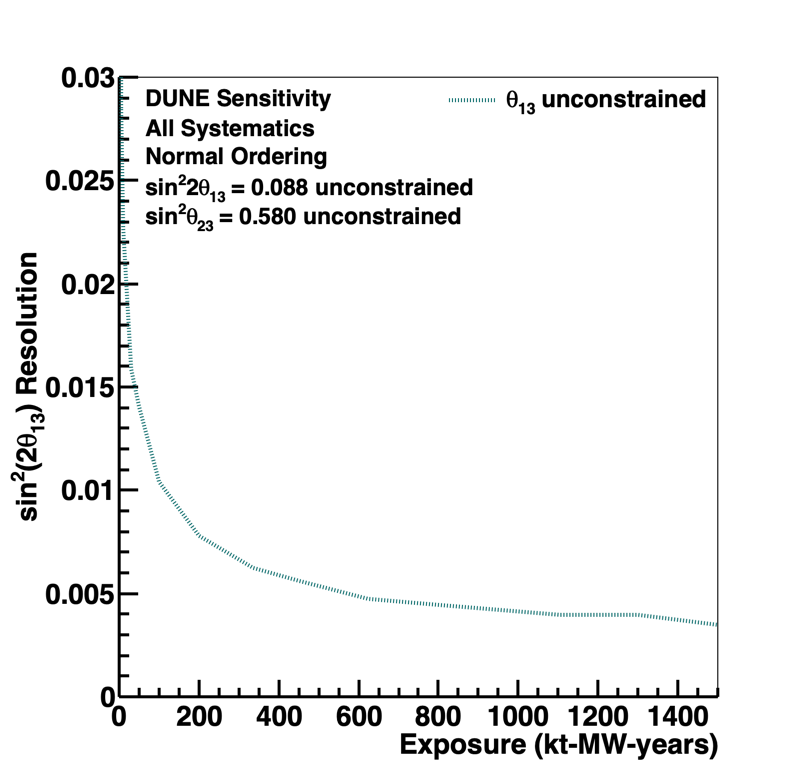} 
	\end{tabular}  
	\caption[Resolution of DUNE measurements of \deltacp and \sinstt{13}, as a function of exposure]{Resolution of DUNE measurements of \deltacp (left) and \sinstt{13} (right), as a function of exposure in kt-MW-years. As seen in Figure~\ref{fig:dcpresvdcp}, the \deltacp resolution has a significant dependence on the true value of \deltacp, so curves for $\deltacp=-\pi/2$ (red) and $\deltacp=0$ (green) are shown. The width of the band shows the impact of applying an external constraint on \sinstt{13}. For the \sinstt{13} resolution, an external constraint does not make sense, so only the unconstrained curve is shown.
For reference, 30, 100, 200, 336, 624, and \SI{1104}{\ktMWyr} correspond to 1.2, 3.1, 5.2, 7, 10, and 15 staged years, respectively.
}
    \label{fig:appres_exp}
\end{figure}

\begin{figure}[h!]
    \centering
    \begin{tabular}{cc}
		\includegraphics[width=0.475\linewidth]{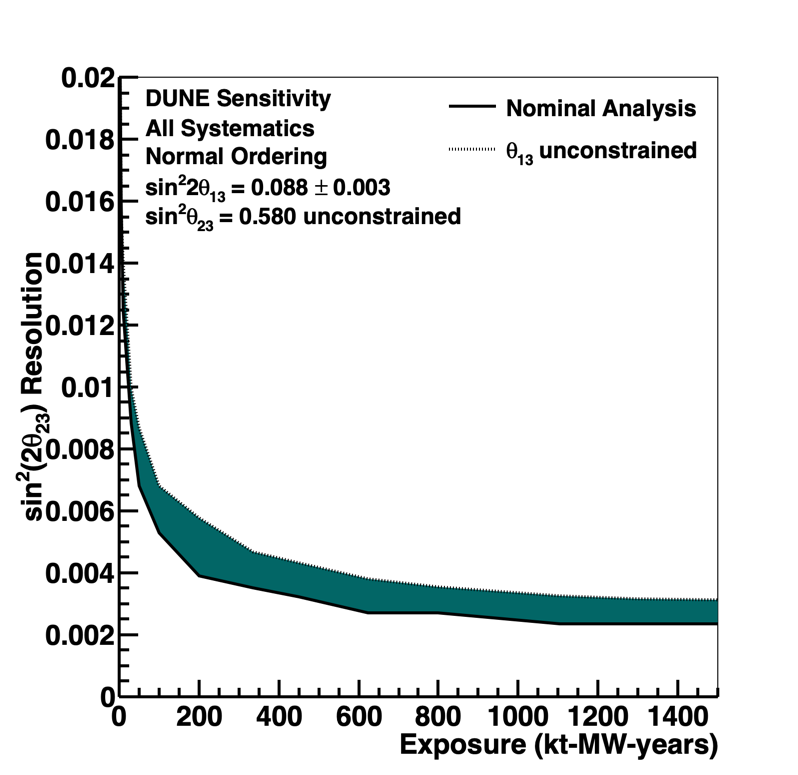} &
		\includegraphics[width=0.475\linewidth]{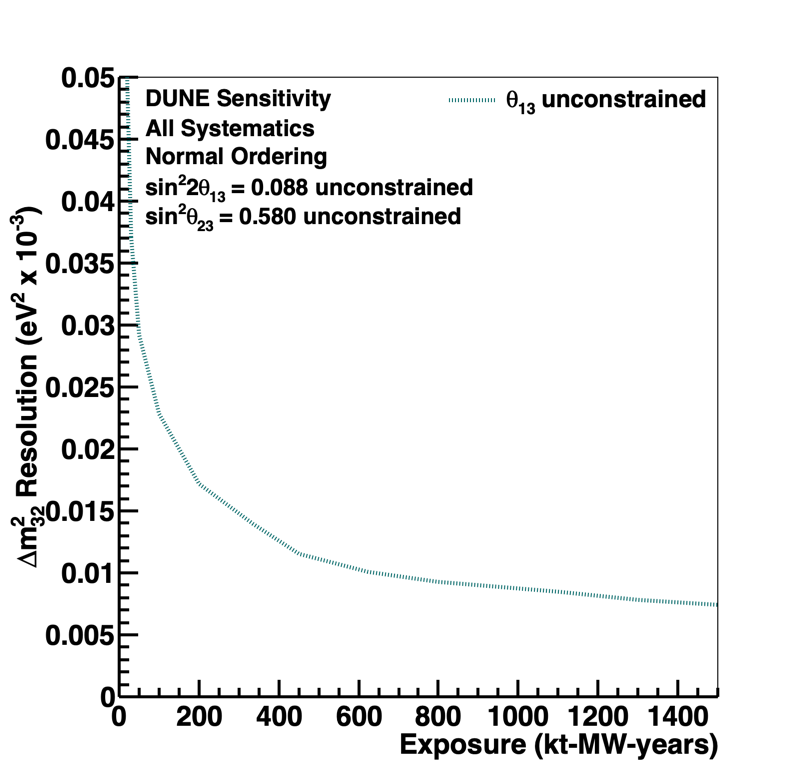} 
	\end{tabular}  
	\caption[Resolution of DUNE measurements of \deltacp and \sinstt{13}, as a function of exposure]{Resolution of DUNE measurements of \sinstt{23} (left) and $\Delta m^{2}_{32}$ (right), as a function of exposure in kt-MW-years. The width of the band for the \sinstt{23} resolution shows the impact of applying an external constraint on \sinstt{13}. For the $\Delta m^{2}_{32}$ resolution, an external constraint does not have a significant impact, so only the unconstrained curve is shown.
For reference, 30, 100, 200, 336, 624, and \SI{1104}{\ktMWyr} correspond to 1.2, 3.1, 5.2, 7, 10, and 15 staged years, respectively.
}
    \label{fig:disres_exp}
\end{figure}

Figures \ref{fig:appres_exp} and  \ref{fig:disres_exp} show the resolution of DUNE's measurements of \deltacp and \sinstt{13} and of \sinstt{23} and $\Delta m^{2}_{32}$, respectively, as a function of exposure in kt-MW-years. As seen in Figure~\ref{fig:dcpresvdcp}, the \deltacp resolution varies significantly with the true value of \deltacp, but for favorable values, resolutions near five degrees are possible for large exposure. The DUNE measurement of \sinstt{13} approaches the precision of reactor experiments for high exposure, allowing a comparison between the two results, which is of interest as a test of the unitarity of the PMNS matrix. 

\begin{figure}[h!]
    \centering
		\includegraphics[width=0.95\linewidth]{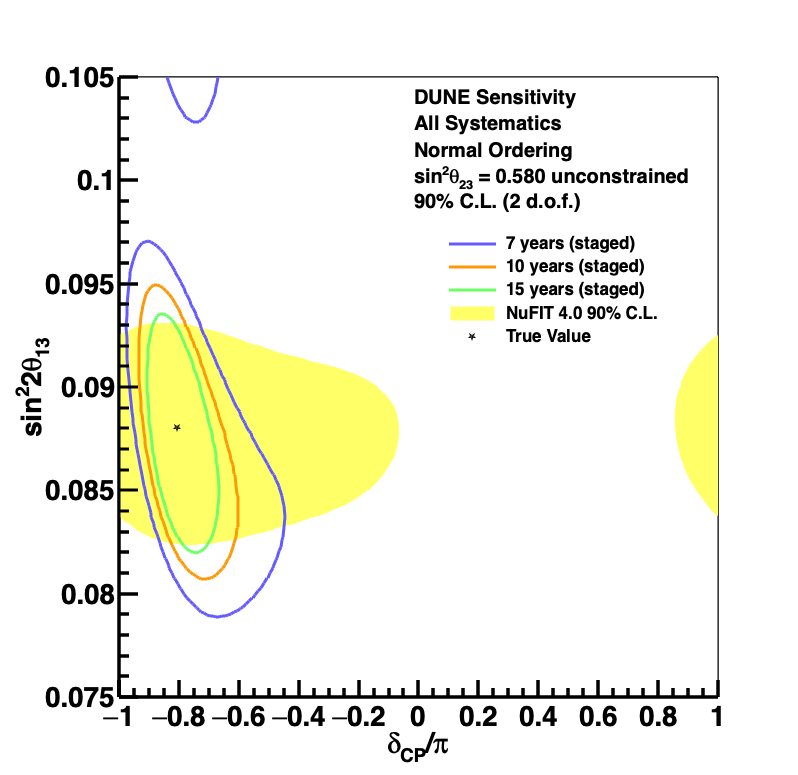}
	\caption[Two-dimensional 90\% C.L. region in \sinstt{13} and \deltacp]{Two-dimensional 90\% C.L. region in \sinstt{13} and \deltacp, for 7, 10, and 15 years of exposure, with equal running in neutrino and antineutrino mode. The 90\% C. L. region for the \dword{nufit} global fit is shown in yellow for comparison. The true values of the oscillation parameters are assumed to be the central values of the \dword{nufit} global fit and the oscillation parameters governing long-baseline oscillation are unconstrained.}
    \label{fig:res_th13vdcp}
\end{figure}

\begin{figure}[h!]
    \centering
		\includegraphics[width=0.95\linewidth]{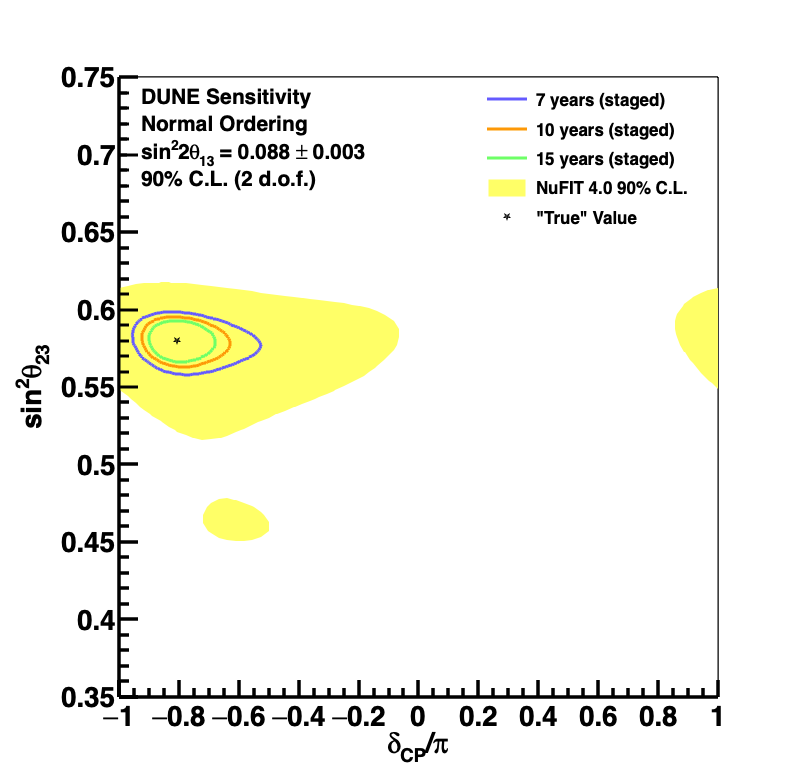}
	\caption[Two-dimensional 90\% C.L. region in \sinst{23} and \deltacp]{Two-dimensional 90\% C.L. region in \sinst{23} and \deltacp, for 7, 10, and 15 years of exposure, with equal running in neutrino and antineutrino mode. The 90\% C. L. region for the \dword{nufit} global fit is shown in yellow for comparison. The true values of the oscillation parameters are assumed to be the central values of the \dword{nufit} global fit and \sinstt{13} is constrained by \dword{nufit}}
    \label{fig:res_th23vdcp}
\end{figure}

\begin{figure}[h!]
    \centering
    \begin{tabular}{cc}
		\includegraphics[width=0.475\linewidth]{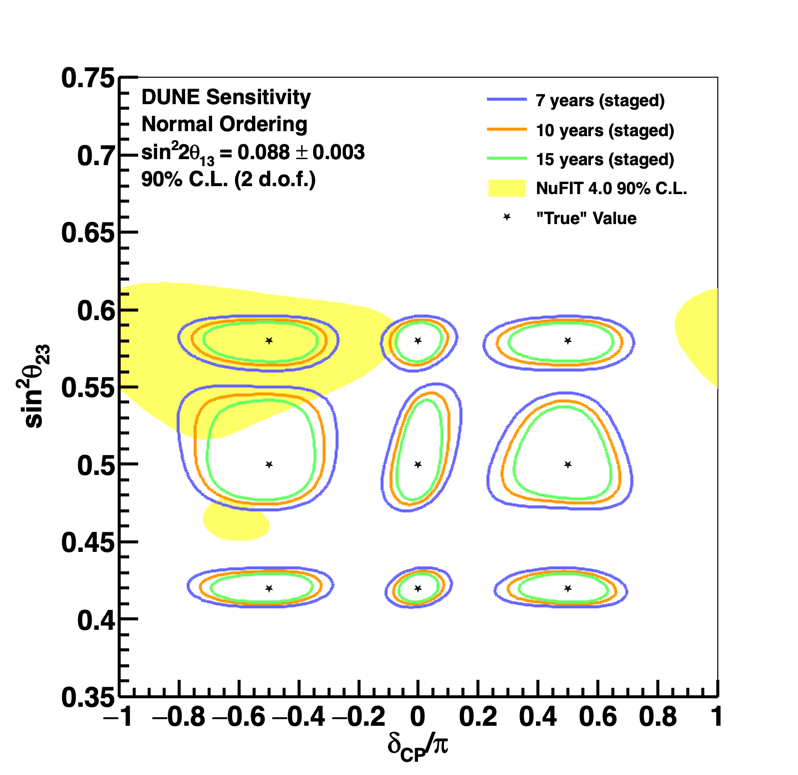} &
		\includegraphics[width=0.475\linewidth]{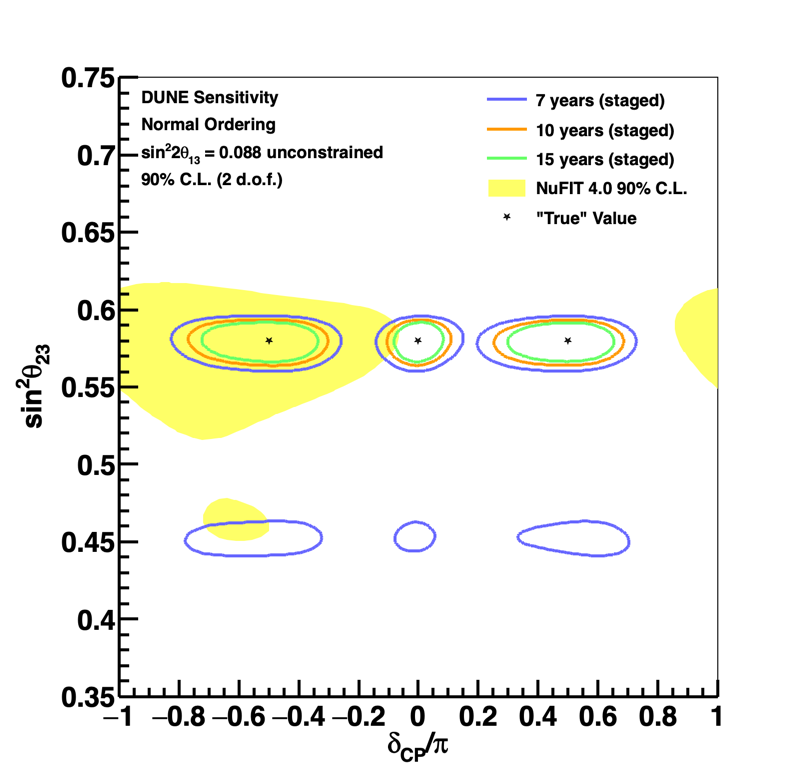} 
	\end{tabular}  
	\caption[Two-dimensional 90\% C.L. region in \sinst{23} and \deltacp]{Two-dimensional 90\% C.L. region in \sinst{23} and \deltacp, for 7, 10, and 15 years of exposure, with equal running in neutrino and antineutrino mode. The 90\% C.L. region for the \dword{nufit} global fit is shown in yellow for comparison. Several possible true values of the oscillation parameters, denoted by stars, are considered, and \sinstt{13} is constrained (left) or unconstrained (right) by \dword{nufit}. In the plot on the right, only one value for \sinst{23} is shown; without the constraint on \sinstt{13}, degenerate regions are allowed for lower exposures.}
    \label{fig:res_th23vdcp_degen}
\end{figure}

One of the primary physics goals for DUNE is the simultaneous measurement of all oscillation parameters governing long-baseline neutrino oscillation, without a need for external constraints. Figure~\ref{fig:res_th13vdcp} shows the 90\% C.L. allowed regions for \sinstt{13} and \deltacp for 7, 10, and 15 years of running, when no external constraints are applied, compared to the current measurements from world data. Note that a degenerate lobe at higher values of \sinstt{13} is present in the 7-year exposure, but is resolved for higher exposures. Figure~\ref{fig:res_th23vdcp} shows the two-dimensional allowed regions for \sinst{23} and \deltacp. Figure~\ref{fig:res_th23vdcp_degen} explores the resolution sensitivity that is expected for values of \sinst{23} different from the \dword{nufit} central value. It is interesting to note that the lower exposure, opposite octant solutions for \sinst{23} are allowed at 90\% C.L. in the absence of an external constraint on \sinstt{13}; however, at the 10 year exposure, this degeneracy is resolved by DUNE data without external constraint.

\begin{figure}[h!]
    \centering
		\includegraphics[width=0.75\linewidth]{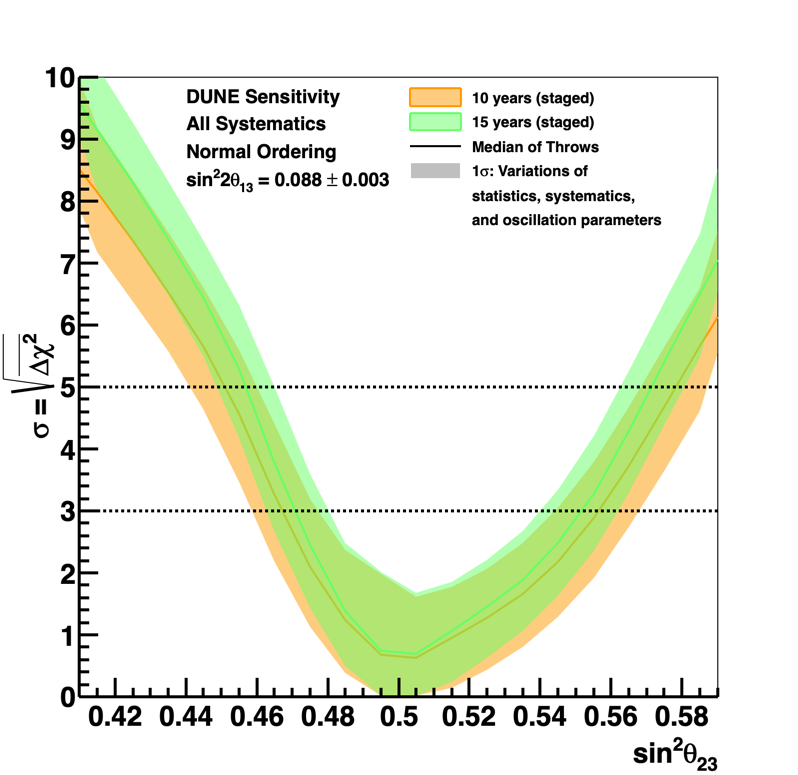}
	\caption[Sensitivity of determination of the $\theta_{23}$ octant as a function of \sinst{23}]{Sensitivity to determination of the $\theta_{23}$ octant as a function of the true value of \sinst{23}, for ten (orange) and fifteen (green) years of exposure. True normal ordering is assumed. The width of the transparent bands cover 68\% of fits in which random throws are used to simulate statistical variations and select true values of the oscillation and systematic uncertainty parameters, constrained by pre-fit uncertainties. The solid lines show the median sensitivity.}
    \label{fig:lbloctant}
\end{figure}

The measurement of $\nu_\mu \rightarrow \nu_\mu$ oscillations is sensitive to $\sin ^2 2 \theta_{23}$, whereas the measurement of $\nu_\mu \rightarrow \nu_e$ oscillations is sensitive to $\sin^2 \theta_{23}$.  A combination of both $\nu_e$ appearance and $\nu_\mu$ disappearance measurements can probe both maximal mixing and
the $\theta_{23}$ octant.  
Figure~\ref{fig:lbloctant} shows the sensitivity to determining the octant as a function of the true value of $\sinst{23}$.

\subsection{Impact of Oscillation Parameter Central Values}
\label{sec:physics-lbnosc-oscvar}
\begin{figure}[h!]
    \centering
    \begin{tabular}{cc}
		\includegraphics[width=0.475\linewidth]{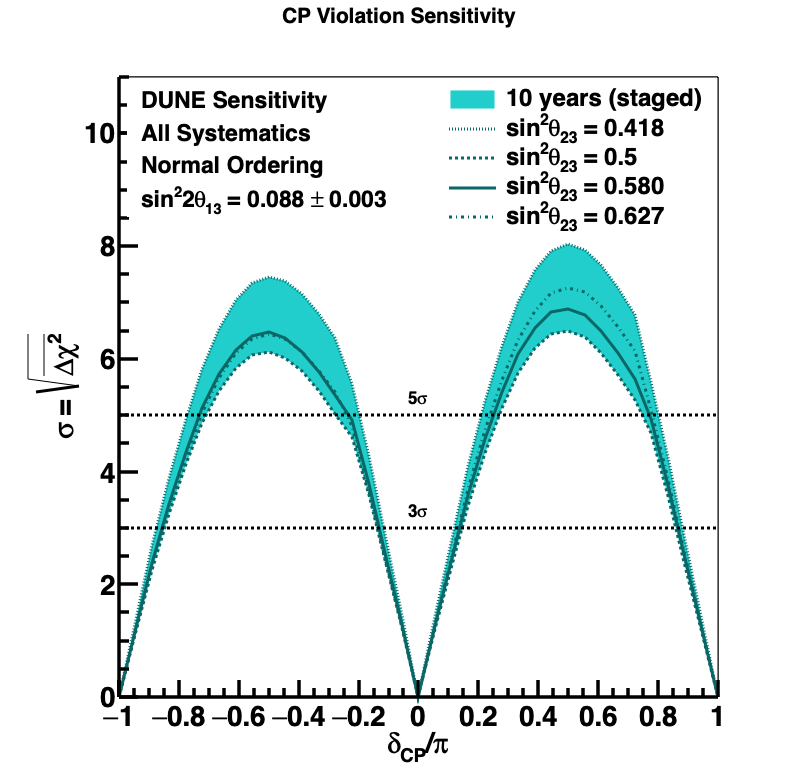} &
		\includegraphics[width=0.475\linewidth]{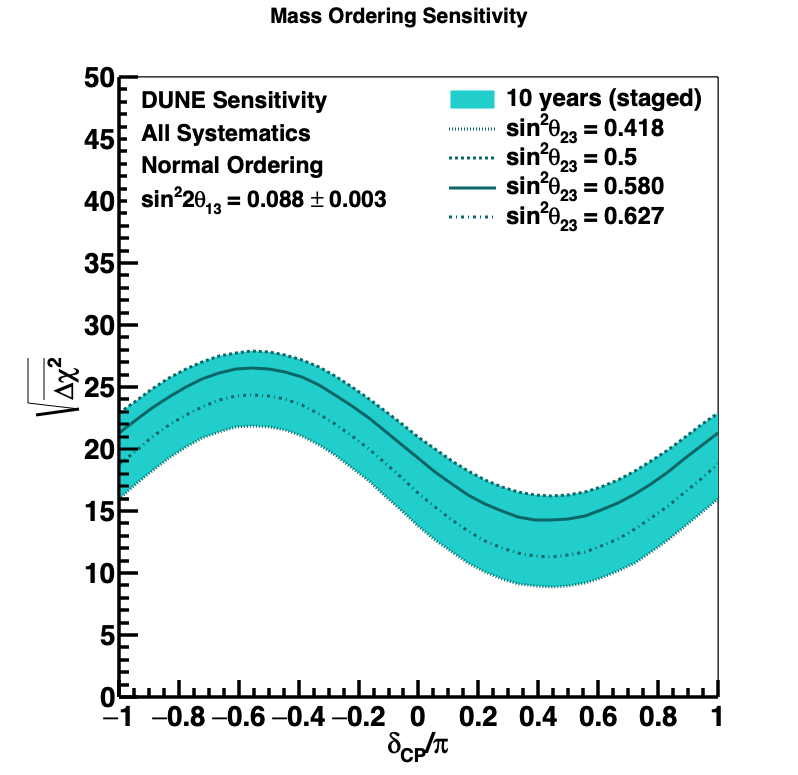}
	\end{tabular}
	\caption[Sensitivity to CP violation and neutrino mass ordering, as a function of $\deltacp$]{Sensitivity to CP violation (left) and neutrino mass ordering (right), as a function of the true value of $\deltacp$, for 10 years of exposure, with equal running in neutrino and antineutrino mode. Curves are shown for true values of $\theta_{23}$ corresponding to the 3$\sigma$ range of values allowed by \dword{nufit}, as well as the \dword{nufit} central value and maximal mixing. The nominal sensitivity analysis is performed.}
    \label{fig:th23var}
\end{figure}

\begin{figure}[h!]
    \centering
    \begin{tabular}{cc}
		\includegraphics[width=0.475\linewidth]{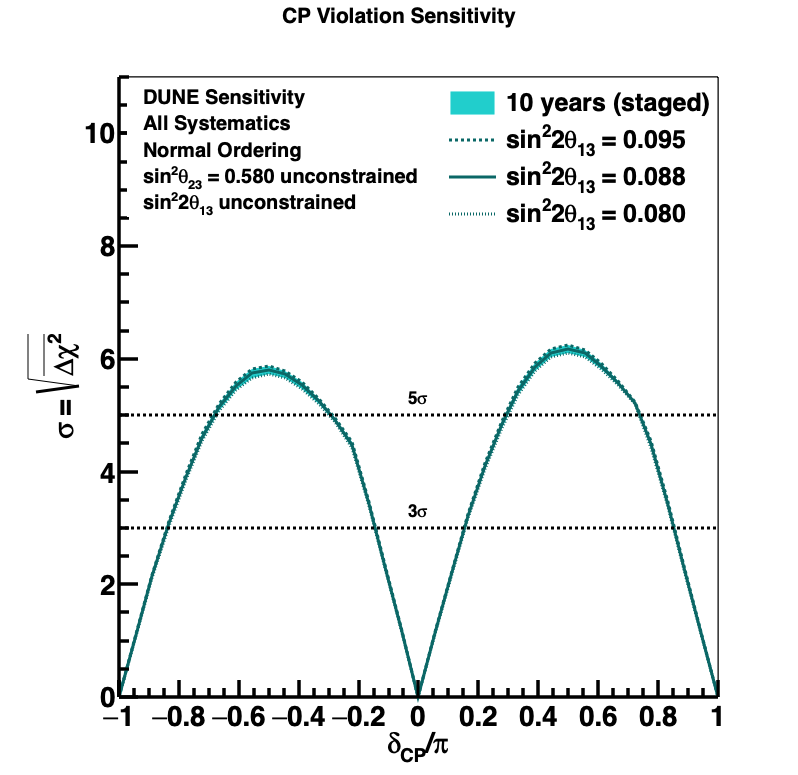} &
		\includegraphics[width=0.475\linewidth]{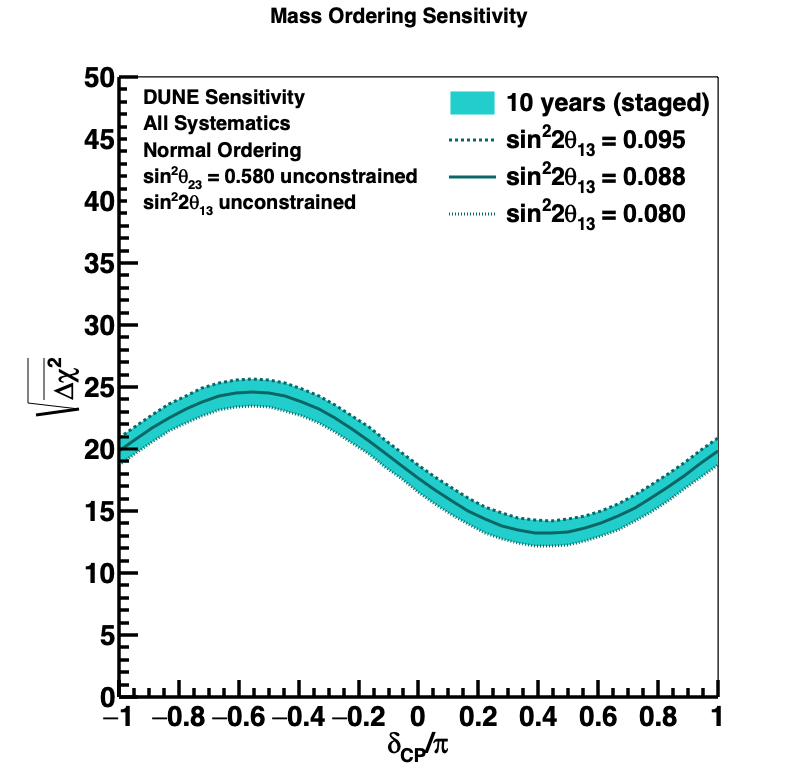}
	\end{tabular}
	\caption[Sensitivity to CP violation and neutrino mass ordering, as a function of $\deltacp$]{Sensitivity to CP violation (left) and neutrino mass ordering (right), as a function of the true value of $\deltacp$, for 10 years of exposure, with equal running in neutrino and antineutrino mode. Curves are shown for true values of $\theta_{13}$ corresponding to the 3$\sigma$ range of values allowed by \dword{nufit}, as well as the \dword{nufit} central value. The nominal sensitivity analysis is performed, with the exception that $\theta_{13}$ is not constrained at the NuFit4.0 central value in the fit.}
    \label{fig:th13var}
\end{figure}

\begin{figure}[h!]
    \centering
    \begin{tabular}{cc}
		\includegraphics[width=0.475\linewidth]{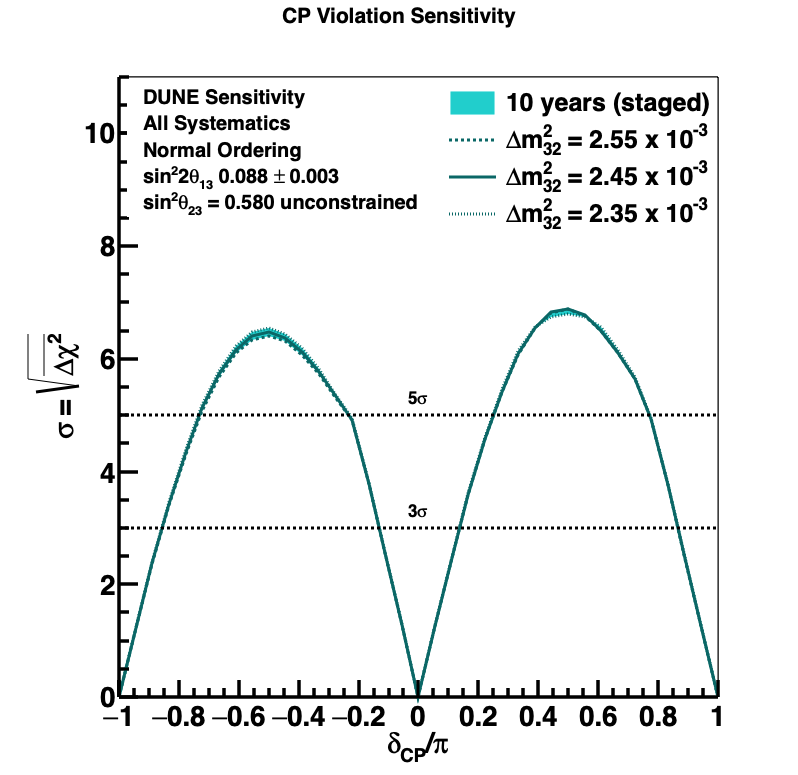} &
		\includegraphics[width=0.475\linewidth]{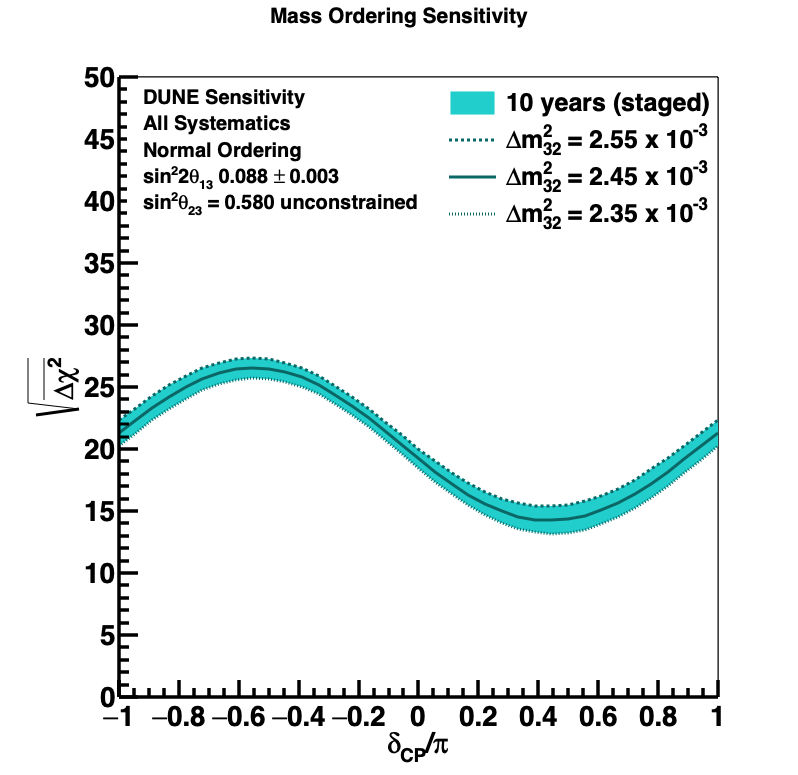}
	\end{tabular}
	\caption[Sensitivity to CP violation and neutrino mass ordering, as a function of $\deltacp$]{Sensitivity to CP violation (left) and neutrino mass ordering (right), as a function of the true value of $\deltacp$, for 10 years of exposure, with equal running in neutrino and antineutrino mode. Curves are shown for true values of $\Delta m^2_{32}$ corresponding to the 3$\sigma$ range of values allowed by \dword{nufit}, as well as the \dword{nufit} central value. The nominal sensitivity analysis is performed.}
    \label{fig:dmsqvar}
\end{figure}

The sensitivity results presented in the preceding sections assume that the true values of the parameters governing long-baseline neutrino oscillation are the central values of the \dword{nufit} global fit, given in Table~\ref{tab:oscpar_nufit}. In this section, variations in DUNE sensitivity with other possible true values of the oscillation parameters are explored.
Figures \ref{fig:th23var}, \ref{fig:th13var}, and \ref{fig:dmsqvar} show DUNE sensitivity to CP violation and neutrino mass ordering when the true values of $\theta_{23}$, $\theta_{13}$, and $\Delta m^{2}_{32}$, respectively, vary within the 3$\sigma$ range allowed by \dword{nufit}. The largest effect is the variation in sensitivity with the true value of $\theta_{23}$, where degeneracy with $\deltacp$ and matter effects are significant. Values of $\theta_{23}$ in the lower octant lead to the best sensitivity to CP violation and the worst sensitivity to neutrino mass ordering, while the reverse is true for the upper octant. DUNE sensitivity for the case of maximal mixing is also shown. The true values of $\theta_{13}$ and $\Delta m^2_{32}$ are highly constrained by global data and, within these constraints, do not have a dramatic impact on DUNE sensitivity.  

\subsection{Impact of Systematic Uncertainties}
\label{sec:physics-lbnosc-systresults}

Implementation of systematic uncertainties in the nominal fits are described in Sections~\ref{sec:nu-osc-04}, \ref{sec:nu-osc-05}, and \ref{sec:physics-lbnosc-syst}. All considered systematic parameters are summarized in Table~\ref{tab:systcheatsheet}. 
In the nominal fits, many systematic uncertainties are constrained by DUNE data, as described in the following section.

\begin{longtable}{ll}\hline \hline
Brief Name & Description of Uncertainty \\  \hline \hline
\textbf{Flux:} & \\
flux[N] & Nth component of flux PCA \\ \hline
\textbf{Interaction Model:} & \\
MaCCQE & Axial mass for CCQE \\
VecFFCCQEshape & Choice of CCQE vector form factors \\
MaCCRES & Axial mass for CC resonance \\
MvCCRES & Vector mass for CC resonance \\
Theta Delta2Npi & $\theta_{\pi}$ distribution in decaying $\Delta$ rest frame \\
AhtBY & $A_{HT}$ higher-twist param in BY model scaling variable $\epsilon_{\omega}$ \\
BhtBY & $B_{HT}$ higher-twist param in BY model scaling variable $\epsilon_{\omega}$ \\
CV1uBY & $C_{V1u}$ valence GRV98 PDF correction param in BY model \\
CV2uBY & $C_{V2u}$ valence GRV98 PDF correction param in BY model \\
MaNCEL & Axial mass for NC elastic \\
MaNCRES & Axial mass for NC resonance \\
MvNCRES & Vector mass for NC resonance \\
FrCEx N & Nucleon charge exchange probability \\
FrElas N & Nucleon elastic reaction probability \\
FrInel N & Nucleon inelastic reaction probability \\
FrAbs N & Nucleon absorption probability \\
FrPiProd N & Nucleon $\pi$-production probability \\
FrCEx pi & $\pi$ charge exchange probability \\
FrElas pi & $\pi$ elastic reaction probability \\
FrInel pi & $\pi$ inelastic reaction probability \\
FrAbs pi & $\pi$ absorption probability \\
FrPiProd pi & $\pi$ $\pi$-production probability \\
BeRPA A & Random Phase Approximation tune: controls low $Q^{2}$ \\
BeRPA B &  Random Phase Approximation tune: controls low-mid $Q^{2}$\\
BeRPA D & Random Phase Approximation tune: controls mid $Q^{2}$ \\
Mnv2p2hGaussEnhancement & Extra strength into 2p2h \\
C12ToAr40 2p2hScaling nu & neutrino 2p2h Ar/C scaling \\
C12ToAr40 2p2hScaling nubar & antineutrino 2p2h Ar/C scaling \\
E2p2h [A,B] [nu,nubar] & 2p2h energy dependence \\
SPPLowQ2Suppression & Low $Q^{2}$ (empirical) suppression \\
MKSPP ReWeight & MK model - alternative strength in W \\
NR nu np CC 1Pi & Norm for $\nu + n/p \rightarrow l + 1\pi$ \\
NR [nu,nubar] [p,n] [CC,NC] [1,2,3]Pi & non-resonant pion production topology norms \\
nuenumu xsec ratio & \nue/\numu uncertainty in \nue unique phase space \\
nuenuebar xsec ratio & Modification of \nue/\numu and \anue/\anumu xsec \\ \hline
\textbf{Detector Effects:} & \\
FVNueFD & FD \nue fiducial volume \\
FVNueFD & FD \numu fiducial volume \\
FDRecoNueSyst & FD \nue selection \\
FDRecoNumuSyst & FD \numu selection \\
ChargedHadResFD & FD charged hadron resolution \\
EMResFD & FD electromagnetic shower resolution \\
MuonResFD & FD muon resolution \\
EMUncorrFD & FD electromagnetic shower energy scale \\
EScaleNFD & FD neutron visible energy scale \\
EScaleMuLArFD & FD muon energy scale \\
EScaleFD & FD overall energy scale \\
\hline
\hline
\caption[Definition of systematic uncertainty parameters]{Definition of systematic uncertainty parameters. The brief names are used in Figures \ref{fig:postfit_unc_ndfd} and \ref{fig:postfit_unc_vsexp}.} \\
\label{tab:systcheatsheet}
\end{longtable}


\subsubsection{Systematic Uncertainty Constraints}

Prefit uncertainties on flux and cross section parameters are at the level of $\sim$10\%. These uncertainties become constrained in the fit, especially by the \dword{nd}. Figure~\ref{fig:postfit_unc_ndfd} shows the level of constraint on each systematic parameter after the fit. The larger band shows the constraint that arises from the far detector alone, while the inner band shows the (much stronger) constraint from the near detector. Figure~\ref{fig:postfit_unc_vsexp} compares the parameter constraints for two different exposures. The wider band shows the \dword{nd}+\dword{fd} constraint expected after 7 years, and the narrower band shows the constraint after 15 years. The effect of increasing the exposure is very small because the \dword{nd} is already systematically limited in the \numu \dword{cc} channel after 7 years. The impact of adding the near detector is significant; flux and cross section parameters are very weakly constrained by the far detector alone. Parameters are implemented in such a way that there are no prefit correlations, but the constraints from the near detector cause parameters to become correlated, which is not shown in the figure.

Some uncertainties are not reduced by the \dword{nd}. For example, the energy scale parameters are treated as uncorrelated between detectors, so naturally the \dword{nd} does not constrain them. Several important cross section uncertainties are not constrained by the near detector. In particular, an uncertainty on the ratio of \numu to \nue cross sections is totally unconstrained. The most significant flux terms are constrained at the level of ~20\% of their \textit{a priori} values.  Less significant principal components have little impact on the observed distributions at either detector, and receive weaker constraints. Most cross section parameters that affect \dword{cc} interactions are well constrained.

\begin{dunefigure}[Post-fit systematic uncertainties]{fig:postfit_unc_ndfd}
{The ratio of post-fit to pre-fit uncertainties for various systematic parameters for a 15-year staged exposure. The red band shows the constraint from the \dword{fd} only in 15 years, while the green shows the \dword{nd}+\dword{fd} constraints. Systematic parameter names are defined in Table \ref{tab:systcheatsheet}.}
  \includegraphics[width=0.8\textwidth]{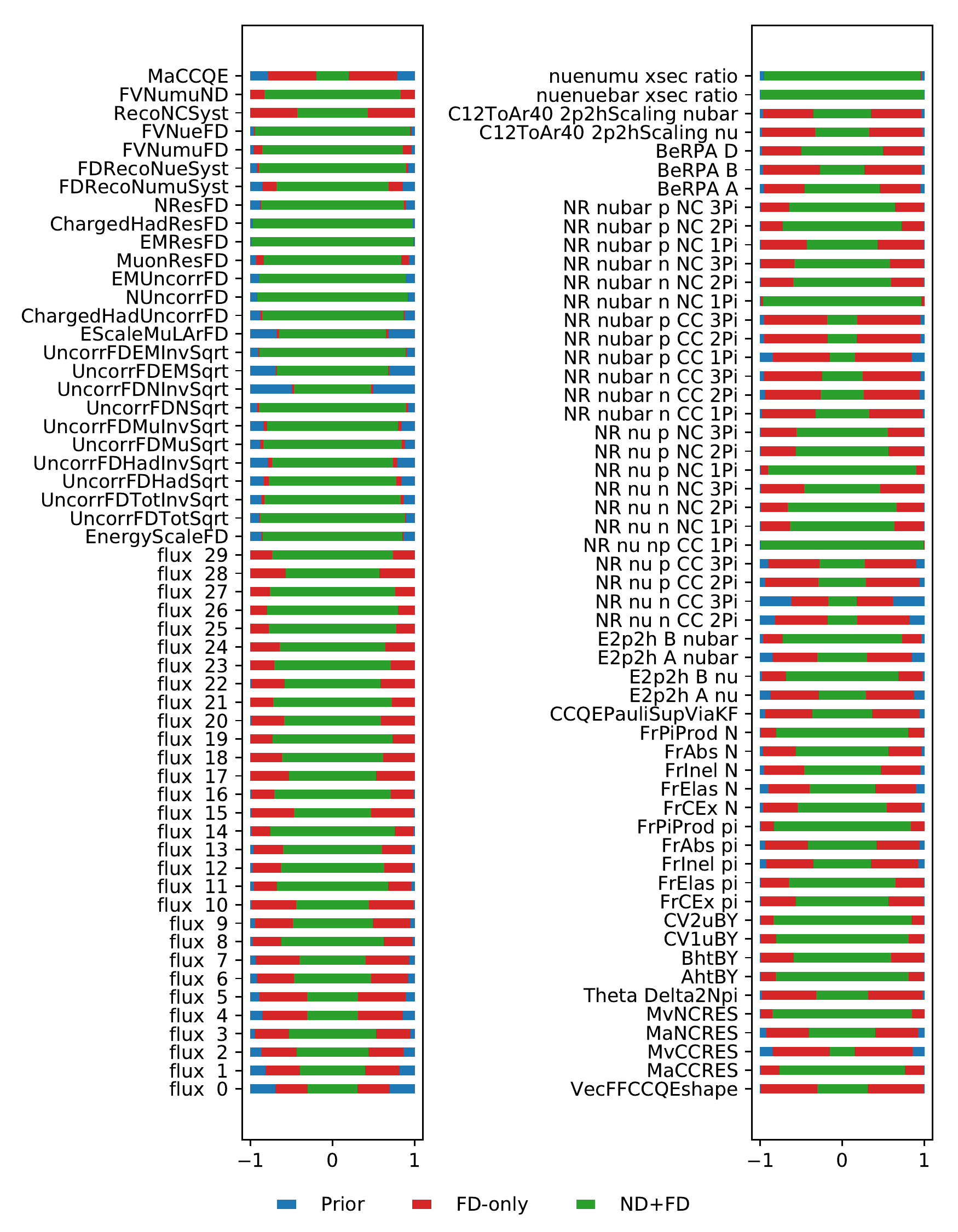}
\end{dunefigure}

\begin{dunefigure}[Post-fit systematic uncertainties]{fig:postfit_unc_vsexp}
{The ratio of post-fit to pre-fit uncertainties for various systematic parameters for a \dword{nd}+\dword{fd} constraint after 7 and 15 years. The difference in parameter constraints due to increasing the exposure is very small. Systematic parameter names are defined in Table \ref{tab:systcheatsheet}.}
  \includegraphics[width=0.8\textwidth]{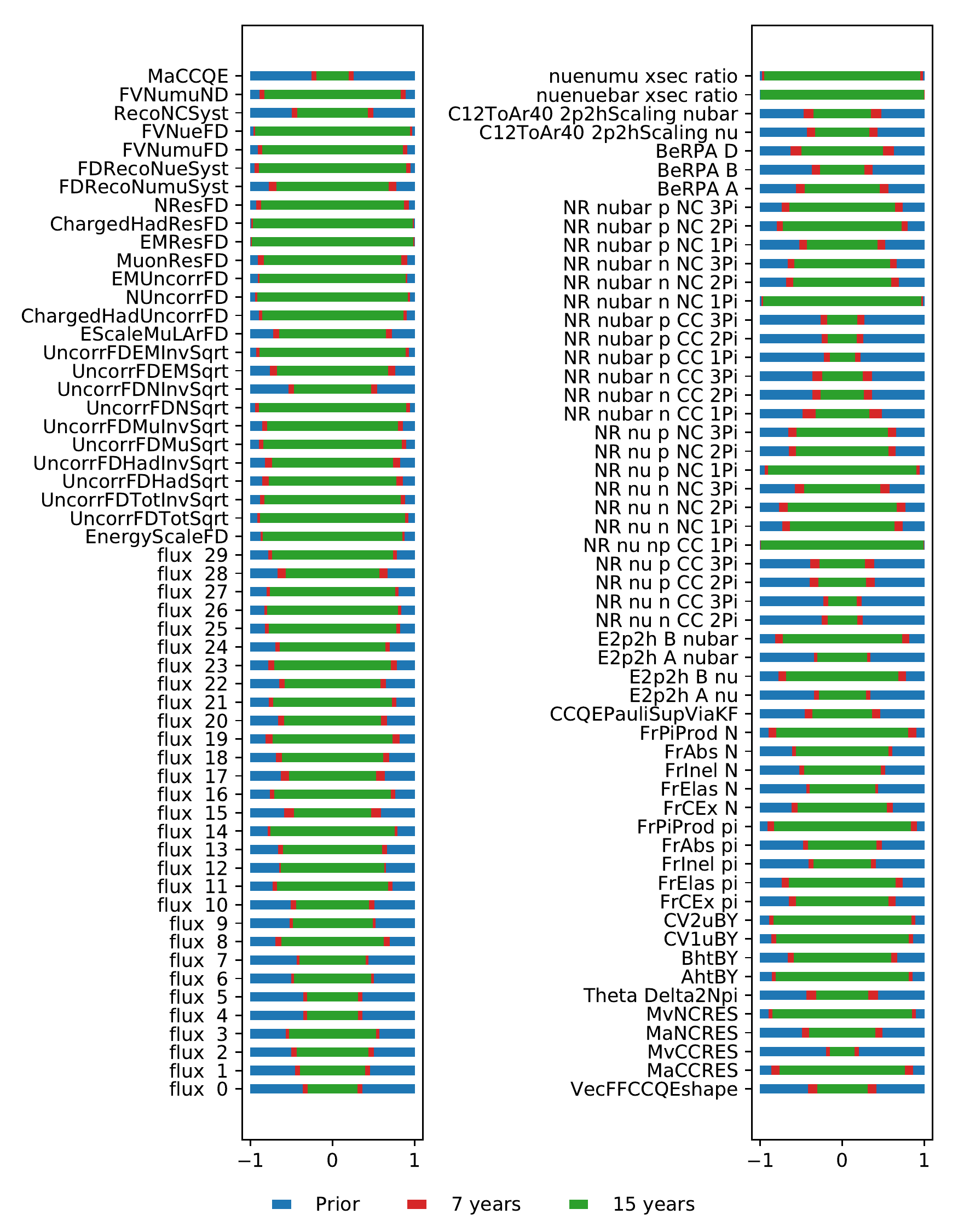}
\end{dunefigure}




\subsection{Impact of the Near Detector}
\label{sec:ndimpact}

The oscillation sensitivity analysis presented in the previous section is intended to demonstrate the full potential of \dword{dune}, with constraints from the full suite of near detectors described in \introchnd, including the \dword{lar} \dword{tpc}, \dword{mpd}, 
\dword{sand}, 
and off-axis measurements. In addition to the \numu and \anumu \dword{cc} spectra used explicitly in this analysis, the \dword{lar} \dword{tpc} is also expected to measure numerous exclusive final-state \dword{cc} channels, including 1$\pi^{\pm}$, 1$\pi^0$, and multi-pion production. Measurements will be made as a function of other kinematic quantities in addition to reconstructed $E_{\nu}$ and $y$, such as four-momentum transfer to the nucleus, lepton angle, or final-state meson kinematics. The \dword{lar} \dword{tpc} will also measure the sum of \nue and \anue \dword{cc} scattering, and \dword{nc} events. Direct flux measurements will be possible with neutrino-electron elastic scattering, and the low-$\nu$ technique.

In addition to the many on-axis \dword{lar} samples, a complementary set of neutrino-argon measurements is expected from the \dword{hpg} \dword{tpc}. This detector will be sensitive to charged tracks at kinetic energies of just a few MeV, enabling the study of nuclear effects in unprecedented detail. It will also sign-select all charged particles, with nearly perfect pion-proton separation from dE/dx out to over 1 GeV/c momentum, so that high-purity measurements of CC1$\pi^{+}$ and CC1$\pi^{-}$ are possible. It may be possible to directly measure neutron energy spectra from time of flight using the \dword{hpg} \dword{tpc} coupled to a high-performance \dword{ecal}. 
%
The \dword{sand} on-axis beam monitor
will measure neutrino-carbon scattering and neutron production while ensuring excellent beam stability.

The \dword{lar} and \dword{mpd} will also move off-axis to measure neutrino-argon interactions in many different fluxes. This will provide a direct constraint on the relationship between neutrino energy and visible energy in \dword{lar}. By taking linear combinations of spectra at many off-axis positions, it is possible to reproduce the expected \dword{fd} energy spectrum for a given set of oscillation parameters and directly measure visible energy.

All of these capabilities of the \dword{nd} benefit the \dword{dune} physics program. However, due to the timing of the \dword{nd} process, design details of the \dword{nd} are not available at the time of preparing this document, and it is not practical to include all of these samples and demonstrate their impact on oscillation sensitivity directly. Instead, we assume a model that implicitly includes these constraints, with further direct demonstration planned for the \dword{nd} \dword{tdr}.

The neutrino interaction model uncertainties shown in Section~\ref{sec:nu-osc-05} represent our current knowledge of neutrino interactions, motivated by measurements wherever possible. The \dword{dune} \dword{nd} is able to constrain these uncertain parameters, as demonstrated in the previous section. However, due to the complexity of modeling neutrino-argon interactions, and the dearth of neutrino-argon measurements in the energy range relevant for \dword{dune}, this is a necessary but insufficient condition for the \dword{nd} program. There are possible variations to the interaction model that cannot be readily estimated, simply because we have yet to observe the inadequacy of the model. While these ``unknown unknowns'' are impossible to predict, guarding against them is critically important to the success of the \dword{dune} physics program. For this reason, the \dword{nd} is designed under the assumption that it must not only constrain some finite list of model parameters, but also be sensitive to general modeling deficiencies.

The sensitivity analysis presented in the previous section assumes the success of the \dword{nd} program. Because of this assumption, in order to estimate the expected sensitivity without a \dword{nd}, it is not sufficient to simply remove the on-axis \dword{lar} \dword{nd} sample that is explicitly included in the analysis. We must also account for other potential biases from the interaction model, the ``unknown unknowns.'' In this section, we consider two simple examples of bias, and evaluate the potential impact on oscillation parameter measurements in a scenario where the \dword{nd} capacity is reduced. In Section~\ref{sec:FDonlyNuWro}, we consider the case where there is no near detector, and show a ``mock data'' sample that results in a high-quality \dword{fd}-only fit with a significant bias in the measured value of $\delta_{CP}$. This bias would be undetectable with a FD-only fit, but easily detected at the \dword{nd}. In Section~\ref{sec:missingProtonMD}, we consider an alternative mock data set that gives a high-quality fit to the \dword{fd} as well as the on-axis \dword{nd} spectra, but has significant biases that are easily detected with off-axis \dword{nd} data. These bias tests are not meant as exact estimates of the reduction in sensitivity that would be expected without a \dword{nd} or with only on-axis \dword{nd}, but they do serve as examples of the kind of bias that is possible. By estimating an additional uncertainty on oscillation parameters to cover the observed bias, it is possible to produce a sensitivity estimate; however, as it is based on one single possible bias, {\em it should be considered a lower bound on the potential reduction in sensitivity.}

\subsubsection{Bias study: \dword{fd}-only fit to NuWro}
\label{sec:FDonlyNuWro}


An alternative Monte Carlo sample is produced by reweighting the \dword{genie} simulated events to \dword{nuwro}. The objective of the reweighting is to reproduce the \dword{nuwro} event spectra as a function of reconstructed neutrino energy, but without re-running the reconstruction. Simple reweighting schemes typically determine weights by taking the ratio between two generators in some limited kinematic space of true quantities. A common shortcoming of such techniques is that the reconstructed energy depends on many true quantities, and perhaps in a complicated way. Defining weights in a limited space effectively projects away any differences in other variables. To overcome this limitation, 18 true quantities that impact the reconstructed neutrino energy are identified: neutrino energy, lepton energy, lepton angle, $Q^{2}$, $W$, $x$, $y$, as well as the number and total kinetic energy carried by protons, neutrons, $\pi^{+}$, $\pi^{-}$, $\pi^{0}$, and the number of electromagnetic particles. A \dword{bdt} is trained on vectors of these 18 quantities in \dword{genie} and \dword{nuwro}. The \dword{bdt} minimizes a logistic loss function between \dword{genie} and \dword{nuwro} in the 18-dimensional space, producing a set of weights. When these weights are applied to \dword{genie} events, the resulting event spectra match the \dword{nuwro} spectra in all 18 quantities.

The resulting selected samples of \dword{fd} \numu and \nue \dword{cc} events in \dword{fhc} and \dword{rhc} beam modes are fit using the nominal \dword{genie}-based model and its uncertainties as described in Sections~\ref{sec:nu-osc-05} and \ref{sec:physics-lbnosc-syst}. The fit quality in the \dword{fd}-only scenario is high, with $\chi^{2}$ per degree of freedom smaller than unity for all oscillation parameters. Systematic nuisance parameters are pulled from their best fit values by more than $\sim$0.6$\sigma$. 

The best-fit value of $\delta_{CP}$ is determined for the full range of possible true $\delta_{CP}$ values between $-\pi$ and $+\pi$. The difference between the best-fit and true values of $\delta_{CP}$ is found to be less than 14 degrees for 68\% of the true values. To estimate the impact of such a bias on CP-violation sensitivity, an uncertainty equal to 14 degrees is added to the $\delta_{CP}$ resolution in quadrature. For a 10-year staged \dword{dune} \dword{fd} exposure, the resulting resolution is shown in the left panel of Figure~\ref{fig:nuwro_bias} compared to the nominal sensitivity with the \dword{nd} included. In the \dword{nd}+\dword{fd} (nominal) fit the bias is excluded, because in the \dword{nd} the bias is easily detected and not attributable to oscillations. To estimate the sensitivity to nonzero CP violation as shown in the right panel of Figure~\ref{fig:nuwro_bias}, the nominal \dword{fd}-only curve is reduced by the fractional increase in the $\delta_{CP}$ resolution at each point. The latter step is necessary because the uncertainty on $\delta_{CP}$ is not Gaussian.

\begin{dunefigure}[Oscillation sensitivities including example FD-only bias]{fig:nuwro_bias}
{The CP violation sensitivity for a \dword{fd}-only scenario with an additional uncertainty added to cover the observed bias from one example variation. The $\delta_{CP}$ resolution (left) and CP violation sensitivity (right) are compared to the results from the nominal \dword{nd}+\dword{fd} analysis.}
  \includegraphics[width=0.48\textwidth]{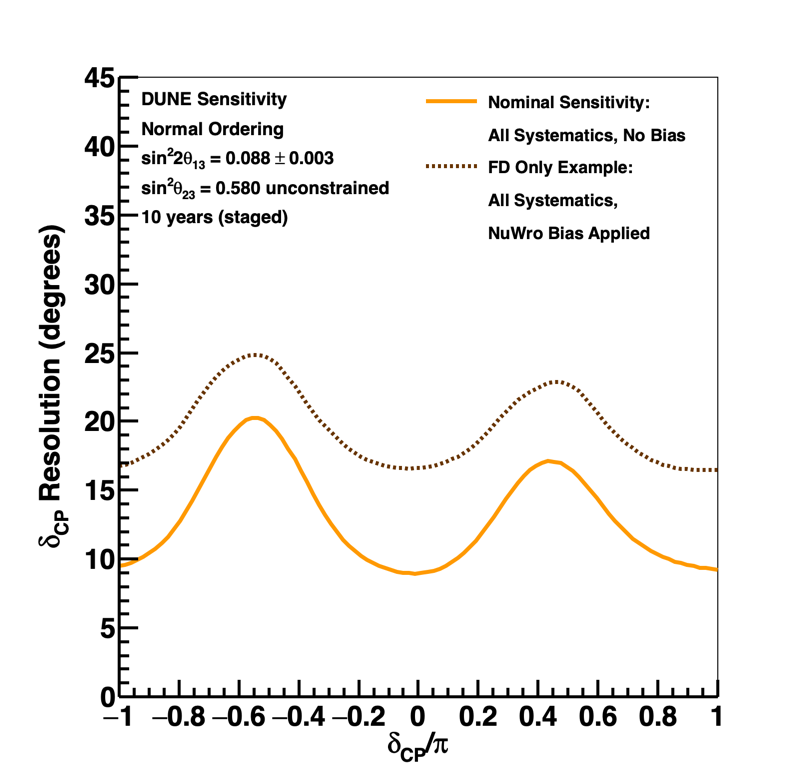}
  \includegraphics[width=0.48\textwidth]{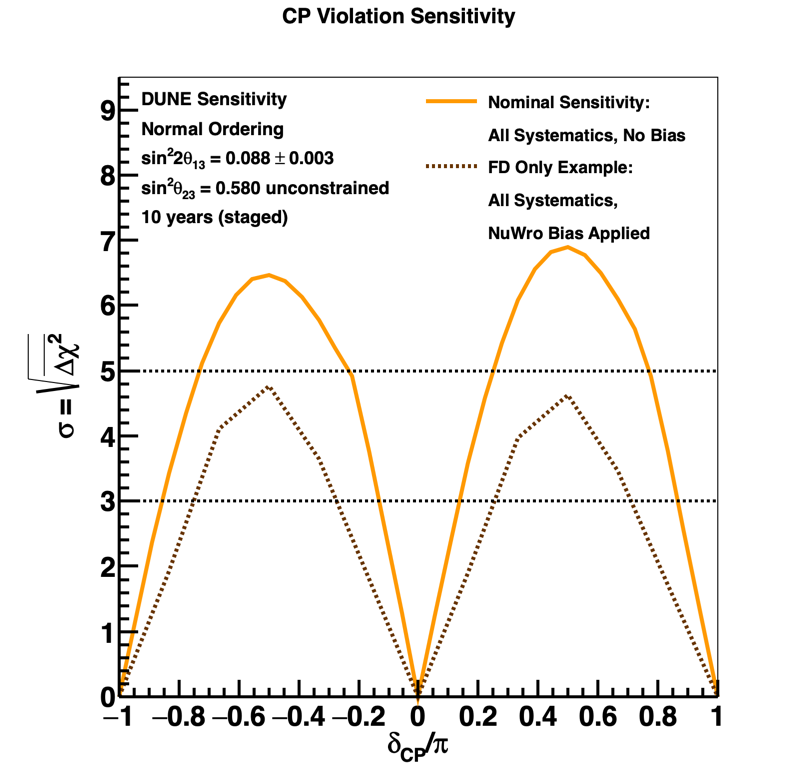}
\end{dunefigure}

As seen in Figure~\ref{fig:nuwro_bias}, the reduction in experimental sensitivity that would result from treating this example bias as a systematic uncertainty, which would be required in the absence of near detector data, is dramatic. Many other reasonable variations of the neutrino interaction model are allowed by world data and would also have to be considered as potential sources of uncertainty without near detector data to observe and resolve model incompatibility. 

\subsubsection{Bias study: shifted visible energy}
\label{sec:missingProtonMD}

As another example, we consider a possible deficiency of the \dword{genie} model, specifically the case where the energy of final-state protons is reduced by 20\%, with the energy going to neutrons instead. As neutrons are generally not observed, this will modify the relationship between neutrino energy and visible energy at the \dword{nd} and \dword{fd}. At the same time, the cross section model is altered so that the distribution of proton kinetic energy is unchanged. This alternate model is perfectly consistent with all available data; there is no reason to prefer our nominal \dword{genie} model to this one.

By construction, this alternate model will not affect the fit at the on-axis near detector, as the cross section shift exactly cancels the loss in hadronic visible energy due to changing protons for neutrons. Nuisance parameters that affect the near detector spectra, namely flux and cross section uncertainties, are not pulled and remain at their nominal values with the same post-fit uncertainties observed in the Asimov sensitivity. At the far detector, however, the different neutrino energy spectrum leads to an observed shift in reconstructed energy with respect to the nominal prediction, visible in Figure~\ref{fig:missingProton_spectra}.

\begin{dunefigure}[Shifted proton energy FD spectra]{fig:missingProton_spectra}
{Predicted distributions of reconstructed neutrino energy for selected \numu (top) and \nue (bottom) events, in \dword{fhc} (left) and \dword{rhc} (right) beam modes in 7 years. The black curve shows the nominal \dword{genie} prediction, while the red points are the mock data, where 20\% of proton energy is shifted to neutrons. The blue curve is the post-fit result, where systematic and oscillation parameters are shifted to match the mock data. The \dword{nd} spectra match the pre-fit prediction by construction and are not shown.}
  \includegraphics[width=0.45\textwidth]{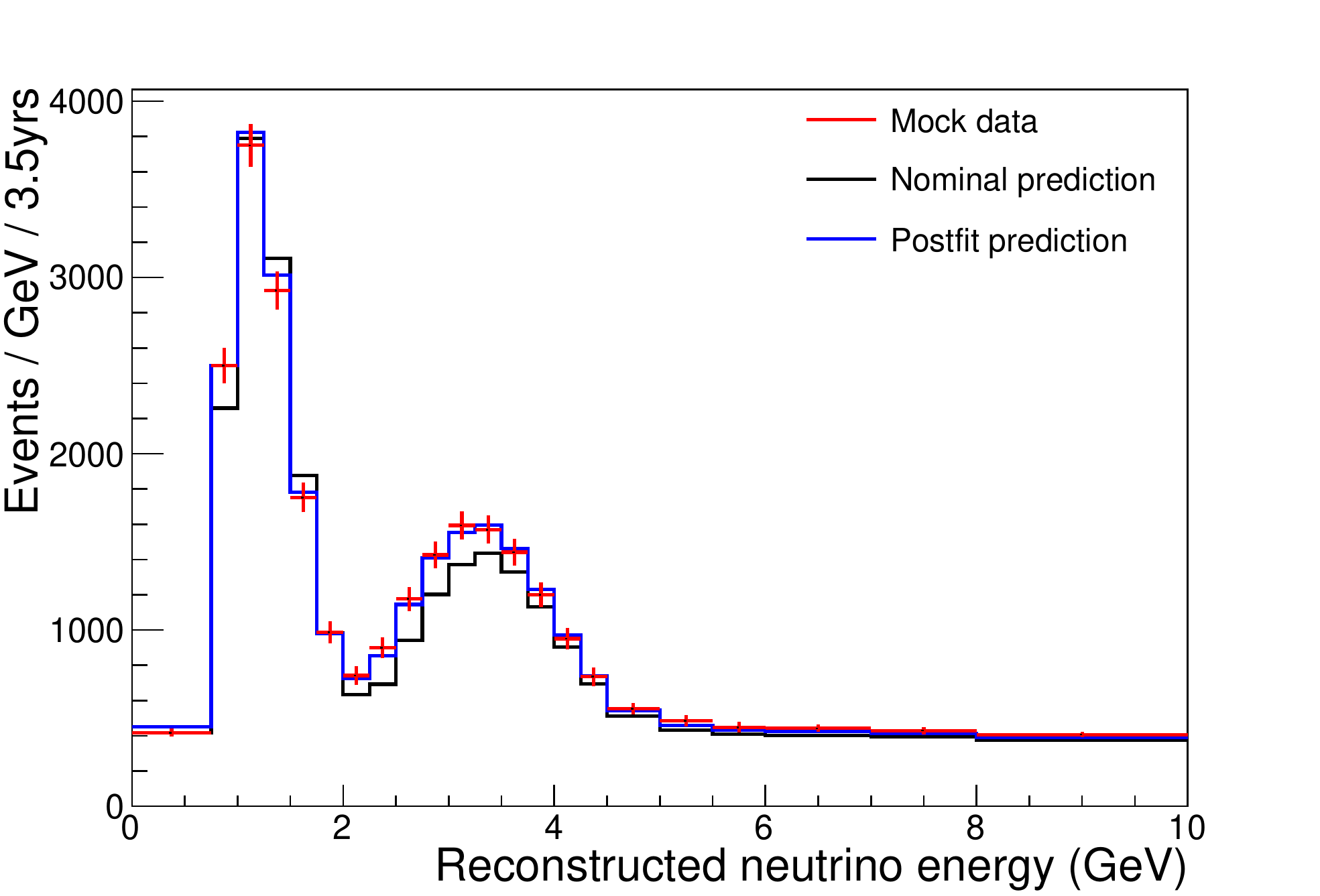}
  \includegraphics[width=0.45\textwidth]{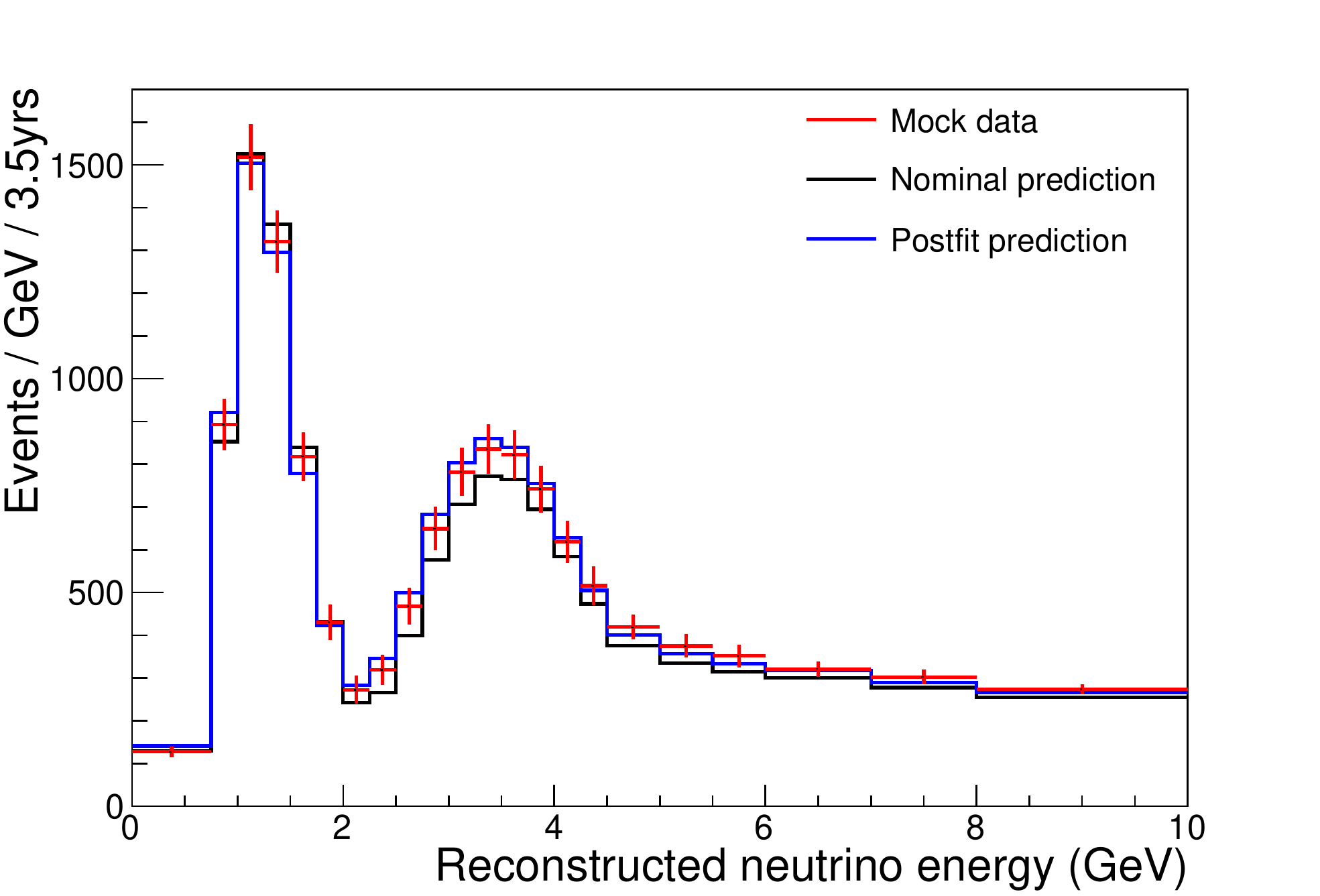}
  \includegraphics[width=0.45\textwidth]{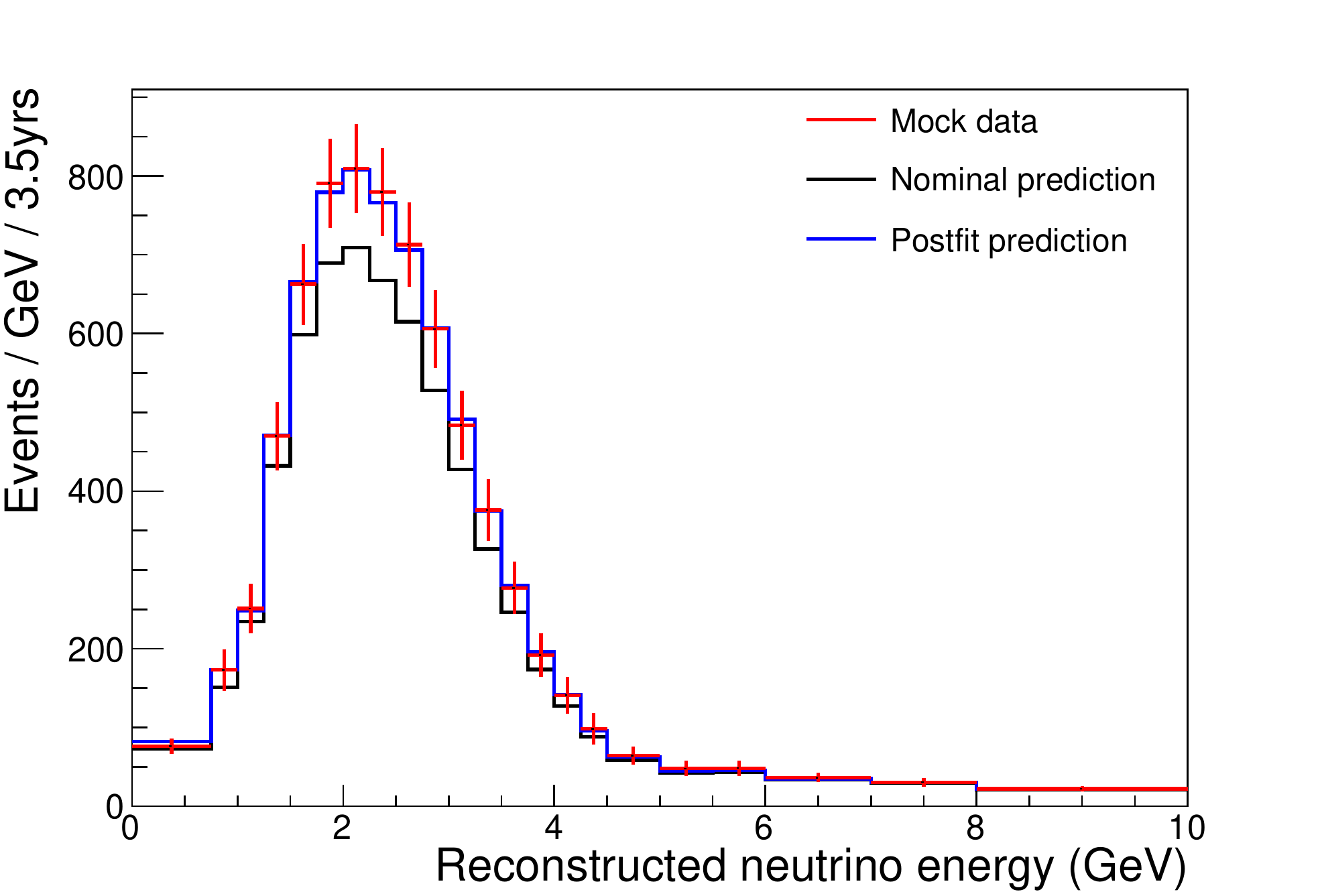}
  \includegraphics[width=0.45\textwidth]{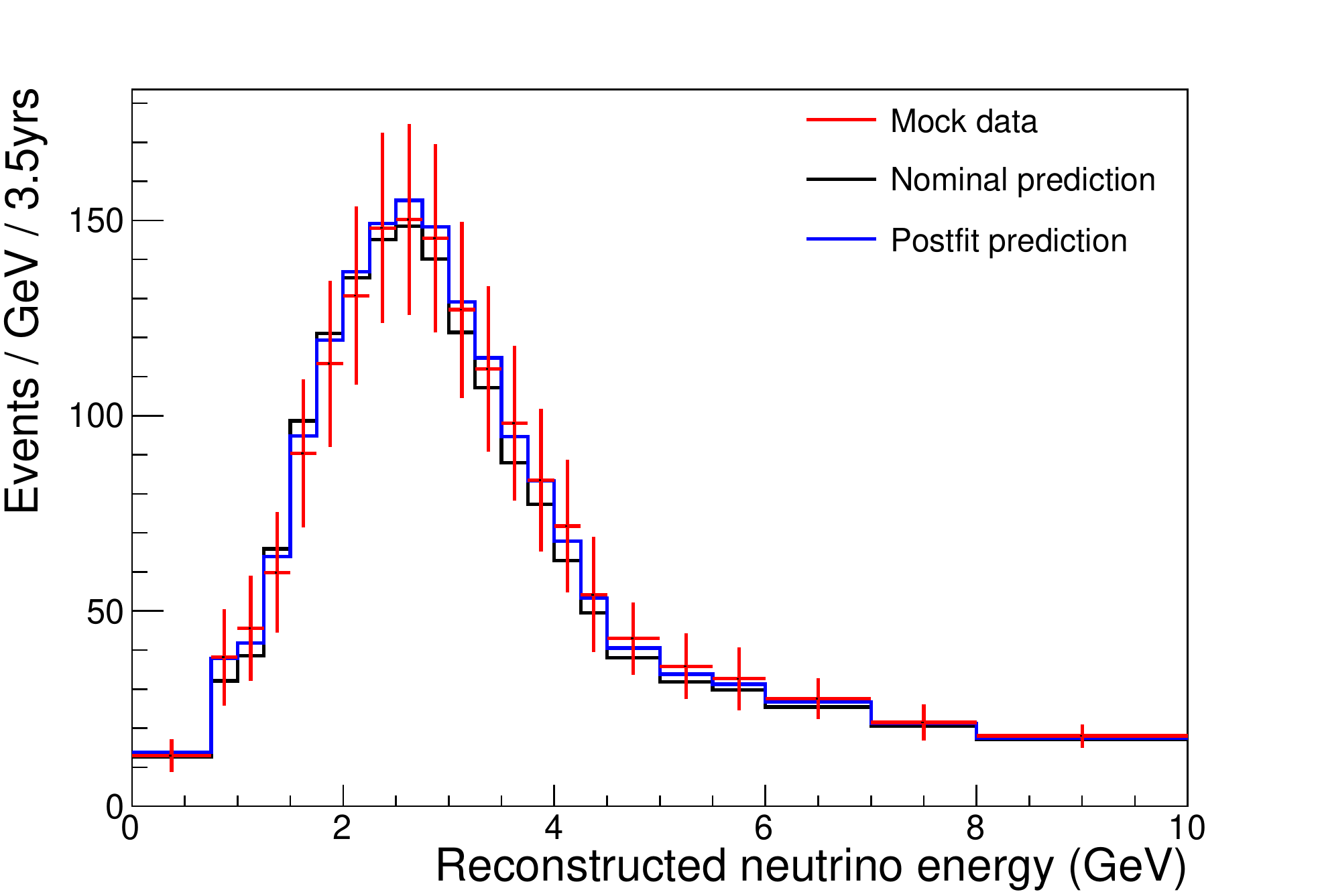}
\end{dunefigure}

Measured oscillation parameters returned by this fit are biased with respect to their true values. In particular, the best-fit values of \dm{32} and sin$^{2}\theta_{23}$ are significantly incorrect, as shown in Figure~\ref{fig:missingProton_dm2th23}. Other parameters, including $\delta_{CP}$, happen not to be pulled significantly from their true values by this particular model variation.

\begin{dunefigure}[\dm{32}-sin$^{2}\theta_{23}$ contour for shifted proton energy]{fig:missingProton_dm2th23}
{Results of a fit to mock data where 20\% of proton energy is shifted to neutrons. The true values of \dm{32} and sin$^{2}\theta_{23}$ are given by the star, while the allowed 90\% C.L. regions are drawn around the best-fit point, for 7, 10, and 15 years of exposure. The solid region shows the result for a fit using the mock data, while the dashed curve shows the result for a fit using nominal simulation, for comparison.}
  \includegraphics[width=0.5\textwidth]{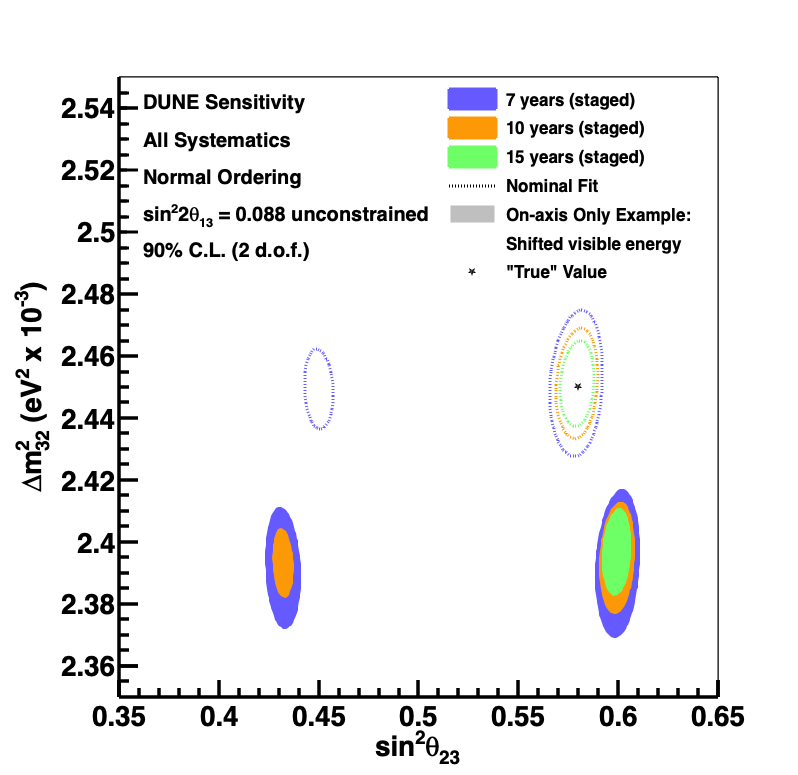}
\end{dunefigure}

While the nominal model gives a good fit to the mock data in the on-axis \dword{nd}, reconstructed spectra from off-axis \dword{nd} data give a poor fit. This occurs because the cancellation between the cross section shift and the final-state proton-to-neutron ratio is dependent on the true neutrino energy spectrum. Off-axis data access different neutrino energy spectra, where the relationship is broken. By combining data at many off-axis positions, it is possible to produce a data-driven prediction of the expected \dword{fd} flux for a given set of oscillation parameters, and directly compare this to the observation. Such a technique is not possible with solely on-axis \dword{nd} data. This example demonstrates the importance of a capable \dword{nd}, including the capability for off-axis measurements, to constrain not only the uncertain parameters of the interaction model, but also the physics in the model itself.

\section{Conclusion}
\label{sec:physics-lbnosc-conclude}

The studies presented in this chapter are based on full, end-to-end simulation, reconstruction, and event selection of \dword{fd} Monte Carlo and parameterized analysis of \dword{nd} Monte Carlo. Detailed uncertainties from flux, the neutrino interaction model, and detector effects have been included in the analysis. Sensitivity results are obtained using a sophisticated, custom fitting framework. These studies demonstrate that DUNE will be able to achieve its primary physics goals of measuring \deltacp to high precision, unequivocally determining the neutrino mass ordering, and making precise measurements of the oscillation parameters governing long-baseline neutrino oscillation. It has also been demonstrated that accomplishing these goals relies upon accumulated statistics from a well-calibrated, full-scale FD,  operation of a 1.2-MW beam upgraded to 2.4~MW, and detailed analysis of data from a highly capable ND.

DUNE will be able to establish the neutrino mass ordering at the 5$\sigma$ level for 100\% of \deltacp values after between two and three years. CP violation can be observed with 5$\sigma$ significance after about 7 years if \deltacp = $-\pi/2$ and after about 10 years for 50\% of \deltacp values. CP violation can be observed with 3$\sigma$ significance for 75\% of \deltacp values after about 13 years of running. For 15 years of exposure, \deltacp resolution between five and fifteen degrees are possible, depending on the true value of \deltacp. The DUNE measurement of \sinstt{13} approaches the precision of reactor experiments for high exposure, allowing measurements that do not rely on an external \sinstt{13} constraint and facilitating a comparison between the DUNE and reactor \sinstt{13}  results, which is of interest as a potential signature for beyond the standard model physics. DUNE will have significant sensitivity to the $\theta_{23}$ octant for values of \sinst{23} less than about 0.47 and greater than about 0.55.

These measurements will make significant contributions to completion of the standard three-flavor 
mixing picture and guide theory in understanding if there are new symmetries in the neutrino sector or whether there is a relationship between the generational structure of quarks and leptons. Observation of CP violation in neutrinos would be an important step in understanding the origin of the baryon asymmetry of the universe. Precise measurements made in the context of the three-flavor paradigm may also yield inconsistencies that point us to physics beyond the standard three-flavor model. 



\cleardoublepage

\chapter{GeV-Scale Non-accelerator Physics Program}
\label{ch:nonaccel}

\section{Nucleon Decay}
\label{sec:nonaccel-ndk}

Unifying three of the fundamental forces in the universe, the strong, 
electromagnetic, and weak interactions, is a shared goal for the current 
world-wide program in particle physics. \Dwords{gut}, extending the standard model of particle physics to include a unified gauge symmetry 
at very high energies  (more than \SI{1e15}{\GeV}), predict a number of observable 
effects at low energies, such as nucleon  decay \cite{Pati:1973rp,Georgi:1974sy,Dimopoulos:1981dw,Langacker:1980js,deBoer:1994dg,Nath:2006ut}. 
Since the early 1980s, supersymmetric \dword{gut} models were preferred for a number of reasons, including gauge-coupling unification, natural embedding in superstring theories, and their ability to solve the fine-tuning problem of the \dword{sm}.  Supersymmetric \dword{gut} models generically predict that the dominant proton decay mode is \ptoknubar, in contrast to non-supersymmetric \dword{gut} models, which typically predict
the dominant decay mode to be \ptoepizero.  Although the LHC has not found evidence for \dword{susy} at the electroweak scale as was expected if \dword{susy} were to solve the gauge hierarchy problem in the \dword{sm}, the appeal of a \dword{gut} still remains. In particular, gauge-coupling unification can still be achieved in non-supersymmetric \dword{gut} models by the introduction of one or more intermediate scales (see, for example, \cite{Altarelli:2013aqa}).
Several experiments have sought signatures of nucleon decay, with the best limits for most decay modes set by the \superk experiment~\cite{Abe:2014mwa,Miura:2016krn,TheSuper-Kamiokande:2017tit}, 
which features the largest sensitive mass and exposure to date. 

Although no evidence for proton decay has been found, lifetime limits from the current generation of experiments already constrain many \dword{gut} models, as shown in Figure~\ref{fig:theoryexplimitsummary}~(updated from~\cite{Babu:2013jba}). In some cases, these limits have eliminated models and approach the upper bounds of what other models will allow. This situation points naturally toward continuing the search with new, highly capable underground detectors, especially those with improved sensitivity to specific proton decay modes favored by \dword{gut} models. Given \superk's long exposure time (more than \SI{30}{years}), extending the lifetime limits will require detectors with long exposure times coupled with larger sensitive mass or improved detection efficiency and background rejection.  

The excellent imaging, as well as calorimetric and particle identification capabilities, of the \dword{lartpc} technology  implemented for the \dword{dune} \dword{fd} will exploit a number of complementary signatures for a broad range of nucleon decay channels.  Should nucleon decay rates lie just beyond current limits, observation of even one or two candidate events with negligible background could constitute compelling evidence.

In the \dword{dune} era, possibly two other large detectors, \hyperk~\cite{Abe:2018uyc} and JUNO~\cite{Djurcic:2015vqa} will be conducting nucleon decay searches. Should a signal be observed in any single experiment, confirmation from experiments using different detector technologies, and therefore different backgrounds, would be very powerful.

As mentioned above, the \dword{gut} models present two benchmark decay modes, \ptoepizero and \ptoknubar.  The decay \ptoepizero arises from gauge boson mediation and is often predicted to have the higher branching fraction of the two key modes. In this mode, the total mass of the proton is converted into the electromagnetic shower energy of the positron and two photons from $\pi^0$ decay with a net momentum vector near zero. 
The second key mode is \ptoknubar. This mode is dominant in most supersymmetric \dword{gut} models,
many of which also favor other modes involving kaons in the final state~\cite{Dimopoulos:1981dw}.
Although significant attention will focus on these benchmark modes, the nucleon decay program at \dword{dune} will be a broad effort, covering many possible decay channels.

\begin{dunefigure}[Summary of nucleon decay experimental limits and model predictions]{fig:theoryexplimitsummary}{Summary of nucleon decay experimental lifetime limits from past or currently running experiments for several modes and a set of  model predictions for the lifetimes in the two benchmark modes.  The limits shown are 90\% \dword{cl} lower limits on the partial lifetimes, $\tau/B$, where $\tau$ is the total mean life and $B$ is the branching fraction. Updated from~\cite{Babu:2013jba}.}
\includegraphics[width=0.9\textwidth]{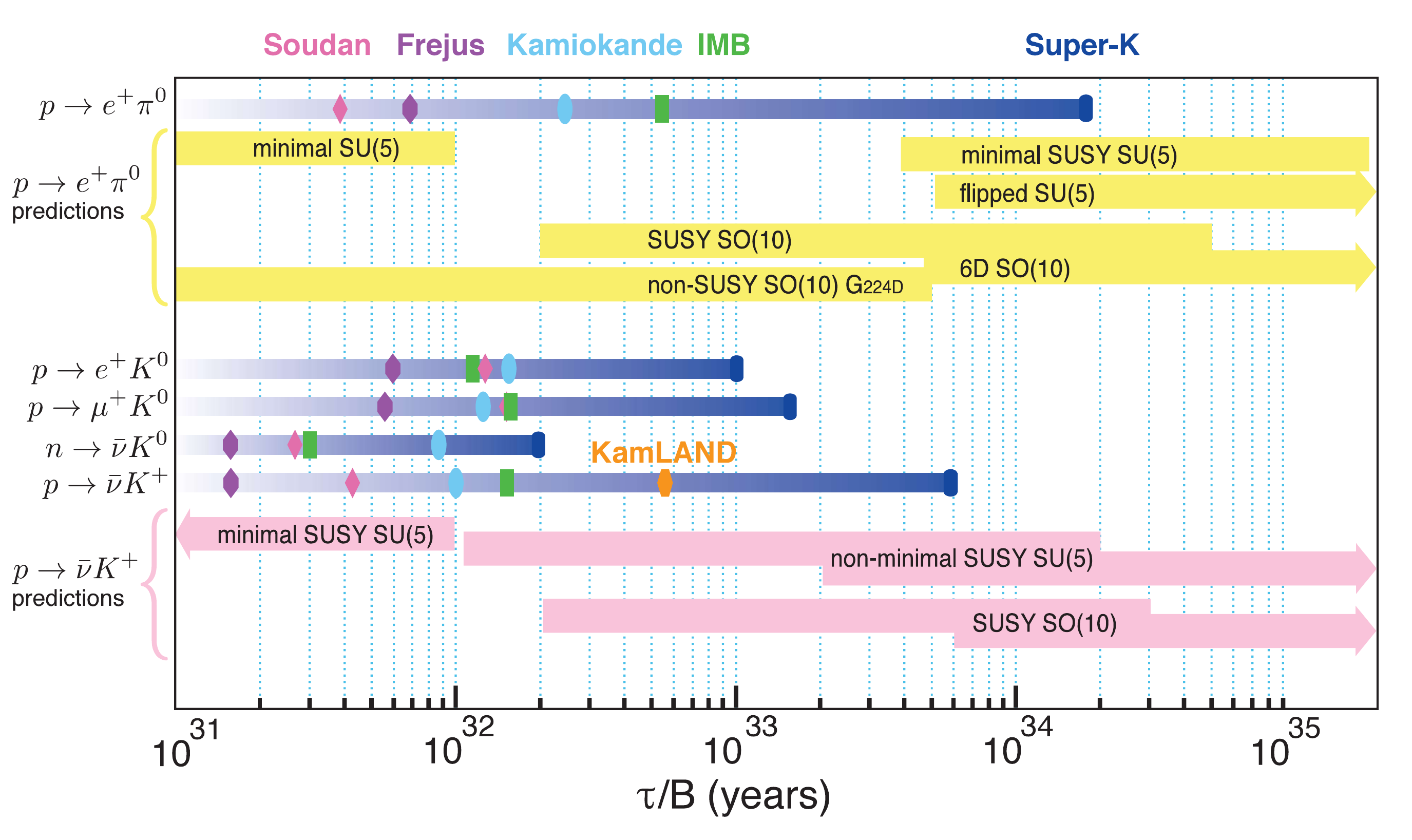}
\end{dunefigure}

\subsection{Experimental Signatures for Nucleon Decay Searches in DUNE}
\label{subsec:nonaccel-ndk-dune}

The \dword{dune} \dword{fd}, with the largest active volume of \dword{lar}, 
will be highly sensitive to several possible nucleon decay modes, 
in many cases complementing the capabilities of large water detectors.
In particular, \dword{lartpc} technology offers the opportunity to observe the entire decay chain for nucleon decays into charged kaons; in \ptoknubar, the kaon is typically below \cherenkov threshold in a water \cherenkov detector, but can be identified by its distinctive $dE/dx$ signature as well as by its decay in a \dword{lartpc}.
Therefore, this mode can be tagged in a \dword{lartpc} if a single kaon within a proper energy/momentum range can be reconstructed with its point of origin lying within the fiducial volume followed by a known decay mode of the kaon.
Background events initiated by cosmic-ray muons can be controlled  by requiring no activity close to the edges of the \dwords{tpc} and by stringent single kaon identification within the energy range of interest~\cite{bib:docdb3384,bib:docdb1752}.  Atmospheric neutrinos make up the dominant background.

Because of the already stringent limits set by \superk on \ptoepizero and the unique ability to track and identify kaons in a \dword{lartpc}, the initial nucleon decay studies in \dword{dune} have focused on nucleon decay modes featuring kaons.  Studies of \ptoepizero have begun (see Section~\ref{subsec:nonaccel-ndk-other}) but are less advanced than the kaon studies.  The remainder of this section describes the background assumptions, signal simulation, particle tracking and identification, and event classification with a focus on nucleon decay involving kaons.

\subsubsection{Background Simulation}
\label{sec:ndkbkgd}

The main background for nucleon decay searches is in the interactions of  atmospheric neutrinos. In this analysis, the Bartol model of atmospheric neutrino flux~\cite{Barr:2004br} is used.
Neutrino interactions in argon are simulated with the \dword{genie} event generator~\cite{Andreopoulos:2009rq}. To estimate the event rate, we integrate the product of the neutrino flux and interaction cross section.
Table \ref{tab:rate} shows the event rate for different neutrino species for an exposure of \SI{10}{\ktyr}, where oscillation effects are not included.

\begin{dunetable}
[Expected rate of atmospheric neutrino interactions]
{cccc}
{tab:rate}
{Expected rate of atmospheric neutrino interactions in \argon40 for a \SI{10}{\ktyr} exposure (not including oscillations).}
  ~\SI{10}{\ktyr}~   &~CC~&~NC~&~Total \\
\numu & \num{1038} & \num{398} &\num{1436} \\
\anumu &\num{280} & \num{169} & \num{449} \\
\nue & \num{597} &  \num{206} &\num{803} \\
\anue & \num{126} & \num{72} & \num{198} \\
Total & \num{2041} & \num{845} & \num{2886} \\
\end{dunetable}

Thus, to suppress atmospheric neutrino background to the level of one event per \si{\Mtyr}, which would yield \num{0.4} events after ten years of operation with a \SI{40}{\kt} fiducial volume, the necessary background rejection is $1 - (1/288600) = 1 - 3\times10^{-6} = 0.999997$, where background rejection is defined as the fraction of background that is not selected.

\subsubsection{Nucleon Decay Simulation}
\label{sec:ndksim}

The simulation of nucleon decay events is performed using GENIE v.2.12.10. 
A total of \num{68} single-nucleon exclusive decay channels listed in the 2016 update of the \dword{pdg}~\cite{Tanabashi:2018oca} 
is available in \dword{genie} (see Table~\ref{tab:genie_ndk}). 
The list includes two-, three-, and five-body decays. 
If a bound nucleon decays, the remaining nucleus can be in an excited state and will typically de-excite by emitting nuclear fission fragments, nucleons, and photons. At present, de-excitation photon emission is simulated only for oxygen~\cite{Andreopoulos:2015wxa}.  However, the \argoneut collaboration~\cite{Acciarri:2018myr} has reported measurements of argon de-excitation photons in \dword{lartpc} detectors,
where energy depositions and positions of these depositions have been compared to those from simulations of neutrino-argon interactions using the FLUKA Monte Carlo generator.

\begin{table} 
  \includegraphics[width=\linewidth]{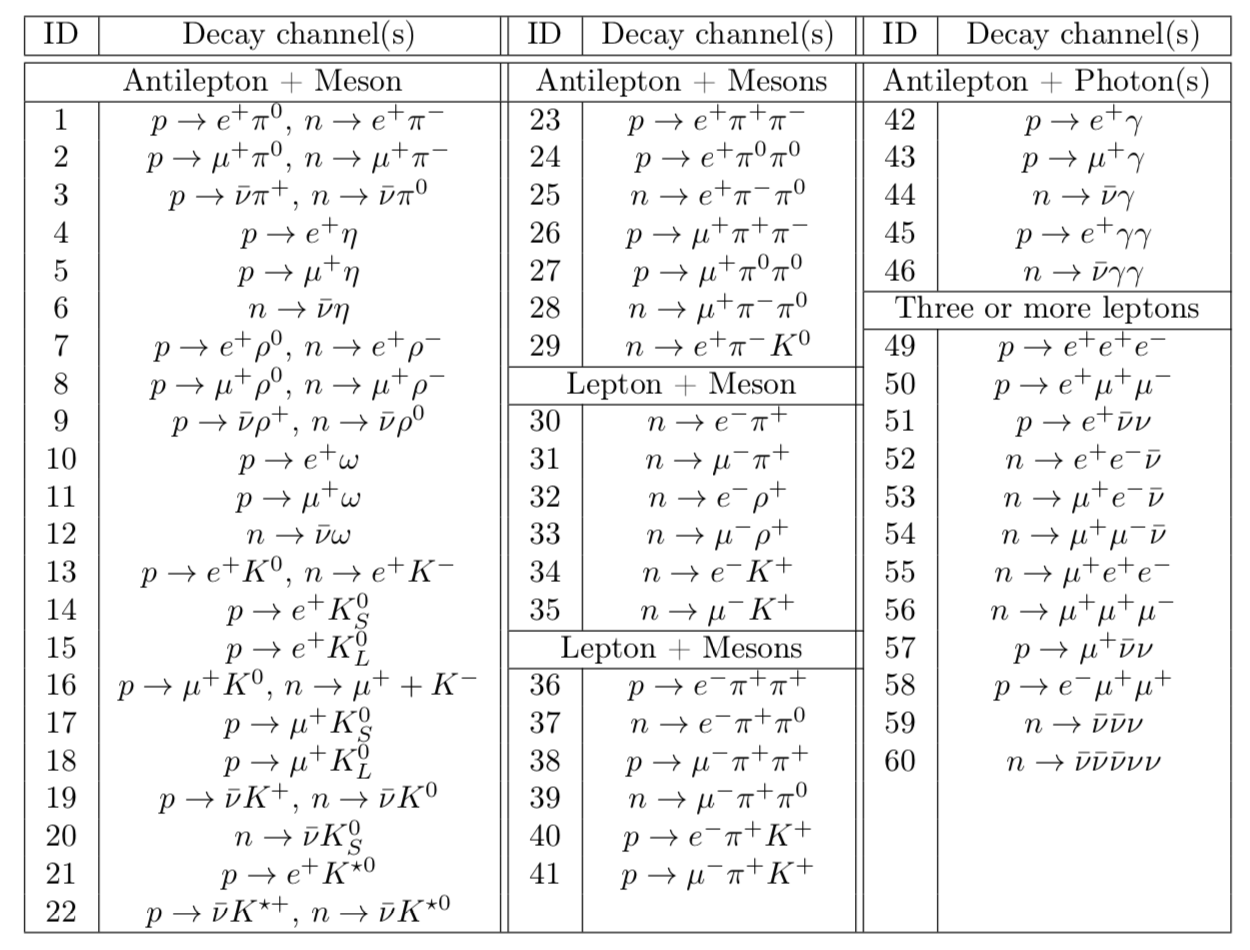}
  \caption[GENIE nucleon decay topologies]{Decay topologies considered in \dword{genie} nucleon decay simulation.}
  \label{tab:genie_ndk}
\end{table}

\subsubsection{Kaon Final State Interactions}
\label{sec:final-state-interactions}

The propagation of the decay products in the nucleus is simulated using an intranuclear cascade \dword{mc}. 
Charged kaons can undergo various scattering processes in the nucleus: elastic scattering, charge exchange, absorption (only $K^{-}$; $K^{+}$ absorption is forbidden), and $K^{+}$ production via strong processes such as $\pi^{+}n \rightarrow K^{+} \Lambda$.  In this analysis, the $hA2015$ model in \dword{genie} is used as the default model for these \dword{fsi}.  $hA2015$ is an empirical, data-driven method that does not model the cascade of hadronic interactions step by step, but instead uses one effective interaction where hadron+nucleus data is used to determine the final state.
For kaons, $K^{+}+C$ data~\cite{Bugg:1968zz,Friedman:1997eq}
is used when available. $hA2015$ only considers kaon-nucleon elastic scattering inside the nucleus.  Charge exchange is not included, nor is $K^+$ production in pion reactions, and therefore a $K^+$ is never added or removed from the final state in this model. 

Other \dword{fsi} models include the full cascade, but there is not enough data to favor one model over the other.  As an example of the limitations of the current data on kaon \dword{fsi}, a recent measurement of kaon production in neutrino interactions shows only a weak preference for including \dword{fsi} as opposed to a model with no \dword{fsi}~\cite{Marshall:2016rrn}.  In this case, the kaon \dword{fsi} have a relatively subtle effect on the differential cross section, and the available statistics are not sufficient to conclusively prefer one model over another.  For nucleon decay into kaons, the \dword{fsi} have a much larger impact, and the differences between models are less significant than the overall effect.  Kaon \dword{fsi} introduce an important uncertainty that is included in this analysis.

\dword{fsi} can significantly modify the observable distributions in the detector.  For example, Figure~\ref{fig:K-wFSI-hA2015} shows the kinetic energy of a kaon from \ptoknubar before and after \dword{fsi}. Because of \dword{fsi} the kaon spectrum becomes softer on average. Of the kaons, \num{31.5}\%  undergo elastic scattering resulting in events with very low kinetic energy;  \num{25}\% of kaons have a kinetic energy of $\le\SI{50}{\MeV}$. When the kaon undergoes elastic scattering, a nucleon can be knocked out of the nucleus. Of decays via this channel, \num{26.7}\%  have one neutron coming from \dword{fsi}, \num{15.3}\% have at least one proton, and \num{10.3}\% have two protons coming from \dword{fsi}. These secondary nucleons are detrimental to reconstructing and selecting  $K^{+}$.

\begin{dunefigure}[Kaon kinetic energy before and after final state interactions]{fig:K-wFSI-hA2015}{Kinetic energy of kaons in simulated proton decay events, \ptoknubar.  The kinetic energy distribution is shown before and after final state interactions in the argon nucleus.}
\includegraphics[width=0.8\textwidth]{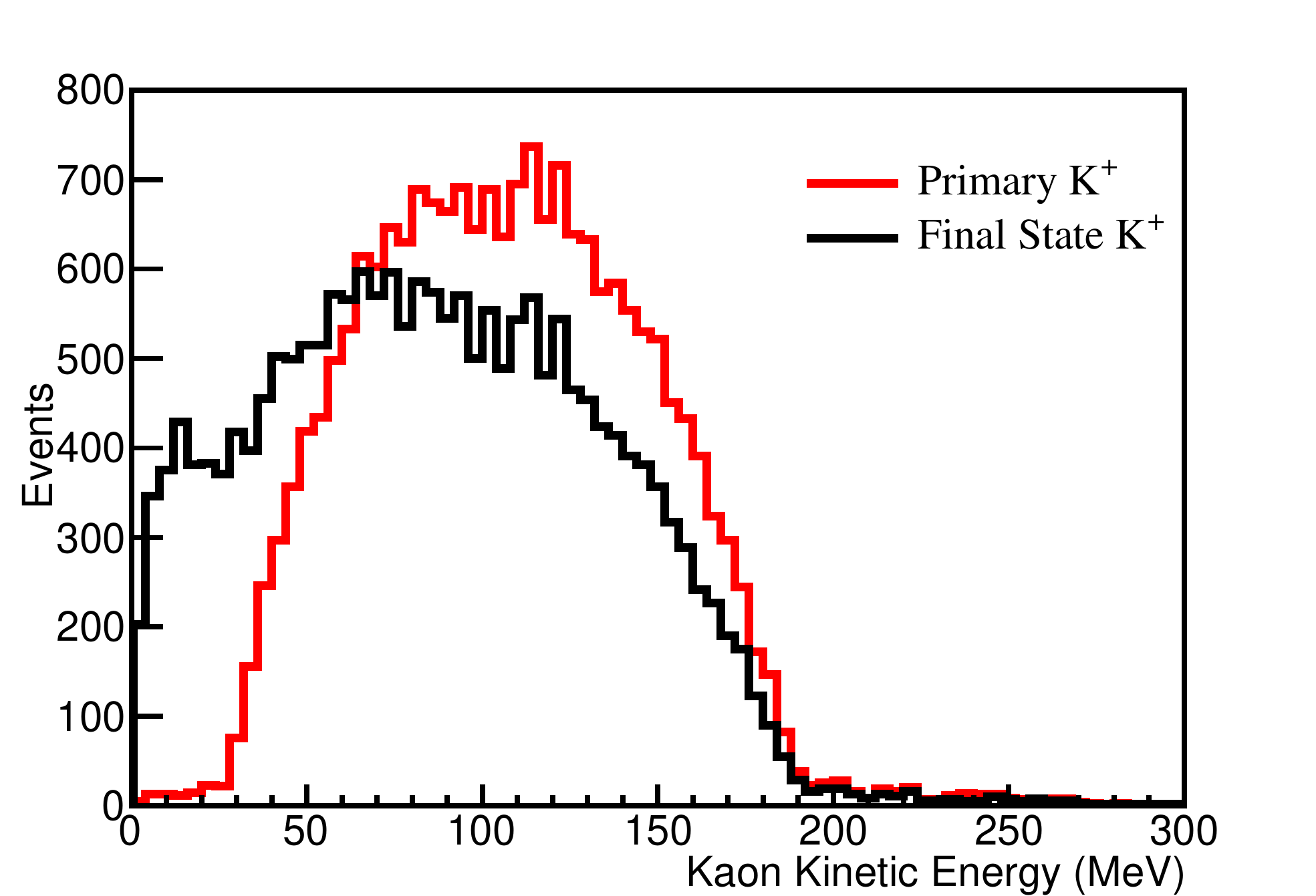}
\end{dunefigure}

The kaon \dword{fsi} in \superk's simulation of \ptoknubar in oxygen seem to have a smaller effect on the outgoing kaon momentum distribution~\cite{Abe:2014mwa} than is seen here with the \dword{genie} simulation on argon.  Some differences are expected due to the different nuclei, but differences in the \dword{fsi} models are under investigation.

\subsubsection{Tracking and Particle Identification}
\label{sec:event-reconstruction}

The \dword{dune} reconstruction algorithms are described in Chapter~\ref{ch:tools}.  This analysis uses \threed track and vertex reconstruction provided by \dword{pma}.

Track reconstruction efficiency for a charged particle $x^{\pm}$ is defined as 
\begin{equation}
\epsilon_{x^{\pm}} = \frac{\mbox{$x^{\pm}$ particles with a reconstructed track}}{\mbox{events with $x^{\pm}$ particle }}.
\end{equation}
The denominator includes events in which an $x^{\pm}$ particle was created and has deposited energy within any of the \dwords{tpc}.  The numerator includes events in which an $x^{\pm}$ particle was created and has deposited energy within any of the \dwords{tpc}, and a reconstructed track can be associated to the $x^{\pm}$ particle based on the number of hits generated by that particle along the track. This efficiency can be calculated as a function of true kinetic energy and true track length.

\begin{dunefigure}[Tracking efficiency of kaons in \ptoknubar]{fig:k-trk-eff}{Tracking efficiency for kaons in simulated proton decay events, \ptoknubar, as a function of kaon kinetic energy (left) and true path length (right).}
\includegraphics[width=0.4\textwidth]{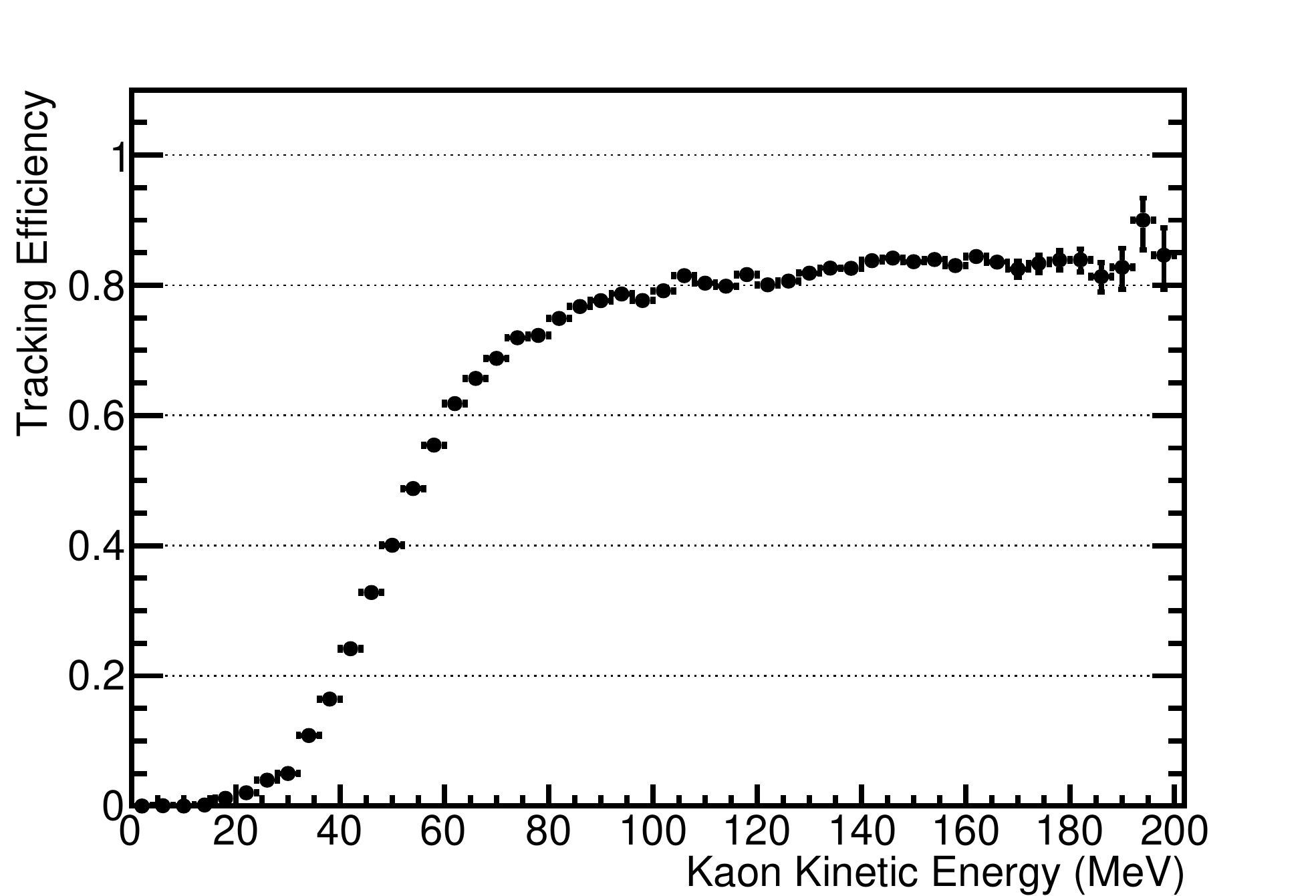}
\includegraphics[width=0.4\textwidth]{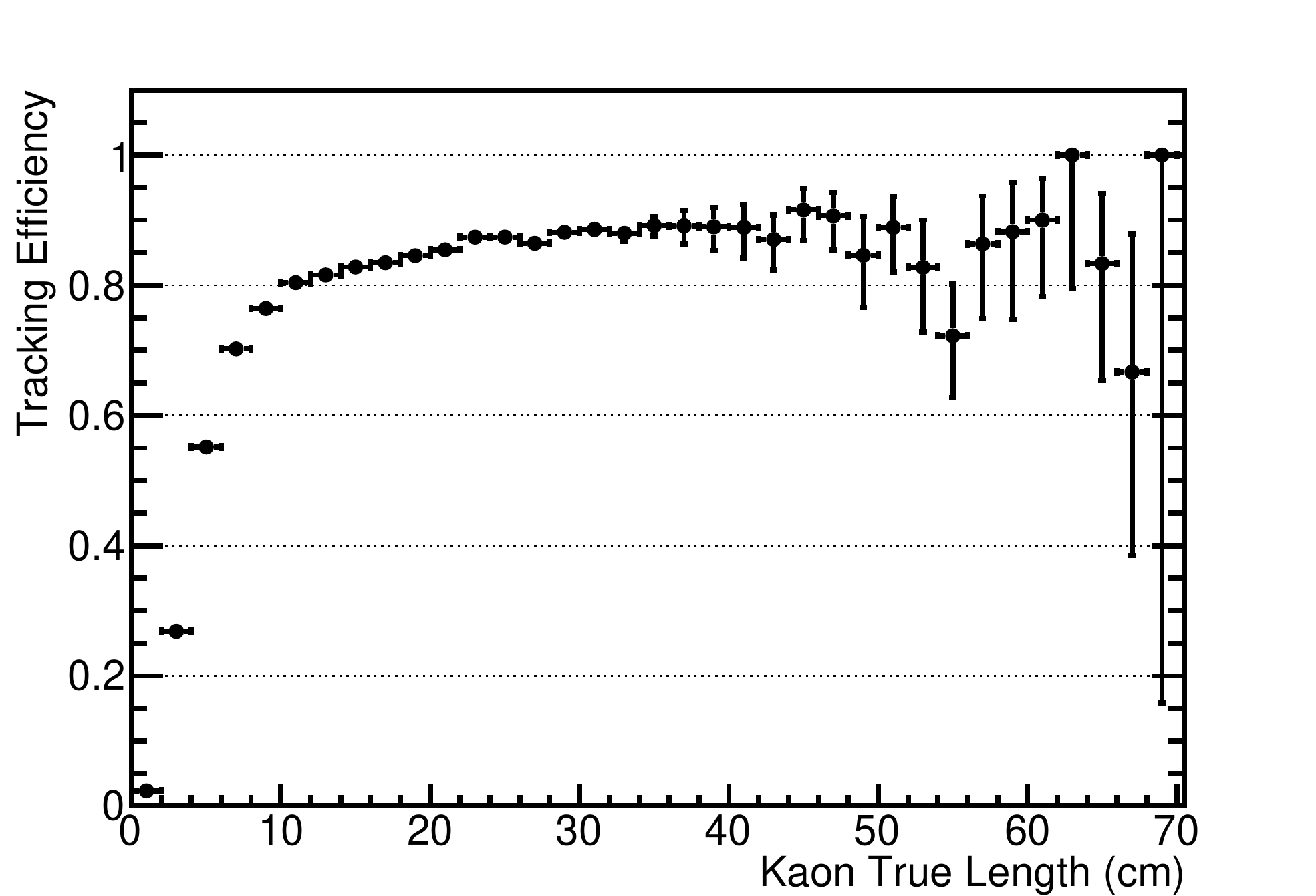}
\end{dunefigure}

Figure~\ref{fig:k-trk-eff} shows the tracking efficiency for $K^{+}$  from proton decay via \ptoknubar as a function of true kinetic energy and true path length. The overall tracking efficiency for kaons is \num{58.0}\%, meaning that \num{58.0}\% of all the simulated kaons are associated with a reconstructed track in the detector.  From Figure~\ref{fig:k-trk-eff}, the tracking threshold is approximately $\sim\SI{40}{\MeV}$ of kinetic energy, which translates to $\sim\SI{4.0}{\cm}$ in true path length.  The biggest loss in tracking efficiency is due to kaons with $<\SI{40}{\MeV}$ of kinetic energy due to scattering inside the nucleus as described in Section~\ref{sec:final-state-interactions}.  The efficiency levels off to approximately \num{80}\% above \SI{80}{\MeV} of kinetic energy.  This inefficiency even at high kinetic energy is due mostly to kaons that decay in flight./footnote{No attempt has been made at this point to recover such events.}
Both kaon scattering in the \dword{lar} and charge exchange are included in the simulation but are relatively small effects (\num{4.6}\% of kaons scatter in the \dword{lar} and \num{1.2}\% of kaons experience charge exchange).   The tracking efficiency for muons from the decay of the $K^{+}$ in \ptoknubar is \num{90}\%.

Charged particles lose energy through ionization and scintillation when traversing the \dword{lar}. This energy loss provides valuable information on particle energy and species. To identify a given particle, the hits associated with a reconstructed track are used.
If the charged particle stops in the \dword{lartpc} active volume, a combination of $dE/dx$ and the reconstructed residual range ($R$, the path length to the end point of the track) is used to define a parameter for \dword{pid}.  The parameter, $PIDA$, 
is defined as~\cite{Acciarri:2013met}  
\begin{equation}
PIDA = \left\langle \left(\frac{dE}{dx}\right)_{i}R^{0.42}_{i}\right\rangle,\label{eqn:PIDA}
\end{equation}
where the median is taken over all track points $i$ for which the residual range $R_i$ is less than \SI{30}{\cm}.

\begin{dunefigure}[Particle identification using $PIDA$ for \ptoknubar]{fig:PIDA}{Particle identification using $PIDA$ for muons and kaons in simulated proton decay events, \ptoknubar, and protons in simulated atmospheric neutrino background events.  The curves are normalized by area.}
\includegraphics[width=0.8\textwidth]{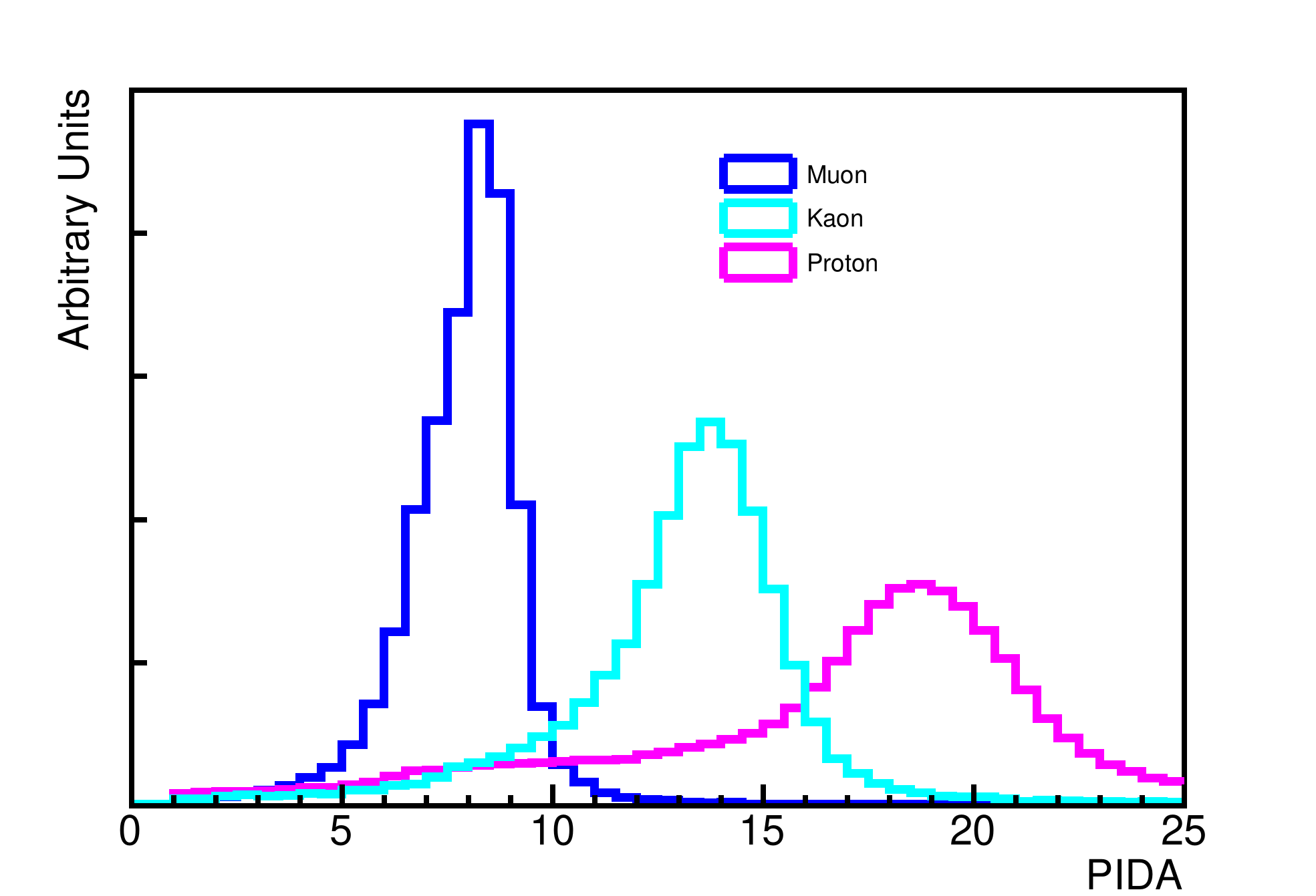}
\end{dunefigure}

Figure~\ref{fig:PIDA} shows the $PIDA$ performance for kaons (from proton decay), muons (from kaon decay), and protons produced by atmospheric neutrino interactions. The tail with lower values in each distribution is due to cases where the decay/stopping point was missed by the track reconstruction. The tail with higher values is caused when a second particle overlaps at the decay/stopping point causing higher values of $dE/dx$ and resulting in higher values of $PIDA$. In addition, ionization fluctuations smear out these distributions.

A complication for \dword{pid} via $dE/dx$ results when ambiguity occurs in reconstructing track direction, which is even more problematic because additional energy deposition may occur at the originating point in events where \dword{fsi} is significant.  The dominant background to \ptoknubar in \dword{dune} is atmospheric neutrino \dword{cc} \dword{qe}
scattering, $\nu_{\mu} n \rightarrow \mu^{-} p$.  When the muon happens to have very close
to the \SI{237}{\MeV$/c$} momentum expected from a $K^{+}$ decay at rest and does not capture, it is indistinguishable from the muon resulting from \ptoknubar followed by $K^{+} \rightarrow \mu^{+}\nu_{\mu}$. When
the proton is also mis-reconstructed as a kaon, this background mimics the signal process.  

The most important difference between signal and this background source is the direction of the hadron track. For an atmospheric neutrino, the proton and muon originate from the same neutrino interaction point, and the characteristic Bragg rise occurs at the end of the proton track farthest from the muon-proton vertex. For signal, the kaon-muon vertex location is where the $K^{+}$ stops and decays at rest, so its ionization energy deposit is highest near the kaon-muon vertex.  To take advantage of this difference, a log-likelihood ratio discriminator is used to distinguish signal from background.  Templates are formed by taking the reconstructed and calibrated energy deposit as a function of the number of wires from both the start and end of the $K^{+}$ candidate hadron track. 
Two log-likelihood ratios are computed separately for each track. The first begins at the hadron-muon shared vertex and moves along the hadron track (the ``backward'' direction). The second begins at the other end of the track, farthest from the hadron-muon shared vertex, moves along the hadron track the other way (the ``forward'' direction). For signal events, this effectively looks for the absence of a Bragg rise at the $K^{+}$ start, and the presence of one at the end, and vice versa for background.  At each point, the probability density for signal and background, $P^{sig}$ and $P^{bkg}$, are determined from the templates. Forward and backward log-likelihood ratios are computed as
\begin{align}
 \mathcal{L}_{fwd(bkwd)} = \sum_{i} \log\frac{P^{sig}_i}{P^{bkg}_i}, 
\end{align}
where the summation is over the wires of the track, in either the forward or backward direction.  Using either the forward or backward log-likelihood ratio alone gives some discrimination between signal and background, but using the sum gives better discrimination. While the probability densities are computed based on the same samples, defining one end of the track instead of the other as the vertex provides more information. The discriminator is the sum of the forward and backward log-likelihood ratios:
\begin{align}
    \mathcal{L} = \mathcal{L}_{fwd} + \mathcal{L}_{bkwd}.\label{eqn:L}
\end{align}
Applying this discriminator to tracks with at least ten wires gives a signal efficiency of roughly \num{0.4} with a background rejection of \num{0.99}.

\subsubsection{Event Classification}

Multivariate classification methods based on machine learning techniques have become a fundamental part of most analyses in high-energy physics. 
To develop an event selection to search for nucleon decay, a \dword{bdt}
classifier is used. The software package Toolkit for Multivariate Data Analysis with ROOT (TMVA4)~\cite{Hocker:2007ht}
was used with AdaBoost as the boosted algorithm.  In the analyses presented here, the \dword{bdt} is trained on a sample of \dword{mc} events (\num{50000} events for signal and background) that is statistically independent from the sample of \dword{mc} events used in the analysis (approximately \num{100000} events for signal and \num{600000} events for background.)  This technique is used for the nucleon decay and neutron-antineutron analyses presented below.

As an independent method of identifying nucleon decay events, image classification using a \dword{cnn}
can be performed using \twod images of \dword{dune} \dword{mc} events. The image classification provides a single score value as a metric of whether any given event is consistent with a proton decay, and this score can be used as a powerful discriminant for event identification.  In the analyses presented here, the \dword{cnn} technique alone does not discriminate between signal and background as well as a \dword{bdt}.  For that reason, the \dword{cnn} score is used as one of the input variables in the \dword{bdt} in each analysis.

\subsection{Sensitivity to \ptoknubar Decay}
\label{subsec:nonaccel-ndk-nubarkplus}

Monte Carlo studies of the \ptoknubar signal and corresponding atmospheric neutrino backgrounds have been carried out with the \dword{dune} multipurpose full event simulation and reconstruction software.  As indicated in Section~\ref{sec:event-reconstruction},
they reveal that one of the main challenges in identifying proton decay candidates is suppressing backgrounds arising from the mis-reconstruction of protons as positive kaons. This happens when a \dword{cc} neutrino interaction produces a muon and a recoiling proton, and the primary vertex for neutrino interaction is mislabeled as a secondary vertex where the kaon decays.  Complicating the ability to reject pathological events of this type is the presence of \dword{fsi}, which can shift the spectrum of kaons toward low energies, with possible concurrent emission of nucleons, which together weaken the otherwise distinct energy and $dE/dx$ signature of the kaon. 

The branching fraction for leptonic decay of charged kaons, $K\rightarrow \mu \nu_{\mu}$, is approximately \num{64}\%. The remaining decay modes are semileptonic or hadronic and include charged and neutral pions. The leptonic decay offers a distinguishable topology with a heavy ionizing particle followed by a minimum ionizing particle. In addition, given the kinematics of a proton decay event, \num{92}\% of kaons decay at rest. Using two-body kinematics, the momentum of the muon is approximately \SI{237}{\MeV$/c$}. The reconstructed momentum of the muon offers a powerful discriminating variable to separate signal from background events.  This analysis includes all modes of kaon decay, but the selection strategy so far has focused on kaon decay to muons.

The proton decay signal and atmospheric neutrino background events are processed using the same reconstruction chain and subject to the same selection criteria. There are two pre-selection cuts to remove obvious background. One cut requires at least two tracks, which aims to select events with a kaon plus a kaon decay product (usually a muon).  The other cut requires that the longest track be less than \SI{100}{\cm}; this removes backgrounds from high energy neutrino interactions.  After these cuts, \num{50}\% of the signal and \num{17.5}\% of the background remain in the sample.  The signal inefficiency at this stage of selection is due mainly to the kaon tracking efficiency.

A \dword{cnn} was developed to classify signal and background events that gives \num{99.9}\% background rejection at \num{6}\% signal efficiency.  Better discriminating power is achieved using a \dword{bdt} with \num{14} input variables, including the \dword{cnn} score as one  variable.  The other variables in the \dword{bdt} include numbers of reconstructed objects (tracks, showers, vertices), variables related to visible energy deposition, \dword{pid} variables ($PIDA$, Equation~\ref{eqn:PIDA}, and $\mathcal{L}$, Equation~\ref{eqn:L}), reconstructed track length, and reconstructed momentum.

Figure~\ref{fig:BDT_response} shows the distribution of the \dword{bdt} output for signal and background.

\begin{dunefigure}
[Boosted Decision Tree response for \ptoknubar ]{fig:BDT_response}
{Boosted Decision Tree response for \ptoknubar for signal (blue) and background (red).}
\includegraphics[width=0.8\textwidth]{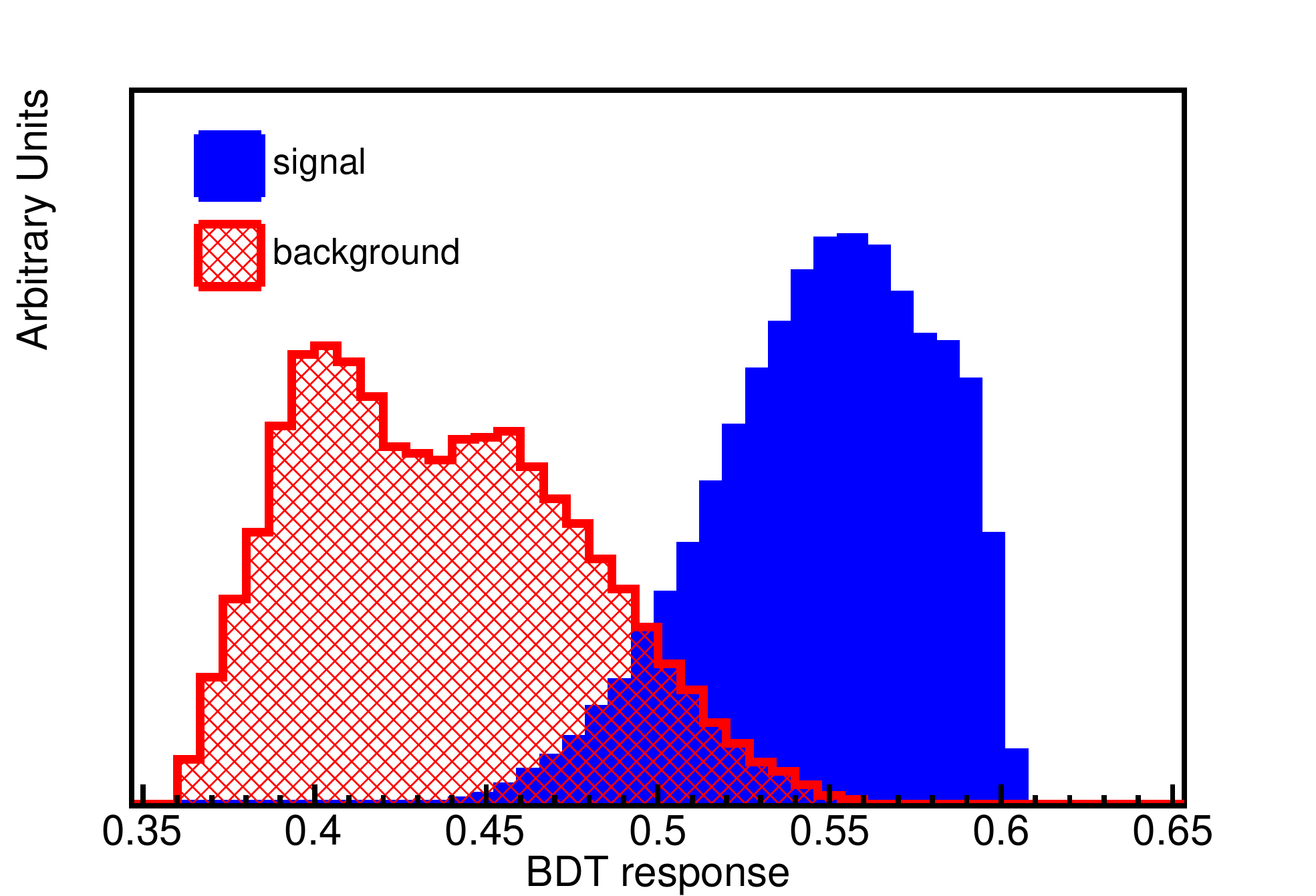}
\end{dunefigure} 

Figure~\ref{fig:event_signal} shows a signal event with high \dword{bdt} response value (\num{0.605}), meaning a well-classified event. The event display shows the reconstructed kaon track in green, the reconstructed muon track from the kaon decay in maroon, and the reconstructed shower from the Michel electron coming from the muon decay in red. Figure~\ref{fig:event_bkgd} shows event displays for atmospheric neutrino interactions.  The left figure (\dword{bdt} response value of \num{0.394}) shows the interaction of an atmospheric electron neutrino, $\nu_{e}+n\rightarrow e^{-}+p+\pi^{0}$.  This event is clearly distinguishable from the signal.  However, the right figure (\dword{bdt} response value \num{0.587}) shows a \dword{cc}\dword{qe} interaction of an atmospheric muon neutrino, $\nu_{\mu}+n \rightarrow \mu^{-}+p$, which is more likely to be mis-classified as a signal interaction. These types of interactions present a challenge if the proton track is misidentified as kaon. A tight cut on \dword{bdt} response can remove most of these events, but this significantly reduces signal efficiency.

\begin{dunefigure}
[\ptoknubar signal event display]{fig:event_signal}
{Event display for a well-classified \ptoknubar signal event.  The vertical axis is time ticks (each time tick corresponds to \SI{500}{\ns}), and the horizontal axis is wire number. The bottom view is induction plane one, middle is induction plane two and top is the collection plane. The color represents the charge deposited in each hit.}
\includegraphics[width=0.8\textwidth]{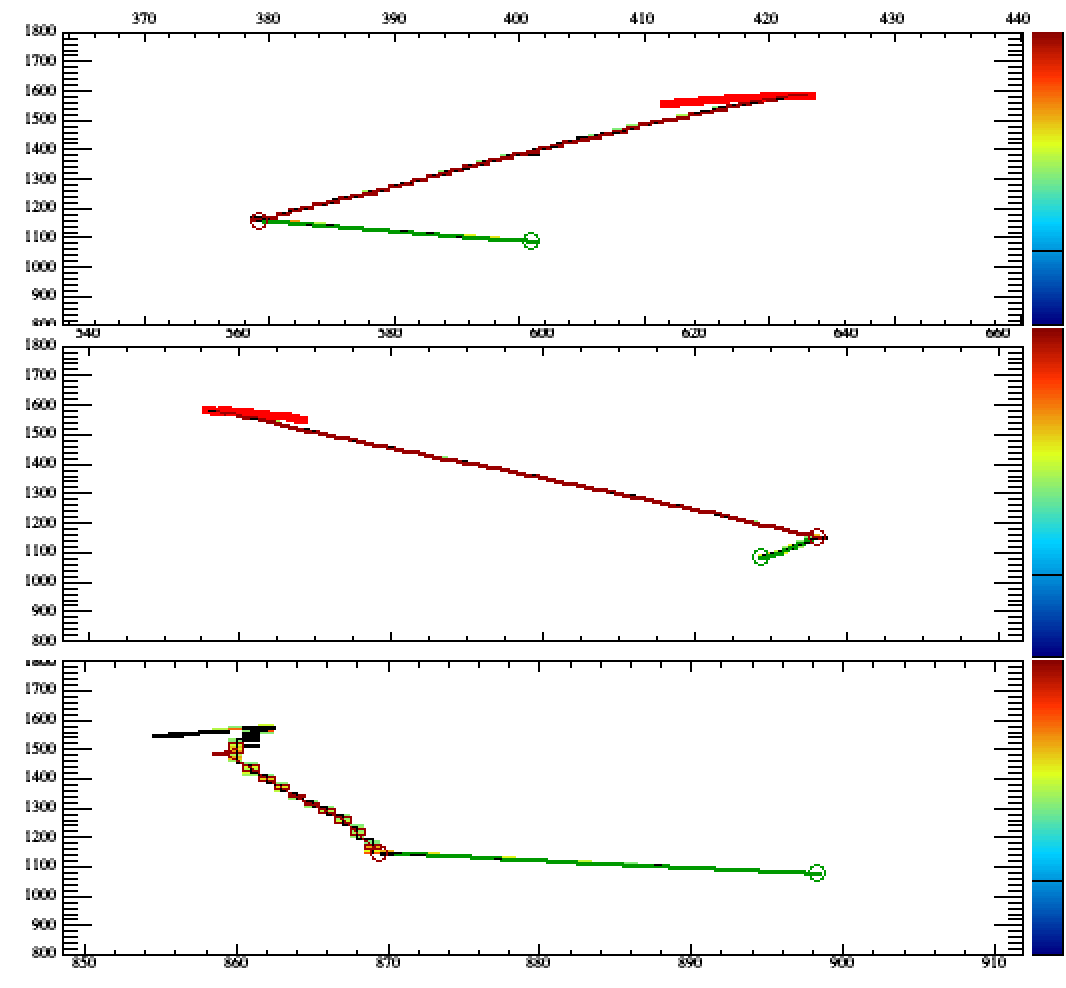}
\end{dunefigure} 

\begin{dunefigure}
[\ptoknubar background event displays]{fig:event_bkgd}
{Event displays for \ptoknubar backgrounds.  The vertical axis is time ticks (each time tick corresponds to \SI{500}{\ns}), and the horizontal axis is wire number. The bottom view is induction plane one, middle is induction plane two and top is the collection plane. The color represents the charge deposited in each hit. The left shows an atmospheric neutrino interaction unlikely to be classified as signal. The right shows an atmospheric neutrino interaction which could make it into the selected sample without a tight cut.}
\includegraphics[width=0.48\textwidth]{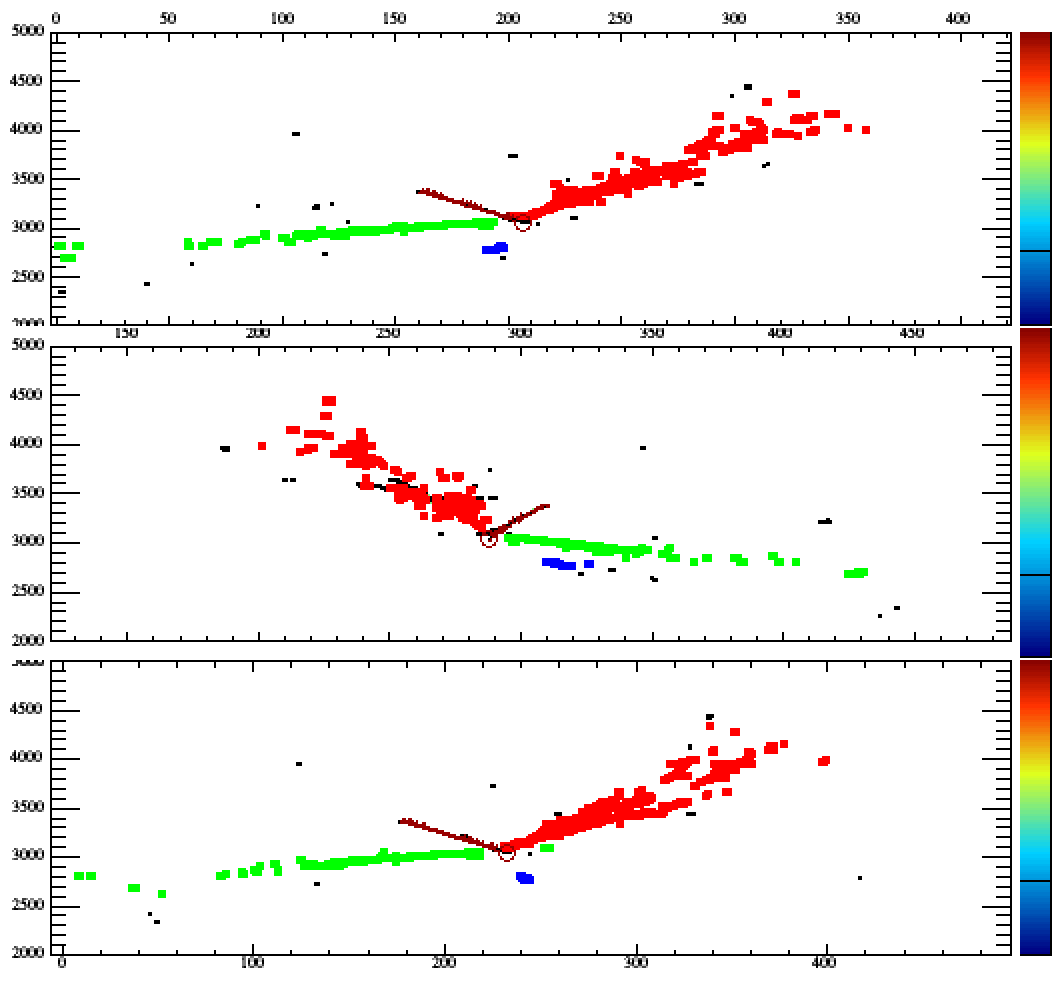}
\includegraphics[width=0.48\textwidth]{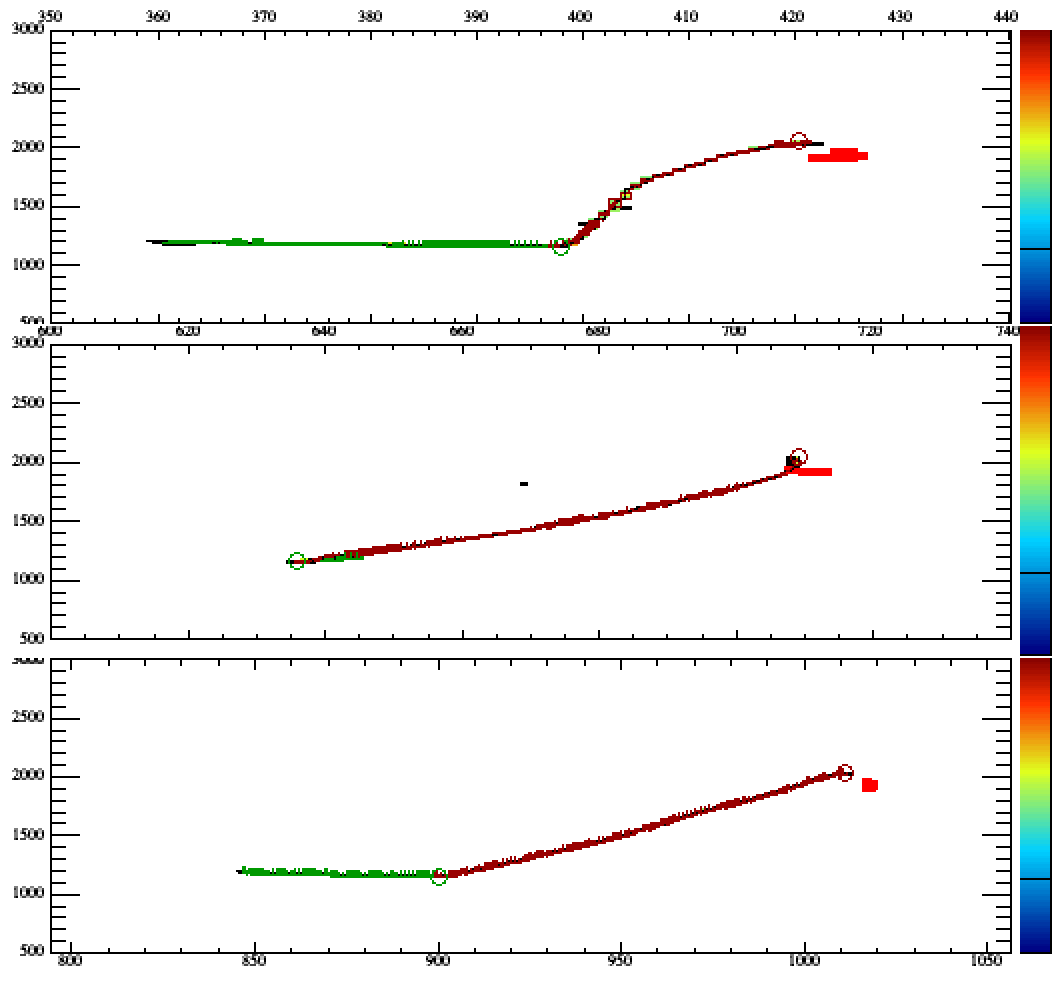}
\end{dunefigure}

Optimal lifetime sensitivity is achieved by combining the pre-selection cuts with a \dword{bdt} cut that gives a signal efficiency of \num{0.15} and a background rejection of 
$0.999997$,
which corresponds to approximately one background event per \si{\Mtyr}. 

The limiting factor in the sensitivity is the kaon tracking efficiency.  With the current reconstruction, 
the overall kaon tracking efficiency is \num{58}\%.
The reconstruction is not yet optimized, and the kaon tracking efficiency should increase with improvements in the reconstruction algorithms.  
To understand the potential improvement, a visual scan of simulated decays of kaons into muons was performed. For this sample of events, with kaon momentum in the \SIrange{150}{450}{\MeV$/c$} range, scanners achieved greater than \num{90}\% efficiency at recognizing the $K^{+} \rightarrow \mu^{+} \rightarrow e^{+}$ decay chain.  The inefficiency came mostly from short kaon tracks (momentum below \SI{180}{\MeV$/c$}) and kaons that decay in flight. Note that the lowest momentum kaons ($<$\SI{150}{\MeV$/c$}) were not included in the study; the path length for kaons in this range would also be too short to track.  Based on this study, the kaon tracking efficiency could be improved to a maximum value of approximately \num{80}\% with optimized reconstruction algorithms, where the remaining inefficiency comes from low-energy kaons and kaons that charge exchange, scatter, or decay in flight.
Combining this tracking performance improvement with some improvement in the $K/p$ separation performance for short tracks, the overall signal selection efficiency improves from \num{15}\% to approximately \num{30}\%.

The analysis presented above is inclusive of all possible modes of kaon decay; however, the current version of the \dword{bdt} preferentially selects kaon decay to muons, which has a branching fraction of roughly \num{64}\%. The second most prominent kaon decay is $K^{+} \rightarrow \pi^{+}\pi^0$, which has a branching fraction of \num{21}\%.  Preliminary studies that focus on reconstructing a $\pi^{+}\pi^0$ pair with the appropriate kinematics indicate that the signal efficiency for kaons that decay via the $K^{+} \rightarrow \pi^{+}\pi^0$ mode is approximately the same as the signal efficiency for kaons that decay via the $K^{+} \rightarrow \mu^{+}\nu_{\mu}$ mode.  This assumption is included in our sensitivity estimates below.

The dominant systematic uncertainty in the signal is expected to be due to the kaon \dword{fsi}. To account for this uncertainty, kaon-nucleon elastic scattering ($K^{+}p(n)\rightarrow K^{+}p(n)$) is re-weighted by $\pm \num{50}\%$ in the simulation. 
The absolute uncertainty on the efficiency with this re-weighting is \num{2}\%, which is taken as the systematic uncertainty on the signal efficiency.
The dominant uncertainty in the background 
is due to the absolute normalization of the atmospheric neutrino rate. The Bartol group has carried out a detailed study of the systematic uncertainties, where the absolute neutrino fluxes have uncertainties of approximately \num{15}\%~\cite{Barr:2006it}.
The remaining uncertainties are due to the cross section models for neutrino interactions.
The uncertainty on the \dword{cc}0$\pi$ cross section in the energy range relevant for these backgrounds is roughly \num{10}\%~\cite{Mahn:2018mai}.
Based on these two effects, a conservative \num{20}\% systematic uncertainty in the background is estimated.

With a \num{30}\% signal efficiency and an expected background of one event per \si{\Mtyr}, a \num{90}\% \dword{cl} lower limit on the proton lifetime in the \ptoknubar channel of \SI{1.3e34}{years} can be set, assuming no signal is observed over ten years of running with a total of \SI{40}{\kt} of fiducial mass. This calculation assumes constant signal efficiency and background rejection over time and for each of the \dword{fd} modules.  Additional running improves the sensitivity proportionately if the experiment remains background-free.

\subsection{Sensitivity to Other Key Nucleon Decay Modes}
\label{subsec:nonaccel-ndk-other}

Another potential mode for a baryon number violation search is the decay of the neutron into a charged lepton plus meson, i.e.~\ntoek. In this mode, $\Delta B = -\Delta L$, where $B$ is baryon number and $L$ is lepton number.  The current best limit on this mode is \SI{3.2e31}{years} from the FREJUS collaboration~\cite{Berger:1991fa}. The reconstruction software for this analysis is the same as for the \ptoknubar analysis; the analysis again uses a \dword{bdt} that includes image classification score as an input. To calculate the lifetime sensitivity for this decay mode the same systematic uncertainties and procedure is used. The selection efficiency for this channel including the expected tracking improvements is \num{0.47}
with a background rejection of 
$0.99995$,
which corresponds to \num{15} background events per \si{\Mtyr}. The lifetime sensitivity for a \SI{400}{\ktyr} exposure is \SI{1.1e34}{years}. 
The \dword{dune} \dword{fd} technology can improve the lifetime limit for this particular channel by more than two orders of magnitude from the current world's best limit.

The sensitivity to the \ptoepizero mode has also been calculated. For this analysis, reconstruction was not applied, and true quantities were used as inputs to a \dword{bdt} to isolate events that contain a positron and two photons from the $\pi^0$ decay.  Energy smearing simulated the effects of reconstruction.  Applying the same selection to the atmospheric neutrino background and calculating the limit yields a sensitivity for an exposure of \SI{400}{\ktyr} in the range of \SIrange{8.7e33}{1.1e34}{years} depending on the level of energy smearing (in the range 5-30\%).  This initial study indicates that with a longer exposure of \SI{800}{\ktyr} \dword{dune} could achieve a sensitivity comparable to \superk's current limit of \SI{1.6e34}{years}~\cite{Miura:2016krn}.

\subsection{Detector Requirements for Nucleon Decay Searches}
\label{subsec:nonaccel-ndk-requirements}

As is the case for the entire \dword{fd} non-accelerator 
based physics program of \dword{dune}, nucleon decay searches require 
efficient triggering and event localization  
capabilities. The nucleon decay search program also relies 
on both the event imaging and particle identification 
(via $dE/dx$) capabilities of the \dword{lartpc} technology.  

Event localization within the \dword{fd} along the ionization 
drift direction is required in order to reject cosmic ray 
backgrounds via fiducial volume cuts. This can be achieved by 
requiring an event time ($t_0$) signal for nucleon decay 
candidates so that \dword{tpc} anode signal times can be used to 
determine the drift time.  Within \dword{dune}, the $t_0$ is provided by 
the \dword{pds}, which must have high detection 
efficiency throughout the \dword{fd} active volume for a 
scintillation photon signal corresponding to 
$>\SI{200}{\MeV}$ of deposited energy.

For nucleon decays into charged kaons, the possibility of using 
the time difference between the kaon scintillation signal and 
the scintillation signal from the muon from the kaon decay has 
been investigated.  
In the \superk analysis of \ptoknubar, the 
corresponding timing difference (between the de-excitation 
photons from the oxygen nucleus and the muon from kaon decay) 
was found to be an effective way to reduce
backgrounds~\cite{Abe:2014mwa}.  
Studies indicate that measuring time differences on the scale 
of the kaon lifetime (\SI{12}{\ns}) is difficult in \dword{dune}, 
independent of photon detector acceptance and timing resolution, 
due to both the scintillation process in argon 
-- consisting of fast (\si{\ns}-scale) and slow (\si{\micro\second}-scale) components --  and 
Rayleigh scattering over long distances.

Given the $\sim \SI{1}{GeV}$ energy release, 
the requirements for tracking 
and calorimetry performance are similar to those for the 
beam-based neutrino oscillation program described 
in Chapter~\ref{ch:osc}.  Especially important are the 
event imaging function and the $dE/dx$ measurement 
capability for particle identification.   
With a well-functioning \dword{lartpc}, nucleon decay search  
capabilities are ultimately limited by physics, namely
complexities arising from 
final state interactions (such as nucleon emission) 
as well as ionization fluctuations for example, 
rather than by detector performance per se. This is 
the case provided that readout noise is small compared to 
the ionization signal expected for minimum-ionizing particles
located anywhere within the active volume of the detector 
(see Sec.~\ref{sec:exec-key-reqs}).

\subsection{Nucleon Decay Summary}
\label{sec:ndksummary}

In summary, projecting from our current analysis of the sensitivity to proton decay via \ptoknubar in \dword{dune} with full simulation and reconstruction, we find that the sensitivity after a \SI{400}{\ktyr} exposure
is roughly twice the current limit from \superk based on an exposure of \SI{260}{\ktyr}~\cite{Abe:2014mwa}.  
An analysis of the sensitivity to neutron decay via \ntoek has also been completed; \dword{dune} could improve the lifetime limits in this mode by more than two orders of magnitude from the current world's best limit.  Future studies of nucleon decay into kaons will focus on potential improvements in track reconstruction, improved methods of particle and event identification, and understanding kaon \dword{fsi} models.  
Analysis of other modes of nucleon decay into kaons is underway, as well as the first investigations of the \ptoepizero with full simulation and reconstruction.


\section{Neutron-Antineutron Oscillations}
\label{sec:nonaccel-nnbar}

Neutron-antineutron (\nnbar) oscillation is a baryon number violating process that
has never been observed but is predicted by a number of \dword{bsm} theories~\cite{Phillips:2014fgb}. 
Discovering baryon
number violation via observation of this process would have implications about the source of matter-antimatter
symmetry in our universe given Sakharov's conditions for such asymmetry to arise~\cite{Sakharov:1967dj}.
In particular, the neutron-antineutron oscillation (\nnbar) process violates
baryon number by two units and, therefore, could also have further implications for
the smallness of neutrino masses~\cite{Phillips:2014fgb}. 
Since the \nnbar transition operator is a six-quark operator, of Maxwellian dimension \num{9}, with a coefficient function of dimension (mass)$^{-5}$, while the proton decay operator is a four-fermion operator, of dimension \num{6}, with a coefficient function of dimension (mass)$^{-2}$, one might naively assume that \nnbar oscillations would always be suppressed relative
to proton decay as a manifestation of baryon number violation.  However, this is not necessarily the case; indeed, there are models~\cite{Nussinov:2001rb} in which proton decay is very strongly suppressed down to an unobservably small level, while \nnbar oscillations occur at a level comparable to present limits. This shows the
value of a search for \nnbar transitions at DUNE.
The \nnbar process is one of many possible baryon number violating processes that can be investigated in \dword{dune}. Searches for this process using
both free neutrons and nucleus-bound neutron states have continued 
since the 1980s. The current best \num{90}\% \dword{cl} limits on the (free) neutron oscillation lifetime are \SI{8.6e7}{\s} from free \nnbar searches and \SI{2.7e8}{\s} from nucleus-bound \nnbar searches~\cite{BaldoCeolin:1994jz,Abe:2011ky}.

Neutron-antineutron oscillations can be detected via the subsequent antineutron annihilation with a neutron or a proton. Table~\ref{tab:nnbar-br} shows the branching ratios for the antineutron annihilation modes applicable to intranuclear searches.  This annihilation event will have a distinct signature of a vertex with several emitted light hadrons, with total energy of twice the nucleon mass and zero net momentum. Reconstructing these hadrons correctly and measuring their energies is key to identifying the signal event. The main background for these \nnbar annihilation events is caused by atmospheric neutrinos. Most common among mis-classified events are \dword{nc} \dword{dis} events without a lepton in the final state. As with nucleon decay, nuclear effects and \dword{fsi} make the picture more complicated.

\begin{table}
\caption[\nnbar annihilation modes]{Effective branching ratios for antineutron annihilation in \argon40, as implemented
in GENIE.}
\begin{tabular}{p{.22\textwidth}p{.22\textwidth}p{.22\textwidth}p{.22\textwidth}}
\rowcolor{dunetablecolor} 
\multicolumn{2}{^c}{$\bar{n}+p$} & \multicolumn{2}{^c}{$\bar{n}+n$}\\
\rowcolor{dunetablecolor}
         Channel & Branching ratio & Channel & Branching ratio \\ \toprowrule
         $\pi^{+}\pi^{0}$ & 1.2\% & $\pi^{+}\pi^{-}$ & 2.0\% \\ \colhline
         $\pi^{+}2\pi^{0}$ & 9.5\% & $2\pi^{0}$ & 1.5\% \\ \colhline
         $\pi^{+}3\pi^{0}$ & 11.9\% & $\pi^{+}\pi^{-}\pi^{0}$ & 6.5\% \\ \colhline
         $2\pi^{+}\pi^{-}\pi^{0}$ & 26.2\% & $\pi^{+}\pi^{-}2\pi^{0}$ & 11.0\% \\ \colhline
         $2\pi^{+}\pi^{-}2\pi^{0}$ & 42.8\% & $\pi^{+}\pi^{-}3\pi^{0}$ & 28.0\% \\ \colhline
         $2\pi^{+}\pi^{-}2\omega$ & 0.003\% & $2\pi^{+}2\pi^{-}$ & 7.1\% \\ \colhline
         $3\pi^{+}2\pi^{-}\pi^{0}$ & 8.4\% & $2\pi^{+}2\pi^{-}\pi^{0}$ & 24.0\% \\ \colhline
          &  & $\pi^{+}\pi^{-}\omega$ & 10.0\% \\ \colhline
          &  & $2\pi^{+}2\pi^{-}2\pi^{0}$ & 10.0\% \\ \colhline
\label{tab:nnbar-br}
\end{tabular}
\end{table}

\subsection{Sensitivity to Intranuclear Neutron-Antineutron Oscillations in DUNE}
\label{subsec:nonaccel-nnbar-dunesensitivity}

The simulation of neutron-antineutron oscillation was developed~\cite{Hewes:2017xtr} and implemented in \dword{genie}. This analysis uses \dword{genie} v.2.12.10.
Implementing this process in \dword{genie} used \dword{genie}'s existing modeling of Fermi momentum and binding energy for both the oscillating neutron and the nucleon with which the resulting antineutron annihilates.   Once a neutron has oscillated to an antineutron in a nucleus, the antineutron has a $18/39$ chance of annihilating with a proton in argon, and a $21/39$ chance of annihilating with a neutron. The energies and momenta of the annihilation products are assigned randomly but consistently with four-momentum conservation. The products of the annihilation process follow the branching fractions (shown in Table~\ref{tab:nnbar-br}) measured in low-energy antiproton annihilation on hydrogen.
Since the annihilation products are produced inside the nucleus, \dword{genie} further models re-interactions of those products as they propagate in the nucleus (until they escape the nucleus).  The \dword{fsi} are simulated using the $hA2015$ model in \dword{genie} as described in Section~\ref{sec:final-state-interactions}.

Figure~\ref{fig:pi_FSI_m} shows the momentum distributions for charged and neutral pions before \dword{fsi} and after \dword{fsi}. These distributions show the \dword{fsi} makes both charged and neutral pions less energetic.  The effect of \dword{fsi} on pion multiplicity is also rather significant; \num{0.9}\% of the events have no charged pions before \dword{fsi}, whereas after \dword{fsi} \num{11.1}\% of the events have no charged pions. In the case of the neutral pion, \num{11.0}\% of the events have no neutral pions before \dword{fsi}, whereas after \dword{fsi}, \num{23.4}\% of the events have no neutral pions. The decrease in pion multiplicity is primarily due to pion absorption in the nucleus. Another effect of \dword{fsi} is nucleon knockout from pion elastic scattering. Of the events, \num{94}\% have at least one proton from \dword{fsi} and \num{95}\% of the events have at least one neutron from \dword{fsi}. Although the kinetic energy for these nucleons peak at a few tens of \si{\MeV}, the kinetic energy can be as large as hundreds of \si{\MeV}.  In summary, the effects of \dword{fsi} in \nnbar become relevant because they modify the kinematics and topology of the event. For instance, even though the decay modes of Table \ref{tab:nnbar-br} do not include nucleons in their decay products, nucleons appear with high probability after \dword{fsi}.

\begin{dunefigure}
[Final state interactions in \nnbar]{fig:pi_FSI_m}
{Momentum of an individual charged pion before and after final state interactions (left): momentum of an individual neutral pion before and after final state interactions (right).}
\includegraphics[width=0.49\textwidth]{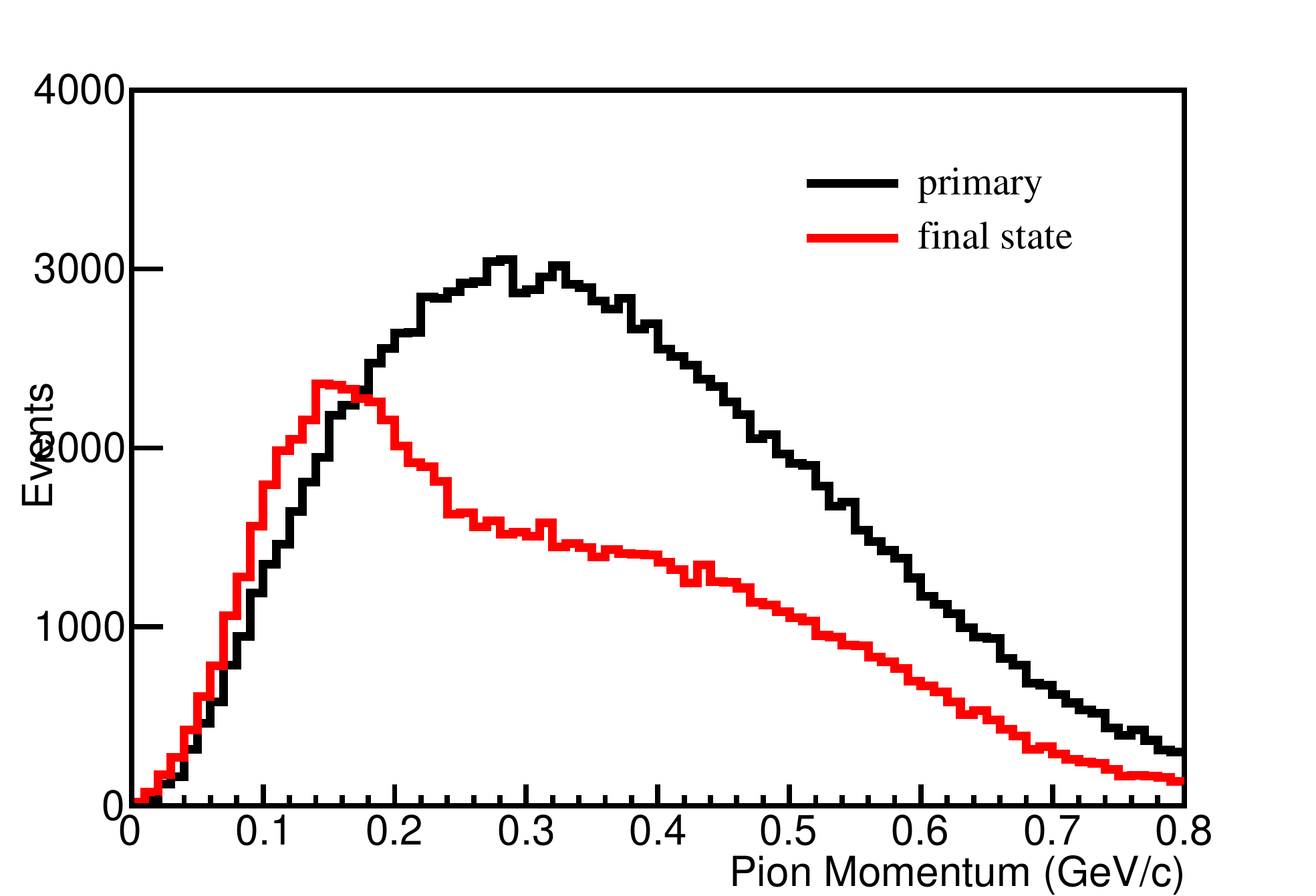}
\includegraphics[width=0.49\textwidth]{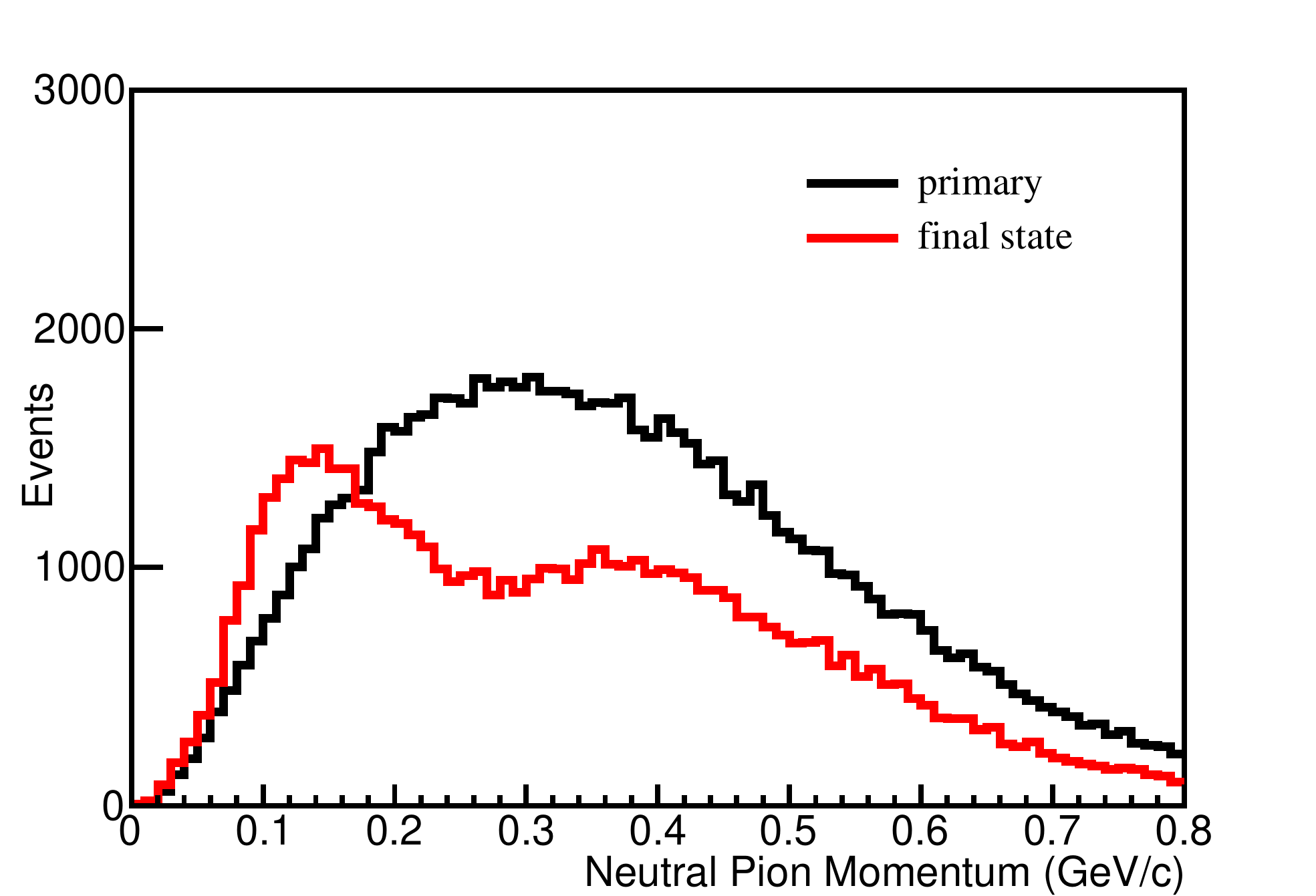}
\end{dunefigure} 

The main background process in search of bound \nnbar oscillation in \dword{dune} is assumed to be atmospheric neutrino interactions in the detector.  This is simulated in \dword{genie} as described in Section~\ref{sec:ndkbkgd}.

As with the \ptoknubar analysis, two distinct methods of reconstruction and event selection have been applied in this search. One involves traditional reconstruction methods (\threed track and vertex reconstruction by \dword{pma}); the other involves image classification 
of \twod images of reconstructed hits (\dword{cnn}). The two methods, combined in the form of a multivariate analysis, uses the image classification score with other physical observables extracted from traditional reconstruction.  A \dword{bdt} classifier is used. Ten variables are used in the \dword{bdt} event selection, including number of reconstructed tracks and showers; variables related to visible energy deposition; $PIDA$ and $dE/dx$; reconstructed momentum; and CNN score.  Figure~\ref{fig:bdt_nnbar} shows the distribution of the \dword{bdt} output for signal and background.

\begin{dunefigure}
[\nnbar Boosted Decision Tree response]{fig:bdt_nnbar}
{Boosted Decision Tree response for \nnbar oscillation for signal (blue) and background (red).}
\includegraphics[width=0.8\textwidth]{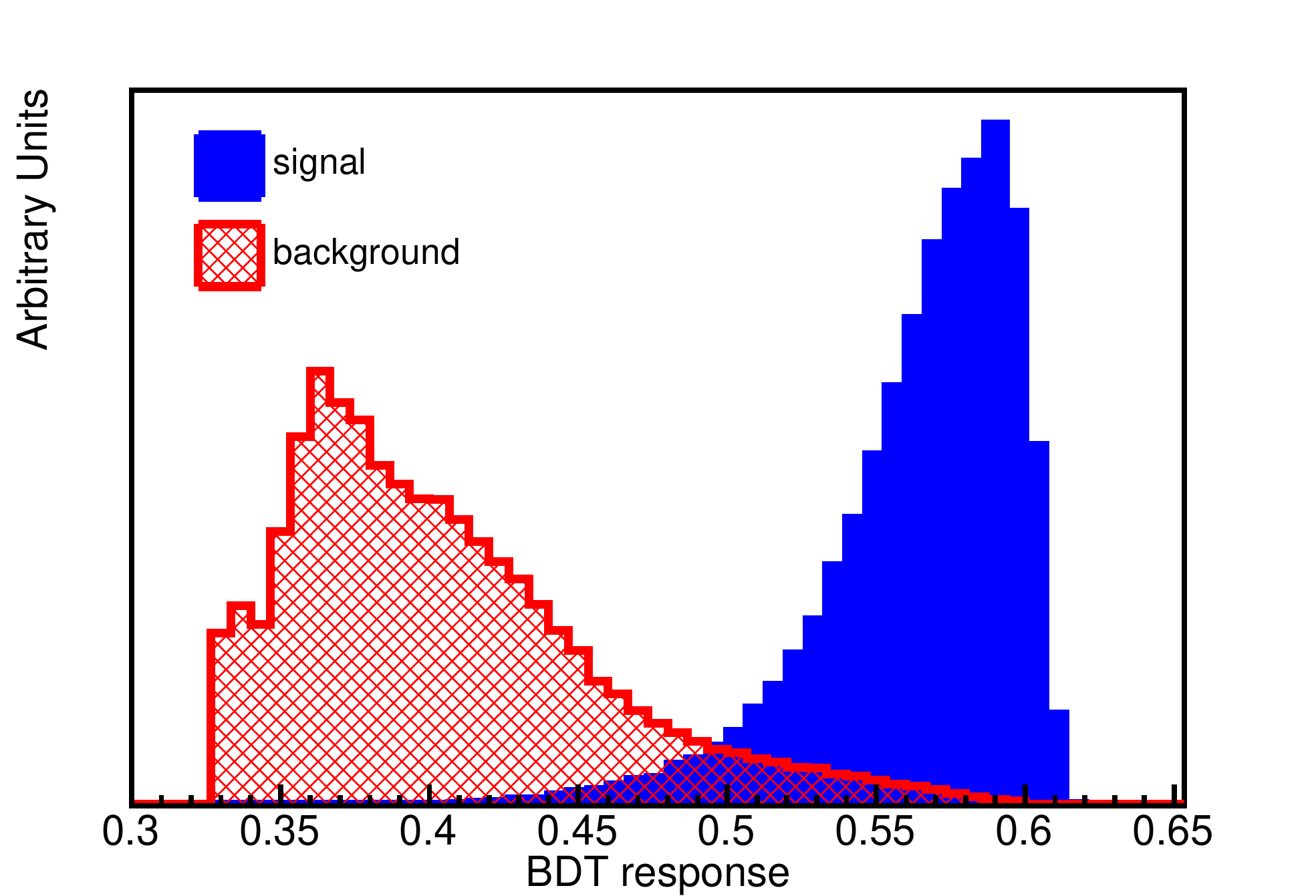}
\end{dunefigure} 

Figure~\ref{fig:nnbar_sig} shows an \nnbar event with high \dword{bdt} response value (\num{0.592}). Showers from neutral pions are shown in red, blue, yellow, and green. The reconstructed charged pion tracks are shown as dark green and maroon lines. The topology of this event is consistent with charged pion and neutral pion production. 

The left side plot in Figure~\ref{fig:nnbar_bkgd} shows a \dword{nc} atmospheric neutrino interaction $\nu_{e}+n\rightarrow \nu_{e}+p+p$ with a low \dword{bdt} response value (\num{0.388}). This type of interaction is easily distinguished from the signal.  The two protons from the \dword{nc} interaction are reconstructed as tracks, and no shower activity is present. However, the right side plot in Figure~\ref{fig:nnbar_bkgd} displays a \dword{cc} atmospheric neutrino interaction $\nu_{e}+n\rightarrow {e}^{-}+p+\pi +p$ with a high \dword{bdt} response value (\num{0.598}). This background event mimics the signal topology by having multi-particle production and an electromagnetic shower. Further improvements in shower reconstruction, especially $dE/dx$, should help in classifying these types of background events in the future because the electron shower $dE/dx$ differs from the $dE/dx$ of a shower induced by a gamma-ray.

\begin{dunefigure}
[Event display for well-classified \nnbar signal event]{fig:nnbar_sig}
{Event display for a well-classified \nnbar signal event.  The vertical axis is time ticks (each time tick corresponds to \SI{500}{\ns}), and the horizontal axis is wire number.  The bottom view is induction plane one, middle is induction plane two, and the top is the collection plane.  The color represents the charge deposited in each hit.}
\includegraphics[width=0.8\textwidth]{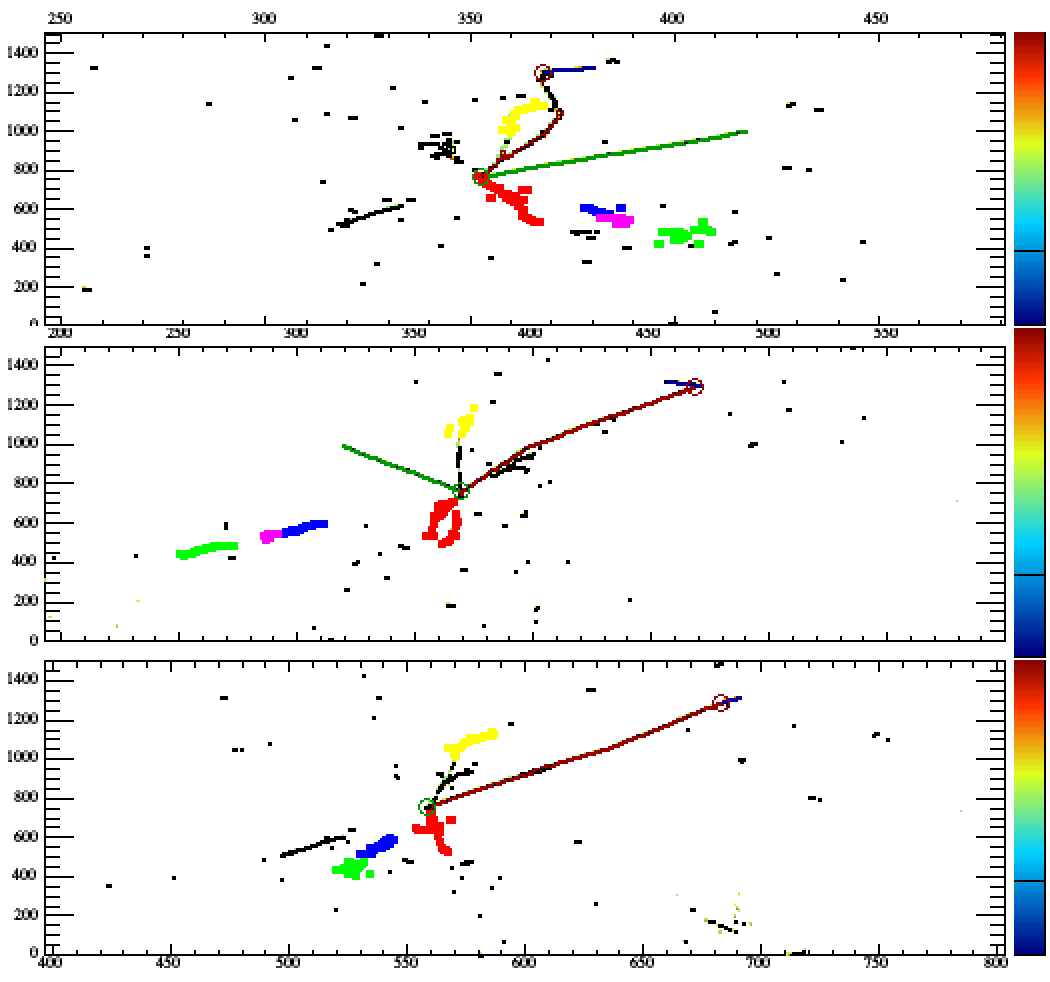}
\end{dunefigure} 

\begin{dunefigure}
[Event displays for \nnbar background events]{fig:nnbar_bkgd}
{Event displays for \nnbar backgrounds.  The vertical axis is time ticks (each time tick corresponds to \SI{500}{\ns}), and the horizontal axis is wire number.  The bottom view is induction plane one, middle is induction plane two, and the top is the collection plane.  The color represents the charge deposited in each hit.  The left plot shows an atmospheric neutrino interaction unlikely to be classified as signal. The right plot shows an atmospheric neutrino interaction which could make it into the selected sample.}
\includegraphics[width=0.49\textwidth]{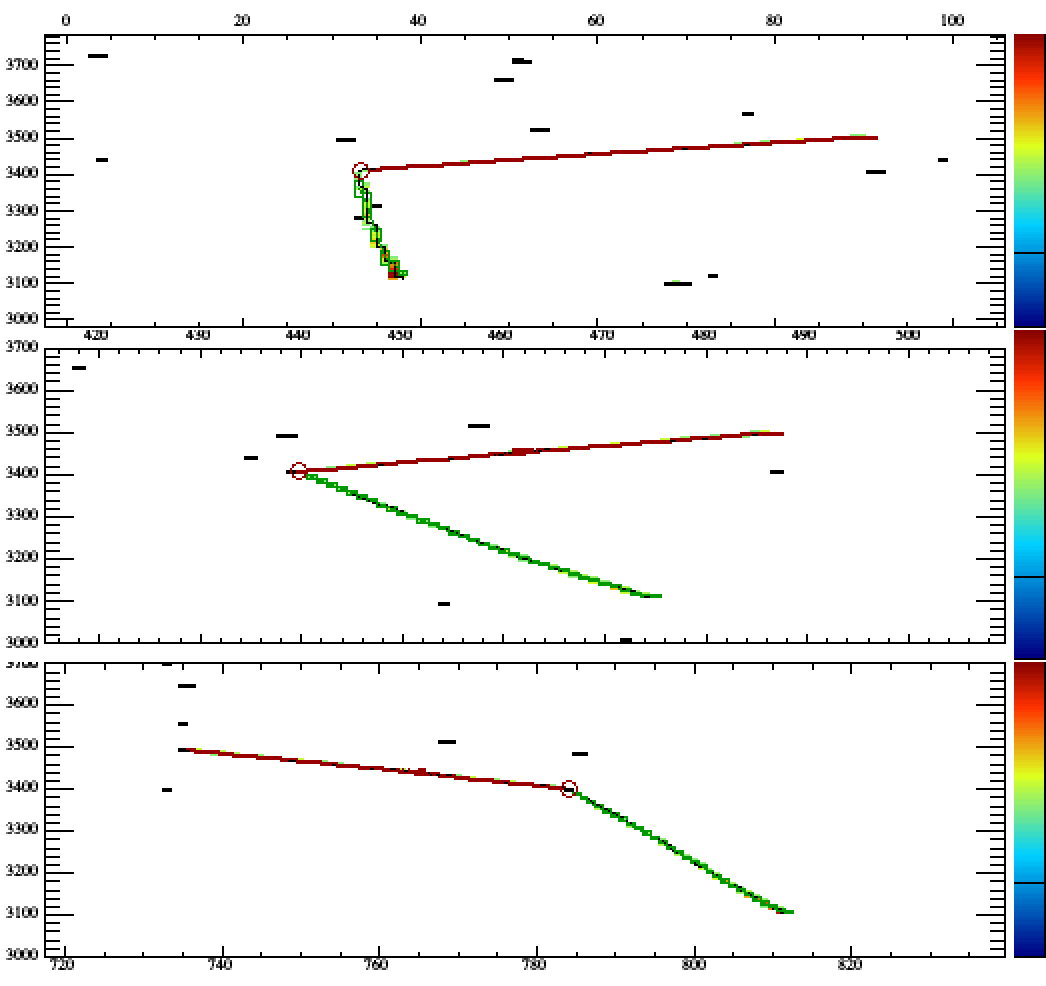}
\includegraphics[width=0.49\textwidth]{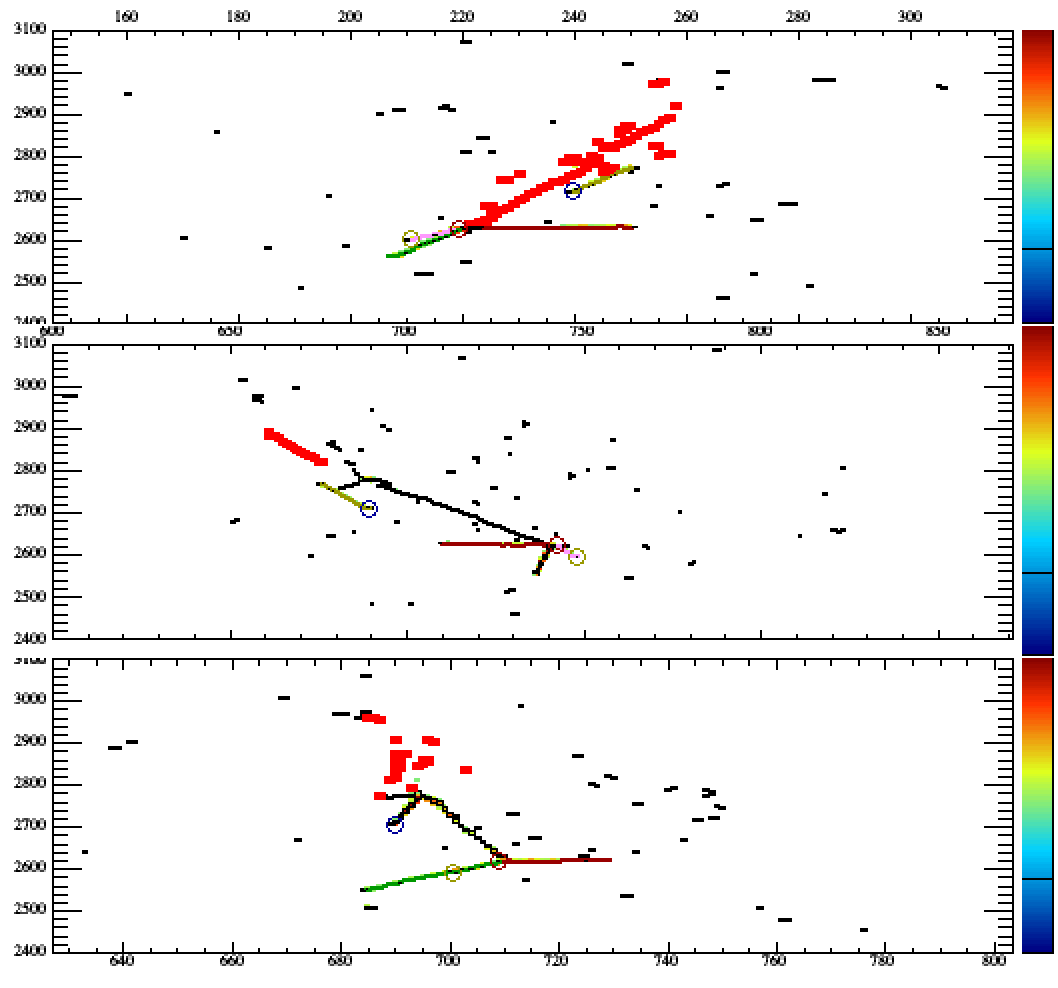}
\end{dunefigure} 

The sensitivity to the \nnbar oscillation lifetime can be calculated for a given exposure, the efficiency of selecting signal events, and the background rate along their uncertainties. The lifetime sensitivity is obtained at \num{90}\% \dword{cl} for the bound neutron. Then, the lifetime sensitivity for a free neutron is acquired using the conversion from nucleus bounded neutron to free neutron \nnbar oscillation~\cite{Friedman:2008es}.  
The uncertainties on the signal efficiency and background rejection are conservatively estimated to be \num{25}\%.  A detailed evaluation of the uncertainties is in progress.

The free \nnbar oscillation lifetime, $\tau_{n-\bar{n}}$, and bounded \nnbar oscillation lifetime, $T_{n-\bar{n}}$, are related to each other through the suppression factor $R$ as
\begin{equation}
    \tau^{2}_{n-\bar{n}} = \frac{T_{n-\bar{n}}}{R} ~.
    \label{eq:tau}
\end{equation}
The suppression factor $R$ varies for different nuclei. This suppression factor was calculated for $^{16}$O and $^{56}$Fe~\cite{Friedman:2008es}. The $R$ for $^{56}$Fe, \SI{0.666e23}{\per\s}, is used in this analysis for \argon40 nuclei.

The best bound neutron lifetime limit is achieved using a signal efficiency of \num{8.0}\% at the background rejection probability of \num{99.98}\%. The \num{90}\% \dword{cl} limit of a bound neutron lifetime is \SI{6.45e32}{years} for a \SI{400}{\ktyr} exposure. The corresponding limit for the oscillation time of free neutrons is calculated to be \SI{5.53e8}{\s}. This is approximately an improvement by a factor of two from the current best limit, which comes from \superk~\cite{Abe:2011ky}.  Planned improvements to this analysis include improved \dword{cnn} performance and better estimates of systematic uncertainties.  As with nucleon decay, searches for \nnbar oscillations performed by \dword{dune} and those performed by \superk or \hyperk are highly complementary.  Should a signal be observed in any one experiment, confirmation from another experiment with a different detector technology and backgrounds would be very powerful. 

\section{Physics with Atmospheric Neutrinos}
\label{sec:nonaccel-atm}

Atmospheric neutrinos are a unique tool for studying neutrino oscillations: the oscillated flux contains all flavors of neutrinos and antineutrinos, is very sensitive to matter effects and to both \dm{} parameters, and covers a wide range of $L/E$. In principle, all oscillation parameters could be measured, with high
complementarity to measurements performed with a neutrino beam. In addition, atmospheric neutrinos are available all the time, in particular before the beam becomes operational. The \dword{dune} \dword{fd}, with its large mass and the overburden to protect it from atmospheric muon background, is an ideal tool for these studies.  Given the strong overlap in event topology and energy scale with beam neutrino interactions, most requirements will necessarily be met by the \dword{fd} design. Additional requirements include a self-trigger because atmospheric neutrino events are asynchronous with accelerator timing and a more stringent demand on neutrino direction reconstruction.

\subsection{Oscillation Physics with Atmospheric Neutrinos}
\label{sec:nonaccel-atm-oscillations}

Sensitivity to oscillation parameters with atmospheric neutrinos in \dword{dune} has been evaluated.
The fluxes of each neutrino species were computed at the \dword{fd} location after 
oscillation. Interactions in the \dword{lar} medium were simulated with the \dword{genie} event 
generator. Detection thresholds and energy resolutions based on full 
simulations were applied to the outgoing particles to take 
detector effects into account. Events were classified as fully contained or partly contained by placing the vertex at a random position inside the 
detector and tracking the lepton until it reaches the edge of the detector.
Partly contained events 
are those where a final state muon exits the detector.  The number of events expected 
for each flavor and category is summarized in Table~\ref{tab:atmos_rates}.

\begin{dunetable}
[Atmospheric neutrino event rates per year in different analysis categories]
{lc}
{tab:atmos_rates}
{Atmospheric neutrino event rates per year in \SI{40}{\kt} of fiducial mass including oscillations in different analysis categories}
Sample   &  Event rate per year \\ \toprowrule
fully contained electron-like   & \num{1600} \\ \colhline
fully contained muon-like       & \num{2400} \\ \colhline
partly contained muon-like   & \num{790} \\ 
\end{dunetable}

Figure~\ref{fig:lovere} shows the expected $L/E$ distribution for high-resolution, muon-like 
events from a \SI{400}{\ktyr} exposure. The data provide excellent resolution of the 
first two oscillation nodes, even with the expected statistical uncertainty.
In performing oscillation fits, the data in each flavor/containment category are 
binned in energy and zenith angle.

\begin{dunefigure}
[Reconstructed $L/E$ Distribution of `High-Resolution' Atmospheric Neutrinos]{fig:lovere}
{Reconstructed $L/E$ Distribution of `High-Resolution'
$\mu$-like atmospheric neutrino events in a \SI{400}{\ktyr} exposure with and
without oscillations (left), and the ratio of the two (right), with the error bars indicating the size of the statistical uncertainty.}
\includegraphics[width=0.49\textwidth]{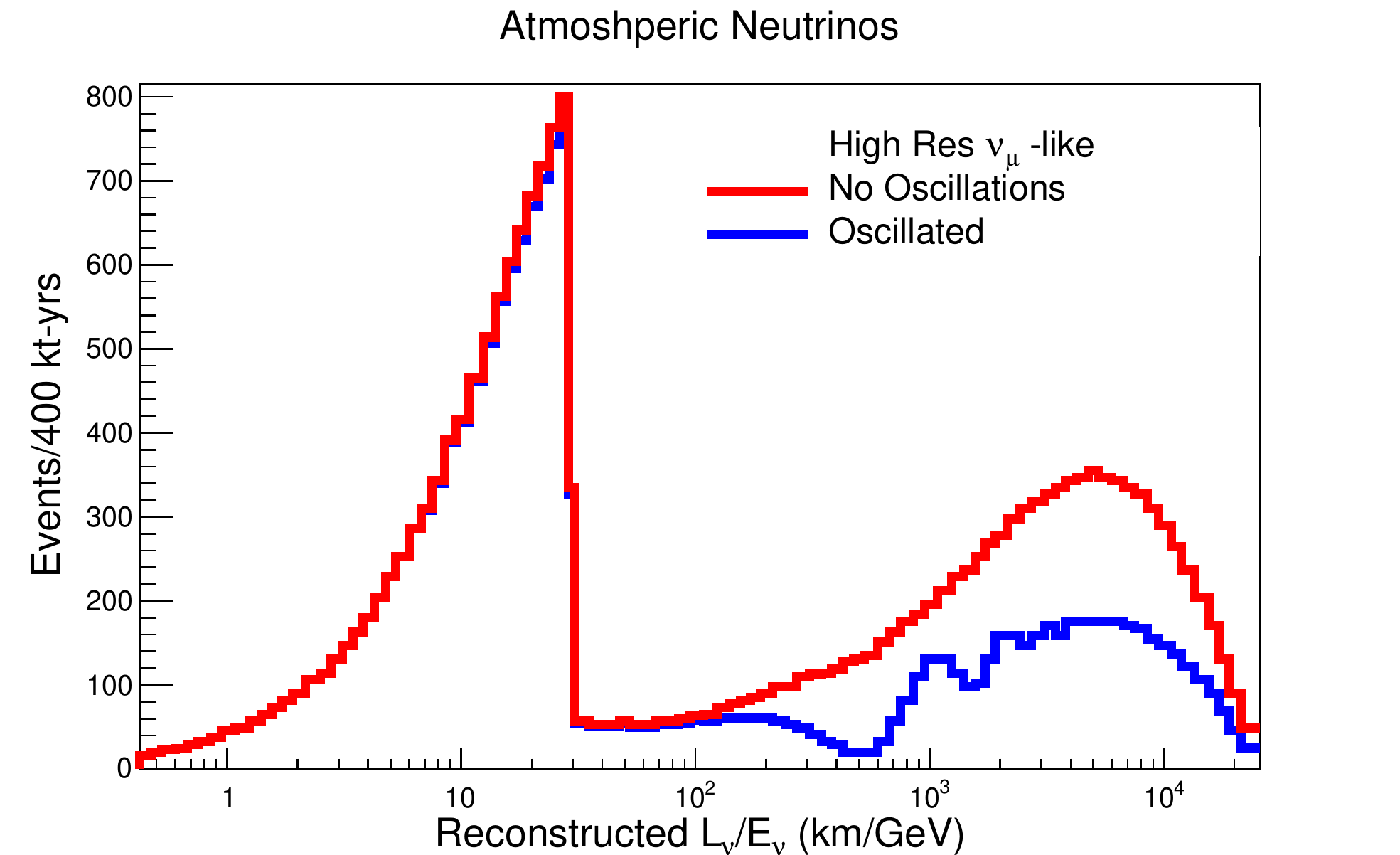}
\includegraphics[width=0.49\textwidth]{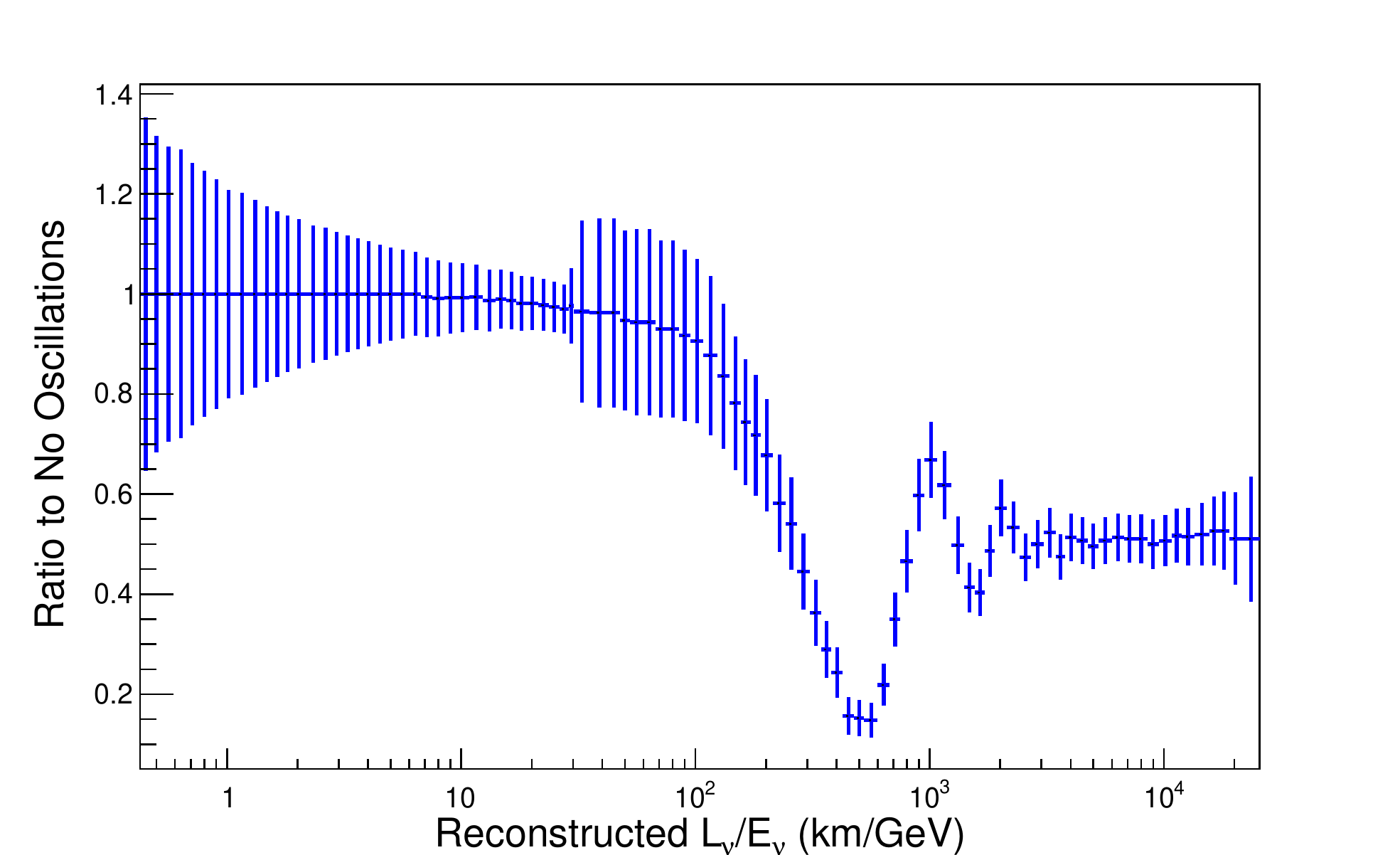}
\end{dunefigure}

When neutrinos travel through the Earth, the \dword{msw} resonance influences 
electron neutrinos in the few-\si{GeV} energy range. More precisely, the resonance 
occurs for \nue in the case of normal ordering and for \anue in the case of inverted ordering.

The mass ordering sensitivity can be greatly enhanced if neutrino and antineutrino events can be 
separated. The \dword{dune} \dword{fd} will not be magnetized, but its high-resolution 
imaging offers possibilities for tagging features of events that provide statistical 
discrimination between neutrinos and antineutrinos. For the sensitivity calculations, 
two such tags were included: a proton tag and a decay electron tag. 

Figure~\ref{fig:atm_mh} shows the mass ordering sensitivity as a function of the fiducial exposure.  Over this range of fiducial exposures, the sensitivity essentially follows the square root of the exposure, indicating that the measurement is not systematics-limited. Unlike beam measurements, the sensitivity to the mass ordering with atmospheric neutrinos is nearly independent of the \dword{cp} violating phase.  The sensitivity comes from both electron neutrino appearance as well as muon neutrino disappearance and depends strongly on the true value of \sinst{23}, as shown in Figure~\ref{fig:atm_mh}. For comparison, the sensitivity for \hyperk atmospheric neutrinos with a \SI{1900}{\ktyr} exposure is also shown.

\begin{dunefigure}
[Mass Ordering Sensitivity vs. Exposure for Atmospheric Neutrinos]{fig:atm_mh}
{Sensitivity to mass ordering using atmospheric neutrinos as a function of fiducial exposure in \dword{dune} (left) and as a function of the true value of \sinst{23} (right).  For comparison, \hyperk sensitivities are also shown \cite{Abe:2018uyc}.}
\includegraphics[width=0.41\textwidth]{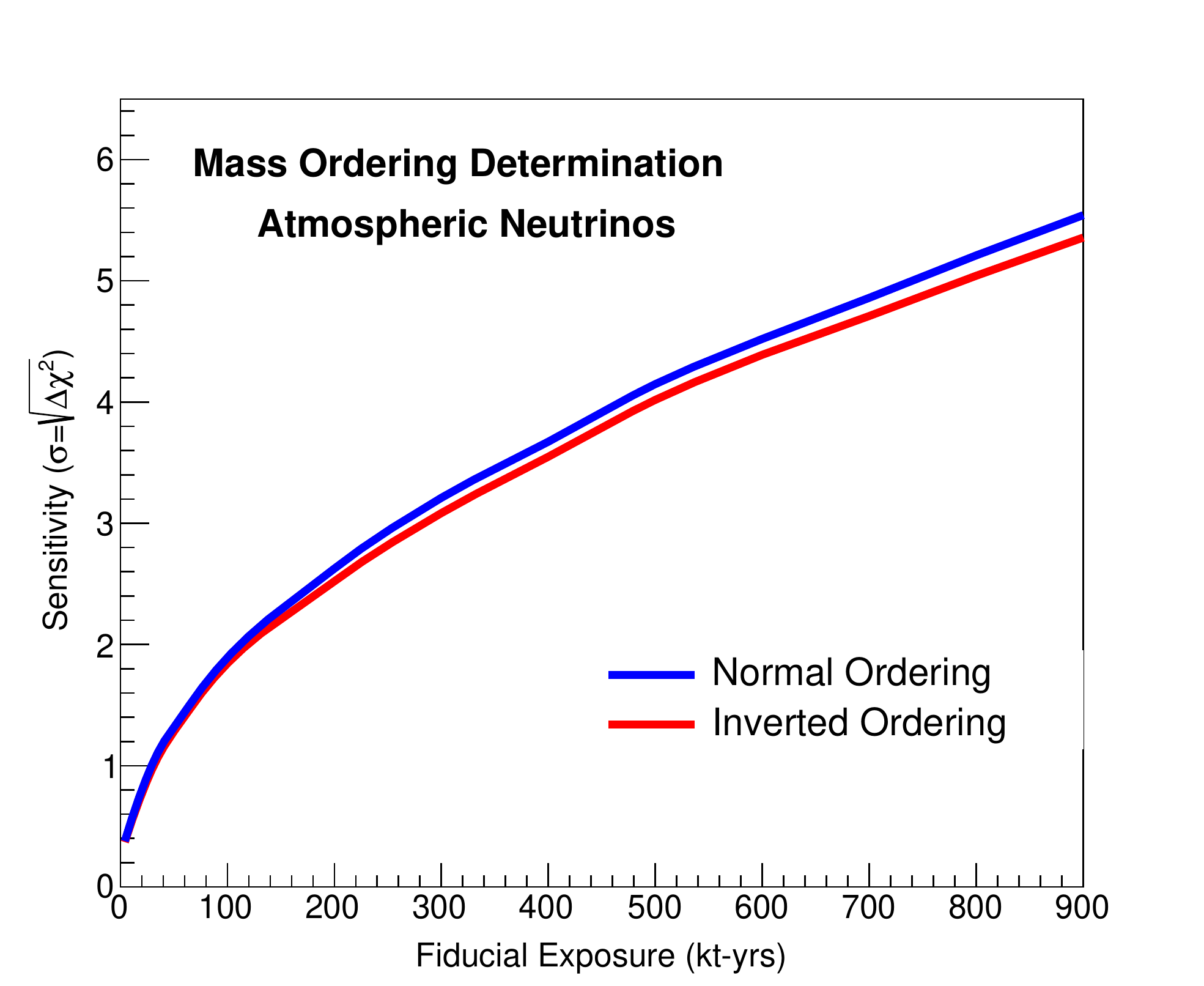}
\includegraphics[width=0.58\textwidth]{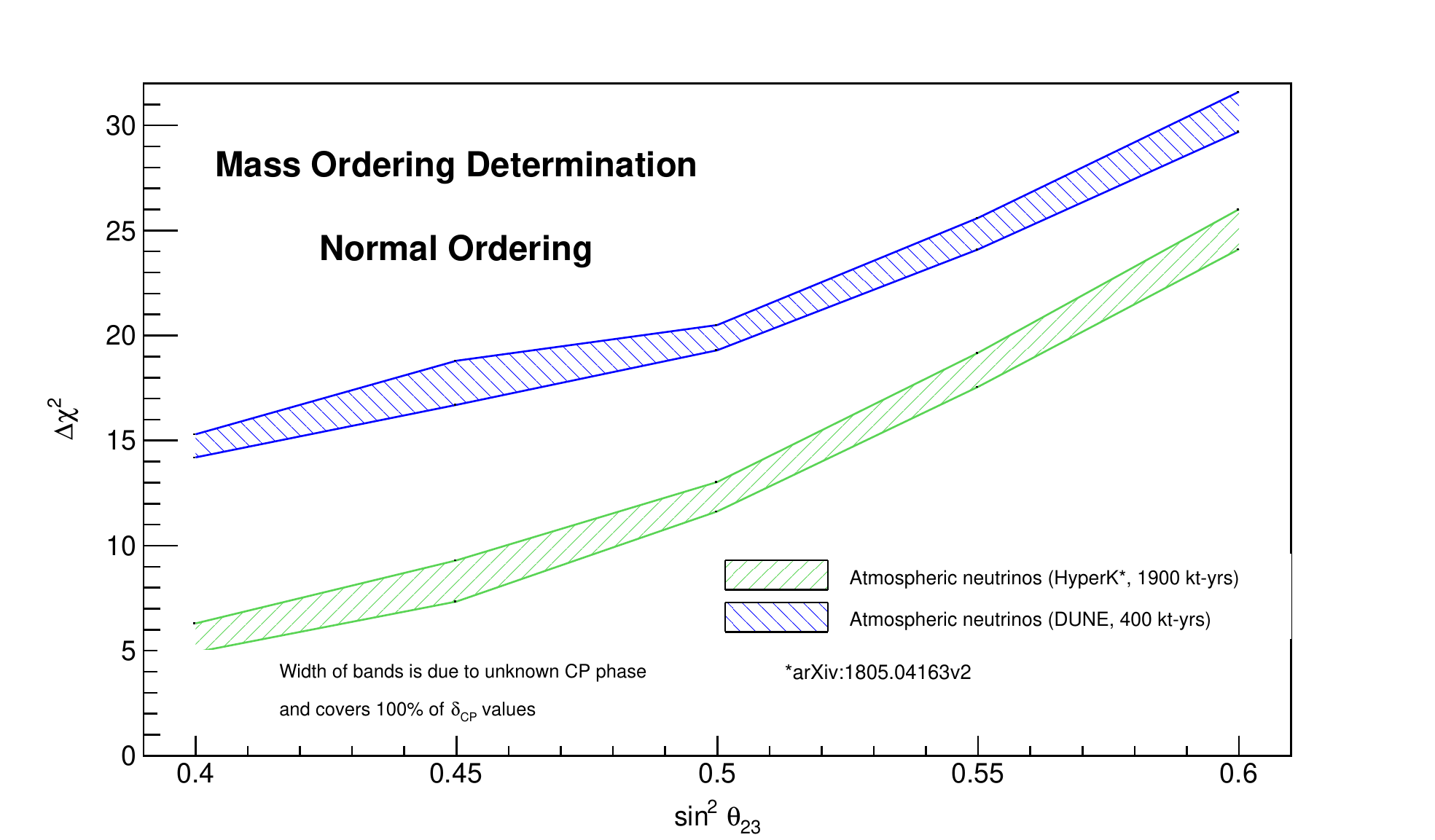}
\end{dunefigure}

In the two-flavor approximation, neutrino oscillation probabilities depend on 
\sinst, which is invariant when changing $\theta$ to $\pi/2-\theta$. In this case, the octant degeneracy remains for $\theta_{23}$ in the leading order terms of the full 
three-flavor oscillation probability, making it impossible to determine whether $\theta_{23}< \pi/4$ or 
$\theta_{23}> \pi/4$. Accessing full three-flavor oscillation with atmospheric neutrinos 
should help solve the ambiguity.

These analyses will provide an approach complementary to the beam neutrino approach. For instance, atmospheric neutrinos should resolve  degeneracies present in beam analyses because the mass ordering sensitivity is essentially independent of \deltacp. Atmospheric neutrino data will be acquired 
even in the absence of the beam and will provide a useful sample for developing reconstruction software, analysis methodologies, and calibrations.  Atmospheric neutrinos provide a window into a range of new physics scenarios, and may allow \dword{dune} to place limits on Lorentz and \dword{cpt} violation (see Section~\ref{sec:nonaccel-atm-bsm}), 
non-standard interactions~\cite{Chatterjee:2014gxa}, mass-varying neutrinos~\cite{Abe:2008zza}, and
sterile neutrinos~\cite{Abe:2014gda}.
Recent studies have also indicated that sub-GeV atmospheric neutrinos could be used to exclude some values of \deltacp independently from the beam neutrino measurements~\cite{Kelly:2019itm}.

\subsection{BSM Physics with Atmospheric Neutrinos}
\label{sec:nonaccel-atm-bsm}

Studying \dword{dune} atmospheric neutrinos is a promising approach
to search for \dword{bsm} effects such as Lorentz and \dword{cpt} violation,
which has been hypothesized
to emerge from an underlying Planck-scale theory like strings~\cite{Kostelecky:1988zi,Kostelecky:1991ak}.
The comprehensive realistic effective field theory
for Lorentz and \dword{cpt} violation,
the \dword{sme}~\cite{Kostelecky:1994rn,Colladay:1996iz,Colladay:1998fq,Kostelecky:2003fs},
is a powerful and calculable framework
for analyzing experimental data.
All \dword{sme} coefficients for Lorentz and \dword{cpt} violation
governing the propagation and oscillation of neutrinos
have been enumerated~\cite{Kostelecky:2003cr,Kostelecky:2011gq},
and many experimental measurements of \dword{sme} coefficients 
have been performed to date~\cite{Kostelecky:2008ts}.
Nonetheless,
much of the available \dword{sme} coefficient space 
in the neutrino sector remains unexplored.

Experimental signals predicted by the \dword{sme} include
corrections to standard neutrino-neutrino 
and antineutrino-antineutrino mixing probabilities,
oscillations between neutrinos and antineutrinos,
and modifications of oscillation-free propagation,
all of which incorporate unconventional dependencies
on the magnitudes and directions of momenta and spin.
For \dword{dune} atmospheric neutrinos,
the long available baselines,
the comparatively high energies accessible,
and the broad range of momentum directions
offer advantages that can make possible great
improvements 
in sensitivities to certain types of Lorentz and \dword{cpt} violation~\cite{Kostelecky:2003cr,Kostelecky:2011gq,Kostelecky:2003xn,Kostelecky:2004hg,Diaz:2009qk,Diaz:2013saa,Diaz:2013wia}.
To date,
experimental searches for Lorentz and \dword{cpt} violation
with atmospheric neutrinos have been published 
by the IceCube and \superk collaborations~\cite{Abbasi:2010kx,Abe:2014wla,Aartsen:2017ibm}.
Similar studies are possible with \dword{dune},
and many \dword{sme} coefficients can be measured that remain unconstrained to date.

An example of the potential reach of studies with \dword{dune} atmospheric neutrinos
is shown in Figure~\ref{fig:atm},
which displays estimated sensitivities
from \dword{dune} atmospheric neutrinos to a subset of coefficients 
controlling isotropic (rotation-invariant) violations 
in the Sun-centered frame~\cite{Kostelecky:2002hh}.
The sensitivities are estimated by requiring that the
Lorentz/\dword{cpt}-violating effects are comparable in size to
those from conventional neutrino oscillations. 
The eventual \dword{dune} constraints will be determined by the
ultimate precision of the experiment (which is set in
part by the exposure).  The gray bars in Figure~\ref{fig:atm} show existing limits.  These conservative sensitivity estimates show that \dword{dune} can achieve first measurements (red) on some coefficients
and improved measurements (green) on others.

To illustrate an \dword{sme} modification of oscillation probabilities,
consider a measurement of the atmospheric neutrino and antineutrino flux
as a function of energy.
For definiteness,
we adopt atmospheric neutrino fluxes~\cite{Honda:2015fha},
evaluated using the NRLMSISE-00 global atmospheric model~\cite{Picone},
that result from a production event at an altitude of \SI{20}{\km}.
Assuming conventional oscillations with standard mass-matrix values from the
\dword{pdg}~\cite{Tanabashi:2018oca},
the fluxes at the \dword{fd} are shown in Figure~\ref{fig:atm2}.
The sum of the \nue and \anue fluxes
is shown as a function of energy as a red dashed line, 
while the sum of the \numu and \anumu fluxes 
is shown as a blue dashed line. 
Adding an isotropic non-minimal coefficient for Lorentz violation
of magnitude $\mathaccent'27 c^{(6)}_{e \mu} = \SI{1e-28}{\per\GeV\square}$
changes the fluxes from the dashed lines to the solid ones.
This coefficient is many times smaller
than the current experimental limit.
Nonetheless,
the flux spectrum is predicted to change significantly 
at energies over approximately \SI{100}{\GeV}. 

\begin{dunefigure}[Sensitivity to Lorenz and CPT violation with atmospheric neutrinos]{fig:atm}{Estimated sensitivity to Lorenz and CPT violation with atmospheric neutrinos in the non-minimal isotropic Standard Model Extension. The sensitivities are estimated by requiring that the Lorentz/CPT-violating effects are comparable in size to
those from conventional neutrino oscillations.}
\includegraphics[width=0.8\textwidth]{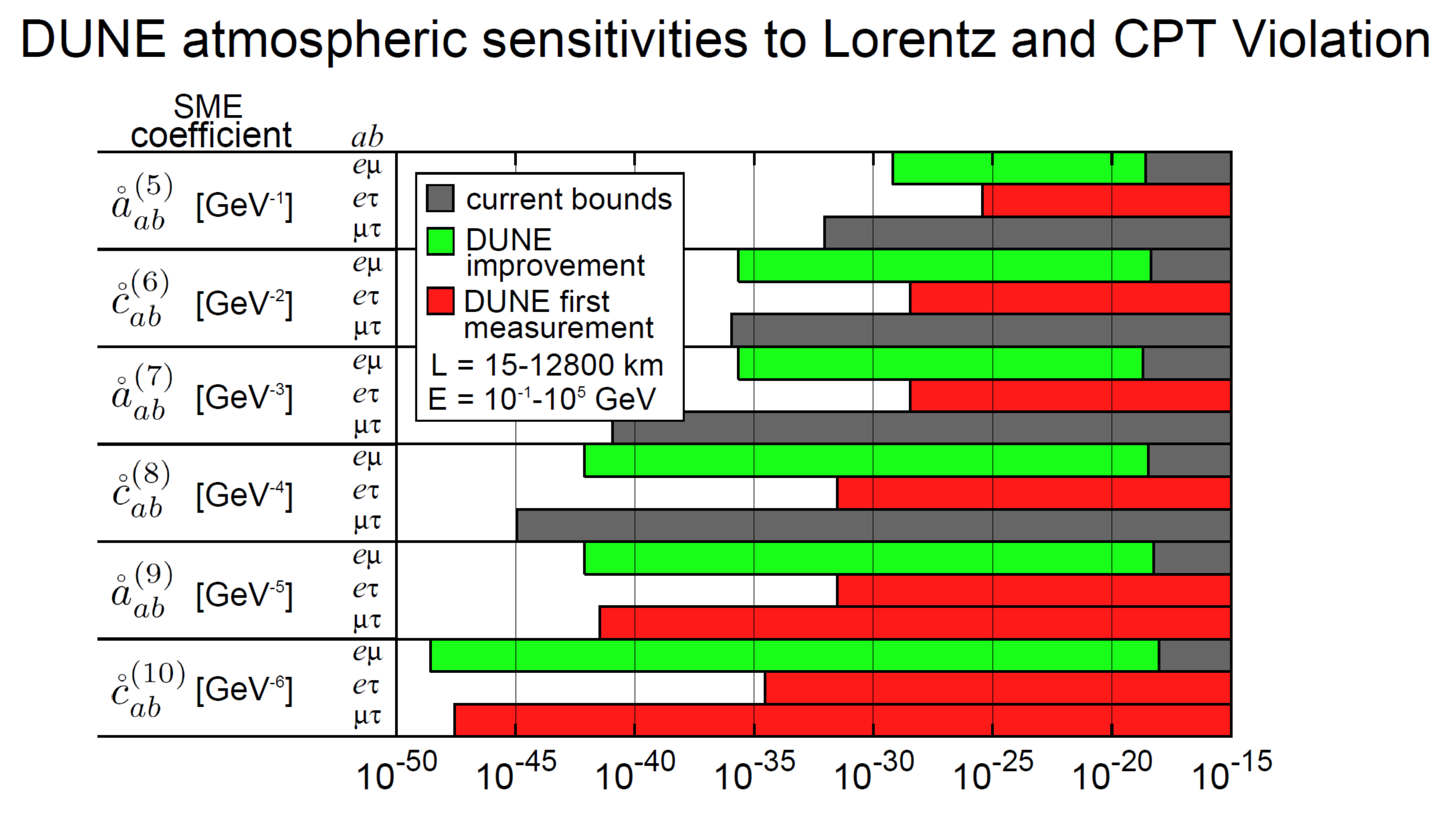}
\end{dunefigure}

\begin{dunefigure}[Atmospheric $\nu$ and $\bar{\nu}$ fluxes in the non-minimal isotropic Standard Model Extension]{fig:atm2}{Atmospheric fluxes of neutrinos and antineutrinos as a function of energy for conventional oscillations (dashed line) and in the non-minimal isotropic Standard Model Extension (solid line).}
\includegraphics[width=0.8\textwidth]{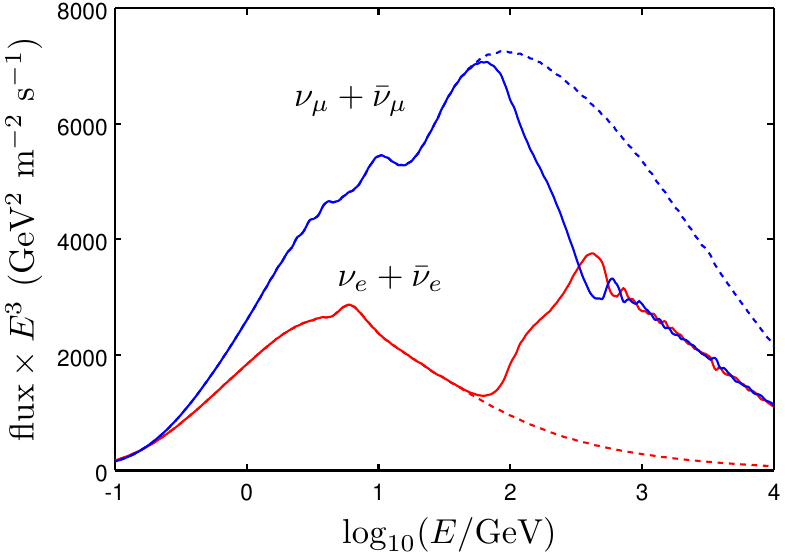}
\end{dunefigure}


\cleardoublepage

\chapter{Supernova neutrino bursts and physics with low-energy neutrinos}
\label{ch:snb-lowe}

The DUNE experiment will be sensitive to neutrinos from around 5 MeV
to a few
tens of MeV.  Charged-current interactions of neutrinos in this range create short electron tracks in liquid argon, potentially accompanied by
gamma ray and other secondary particle signatures.   
This regime is of
particular interest for detection of the burst of neutrinos from a galactic
core-collapse supernova (the primary focus of this section). 
The sensitivity of DUNE is primarily to \textit{electron flavor} supernova neutrinos, and this capability is unique among existing and proposed supernova neutrino detectors for the next decades.  
Neutrinos and antineutrinos from other astrophysical sources are also potentially detectable.  
The low-energy event regime has particular reconstruction, background and triggering challenges.

In this section we will describe studies done in the DUNE 
\dword{snble}
physics working group so far towards
understanding of DUNE's sensitivity to low-energy neutrinos, with an
emphasis on supernova burst signals.   In Sec.~\ref{sec:snb-lowe-snb},
we describe basic supernova neutrino physics.  In
Sec.~\ref{sec:lowe-events} we describe the general properties of
low-energy events in DUNE including interaction channels, the tools we
have developed so far, and backgrounds.  The tools include a neutrino event generator
specifically for this energy regime, and the \dword{snowglobes} fast event-rate calculation tool.  Some of the subsequent studies are done using
a full simulation and reconstruction, whereas others make use of
 \dword{snowglobes}.
Section~\ref{sec:sn-signals} describes the
expected supernova signal: event rates, and pointing properties.
Section~\ref{sec:physics-snblowe-astrophysics} describes astrophysics
of the collapse, explosion and remnant that we will learn from the burst.
Section~\ref{sec:physics-snblowe-neutrino-physics} describes neutrino
physics that can be extracted from a \dword{snb} observation.
Section~\ref{sec:physics-snblowe-other} mentions
some other possible astrophysical neutrinos, including solar and
diffuse supernova background neutrinos.
Section~\ref{sec:physics-snblowe-detector-requirements} summarizes the
detector requirements.

\section{Supernova neutrino bursts}
\label{sec:snb-lowe-snb}

\subsection{Neutrinos from collapsed stellar cores: basics}

A core-collapse supernova\footnote{``Supernova'' always
  refers to a ``core-collapse supernova'' in this chapter.} occurs when a massive star reaches the end of its
life. As a result of nuclear burning throughout the star's life, the
central region of such a star gains an ``onion'' structure, with an iron core at the center surrounded by concentric shells of lighter elements (silicon, oxygen, neon, magnesium, carbon, etc). At temperatures of $T\sim 10^{10}$ K and densities of $\rho \sim 10^{10}$ g/cm$^{3}$, the Fe core continuously loses energy by neutrino emission (through pair annihilation and plasmon decay). Since iron cannot be further burned, the lost energy cannot be replenished throughout the volume and the core continues to contract and heat up, while also growing in mass thanks to the shell burning. Eventually, the critical mass of about $1.4 M_{\odot}$ of Fe is reached, at which point a stable configuration is no longer possible. As electrons are absorbed by the protons and some iron is disintegrated by thermal photons, the pressure support is suddenly removed and the core collapses essentially in free fall, reaching speeds of about a quarter of the speed of light.\footnote{Other collapse mechanisms are possible: an ``electron-capture'' supernova does not reach the final burning phase before highly degenerate electrons break apart nuclei and trigger a collapse.}

The collapse of the central region is suddenly halted after $\sim 10^{-2}$ seconds, as the density reaches nuclear (and up to supra-nuclear)  values. The central core bounces and a shock wave is formed. The extreme physical conditions of this core, in particular the densities of order $10^{12}-10^{14}$ g/cm$^{3}$, create a medium that is opaque even for neutrinos. As a consequence, the core initially has a trapped lepton number. The gravitational energy of the collapse at this stage is stored mostly in the degenerate Fermi sea of electrons ($E_{F}\sim 200$ MeV) and electron neutrinos, which are in equilibrium with the former. The temperature of this core is not more than 30 MeV, which means the core is relatively \emph{cold}. 

At the next stage, the trapped energy and lepton number both escape from the core, carried by the least interacting particles, which in the Standard Model are neutrinos.  Neutrinos and antineutrinos of all flavors are emitted in a time span of a few seconds (their diffusion time). The resulting central object then settles to a neutron star, or a black hole. A tremendous amount of energy, some~$10^{53}$ ergs, is released in $10^{58}$ neutrinos with energies $\sim 10$~MeV. A fraction of this energy is absorbed by beta reactions behind the shock wave that blasts away the rest of the star, creating a spectacular explosion.
Yet, from the energetics point of view, this visible explosion is but a tiny perturbation. Over 99\% of all gravitational binding energy of the $1.4 M_{\odot}$ collapsed core -- some 10\% of its rest mass -- is emitted in neutrinos. 

\subsection{Stages of the explosion}

Electron antineutrinos from the celebrated SN1987A core
collapse~\cite{Bionta:1987qt,Hirata:1987hu} in the Large Magellanic
Cloud outside the Milky Way were reported in water Cherenkov and
scintillator detectors. This observation provided qualitative validation of the basic physical picture outlined above and provided powerful constraints on numerous models of new physics. At the same time, the
statistics 
were sparse 
and a great many questions remain.  A high-statistics observation of a
nearby supernova neutrino burst will be possible with the current
generation of detectors. Such an observation will shed light
on 
the nature of the astrophysical event, as well as on the nature of
neutrinos themselves.  Sensitivity to the \textit{different flavor components}
of the flux is highly desirable.

The core-collapse neutrino signal starts with a short, sharp
\emph{neutronization} (or \emph{break-out}) burst primarily composed of
$\nu_e$. These neutrinos are messengers of the shock front breaking through the neutrinosphere (the surface of neutrino trapping): when this happens, iron is disintegrated, the neutrino scattering cross section drops and the lepton number trapped just below the original neutrinosphere is suddenly released. This quick and intense burst is followed by an
\emph{``accretion'' phase} lasting some hundreds of milliseconds, depending on the progenitor star mass, as matter falls onto the collapsed core and the shock is stalled at the distance of perhaps $\sim 200$ km. The gravitational binding energy of the accreting material is powering the neutrino luminosity during this stage. The later
\emph{``cooling'' phase} over $\sim$10~seconds represents the main part of
the signal, over which the proto-neutron star sheds its trapped energy.  

The flavor content and spectra of the neutrinos emitted from the neutrinosphere change
throughout these phases, and the supernova's evolution can
be followed with the neutrino signal. 
Some fairly generic features of these emitted neutrino fluxes are
illustrated in Figures~\ref{params},~\ref{3timescales}.

\begin{dunefigure}[Expected core-collapse parameters]{params}{Expected
  time-dependent signal for a specific flux model for an
  electron-capture supernova~\cite{Huedepohl:2009wh} at 10~kpc.  No oscillations are assumed. The
  top plot shows the luminosity as a function of time, the second plot
  shows average neutrino energy, and the third plot shows the $\alpha$
  (pinching) parameter.  The vertical dashed line at 0.02 seconds indicates
  the time of core bounce, and the vertical lines indicate different
  eras in the supernova evolution.  The leftmost time interval
  indicates the infall period.  The next interval, from core bounce to
  50~ms, is the neutronization burst era, in which the flux is
  composed primarily of $\nu_e$.  The next period, from 50 to 200~ms,
  is the accretion period. The final era, from 0.2 to 9~seconds, is
  the proto-neutron-star cooling period.  The general features are
  qualitatively similar for most core-collapse supernovae.}
\includegraphics[width=0.9\textwidth]{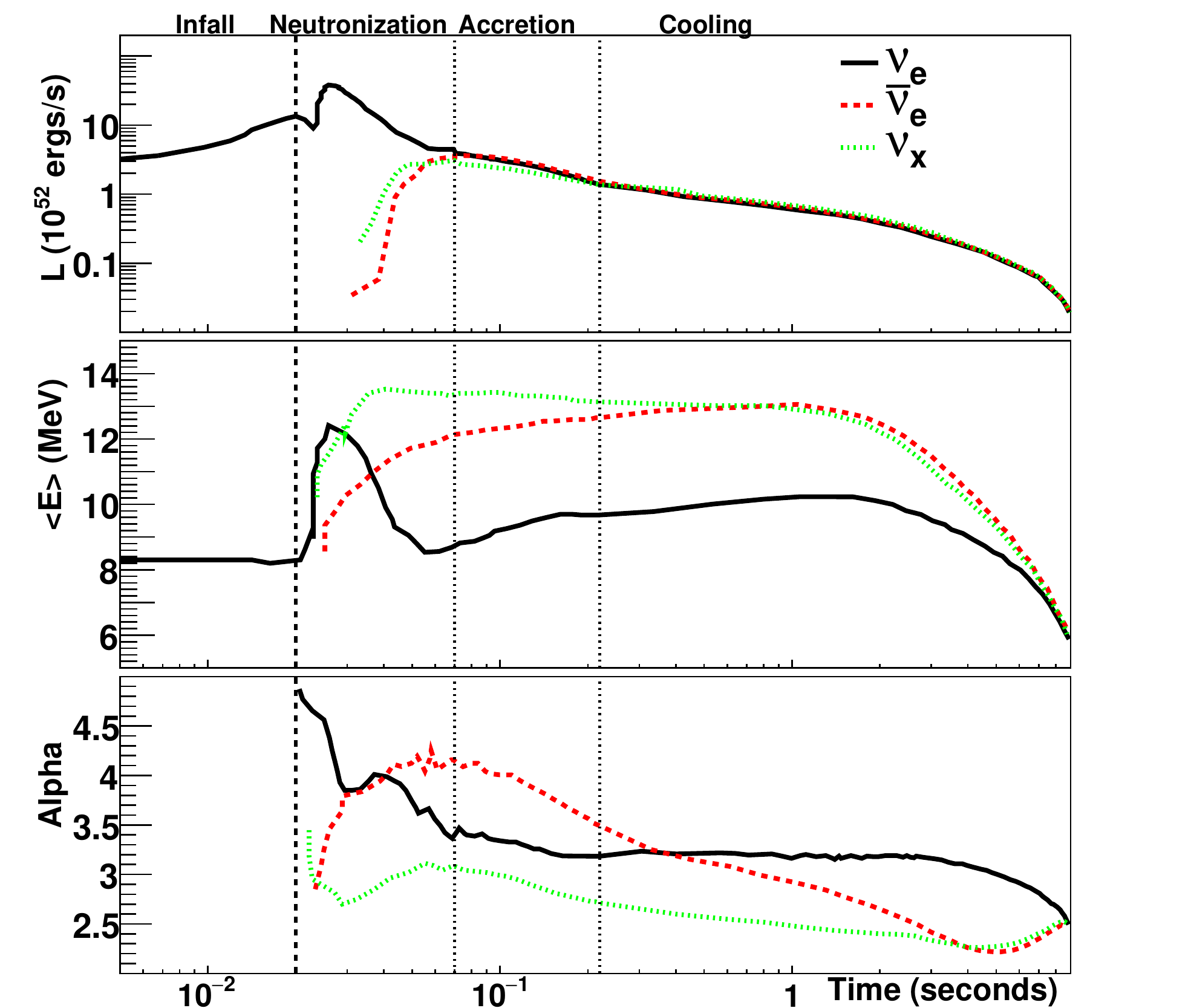}
\end{dunefigure}

\begin{dunefigure}[Expected fluxess]{3timescales}{Example of time-dependent spectra for the electron-capture supernova model~\cite{Huedepohl:2009wh} parameterized in Figure~\ref{params}, on three different timescales.   The z-axis units are neutrinos per cm$^2$ per millisecond per 0.2~MeV.  Top: $\nu_e$.  Center: $\bar{\nu}_e$.  Bottom: $\nu_x$.  Oscillations are not included here; note they can have dramatic effects on the spectra.}
\centerline{\includegraphics[width=10cm]{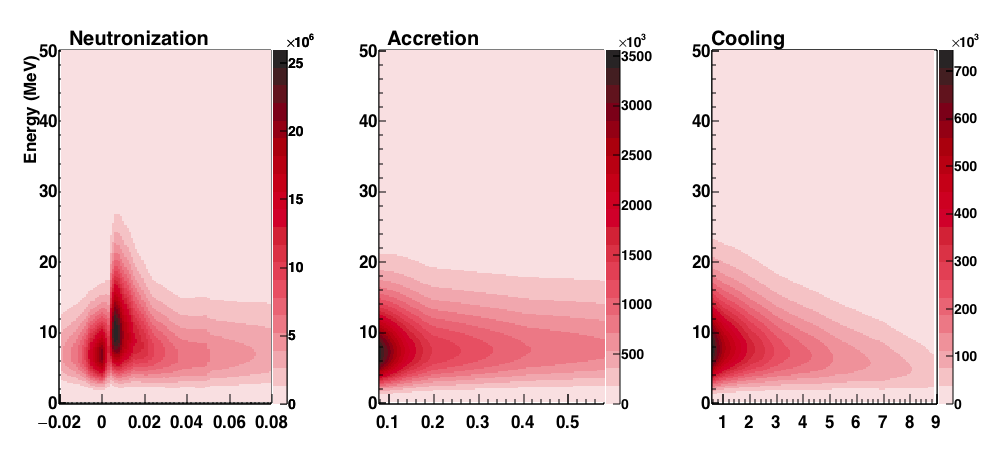}}
\centerline{\includegraphics[width=10cm]{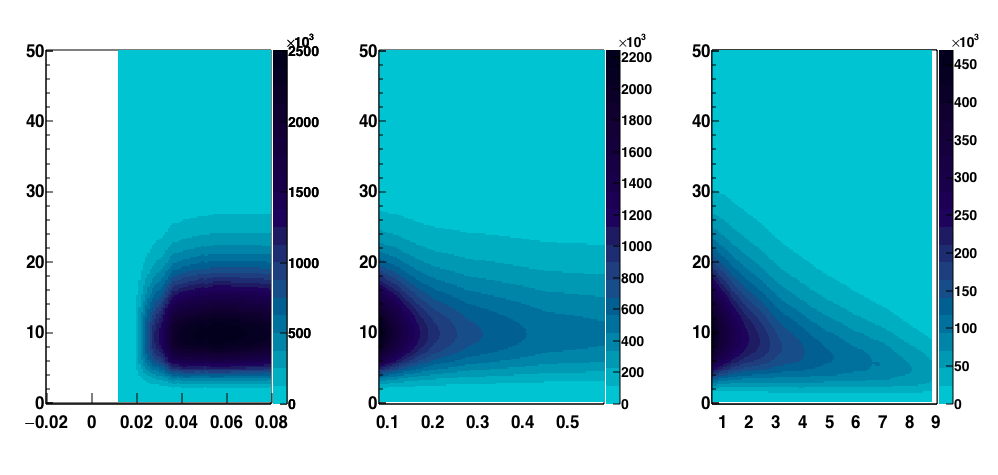}}
\centerline{\includegraphics[width=10cm]{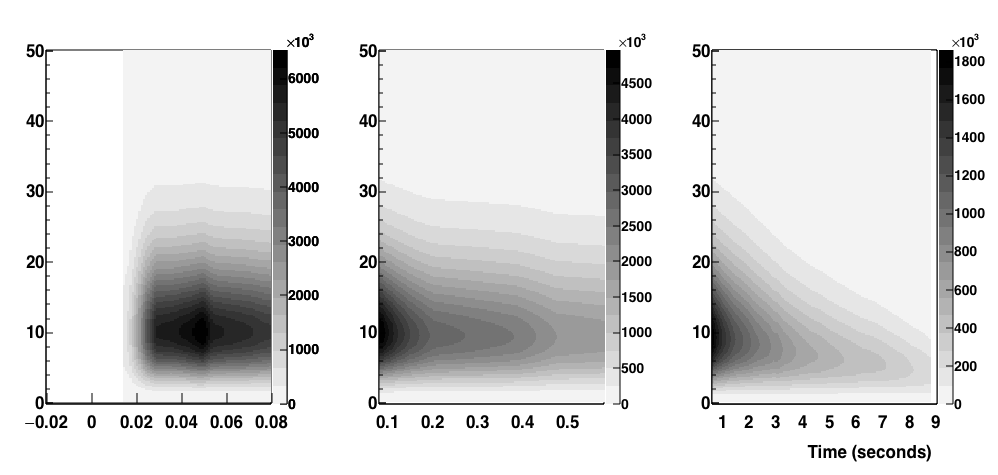}}
\end{dunefigure}

The physics of neutrino decoupling and spectra formation is far from trivial, owing to the energy dependence of the cross sections and the roles played by both charged- and neutral-current reactions.
Detailed transport calculations using methods such as \dword{mc} or Boltzmann solvers have been employed. It has been observed that spectra coming out of such simulations can typically be parameterized at a given moment in time by the following ansatz (e.g.,~\cite{Minakata:2008nc,Tamborra:2012ac}):
\begin{equation}
        \label{eq:pinched}
        \phi(E_{\nu}) = \mathcal{N} 
        \left(\frac{E_{\nu}}{\langle E_{\nu} \rangle}\right)^{\alpha} \exp\left[-\left(\alpha + 1\right)\frac{E_{\nu}}{\langle E_{\nu} \rangle}\right] \ ,
\end{equation}
where $E_{\nu}$ is the neutrino energy, $\langle E_\nu \rangle$ is the
mean neutrino energy, $\alpha$ is a ``pinching parameter'', and
$\mathcal{N}$ is a normalization constant.
Large $\alpha$ corresponds to a more ``pinched'' spectrum (suppressed
high-energy tail). This parameterization is referred to as a
``pinched-thermal'' form. The different $\nu_e$, $\overline{\nu}_e$ and
$\nu_x, \, x = \mu, \tau$ flavors are expected to have different
average energy and $\alpha$ parameters and to evolve differently in
time. 

The initial spectra get further processed (permuted) by flavor oscillations and understanding these oscillations is very important for extracting physics from the detected signal.

\section{Low-Energy Events in DUNE}\label{sec:lowe-events}

\subsection{Detection Channels and Interaction Rates}

Liquid argon should have a particular sensitivity to the $\nu_e$
component of a supernova neutrino burst, via the dominant interaction,
charged-current
absorption of $\nu_e$ on $^{40}$Ar,
\begin{equation}
\nu_e + ^{40}{\rm Ar} \rightarrow e^- + ^{40}{\rm K^*},
\label{eq:nueabs}
\end{equation}
for which the observable is the $e^-$ plus deexcitation products from
the excited $K^*$ final state.  Additional channels include a
$\bar{\nu}_e$ CC interaction and elastic scattering on electrons.
Cross sections for the most
relevant interactions are shown in Figure~\ref{xscns}.  It is worth
noting that none of the neutrino-$^{40}$Ar cross sections in this
energy range have been experimentally measured; theoretical
calculations may have large uncertainties.

Another process of interest for supernova detection in \dword{lar} detectors,
not yet fully studied,  is neutral-current  scattering on Ar nuclei by
any type of neutrino: $\nu_x + {\rm Ar} \rightarrow \nu_x + {\rm
  Ar}^*$,  for which the signature is given by the cascade of
deexcitation $\gamma$s from the final state Ar nucleus.  A dominant
9.8-MeV Ar$^*$ decay line has been recently identified as a spin-flip
M1 transition~\cite{Hayes}.   At this energy the probability of
$e^+e^-$ pair production is relatively high, offering a potentially
interesting neutral-current tag.  Other transitions are under investigation.

\begin{dunefigure}[Cross sections for supernova-relevant interactions in argon]{xscns}{Cross sections for supernova-relevant interactions in argon~\cite{snowglobes,GilBotella:2003sz}.}
\includegraphics[width=0.6\textwidth]{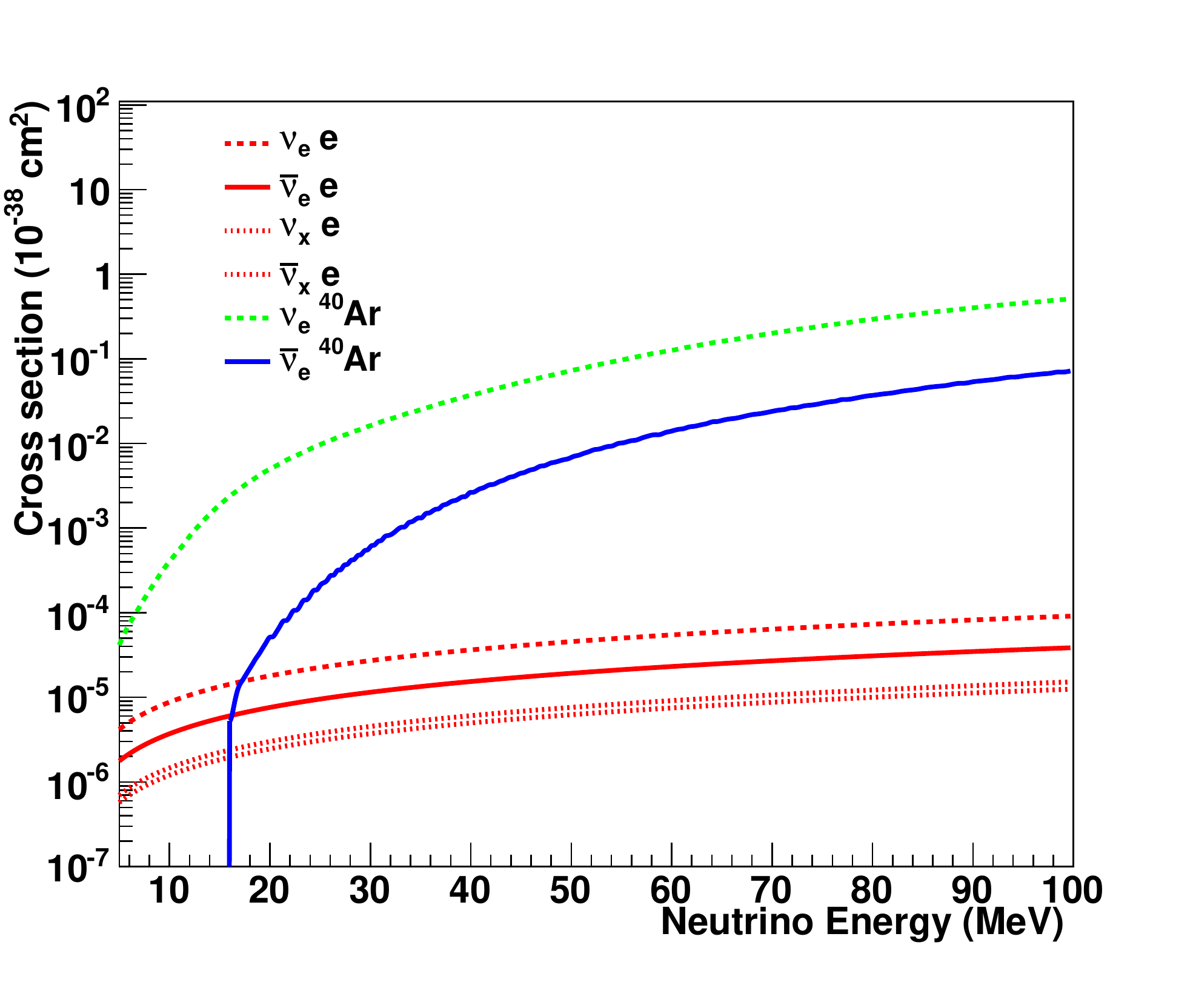}
\end{dunefigure}

The predicted event rate from a supernova burst may be calculated by
folding expected neutrino differential energy spectra with cross
sections for the relevant channels, and with detector response; we do
this using \dword{snowglobes}~\cite{snowglobes} (see Sec.~\ref{snowglobes}.)

\subsection{Event Simulation and Reconstruction}

Supernova neutrino events, due to their low energies, will manifest
themselves primarily as spatially small events, perhaps up to a few tens of cm
scale, with stub-like tracks from
electrons (or positrons from the rarer $\bar{\nu}_e$ interactions).
Events from $\nu_e $ charged-current interactions, $\nu_e+{}^{40}{\rm
  Ar}\rightarrow e^{-}+{}^{40}{\rm K}^{*}$, are likely to be
accompanied by de-excitation products-- gamma rays and/or ejected
nucleons. Gamma-rays are in principle observable via energy deposition
from Compton scattering, which will show up as small charge blips in
the \dword{tpc}.  
Ejected nucleons may result in loss of observed energy for
the event.  Elastic scattering on electrons will result in single
scattered electrons, and single gamma rays may result from \dword{nc} 
excitations of the argon nucleus.   Each event category has, in principle, a
distinctive signature.

The canonical reconstruction task is to identify the interaction
channel, the neutrino flavor for \dword{cc} events, and to determine the
4-momentum of the incoming neutrino; this overall task is the same for
low-energy events as for high-energy ones.  The challenge is to
reconstruct the properties of the lepton (if present), and to the extent
possible, to tag the interaction channel by the pattern of final-state
particles.

While
some physics studies in the \dword{snble} group
 use a fast event-rate calculation tool called \dword{snowglobes}, most
 activity is towards development of realistic and comprehensive
 simulation and reconstruction tools, from neutrino interaction event
 generators through full event reconstruction, in both single and
 dual-phase detectors, with \dword{larsoft}.

\subsubsection{\dshort{marley}}

\dword{marley}~\cite{marley} simulates tens-of-MeV
neutrino-nucleus interactions in liquid argon. Currently, \dword{marley} can only
simulate charged-current $\nu_e$ scattering on $^{40}$Ar, but other
reaction channels will be added in the future.

\dword{marley} weights the incident neutrino spectrum, selects an initial excited state
of the residual $^{40}$K$^*$ nucleus, and samples an outgoing electron
direction using the allowed approximation for the $\nu_e$ \dword{cc} differential cross
section.\footnote{That is, the zero momentum transfer and zero nucleon velocity
limit of the tree-level $\nu_e$ \dword{cc} differential cross section, which may be
written as
\[
\frac{d\sigma}{d\cos \theta}
= \frac{G_F^2 |V_{ud}|^2}{2\pi} |\mathbf{p}_e|\, E_e \,F(Z_f, \beta_e)
\left[(1+\beta_e \cos\theta)B(F) + \left(\frac{3 - \beta_e \cos\theta}
{3}\right)B(GT)\right].
\]
In this expression, $\theta$ is the angle between the incident neutrino and the
outgoing electron, $G_F$ is the Fermi constant, $V_{ud}$ is the quark mixing
matrix element, $F(Z_f, \beta_e)$ is the Fermi function, and $|\mathbf{p}_e|$,
$E_e$, and $\beta_e$ are the outgoing electron's three momentum, total energy,
and velocity, respectively. $B(F)$ and $B(GT)$ are the Fermi and Gamow-Teller
matrix elements.
}
\dword{marley} computes this cross section using a table of Fermi and Gamow-Teller
nuclear matrix elements. Their values are taken from experimental measurements
at low excitation energies and a quasiparticle random phase approximation
(QRPA) calculation at high excitation energies. As the code develops, a more
sophisticated treatment of this cross section will  be included.

After simulating the initial two-body $^{40}${Ar}($\nu_e$,
$e^{-}$)$^{40}$K$^*$ reaction for an event, \dword{marley}
also handles the subsequent nuclear de-excitation. For bound nuclear
states, the de-excitation $\gamma$-rays are sampled using tables of
experimental branching ratios. These tables are supplemented with
theoretical estimates when experimental data are unavailable. For
particle-unbound nuclear states, \dword{marley} simulates the competition between
$\gamma$-ray and nuclear fragment\footnote{ Nucleons and light nuclei up to
$^{4}${He} are considered.} emission using the Hauser-Feshbach
statistical model.   Figure~\ref{fig:marleydist} shows an example
visualization of a simulated \dword{marley} event.

Although many refinements remain to be made, \dword{marley}'s treatment of high-lying
Gamow-Teller strength and nuclear de-excitations represents a significant
improvement over existing tools for simulating supernova $\nu_e$ \dword{cc} events.  \dword{marley} has been now been fully incorporated into the \dword{larsoft} code base.

\begin{dunefigure}[\dword{marley} event]{fig:marleydist}{Visualization of an
    example \dword{marley}-simulated $\nu_e$\dword{cc} event, showing the trajectories and energy
    deposition points of the interaction products. }
\includegraphics[width=0.6\textwidth]{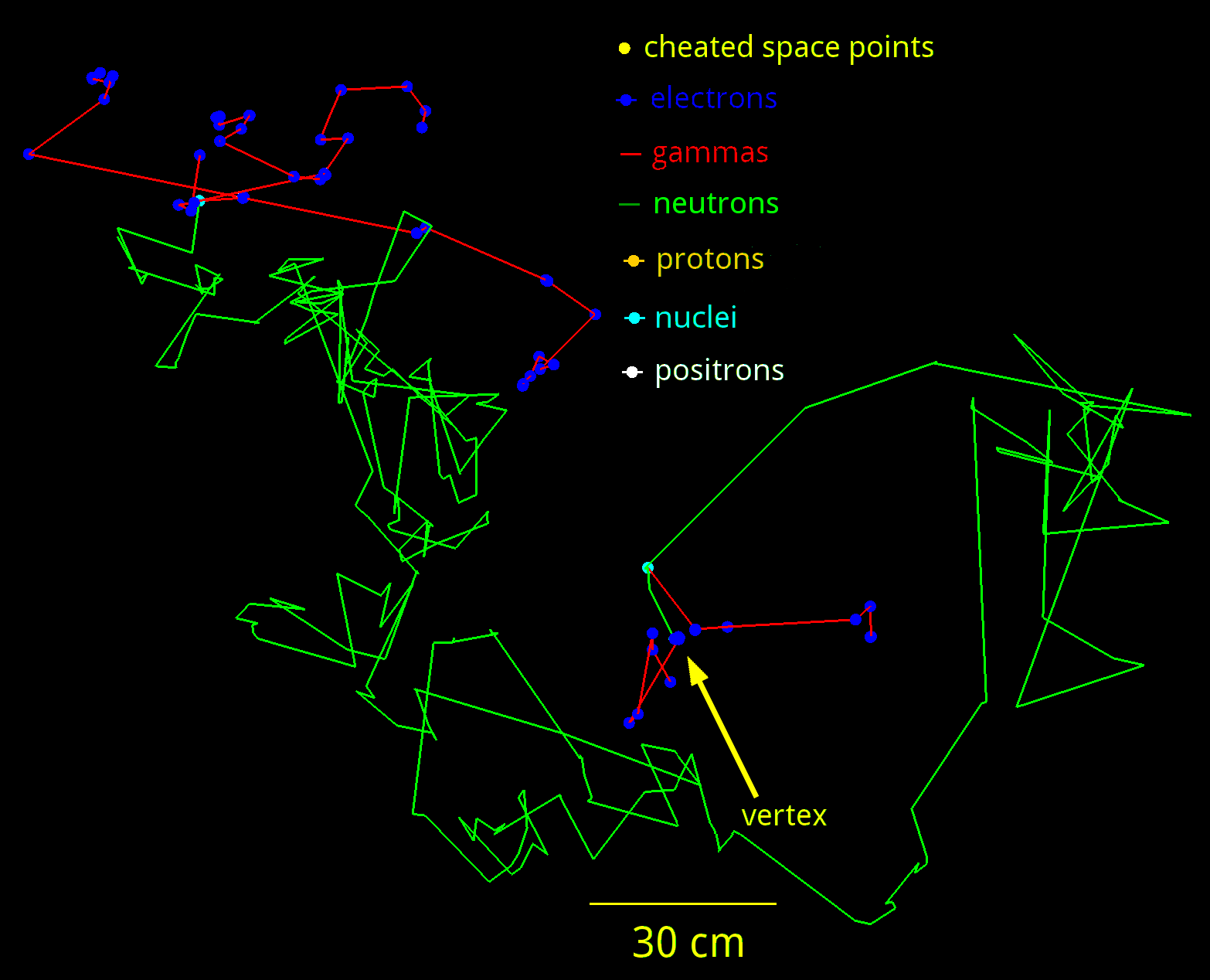}
\end{dunefigure}

\subsubsection{Low-energy Event Reconstruction Performance}

The standard DUNE reconstruction tools in \dword{larsoft} provide
energy and track reconstruction for
low energy events.  Photons may also be used for calorimetry.
Figure~\ref{reseff} shows summarized resolution and efficiency for
\dword{marley} events.

\begin{dunefigure}[Resolution and efficiency]{reseff}{Left:
    reconstruction efficiency as a function of neutrino energy for
    MARLEY events, for different minimum required reconstructed
    energy. Right: fractional energy resolution as a function of
    neutrino energy for TPC tracks (black) and photon detector
    calorimetry (blue). The red ``physics-limited resolution'' assumes
  all energy deposited by final-state particles is reconstructed; the
  finite resolution represents loss of energy from escaping particles.}
\includegraphics[width=0.45\textwidth]{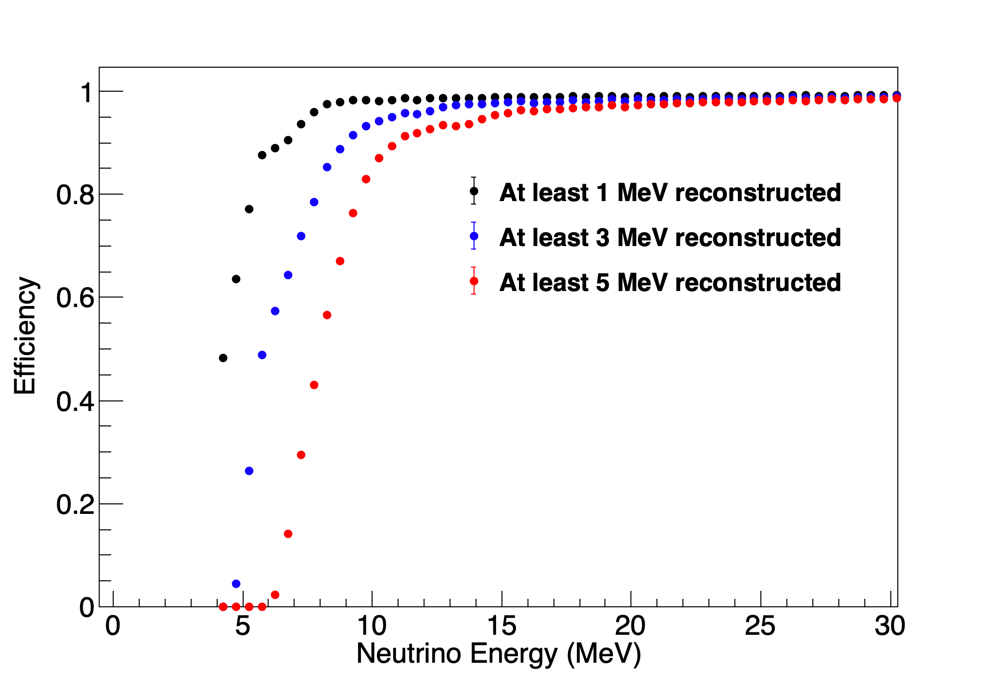}
\includegraphics[width=0.45\textwidth]{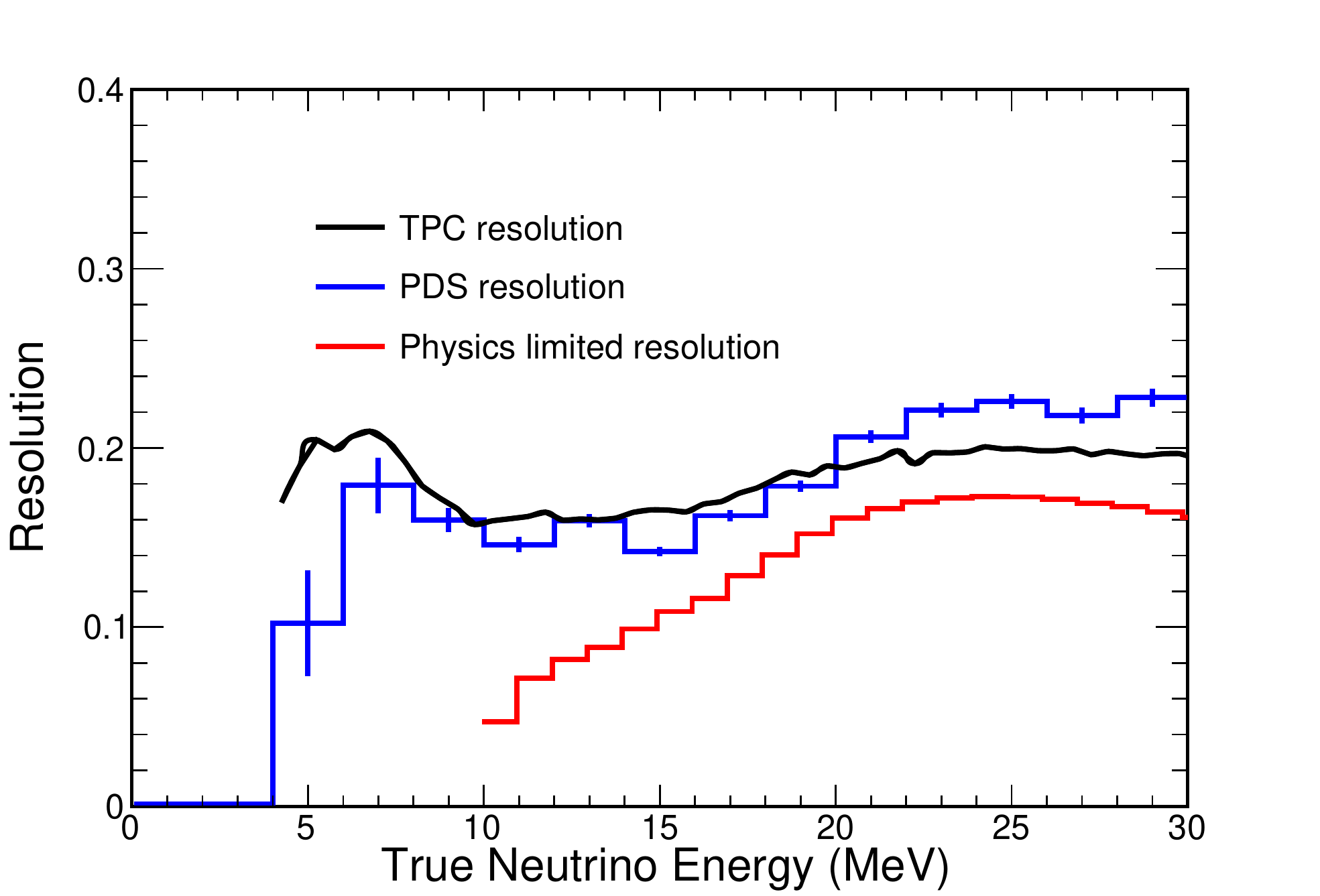}
\end{dunefigure}

\subsubsection{ \dword{snowglobes}}\label{snowglobes}

Most supernova neutrino studies done for DUNE so far, including of the
plots included in the \dword{cdr}~\cite{Acciarri:2015uup}, have employed
 \dword{snowglobes}\cite{snowglobes}, a fast event-rate computation tool.  This
uses 
\dword{globes} front-end software~\cite{Huber:2004ka} to
convolve fluxes with cross-sections and detector parameters.  The
output is in the form of interaction rates for each channel as a
function of neutrino energy, and ``smeared'' rates as a function of
detected energy for each channel (i.e., the spectrum that
actually would be observed in a detector).  
The smearing (transfer) matrices incorporate both
interaction product spectra for a given neutrino energy, and detector
response.   Figure~\ref{marleysmearing} shows such a transfer matrix
created
using \dword{marley}, by determining the distribution of observed charge, and
 a full simulation of the detector response (including the generation,
 transport, and detection of ionization signals and the electronics)
 as a function of neutrino energy in 0.5-MeV neutrino energy steps.
Time dependence in  \dword{snowglobes} can be straightforwardly
handled by providing multiple files with fluxes divided into different
time bins. \footnote{Note that  \dword{snowglobes} is \textit{not} a \dlong{mc}
code--- it calculates mean event rates using a transfer matrix to
convert neutrino spectra to observed spectra.  This will produce
equivalent results to reweighting Monte Carlo.}

\begin{dunefigure}[\dword{marley} event]{marleysmearing}{Smearing matrix for
     \dword{snowglobes} created with monochromatic \dword{marley} samples run though
    \dword{larsoft}, describing detected charge distribution as a function of
    neutrino energy.  The effects of interaction product distributions
  and detector smearing are both incorporated in this matrix.  The
  right hand plot 
  incorporates an assumed correction for charge attenuation due to
  electron drift,  based on \dlong{mc} truth position of the
  interaction.      The drift correction improves resolution.}
\includegraphics[width=0.45\textwidth]{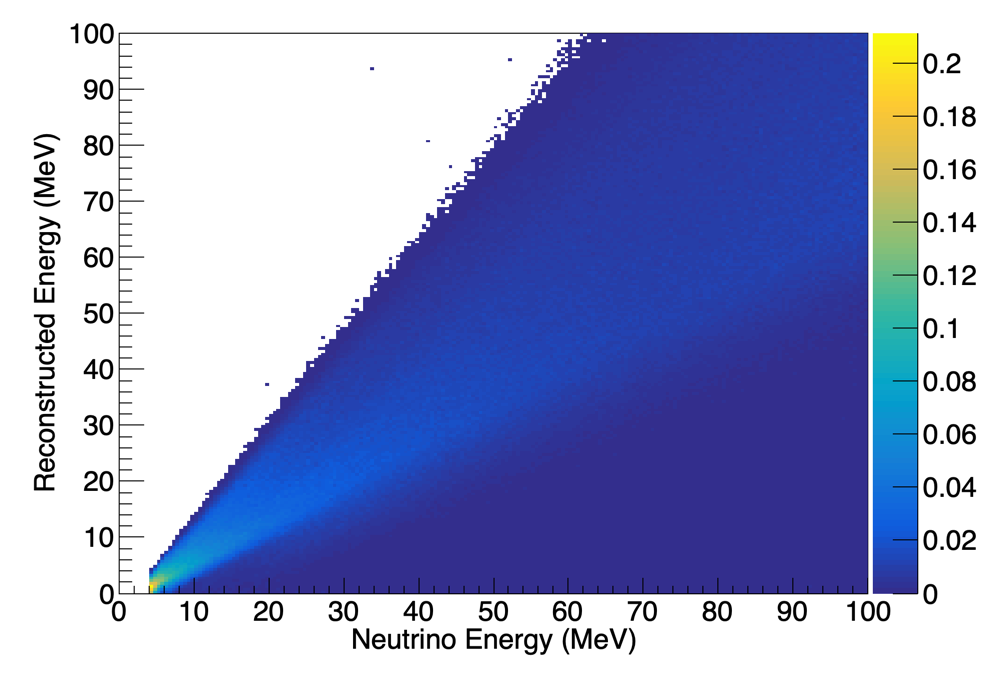}
\includegraphics[width=0.45\textwidth]{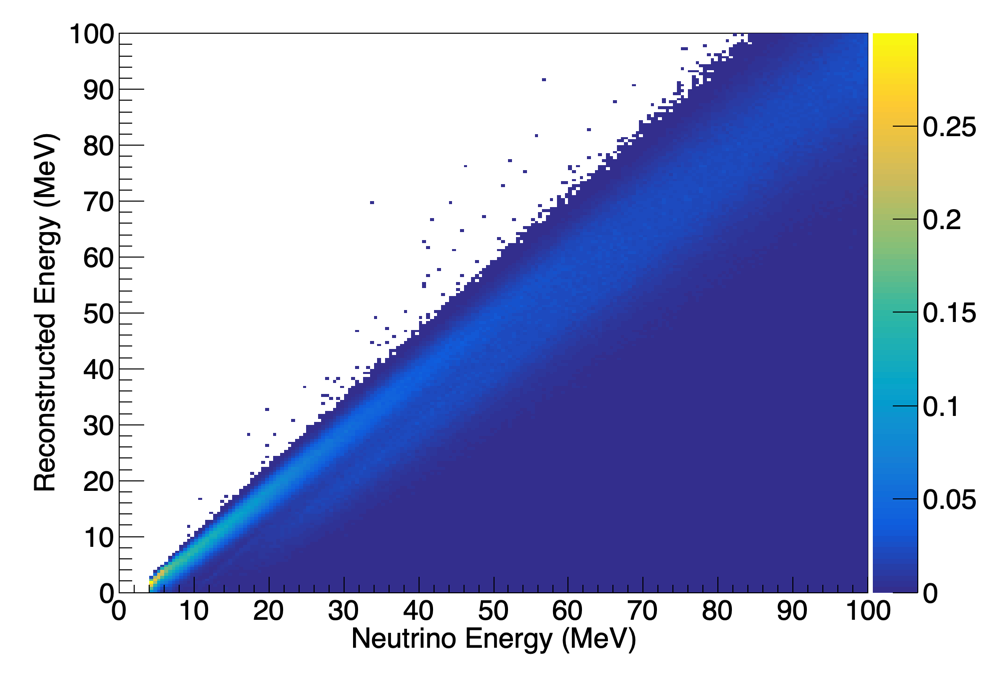}

\end{dunefigure}

While \dword{snowglobes} is, and will continue to be, a fast, useful tool,
it has limitations with respect to a full simulation.  One loses correlated
event-by-event angular and energy information, for example; some
studies, such as the directionality study in
Section~\ref{sec:pointing} require such complete event-by-event information.
Nevertheless, transfer matrices generated with the best available
simulations can be used to compute observed event rates and energy distributions and draw useful conclusions.

\subsubsection{Backgrounds}

Understanding of cosmogenic and radiological
backgrounds is also important for understanding of how
well we can reconstruct low energy events, and for setting detector
requirements.  Small single-hit blips from $^{39}$Ar or other
impurities may fake de-excitation gammas.  While preliminary studies
show that backgrounds will have a minor effect on reconstruction of
triggered supernova burst events, their effects on a \dword{daq} and
triggering system that satisfies supernova burst triggering
requirements requires separate consideration.
These issues are addressed in the DAQ and backgrounds sections of this
TDR.

\section{Expected Supernova Burst Signal Properties}\label{sec:sn-signals}

Table~\ref{argon_events} shows rates calculated  for the dominant interactions in argon for
the ``Livermore'' model~\cite{Totani:1997vj} (out of date, but included for comparison with literature), and the ``GKVM''
model~\cite{Gava:2009pj}; for the former, no oscillations are assumed
in the supernova or Earth; the latter assumes collective effects in
the supernova.  In general, there is a rather wide variation--- up to an order of magnitude --- in event rate for different models, due to different numerical treatment (e.g., neutrino transport, dimensionality), physics input (nuclear equation of state, nuclear correlation and impact on neutrino opacities, neutrino-nucleus interactions) and oscillation effects. In addition, there is intrinsic variation in the nature of the progenitor and collapse mechanism.  Neutrino emission from the supernova may furthermore have an emitted lepton-flavor asymmetry~\cite{Tamborra:2014aua}, so that observed rates may be dependent on the supernova direction.
\begin{dunetable}[Event numbers \SI{40}{\kt} of \dword{lar} at 10~kpc]{lcc}{argon_events}{Event counts for different
    supernova models in \SI{40}{\kt} of liquid argon for a core collapse at 10~kpc, for $\nu_e$ and $\bar{\nu}_e$ charged-current channels and \dword{es} on electrons.
    Event rates will simply scale by active detector mass and inverse
    square of supernova distance.   No oscillations are assumed; we
    note that oscillations (both standard and ``collective'') will
    potentially have a large, model-dependent effect, discussed in Sec.~\ref{sec:mh}.}
Channel & Events & Events \\
\rowtitlestyle
& ``Livermore'' model & ``GKVM'' model  \\ 
\toprowrule

$\nu_e + ^{40}{\rm Ar} \rightarrow e^- + ^{40}{\rm K^*}$ & 2720  & 3350 \\ \colhline

$\overline{\nu}_e + ^{40}{\rm Ar} \rightarrow e^+ + ^{40}{\rm Cl^*}$ & 230 & 160\\ \colhline

$\nu_x + e^- \rightarrow \nu_x + e^-$                           & 350 &  260\\ \colhline

Total &  3300 & 3770 \\ 
\end{dunetable}

Clearly, the $\nu_e$
flavor dominates.  Although water and scintillator detectors will record $\nu_e$ events~\cite{Laha:2013hva,Laha:2014yua}, liquid argon is the only future prospect for a large, clean supernova $\nu_e$ sample~\cite{Scholberg:2012id}.

The number of signal events scales with mass and inverse square of
distance as shown in Figure~\ref{ratesvsdist}.  For a collapse in the
Andromeda galaxy, 780~kpc away, a 40-kton detector would observe a few events.

\begin{dunefigure}[Supernova neutrino rates vs distance]{ratesvsdist}{Estimated numbers of supernova neutrino interactions in DUNE as a function of distance to the supernova, for different detector masses ($\nu_e$ events dominate). The red dashed lines represent expected events for a 40-kton detector and the green dotted lines represent expected events for a 10-kton detector. The lines limit a fairly wide range of possibilities for ``Garching-parameterized'' supernova flux spectra (Equation~\ref{eq:pinched}) with luminosity $0.5\times 10^{52}$ ergs over ten seconds. The optimistic upper line of a pair gives the number of events for average $\nu_e$ energy of $\langle E_{\nu_e}\rangle =12$~MeV, and ``pinching'' parameter $\alpha=2$; the pessimistic lower line of a pair gives the number of events for $\langle E_{\nu_e}\rangle=8$~MeV and $\alpha=6$. (Note that the luminosity, average energy and pinching parameters will vary over the time frame of the burst, and these estimates assume a constant spectrum in time. Oscillations will also affect the spectra and event rates.) The solid lines represent the integrated number of events for the specific time-dependent neutrino flux model in~\cite{Huedepohl:2009wh} (see Figures~\ref{params} and \ref{3timescales}; this model has relatively cool spectra and low event rates). Core collapses are expected to occur a few times per century, at a most-likely distance of around 10 to 15 kpc.}
\includegraphics[width=5in]{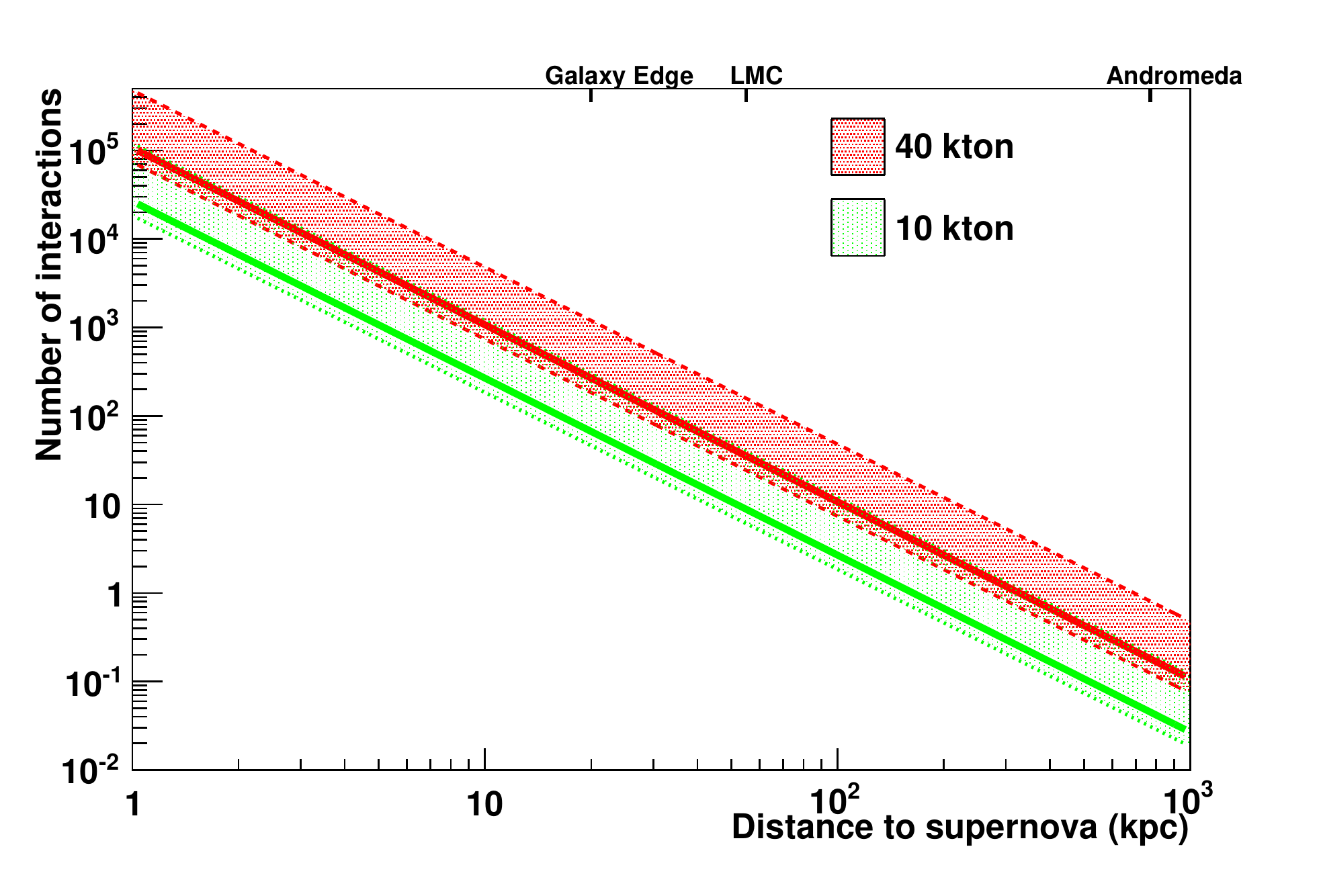}
\end{dunefigure}

\subsection{Directionality: pointing to the supernova}\label{sec:pointing}


It will be valuable to use DUNE's tracking ability to reconstruct the direction of the incoming neutrinos to the extent possible.  Reconstruction of direction to a supernova (or other astrophysical event) will be of obvious use to astronomers for prompt detection of the early turn-on of the light.  Furthermore, some core collapse events may not yield bright electromagnetic fireworks, in which case directional information may help in location of a dim supernova or even a ``disappeared'' progenitor~\cite{Kochanek:2008mp}.
Directional information can be used for correlation with gravitational
wave observations, which also have some directionality.  Pointing
resolution for low-energy events will also be helpful for selecting
signal from background for solar neutrinos or other sources with known
angular distribution.  The directional information could also
potentially be used in a high-level trigger.

The pointing resolution incorporates the intrinsic angular spread of the
interaction products of the neutrino interaction, as well as
resolution for detector reconstruction.  A large fraction of the
events expected from the supernova will not point well; in particular,
the expected angular distribution of the $\nu_e$ \dword{cc} absorption
events which will make up the bulk of the signal events are expected
to have relatively weak, but usable,  anisotropy, with intrinsic physics-related
(not detector-related)
head-tail ambiguity.   Fermi transitions to the final state are
described by a $\propto (1+\cos\theta)$ angular distribution and Gamow-Teller
transitions are
described by a $\propto (1-\frac{1}{3}\cos\theta)$ angular
distribution; these are modeled in MARLEY.
In contrast, the elastic
scattering component of the signal should point more sharply.
Directionality depends also on event energy.  See Figure~\ref{fig:escc}. 

We describe here a study of the ability of DUNE to point to a
supernova using the \dword{tpc} tracks.  This study makes use of full
simulation and reconstruction tools.  We have studied single
electrons, neutrino-electron elastic scattering events, and the full
expected supernova signal, looking at both elastic scattering events
and $\nu_e$CC events.  Future studies will
incorporate additional interaction channels, as well as backgrounds.

\begin{dunefigure}[\dword{es} event]{fig:1eevd}{Example event display for a
    single simulated 10.25~MeV electron, with track reconstruction, in
    time vs wire, with color representing charge.
    The top panel shows the collection plane and the bottom panels
    show induction planes.  The boxes represent reconstructed hits. }
 \includegraphics[width=0.4\textwidth]{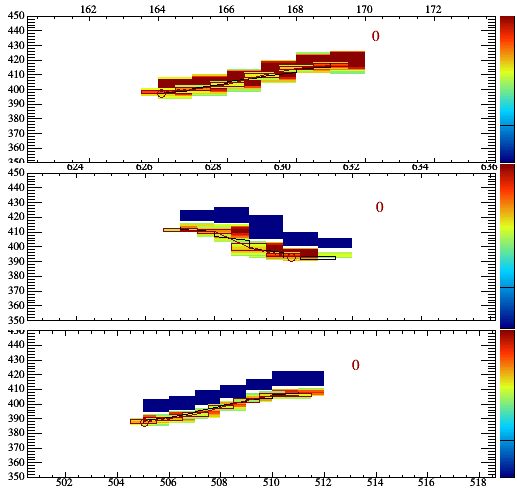}
\end{dunefigure}

\begin{dunefigure}[\dword{es} and CC events]{fig:escc}{Example
   distribution of reconstructed directions of \dword{es} and CC supernova neutrino
    events in a 12-14 MeV reconstructed energy bin (left) and a 24-26
    MeV reconstructed energy bin (right). }
 \includegraphics[width=0.4\textwidth]{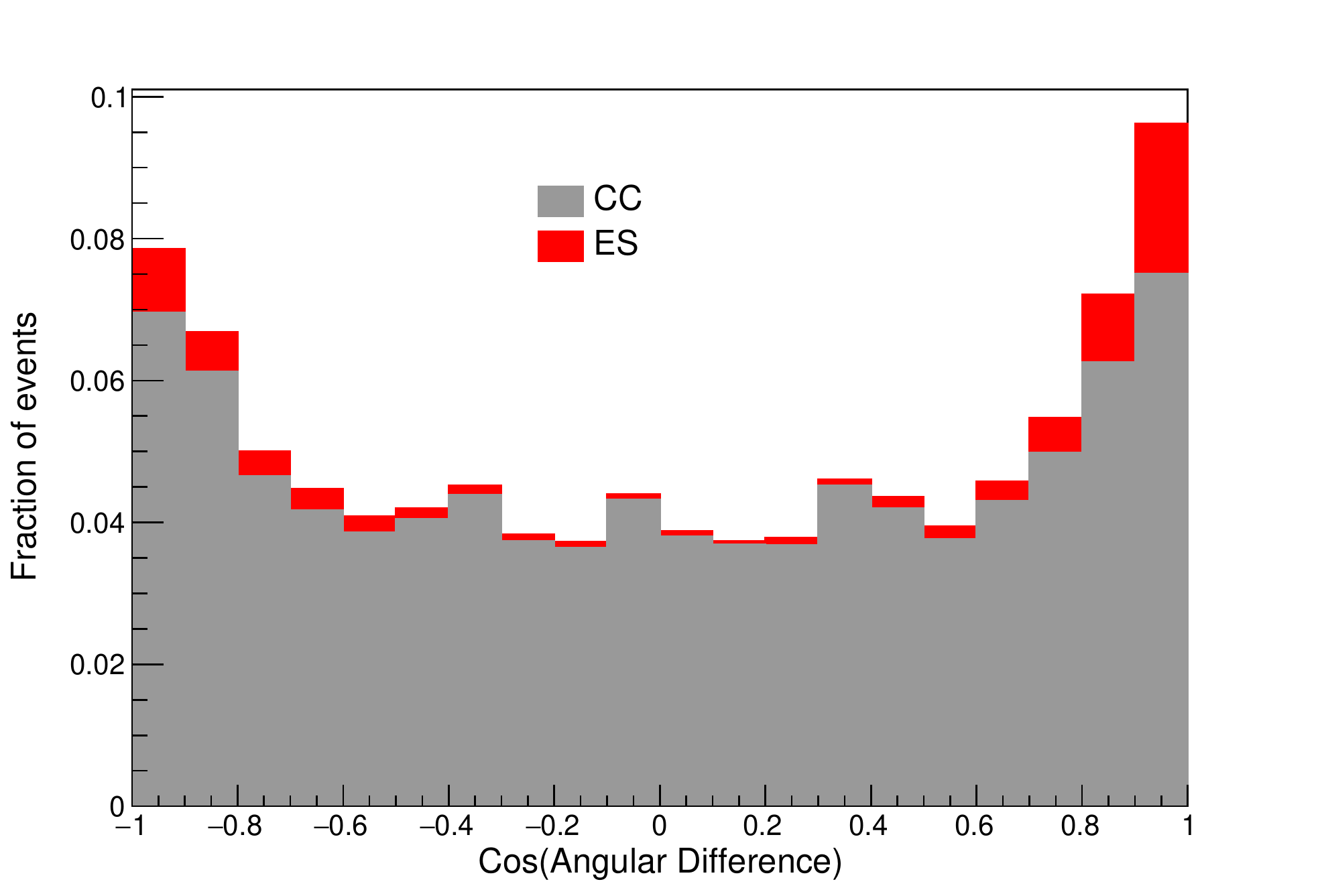}
 \includegraphics[width=0.4\textwidth]{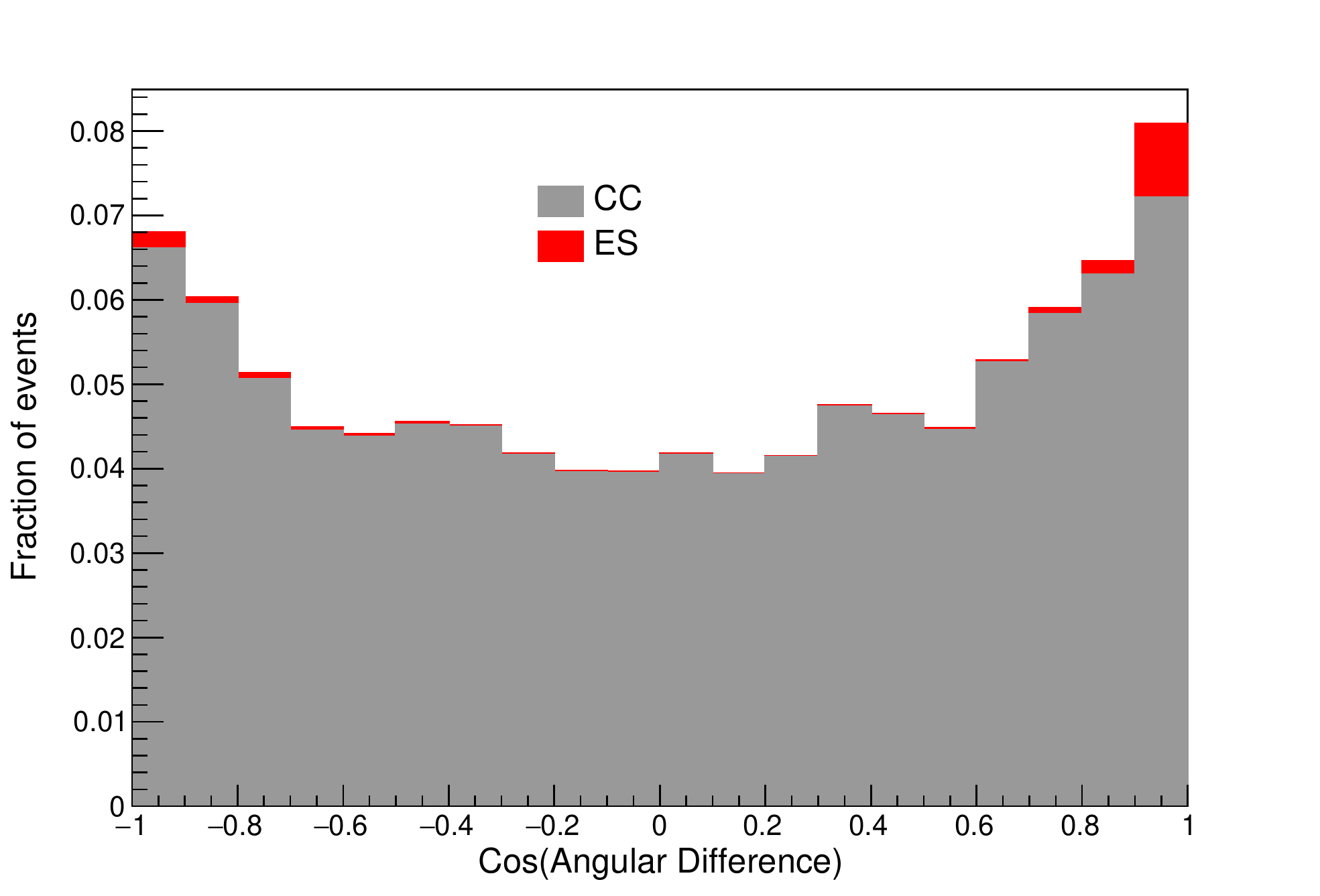}
\end{dunefigure}

\begin{dunefigure}[Pointing vs neutrino
  energy]{fig:ESpointres}{ Left: Pointing resolution for electron
    tracks, showing effect of direction ambiguity, which can be
    partially resolved using bremsstrahlung directionality.  The black
    line shows the angle at which 68\% of angular differences are
    closer to truth, given the entire distribution including events
    misreconstructed $\sim$180$^\circ$ away from the true direction.
    The red line shows the same when the absolute value of the cosine
    of the angle with respect to the true direction is used,
    effectively disambiguating head-tail using truth.  The blue line uses a
    ``daughter flipping'' algorithm which preferentially selects the
    track's    forward direction using the relative positions with respect to the
    track of Compton-scatter blips from bremsstrahlung gamma
    daughters.
    Right: Pointing resolution of elastic scattering
    events versus neutrino energy for each neutrino flavor.}
\includegraphics[width=0.4\textwidth]{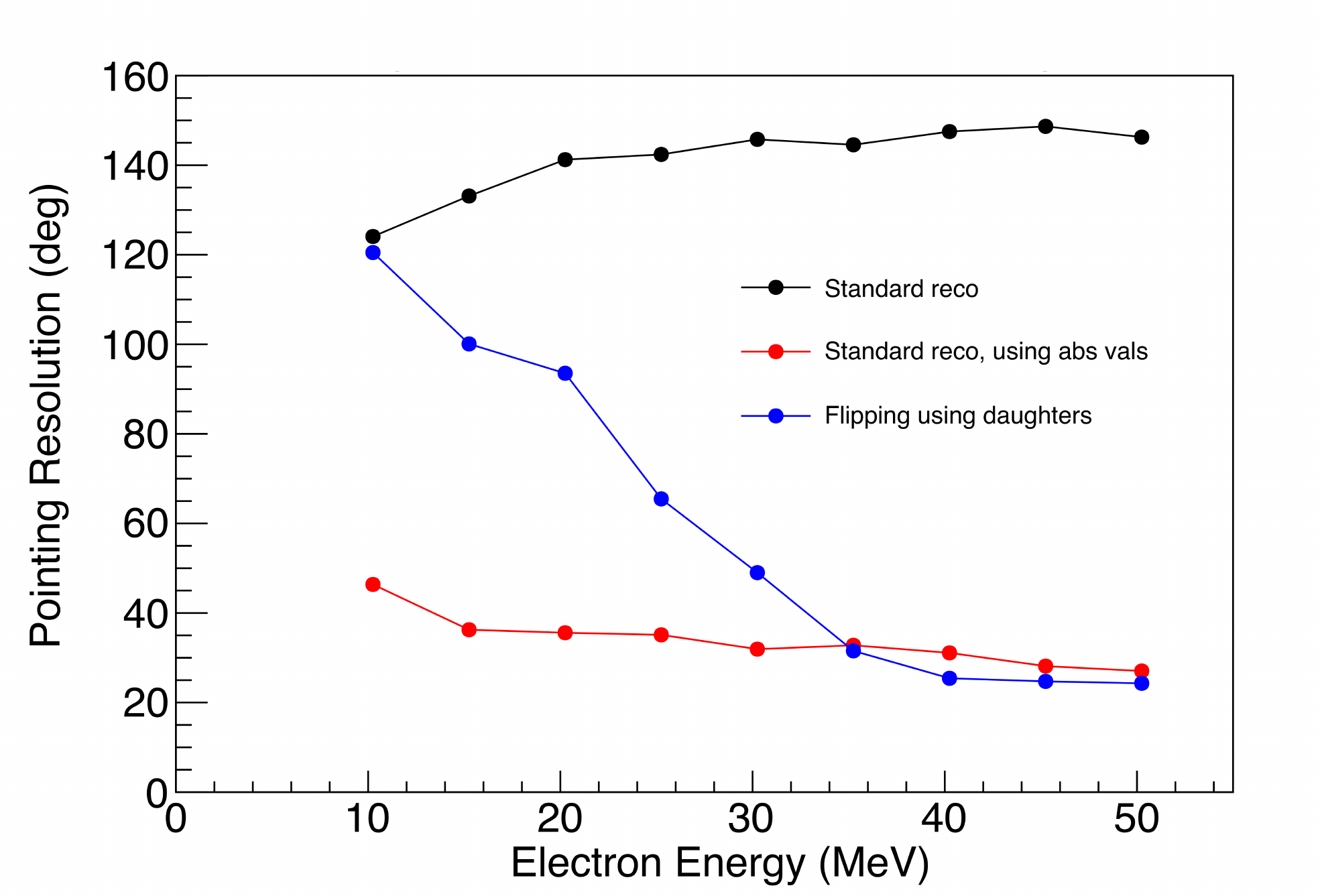}
\includegraphics[width=0.4\textwidth]{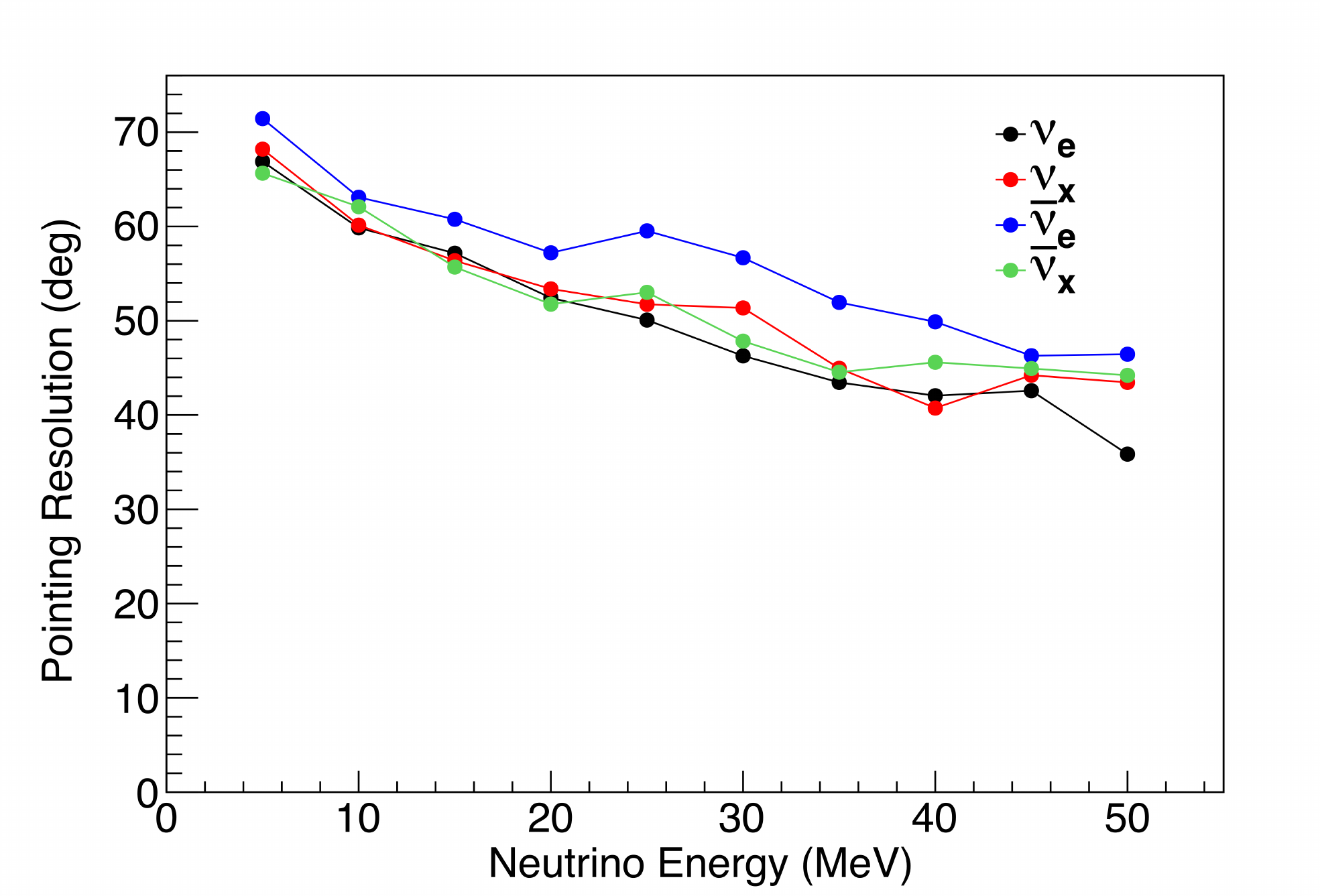}
\end{dunefigure}

The pointing resolution of the reconstructed electron direction with
respect to the true neutrino direction, defined as the angle at which
68\% of angular differences are closer to truth, is plotted in
Figure~\ref{fig:ESpointres} on the right.
The absolute values of cosines of the
angular differences are used, which does not capture the head-tail
directional ambiguity of the electron track.
This pointing resolution is a result of both the neutrino-electron
angle spread, electron scattering and the error in reconstruction.  The left plot shows the
pointing resolution for electrons only, and the effect of a head-tail disambiguation
using bremsstrahlung directionality (``daughter flipping'').  Work is
continuing to improve the directional disambiguation algorithm,
including use of increased multiple
scattering towards the end of a track.

\begin{dunefigure}[Pointing for full supernova]{fig:fullSN}{Left:
    Negative log
    likelihood values as a function of direction for a 10-kpc
    supernova sample.  The sample used to compute the likelihood
    includes also the dominant $\nu_e$CC interactions.  Right: Distribution of angular differences for
    directions to a 10-kpc supernovae using a maximum likelihood
    method. The supernovae incorrectly reconstructed in the backwards direction,
    shown in red,
  have the distribution of absolute value of $\cos \theta$ plotted for display purposes. }
  \includegraphics[width=0.44\textwidth]{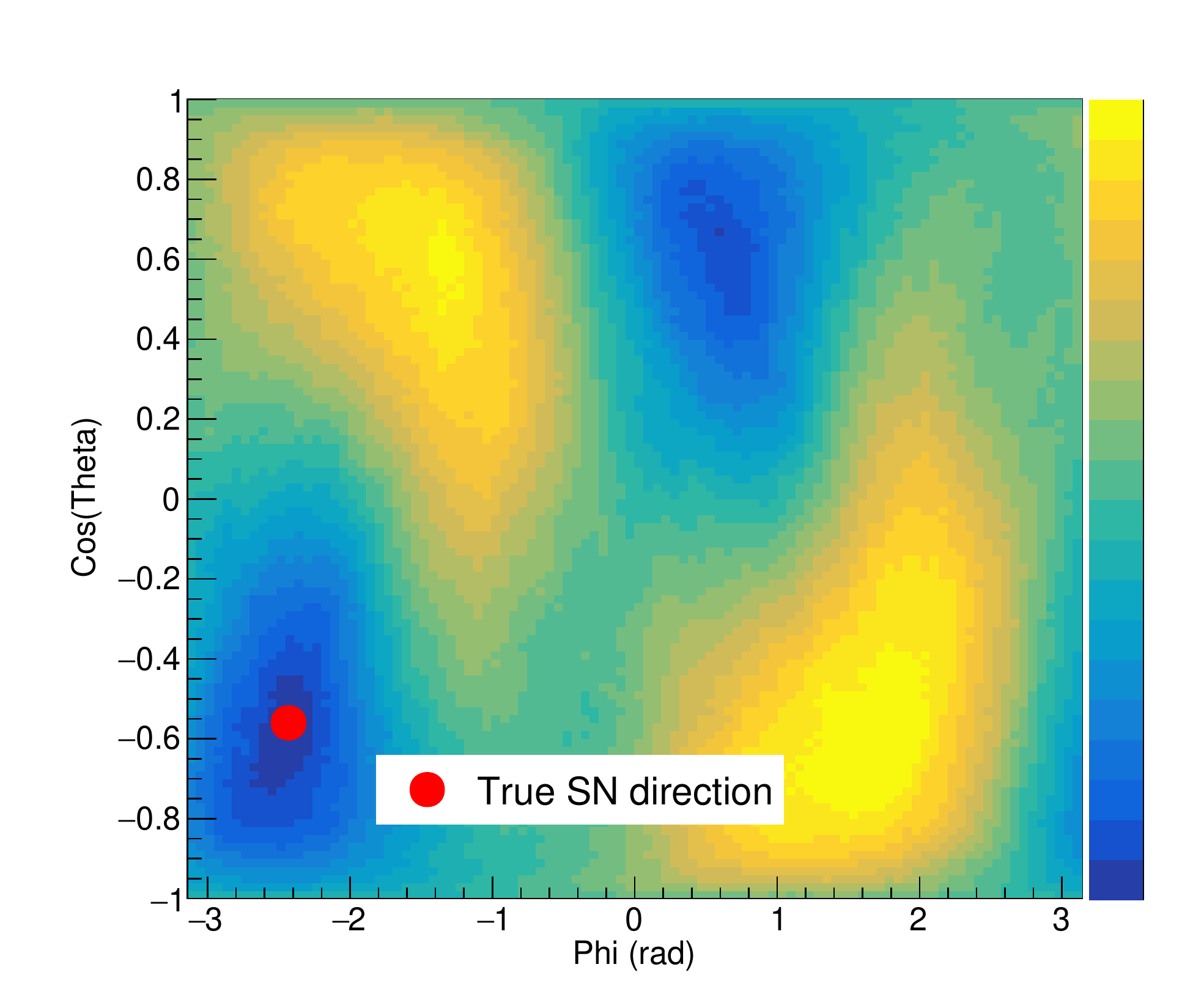}
  \includegraphics[width=0.44\textwidth]{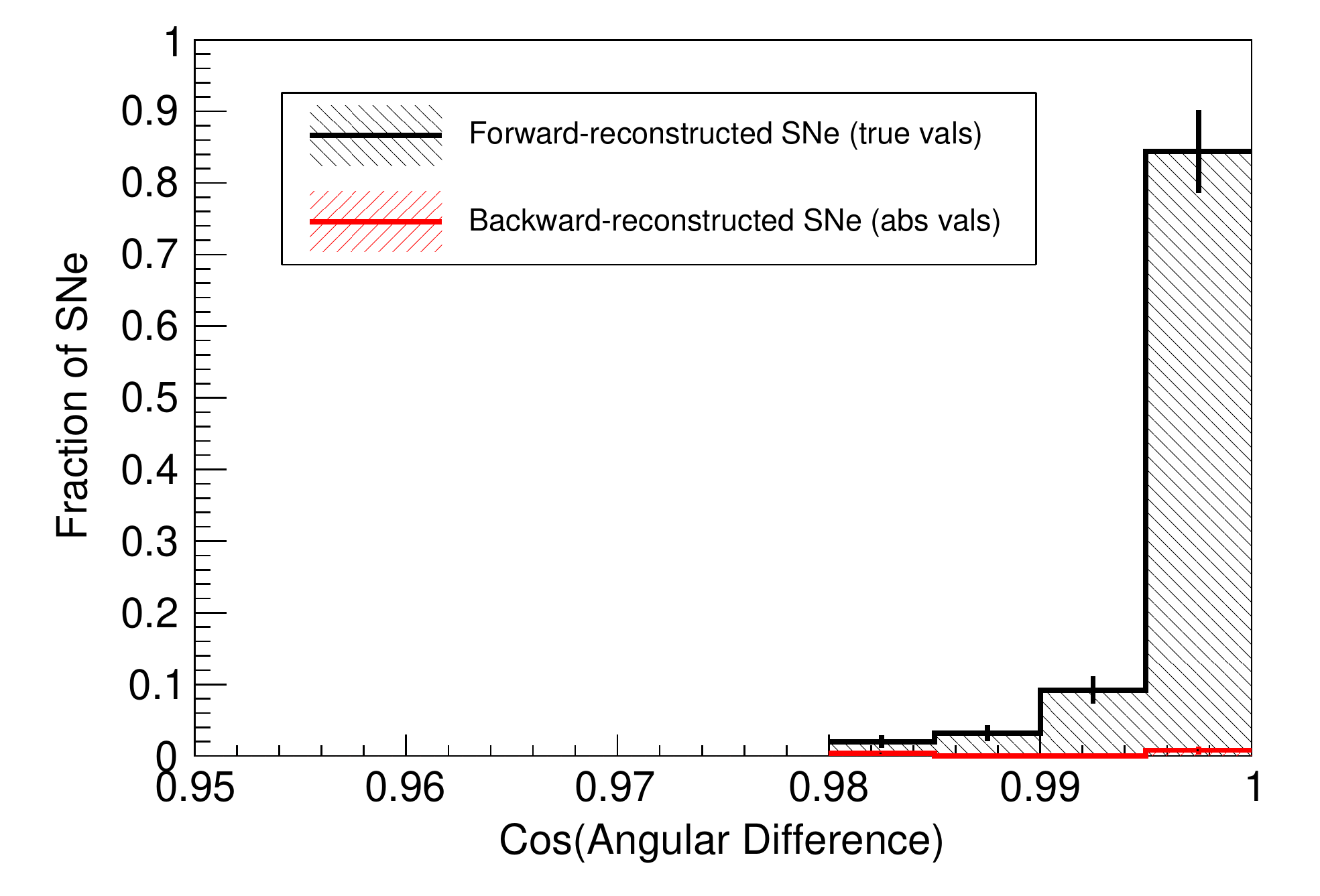}
\end{dunefigure}

If the direction can be disambiguated for $>$50\% of the individual elastic
scatters, the overall direction to the supernova can be disambiguated
with sufficient statistics.  Since the angular
distribution depends on event energy, we can also make use of measured
electron energy to improve the pointing from an ensemble of events.
We employ a maximum likelihood algorithm to estimate the overall
pointing resolution to a supernova, given a mean of 260
neutrino-electron elastic scatters and 3350 $\nu_e$CC at $\sim$10~kpc .
Using 16 energy bins
in the likelihood, the results are shown in
Figure~\ref{fig:fullSN}.  Overall resolution is about 4.5 degrees.

The result shown in the figure includes both \dword{es} and $\nu_e$CC interactions in the
likelihood, without radiological backgrounds or noise.
The addition of $\nu_e$CC events improves the
pointing resolution, even without \dword{es} vs $\nu_e$CC channel tagging.
We will likely be able to improve the pointing further by making use of
channel-tagging algorithms.

\section{Astrophysics of Core Collapse}
\label{sec:physics-snblowe-astrophysics}

A number of astrophysical phenomena associated with supernovae are expected to be observable
in the supernova neutrino signal, providing a remarkable window into the event.  In particular, the supernova explosion mechanism, which in the current paradigm involves energy deposition via neutrinos, is still not well understood, and the neutrinos themselves will bring the insight needed to confirm or refute the paradigm.

There are many other examples of astrophysical observables.
\begin{itemize}
\item The initial burst, primarily composed of $\nu_e$ and called the
  ``neutronization'' or ``breakout''
  burst, 
  represents only a small component of the total signal.  However,
  oscillation effects can manifest themselves in an observable manner
  in this burst, and flavor transformations can be modified by the
  ``halo'' of neutrinos generated in the supernova envelope by
  scattering~\cite{Cherry:2013mv}.
\item The formation of a black hole would cause a sharp signal cutoff
  (e.g.,~\cite{Beacom:2000qy,Fischer:2008rh}).
\item Shock wave effects (e.g.,~\cite{Schirato:2002tg}) would cause a
  time-dependent change in flavor and spectral composition as the
  shock wave propagates.
\item The standing accretion shock instability
  (SASI)~\cite{Hanke:2011jf,Hanke:2013ena}, a ``sloshing'' mode
  predicted by three-dimensional neutrino-hydrodynamics simulations of
  supernova cores, would give an oscillatory flavor-dependent
  modulation of the flux.
\item Turbulence effects~\cite{Friedland:2006ta,Lund:2013uta} would
  also cause flavor-dependent spectral modification as a function of
  time.
\end{itemize}

Observation of a supernova neutrino burst in coincidence with gravitational waves (which would also be prompt, and could indeed provide a time reference for a a time-of-flight analysis) would be especially interesting~\cite{Arnaud:2003zr,Ott:2012jq, Mueller:2012sv, Nishizawa:2014zna}.

The supernova neutrino burst is prompt with respect to the
electromagnetic signal and therefore can be exploited to provide an
early warning to astronomers~\cite{Antonioli:2004zb,Scholberg:2008fa}.  
Additionally, a liquid argon signal~\cite{Bueno:2003ei} is expected to
provide some pointing information, primarily from elastic scattering
on electrons.
We note that not every core collapse will produce an observable supernova, and observation of a neutrino burst in the absence of an electromagnetic event would be very interesting. 

Even non-observation of a burst, or non-observation of
a $\nu_e$ component of a burst in the presence of supernovae (or other
astrophysical events) observed in electromagnetic or gravitational
wave channels, would still provide valuable information about the
nature of the sources.  Further, a long-timescale, sensitive search
yielding no bursts will also provide limits on the rate of
core-collapse supernovae.

We note that the better one can understand the astrophysical nature of core-collapse supernovae, the easier it will be to extract information about particle physics.  DUNE's capability to characterize the $\nu_e$ component of the signal is unique and critical.

\subsection{Supernova Spectral Parameter Fits}


We have investigated how well it will be possible to fit to the supernova
spectral parameters, to determine, for example, the $\epsilon$
parameter related to the total binding energy release of the supernova.  We 
use  \dword{snowglobes} to model signals described by the pinched-thermal form.

We have developed a
forward fitting algorithm requiring a  \dword{snowglobes} binned energy
spectrum for a supernova at a given distance and a ``true'' set of
pinched-thermal parameters $(\alpha^0, \langle E_\nu \rangle^0,
\varepsilon^0)$. As an example, we define the true parameter values as
$(\alpha^0, \langle E_\nu \rangle^0, \varepsilon^0) = (2.5, 9.5,
5\times 10^{52})$, with  $\langle E_\nu \rangle^0$ in MeV and $\varepsilon$ in ergs, assumed
integrated over a ten-second burst.
We focus on the electron neutrino flux. The algorithm uses this
spectrum as a ``test spectrum" to compare against a grid of energy
spectra generated with many different combinations of $(\alpha,
\langle E_\nu \rangle, \varepsilon)$. To quantify these comparisons,
the algorithm employs $\chi^2$ minimization technique to find the
best-fit spectrum.

A test spectrum input into the forward fitting algorithm produces a set of $\chi^2$ values for every element in a grid. While the smallest $\chi^2$ value determines the best fit to the test spectrum, there exists other grid elements that reasonably fit the test spectrum according to their $\chi^2$ values. The collection of these grid elements help determine the parameter measurement uncertainty, and we represent this using ``sensitivity regions'' in \twod spectral parameter space. We can use three sets of \twod parameter spaces: $(\langle E_\nu \rangle, \alpha)$, $(\langle E_\nu \rangle, \varepsilon)$, and $(\alpha, \varepsilon)$.

One ``point" in 2D parameter space encompasses several grid elements,
e.g., the $(\langle E_\nu \rangle, \alpha)$ space contains different
$\varepsilon$ values for a given values of $\langle E_\nu \rangle$ and
$\alpha$. To determine the $\chi^2$ value, we profile over
$\varepsilon$ to select the grid element with the smallest
$\chi^2$. We determine the sensitivity regions by placing a cut of
$\chi^2 = 4.61$ corresponding to a 90\% coverage probability for three
free parameters.
Figure~\ref{fig:example3params} shows an example of a resulting fit,
with the approximate parameters for some specific models
superimposed.  Figure~\ref{fig:exampleAsimovDistance} shows the
statistical effect of supernova distance.

\begin{dunefigure}[Fit to three pinching
  parameters]{fig:example3params}{Sensitivity regions for the three
    pinched-thermal parameters (90\% C.L.).
  \dword{snowglobes} assumed a cross section
    model from \dword{marley}, realistic detector smearing and a step efficiency function with a 5 MeV
    detected energy threshold, for a supernova at 10~kpc. Superimposed
  are parameters corresponding to the time-integrated flux for three different sets of models:
  Nakazato~\cite{Nakazato:2012qf}, Huedepohl black hole formation models, and Huedepohl
  cooling models~\cite{huedepohldb}.  For the Nakazato parameters (for which there is no
  explicit pinching, corresponding to $\alpha=2.3$), the parameters are
  taken directly from the reference; for the Huedepohl models, they are fit to a
  time-integrated flux.}
	\includegraphics[scale = 0.37]{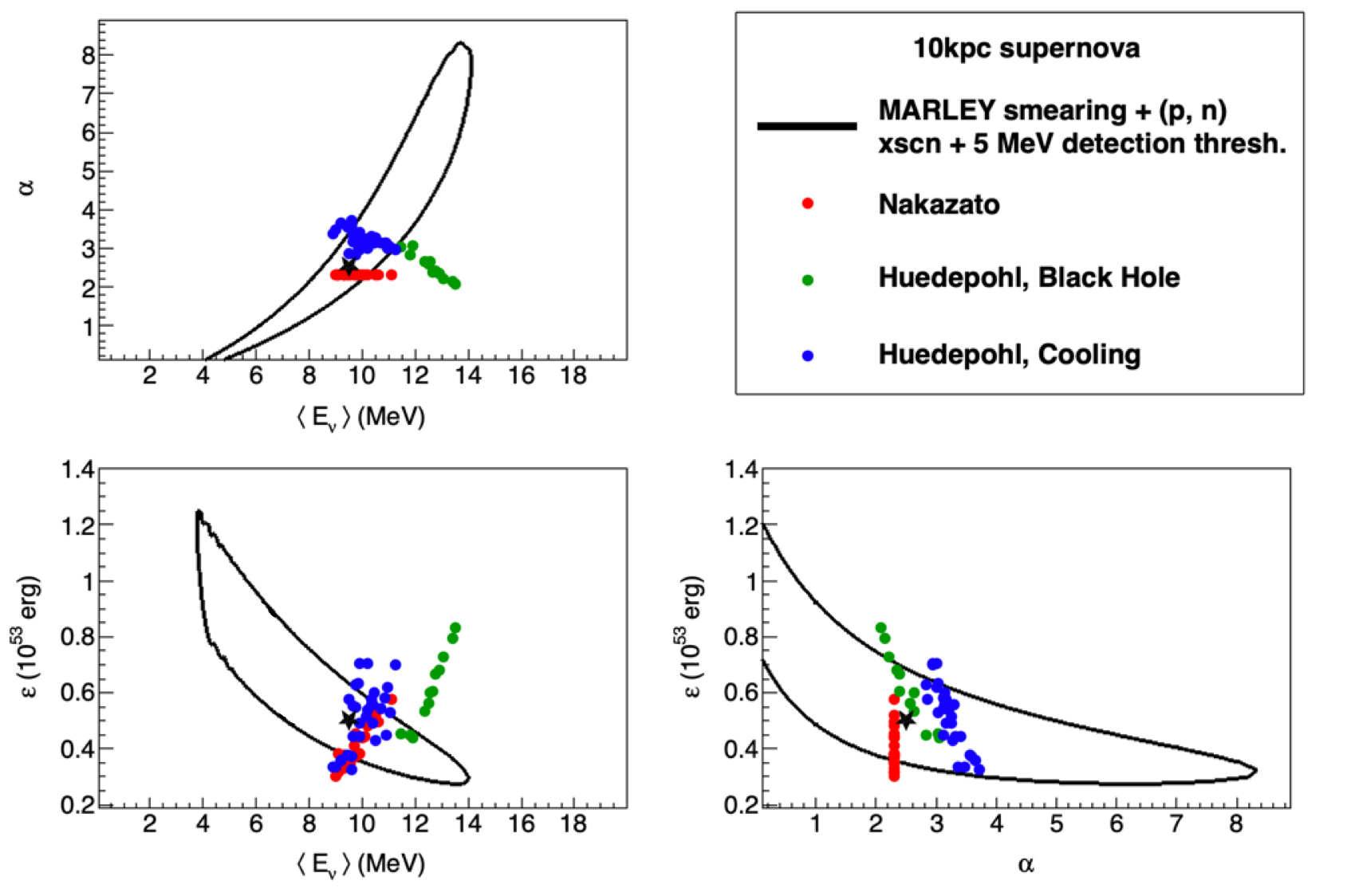}
  \end{dunefigure}

\begin{dunefigure}[Example of distance
  effect]{fig:exampleAsimovDistance}{Sensitivity regions generated
    in $(\langle E_\nu \rangle, \varepsilon)$ space
    for three different supernova distances (90\% C.L.).  \dword{snowglobes} used a
    smearing matrix with 15\% Gaussian resolution, a cross section
    model from \dword{marley}, and a step efficiency function with a 5 MeV
    detected energy threshold.}
	\includegraphics[scale = 0.37]{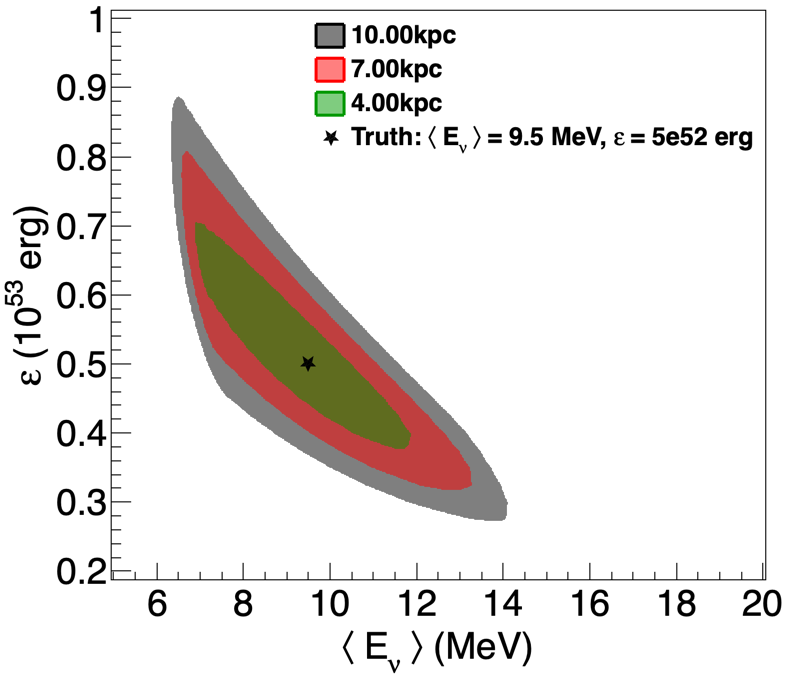}
  \end{dunefigure}

  We have used this method to study the effect of detector parameters,
  and required \textit{knowledge} of detector parameters, on ability
  to extract the flux parameters. While the size of the sensitivity
  regions is highly dependent on statistics (and hence distance), we
  find biases in the best-fit physics parameters
  if assumed understanding of detector parameters such as
 energy resolution, energy threshold, and energy scale does not match
 the truth.  We have
also studied the effect of imperfect knowledge of the $\nu_e$ cross
section on argon.  The results of these studies are extensive and
documented in~\cite{bib:docdb14068}.

\section{Neutrino Physics and Other Particle Physics}
\label{sec:physics-snblowe-neutrino-physics}

A core-collapse supernova is essentially a gravity-powered neutrino bomb: the energy of the collapse is initially stored in the Fermi seas of electrons and neutrinos and then gradually leaked out by neutrino diffusion. The key property of neutrinos that makes them play such a dominant role in the supernova dynamics is the feebleness of their interactions. It then follows that should there be new light ($< 100$ MeV) particles with even weaker interactions, they could alter the energy transport process and the resulting evolution of the nascent proto-neutron star. Moreover, additional interactions or properties of neutrinos could also be manifested in this way. 

Thus, a core-collapse supernova can therefore be thought of as an extremely hermetic system, which can be used to search for numerous types of new physics (e.g.,~\cite{Schramm:1990pf,Raffelt:1999tx}). The list includes various Goldstone bosons (e.g., Majorons), neutrino magnetic moments, new gauge bosons (``dark photons''), ``unparticles'', and extra-dimensional gauge bosons. The existing data from SN1987A already provides significant constraints on these scenarios, by confirming the basic energy balance of the explosion. At the same time, more precision is highly desirable and should be provided with the next galactic supernova.

Such energy-loss-based analysis will make use of two types of information. First, the total energy of the emitted neutrinos should be compared with the expected release in the gravitational collapse.  Note that measurements of all flavors, including $\nu_e$, are needed for the best estimate of the energy release.
Second, the rate of cooling of the protoneutron state should be measured and compared with what is expected from diffusion of the standard neutrinos.

Because DUNE is mostly sensitive to $\nu_e$, complementary data from
water Cherenkov detector and scintillator for the measurement of
$\bar\nu_{e}$ and a careful analysis of the oscillation pattern (see
below) will enable inference of the fluxes of $\mu$ and $\tau$
flavors. As for measuring the energy loss rate, it will require
sufficient statistics at late times.

The flavor oscillation physics and its signatures are a major part of
the physics program. Compared to the well-understood case of solar
neutrinos, in a supernova, neutrino flavor transformations are much
more involved. For supernovae, there are both neutrinos and antineutrinos, and
two mass splittings---``solar" and ``atmospheric" to worry about.
While flavor transitions can be reasonably well understood during 
early periods of the neutrino emission as standard \dword{msw} 
transitions in the varying density profile of the overlying material, during
later periods the physics of the transformations is significantly richer.
For example, several seconds after the onset of the explosion, the
flavor conversion probability is affected by the expanding shock front
and the turbulent region behind it. The conversion process in such a
stochastic profile is qualitatively different from the adiabatic \dword{msw}
effect in the smooth, fixed density profile of the Sun. 

Even more complexity is brought about by the coherent scattering of neutrinos off each other. This neutrino ``self-refraction'' 
 results in highly nontrivial flavor transformations close to the neutrinosphere, typically within a few hundred kilometers from the center, where the density of streaming neutrinos is very high. Since the evolving flavor composition of the neutrino flux feeds back into the oscillation Hamiltonian, the problem is \emph{nonlinear}. Furthermore, as the interactions couple neutrinos and antineutrinos of different flavors and energies, the oscillations are characterized by \emph{collective} modes.    This leads to very rich physics that has been the subject of intense interest over the last decade and a voluminous literature exists exploring these collective phenomena,
e.g.,~\cite{Duan:2005cp,Fogli:2007bk,Raffelt:2007cb,Raffelt:2007xt,EstebanPretel:2008ni,Duan:2009cd,Dasgupta:2009mg,Duan:2010bg,Duan:2010bf,Wu:2014kaa}.  This is an active theoretical field and the effects are not yet fully understood. A supernova burst is the only opportunity to study neutrino-neutrino interactions experimentally.

One may wonder whether all this complexity will impede the extraction
of useful information from the future signal. In fact, the opposite is
true: the new effects can \emph{imprint} information about the inner
workings of the explosion on the signal. The oscillations can modulate
the characteristics of the signal (both event rates and spectra as a
function of time).
Moreover, the oscillations can imprint \emph{non-thermal} features on the energy spectra, potentially making it possible to disentangle the effects of flavor transformations and the physics of neutrino spectra formation. This in turn should help us learn about the development of the explosion during the crucial first 10 seconds.   It is important to note that the features depend on the unknown mass ordering, and so can potentially tell us what the ordering is.

In the following, we examine quantitatively two examples of particle
physics that can be accessed: neutrino mass ordering and Lorentz
invariance violation.

\subsection{Neutrino Mass Ordering}\label{sec:mh}

As described above, flavor transitions
in the supernova can be fairly complex, and the rich phenomenology is
at this time still under active investigation.  The neutrino mass
ordering affects the specific flavor composition in multiple ways
during the different eras of neutrino emission.  
References~\cite{Mirizzi:2015eza,Scholberg:2017czd} survey in some detail the
multiple signatures of mass ordering that will imprint themselves on
the flux.  Table~\ref{tab:snmo} summarized several of them. For many of these, the $\nu_e$ component of the signal will
be critical to measure.    Some signatures of mass ordering are more robust than
others, in the sense that the assumptions are less subject to
theoretical uncertainties.  One of the more robust of these is the
early-time signal, including the \textit{neutronization burst}.   At
early times, the matter potential is dominant over the
neutrino-neutrino potential, which means that standard \dword{msw} effects are
in play.  In this case, for the \dword{no}, the neutronization burst, which is
emitted as nearly pure $\nu_e$, is strongly suppressed, whereas for
the \dword{io}, the neutronization burst is only partly suppressed.  
Figure~\ref{fig:early} gives an example for a specific model, but which
shows typical features.  The same \dword{msw}-dominated transitions also
affect
the subsequent rise of the signal over a fraction of a second; here
the time profile will depend on the turn-on of the non-$\nu_e$ flavors.

\begin{dunetable}[Supernova mass ordering
  signatures]{|c|c|c|c|c|}{tab:snmo}{Table taken from
    ~\cite{Scholberg:2017czd} qualitatively summarizing different neutrino mass ordering
    signatures that will manifest themselves in the supernova neutrino
  time, energy and flavor structure of the burst.}

Signature & Normal &  Inverted & Robustness & Observability\\ 

 Neutronization & \small Very suppressed & \small Suppressed & \small
 Excellent & \small Good, need $\nu_e$ \\  \colhline
Early time profile & \small Low then high & \small Flatter & \small Good,
possibly some self-interaction & \small Good \\ \colhline 
Shock wave & \small Time- & \small Time- & \small Fair,  & \small May be \\ 
        &\small dependent & \small dependent & \small entangled with & \small statistics\\ 
       &                 &                    & \small  self-interaction     &  \small limited\\ 

  & & & \small effects& \\  \colhline
   
Collective effects &\multicolumn{2}{|c|} {\small Various time- and energy-} & \small Unknown, but & \small Want all\\ 
 & \multicolumn{2}{|c|} {\small dependent signatures} & \small multiple signatures &\small flavors\\  \colhline
Earth matter & \small Wiggles in $\bar{\nu}_e$ & \small Wiggles in $\nu_e$ & \small Excellent & \small Difficult, need\\ 
& & & & \small energy resolution,\\ 
& & & &  \small Earth shadowing\\   \colhline
Type Ia &  \small Lower flux & \small Higher flux& \small Moderate & \small Need large detectors,  \\
  & & & & \small very close SN\\  \colhline

\end{dunetable}

\begin{dunefigure}[Event rates with ordering effects at early times]{fig:early}
{Expected event rates as a function of time for the electron-capture model in~\cite{Huedepohl:2009wh} for \SI{40}{kt} of argon during early stages of the event -- the neutronization burst and early accretion phases, for which self-induced effects are unlikely to be important.  Shown is the event rate for the unrealistic case of no flavor transitions (blue), the event rate including the effect of matter transitions for the normal (red)  and inverted (green) orderings.  Error bars are statistical, in unequal time bins.}
\includegraphics[width=0.8\textwidth]{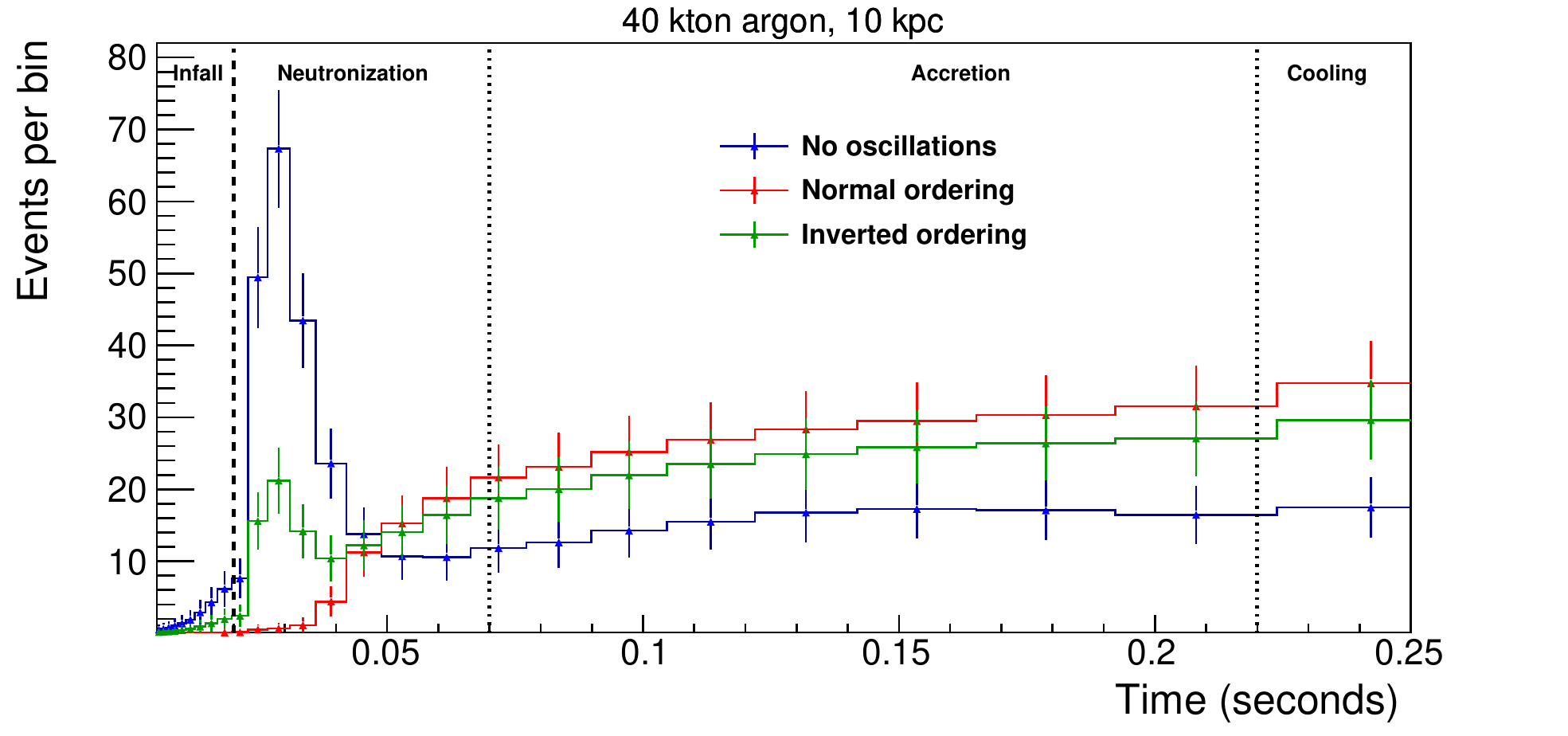}
\end{dunefigure}

Of course, if the mass ordering is already known, we can turn it around 
and use the terrestrial determination to better disentangle the other 
particle physics and astrophysics knowledge from the observed signal.


Figure~\ref{fig:neutronization_mh} shows the results of a simple quantitative study
based in counting observed events in DUNE in the first 50 milliseconds
of the burst.  We expect this early neutronization-burst period to be
dominated by adiabatic \dword{msw} transitions driven by the ``H-resonance''
for $\Delta m^2_{3\ell}$, for which the following
neutrino-energy-independent relations apply:

\begin{eqnarray}  
 F_{\nu_e} &=& F^0_{\nu_x} \,\ \,\ \,\ \,\ \,\ \,\ \,\ \,\  \,\ \,\ \,\ \,\   \,\ \,\  \,\ \,\ \,\ \,\ \,\ \,\  \,\ \,\ \textrm{(\dword{no})} \,\ , \label{eq:msw_nmo}\\
 F_{\nu_e} &=&  \sin^2 \theta_{12} F^0_{\nu_e} +
\cos^2 \theta_{12} F^0_{\nu_x}  \,\ \,\ \,\ \,\ \textrm{(\dword{io})} \,\,
\label{eq:msw_imo}
\end{eqnarray} 
 and 
\begin{eqnarray}  
 F_{\bar\nu_e} &=& \cos^2 \theta_{12} F^0_{\bar\nu_e} + \sin^2 \theta_{12} F^0_{\bar\nu_x}   \,\   \,\ \,\  \,\ \,\ \,\ \,\ \,\ \,\  \,\ \,\ \textrm{(\dword{no})} \,\ , \label{eq:msw_nmo_anti}\\
 F_{\bar\nu_e} &=&   F^0_{\bar\nu_x}  \,\ \,\ \,\ \,\ \,\ \,\ \,\ \,\ \,\ \,\ \,\ \,\ \,\ \,\ \,\ \,\ 
 \,\ \,\ \,\ \,\ \,\ \,\ \,\ \,\ \,\ \,\ \,\ \,\ \,\  \,\
\textrm{(\dword{io})} \,\,\label{eq:msw_imo_anti}
\end{eqnarray} 

where $F$'s are the fluxes corresponding to the respective flavors,
and the $^o$ subscript represents flux before transition.

\begin{dunefigure}[MH1]{fig:neutronization_mh}{Event counts in the
    first 50 milliseconds for the model in~\cite{Huedepohl:2009wh}, under
    assumptions of no oscillations, normal ordering and inverted
    ordering, assuming adiabatic \dword{msw} transitions.  The left plot
    shows the event number as a function of distance with statistical
    errors.  The right plot
    shows the event number scaled by square of distance, under the
    assumption of a 20\% uncertainty on distance. }
\includegraphics[width=3.2in]{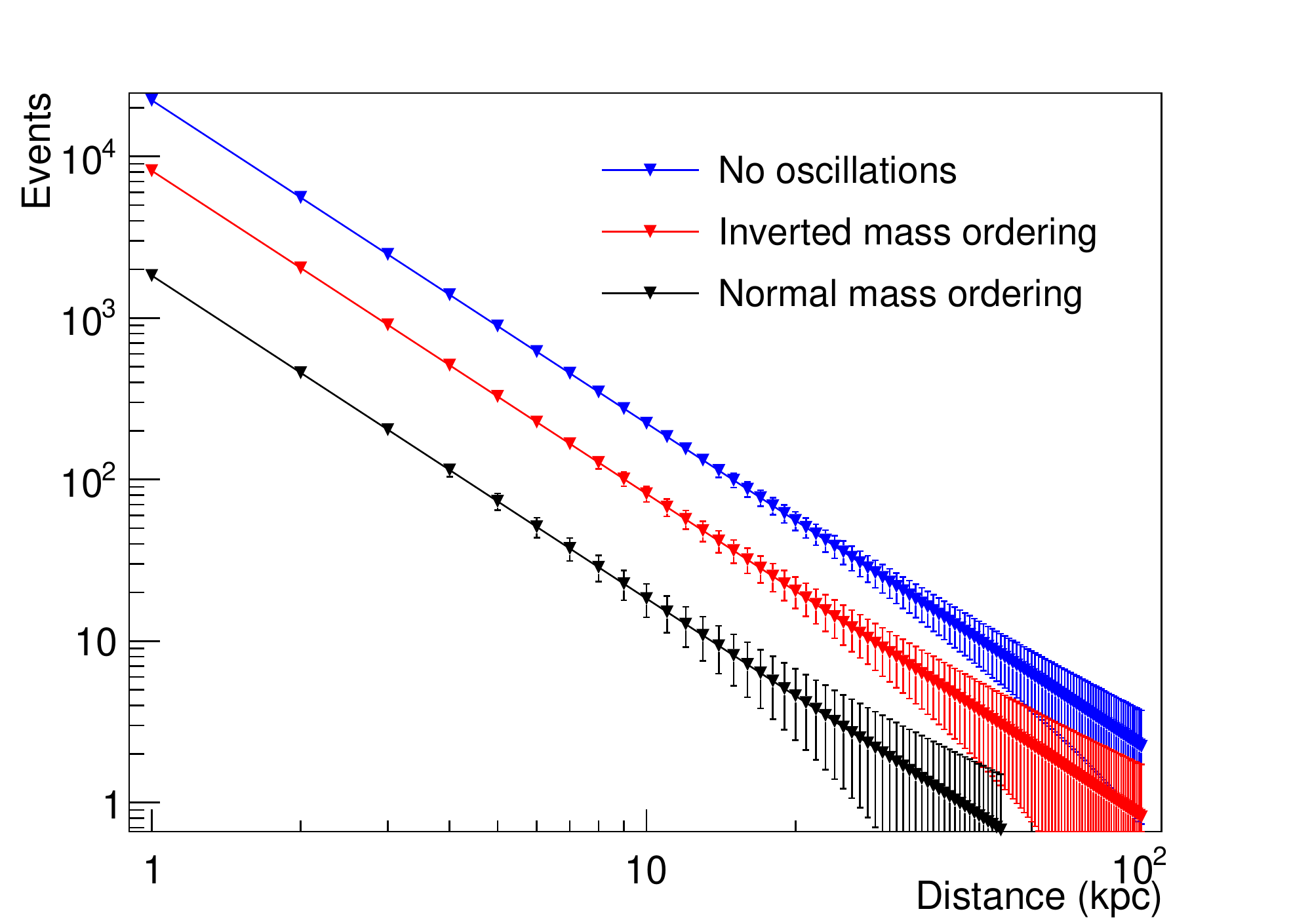}
\includegraphics[width=3.2in]{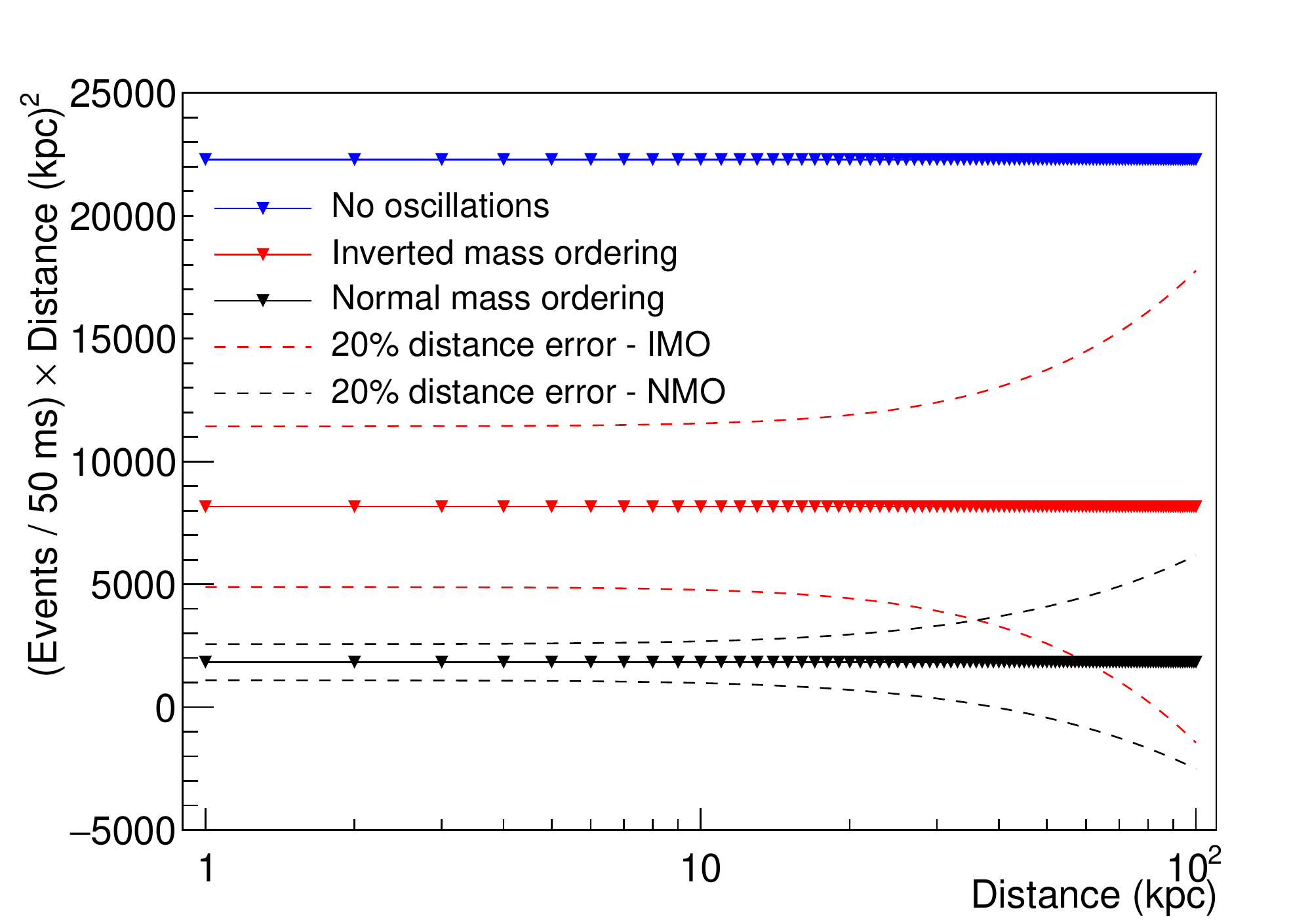}
\end{dunefigure}

Figure~\ref{fig:neutronization_mh} shows that for this model, the event count will be
well separated under the two different assumptions, out to the edge of
the Galaxy.  The right hand plot shows also the effect of uncertainty
on the distance to the supernova, in the scenario of evaluating the
mass ordering based on absolute neutronization-burst counts.
Note that while the neutronization burst is thought to be a ``standard
candle''~\cite{Mirizzi:2015eza}, there will likely be some model
dependence, and early-time-window event count by
itself is not likely a sufficiently robust discriminant.
There will, however, be additional information from other time eras of the burst signal.  Further studies for a
range of additional models and making use of the full burst
time information are underway.

\subsection{Lorentz Invariance Violation}

As another example of a probe of new physics with supernova neutrinos or antineutrinos,
a class of tests of Lorentz and \dword{cpt} violation involves comparing the propagation of neutrinos with other species of neutrinos of the same flavor but different energies~\cite{Kostelecky:2003cr,Kostelecky:2003xn,Kostelecky:2011gq,Diaz:2009qk}. These amount to time-of-flight or dispersion studies.
Time-of-flight and dispersion effects lack the interferometric resolving power available to neutrino oscillations, but they provide instead sensitivity to Lorentz- and \dword{cpt}-violating effects that  leave unaffected neutrino oscillations
and so cannot be measured using atmospheric or long-baseline neutrinos.
The corresponding \dword{sme} coefficients controlling these effects are called oscillation-free coefficients~\cite{Kostelecky:2011gq}.

Supernova neutrinos are of particular interest in this context because of the long baseline, which implies sensitivities many orders of magnitude better than available from time-of-flight measurements in beams. Observations of the supernova SN1987A yield constraints on the difference between the speed of light and the speed of antineutrinos, which translates into constraints on isotropic and anisotropic coefficients in both the minimal and nonminimal sectors of the \dword{sme}. Knowledge of the spread of arrival times constrains the maximum speed difference between SN1987A antineutrinos of different energies in the approximate range 10--40 MeV, which restricts the possible antineutrino dispersion and yields further constraints on \dword{sme} coefficients~\cite{Kostelecky:2011gq}.

Analyses of this type would be possible with DUNE if supernova neutrinos are observed. Key features to maximize sensitivity would include absolute timing information to compare with photon spectral observations (and perhaps ultimately with gravitational-wave data~\cite{Kostelecky:2016kfm})  along with relative timing information for different components of the neutrino energy spectrum. Significant improvements over existing limits are possible.

Figure \ref{fig:livsn} displays DUNE supernova sensitivities 
to these relevant oscillation-free coefficients 
for Lorentz and \dword{cpt} violation.
The estimated sensitivities are obtained using
the general expression (125) in ~\cite{Kostelecky:2011gq}
for the neutrino velocity in oscillation-free models
and its application (132) to dispersion studies.
The figure assumes a supernova comparable to SN1987A
and at the same location on the sky.
Enhancements of the displayed sensitivities 
from angular factors can occur for different sky locations.
Studies of supernova neutrinos using DUNE 
can measure many coefficients (green) 
at levels improving over existing limits (gray).

\begin{dunefigure}[DUNE SN sensitivity to Lorentz, \dword{cpt} violation]{snliv}{DUNE supernova sensitivities to oscillation-free coefficients for Lorentz and \dword{cpt} violation. Studies of DUNE supernova neutrinos can measure many coefficients (green) at levels improving over existing limits (grey). These Lorentz- and \dword{cpt}- violating effects leave oscillations unchanged and so are challenging to detect in atmospheric or long-baseline measurements~\cite{kostelecky}.\label{fig:livsn}}
\includegraphics[width=0.9\textwidth]{liv-sn.pdf}
\end{dunefigure}

Finally, via detection of time-of-flight delayed $\nu_e$ from the  neutronization burst,  DUNE will be able to probe neutrino mass bounds of $\mathcal{O}(1)$~eV for a 10-kpc supernova~\cite{Rossi-Torres:2015rla} (although likely not competitive near-future terrestrial kinematic limits).  If eV-scale sterile neutrinos exist, they will likely have an impact on astrophysical and oscillation aspects of the signal (e.g.,~\cite{Keranen:2007ga,Tamborra:2011is,Esmaili:2014gya}), as well as time-of-flight observables. \\

\section{Additional Astrophysical Neutrinos}
\label{sec:physics-snblowe-other}

\subsection{Solar Neutrinos}

Intriguing questions in solar neutrino physics remain, even after data
from the \dword{sk} and \dword{sno}~\cite{Fukuda:2001nj,Ahmad:2001an}
experiments explained the long-standing mystery of missing solar
neutrinos~\cite{Cleveland:1998nv} as due to flavor transformations.
Some unknowns, such as the fraction of energy production via the
\dword{cno} cycle in the Sun, flux variation due to
helio-seismological modes that reach the solar core, or long-term
stability of the solar core temperature, are astrophysical in
nature. Others directly impact particle physics.  Can the \dword{msw}
model explain the amount of flavor transformation as a function of
energy, or are non-standard neutrino interactions required?  Do solar
neutrinos and reactor antineutrinos oscillate with the same
parameters?  There is a modest tension between the $\Delta m^2_{21}$
values indicated by current global solar neutrino measurements and the
KamLAND reactor measurement~\cite{Capozzi:2018dat}, and further solar
neutrino measurements could help to resolve this.  Interesting
observables are the day/night effect, and potentially the hep flux at
higher energies.

Detection of solar and other low-energy neutrinos is challenging in
a \dword{lartpc} because of relatively high intrinsic detection energy thresholds for
the charged-current interaction on argon ($>$\SI{5}{\MeV}). 
However, compared with other technologies, a \dword{lartpc} offers a large
cross section and unique potential channel-tagging signatures from deexcitation
photons.  Furthermore, observed energy from the final state $\nu_e$CC
interaction follows neutrino energy more closely on an event-by-event
basis (see Figure~\ref{marleysmearing}) with respect to the recoil
spectrum from the ES channel that has been used for most solar neutrino
observations so far. This feature of DUNE enables more precise spectral measurements.
The solar neutrino event rate in a
\ktadj{40} \dword{lartpc} is $\sim$100 per day.
Reference~\cite{Capozzi:2018dat} explores the solar neutrino potential
of DUNE, with somewhat optimistic energy resolution assumptions.

Detailed simulation studies
making use of both TPC and photon information are underway, and
preliminary event selection criteria for solar neutrinos are under development.
Figure~\ref{solareff} shows an example of event selection efficiency
as a function of neutrino energy.  These preliminary cuts require three nearby
\dword{tpc} hits, an associated optical photon flash, and nearby
\dword{tpc} activity associated with deexcitation gammas.  For these
cuts,
background in \SI{10}{\kt} is reduced to less than \SI{0.1}{Hz}.

\begin{dunefigure}[Solar neutrino efficiency]{solareff}{Efficiency for
  selection of neutrinos as a function of neutrino energy, for
  preliminary event selection cuts.
  Normalized spectra for solar neutrino signals are superimposed. }
\includegraphics[width=0.6\textwidth]{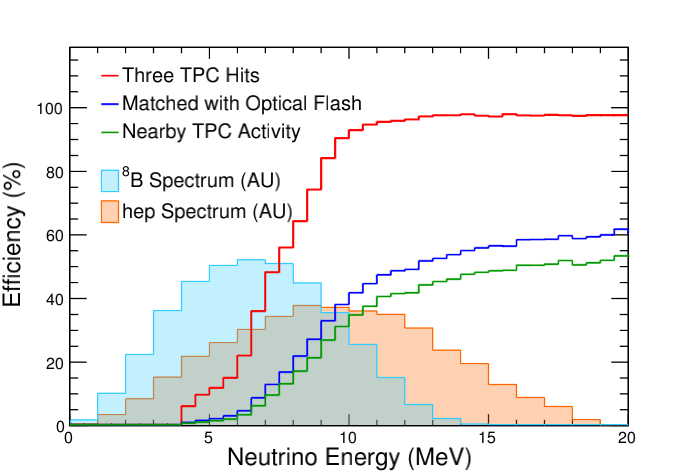}
\end{dunefigure}

Backgrounds for triggering and reconstruction are the most
serious issue.  Cosmogenic backgrounds are likely tractable, but
radiological backgrounds are more troublesome.  In particular, neutron
capture on argon is troublesome, and a relatively rare (and poorly
known) alpha-capture channel on argon may contribute to the background
in the solar neutrino energy regime. 
It is plausible that with sophisticated event selection, and possibly
with additional shielding, a high-statistics solar neutrino sample may be selected.

\subsection{Diffuse Supernova Background Neutrinos}

Galactic supernovae are relatively rare, occurring somewhere between
once and four times a century. In the universe
at large, however, thousands of neutrino-producing explosions occur
every hour.  The resulting neutrinos --- in fact most of the neutrinos
emitted by all the supernovae since the onset of stellar formation ---
suffuse the universe.  Known as the 
\dword{dsnb}, their energies are in the few-to-\MeVadj{30} range.  \dword{dsnb}
have not yet been observed, but an observation would greatly enhance
our understanding of supernova-neutrino emission and the overall
core-collapse rate~\cite{Beacom:2010kk, Lunardini:2012ne, Moller:2018kpn}.

A liquid argon detector such as DUNE's far detector is sensitive to
the $\nu_e$ component of the diffuse relic supernova neutrino flux,
whereas water Cherenkov and scintillator detectors are sensitive to
the antineutrino component.

Background is a serious issue for \dword{dsnb} detection.
The solar {\em hep} neutrinos, which have an                
endpoint at \SI{18.8}{\MeV}, will determine the lower bound of the \dword{dsnb}.
The upper bound is determined
by the atmospheric ${\nu}_{e}$ flux and
is around \SI{40}{MeV}.
Although the \dword{lartpc} provides a unique sensitivity to the
electron-neutrino component of the \dword{dsnb}
flux~\cite{Cocco:2004ac}, event rates are very low.
The expected number of relic
supernova neutrinos, $N_{\rm \dword{dsnb}}$, that could be observed is
1-2 per MeV per 20 years in \SI{10}{\kt}~\cite{Moller:2018kpn}
within the 19-31~MeV window.  For this low signal rate, even rare
radiological cosmogenic backgrounds will be challenging, and are
under study.

\subsection{Other Low-Energy Neutrino Sources}

We note some other potential sources of signals in the tens-of-MeV
range which may be observable in DUNE.  A small flux is expected from
Type I (thermonuclear) supernovae~\cite{Wright:2016gar,
  Wright:2016xma}, with potential detectability by DUNE within a few
kpc.   Other signals include neutrinos from accretion disks~\cite{Caballero:2011dw} and black-hole/neutron star mergers~\cite{Caballero:2009ww}.  These will create spectra not unlike those from core-collapse events, and with potentially large fluxes.  However they are expected to be considerably rarer than core-collapse supernovae within an observable distance range.  There may also be signatures of dark-matter \dword{wimp} annihilations in the low-energy signal range~\cite{Rott:2012qb, Bernal:2012qh}.

\section{Burst Detection and Alert}
\label{sec:physics-snblowe-detector-requirements}

For supernova burst physics, the detector must be able to detect and
reconstruct as well as possible events in the range 5--100~MeV.  As
for proton decay and atmospheric neutrinos, no beam trigger will be
available; therefore there must be special triggering and \dword{daq}
requirements that take into account the short, intense nature of the
burst, and the need for prompt propagation of information in a
worldwide context.  The trigger requirement is for 90\% trigger efficiency for a
supernova burst at 100~kpc.

Events are expected within a time window of approximately 30 seconds,
but possibly over an interval as long as a few hundred seconds; a
large fraction of the events are expected within approximately the 1-2
seconds of the burst.  The data acquisition buffers must be
sufficiently large and the data acquisition system sufficiently robust
to allow full capture of neutrino event information for a supernova as
close as 0.1 kpc.  At 10~kpc, one expects thousands of events within
approximately 10 seconds, but a supernova at a distance of less than
1~kpc would result in $10^5-10^7$ events over 10 seconds.

The far detector must have high uptime to allow the capture of
low-probability astrophysical events that could occur at any time with
no external trigger.  Supernova events are expected to occur a few
times per century within the Milky Way galaxy. For any 10-year period,
the probability of a supernova could be 20 to 30\%.  Capturing such an
event at the same time as many of the other detectors around the Earth
is very important.

The DUNE detector systems must be configured to provide information
to other observatories on possible astrophysical events (such as a
galactic supernova) in a short enough time to allow global
coordination.  This interval should be less than 30 minutes, and
preferably
on a few-minute timescale.  To obtain maximum scientific value out of a singular
astronomical event, it is very important to inform all other
observatories (including optical ones) immediately via SNEWS~\cite{Antonioli:2004zb,Scholberg:2008fa}, so that they can
begin observation of the evolution of the event. Pointing information
should also be made available as promptly as possible.

\spchdaq describes the DUNE triggering and
\dword{daq} configurations designed to meet these challenges for the \dword{spmod}, and similarly for the \dword{dpmod}.

\cleardoublepage

\chapter{Beyond the Standard Model  Physics Program }
\label{ch:bsm}
\section{Executive Summary}
\label{phys:bsm:execsumm}

The unique combination of the high-intensity LBNF neutrino beam with DUNE's \dword{nd}  and massive LArTPC \dword{fd} modules at a \SI{1300}{km} baseline enables a variety of probes of beyond the standard model \dword{bsm} physics, either novel or with unprecedented sensitivity. This section describes a selection of such topics, and briefly summarizes how DUNE can make leading contributions in this arena.

\textit{Search for active-sterile neutrino mixing:} Experimental results in tension with the three-neutrino-flavor paradigm, which may be interpreted as mixing between the known active neutrinos and one or more sterile states, have led to a rich and diverse program of searches for oscillations into sterile neutrinos~\cite{ref:tension,Gariazzo:2017fdh}. DUNE is sensitive over a broad range of potential sterile neutrino mass splittings by looking for disappearance of \dword{cc} and \dword{nc}  interactions over the long distance separating the \dword{nd} and \dword{fd}, as well as over the short baseline of the \dword{nd} . With a longer baseline, a more intense beam, and a high-resolution large-mass \dword{fd}, compared to previous experiments, DUNE provides a unique opportunity to improve significantly on the sensitivities of the existing probes, and greatly enhance the ability to map the extended parameter space if a sterile neutrino is discovered.

\textit{Searches for non-unitarity of the \dword{pmns} matrix:} A generic characteristic of most models explaining the neutrino mass pattern is the presence of heavy neutrino states, additional to the three light states of the \dword{sm} of particle physics~\cite{Minkowski:1977sc,Mohapatra:1979ia,Yanagida:1979as,GellMann:1980vs}. This implies a deviation from 
unitarity of the $3 \times 3$ \dword{pmns} matrix that can 
become particularly sizable the lower the mass of the extra states are.
For values of the unitarity deviations of order $10^{-2}$, this would decrease the expected reach of DUNE to the standard parameters, although stronger bounds existing from charged leptons would be able to restore its expected performance.

\textit{Searches for \dword{nsi}:} 
\Dword{nsi} affecting neutrino propagation through the Earth, can significantly modify the data to be collected by DUNE as long as the new physics parameters are large enough~\cite{Farzan:2017xzy}. Leveraging its very long baseline and wide-band beam, DUNE is uniquely sensitive to these probes. If the DUNE data are consistent with standard oscillations for three massive neutrinos, interaction effects of order 0.1 $G_{F}$ can be ruled out at DUNE. We note that DUNE will improve current constraints on $\epsilon_{\tau e}$ and $\epsilon_{\mu e}$, the magnitude of the \dword{nsi} relative to standard weak interactions, by a factor of 2 to 5.

\textit{Searches for violation of Lorentz or \dword{cpt} Symmetry:} \dword{cpt} symmetry, the combination of charge conjugation, parity and time reversal, is a cornerstone of our model-building strategy and therefore the repercussions of its potential violation will severely threaten the \dword{sm} of particle physics~\cite{Streater:1989vi,Barenboim:2002tz,Kostelecky:2003cr,Diaz:2009qk,Kostelecky:2011gq,Barenboim:2017ewj}. DUNE can improve the present limits on Lorentz and \dword{cpt} violation by several orders of magnitude, contributing as a very important experiment to test these fundamental assumptions underlying quantum field theory.

\textit{Searches for neutrino trident production:} The intriguing possibility that neutrinos may be charged under new gauge symmetries beyond the \dword{sm}  $SU(3)_{c} \times SU(2)_{L} \times U(1)_{Y}$, and interact with the corresponding new gauge bosons can be tested with unprecedented precision by DUNE through \dword{nd}  measurements of neutrino-induced di-lepton production in the Coulomb field of a heavy nucleus, also known as neutrino trident interactions~\cite{Czyz:1964zz,Lovseth:1971vv,Fujikawa:1971nx,Koike:1971tu,Koike:1971vg,Brown:1973ih,Belusevic:1987cw}. Although this process is extremely rare (\dword{sm} rates are suppressed by a factor of $10^{-5} \-- 10^{-7}$ with respect to \dword{cc} interactions), the CHARM-II collaboration and the CCFR collaboration both reported detection of several trident events ($\sim40$ events at CCFR) and quoted cross-sections in good agreement with the \dword{sm} predictions. With a predicted annual rate of over 100 dimuon neutrino trident interactions at the \dword{nd}, DUNE will be able to measure deviations from the \dword{sm} rates and test the presence of new gauge symmetries~\cite{Altmannshofer:2019zhy,Ballett:2018uuc,Ballett:2019xoj}.

\textit{Search for \dword{ldm}:} Various cosmological and astrophysical observations strongly support the existence of \dword{dm} representing $\approx$27\% of the mass-energy of the universe, but its nature and potential non-gravitational interactions with regular matter remain undetermined~\cite{Aghanim:2018eyx}. The lack of evidence for \dwords{wimp} at direct detection and the LHC experiments has resulted in a reconsideration of the \dword{wimp} paradigm. For instance, if \dword{dm} has a mass that is much lighter than the electroweak scale (e.g., below GeV level), it motivates theories for \dword{dm} candidates that interact with ordinary matter through a new ``vector portal'' mediator. High-flux neutrino beam experiments, such as DUNE, have been shown to provide coverage of \dword{dm}+mediator parameter space that cannot be covered by either direct detection or collider experiments~\cite{Alexander:2016aln, Battaglieri:2017aum, LoSecco:1980nf, Acciarri:2015uup}. \dword{dm} particles can be detected in the \dword{nd} through neutral-current-like interactions either with electrons or nucleons in the detector material. The neutrino-induced backgrounds can be suppressed using timing and the kinematics of the scattered electron. These enable DUNE's search for \dword{ldm} to be competitive and complementary to other experiments.

\textit{Search for \dword{bdm}:} Using its large \dword{fd}, DUNE will be able to search for  \dword{bdm}~\cite{Agashe:2014yua,Belanger:2011ww}. A representative model is composed of heavy and light \dword{dm} components and the lighter one can be produced from the annihilation of the heavier one in, e.g., the nearby sun or galactic centers. Due to the large mass difference between the two \dword{dm} components, the lighter one is produced relativistically. The incoming energy of the lighter \dword{dm} component can be high enough above the expected energy thresholds of DUNE in a wide range of parameter space. A first attempt at observing the inelastic \dword{bdm} signal with \dword{protodune} prior to running DUNE is proposed in Ref.~\cite{Chatterjee:2018mej}.
Further, a significant \dword{bdm} flux can arise from \dword{dm} annihilation in the core of the sun~\cite{Huang:2013xfa,Berger:2014sqa,Kong:2014mia,Kim:2018veo}. \dword{dm} particles can be captured by the sun through their scattering with the nuclei in the sun, mostly hydrogen and helium. This makes the core of the sun a region with concentrated \dword{dm} distribution. Through various processes, this \dword{dm} can then be emitted as \dword{bdm} and its flux probed on Earth by DUNE.

Section~\ref{sec:otheropps} details several other compelling BSM Physics scenarios DUNE will be sensitive to.

\section{Common Tools: Simulation, Systematics, Detector Components}
\label{sec:tools}

DUNE will be the future leading-edge neutrino experiment. The DUNE detector-beam configuration provides an excellent opportunity to study the physics beyond standard neutrino
oscillations. It utilizes a megaWatt class proton accelerator (with beam power of up to \SI{2.4}{MW}), a massive (\fdfiducialmass) liquid argon time-projection chamber (LArTPC) \dword{fd}, and a high-resolution near detector. The neutrino beam, \dword{nd} and \dword{fd} configurations used for the \dword{bsm} searches are discussed in the following sections.

\subsection{Neutrino Beam Simulation}
\label{Nusim}

The DUNE experiment will use an optimized neutrino beam designed to provide maximum sensitivity to leptonic \dword{cp} violation. The optimized beam includes a three-horn system with a longer target embedded within the first horn and a decay pipe with \SI{194}{m} length and \SI{4}{m} diameter. In this design, a genetic algorithm is used to determine values for 20 beamline parameters describing the primary proton momentum and the target dimensions, along with the horn shapes, horn positions, and horn current values that maximize DUNE's sensitivity to \dword{cpv}. The optimized neutrino beam is further described in~\cite{bib:docdb4559}. We discuss the \dword{nd} and \dword{fd} flux used for the \dword{bsm} searches below. 

The neutrino flux for the \dword{nd} is generated at a distance of \SI{574}{m} downstream of the start of horn 1. Fluxes have been generated for both neutrino mode and antineutrino mode. The detailed beam configuration used for the \dword{nd}  analysis is given in Table~\ref{tabBC}.

Unless otherwise noted, the neutrino fluxes used in the \dword{bsm} physics analysis are the same as those used in the long-baseline three-flavor analysis, introduced in Section~\ref{sec:physics-lbnosc-simreco}. These fluxes were produced using G4LBNF, a \dword{geant4}-based simulation. The fluxes are weighted  at the \dword{fd}, located \SI{1297}{km} downstream of the start of horn 1. The flux files  contain \dword{nc} and \dword{cc} spectra, which are obtained by multiplying the flux by inclusive cross sections supplied by \dword{genie} version 2.8.4. 
Note that these histograms have variable bin widths, so discontinuities in the number of events per bin are expected. \\
The beam power configuration used for both \dword{nd} and \dword{fd} is given in Table~\ref{tabBC}.
\begin{dunetable}
[Beam power configuration assumed for the LBNF neutrino beam.]
{ | c | c | c | c |} 
{tabBC}
{Beam power configuration assumed for the LBNF neutrino beam.}
               {\bf Energy (GeV)} & {\bf Beam Power (MW)} & {\bf Uptime Fraction} & {\bf POT/year}\\ \toprowrule
         
              120 &1.2 &0.56& 1.1$\times10^{21}$\\ 
\end{dunetable}

\subsection{Detector Properties}
\label{sec:ndprops}

The \dword{nd} configuration is not yet finalized, so we have adopted an overall structure for the \dword{lartpc} component of the detector and its fiducial volume. 
The \dword{nd} will be located at a distance of \ndfromtarget from the target. The \dword{nd} dimensions and properties used for the \dword{bsm} searches are given below.
The \dword{nd} concept 
consists of a modular \lartpc and a magnetized high-pressure gas argon TPC. In the \dword{bsm} physics analysis, 
the \dword{lartpc} is assumed to be \SI{7}{m} wide, \SI{3}{m} high, and \SI{5}{m} long. The fiducial volume is assumed to include the detector volume up to 50 cm of each face of the detector.
The \dword{nd} properties are given in Table~\ref{tabND}. The signal and background efficiencies for different physics models are different. 
Detailed signal and background efficiencies for each physics topic are discussed along with each analysis.

\begin{dunetable}
[ND properties used in the BSM physics analyses.]
{ | c | c |}
{tabND}
{ND properties used in the BSM physics analyses.}
   {\bf \dword{nd} Properties} & {\bf Values}\\ \toprowrule  
    Dimensions &  \SI{7}{m} wide, \SI{3}{m} high, and \SI{5}{m} long \\ \colhline
    Dimensions of fiducial volume & \SI{6}{m} wide, \SI{2}{m} high, and \SI{4}{m} long\\ \colhline
    Total mass  & 147 ton \\ \colhline
    Fiducial mass & 67.2 ton \\ \colhline
    Distance from target & \ndfromtarget \\
 \end{dunetable} 

The DUNE \dword{fd} will consist of four \nominalmodsize \lartpc modules located at Sanford Underground Research Facility (\surf), 
either \dword{sp} or \dword{dp} with integrated \dwords{pds}.  
The effective active mass of the detector used for the analysis is \fdfiducialmass. The \dword{fd} dimensions and \dword{globes} configurations are given below.
The \dword{gdml} files for the two \dword{fd} workspace geometries described here, with and without the \dword{apa}
sense wires, are the same used in the long-baseline three-flavor analysis, as described in Section~\ref{sec:physics-lbnosc-simreco}.
The single-particle detector responses used for the analyses are listed in Table~\ref{tabFD}.
\begin{dunetable}
[FD properties used in the BSM physics analyses.]
{ | c | c | c | c|} 
{tabFD}
{FD properties used in the BSM physics analyses.}
            {\bf Particle Type} & {\bf Threshold} & {\bf Energy Resolution} & {\bf Angular Resolution}\\ \toprowrule 
            $\mu^{\pm}$ & 30 MeV &Contained track: track length&$1^{o}$\\ 
            \colhline
            $e^{\pm}$ &30 MeV&2$\%$&$1^{o}$\\ 
            \colhline
            $\pi^{\pm}$ & 100 MeV&30$\%$&$5^{o}$\\ 
\end{dunetable} 
    
 \subsubsection{ \dword{globes} Configuration for the \dword{fd} analysis}
 
The  \dword{globes} configuration files reproduce the \dword{fd} simulation used in the long-baseline three-flavor analysis, introduced in Section~\ref{sec:physics-lbnosc-simreco}.
The flux normalization factor is included using a \dword{globes} Abstract Experiment Definition Language (AEDL) 
file to ensure that all variables have the proper units; its value is $@norm = 1.017718\times 10^{17}$. \fixme{I don't know what @norm is. Anne} Cross-section files describing \dword{nc} and \dword{cc} interactions with argon, generated using \dword{genie} 2.8.4, are included in the configuration. The true-to-reconstructed smearing matrices and the selection efficiency as a function of energy for various signal and background modes are included within \dword{globes}. The  \dword{globes} configuration provided in the ancillary files corresponds to \SI{300}{\ktMWyr} of exposure, with 3.5 years each of running in neutrino and antineutrino mode. A \fdfiducialmass fiducial mass is assumed for the \dword{fd}, exposed to a \SI{120}{GeV}, \SI{1.2}{MW} beam.The $\nu_{e}$ and $\bar\nu_{e}$ signal modes have independent normalization uncertainties of $2\%$ each, while $\nu_{\mu}$ and $\bar{\nu}_{\mu}$ signal modes have independent normalization uncertainties of $5\%$. The background normalization uncertainties range from $5\%$ to $20\%$ and include
correlations among various sources of background; the correlations among the background normalization parameters are given in the AEDL file of Ref.~\cite{Alion:2016uaj}. The \dword{fd} response for the different particles used are the same as used in Section~\ref{sec:physics-lbnosc-simreco}.

\section{Sterile Neutrino Searches}
Experimental results in tension with the three-neutrino-flavor paradigm~\cite{LSNDSterile,MiniBooNESterile,GalliumSummary,ReactorSummary, ref:tension,Gariazzo:2017fdh}, which may be interpreted as mixing between the known active neutrinos and one or more \textit{sterile} states, have led to a rich and diverse program of searches for oscillations into sterile neutrinos. Having a longer baseline, a more intense beam, and a high-resolution large-mass \dword{fd}, 
compared to previous experiments, DUNE provides a unique opportunity to improve significantly on the sensitivities of existing probes, and to enhance the ability to map the extended parameter space if a sterile neutrino is discovered. Conversely, the presence of light sterile neutrino mixing would impact the interpretation of the DUNE physics results~\cite{Dutta:2016glq}, so studying sterile neutrinos within DUNE is essential.
	
\subsection{Probing Sterile Neutrino Mixing with DUNE}

Long-baseline experiments like DUNE can look for sterile neutrino oscillations by measuring disappearance of the beam neutrino flux between the \dword{nd} and \dword{fd}. This results from the quadratic suppression of the sterile mixing angle measured in appearance experiments, $\theta_{\mu e}$, with respect to its disappearance counterparts, $\theta_{\mu\mu}\approx\theta_{24}$ for \dword{lbl} experiments, and $\theta_{ee}\approx\theta_{14}$ for reactor experiments. These disappearance effects have not yet been observed and are in tension with appearance results~\cite{ref:tension,Gariazzo:2017fdh} when global fits of all available data are carried out. The exposure of DUNE's high-resolution \dword{fd} to the high-intensity LBNF beam will also allow direct probes of nonstandard electron (anti)neutrino appearance. 

DUNE will look for active-to-sterile neutrino mixing using the reconstructed energy spectra of both \dword{nc} and \dword{cc}  neutrino interactions  in the \dword{fd}, and their comparison to the extrapolated predictions from the \dword{nd} measurement. Since \dword{nc} cross sections and interaction topologies are the same for all three active neutrino flavors, the \dword{nc} spectrum is insensitive to standard neutrino mixing. However, should there be oscillations into a fourth light neutrino, an energy-dependent depletion of the neutrino flux would be observed at the \dword{fd}, as the sterile neutrino would not interact in the detector volume. Furthermore, if sterile neutrino mixing is driven by a large mass-square difference $\Delta m^2_{\rm{41}}$ $\sim$1\,eV$^{2}$, the \dword{cc} spectrum will be distorted at energies higher than the energy corresponding to the standard oscillation maximum. Therefore, \dword{cc} disappearance is also a powerful probe of sterile neutrino mixing at long baselines. 

At long baselines, the \dword{nc} disappearance probability to first order in small mixing angles is given by:
\begin{equation}\label{eq:numu_nus}
\begin{aligned}
1 - P(\nu_{\mu} \rightarrow \nu_s) & \approx 1 - \cos^4\theta_{14}\cos^2\theta_{34}\sin^{2}2\theta_{24}\sin^2\Delta_{41} \\
& - \sin^2\theta_{34}\sin^22\theta_{23}\sin^2\Delta_{31} \\
& + \frac{1}{2}\sin\delta_{24}\sin\theta_{24}\sin2\theta_{23}\sin\Delta_{31},
\end{aligned}
\end{equation}
where $\Delta_{ji} = \frac{\Delta m^2_{ji}L}{4E}$. 
The relevant oscillation probability for \numu~\dword{cc} disappearance is the \numu~survival probability, similarly approximated by:
\begin{equation}
\begin{aligned}
P(\nu_{\mu} \rightarrow \nu_{\mu}) &\approx 1 - \sin^22\theta_{23}\sin^2\Delta_{31} \\
& + 2\sin^22\theta_{23}\sin^2\theta_{24}\sin^2\Delta_{31} \\ 
& - \sin^22\theta_{24}\sin^2\Delta_{41}.
\label{eq:NuMuDisFull}
\end{aligned}
\end{equation}
Finally, the disappearance of $\overset{(-)}\nu\!\!_e$~\dword{cc} is described by: 
\begin{equation}
\begin{aligned}
P(\overset{(-)}\nu\!\!_e \rightarrow \overset{(-)}\nu\!\!_e) &\approx 1 - \sin^22\theta_{13}\sin^2\Delta_{31} \\
& - \sin^22\theta_{14}\sin^2\Delta_{41}.
\label{eq:NueDisFull}
\end{aligned}
\end{equation}

\begin{figure}[!htb]
	\begin{center}
	  	\includegraphics[width=0.49\textwidth]{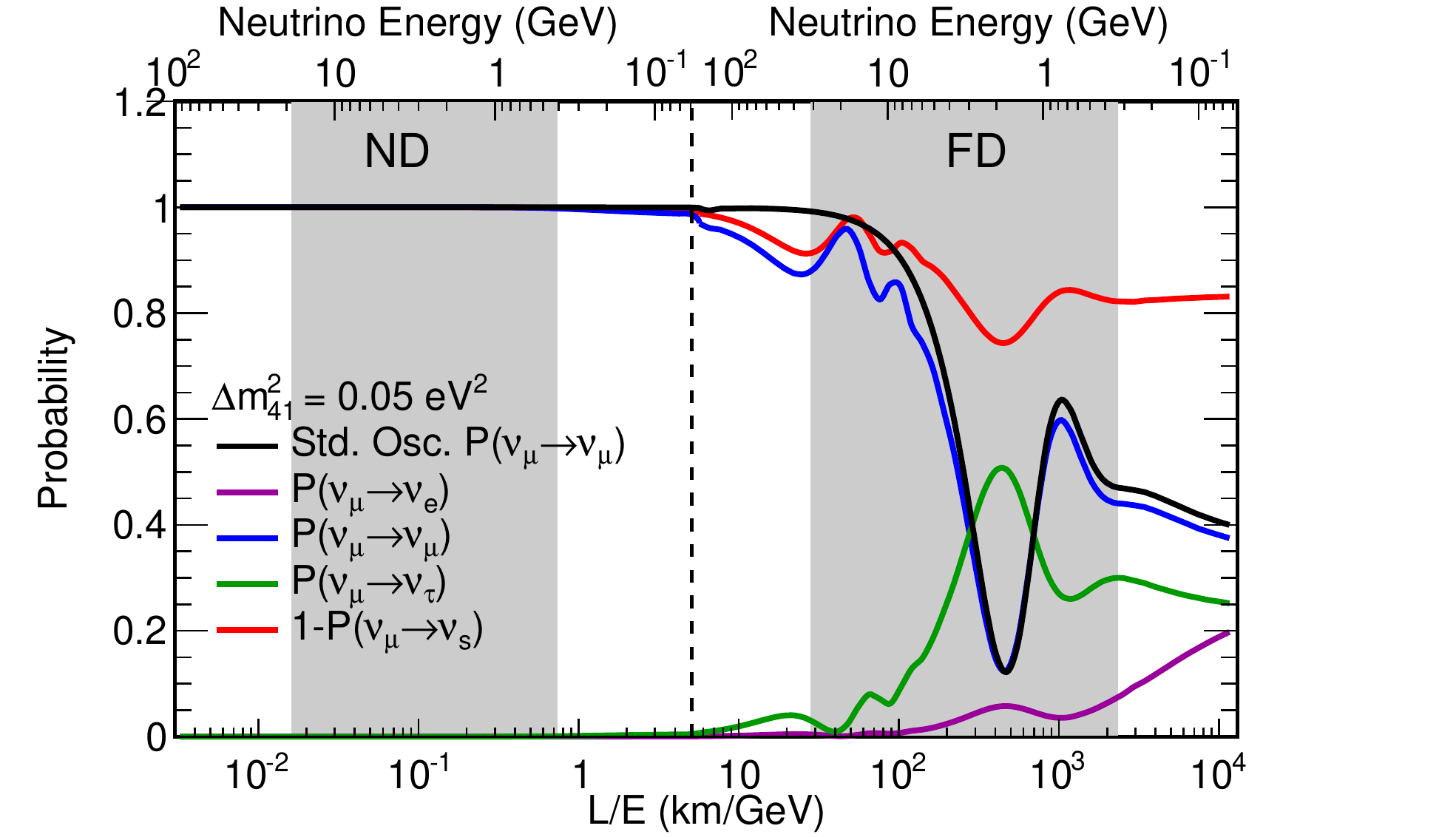}
        \includegraphics[width=0.49\textwidth]{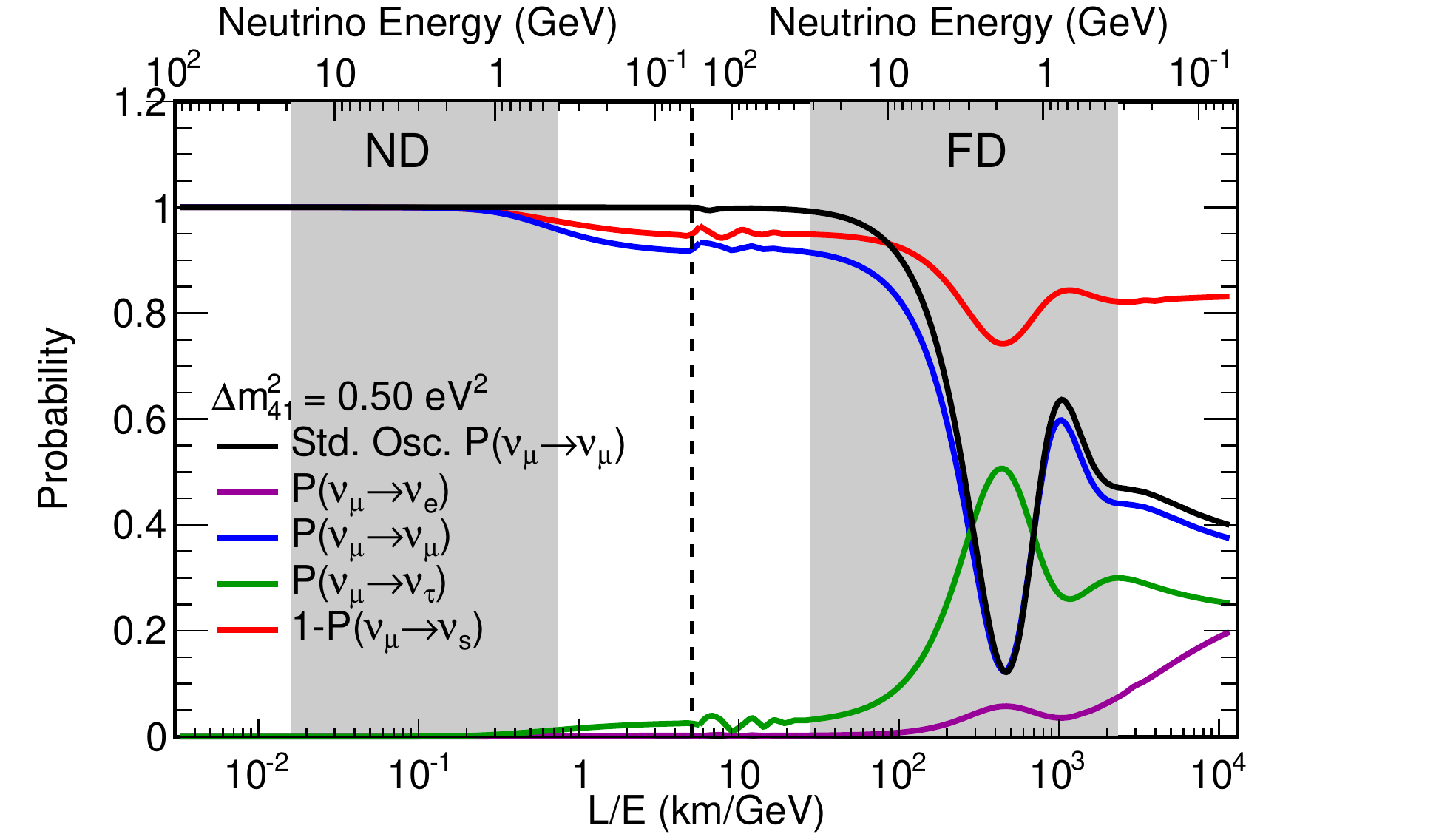}
        \includegraphics[width=0.49\textwidth]{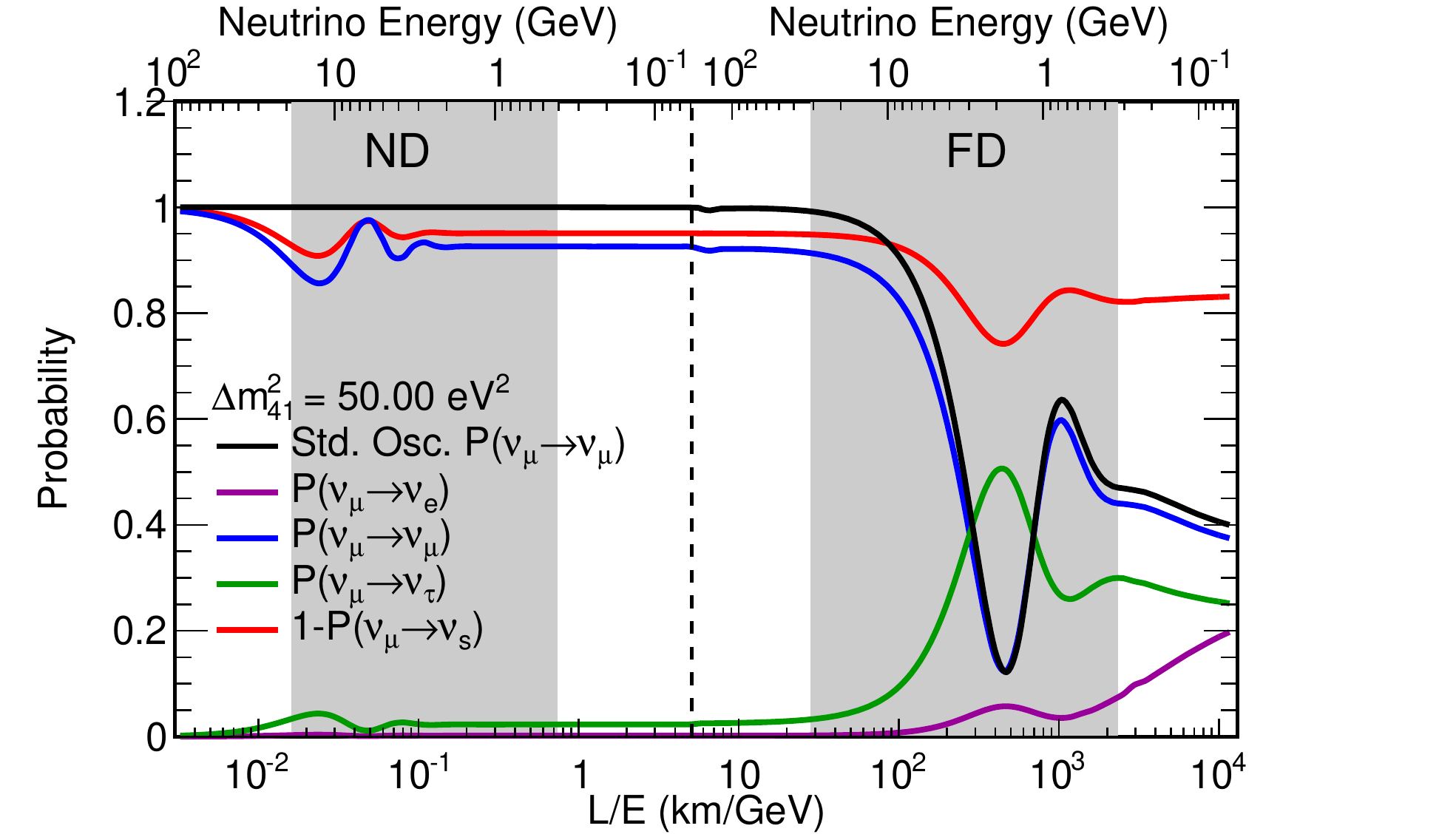}
	\end{center}
\caption[Regions of $L/E$ probed by DUNE for 
3-flavor and 3+1-flavor $\nu$ oscillations]
{Regions of $L/E$ probed by the DUNE detector compared to 3-flavor and 3+1-flavor neutrino disappearance and appearance probabilities. The gray-shaded areas show the range of true neutrino energies probed by the \dword{nd} and \dword{fd}. The top axis shows true neutrino energy, increasing from right to left. The top-left plot shows the probabilities assuming mixing with one sterile neutrino with $\Delta m^2_{\rm{41}}=0.05$~eV$^2$, corresponding to the slow oscillations regime. The top-right plot assumes mixing with one sterile neutrino with $\Delta m^2_{\rm{41}}=0.5$~eV$^2$, corresponding to the intermediate oscillations regime. The bottom plot includes mixing with one sterile neutrino with $\Delta m^2_{\rm{41}}=50$~eV$^2$, corresponding to the rapid oscillations regime. As an example, the slow sterile oscillations cause visible distortions in the three-flavor \numu~survival probability (blue curve) for neutrino energies $\sim10$\,GeV, well above the three-flavor oscillation minimum.}
\label{fig:regimes}
\end{figure}

Figure~\ref{fig:regimes} shows how the standard three-flavor oscillation probability is distorted at neutrino energies above the standard oscillation peak when oscillations into sterile neutrinos are included.

\subsection{Setup and Methods}
The simulation of the DUNE experimental setup was performed with the \dword{globes} software~\cite{Huber:2004ka,Huber:2007ji} using the same flux and equivalent detector definitions used by the three-neutrino flavor analysis presented in Section~\ref{sec:physics-lbnosc-simreco}. Specifically, the neutrino flux used assumes 120 GeV protons incident on the LBNF target, with $1.1\times 10^{21}$~\dword{pot} collected per year. A total exposure of 300~kton.MW.year is used in assessing DUNE's physics reach in probing the relevant sterile neutrino mixing parameter space.

The sterile neutrino effects have been implemented in  \dword{globes} via the existing plug-in for sterile neutrinos and \dword{nsi}~\cite{Joachim}. As described above, the \dword{nd} will play a very important role in the sensitivity to sterile neutrinos both directly, for rapid oscillations with $\Delta m_{41}^2 > 1$~eV$^2$ where the sterile oscillation matches the \dword{nd} baseline, and indirectly, at smaller values of $\Delta m_{41}^2$ where the \dword{nd} is crucial to reduce the systematics affecting the \dword{fd} to increase its sensitivity. To include these \dword{nd} effects in these studies, the latest \dword{globes} DUNE \dword{tdr} configuration files describing the detectors were modified by adding a \dword{nd} with correlated systematic errors with the \dword{fd}. As a first approximation, the \dword{nd} is assumed to be an identical scaled-down version of the TDR \dword{fd} where the same efficiencies, backgrounds and energy reconstruction as presented in Section~\ref{sec:physics-lbnosc-simreco} have been assumed, with detector properties the same as described in Section~\ref{sec:ndprops}. The systematic uncertainties originally defined in the \dword{globes} DUNE \dword{cdr} configuration already took into account the effect of the \dword{nd} constraint. Thus, since we are now explicitly simulating the \dword{nd}, larger uncertainties have been adopted but partially correlated between the different channels in the \dword{nd} and \dword{fd}, so that their impact is reduced by the combination of both data sets. The full list of systematic uncertainties considered and their values is summarized in a technical note~\cite{ref:dune-sterile-note}.

Finally, for oscillations observed at the \dword{nd}, the uncertainty on the production point of the neutrinos can play an important role. We have included an additional $20\%$ energy smearing, which produces a similar effect given the $L/E$ dependence of oscillations. We implemented this smearing in the \dword{nd} through multiplication of the migration matrices provided with the \dword{globes} files by an additional matrix with the $20\%$ energy smearing obtained by integrating the Gaussian
\begin{equation}
R^c(E,E')\equiv\frac{1}{\sigma(E)\sqrt{2\pi}}e^{-\frac{(E-E')^2}{2\sigma(E)}},
\label{R_mat}
\end{equation}
with $\sigma(E)=0.2 E$ in reconstructed energy $E'$.

\subsection{Results}
By default, \dword{globes} treats all systematic uncertainties included in the fit as normalization shifts. However, depending on the value of $\Delta m^2_{41}$, sterile mixing will induce shape distortions in the measured energy spectrum beyond simple normalization shifts. As a consequence, shape uncertainties are very relevant for sterile neutrino searches, particularly in regions of parameter space where the \dword{nd}, with virtually infinite statistics, has a dominant contribution. The correct inclusion of systematic uncertainties affecting the shape of the energy spectrum in the two-detector fit \dword{globes} framework used for this analysis posed technical and computational challenges beyond the scope of the study.
Therefore, for each limit plot, we present two limits bracketing the expected DUNE sensitivity limit, namely: the black limit line, a best-case scenario, where only normalization shifts are considered in a \dword{nd}+\dword{fd} fit, where the ND statistics and shape have the strongest impact; and the grey limit line, corresponding to a worst-case scenario where only the \dword{fd} is considered in the fit, together with a rate constraint from the \dword{nd}. 

Studying the sensitivity to $\theta_{14}$, the dominant channels are those regarding $\nu_e$ disappearance. Therefore, only the $\nu_e$ \dword{cc} sample is analyzed and the channels for \dword{nc} and $\nu_{\mu}$ \dword{cc} disappearance are not taken into account, as they do not influence greatly the sensitivity and they slow down the simulations. The sensitivity at the 90\% \dword{cl}, taking into account the systematics mentioned above, is shown in Figure~\ref{fig:th_14+th_24}, along with a comparison to current constraints.

For the $\theta_{24}$ mixing angle, we analyze the $\nu_{\mu}$ \dword{cc} disappearance and the \dword{nc} samples, which are the main contributors to the sensitivity. 
The results are shown in Figure~\ref{fig:th_14+th_24}, along with comparisons with present constraints.

\begin{dunefigure}[Sensitivities to $\theta_{14}$ from $\nu_e$ CC samples, and to $\theta_{24}$ using $\nu_\mu$ CC and NC samples] 
{fig:th_14+th_24}
{The left-hand plot shows the DUNE sensitivities to $\theta_{14}$ from the $\nu_e$ \dword{cc} samples at the \dword{nd} and \dword{fd}, along with a comparison with the combined reactor result from Daya Bay and Bugey-3. The right-hand plot displays sensitivities to $\theta_{24}$ using the $\nu_\mu$ \dword{cc} and \dword{nc} samples at both detectors, along with a comparison with previous and existing experiments. In both cases, regions to the right of the contours are excluded.}
\includegraphics[width=0.48\columnwidth]{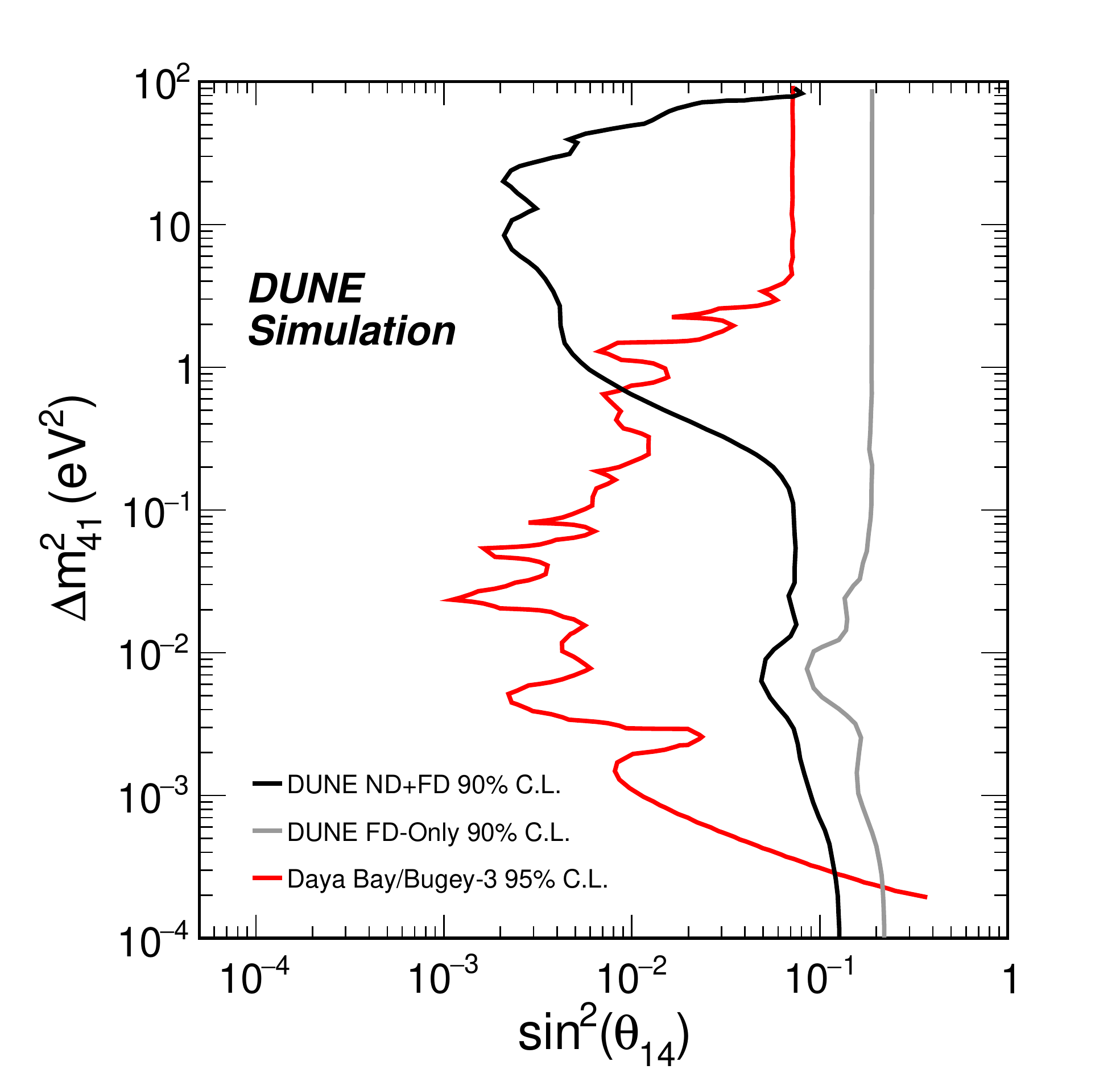}
\includegraphics[width=0.48\columnwidth]{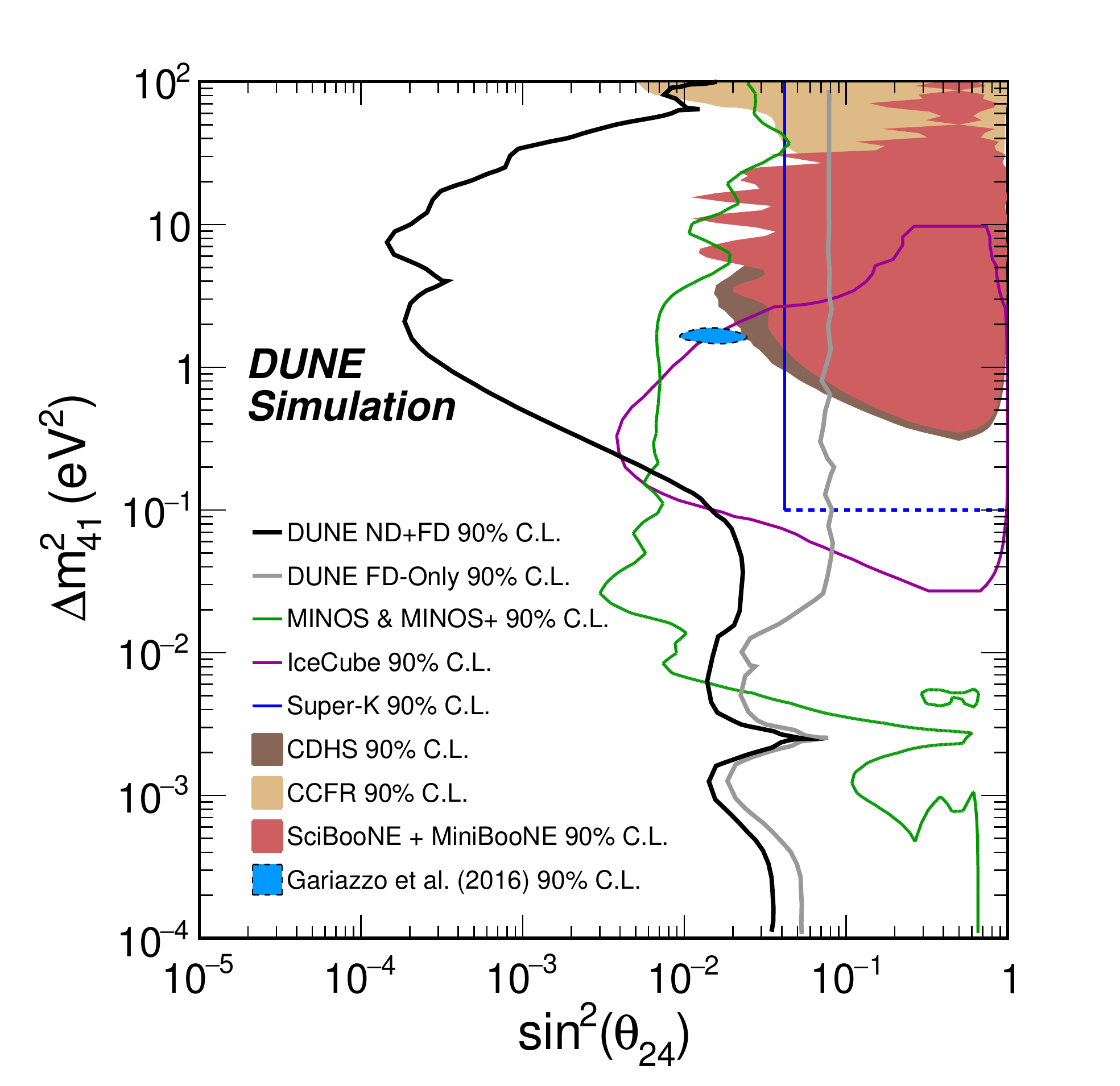}
\end{dunefigure}

In the case of the $\theta_{34}$ mixing angle, we look for disappearance in the \dword{nc} sample, the only contributor to this sensitivity. The results are shown in Figure~\ref{fig:th_34}. Further, a comparison with previous experiments sensitive to \numu, \nutau~mixing with large mass-squared splitting is possible by considering an effective mixing angle $\theta_{\mu\tau}$, such that $\sin^2{2\theta_{\mu\tau}}\equiv 4|U_{\tau4}|^2|U_{\mu 4}|^2=\cos^4\theta_{14}\sin^22\theta_{24}\sin^2\theta_{34}$, and assuming conservatively that $\cos^4\theta_{14}=1$, and $\sin^22\theta_{24}=1$. This comparison with previous experiments is also shown in Figure~\ref{fig:th_34}.
The sensitivity to $\theta_{34}$ is largely independent of 
$\Delta m^2_{41}$, since the term with $\sin^2\theta_{34}$ in the expression describing $P(\nu_{\mu} \rightarrow \nu_s)$ Eq.~\ref{eq:numu_nus}, depends solely on the $\Delta m^2_{31}$ mass splitting.

\begin{dunefigure}[Sensitivity to $\theta_{34}$ using the NC samples at the ND and FD] {fig:th_34} 
{DUNE sensitivity to $\theta_{34}$ using the \dword{nc} samples at the \dword{nd} and \dword{fd} compared to previous and existing experiments. Regions to the right of the contour are excluded.}
\includegraphics[width=0.48\columnwidth]{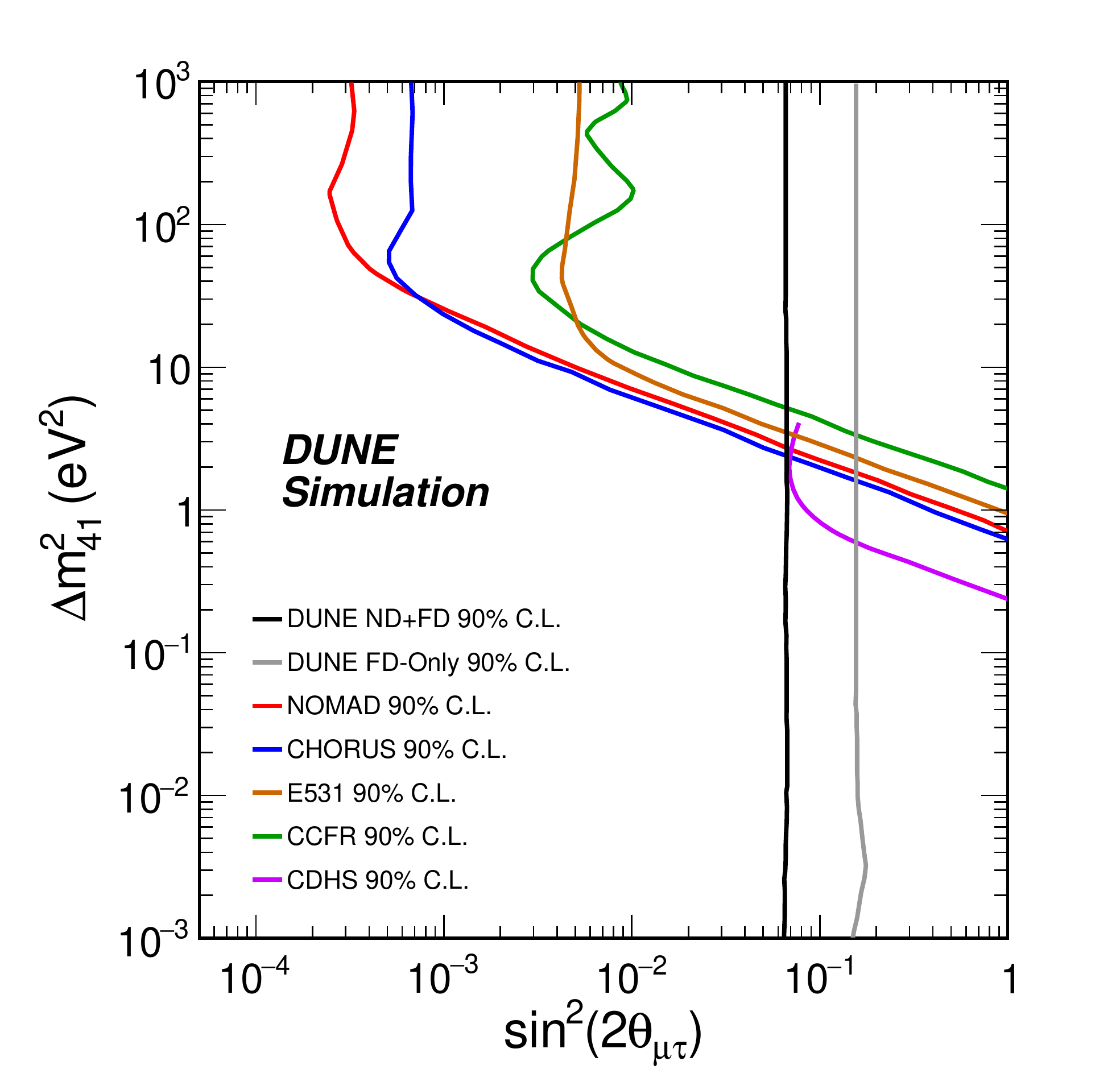}
\end{dunefigure}

Another quantitative comparison of our results for $\theta_{24}$ and $\theta_{34}$ with existing constraints can be made for projected upper limits on the sterile mixing angles assuming no evidence for sterile oscillations is found, and picking the value of  $\Delta m^2_{41} = 0.5$~eV$^2$ corresponding to the simpler counting experiment regime. For the $3+1$ model, upper limits of $\theta_{24}$\,$<$\,$1.8^{\circ}(15.1^{\circ})$ and $\theta_{34}$\,$<$\,$15.0^{\circ}(25.5^{\circ})$ are obtained at the 90\% \dword{cl} from the presented best(worst)-case scenario DUNE sensitivities. If expressed in terms of the relevant matrix elements
\begin{align}
\begin{split}
|U_{\mu4}|^2 =&\,\,\cos^2\theta_{14}\sin^2\theta_{24} \\
|U_{\tau4}|^2= & \,\,\cos^2\theta_{14}\cos^2\theta_{24}\sin^2\theta_{34},
\end{split}
\label{eq:DisapToApp}
\end{align}
these limits become $|U_{\mu4}|^{2}$\,$<$\,0.001(0.068) and $|U_{\tau4}|^{2}$\,$<$\,0.067(0.186) at the 90\% \dword{cl}, where we conservatively assume $\cos^2\theta_{14}$\,=\,1 in both cases, and additionally $\cos^2\theta_{24}$\,=\,1 in the second case.
\begin{dunetable}
[Projected 90\% \dword{cl} upper limits on sterile mixing angles and matrix elements]
{c c c c c}
{tab:limits}
{The projected DUNE 90\% \dword{cl} upper limits on sterile mixing angles and matrix elements compared to the equivalent 90\% \dword{cl} upper limits from \nova~\cite{ref:novasterile}, MINOS/MINOS+~\cite{Adamson:2017uda}, \superk~\cite{ref:superksterile}, IceCube~\cite{ref:IceCube}, and IceCube-DeepCore~\cite{ref:DeepCore}. The limits are shown for $\Delta m^2_{41} = 0.5$~eV$^2$ for all experiments, except for IceCube-DeepCore, where the results are reported for $\Delta m^2_{41} = 1.0$~eV$^2$.}
& $\theta_{24}$ & $\theta_{34}$ & $|U_{\mu4}|^2$ &  $|U_{\tau4}|^2$  \\ \toprowrule
DUNE Best-Case  & $1.8^{\circ}$ & $15.0^{\circ}$ & 0.001 & 0.067  \\ \colhline
DUNE Worst-Case  & $15.1^{\circ}$ & $25.5^{\circ}$ & 0.068 & 0.186  \\ \colhline
NOvA  & $20.8^{\circ}$ & $31.2^{\circ}$ & 0.126 & 0.268  \\ \colhline
MINOS/MINOS+ & $4.4^{\circ}$ & $23.6^{\circ}$ & 0.006 & 0.16  \\ \colhline
\superk & $11.7^{\circ}$ & $25.1^{\circ}$ & 0.041 & 0.18  \\ \colhline
IceCube & $4.1^{\circ}$ & \-- & 0.005 & \--   \\ \colhline 
IceCube-DeepCore & $19.4^{\circ}$ & $22.8^{\circ}$ & 0.11 & 0.15 \\
\end{dunetable}  
  
Finally, sensitivity to the $\theta_{\mu e}$ effective mixing angle, defined above as $\sin^2{2\theta_{\mu e}}\equiv 4|U_{e4}|^2|U_{\mu 4}|^2=\sin^22\theta_{14}\sin^2\theta_{24}$, is shown in Figure~\ref{fig:th_me}, which also displays a comparison with the allowed regions from \dword{lsnd} and MiniBooNE, as well as with present constraints and projected constraints from the \fnal \dword{sbn} program.

Further, to illustrate that DUNE would not be limited to constraining active-sterile neutrino mixing, we have produced a discovery potential plot, for a scenario with one sterile neutrino governed by the \dword{lsnd} best-fit parameters: $\left(\Delta m_{14}^2= 1.2\;\text{eV}^2;\,\,\sin^2{2\theta_{\mu e}}=0.003\right)$~\cite{LSNDSterile}. 
A small 90\% \dword{cl} allowed region, shown in Figure~\ref{fig:th_me}, is obtained, which can be compared with the \dword{lsnd} allowed region in the same figure. 
\begin{dunefigure}
[Sensitivity to $\theta_{\mu e}$ from  (dis)appearance samples and discovery potential at \dword{lsnd} best fit]
{fig:th_me}
{DUNE sensitivities to $\theta_{\mu e}$ from the appearance and disappearance samples at the \dword{nd} and \dword{fd} is shown on the left-hand plot, along with a comparison with previous existing experiments and the sensitivity from the future \dword{sbn} program. Regions to the right of the DUNE contours are excluded. The right-hand plot displays the discovery potential assuming $\theta_{\mu e}$ and $\Delta m_{41}^2$ set at the best-fit point determined by \dword{lsnd}~\cite{LSNDSterile} for the best-case scenario referenced in the text.}
$\vcenter{\hbox{\includegraphics[width=0.48\columnwidth]{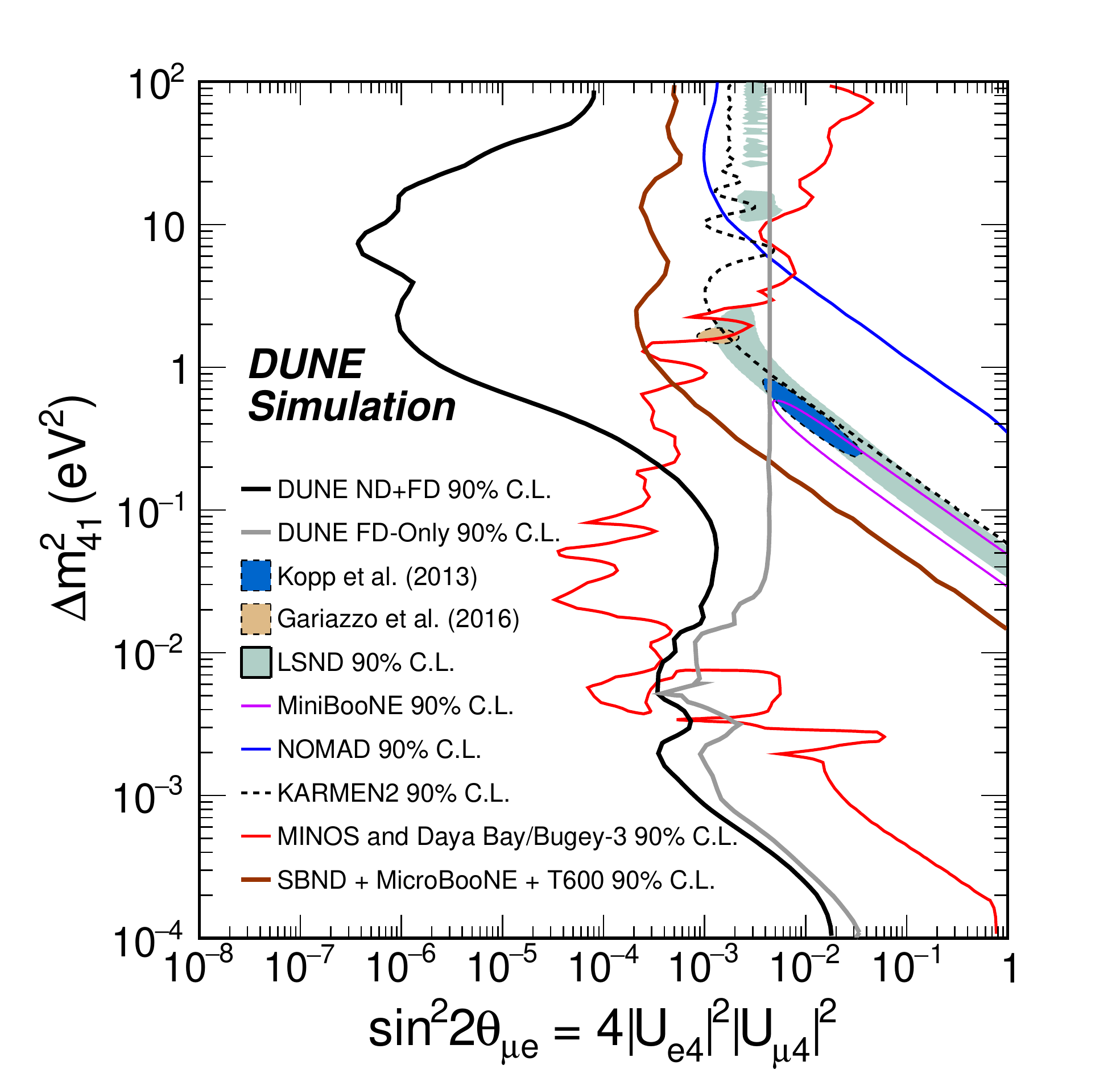}}}$
$\vcenter{\hbox{\includegraphics[width=0.42\columnwidth]{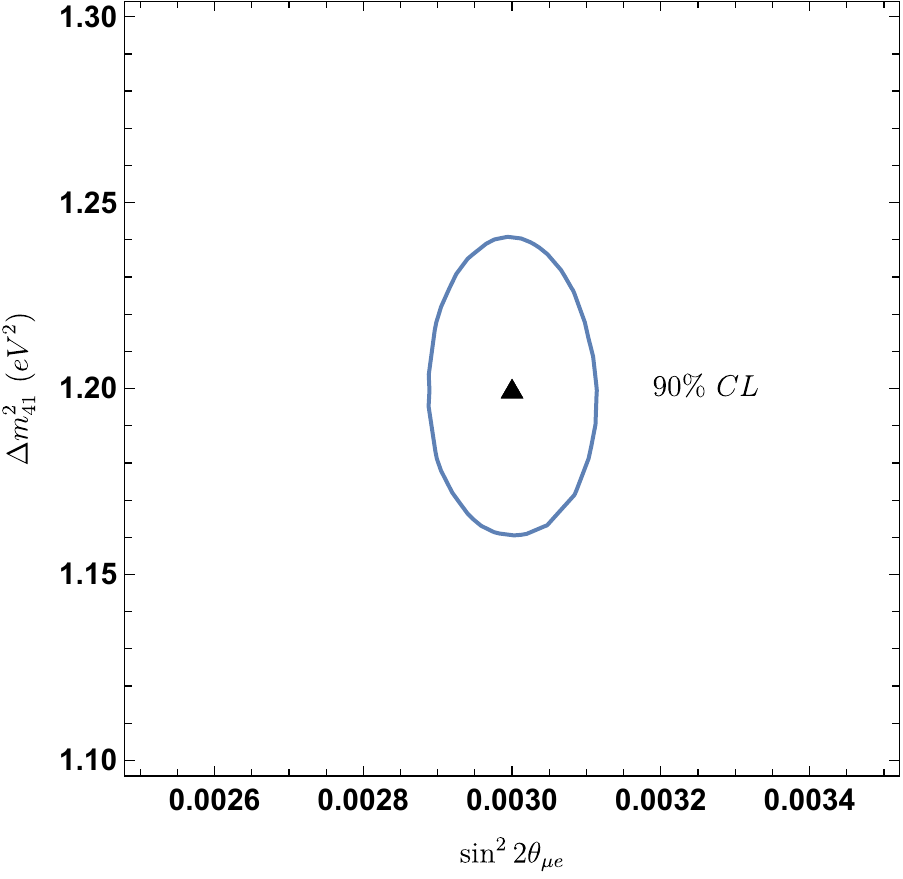}}}$
\end{dunefigure}

The physics reach plots shown above illustrate the excellent potential of DUNE to discover or constrain mixing with sterile neutrinos. Notably, in the case of sterile-mediated \numu~to \nue~transitions, DUNE can place very competitive constraints on its own, without requiring a combination with reactor experiments. 

These studies show compelling motivation for DUNE to deploy a highly-capable \dword{nd}  given its high potential for discovery or constraining of new physics, including mixing with sterile neutrino species. These capabilities can be further improved by a high-precision muon monitor system for the LBNF beam, which would provide an independent constraint on the neutrino flux through measurements of the associated muon flux, not susceptible to mixing with sterile neutrinos.


\section{Non-Unitarity of the Neutrino Mixing Matrix}
\label{sec:nonUnitarity}
A generic characteristic of most models explaining the neutrino mass
pattern is the presence of heavy neutrino states, beyond the
three light states of the \dword{sm}  of particle
physics~\cite{Minkowski:1977sc,Mohapatra:1979ia,Yanagida:1979as,GellMann:1980vs}. This implies a deviation from unitarity of the $3\times3$ \dword{pmns} matrix, which can be particularly sizable 
as the mass of the extra states becomes lower~\cite{Lee:1977tib,Schechter:1980gr,Mohapatra:1986bd,Akhmedov:1995vm,Akhmedov:1995ip,Malinsky:2005bi}.
For values of non-unitarity parameter deviations of order $10^{-2}$, this would decrease the expected reach of DUNE to the standard parameters, although stronger bounds existing from charged leptons would be able to restore its expected performance~\cite{Blennow:2016jkn,Escrihuela:2016ube}.

A generic characteristic of most models explaining the neutrino mass
pattern is the presence of heavy neutrino states, additional to the
three light states of the \dword{sm}  of particle
physics~\cite{Mohapatra:1998rq,Valle:2015pba,Fukugita:2003en}. These
types of models will imply that the $3 \times 3$ \dword{pmns} matrix is not unitary due to the mixing with the additional states.  Besides the type-I seesaw
mechanism~\cite{GellMann:1980vs,Yanagida:1979as,Mohapatra:1979ia,Schechter:1980gr},
different low-scale seesaw models include right-handed neutrinos that are relatively not-so-heavy~\cite{Mohapatra:1986bd} and perhaps detectable at collider experiments.

These additional heavy leptons would mix with the light neutrino states and, as a result, the complete unitary mixing matrix would be a squared $n \times n$ matrix, with $n$ the total number of neutrino
states. As a result, the usual $3 \times 3$ \dword{pmns} matrix, which we dub $N$ to stress its non-standard nature, will be
non-unitary. One possible general way to parameterize these unitarity deviations in $N$ is through a triangular matrix~\cite{Escrihuela:2015wra}\footnote{For a similar parameterization corresponding to a $(3+1)$ and a $(3+3)$-dimensional mixing matrix,  see Refs.~\cite{Xing:2007zj,Xing:2011ur}}
 \begin{equation}
  N = 
 \left\lgroup
 \begin{array}{ccc} 
 1-\alpha_{ee} & 0 & 0 \\
 \alpha_{\mu e} & 1-\alpha_{\mu \mu} & 0 \\
  \alpha_{\tau e} & \alpha_{\tau \mu} & 1-\alpha_{\tau \tau}
 \end{array}
 \right \rgroup U \,,
 \label{eq:triangular}
 \end{equation}
with $U$ a unitary matrix that tends to the usual \dword{pmns} matrix when the non-unitary parameters $\alpha_{ij} \rightarrow 0$\footnote{The original parameterization in Ref.~\cite{Escrihuela:2015wra} uses $\alpha_{ii}$ instead of $\alpha_{\beta\gamma}$. The equivalence between the two notations is as follows: $\alpha_{ii} = 1-\alpha_{\beta\beta}$ and $\alpha_{ij} = \alpha_{\beta\gamma}$.} .
The triangular matrix in this equation accounts for the non-unitarity of the $3 \times 3$ matrix for any number of extra neutrino species. This pasteurization has been shown to be particularly well-suited for oscillation searches~\cite{Escrihuela:2015wra,Blennow:2016jkn} since, compared to other alternatives, it minimizes the departures of its unitary component $U$ from the mixing angles that are directly measured in neutrino oscillation experiments when unitarity is assumed.

The phenomenological implications of a non-unitary leptonic mixing matrix have been extensively studied in flavor and electroweak precision observables as well as in the neutrino oscillation phenomenon~\cite{Shrock:1980vy,Schechter:1980gr,Shrock:1980ct,Shrock:1981wq,Langacker:1988ur,Bilenky:1992wv,Nardi:1994iv,Tommasini:1995ii,Antusch:2006vwa,FernandezMartinez:2007ms,Antusch:2008tz,Biggio:2008in,Antusch:2009pm,Forero:2011pc,Alonso:2012ji,Antusch:2014woa,Abada:2015trh,Fernandez-Martinez:2015hxa,Escrihuela:2015wra,Parke:2015goa,Miranda:2016wdr,Fong:2016yyh,Escrihuela:2016ube}. For recent global fits to all flavor and electroweak precision data summarizing present bounds on non-unitarity see Refs.~\cite{Antusch:2014woa,Fernandez-Martinez:2016lgt}.

\begin{dunetable}
[Expected 90\%~CL constraints on the non-unitarity parameters $\alpha$]
{|c|c|}
{tab:bounds}
{Expected $90\%$~\dword{cl} constraints on the non-unitarity parameters $\alpha$ from DUNE.}
{\bf Parameter} & {\bf Constraint} \\ \toprowrule
$\alpha_{ee}$ & $0.3$   \\ \colhline
$\alpha_{\mu\mu}$ & $0.2$ \\ \colhline
$\alpha_{\tau\tau}$ & $0.8$ \\ \colhline
$\alpha_{\mu e}$ & $0.04$ \\ \colhline
$\alpha_{\tau e}$ & $0.7$ \\ \colhline
$\alpha_{\tau\mu}$ & $0.2$ \\
\end{dunetable}

\subsection{NU constraints from DUNE}
Recent studies have shown that DUNE can constrain the non-unitarity parameters~\cite{Blennow:2016jkn, Escrihuela:2016ube}. The summary of the $90 \%$~\dword{cl}  bounds on the different $\alpha_{ij}$ elements profiled over all other parameters is given in Table~\ref{tab:bounds}. 
These bounds are comparable with other constraints from present oscillation experiments, although they are not competitive with those obtained from flavor and electroweak precision data.
For this analysis, and 
those presented below, we have used the \dword{globes} software~\cite{Huber:2004ka,Huber:2007ji} with the DUNE \dword{cdr} configuration presented in Ref.~\cite{Alion:2016uaj}, and assuming a data exposure of 300~kton.MW.year. The standard (unitary) oscillation parameters have also been treated as in~\cite{Alion:2016uaj}. The unitarity deviations have been included both by an independent code (used to obtain the results shown in Ref.~\cite{Escrihuela:2016ube}) and via the MonteCUBES~\cite{Blennow:2009pk} plug-in to cross validate our results.

\subsection{NU impact on DUNE standard searches}
Conversely, the presence of non-unitarity may affect the determination of the
Dirac \dword{cp}-violating phase $\delta_{CP}$ in long-baseline experiments~\cite{Miranda:2016wdr,Fernandez-Martinez:2016lgt,Escrihuela:2016ube}.
Indeed, when allowing for unitarity deviations, the expected \dword{cp} discovery potential for DUNE could be significantly reduced.
However, the situation is alleviated when a combined analysis with the constraints on non-unitarity from other experiments is considered. This is illustrated in Figure~\ref{fig:CPsens}. In the left panel, the discovery potential for \dword{cpv} is computed when the non-unitarity parameters introduced in Eq.~(\ref{eq:triangular}) are allowed in the fit. While for the Asimov data all $\alpha_{ij}=0$, the non-unitary parameters are allowed to vary in the fit with $1 \sigma$ priors of $10^{-1}$, $10^{-2}$ and $10^{-3}$ for the dotted green, dashed blue and solid black lines respectively. For the dot-dashed red line no prior information on the non-unitarity parameters has been assumed. As can be observed, without additional priors on the non-unitarity parameters, the capabilities of DUNE to discover \dword{cpv} from $\delta_{CP}$ would be seriously compromised~\cite{Escrihuela:2016ube}. However, with priors of order $10^{-2}$ matching the present constraints from other neutrino oscillation experiments~\cite{Escrihuela:2016ube,Blennow:2016jkn}, the standard sensitivity is almost recovered. If the more stringent priors of order $10^{-3}$ stemming from flavor and electroweak precision observables are added~\cite{Antusch:2014woa,Fernandez-Martinez:2016lgt}, the standard sensitivity is obtained.   

The right panel of Figure~\ref{fig:CPsens} concentrates on the impact of the phase of the element $\alpha_{\mu e}$ in the discovery potential of \dword{cpv} from $\delta_{CP}$, since this element has a very important impact in the $\nu_e$ appearance channel. In this plot the modulus of $\alpha_{ee}$, $\alpha_{\mu \mu}$ and $\alpha_{\mu e}$ have been fixed to $10^{-1}$, $10^{-2}$, $10^{-3}$ and 0 for the dot-dashed red, dotted green, dashed blue and solid black lines respectively. All other non-unitarity parameters have been set to zero and the phase of $\alpha_{\mu e}$ has been allowed to vary both in the fit and in the Asimov data, showing the most conservative curve obtained. As for the right panel, it can be seen that a strong deterioration of the \dword{cp} discovery potential could be induced by the phase of $\alpha_{\mu e}$ (see Ref.~\cite{Escrihuela:2016ube}). However, for unitarity deviations of order $10^{-2}$, as required by present neutrino oscillation data constraints, the effect is not too significant in the range of $\delta_{CP}$ for which a $3 \sigma$ exclusion of \dword{cp} conservation would be possible and it becomes negligible if the stronger $10^{-3}$ constraints from flavor and electroweak precision data are taken into account.  

Similarly, the presence of non-unitarity worsens degeneracies involving $\theta_{23}$, making the determination of the octant or even its maximality challenging.
This situation is shown in Figure~\ref{fig:octant} where an input value of $\theta_{23} = 42.3^\circ$ was assumed. As can be seen, the fit in presence of non-unitarity (solid lines) introduces degeneracies for the wrong octant and even for maximal mixing~\cite{Blennow:2016jkn}. However, these degeneracies are solved upon the inclusion of present priors on the non-unitarity parameters from other oscillation data (dashed lines) and a clean determination of the standard oscillation parameters following DUNE expectations is again recovered.   

\begin{dunefigure}
[Impact of non-unitarity on the CPV discovery potential]
{fig:CPsens}
{The impact of non-unitarity on the DUNE \dword{cpv} discovery potential. See the text for details.}
 \includegraphics[width=0.8\columnwidth]{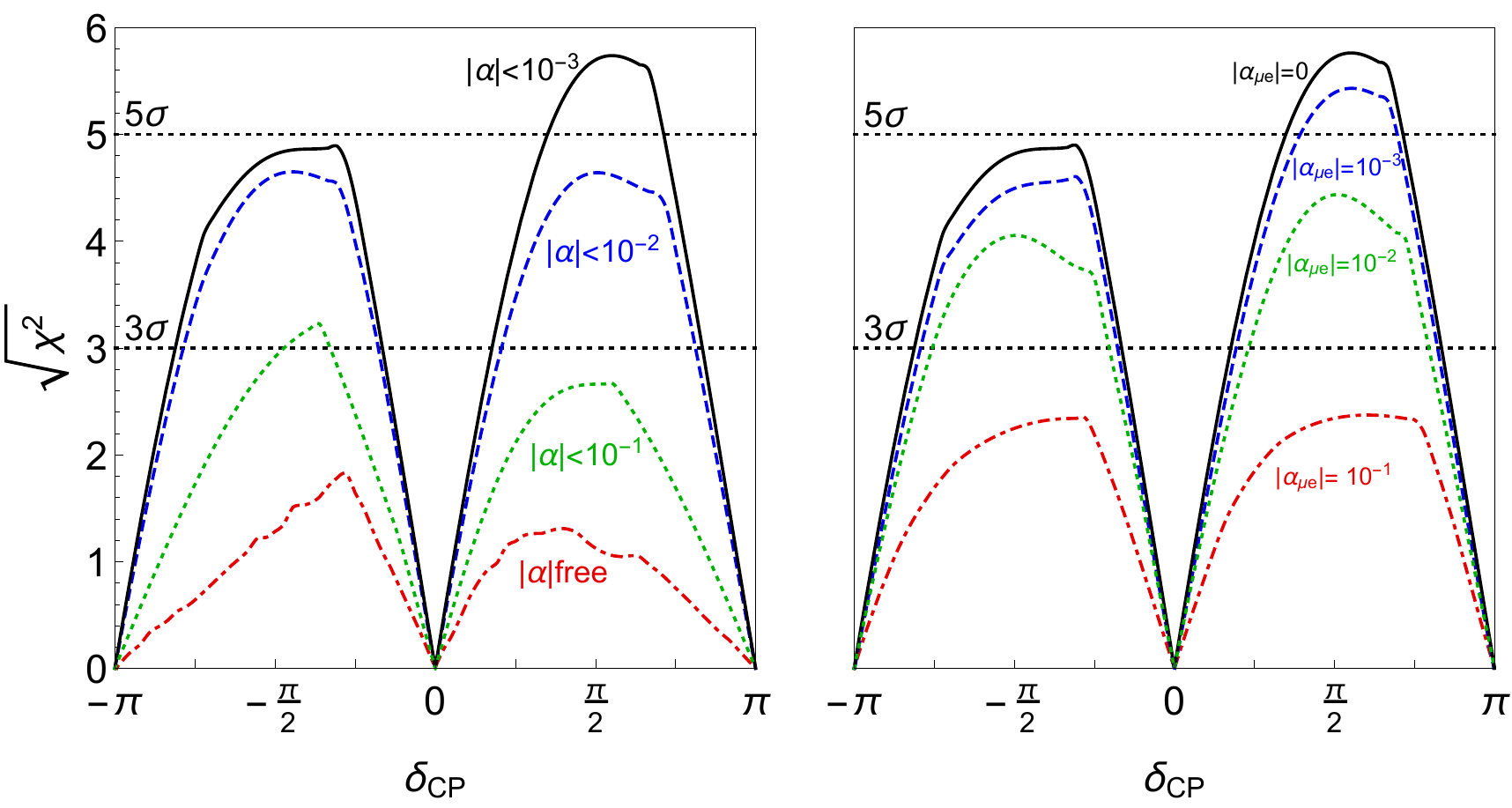}
\end{dunefigure}

\subsubsection{Conclusions}
A non-unitary lepton mixing matrix, as generally expected from the most common extensions of the \dword{sm}  explaining neutrino masses, would affect the neutrino oscillations to be measured by DUNE. The sensitivity that DUNE would provide to the non-unitarity parameters is comparable to that from present oscillation experiments, while not competitive to that from flavor and electroweak precision observables, which is roughly an order of magnitude more stringent. Conversely, the capability of DUNE to determine the standard oscillation parameters such as \dword{cpv} from $\delta_{CP}$ or the octant or maximality of $\theta_{23}$ would be seriously compromised by unitarity deviations in the \dword{pmns}. This negative impact is however significantly reduced when priors on the size of these deviations from other oscillation experiments are considered and disappears altogether if the more stringent constraints from flavor and electroweak precision data are added instead.

\begin{dunefigure}
[Expected frequentist allowed regions at the $1 \sigma$, $90\%$ and $2\sigma$ \dword{cl}]
{fig:octant}
{Expected frequentist allowed regions at the $1 \sigma$, $90\%$ and $2\sigma$ \dword{cl}\ for DUNE. All new physics parameters are assumed to be zero so as to obtain the expected non-unitarity sensitivities. The solid lines correspond to the analysis of DUNE data alone, while the dashed lines include the present constraints on non-unitarity.}
 \includegraphics[width=0.8\columnwidth]{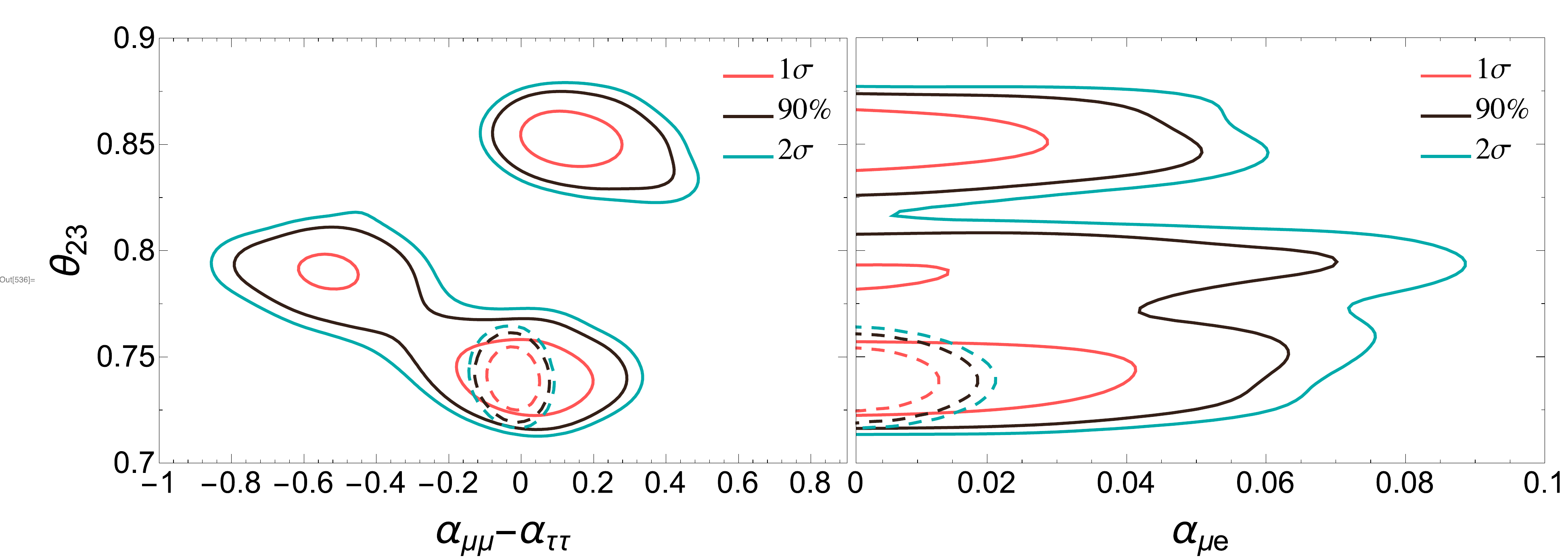} 
\end{dunefigure}

\section{Non-Standard Neutrino Interactions}
\label{sec:nsi}
\dword{nsi} can significantly modify the data to be collected by DUNE as long as the new physics parameters are large enough. \dword{nsi} may impact the determination of current unknowns such as \dword{cpv}~\cite{Masud:2015xva,Masud:2016bvp}, mass hierarchy~\cite{Masud:2016gcl} and octant of $\theta_{23}$~\cite{Agarwalla:2016fkh}. If the DUNE data are consistent with the standard oscillation for three massive neutrinos, \dword{nc} \dword{nsi} effects of order 0.1 $G_F$, affecting neutrino propagation through the Earth, can be ruled out at DUNE~\cite{deGouvea:2015ndi,Coloma:2015kiu}. We notice that DUNE might improve current constraints on $|\epsilon^m_{e \tau}|$ and $|\epsilon^m_{e \mu}|$ by a factor 2-5~\cite{Ohlsson:2012kf,Miranda:2015dra,Farzan:2017xzy}. New  \dword{cc}  interactions can also lead to modifications in the production and the detection of neutrinos. The findings on source and detector \dword{nsi} studies at DUNE are presented in~\cite{Blennow:2016etl,Bakhti:2016gic}. In particular, the simultaneous impact on the measurement of $\delta_{\rm CP}$ and $\theta_{23}$ is investigated in detail. Depending on the assumptions, such as the use of the \dword{nd}  and whether \dword{nsi} at production and detection are the same, the impact of source/detector \dword{nsi} at DUNE may be relevant. We are assuming the results from~\cite{Blennow:2016etl}, in which DUNE does not have sensitivity to discover or to improve bounds on source/detector \dword{nsi}, and focus our attention in the propagation.

\subsection{NSI in propagation at DUNE}
\dword{nc} \dword{nsi} can be understood as non-standard
matter effects that are visible only in a \dword{fd} at a sufficiently long baseline. They can be parameterized as new contributions
to the \dword{msw} matrix in the neutrino-propagation Hamiltonian:
\begin{equation}
  H = U \left( \begin{array}{ccc}
           0 &                    & \\
             & \Delta m_{21}^2/2E & \\
             &                    & \Delta m_{31}^2/2E
         \end{array} \right) U^\dag + \tilde{V}_{\rm MSW} \,,
\end{equation}
with
\begin{equation}
  \tilde{V}_{\rm MSW} = \sqrt{2} G_F N_e
\left(
  \begin{array}{ccc}
    1 + \epsilon^m_{ee}       & \epsilon^m_{e\mu}       & \epsilon^m_{e\tau}  \\
        \epsilon^{m*}_{e\mu}  & \epsilon^m_{\mu\mu}     & \epsilon^m_{\mu\tau} \\
        \epsilon^{m*}_{e\tau} & \epsilon^{m*}_{\mu\tau} & \epsilon^m_{\tau\tau}
  \end{array} 
\right)
\label{eq:epsmatrix}
\end{equation}
Here, $U$ is the standard \dword{pmns} leptonic mixing matrix, for which we use the standard parameterization found, e.g., in~\cite{Agashe:2014yua}, 
and the $\epsilon$-parameters give the
magnitude of the \dword{nsi} relative to standard weak interactions.  For new physics scales of a few hundred GeV,  a value of $|\epsilon|$ of the order
0.01 or less is
expected~\cite{Davidson:2003ha,GonzalezGarcia:2007ib,Biggio:2009nt}. The DUNE
baseline provides an advantage in the detection of \dword{nsi} relative
to existing beam-based experiments with shorter baselines.
Only atmospheric-neutrino experiments have longer baselines, but the sensitivity of these experiments to \dword{nsi} is limited by systematic effects~\cite{Adams:2013qkq}.

To assess DUNE sensitivity to \dword{nc}  \dword{nsi}, the \dword{nsi} discovery reach is defined in the following way: the expected event spectra are simulated using  \dword{globes}~\cite{Huber:2004ka,Huber:2007ji}, assuming \textit{true} values for the  \dword{nsi} 
parameters, and a fit is then attempted assuming no \dword{nsi}. If the fit is
incompatible with the simulated data at a given confidence level,
the chosen \emph{true} values of the \dword{nsi} parameters are considered to be within the experimental discovery reach.

In this analysis, we use \dword{globes} with the Monte Carlo Utility Based Experiment Simulator (MonteCUBES) C library~\cite{Blennow:2009pk}, a plugin that replaces the deterministic  \dword{globes} minimizer by a Markov Chain Monte Carlo (MCMC) method that is able to handle higher dimensional parameter spaces. In the simulations we use the configuration for the DUNE \dword{cdr}~\cite{Alion:2016uaj}. Each point scanned by the MCMC is stored and a frequentist $\chi^2$ analysis is performed with the results. The analysis assumes an exposure of 300~kton.MW.year.

Considering that  \dword{nsi} exists, conducting the analysis with all the  \dword{nsi} parameters free to vary, we obtain the sensitivity regions in Figure~\ref{fig:nsi}. We omit the superscript $m$ that appears in Eq.~\ref{eq:epsmatrix}. 
The credible regions are shown for different confidence level intervals.
\begin{dunefigure}
[Allowed regions for NSI parameters]
{fig:nsi}
{Allowed regions of the non-standard oscillation parameters in which we see important degeneracies (top) and the complex non-diagonal ones (bottom). We conduct the analysis considering all the \dword{nsi} parameters as non-negligible. The sensitivity regions are for 68\% CL [red line (left)], 90\% CL [green dashed line (middle)], and 95\% CL [blue dotted line (right)]. Current bounds are taken from~\cite{Gonzalez-Garcia:2013usa}.}
\includegraphics[width=0.8\columnwidth]{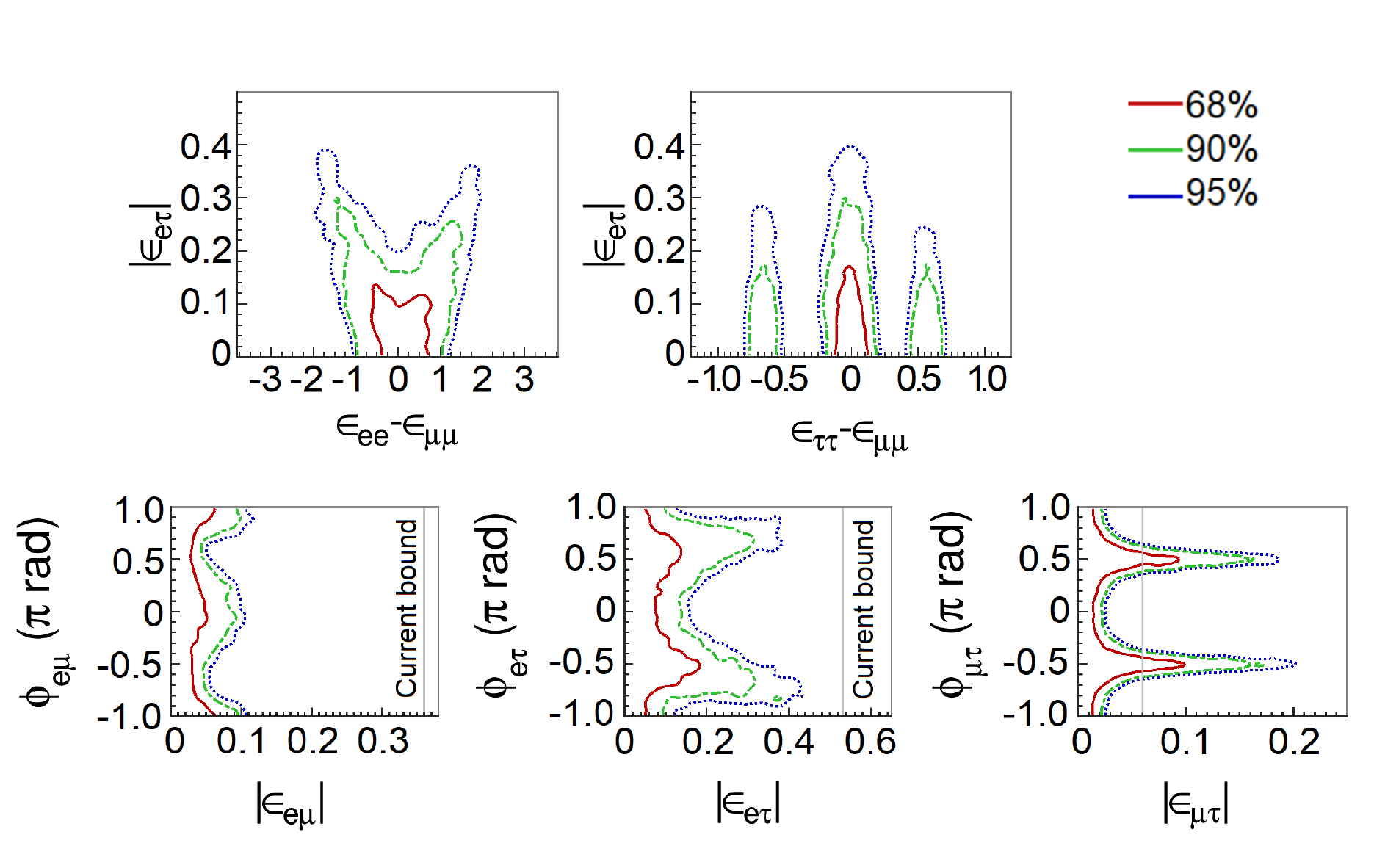}
\end{dunefigure}
We note, however, that constraints on $\epsilon_{\tau\tau}-\epsilon_{\mu\mu}$ coming from global fit analysis~\cite{Gonzalez-Garcia:2013usa,Miranda:2015dra,Farzan:2017xzy,Esteban:2018ppq} can remove the left and right solutions of $\epsilon_{\tau\tau}-\epsilon_{\mu\mu}$ in Figure~\ref{fig:nsi}.

In order to constrain the standard oscillation parameters when  \dword{nsi} are 
present, we use the fit for three-neutrino mixing from~\cite{Gonzalez-Garcia:2013usa} and implement prior constraints to restrict the region sampled by the MCMC. The sampling of the parameter space is explained in~\cite{Coloma:2015kiu} and the priors that we use can be found in table~\ref{tab:priors1}.
\begin{dunetable}
[Oscillation parameters and priors implemented in MCMC.]
{| c | c | c |}
{tab:priors1}
{Oscillation parameters and priors implemented in MCMC for calculation of Figure~\ref{fig:nsi}.} 
{\bf Parameter} & {\bf Nominal} & {\bf 1$\sigma$ Range ($\pm$) }\\ \toprowrule
$\theta_{12}$ &0.19$\pi~\textrm{rad}$&2.29\%\\ \colhline
$\sin^2(2\theta_{13})$ &0.08470&0.00292\\ \colhline
$\sin^2(2\theta_{23})$ &0.9860&0.0123\\ \colhline
$\Delta m^2_{21} $ &7.5 $\times10^{-5}\textrm{eV}^2$&2.53\%\\ \colhline
$\Delta m^2_{31} $ &2.524 $\times10^{-3}\textrm{eV}^2$&free\\ \colhline
$\delta_{\rm CP} $ &1.45$\pi~\textrm{rad}$&free\\
\end{dunetable}

Then we can observe the effects of \dword{nsi} on the measurements of the standard oscillation parameters at DUNE. In Figure~\ref{fig:standar-nsi}, we superpose the allowed regions with non-negligible  \dword{nsi} and the standard-only credible regions at 90\% \dword{cl}. 
An important degeneracy appears in the measurement of the mixing angle $\theta_{23}$. We also see that the sensitivity of the \dword{cp} phase is strongly affected.
\begin{dunefigure}
[Projections of the standard oscillation parameters with nonzero NSI]
{fig:standar-nsi}
{Projections of the standard oscillation parameters with nonzero \dword{nsi}. The sensitivity regions are for 68\%, 90\%, and 95\% \dword{cl}. The allowed regions considering negligible \dword{nsi} (standard oscillation (SO)) are superposed to the SO+NSI at 90\% \dword{cl}.}
\includegraphics[width=0.6\columnwidth]{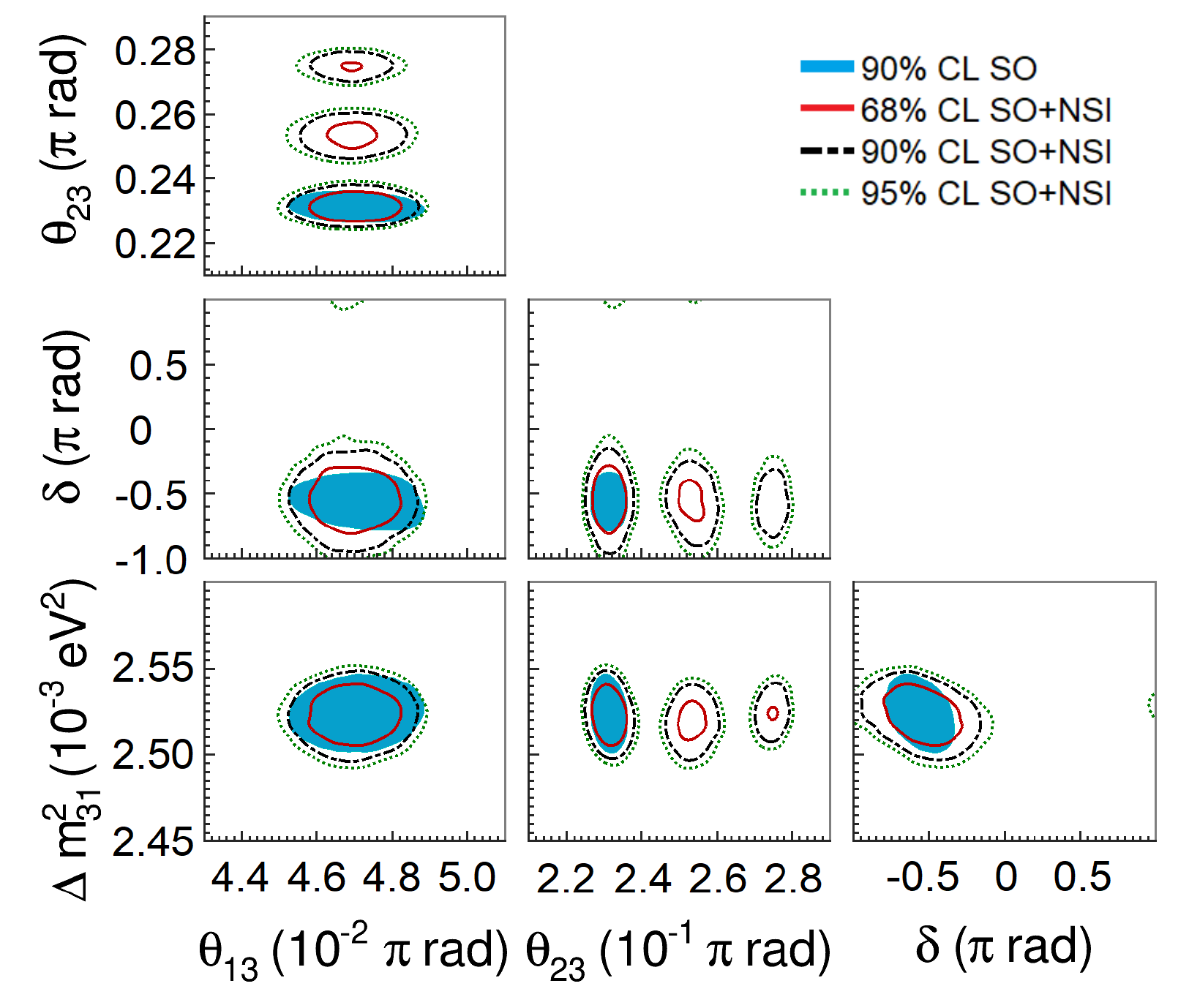}
\end{dunefigure}

\subsection{Effects of baseline and matter-density variation on  NSI measurements}\label{ssec:matter}
The effects of matter density variation and its average along the beam path from \fnal to \surf  were studied considering the standard neutrino oscillation framework with three flavors~\cite{Roe:2017zdw,Kelly:2018kmb}. In order to obtain the results of Figures~\ref{fig:nsi} and~\ref{fig:standar-nsi}, we use a high-precision calculation for the baseline of \SI{1284.9}{km} and the average density of \SI{2.8482}{g/cm^3}~\cite{Roe:2017zdw}.

The DUNE collaboration has been using the so-called PREM~\cite{Dziewonski:1981xy,PREM2} density profile to consider matter density variation. With this assumption, the neutrino beam crosses a few constant density layers.
However, a more detailed density map is available for the USA with more than 50 layers and $0.25 \times 0.25$ degree cells of latitude and longitude: The Shen-Ritzwoller or S.R. profile~\cite{SR:2016,Roe:2017zdw}. Comparing the S.R. with the PREM profiles, Kelly and Parke~\cite{Kelly:2018kmb} show that, in the standard oscillation paradigm, DUNE is not highly sensitive to the density profile and that the only oscillation parameter with its measurement slightly impacted by the average density true value is \deltacp{}.
\dword{nsi}, however, may be sensitive to the profile, particularly considering the phase $\phi_{e\tau}$~\cite{Chatterjee:2018dyd}, to which DUNE will have a high sensitivity~\cite{Ohlsson:2012kf,Miranda:2015dra,deGouvea:2015ndi,Coloma:2015kiu,Farzan:2017xzy}, as we also see in Figure~\ref{fig:nsi}.

In order to compare the results of our analysis predictions for DUNE with the constraints from other experiments we use the results from~\cite{Farzan:2017xzy}. There are differences in the parameter nominal values used for calculating the $\chi^2$ function and other assumptions. This is the reason why the regions in Figure~\ref{fig:bars} do not have the same central values, but this comparison gives a good view of how DUNE can substantially improve the bounds on, for example, $\varepsilon_{\tau\tau}-\varepsilon_{\mu\mu}$, $\Delta m^2_{31}$, and the non-diagonal \dword{nsi} parameters.

\begin{dunefigure}
[1D DUNE constraints versus current constraints]
{fig:bars}
{One-dimensional DUNE constraints compared with current constraints calculated in \cite{Farzan:2017xzy}. See text for details.}
\includegraphics[width=1.0\columnwidth]{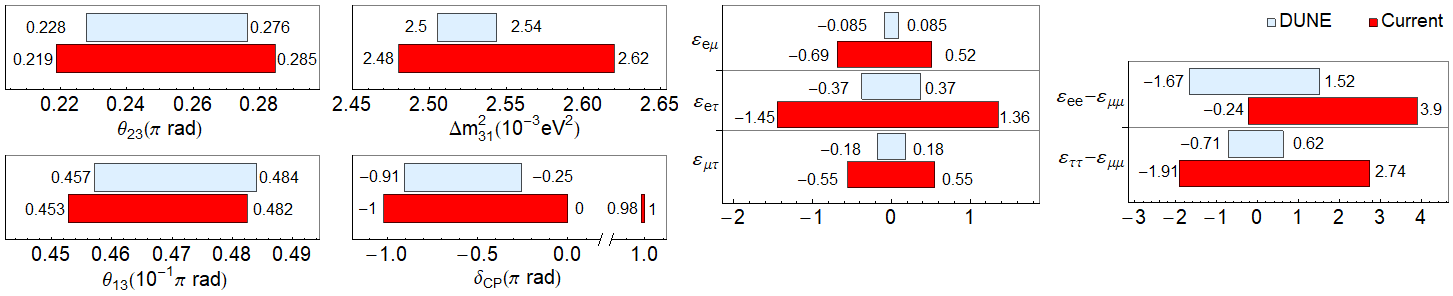}
\end{dunefigure}

\subsubsection{Conclusions and prospects}
\dword{nsi} can significantly impact the determination of current unknowns such as \dword{cpv} and the octant of $\theta_{23}$. Clean determination of the intrinsic \dword{cp} phase at long-baseline experiments such as DUNE is a formidable task~\cite{Rout:2017udo}. A feasible strategy to extricate physics scenarios at DUNE using high-energy beams was suggested in~\cite{Masud:2017bcf}.

\section{CPT Symmetry Violation}

\dword{cpt} symmetry, the combination of charge conjugation, parity and time reversal, is a cornerstone of our model-building strategy. 
DUNE can improve the present limits on Lorentz and \dword{cpt} violation by several orders of magnitude~\cite{Streater:1989vi,Barenboim:2002tz,Kostelecky:2003cr,Diaz:2009qk,Kostelecky:2011gq,Barenboim:2017ewj}, contributing as a very important experiment to test these fundamental assumptions underlying quantum field theory.

\dword{cpt} invariance is one of the predictions of major importance of local, relativistic quantum field theory. One of the predictions of \dword{cpt} invariance is that particles and antiparticles have the same masses and, if unstable, the same lifetimes. To prove the \dword{cpt} theorem one needs only three ingredients~\cite{Streater:1989vi}: Lorentz invariance, hermeticity of the Hamiltonian, and locality.

Experimental bounds on \dword{cpt} invariance can be derived using the neutral kaon system~\cite{Schwingenheuer:1995uf}:
\begin{equation}
  \frac{|m(K^0) - m(\overline{K}^0)|}{m_K} < 0.6 \times 10^{-18}\,. 
  \label{eq:mK}
\end{equation}
This result, however, should be interpreted very carefully for two reasons. First, we do not have a complete theory of \dword{cpt} violation, and it is therefore arbitrary to take the kaon-mass as a scale. Second, since kaons are bosons, the term entering the Lagrangian is the mass squared and not the mass itself. With this in mind, we can rewrite the previous bound as:
%
 $ |m^2(K^0) - m^2(\overline{K}^0)| < 0.3~\mbox{eV}^2 \, $.
%
Here we 
see that neutrinos can test the predictions of the \dword{cpt} theorem to an unprecedented extent and could, therefore, provide stronger limits than the ones regarded as the most stringent ones 
to date\footnote{\dword{cpt} was tested also using charged leptons. However, these measurements involve a combination
of mass and charge and are not a direct \dword{cpt} test. Only neutrinos can provide \dword{cpt} tests on an elementary mass not contaminated by charge.}. 
In the absence of a solid model of flavor, not to mention one of \dword{cpt} violation, the spectrum  of neutrinos and antineutrinos can differ both  in the mass eigenstates themselves as well as in the flavor composition of each of these states. It is important to notice then that neutrino oscillation experiments can only test \dword{cpt} in the mass differences and mixing angles. An overall shift between the neutrino and antineutrino spectra will be missed by oscillation experiments.  Nevertheless, such a pattern can be bounded by cosmological data~\cite{Barenboim:2017vlc}. Unfortunately direct searches for neutrino mass (past, present, and future) involve only antineutrinos and hence cannot be used to draw any conclusion on \dword{cpt} invariance on the absolute mass scale, either.
Therefore, using neutrino oscillation data, we will compare the mass splittings and mixing angles of  neutrinos with those of antineutrinos. Differences in the neutrino and antineutrino spectrum would imply the violation of the \dword{cpt} theorem.

In Ref.~\cite{Barenboim:2017ewj} the authors derived the most up-to-date bounds on \dword{cpt} invariance from the neutrino sector 
using the same data that was used in the global fit to neutrino oscillations in Ref.~\cite{deSalas:2017kay}. 
Of course, experiments that cannot distinguish between neutrinos and antineutrinos, such as atmospheric data from \superk~\cite{Abe:2017aap}, IceCube-DeepCore~\cite{Aartsen:2014yll,Aartsen:2017nmd} and ANTARES~\cite{AdrianMartinez:2012ph} were not included. The complete data set used, as well as the parameters to which they are sensitive, are 
%
 (1) from solar neutrino data~\cite{Cleveland:1998nv,Kaether:2010ag,Abdurashitov:2009tn,hosaka:2005um,Cravens:2008aa,Abe:2010hy,Nakano:PhD,Aharmim:2008kc,Aharmim:2009gd,Bellini:2013lnn}:  $\theta_{12}$, $\Delta m_{21}^2$, and $\theta_{13}$;
 (2) from neutrino mode in long-baseline experiments K2K~\cite{Ahn:2006zza}, MINOS~\cite{Adamson:2013whj,Adamson:2014vgd}, T2K~\cite{Abe:2017uxa,Abe:2017bay}, and NO$\nu$A~\cite{Adamson:2017qqn,Adamson:2017gxd}:  $\theta_{23}$, $\Delta m_{31}^2$, and $\theta_{13}$;
 (3) from KamLAND reactor antineutrino data~\cite{Gando:2010aa}: $\overline{\theta}_{12}$, $\Delta \overline{m}_{21}^2$, and $\overline{\theta}_{13}$;
 (4) from short-baseline reactor antineutrino experiments Daya Bay~\cite{An:2016ses}, RENO~\cite{RENO:2015ksa}, and Double Chooz~\cite{Abe:2014bwa}:     $\overline{\theta}_{13}$ and $\Delta \overline{m}_{31}^2$; and 
 (5) from antineutrino mode in long-baseline experiments MINOS~\cite{Adamson:2013whj,Adamson:2014vgd} and T2K~\cite{Abe:2017uxa,Abe:2017bay}: $\overline{\theta}_{23}$, $\Delta \overline{m}_{31}^2$, and
$\overline{\theta}_{13}$\footnote{The K2K experiment took  data only in neutrino mode, while the \nova experiment had not published data in the antineutrino mode when these bounds were calculated.}. 

From the analysis of all previous data samples, one can derive the most up-to-date bounds on \dword{cpt} violation:
%
$ |\Delta m_{21}^2-\Delta \overline{m}_{21}^2| < 4.7\times 10^{-5} \,  \text{eV}^2,\,\,
 |\Delta m_{31}^2-\Delta \overline{m}_{31}^2| < 3.7\times 10^{-4} \, \text{eV}^2,\,\,
 |\sin^2\theta_{12}-\sin^2\overline{\theta}_{12}| < 0.14\,,\,\,
 |\sin^2\theta_{13}-\sin^2\overline{\theta}_{13}| < 0.03\,, \,\,
 {\rm and}~|\sin^2\theta_{23}-\sin^2\overline{\theta}_{23}| < 0.32\,.
 $  

At the moment it is not possible to set any bound on $|\delta-\overline{\delta}|$, since all possible values of
$\delta$ or $\overline{\delta}$ are allowed by data. The preferred intervals of $\delta$ obtained in Ref.~\cite{deSalas:2017kay} can only be obtained after combining the neutrino and antineutrino data samples. 
The limits  on $\Delta(\Delta m_{31}^2)$ and $\Delta(\Delta m_{21}^2)$  are already better than the one derived from the neutral kaon system and should be regarded as the best current bounds on \dword{cpt} violation on the mass squared. 
Note that these results were derived assuming the same mass ordering for neutrinos and antineutrinos. If the ordering was different for neutrinos and antineutrinos, this would be an indication for \dword{cpt} violation on its own. In the following we show how DUNE could improve this bound.

\subsubsection{Sensitivity to \dword{cpt} symmetry violation in the neutrino sector}
\label{sec:sensitivity}

\begin{dunetable}
[Oscillation parameters used to simulate (anti)neutrino data.]
{| c | c |}
{tab:par2}
{Oscillation parameters used to simulate neutrino and antineutrino data analyzed in Section~\ref{sec:sensitivity}.}
    {\bf Parameter} & {\bf Value}  \\ \toprowrule
    $\Delta m^2_{21}$& $7.56\times 10^{-5}\text{eV}^2$\\  \colhline
    $\Delta m^2_{31}$&  $2.55\times 10^{-3}\text{eV}^2$\\  \colhline
    $\sin^2\theta_{12}$ & 0.321\\  \colhline
    $\sin^2\theta_{23}$ &  0.43, 0.50, 0.60\\  \colhline
    $\sin^2\theta_{13}$ & 0.02155\\  \colhline
    $\delta$ & 1.50$\pi$\\
\end{dunetable}

Here we study the sensitivity of the DUNE experiment to measure \dword{cpt} violation in the neutrino sector by analyzing neutrino and antineutrino oscillation parameters separately. We assume the neutrino oscillations being parameterized by the usual \dword{pmns} matrix $U_{\text{PMNS}}$, with parameters $\theta_{12},\theta_{13},\theta_{23},\Delta m_{21}^2,\Delta m_{31}^2, {\rm and}~\delta$, while the antineutrino oscillations are parameterized by a matrix $\overline{U}_{\text{PMNS}}$ with parameters $\overline{\theta}_{12},\overline{\theta}_{13},\overline{\theta}_{23},\Delta \overline{m}_{21}^2,\Delta \overline{m}_{31}^2, {\rm and}~\overline{\delta}$. Hence, antineutrino oscillation is described  by the same probability functions as neutrinos with the neutrino parameters replaced by their antineutrino counterparts\footnote{Note that the antineutrino oscillation probabilities also include the standard change of sign in the \dword{cp} phase.}. 
To simulate the future neutrino data signal in DUNE, we assume the true values for neutrinos and antineutrinos to be as listed in Table~\ref{tab:par2}.
Then, in the statistical analysis, we vary freely all the oscillation parameters, except the solar ones, which are fixed to their best fit values throughout the simulations. Given the great precision in the determination of the reactor mixing angle by the short-baseline reactor experiments~\cite{An:2016ses,RENO:2015ksa,Abe:2014bwa}, in our analysis we use a prior on $\overline{\theta}_{13}$, but not on $\theta_{13}$. We also consider three different values for the atmospheric angles, as indicated in Table~\ref{tab:par2}. The exposure considered in the analysis corresponds to 300~kton.MW.year.

\begin{figure}[!htb]
 \centering
        \includegraphics[width=0.55\columnwidth]{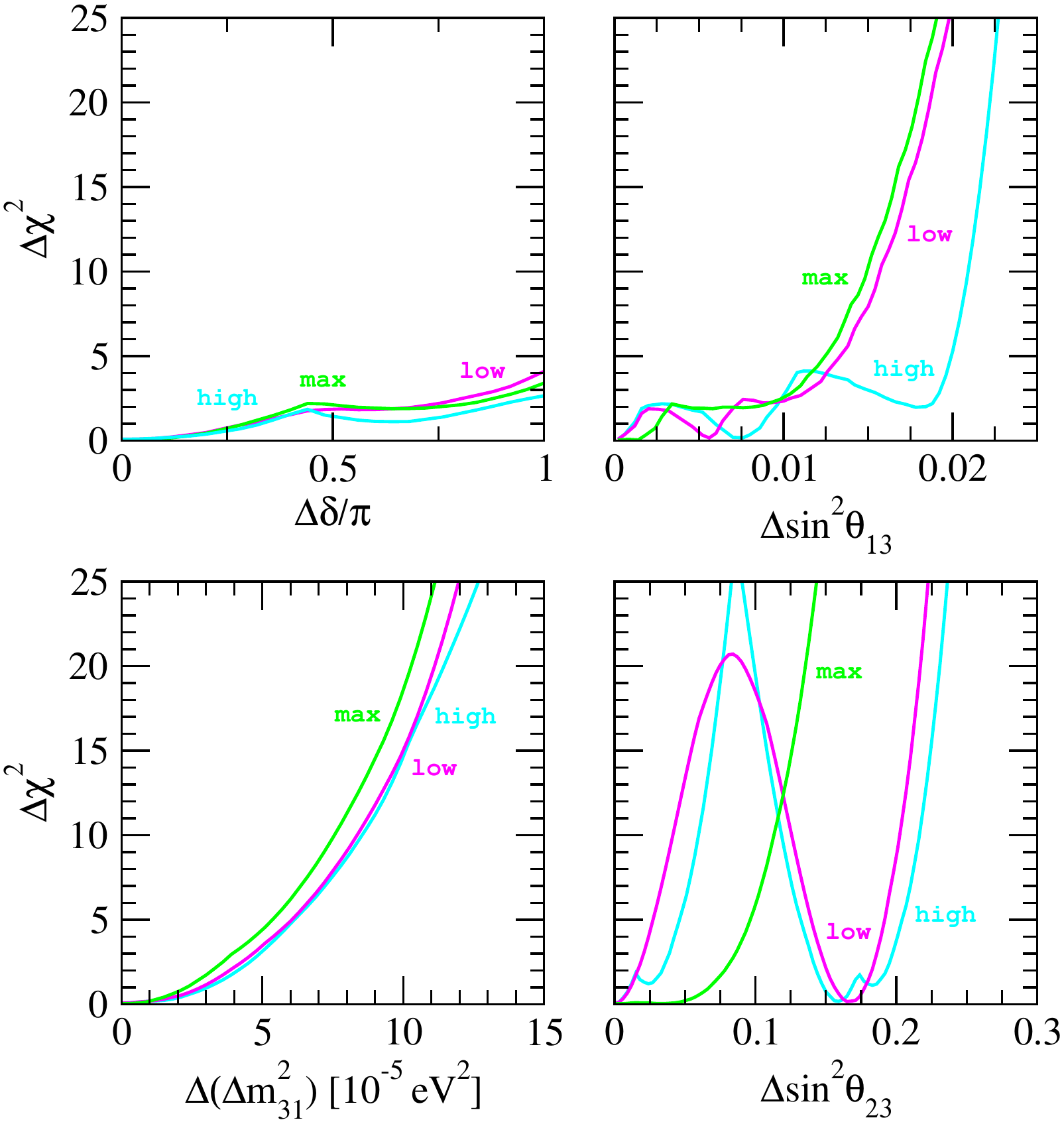}
        \caption[Sensitivities to the difference of neutrino and antineutrino parameters]{The sensitivities of DUNE to the difference of neutrino and antineutrino parameters: 
        $\Delta\delta$, $\Delta(\Delta m_{31}^2)$, $\Delta(\sin^2\theta_{13})$ and $\Delta(\sin^2\theta_{23})$  
        for the atmospheric angle in the lower octant (magenta line),  in the upper octant (cyan line) and for maximal mixing (green line).}
	\label{fig:sensitivity-CPT}
\end{figure}

Therefore, to test the sensitivity at DUNE we perform the simulations assuming $\Delta x = |x-\overline{x}| = 0$, where $x$ is any of the oscillation parameters. Then we estimate the sensitivity to $\Delta x\neq 0$. To do so we calculate two $\chi^2$-grids, one for neutrinos and one for antineutrinos, varying the four parameters of interest. After minimizing over all parameters except $x$ and $\overline{x}$, we calculate 
\begin{equation}
 \chi^2(\Delta x) = \chi^2(|x-\overline{x}|) = \chi^2(x)+\chi^2(\overline{x}),
 \label{eq:chi2-nu-nubar}
\end{equation}
where we have considered all the possible combinations of $|x-\overline{x}|$. The results are presented in Figure~\ref{fig:sensitivity-CPT}, where we plot three different lines, labeled as ``high'', ``max'' and ``low.'' These refer to the assumed value for the atmospheric angle: in the lower octant (low), maximal mixing (max) or in the upper octant (high). Here we can see that there is sensitivity neither to $\Delta(\sin^2\theta_{13})$, where the 3$\sigma$ bound would be of the same order 
as the current measured value for $\sin^2\overline{\theta}_{13}$, nor to $\Delta\delta$, where no single value of the parameter would be excluded at more than 2$\sigma$.

On the contrary, interesting results for $\Delta(\Delta m_{31}^2)$ and $\Delta(\sin^2\theta_{23})$ are obtained. First, we see that DUNE can put stronger bounds on the difference of the atmospheric mass splittings, namely $\Delta(\Delta m_{31}^2) < 8.1\times 10^{-5}$, improving the current neutrino bound by one order of magnitude. For the atmospheric angle, we obtain different results depending on the true value assumed in the simulation of DUNE data. In the lower right panel of Figure~\ref{fig:sensitivity-CPT} we see the different behavior obtained for $\theta_{23}$ with the values of $\sin^2\theta_{23}$ from table~\ref{tab:par2}, i.e., lying in the lower octant, being maximal, and lying in the upper octant.
As one might expect, the sensitivity increases with $\Delta\sin^2\theta_{23}$ in the case of maximal mixing. However, if the true value lies in the lower or upper octant, a degenerate solution appears in the complementary octant.

\subsection{Imposter solutions}
\label{sec:impost}
In 
some types of neutrino oscillation experiments, e.g., accelerator experiments, neutrino and antineutrino data are obtained in separate experimental runs. The usual procedure followed by the experimental collaborations, as well as the global oscillation fits as for example Ref.~\cite{deSalas:2017kay}, assumes \dword{cpt} invariance and analyzes the full data sample in a joint way.
However, if \dword{cpt} is violated in nature, the outcome of the joint data analysis might give rise to what we call an ``imposter'' solution, i.e., one that does not correspond to the true solution of any channel. 

\begin{figure}[!htb]
 \centering
        \includegraphics[width=0.45\columnwidth]{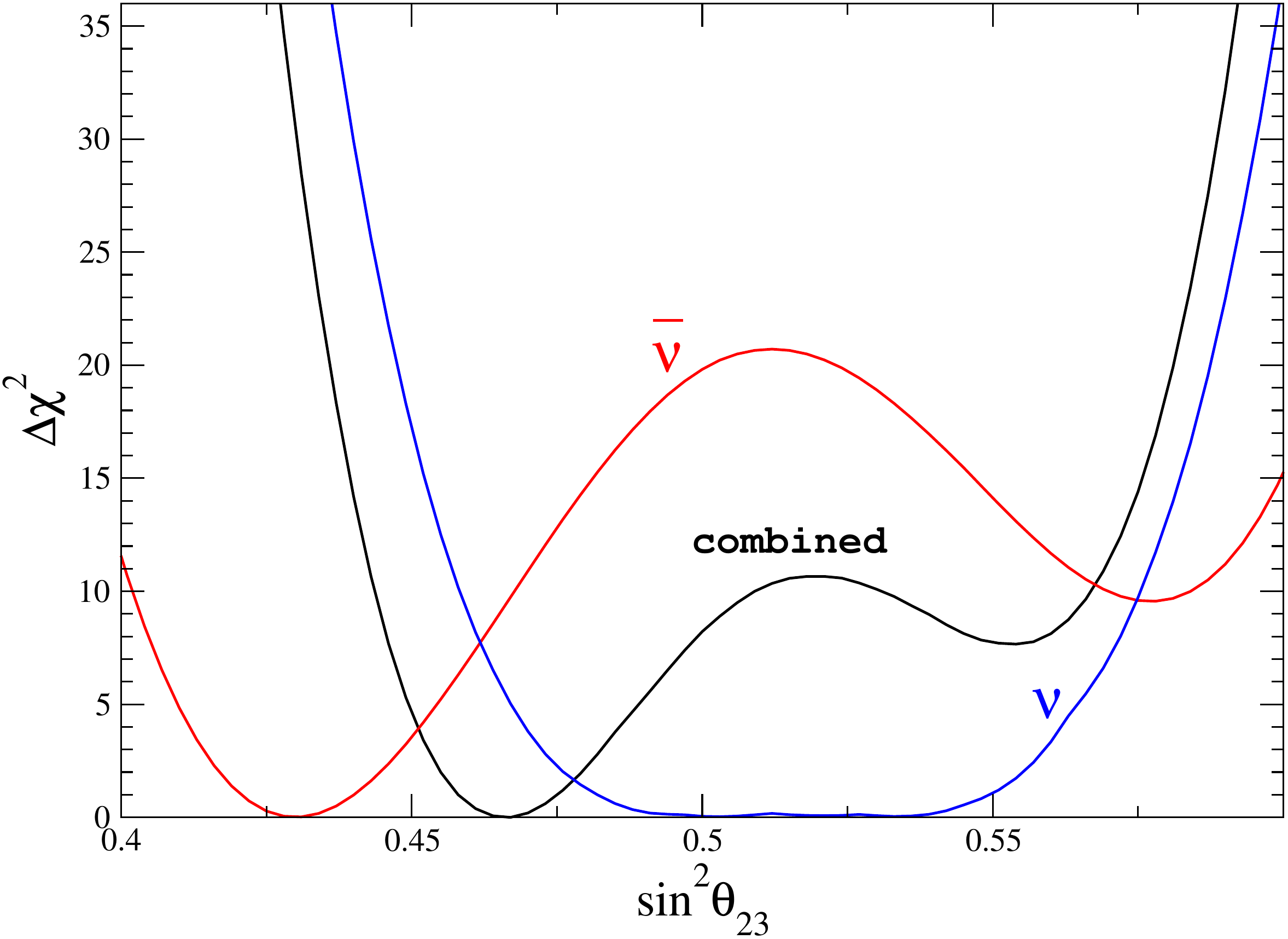}
        \caption[Sensitivity to $\theta_{23}$ for (anti)neutrinos, and combination under CPT conservation]{DUNE sensitivity to the atmospheric angle for neutrinos (blue), antineutrinos (red), and to the combination of both under the assumption of \dword{cpt} conservation (black).
         }
	\label{fig:imposter-sq23}
\end{figure}

Under the assumption of \dword{cpt} conservation, the $\chi^2$ functions are computed according to
\begin{equation}
 \chi^2_{\text{total}}=\chi^2(\nu)+\chi^2(\overline{\nu})\, ,
 \label{eq:CPT-cons}
\end{equation}
and assuming that the same parameters describe neutrino and antineutrino flavor oscillations. In contrast, in Eq.~(\ref{eq:chi2-nu-nubar}) we first profiled over the parameters in neutrino and antineutrino mode separately and then added the profiles. Here, we shall assume \dword{cpt} to be violated in nature, but perform our analysis as if it were conserved. As an example, we assume that the true value for the atmospheric neutrino mixing is $\sin^2\theta_{23}=0.5$, while the antineutrino mixing angle is given by $\sin^2\overline{\theta}_{23}=0.43$. The rest of the oscillation parameters are set to the values in Table~\ref{tab:par2}. Performing the statistical analysis in the \dword{cpt}-conserving way, as indicated in Eq.~(\ref{eq:CPT-cons}), we obtain the profile of the atmospheric mixing angle presented in Figure~\ref{fig:imposter-sq23}. The profiles for the individual reconstructed results (neutrino and antineutrino) are also shown in the figure for comparison.
The result is a new best fit value at $\sin^2\theta^\text{comb}_{23}=0.467$, disfavoring the true values for neutrino and antineutrino parameters at approximately 3$\sigma$ and more than 5$\sigma$, respectively. 

\section{Search for Neutrino Tridents at the Near Detector}
\begin{figure}[!hb]
\centering
\includegraphics[width=0.27\textwidth]{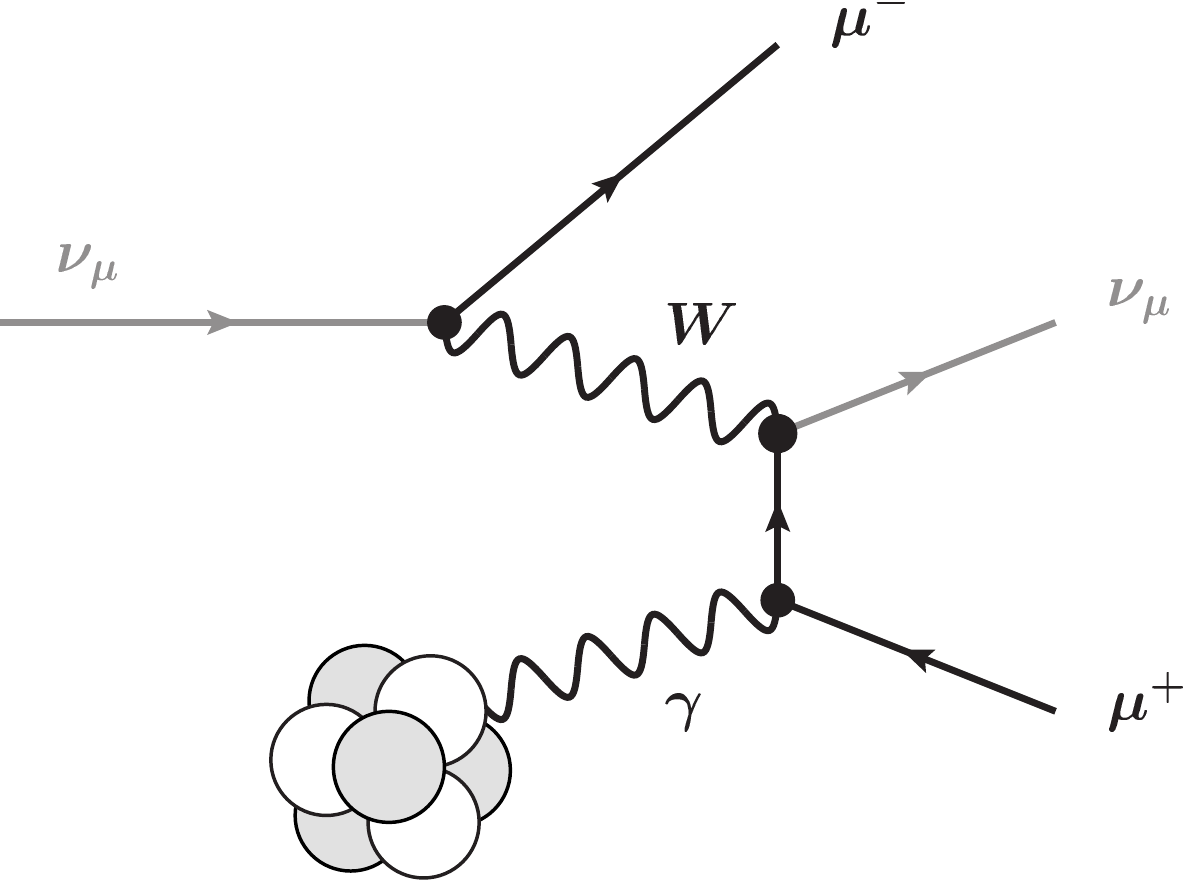} \qquad
\includegraphics[width=0.27\textwidth]{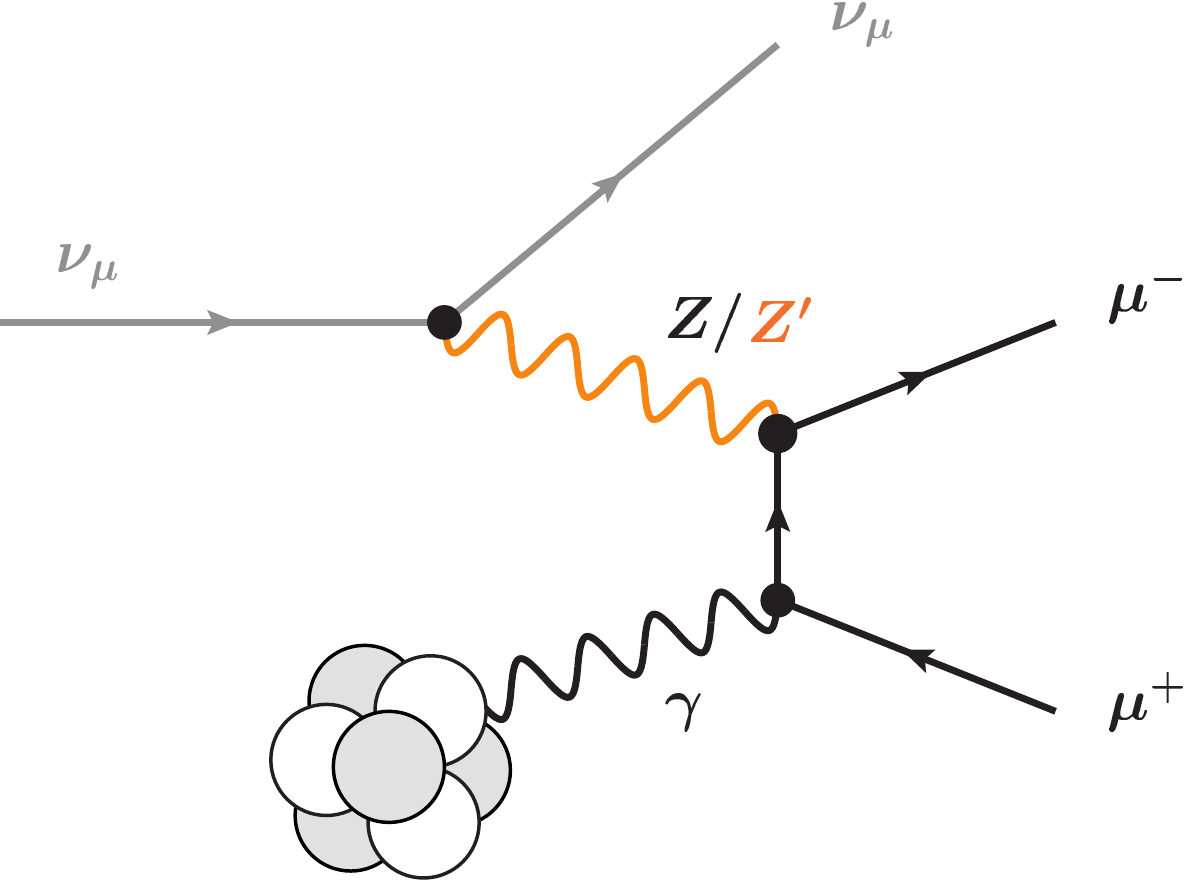} \\[\baselineskip]
\includegraphics[width=0.27\textwidth]{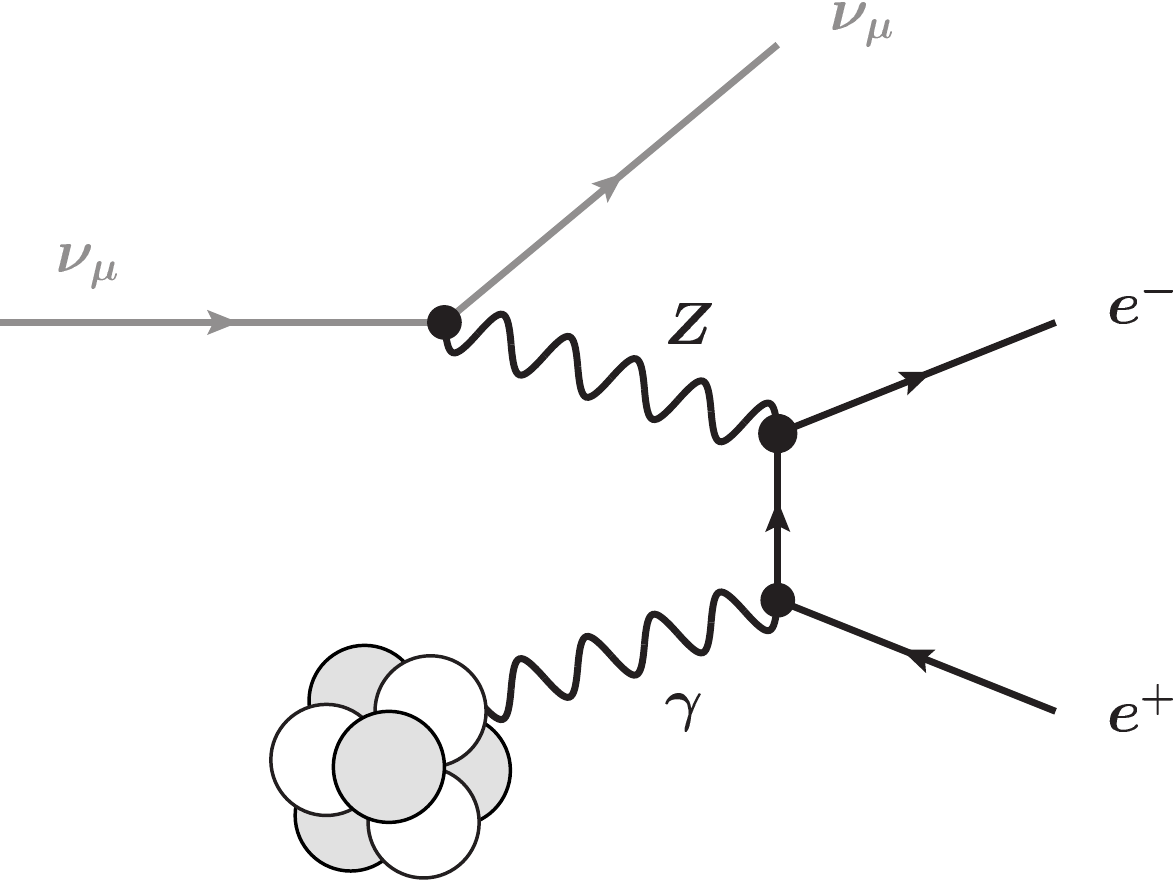} \qquad
\includegraphics[width=0.27\textwidth]{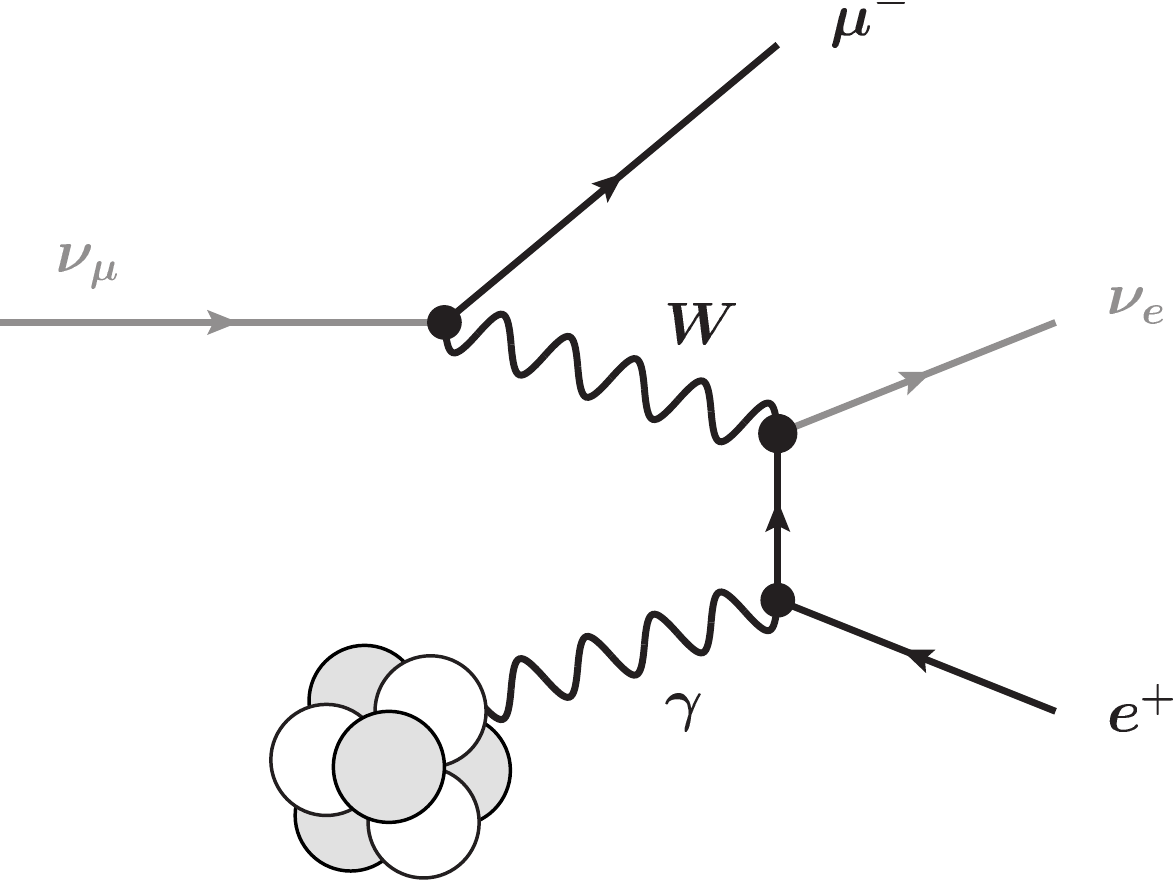} \\[\baselineskip]
\caption[Example diagrams for $\numu$-induced trident processes in the SM]{Example diagrams for muon-neutrino-induced trident processes in the Standard Model. A second set of diagrams where the photon couples to the negatively charged leptons is not shown. 
Analogous diagrams exist for processes induced by different neutrino flavors and by anti-neutrinos. A diagram illustrating trident interactions mediated by a new $Z'$ gauge boson, discussed in the text, is shown on the top right.}
\label{fig:diagrams}
\end{figure}

Neutrino trident production is a weak process in which a neutrino, scattering off the Coulomb field of a heavy nucleus, generates a pair of charged leptons, as shown in Fig.~\ref{fig:diagrams}~\cite{Czyz:1964zz,Lovseth:1971vv,Fujikawa:1971nx,Koike:1971tu,Koike:1971vg,Brown:1973ih,Belusevic:1987cw}.
Measurements of muonic neutrino tridents ($\nu_\mu \to \nu_\mu \mu^+\mu^-$) were carried out at the CHARM-II~\cite{Geiregat:1990gz}, CCFR~\cite{Mishra:1991bv} and NuTeV~\cite{Adams:1999mn} experiments:
\[
\frac{\sigma(\nu_\mu \to \nu_\mu \mu^+\mu^-)_\text{exp}}{\sigma(\nu_\mu \to \nu_\mu \mu^+\mu^-)_\text{SM}} = 
\begin{cases}
1.58 \pm 0.64         & \text{(CHARM-II)} \\ 
0.82 \pm 0.28         & \text{(CCFR)} \\
0.72 ^{+1.73}_{-0.72} & \text{(NuTeV)} 
\end{cases}
\]
The high-intensity muon-neutrino beam at the DUNE \dword{nd}  will lead to a sizable production rate of trident events (see Table~\ref{tab:trident_rates}), offering excellent prospects to improve the above measurements~\cite{Altmannshofer:2019zhy,Ballett:2018uuc,Ballett:2019xoj}. A deviation from the event rate predicted by the \dword{sm} could be an indication of new interactions mediated by the corresponding new gauge bosons~\cite{Altmannshofer:2014pba}. 

\begin{dunetable}
[Expected number of SM $\nu_\mu$ and $\bar\nu_\mu$-induced trident events at ND per ton of Ar per year]
{|l|c|c|}
{tab:trident_rates}
{Expected number of \dword{sm} $\nu_\mu$ and $\bar\nu_\mu$-induced trident events at the LArTPC of the DUNE \dword{nd} per metric ton of argon and year of operation.}
{\bf Process} & {\bf Coherent} & {\bf Incoherent} \\ \toprowrule
\midrule
$\nu_\mu \to \nu_\mu \mu^+\mu^-$ & $1.17 \pm 0.07$ & $0.49 \pm 0.15$ \\
$\nu_\mu \to \nu_\mu e^+e^-$ & $2.84 \pm 0.17$ & $0.18 \pm 0.06$\\
$\nu_\mu \to \nu_e e^+\mu^-$ & $9.8 \pm 0.6$ & $1.2 \pm 0.4$ \\
$\nu_\mu \to \nu_e \mu^+e^-$ & $0$ & $0$ \\
\midrule
$\bar\nu_\mu \to \bar\nu_\mu \mu^+\mu^-$ & $0.72 \pm 0.04$ & $0.32 \pm 0.10$ \\
$\bar\nu_\mu \to \bar\nu_\mu e^+e^-$ & $2.21 \pm 0.13$ & $0.13 \pm 0.04$ \\
$\bar\nu_\mu \to \bar\nu_e e^+\mu^-$ & $0$ & $0$ \\
$\bar\nu_\mu \to \bar\nu_e \mu^+e^-$ & $7.0 \pm 0.4$ & $0.9 \pm 0.3$ \\
\end{dunetable}

The main challenge in obtaining a precise measurement of the muonic trident cross section will be the copious backgrounds, mainly consisting of \dword{cc} single-pion production events, $\nu_\mu N \to \mu \pi N^\prime$, as muon and pion tracks can be easily confused in LArTPC detectors. The discrimination power of the DUNE \dword{nd} LArTPC was evaluated using large simulation datasets of signal and background. Each simulation event represents a different neutrino-argon interaction in the active volume of the detector. Signal events were generated using a standalone code \cite{Altmannshofer:2019zhy} that simulates trident production of muons and electrons through the scattering of $\nu_{\mu}$ and $\nu_e$ on argon nuclei (or iron nuclei, for comparison with CCFR and NuTeV results). The generator considers both the coherent scattering on the full nucleus (the dominant contribution) and the incoherent scattering on individual nucleons. Background events, consisting of several \dword{sm} neutrino interactions, were generated using \dword{genie}. Roughly $38\%$ of the generated events have a charged pion in the final state, leading to two charged tracks with muon-like energy deposition pattern ($\mathrm{d}E/\mathrm{d}x$), as in our trident signal. All final-state particles produced in the interactions were propagated through the detector geometry using the Geant4-based \cite{Agostinelli:2002hh,Allison:2006ve,Allison:2016lfl} simulation of the DUNE \dword{nd}. Charge collection and readout were not simulated, and possible inefficiencies due to misreconstruction effects or event pile-up were disregarded for simplicity.

\begin{figure}[!tb]
\centering
\includegraphics[height=5cm]{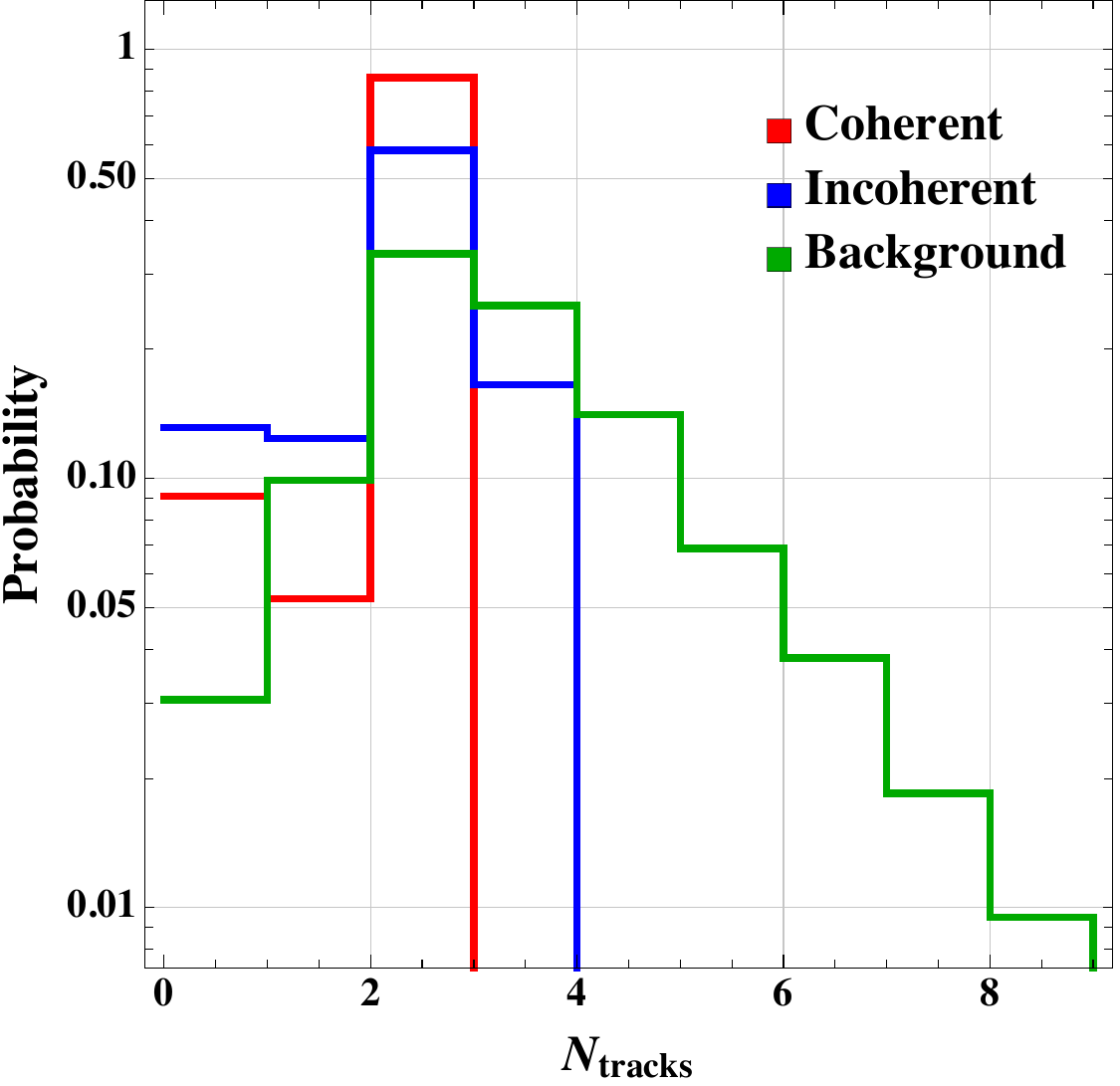}
\includegraphics[height=5cm]{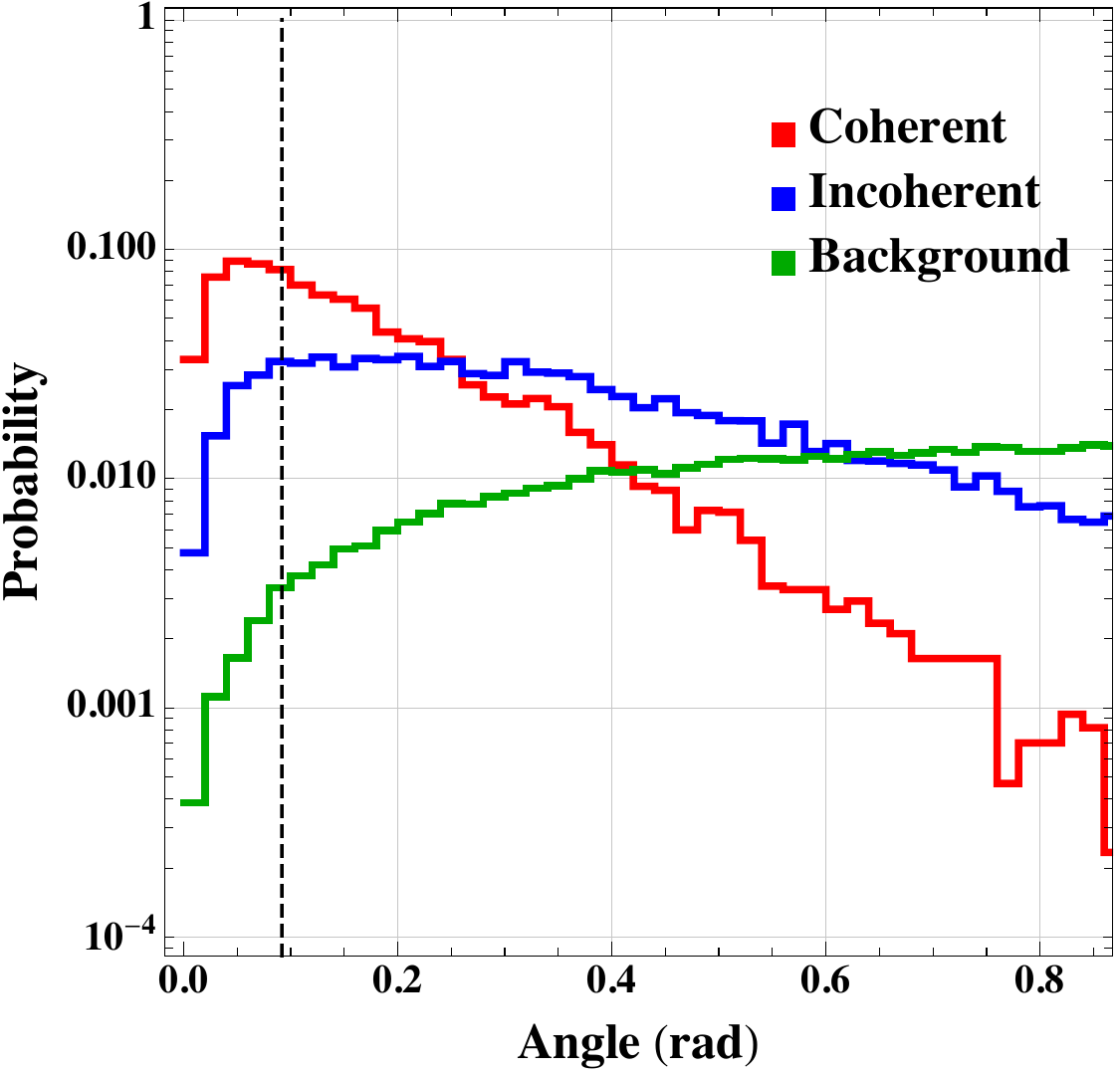} \\[0.75\baselineskip]
\includegraphics[height=5cm]{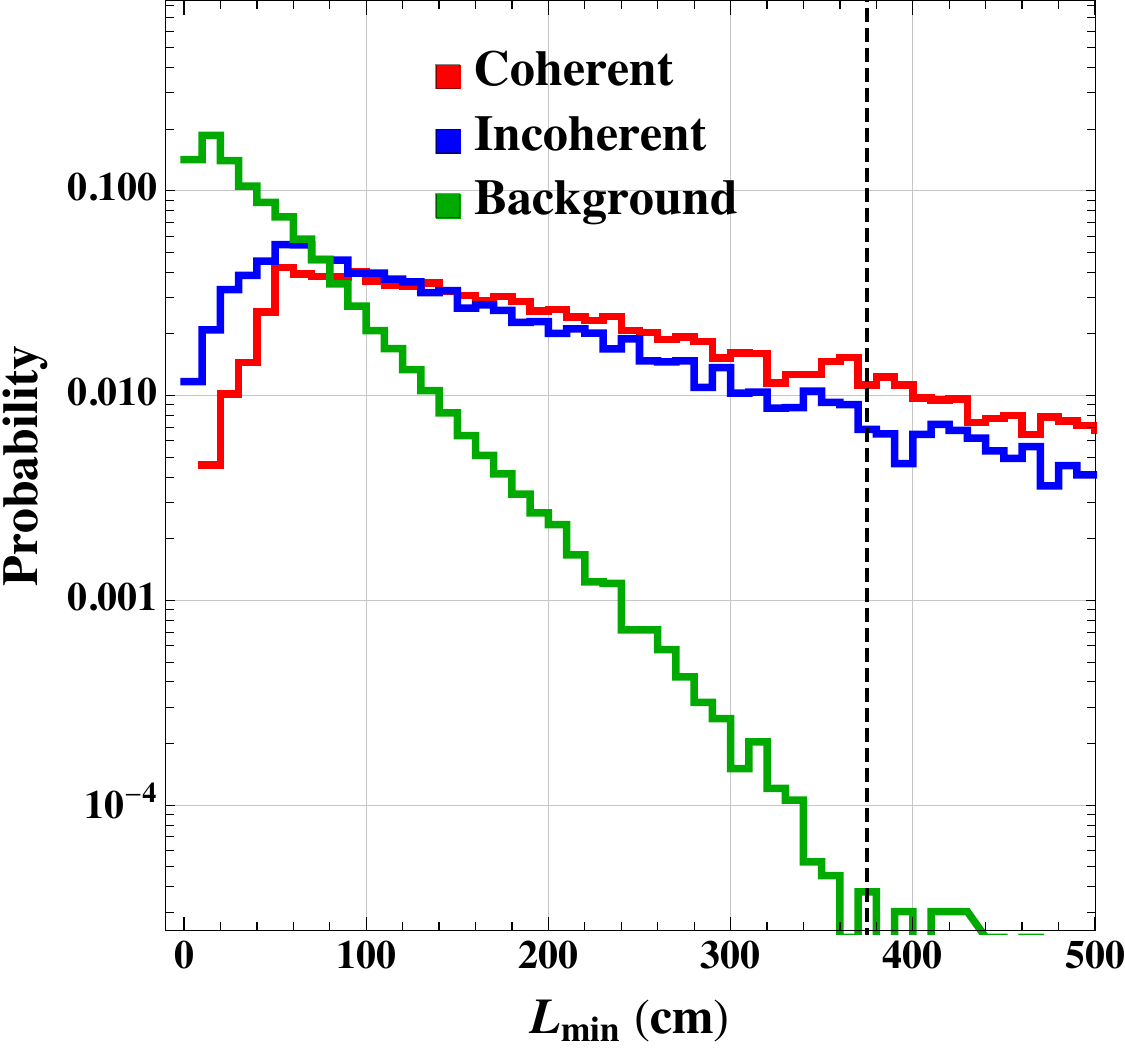}
\includegraphics[height=5cm]{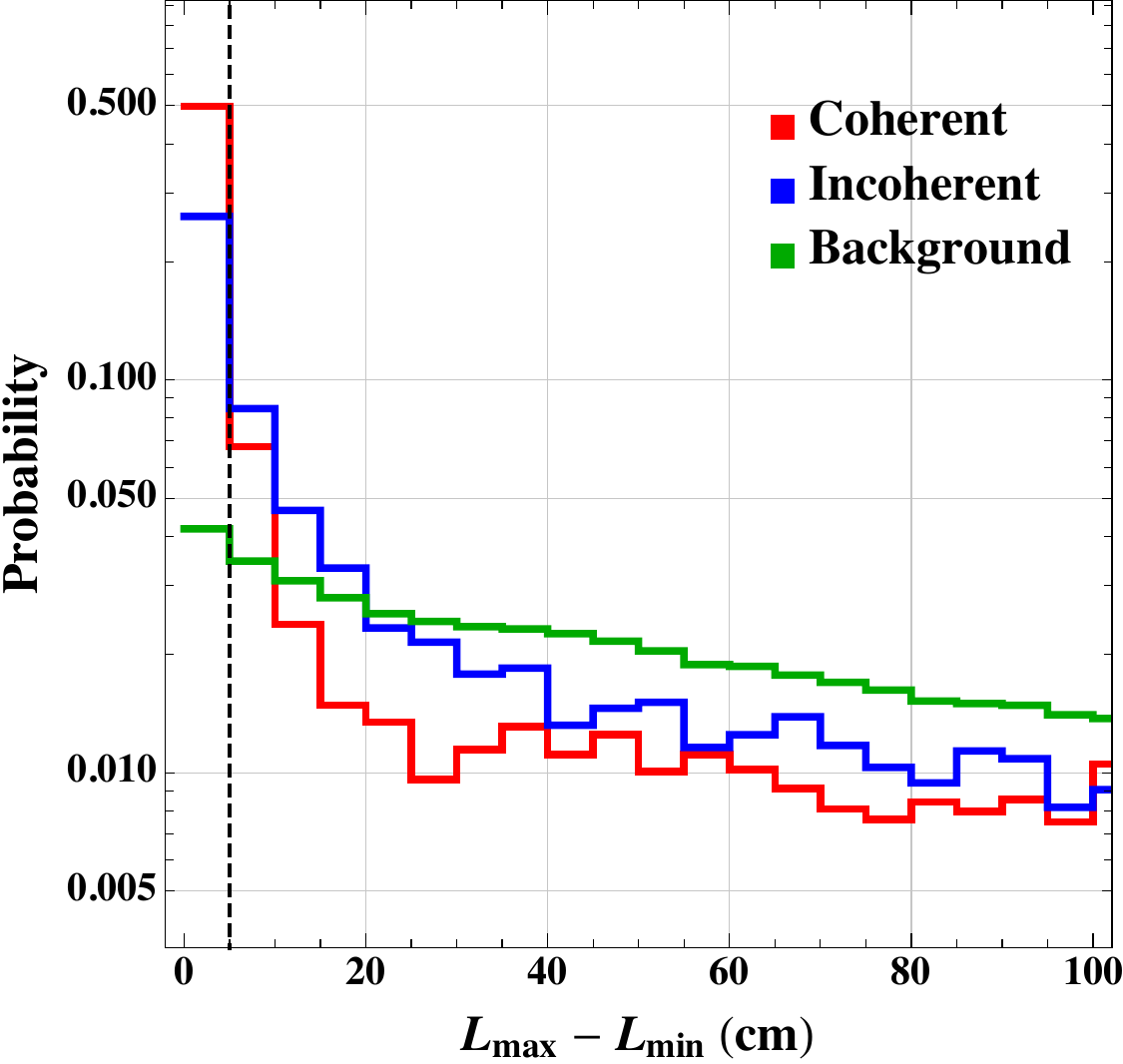}
\caption[
Signal and background 
for selecting 
muonic trident interactions in ND LArTPC]{Event kinematic distributions of signal and background considered for the selection of muonic trident interactions in the \dword{nd} \dword{lartpc}: number of tracks (top left), angle between the two main tracks (top right), length of the shortest track (bottom left), and the difference in length between the two main tracks (bottom right). The dashed, black vertical lines indicate the optimal cut values used in the analysis.} \label{fig:trident_kinematics}
\end{figure}

Figure~\ref{fig:trident_kinematics} shows the distribution (area normalized) for signal and background of the different kinematic variables used in our analysis for the discrimination between signal and background. As expected, background events tend to contain a higher number of tracks than the signal. The other distributions also show a clear discriminating power: the angle between the two tracks is typically much smaller in the signal than in the background. Moreover, the signal tracks (two muons) tend to be longer than tracks in the background (mainly one muon plus one pion).

\subsection{Sensitivity to new physics} 
\begin{figure}[tb!]
\centering
\includegraphics[width=0.4\textwidth]{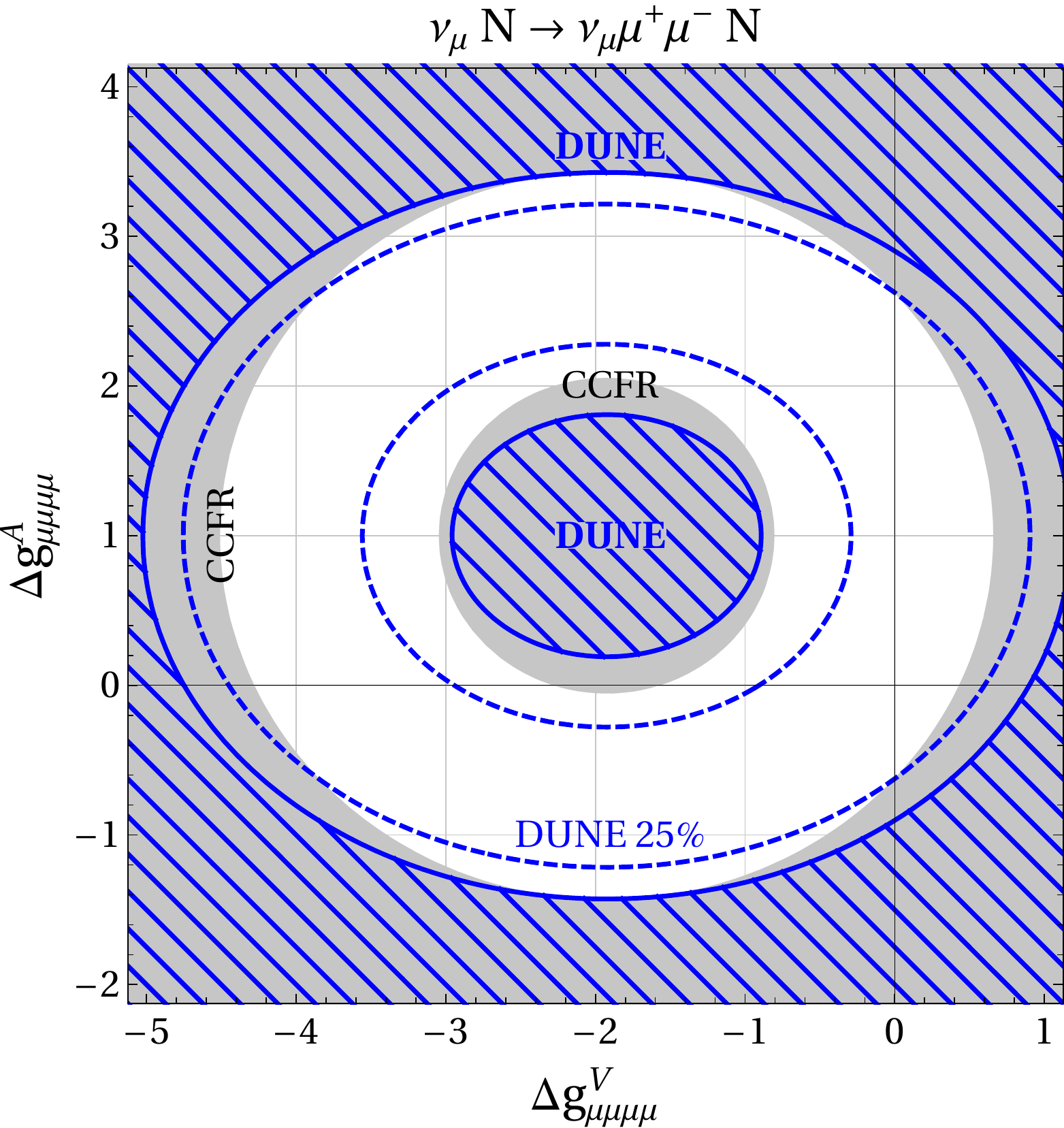}
\caption[$\nu_\mu N \to \nu_\mu \mu^+\mu^- N$ cross section at ND and (axial-)vector couplings of \numu{} to muons]
{Projected sensitivity (95\% \dword{cl}) of a measurement of the $\nu_\mu N \to \nu_\mu \mu^+\mu^- N$ cross section at the DUNE \dword{nd} to modifications of the vector and axial-vector couplings of muon-neutrinos to muons (blue hashed regions). The gray regions are excluded at 95\% \dword{cl} by existing measurements of the cross section by the CCFR collaboration. The intersection of the black lines indicates the \dword{sm} point.}
\label{fig:trident_gVgA}
\end{figure}

The sensitivity of neutrino tridents to heavy new physics (i.e., heavy compared to the momentum transfer in the process) can be parameterized in a model-independent way using a modification of the effective four-fermion interaction Hamiltonian. Focusing on the case of muon-neutrinos interacting with muons, the vector and axial-vector couplings can be written as
\begin{equation}
g_{\mu\mu\mu\mu}^V = 1 + 4 \sin^2\theta_W + \Delta g_{\mu\mu\mu\mu}^V \quad \mathrm{and} \quad g_{\mu\mu\mu\mu}^A = -1 + \Delta g_{\mu\mu\mu\mu}^A ~,
\end{equation}
where $\Delta g_{\mu\mu\mu\mu}^V$ and $\Delta g_{\mu\mu\mu\mu}^A$ parameterize possible new physics contributions. Couplings involving other combinations of lepton flavors can be modified analogously. Note, however, that for interactions that involve electrons, very strong constraints can be derived from LEP bounds on electron contact interactions~\cite{Schael:2013ita}. The modified interactions of the muon-neutrinos with muons alter the cross section of the $\nu_\mu N \to \nu_\mu \mu^+\mu^- N$ trident process. In Figure~\ref{fig:trident_gVgA} we show the regions in the $\Delta g^V_{\mu\mu\mu\mu}$ vs.\ $\Delta g^A_{\mu\mu\mu\mu}$ plane that are excluded by the existing CCFR measurement $\sigma_\text{CCFR} / \sigma_\text{CCFR}^\text{SM} = 0.82 \pm 0.28$~\cite{Mishra:1991bv} at the 95\% \dword{cl} in gray. A measurement of the $\nu_\mu N \to \nu_\mu \mu^+\mu^- N$ cross section with $40\%$ uncertainty at the DUNE \dword{nd}  could cover the blue hashed regions. Our baseline analysis does not extend the sensitivity into parameter space that is unconstrained by the CCFR measurement. However, It is likely that the use of a magnetized spectrometer, as it is being considered for the DUNE ND, able to identify the charge signal of the trident final state, along with a more sophisticated  event selection (e.g.\ deep-learning-based), will significantly improve separation between neutrino trident interactions and backgrounds. Therefore, we also present the region that could be probed by a 25\% measurement of the neutrino trident cross section at DUNE, which would extend the coverage of new physics parameter space substantially.

We consider a class of models that modify the trident cross section through the presence of an additional neutral gauge boson, $Z'$, that couples to neutrinos and charged leptons. A consistent way of introducing such a $Z'$ is to gauge an anomaly-free global symmetry of the \dword{sm}. Of particular interest is the $Z'$ that is based on gauging the difference of muon-number and tau-number, $L_\mu - L_\tau$~\cite{He:1990pn,He:1991qd}. Such a $Z'$ is relatively weakly constrained and can for example address the longstanding discrepancy between \dword{sm} prediction and measurement of the anomalous magnetic moment of the muon, $(g-2)_\mu$~\cite{Baek:2001kca,Harigaya:2013twa}. The $L_\mu - L_\tau$ $Z'$ has also been used in models to explain $B$ physics anomalies~\cite{Altmannshofer:2014cfa} and as a portal to \dword{dm}~\cite{Baek:2008nz,Altmannshofer:2016jzy}. The $\nu_\mu N \to \nu_\mu \mu^+\mu^- N$ trident process has been identified as important probe of gauged $L_\mu - L_\tau$ models over a broad range of $Z^\prime$ masses~\cite{Altmannshofer:2014cfa,Altmannshofer:2014pba}.

\begin{figure}[tb!] \centering
\includegraphics[width=0.75\textwidth]{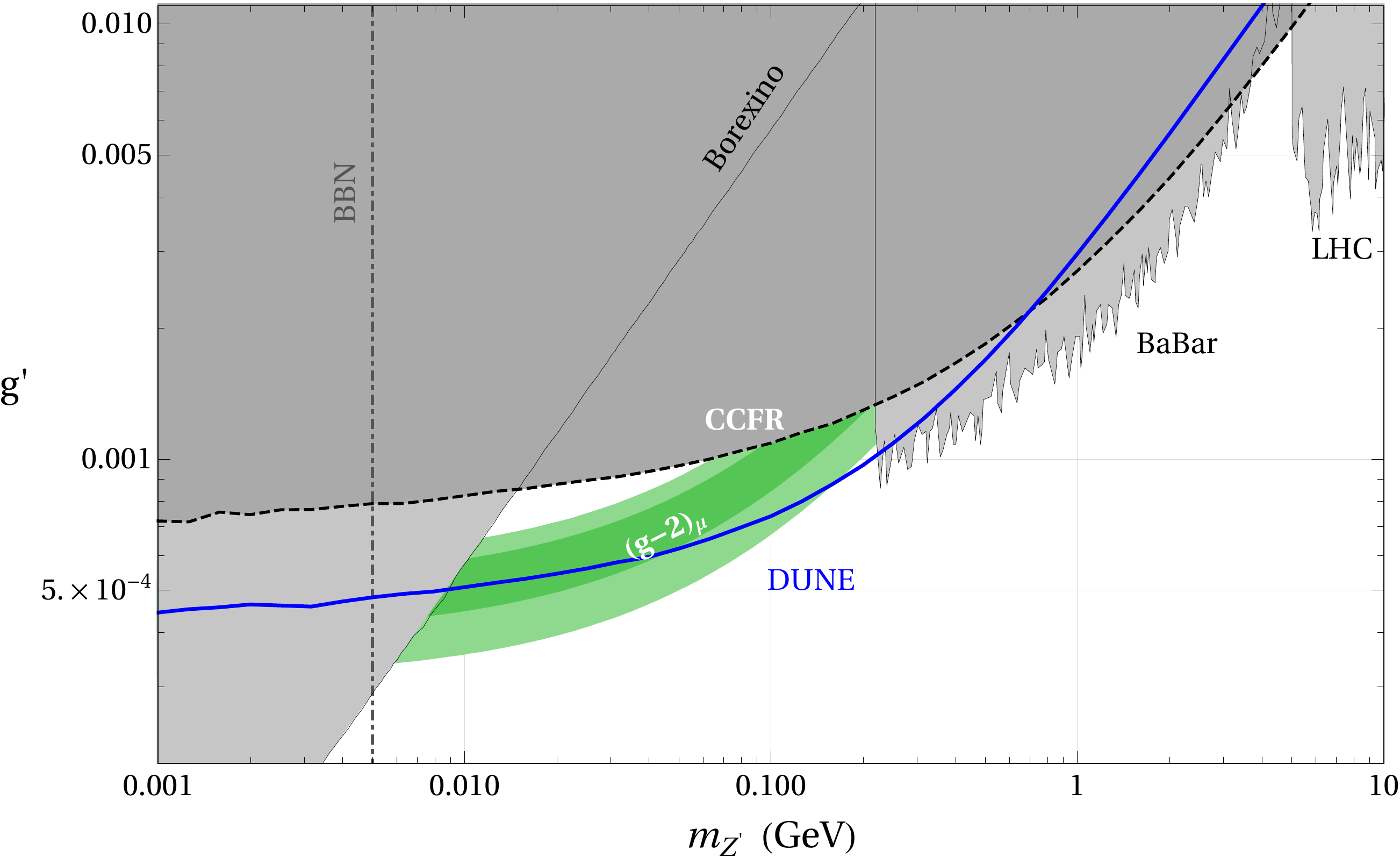}
\caption[Existing constraints and projected sensitivity in the $L_\mu - L_\tau$ parameter space]{Existing constraints and projected DUNE sensitivity in the $L_\mu - L_\tau$ parameter space. Shown in green is the region where the $(g-2)_\mu$ anomaly can be explained at the $2\sigma$ level. The parameter regions already excluded by existing constraints are shaded in gray and correspond to a CMS search for $pp \to \mu^+\mu^- Z' \to \mu^+\mu^-\mu^+\mu^-$~\cite{Sirunyan:2018nnz} (``LHC''), a BaBar search for $e^+e^- \to \mu^+\mu^- Z' \to \mu^+\mu^-\mu^+\mu^-$~\cite{TheBABAR:2016rlg} (``BaBar''), precision measurements of $Z \to \ell^+ \ell^-$ and $Z \to \nu\bar\nu$ couplings~\cite{ALEPH:2005ab,Altmannshofer:2014cfa} (``LEP''), a previous measurement of the trident cross section~\cite{Mishra:1991bv,Altmannshofer:2014pba} (``CCFR''), a measurement of the scattering rate of solar neutrinos on electrons~\cite{Bellini:2011rx,Harnik:2012ni,Agostini:2017ixy} (``Borexino''), and bounds from big bang nucleosynthesis~\cite{Ahlgren:2013wba,Kamada:2015era} (``BBN''). The DUNE sensitivity shown by the solid blue line assumes a measurement of the trident cross section with $40\%$ precision.}
\label{fig:LmuLtau}
\end{figure}

In Figure~\ref{fig:LmuLtau} we show the existing CCFR constraint on the model parameter space in the $m_{Z'}$ vs. $g'$ plane and compare it to the region of parameter space where the anomaly in $(g-2)_\mu = 2 a_\mu$ can be explained. The green region shows the $1\sigma$ and $2\sigma$ preferred parameter space corresponding to a shift $\Delta a_\mu = a_\mu^\text{exp}-a_\mu^\text{SM} = (2.71 \pm 0.73) \times 10^{-9}$~\cite{Keshavarzi:2018mgv}.
Shown are in addition constraints from LHC searches for the $Z'$ in the $pp \to \mu^+\mu^- Z' \to \mu^+\mu^-\mu^+\mu^-$ process~\cite{Sirunyan:2018nnz} (see also~\cite{Altmannshofer:2014pba}), direct searches for the $Z'$ at BaBar using the $e^+e^- \to \mu^+\mu^- Z' \to \mu^+\mu^-\mu^+\mu^-$ process~\cite{TheBABAR:2016rlg}, and constraints from LEP precision measurements of leptonic $Z$ couplings~\cite{ALEPH:2005ab,Altmannshofer:2014cfa}.  
Also a Borexino bound on non-standard contributions to neutrino-electron scattering~\cite{Harnik:2012ni,Bellini:2011rx,Agostini:2017ixy} has been used to constrain the $L_\mu - L_\tau$ gauge boson~\cite{Kamada:2015era,Araki:2015mya,Kamada:2018zxi}. Our reproduction of the Borexino constraint is shown. 
For very light $Z'$ masses of $O$(few MeV) and below, strong constraints from measurements of the effective number of relativistic degrees of freedom during big bang nucleosynthesis (BBN) apply~\cite{Ahlgren:2013wba,Kamada:2015era}.
Taking into account all relevant constraints, parameter space to explain $(g-2)_\mu$ is left below the di-muon threshold $m_{Z'} \lesssim 210$~MeV.


\section{Dark Matter Probes}\label{sec:DM}
Dark matter (\dword{dm}) is a crucial ingredient to understand the cosmological history of the universe, and the most up-to-date measurements suggests the existence of \dword{dm} with an abundance of 27\%~\cite{Aghanim:2018eyx}. 
In light of this situation, a tremendous amount of experimental effort has gone into 
the search for \dword{dm}-induced signatures, for example, \dword{dm} direct and indirect detections and collider searches. However, no ``smoking-gun'' signals have been discovered thus far while more parameter space in relevant \dword{dm} models is simply ruled out. 
It is noteworthy that most conventional \dword{dm} search strategies are designed to be sensitive to signals from the \dword{wimp}, one of the well-motivated \dword{dm} candidates, whose mass range is from a few GeV to tens of TeV. 
The null observation of \dword{dm} via non-gravitational interactions actually motivates unconventional or alternative \dword{dm} search schemes. 
One such possibility is 
a search for experimental signatures induced by boosted, hence relativistic, \dword{dm} for which 
a mass range smaller than that of the weak scale is often motivated. 

One of the possible ways to produce and then detect relativistic \dword{dm} particles can be through accelerator experiments, 
for example, neutrino beam experiments~\cite{Alexander:2016aln, Battaglieri:2017aum, LoSecco:1980nf, Acciarri:2015uup}. 
By construction, large signal statistics is expected so that this sort of search strategy can allow for significant
sensitivity to \dword{dm}-induced signals despite the feeble interaction of \dword{dm} with \dword{sm} particles. 
DUNE will perform a signal search in the relativistic scattering of \dword{ldm} at the \dword{nd}, as it is close enough to the beam source to sample a substantial level of \dword{dm} flux, assuming that \dword{dm} is produced.

Alternatively, it is possible that \dword{bdm} particles are created in the universe under non-minimal dark-sector scenarios~\cite{Agashe:2014yua,
Belanger:2011ww}, and can reach terrestrial detectors. 
For example, one can imagine a two-component \dword{dm} scenario in which a lighter component is usually a subdominant relic with direct coupling to \dword{sm} particles, while the heavier is the cosmological \dword{dm} that pair-annihilates directly to a lighter \dword{dm} pair, not to \dword{sm} particles. Other mechanisms such as semi-annihilation in which a \dword{dm} particle pair-annihilates to a lighter \dword{dm} particle and a dark sector particle that may decay away are also possible~\cite{Carlson:1992fn, Hochberg:2014dra,Huang:2013xfa,Berger:2014sqa,Kong:2014mia}.
In typical cases, the \dword{bdm} flux is not large and thus large-volume neutrino detectors are desirable 
to overcome the challenge in statistics (for an  exception, see~\cite{Cherry:2015oca, Cui:2017ytb}).

Indeed, a (full-fledged) DUNE \dword{fd} with a fiducial mass of \fdfiducialmass and quality detector performance is expected to possess competitive sensitivity to \dword{bdm} signals from various sources in the current universe such as the galactic halo~\cite{Agashe:2014yua,
Alhazmi:2016qcs,Kim:2016zjx,Giudice:2017zke,Chatterjee:2018mej,Kim:2018veo}, the sun~\cite{Huang:2013xfa,Berger:2014sqa,Kong:2014mia,Kim:2018veo}, and dwarf spheroidal galaxies~\cite{Necib:2016aez}.
Furthermore, the \dword{protodune} detectors 
are operational, and we anticipate preliminary studies with their cosmic data. Interactions of \dword{bdm} with electrons~\cite{Agashe:2014yua} 
and with hadrons (protons)~\cite{Berger:2014sqa}, were investigated for Cherenkov detectors, such as \superk, which recently published a dedicated search for \dword{bdm} in the electron channel~\cite{Kachulis:2017nci}. However, in such detectors the \dword{bdm} signal rate is shown to often be significantly attenuated due to Cherenkov threshold, in particular for hadronic channels.  \lar detectors, such as DUNE's, have the potential to greatly improve the sensitivity for \dword{bdm} compared to Cherenkov detectors. This is due to improved particle identification techniques, as well as a significantly lower energy threshold for proton detection. Earlier studies have shown an improvement with DUNE for\dword{bdm}-electron interaction~\cite{Necib:2016aez}.

\subsection{Benchmark Dark Matter Models}
\label{sec:model}

The benchmark ``\dword{dm} models'' defined in this section describe only couplings of dark-sector states including \dword{ldm} particles.
We consider two example models: i) a vector portal-type scenario where a (massive) dark-sector photon $V$ mixes with the \dword{sm} photon and ii) a leptophobic $Z'$ scenario.
The former is used in Sections~\ref{sec:ND} and \ref{sec:FD}, while the latter features in Section~\ref{sec:FDsun}.
\dword{dm} and other dark-sector particles are assumed to be fermionic for convenience.

\paragraph{Benchmark Model i)}

The relevant interaction Lagrangian is given by
\bea
\mathcal{L}_{\rm int} \supset -\frac{\epsilon}{2}V_{\mu\nu}F^{\mu\nu}+g_{11} \bar{\chi}_1\gamma^\mu \chi_1 V_\mu+g_{12} \bar{\chi}_2\gamma^\mu \chi_1 V_\mu +h.c.\,, 
\label{eq:lagrangian}
\eea
where $V^{\mu\nu}$ and $F^{\mu\nu}$ are the field strength tensors for the dark-sector photon and the \dword{sm} photon, respectively. 
Here we have introduced the kinetic mixing parameter $\epsilon$, while $g_{11}$ and $g_{12}$ parameterize the interaction strengths for flavor-conserving (second operator) and flavor-changing (third operator) couplings, respectively.  
Here $\chi_1$ and $\chi_2$ denote a dark matter particle and a heavier, \textit{un}stable dark-sector state, respectively (i.e., $m_{\chi_2}>m_{\chi_1}$), and the third term allows (boosted) $\chi_1$ to up-scatter to this $\chi_2$ (i.e., an ``inelastic'' scattering process).

This model introduces five new free parameters that may be varied for our sensitivity analysis: dark photon mass $m_V$, \dword{dm} mass $m_{\chi_1}$, heavier dark-sector state mass $m_{\chi_2}$, kinetic mixing parameter $\epsilon$, dark-sector diagonal coupling $\alpha_{11} =g_{11}^2/(4\pi)$, and dark-sector off-diagonal coupling $\alpha_{12} =g_{12}^2/(4\pi)$. 
We shall perform our analyses with some of the parameters fixed to certain values for illustration.

\paragraph{Benchmark Model ii)}
This model employs a leptophobic $Z^\prime$ mediator for interactions with the nucleons. The interaction lagrangian for this model is
\bea
\mathcal{L}_{\rm int} \supset - g_{\rm Z^\prime} \sum_f Z^\prime_\mu \bar{q}_f \gamma^\mu \gamma^5 q_f - g_{\rm Z^\prime} Z^\prime_\mu \bar{\chi} \gamma^\mu \gamma^5 \chi - Q_\psi g_{\rm Z^\prime} Z^\prime_\mu \bar{\psi} \gamma^\mu \gamma^5 \psi. 
\label{eq:zprimelag}
\eea
Here, all couplings are taken to be axial. $f$ denotes the quark flavors in the \dword{sm} sector. The dark matter states are denoted by $\chi$ and $\psi$ with $m_\chi < m_\psi$. The coupling $g_{\rm Z^\prime}$ and the masses of the dark matter states are free parameters. $Q_\psi$ is taken to be less than 1 and determines the abundance of dark matter in the universe. The hadronic interaction model study presented here is complementary to and has different phenomenology compared to others such as Benchmark Model i).
The study of this benchmark model and the result are discussed in Section~\ref{sec:FDsun}.

\subsection{Search for Low-Mass Dark Mater at the Near Detector} \label{sec:ND}
\subsubsection{Dark Matter Production and Detection}
\label{sec:DMProd}

\begin{figure}[t]
\centering
 \includegraphics[width=0.30\linewidth]{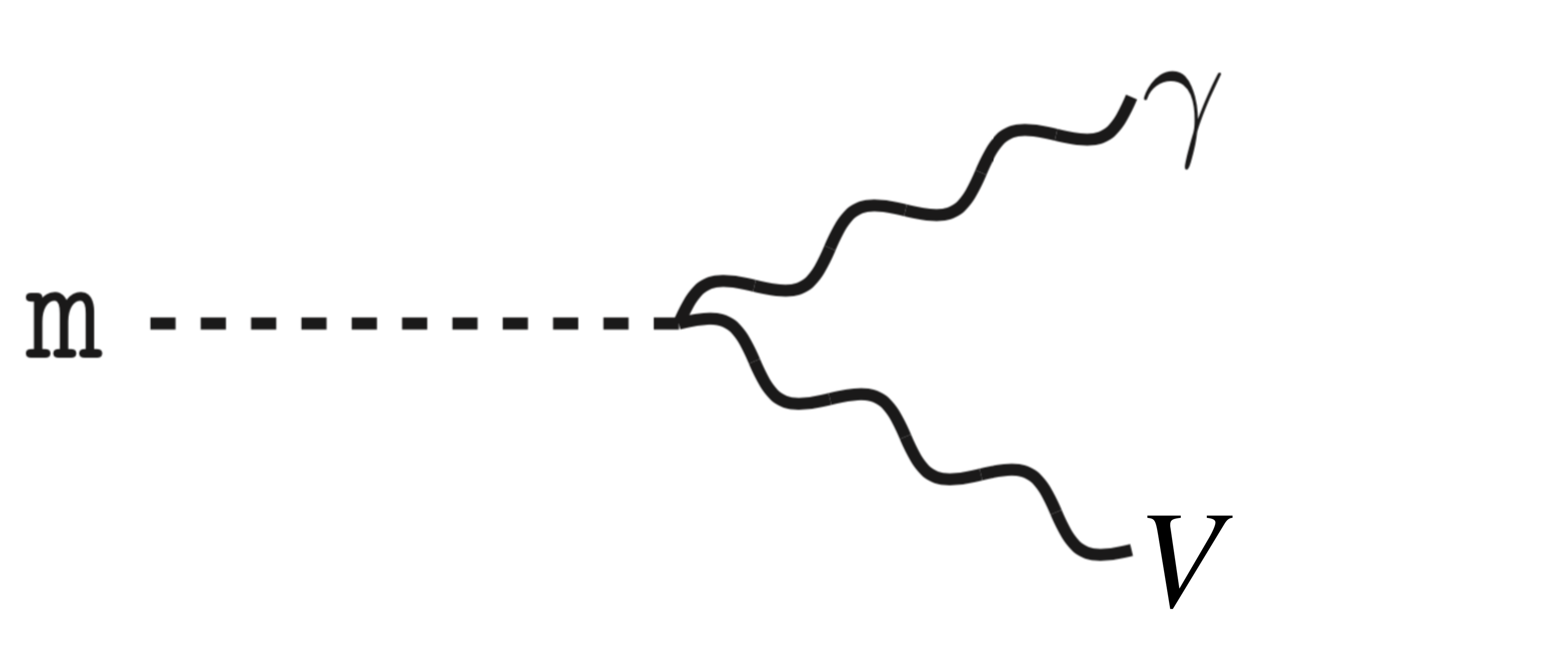}
    \includegraphics[width=0.30\linewidth]{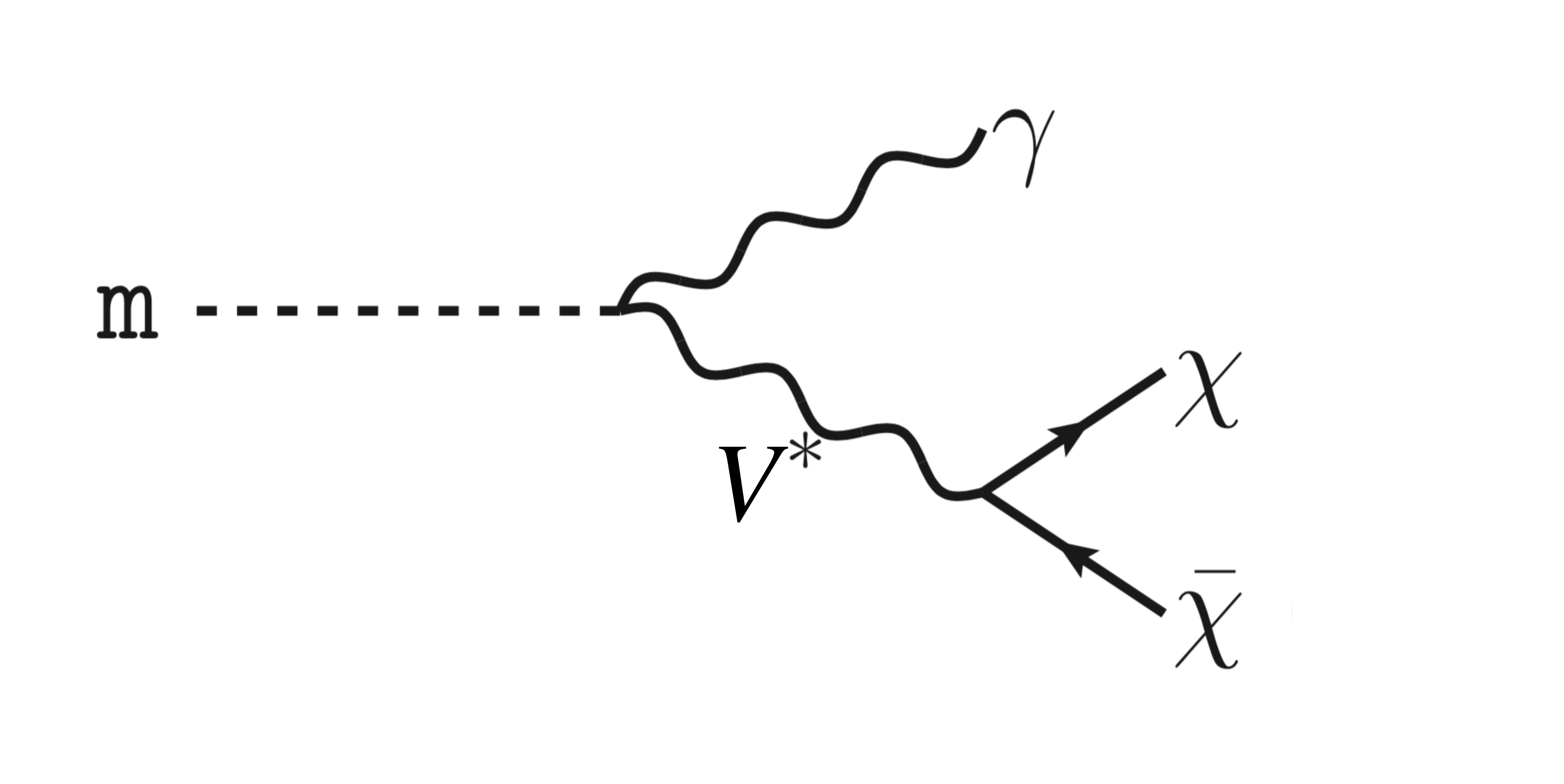}
\includegraphics[width=0.30\textwidth]{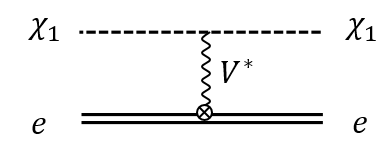}
\caption[DM production via meson decays and DM-e$^-$ elastic scattering]{ 
Production of fermionic \dword{dm} via two-body pseudoscalar meson decay $\mathfrak{m} \to \gamma V$, when $M_{V} < m_\mathfrak{m}$ (left) or via three-body decay $\mathfrak{m} \to \gamma \chi \overline{\chi}$ (center) and \dword{dm}-electron elastic scattering (right panel).}
\label{fig:dm_prod}
\end{figure}

Here, we focus on Benchmark Model i) from Eq.~(\ref{eq:lagrangian}), specifically where only one \dword{dm} particle $\chi \equiv \chi_1$ exists. We also define the dark fine structure constant $\alpha_D \equiv g_{11}^2/4\pi$. We assume that $\chi$ is a fermionic thermal relic -- in this case, the \dword{dm}/dark photon masses and couplings will provide a target for which the relic abundance matches the observed abundance in the universe. Here, the largest flux of dark photons $V$ and \dword{dm} to reach the DUNE \dword{nd} will come from the decays of light pseudoscalar mesons (specifically $\pi^0$ and $\eta$ mesons) that are produced in the DUNE target, as well as proton bremsstrahlung processes $p + p \to p + p + V$.
For the entirety of this analysis, we will fix $\alpha_D = 0.5$ and assume that the DM mass $M_{\chi}$ is lighter than half the mass of a pseudoscalar meson $\mathfrak{m}$ that is produced in the DUNE target. In this scenario, $\chi$  is produced via two decays, those of on-shell $V$ and those of off-shell $V$. This production is depicted in Figure~\ref{fig:dm_prod}. 

The flux of \dword{dm} produced via meson decays -- via on-shell $V$ -- may be estimated by\footnote{See Ref.~\cite{DeRomeri:2019kic} for a complete derivation of these expressions, including those for meson decays via off-shell $V$.}
\begin{equation}
    N_\chi = 2 N_\mathrm{POT} c_\mathfrak{m} \mathrm{Br}(\mathfrak{m}\to \gamma\gamma) \left[ 2 \varepsilon^2 \left(1 - \frac{M_{V}^2}{m_\mathrm{m}^2}\right)^3\right] \mathrm{Br}(V \to \chi\bar{\chi}) g(M_\chi, M_{V}),
\end{equation}
where $N_\mathrm{POT}$ is the number of protons on target delivered by the beam, $c_\mathfrak{m}$ is the average number of meson $\mathfrak{m}$ produced per POT, the term in braces is the relative branching fraction of $\mathfrak{m} \to \gamma V$ relative to $\gamma\gamma$, and $g(x, y)$ characterizes the geometrical acceptance fraction of \dword{dm} reaching the DUNE \dword{nd}. $g(x, y)$ is determined given model parameters using Monte Carlo techniques. For the range of dark photon and \dword{dm} masses in which DUNE will set a competitive limit, the \dword{dm} flux due to meson decays will dominate over the flux due to proton bremsstrahlung. Considering \dword{dm} masses in the $\sim$1-300 MeV range, this will require production via the $\pi^0$ and $\eta$ mesons. Our simulations using {\sc Pythia} determine that $c_{\pi^0} \approx 4.5$ and $c_\eta \approx 0.5$.

If the \dword{dm} reaches the near detector, it may scatter elastically off nucleons or electrons in the detector, via a $t$-channel dark photon. Due to its smaller backgrounds, we focus on scattering off electrons, depicted in the right panel of Figure~\ref{fig:dm_prod}. The differential cross section of this scattering, as a function of the recoil energy of the electron $E_e$, is
\begin{equation}
\frac{d\sigma_{{\chi}e}}{dE_{e}} 
= 4\pi \epsilon^{2}\alpha_D\alpha_{EM} \frac{2m_{e}E_{\chi}^{2} - (2m_{e}E_{\chi} + m_{\chi}^{2})(E_e-m_{e})}{(E_e^{2}-m_{\chi}^{2})(m_{V}^{2}+2m_{e}E_{e}-2m_{e}^{2})^{2}}\,,
\end{equation}
where $E_{\chi}$ is the incoming \dword{dm} $\chi$ energy. The signal is an event with only one recoil electron in the final state. We may use the scattering angle and energy of the electron to distinguish between signal and background (discussed in the following) events.

\subsubsection{Background Considerations}
 The background to the process shown in the right panel of Figure~\ref{fig:dm_prod} consists of any processes involving an electron recoil. As the \dword{nd} is located near the surface, background events, in general, can be induced by cosmic rays as well as by neutrinos generated from the beam. Since majority of cosmic-induced, however, will be vetoed by triggers and timing information, the dominant background will be from neutrinos coming in the DUNE beam.

The two neutrino-related backgrounds are $\nu_\mu -e^-$ scattering, which looks nearly identical to the signal, and $\nu_e$ CCQE scattering, which does not. The latter has a much larger rate ($\sim$ 10 times higher) than the former, however, we expect that using the kinematical variable $E_e \theta_e^2$ of the final state, where $\theta_e$ is the direction of the outgoing electron relative to the beam direction, will allow the $\nu_e$ CCQE background to be vetoed effectively.

While spectral information regarding $E_e$ could allow a search to distinguish between $\chi e$ and $\nu_\mu e$ scattering, we expect that uncertainties in the $\nu_\mu$ flux (both in terms of overall normalization and shape as a function of neutrino energy) will make such an analysis very complicated. For this reason, we include a normalization uncertainty of $10\%$ on the expected background rate and perform a counting analysis. Studies are ongoing to determine how such an analysis may be improved.

For this analysis we have assumed $3.5$ years of data collection each in neutrino and antineutrino modes, analyzing events that occur within the fiducial volume of the DUNE near detector. We compare results assuming either all data is collected with the ND on-axis, or data collection is divided equally among all off-axis positions, $0.7$ yr at each position  $i$, between $0$ and $24$ m transverse to the beam direction (in steps of 6 meters).
We assume three sources of uncertainty: statistical, correlated systematic, and an uncorrelated systematic in each bin. 
For a correlated systematic uncertainty, we include a nuisance parameter $A$ that modifies the number of neutrino-related background events in all bins -- an overall normalization uncertainty across all off-axis locations. 
We further include an additional term in our test statistic for $A$, a  Gaussian probability with width $\sigma_A = 10\%$. 
We also include an uncorrelated uncertainty in each bin, which we assume to be much narrower than $\sigma_A$. 
We assume this uncertainty to be parameterized by a Gaussian with width $\sigma_{f_i} = 1\%$. 
After marginalizing over the corresponding uncorrelated nuisance parameters, the test statistic reads
\begin{eqnarray}\label{eq:chisqfull}
-2\Delta \mathcal{L} = \sum_i \frac{r_i^m\left( \left(\frac{\varepsilon}{\varepsilon_0}\right)^4 N_i^\chi + (A-1)N_i^\nu\right)^2}{A\left(N_i^\nu + (\sigma_{f_i} N_i^\nu)^2 \right)} + \frac{\left(A-1\right)^2}{\sigma_A^2}.
\end{eqnarray}

In Eq.~(\ref{eq:chisqfull}), $N_i^\chi$ is the number of \dword{dm} scattering events, calculated assuming $\varepsilon$ is equal to some reference value $\varepsilon_0 \ll 1$. $N_i^\nu$ is the number of $\nu_\mu e^-$ scattering events expected in detector position $i$, and $r_i^m$ is the number of years of data collection in detector position $i$ during beam mode $m$ (neutrino or antineutrino mode). If data are only collected on-axis, then this test statistic will be dominated by the systematic uncertainty associated with $\sigma_A$. If on- and off-axis measurements are combined, then the resulting sensitivity will improve significantly.

\subsubsection{Sensitivity Calculation and Results}

\begin{figure}[t]
 \centering
 \includegraphics[width=\linewidth]{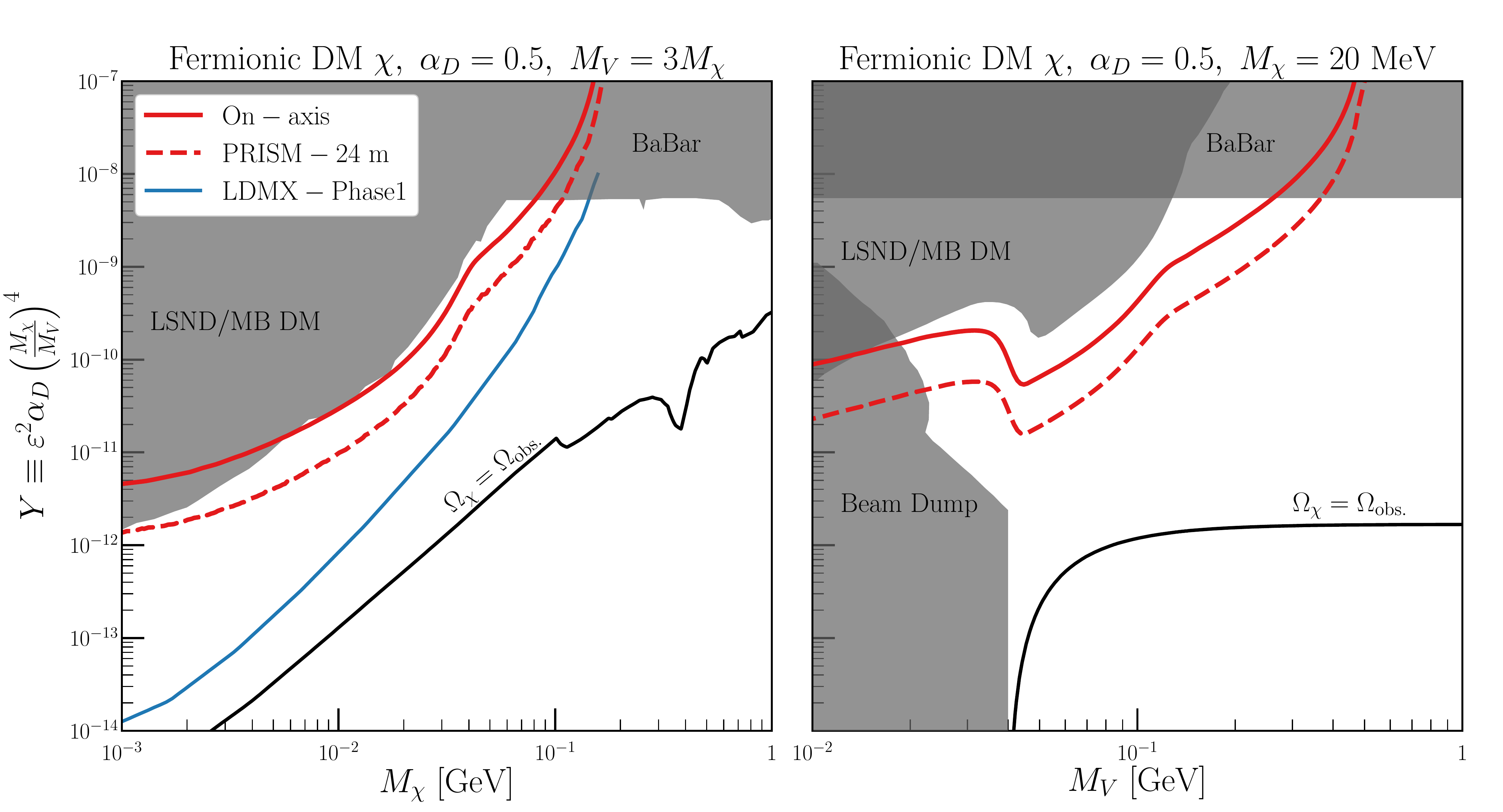}
 \caption[90$\%$ \dword{cl} limit for Y as a function of $m_{\chi}$ at the ND]{\label{fig:chisq} Expected DUNE On-axis (solid red) and PRISM (dashed red) sensitivity using $\chi e^- \to \chi e^-$ scattering. We assume $\alpha_D = 0.5$ in both panels, and $M_V = 3M_\chi$ ($M_\chi = 20$ MeV) in the left (right) panel, respectively. Existing constraints are shown in grey, and the relic density target is shown as a black line. We also show for comparison the sensitivity curve expected for LDMX-Phase I (solid blue)~\cite{Akesson:2018vlm}.
 }
 \end{figure}

We compute the expected DUNE sensitivity assuming all data collected with the ND on-axis (DUNE On-axis) or equal times at each ND off-axis position (DUNE-PRISM). We present results in terms of the \dword{dm} or dark photon mass and the parameter $Y$, where
\begin{equation}
Y \equiv \varepsilon^2 \alpha_D \left(\frac{M_\chi}{M_V}\right)^4.    
\end{equation}
Assuming $M_V \gg M_\chi$, this parameter determines the relic abundance of \dword{dm} in the universe today, and sets a theoretical goal in terms of sensitivity reach. We present the 90\% CL sensitivity reach of the DUNE \dword{nd} in Figure~\ref{fig:chisq}. 
We assume $\alpha_D = 0.5$ in our simulations and we display the results fixing $M_V = 3M_\chi$ (left panel) and $M_\chi = 20$ MeV (right panel).
We also compare the sensitivity reach of this analysis with other existing experiments, shown as grey shaded regions. We further show for comparison the sensitivity curve expected for a proposed dedicated experiment to search for \dword{ldm}, LDMX-Phase I~\cite{Akesson:2018vlm} (solid blue).

 From our estimates, we see that DUNE can significantly improve the constraints from LSND~\cite{deNiverville:2018dbu} and the MiniBooNE-DM search~\cite{Aguilar-Arevalo:2018wea}, as well as BaBar~\cite{Lees:2017lec} if $M_V \lesssim 200$ MeV. We also show limits in the right panel from beam-dump experiments (where the dark photon is assumed to decay visibly if $M_V < 2 M_\chi$)~\cite{Davier:1989wz,Batley:2015lha,Bjorken:1988as,Riordan:1987aw,Bjorken:2009mm,Bross:1989mp}, as well as the lower limits obtained from matching the thermal relic abundance of $\chi$ with the observed one (black).

The features in the sensitivity curve in the right panel can be understood by looking at the DM production mechanism.
For a fixed $\chi$ mass, as $M_V$ grows, the DM production goes from off-shell to on-shell and back to off-shell. The first transition explains the strong feature near $M_V=2M_\chi = 40$~MeV, while the second is the source for the slight kink around $M_V=m_{\pi^0}$ (which appears also in the left panel).

\subsection{Inelastic Boosted Dark Matter Search at the DUNE FD 
\label{sec:FD}}

\subsubsection{\dshort{bdm} Flux from the Galactic Halo \label{sec:flux}}

As we mentioned in Section~\ref{phys:bsm:execsumm}, 
we look at an annihilating two-component \dword{dm} scenario~\cite{Belanger:2011ww} in this study. 
The heavier \dword{dm} (denoted $\chi_0$) plays a role of cosmological \dword{dm} and pair-annihilates to a pair of lighter \dword{dm} particles (denoted $\chi_1$) in the universe today. 
The expected flux near the Earth is given by~\cite{Agashe:2014yua,
Giudice:2017zke, Kim:2018veo}
\bea 
\mathcal{F}_1= & 1.6 \times 10^{-6} {\rm cm}^{-2}{\rm s}^{-1}\times \left( \frac{\langle \sigma v\rangle_{0\rightarrow 1}}{5\times 10^{-26}{\rm cm}^3{\rm s}^{-1}}\right) 
 \times \left( \frac{10\, {\rm GeV}}{m_{\chi_0}}\right)^2\,,
\label{eq:flux}
\eea
where $m_{\chi_0}$ is the mass of $\chi_0$ and $\langle \sigma v\rangle_{0\rightarrow 1}$ stands for the velocity-averaged annihilation cross section of $\chi_0\bar{\chi}_0 \to \chi_1\bar{\chi}_1$ in the current universe.
To evaluate the reference value shown as the first prefactor, we take $m_{\chi_0} = 10$ GeV and $\langle \sigma v\rangle_{0\rightarrow 1}=5\times 10^{-26}{\rm cm}^3{\rm s}^{-1}$, the latter of which is consistent with the current observation of \dword{dm} relic density assuming $\chi_0$ and its anti-particle $\bar{\chi}_0$ are distinguishable. 
To integrate all relevant contributions over the entire galaxy, we assume the Navarro-Frenk-White (NFW) \dword{dm} halo profile~\cite{Navarro:1995iw, Navarro:1996gj}.
In this section we assume the \dword{bdm} flux with a $m_{\chi_0}$ dependence given by Eq.~(\ref{eq:flux}) for the phenomenological analysis.

\subsubsection{Experimental Signatures}

\begin{figure}[t]
\centering
\includegraphics[width=9cm]{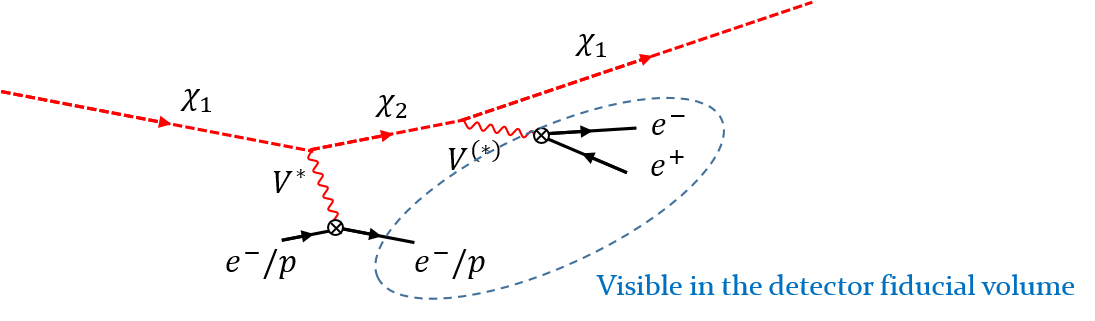}
\caption{\label{fig:sig} The inelastic BDM signal under consideration.}
\end{figure}

The \dword{bdm} that is created, e.g., at the galactic center, reaches the DUNE \dword{fd} 
detectors and scatters off either electrons or protons energetically. 
In this study, we focus on electron scattering signatures for illustration, under Benchmark Model i) defined in Eq.~\eqref{eq:lagrangian}. 
The overall process is summarized as follows:
\bea 
\chi_1 + e^- \to e^- + \chi_2 (\to \chi_1 + V^{(*)} \to \chi_1 + e^+ +e^-)\,,
\eea
and a diagrammatic description is shown in Figure~\ref{fig:sig} where 
particles visible by the detector are circled in blue. 
In the final state, there exist three visible particles that usually leave sizable ($e$-like) tracks in the 
detectors.  
Note that we can replace $e^-$ in the left-hand side and the first $e^-$ in the right-hand side of the above process to $p$ for the $p$-scattering case.
In the basic model, Eq.~\eqref{eq:lagrangian}, and given the source of \dword{bdm} at the galactic center,  the primary signature is quasi-elastic proton recoiling~\cite{pscattering} in this case.

\subsubsection{Background Estimation}

As we have identified a possible $i$\dword{bdm} signature, we are now in a position to discuss potential \dword{sm} background events.

For the DUNE \dwords{detmodule} located $\sim 1480$ m deep underground, the cosmic-induced background discussed earlier is not an issue. 
The most plausible scenario for background production is the creation of multiple pions that subsequently decay to electrons, positrons, and neutrinos. 
Relevant channels are the resonance production and/or \dword{dis} by the \dword{cc} $\nu_e$ or $\bar \nu_e$ scattering with a nucleon in the \lar target.
Summing up all the resonance production and \dword{dis} events that are not only induced by $\nu_e$ or $\bar \nu_e$ 
but relevant to production of a few pions, we find that the total number of multi-pion production events is at most $\sim 12$ kt$^{-1}$yr$^{-1}$ based on the neutrino flux in Ref.~\cite{Honda:2015fha} and the cross section in Ref.~\cite{Formaggio:2013kya}.
In addition, the charged pions often leave appreciable tracks inside the detector so that the probability of misidentifying the $e^\pm$ from the decays of $\pi^\pm$ with the \textit{i}\dword{bdm} signal events would be very small.
Hence, we conclude that it is fairly reasonable to assume that almost no background events exist.

\subsubsection{Phenomenology}\label{Sec:Pheno}

We finally present the expected experimental sensitivities at 
DUNE, in the searches for $i$\dword{bdm}. 
We closely follow the strategies illustrated in Refs.~\cite{Giudice:2017zke, Chatterjee:2018mej, Kim:2018veo} to represent phenomenological interpretations. 

\begin{figure}[t]
\centering
\includegraphics[width=6cm]{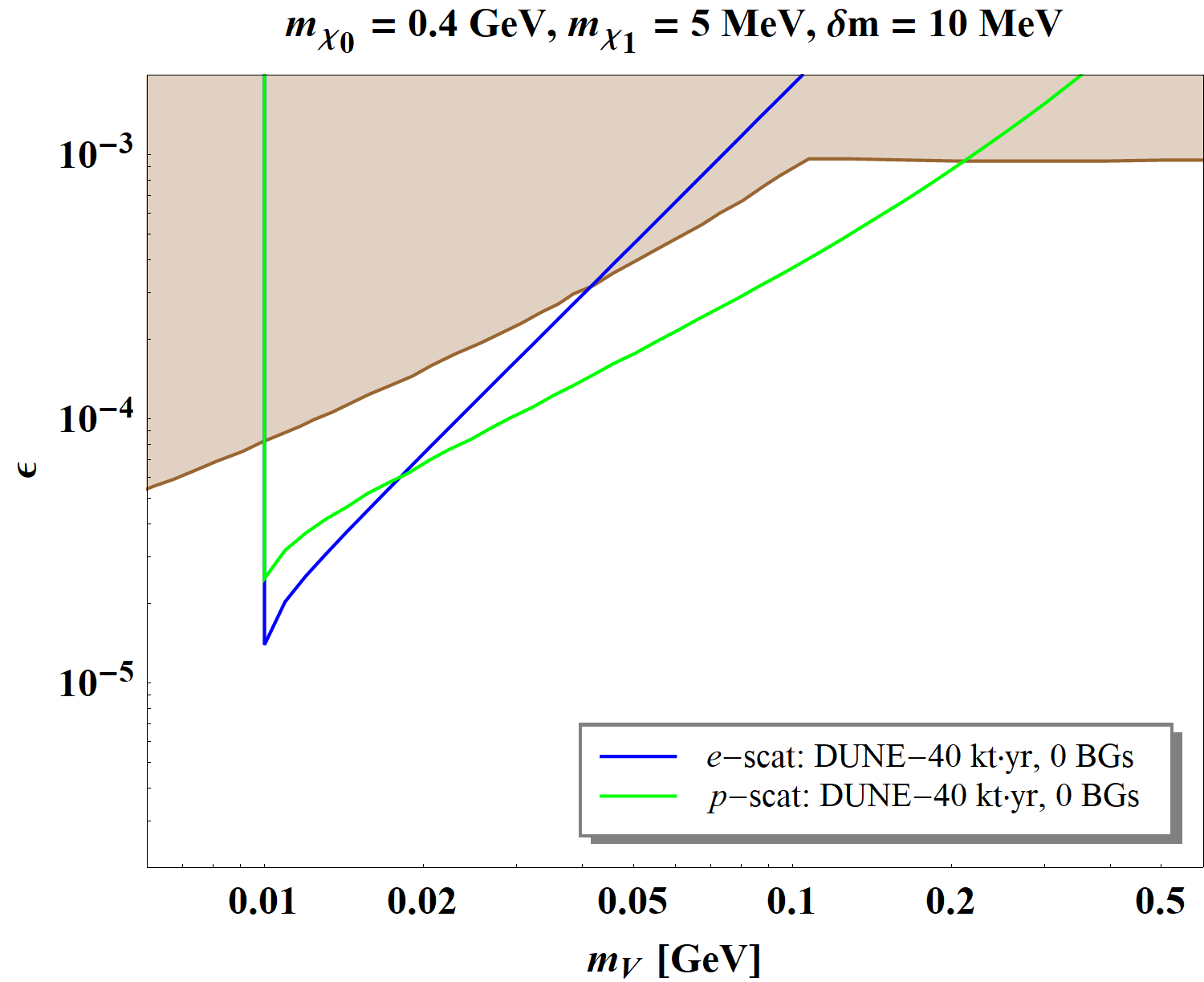} 
\includegraphics[width=6cm]{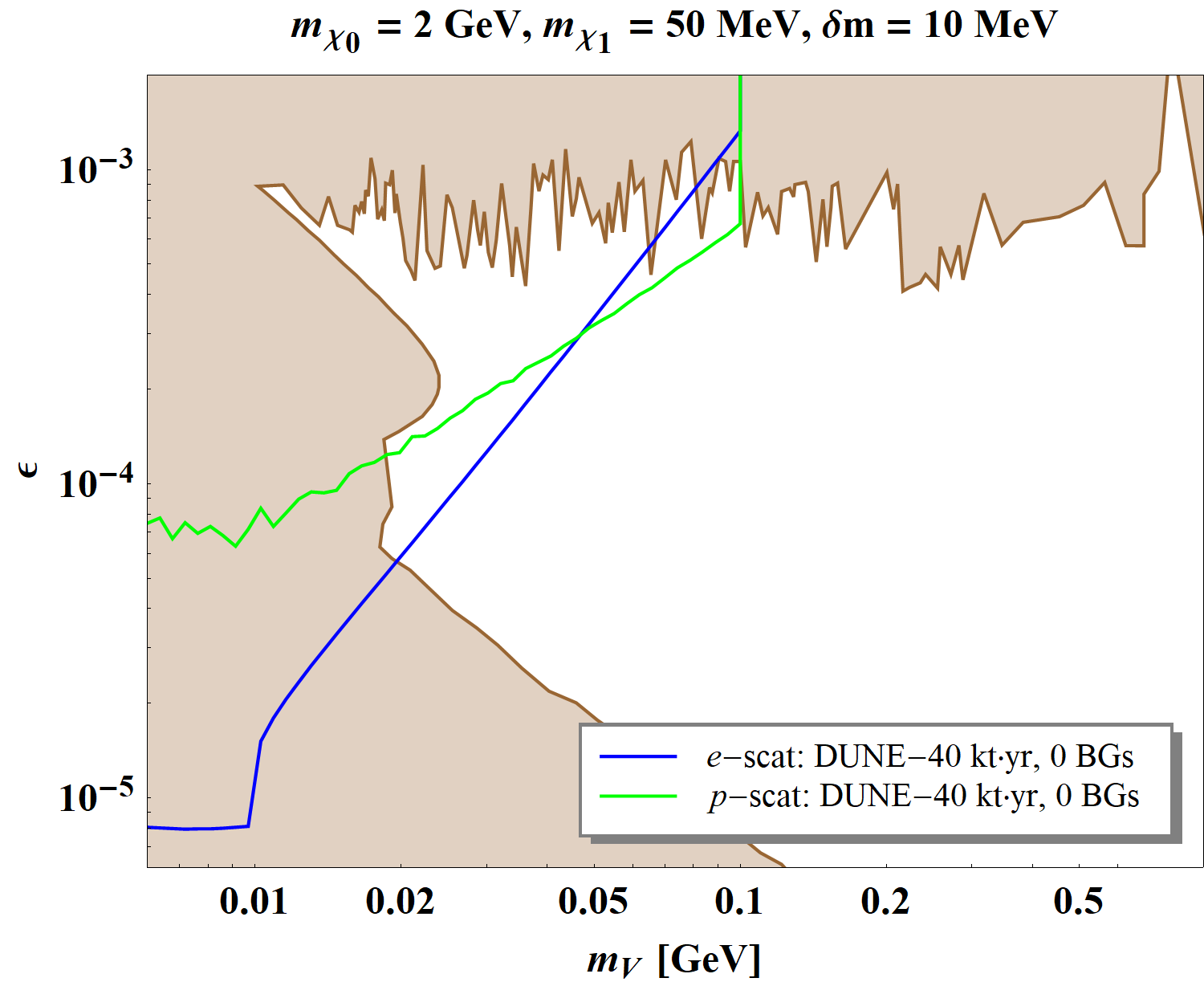} \\
\vspace{0.3cm}
\includegraphics[width=6cm]{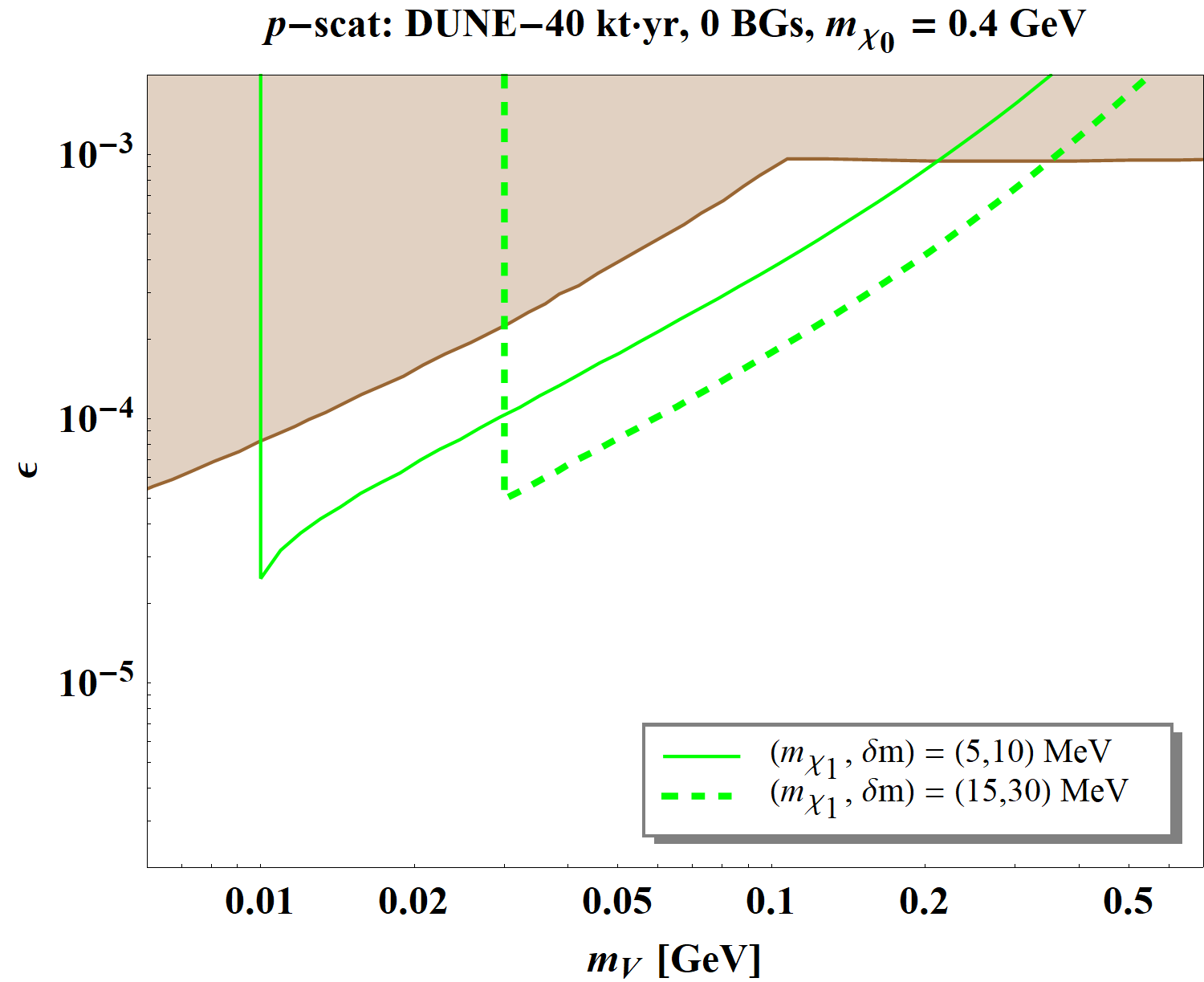}
\includegraphics[width=6cm]{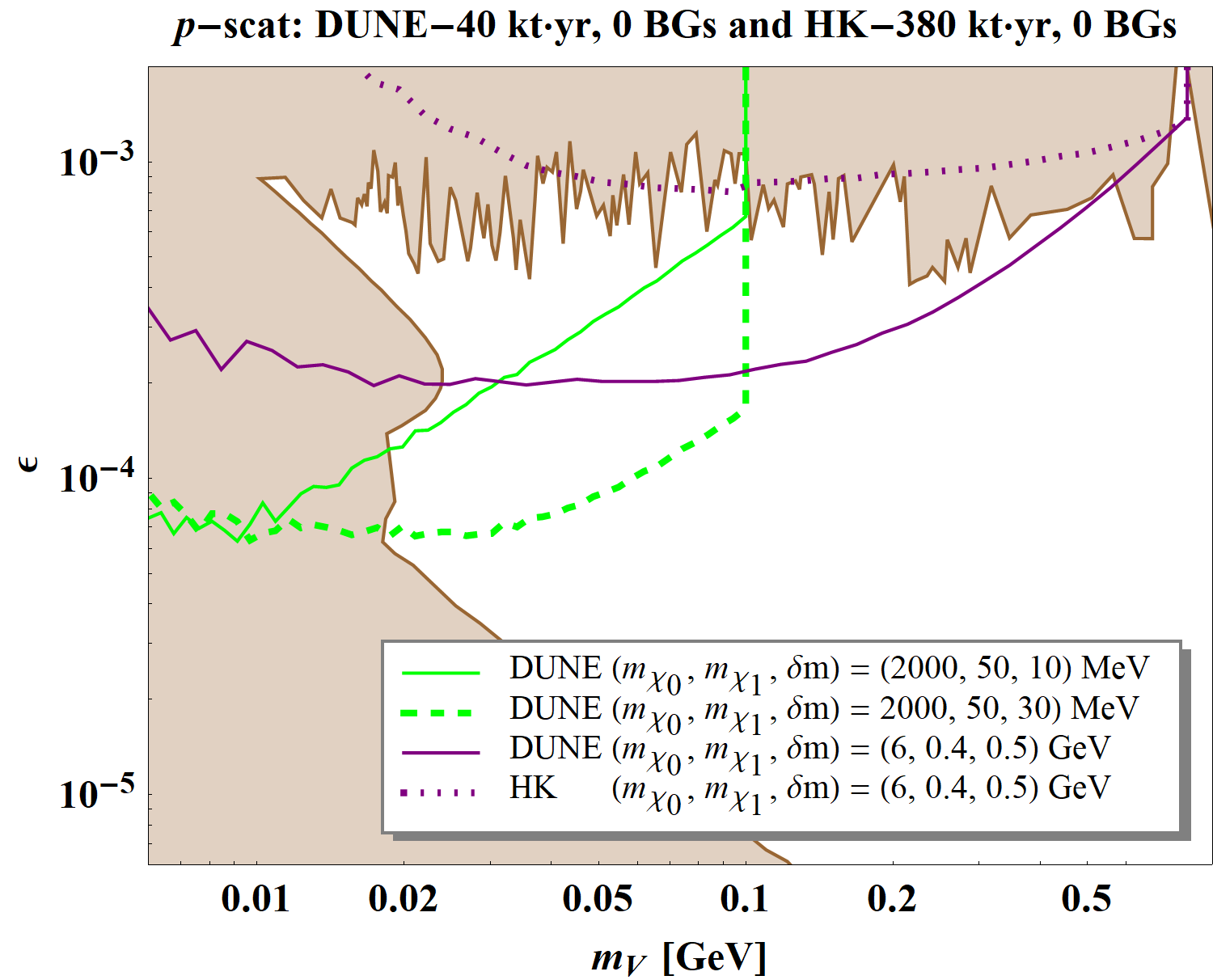}
\caption[Experimental sensitivities for $m_{\chi_n}$ values in terms of $m_V - \epsilon$]{
The experimental sensitivities in terms of reference model parameters $m_V - \epsilon$ 
for $m_{\chi_0} = 0.4$ GeV, $m_{\chi_1} = 5$ MeV, and $\delta m = m_{\chi_2} - m_{\chi_1} = 10$ MeV (upper-left panel) and $m_{\chi_0} = 2$ GeV, $m_{\chi_1} = 50$ MeV, and $\delta m = 10$ MeV (upper-right panel).
The left panels are for Scenario 1 and the right ones are for Scenario 2.
The lower panels compare different reference points in the $p$-scattering channel.
See the text for the details.
\label{fig:darkphotonparameter} }
\end{figure}

In displaying the results, we separate the signal categories into 
\begin{itemize}
\item Scenario 1: $m_V > 2 m_{\chi_1}$, experimental limits for $V \to$ invisible  applied.
\item Scenario 2: $m_V \le 2 m_{\chi_1}$, experimental limits for $V \to e^+ e^-$ invisible  applied.
\end{itemize}

The brown-shaded region shows the latest limits set by various experiments such as the fixed-target experiment NA64 at the CERN SPS and the B-factory experiment BaBar~\cite{Banerjee:2017hhz}.
The blue solid line describes the experimental sensitivity\footnote{This is defined as the boundary of parameter space that can be probed by the dedicated search in a given experiment at 90\% \dword{cl}, practically obtained from Eq.~(\ref{eq:MIsensitivity}).} at DUNE \dword{fd} under a zero background assumption.
The associated exposure is \fdfiducialmass $\cdot$ yr, i.e., a total fiducial volume of 40 kilo-ton times 1-year running time.
For comparison, we also show the sensitivities of DUNE to the $p$-scattering signal as a green solid line. 

Inspired by this potential of searching for the proton scattering channel, we study another reference parameter and compare it with the original one in the lower-left panel of Figure~\ref{fig:darkphotonparameter}. 
We see the reachable $\epsilon$ values rise, as $m_V$ increases.

For Scenario 2 (the right panels of Figure~\ref{fig:darkphotonparameter}), we choose a different reference parameter set: $m_{\chi_0} = 2$ GeV, $m_{\chi_1} = 50$ MeV, $\delta m = 10$ MeV. 
The current limits (brown shaded regions), from various fixed target experiments, B-factory experiments, and astrophysical observations, are taken from Ref.~\cite{Banerjee:2018vgk}.

We next discuss model-independent experimental sensitivities. 
The experimental sensitivities are determined by the number of signal events excluded at 90\% \dword{cl} in the absence of an observed signal.
The expected number of signal events, $N_{\rm sig}$, is given by
\begin{align}
N_{\rm sig} = \sigma_\epsilon \mathcal F A(\ell_{\rm lab}) t_{\rm exp} N_T\,,
\label{eq:NS}
\end{align}
where $T$ stands for the target that $\chi_1$ scatters off, $\sigma_\epsilon$ is the cross section of the primary scattering $\chi_1 T \to \chi_2 T$, $\mathcal F$ is the flux of $\chi_1$, $t_{\rm exp}$ is the exposure time, and $A(\ell_{\rm lab})$ is the acceptance that is defined as 1 if the event occurs within the fiducial volume and 0 otherwise.
Here we determine the acceptance for an $i$\dword{bdm} signal by the distance between the primary and secondary vertices in the laboratory frame, $\ell_{\rm lab}$, so $A(\ell_{\rm lab}) = 1$ when both the primary and secondary events occur inside the fiducial volume. (Given this definition, obviously, $A(\ell_{\rm lab}) = 1$ for elastic \dword{bdm}.)
Our notation $\sigma_\epsilon$ includes additional realistic effects from cuts, threshold energy, and the detector response, hence it can be understood as the fiducial cross section.

\begin{figure}[t]
\centering
\includegraphics[width=6cm]{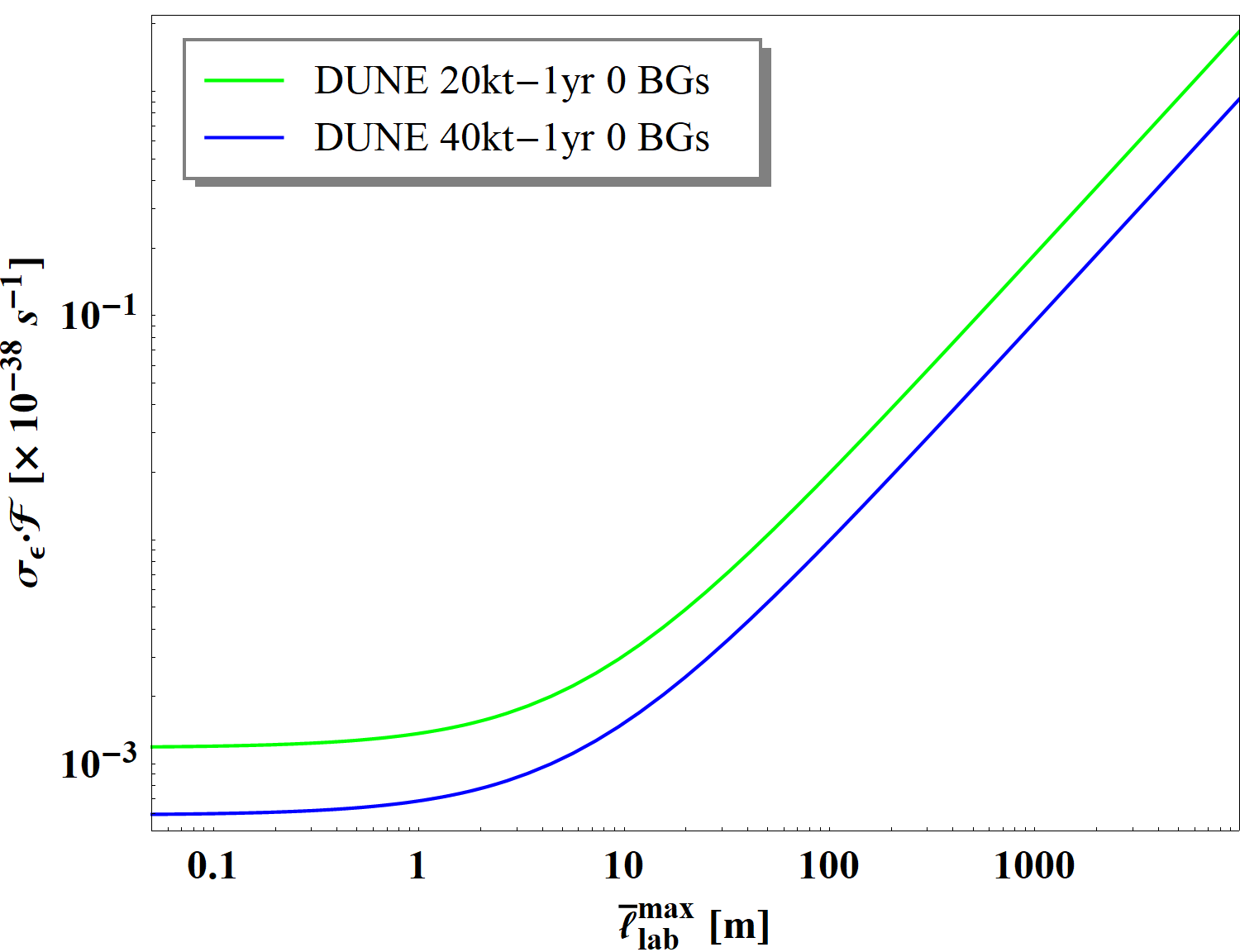}
\includegraphics[width=6cm]{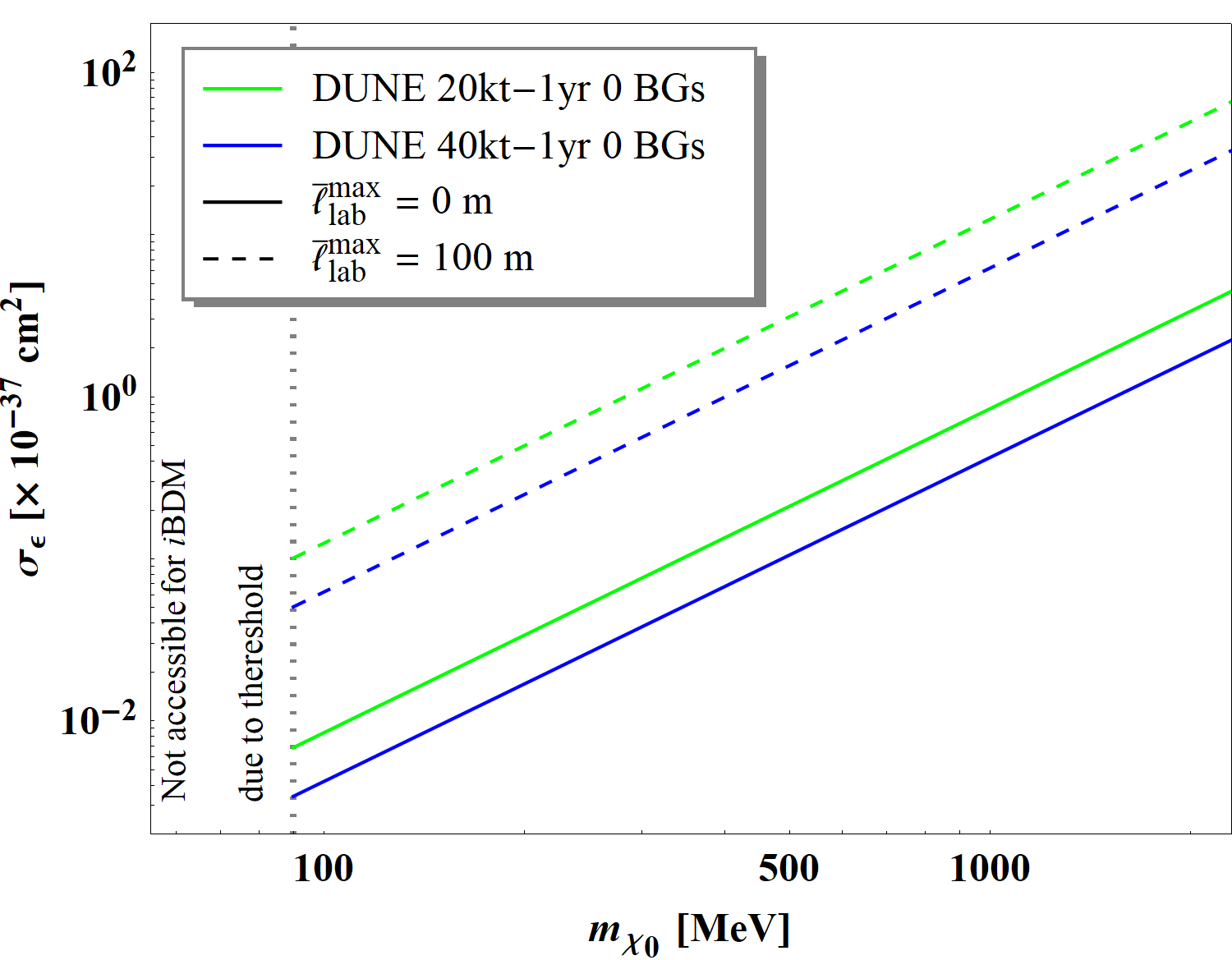}
\caption[Model-independent experimental sensitivities of $i$BDM search]{
Left: model-independent experimental sensitivities of $i$\dword{bdm} search in $\bar{\ell}_{\rm lab}^{\rm max} - \sigma_\epsilon \cdot \mathcal F$ plane. 
The reference experiments are
DUNE 20kt (green), and DUNE 40kt (blue) with zero-background assumption for 1-year time exposure. 
Right: Experimental sensitivities of $i$\dword{bdm} search in $m_{\chi_0} - \sigma_\epsilon$ plane. The sensitivities for $\bar{\ell}_{\rm lab}^{\rm max} = 0$ m and 100 m are shown as solid and dashed lines for each reference experiment in the left panel.
\label{fig:modelindependent} }
\end{figure}

The 90\% \dword{cl} exclusion limit, $N_s^{90}$, can be obtained with a modified frequentist construction~\cite{cls1,cls2}. We follow the methods in Refs.~\cite{Dermisek:2013cxa,Dermisek:2014qca,Dermisek:2016via} in which the Poisson likelihood is assumed. 
An experiment becomes sensitive to the signal model independently if $N_{\rm sig} \ge N_s^{90}$.
Plugging Eq.~\eqref{eq:NS} here, we find the experimental sensitivity expressed by 
\begin{align}
\sigma_\epsilon \mathcal F \ge \frac{N_s^{90}}{A(\ell_{\rm lab}) t_{\rm exp} N_T}\,. 
\label{eq:MIsensitivity}
\end{align}
Since $\ell_{\rm lab}$ differs event-by-event, we take the maximally possible value of laboratory-frame mean decay length, i.e., $\bar{\ell}_{\rm lab}^{\rm max} \equiv \gamma_{\chi_2}^{\max} \bar{\ell}_{\rm rest}$ where $\gamma_{\chi_2}^{\max}$ is the maximum boost factor of $\chi_2$ and $\bar{\ell}_{\rm rest}$ is the rest-frame mean decay length. 
We emphasize that this is a rather conservative approach, because the acceptance $A$ is inversely proportional to $\ell_{\rm lab}$.
We then show the experimental sensitivity of any kind of experiment for a given background expectation, exposure time, and number of targets, in the plane of $\bar{\ell}_{\rm lab}^{\rm max} - \sigma_\epsilon \cdot \mathcal F$. 
The left panel of Figure~\ref{fig:modelindependent} demonstrates the expected model-independent sensitivities at the DUNE experiment.
The green (blue) line is for the DUNE \dword{fd} with a background-free assumption and 20 (40) kt$\cdot$yr exposure.

The right panel of Figure~\ref{fig:modelindependent} reports model-dependent sensitivities for $\bar{\ell}_{\rm lab}^{\rm max} = 0$ m and 100 m corresponding to the experiments in the left panel.
Note that this 
method of presentation is reminiscent of the widely known scheme for showing the experimental reaches in various \dword{dm} direct detection experiments, i.e., $m_{\rm DM} - \sigma_{\rm DM - target}$ where $m_{\rm DM}$ is the mass of \dword{dm} and $\sigma_{\rm DM - target}$ is the cross section between the \dword{dm} and target. 
For the case of non-relativistic \dword{dm} scattering in the direct-detection experiments, $m_{\rm DM}$ determines the kinetic energy scale of the incoming \dword{dm}, just like $m_{\chi_0}$ sets out the incoming energy of boosted $\chi_1$ in the $i$\dword{bdm} search. 

\subsection{Elastic Boosted Dark Matter from the Sun \label{sec:FDsun}}

\subsubsection{\label{sec:level2} Introduction and theoretical framework}

In this section, we focus on the Benchmark Model ii) discussed in Section~\ref{sec:model}. This study represents the first assessment of sensitivity to this model in DUNE using DUNE's full event generation and detector simulation. We focus on \dword{bdm} flux sourced by \dword{dm} annihilation in the core of the sun. \dword{dm} particles can be captured through their scattering with the nuclei within the sun, mostly hydrogen and helium. This makes the core of the sun a region with concentrated \dword{dm} distribution. The \dword{bdm} flux is
\begin{eqnarray} \label{eq:fluxbdm}
\Phi= f \frac{A}{4\pi D^2},
\end{eqnarray}
where $A$ is the annihilation rate, and $D = 1\,\rm{\dword{au}}$ is the distance from the sun. $f$ is a model-dependent parameter, where $f = 2$ for two-component \dword{dm} as considered here.

\begin{figure}[h!]
  \centering
  \includegraphics[width=0.7\textwidth]{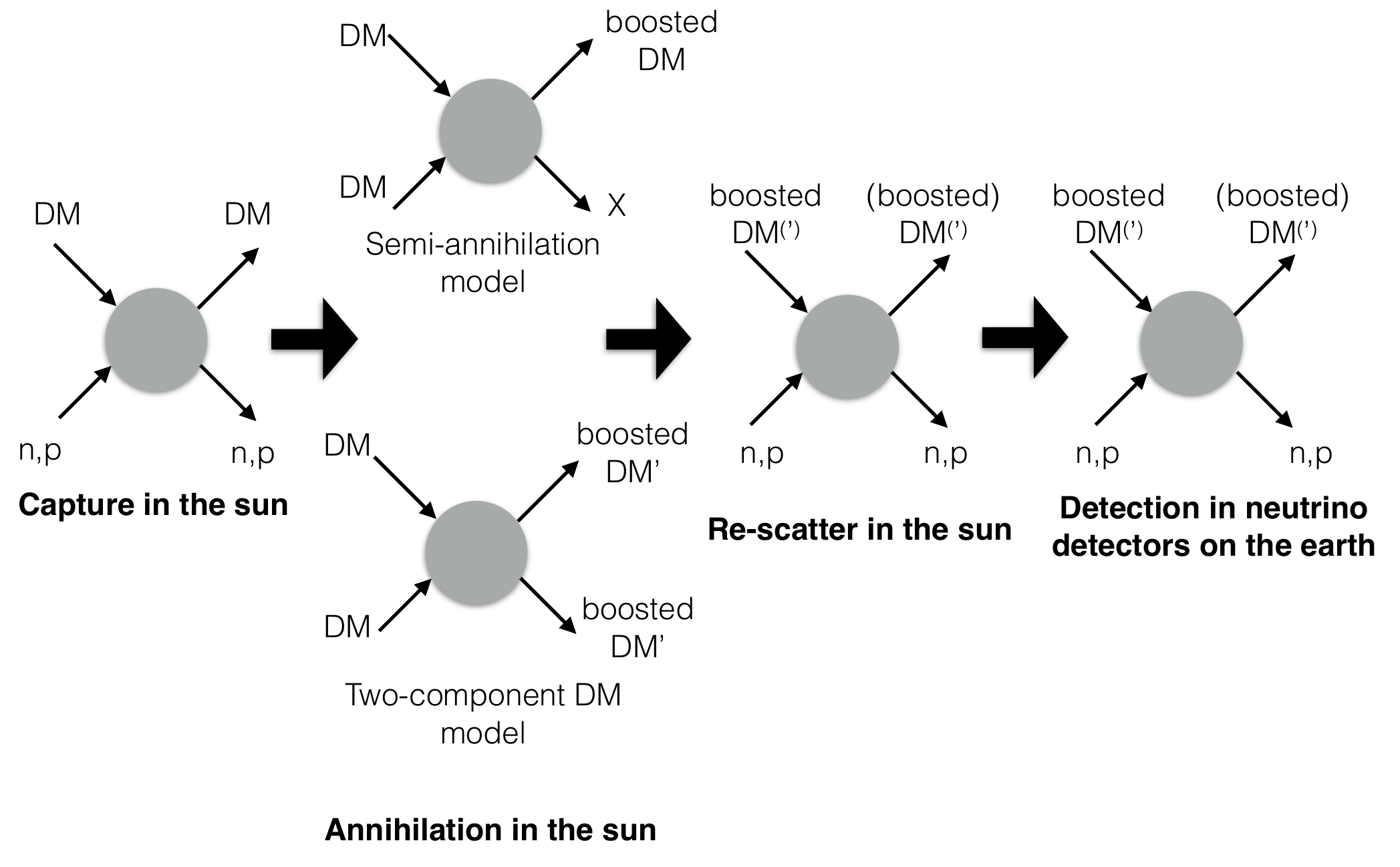}
  \caption[Processes leading to boosted DM signal from the sun]{The chain of processes leading to boosted DM signal from the sun. The semi-annihilation and two-component DM models refer to the two examples of the non-minimal dark-sector scenarios introduced in the beginning of Section~\ref{sec:DM}. DM' denotes the lighter DM in the two-component DM model. $X$ is a lighter dark sector particle that may decay away.}
    \label{fig:processes}
\end{figure}

For the parameter space of interest, 
assuming that the 
\dword{dm} annihilation cross section is not too small, the \dword{dm} distribution in the sun has reached an equilibrium between capture and annihilation. This helps to eliminate the annihilation cross section dependence in our study. The chain of processes involved in giving rise to the boosted DM signal from the Sun is illustrated in Fig.~\ref{fig:processes}.

Two additional comments are in order. First, the \dword{dm} particles cannot be too light, i.e.,  lighter than 4\,GeV~\cite{Griest:1986yu,Gould:1987ju}, otherwise we will lose most of the captured \dword{dm} through evaporation rather than annihilation; this would dramatically reduce the \dword{bdm} flux. Additionally, one needs to check that \dword{bdm} particles cannot lose energy and potentially be recaptured by scattering with the solar material when they escape from the core region after production. Rescattering is found to be rare for the benchmark models considered in this study and we consider the \dword{bdm} flux to be monochromatic at its production energy.

The event rate to be observed at DUNE is 
\begin{equation}
R = \Phi \times \sigma_{\rm{SM} - \chi} \times \varepsilon \times N,
\end{equation}
 where $\Phi$ is the flux given by Eq. \eqref{eq:fluxbdm}, $\sigma_{\rm{SM} - \chi}$ is the scattering cross section of the \dword{bdm} off of \dword{sm} particles, $\varepsilon$ is the efficiency of the detection of such a process, and $N$ is the number of target particles in DUNE. The computation of the flux of \dword{bdm} from the sun can be found in~\cite{Berger:2014sqa}. 

\begin{figure}[h!]
  \centering
  \includegraphics[width=0.256\textwidth]{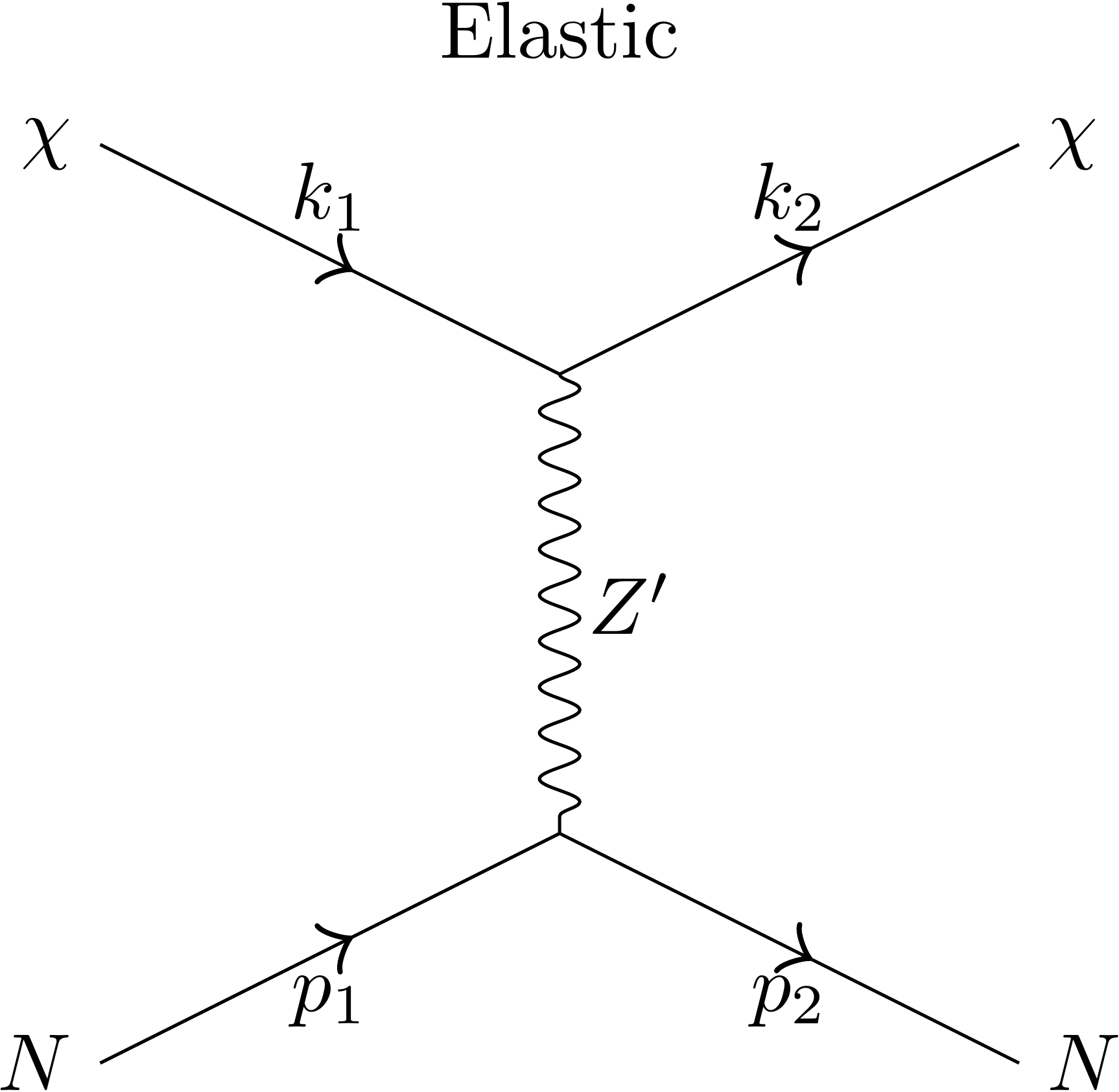}
  \includegraphics[width=0.33\textwidth]{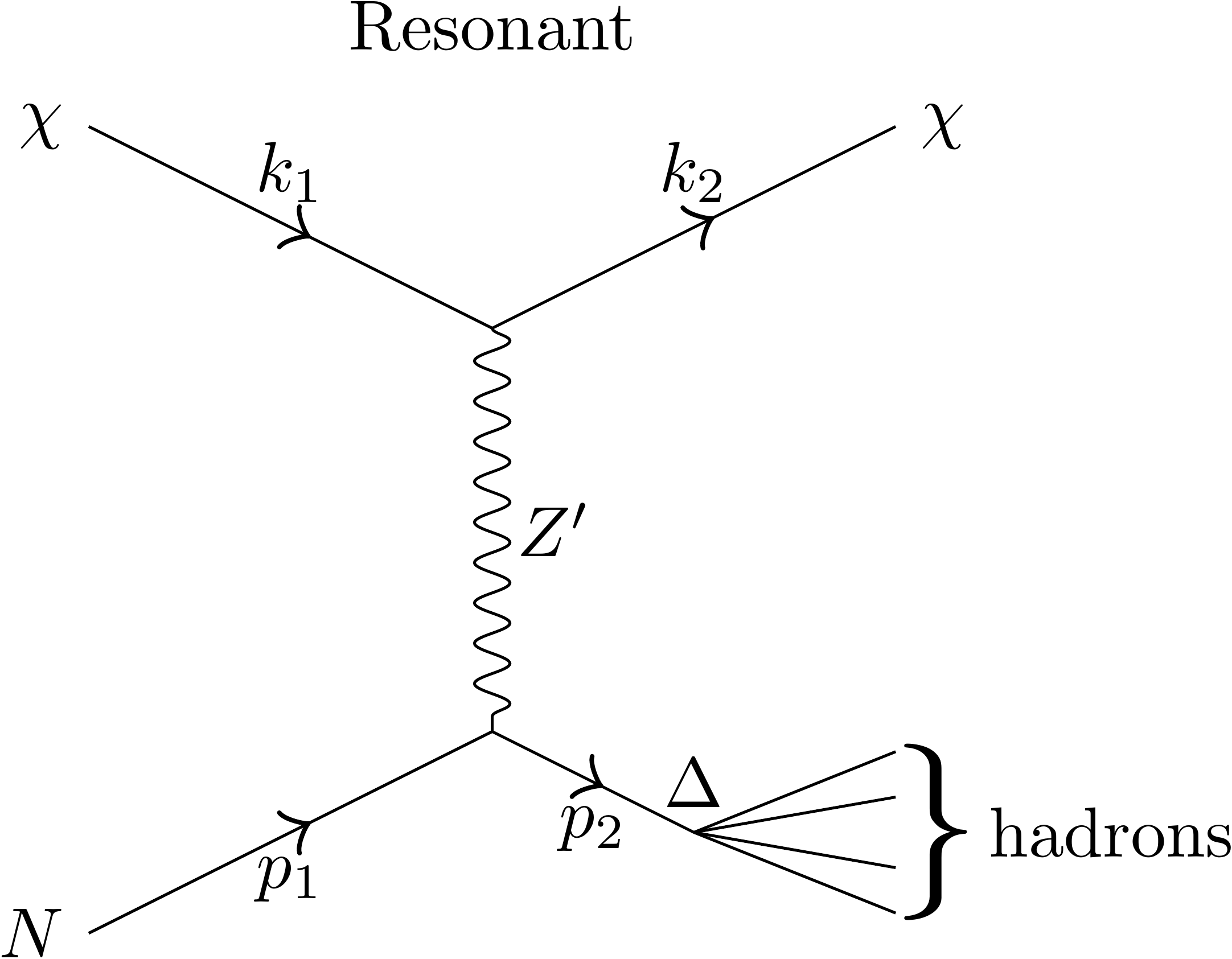}
  \includegraphics[width=0.29\textwidth]{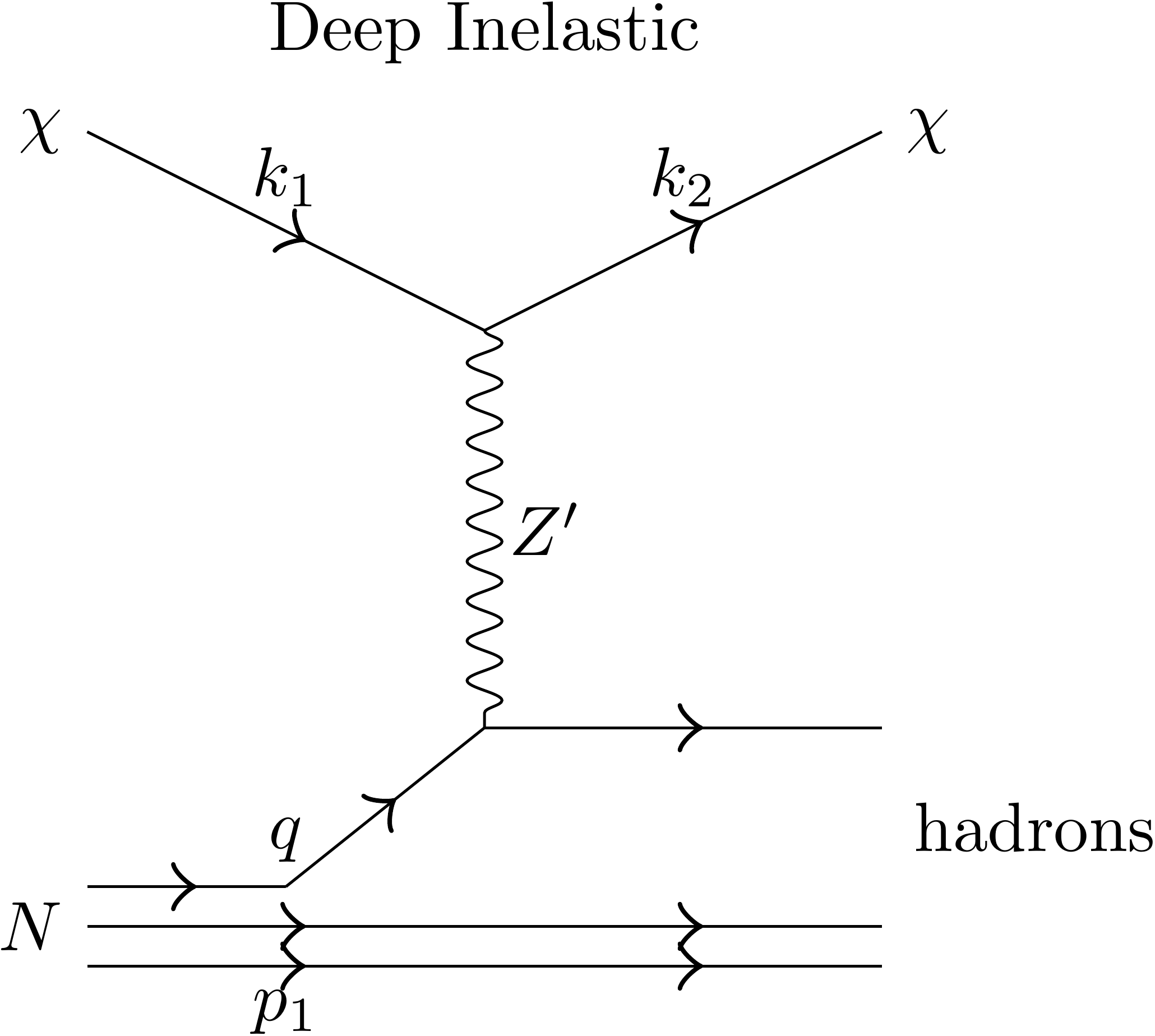}
  \caption{Diagram illustrating each of the three processes contributing to dark matter scattering in argon: elastic (left), baryon resonance (middle), and deep inelastic (right).}
  \label{fig:BDM-argon}
\end{figure}

The processes of typical BDM scattering in argon are illustrated in Fig.~\ref{fig:BDM-argon}.
We generate the signal events and calculate interaction cross sections in the detector using a newly developed \dword{bdm} module~\cite{Andreopoulos:2009rq,Andreopoulos:2015wxa,Berger:2018} that includes elastic and deep inelastic scattering, as well as a range of nuclear effects. This conservative event generation neglects the dominant contributions from baryon resonances in the final state hadronic invariant mass range of 1.2 to 1.8 GeV, which should not have a major effect on our main results. The interactions are taken to be mediated by an axial, flavor-universal $Z^\prime$ coupling to both the \dword{bdm} and with the quarks. The axial charge is taken to be 1. 
The events are generated for the \nominalmodsize DUNE detector module~\cite{dunetpc_code}, though we only study the dominant scattering off of the \argon40 atoms therein. The method for determining the efficiency $\varepsilon$ is described below. The number of target argon atoms is $N = 1.5  \times 10^{32}$ assuming a target mass of \nominalmodsize{}.

\subsubsection{Background Estimation}
\label{sec:background}

The main background in this process comes from the \dword{nc} 
interactions of atmospheric neutrinos and argon,
as they share the features that the timing of events is unknown in advance,
and that the interactions with argon produce hadronic activity in the detector.
We use \dword{genie}~\cite{Andreopoulos:2009rq,Andreopoulos:2015wxa}
interfaced by the \dword{larsoft} toolkit to generate the \dword{nc} atmospheric
neutrino events, and obtain 845 events in a \nominalmodsize{} module for one year of
exposure.

\subsubsection{Detector Response}
\label{sec:detector_resp}

\begin{figure*}[!htb]
\centering
\includegraphics[width=0.45\textwidth]{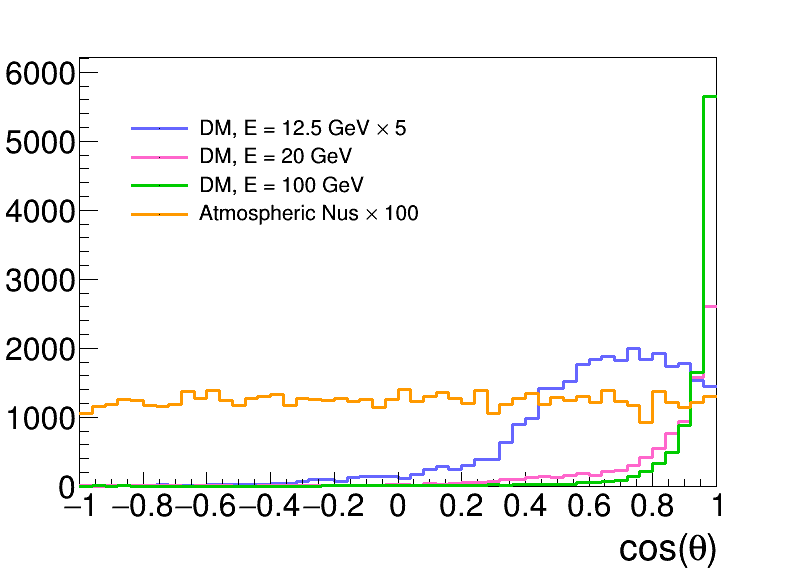}
\includegraphics[width=0.45\textwidth]{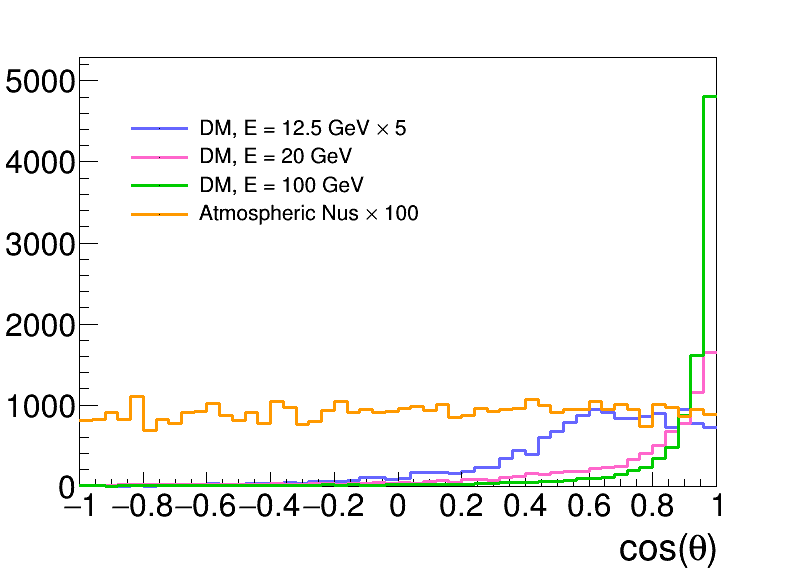}
\caption[Angular distribution of the BDM signal events for a BDM mass of 10\,GeV]{Angular distribution of the \dword{bdm} signal events for a \dword{bdm} mass of 10\,GeV
and different boosted factors, $\gamma$, and of the atmospheric neutrino NC
background events.
$\theta$ represents the angle of the sum over all the stable final state
particles as detailed in the text.
The amount of background represents one-year data collection, magnified by a factor 100,
while the amount of signal reflects the detection efficiency of 10,000 \dword{mc} events, as
described in this note.
The left plot shows the scenario where neutrons can be reconstructed,
while the right plot represents the scenario without neutrons.}
\label{fig:m10_SmearedReconstructableAngle}
\end{figure*}

The finite detector resolution is taken into
account by smearing the direction of the stable final state particles, 
including protons, neutrons, charged pions, muons, electrons, and photons,
with the expected angular resolution,
and by ignoring the ones with kinetic energy below detector threshold,
using the parameters reported in the DUNE \dword{cdr}~\cite{Acciarri:2015uup}.
We form as the observable the total momentum from all the stable final state particles,
and obtain its angle with respect to the direction of the sun.
The sun position is simulated with the SolTrack package~\cite{SolTrack}
including the geographical coordinates of the DUNE \dword{fd}~\cite{DUNE_DocDB136}.
We consider both the scenarios in which we can reconstruct neutrons and in which 
neutrons will not be reconstructed.
Figure~\ref{fig:m10_SmearedReconstructableAngle} shows the angular distributions of
the \dword{bdm} signals with mass of 10\,GeV and different boost factors,
and of the background events.

To increase the signal fraction in our samples, we select events with $\cos\theta > 0.6$,
and obtain the selection efficiency $\varepsilon$ for different \dword{bdm} models.
We predict that $104.0 \pm 0.7$ and $79.4 \pm 0.6$ background events per year, in the scenarios with and without neutrons respectively, survive the selection in a DUNE \nominalmodsize module.

\subsubsection{Results}

\begin{figure*}[!htb]
\centering
\includegraphics[width=0.45\textwidth]{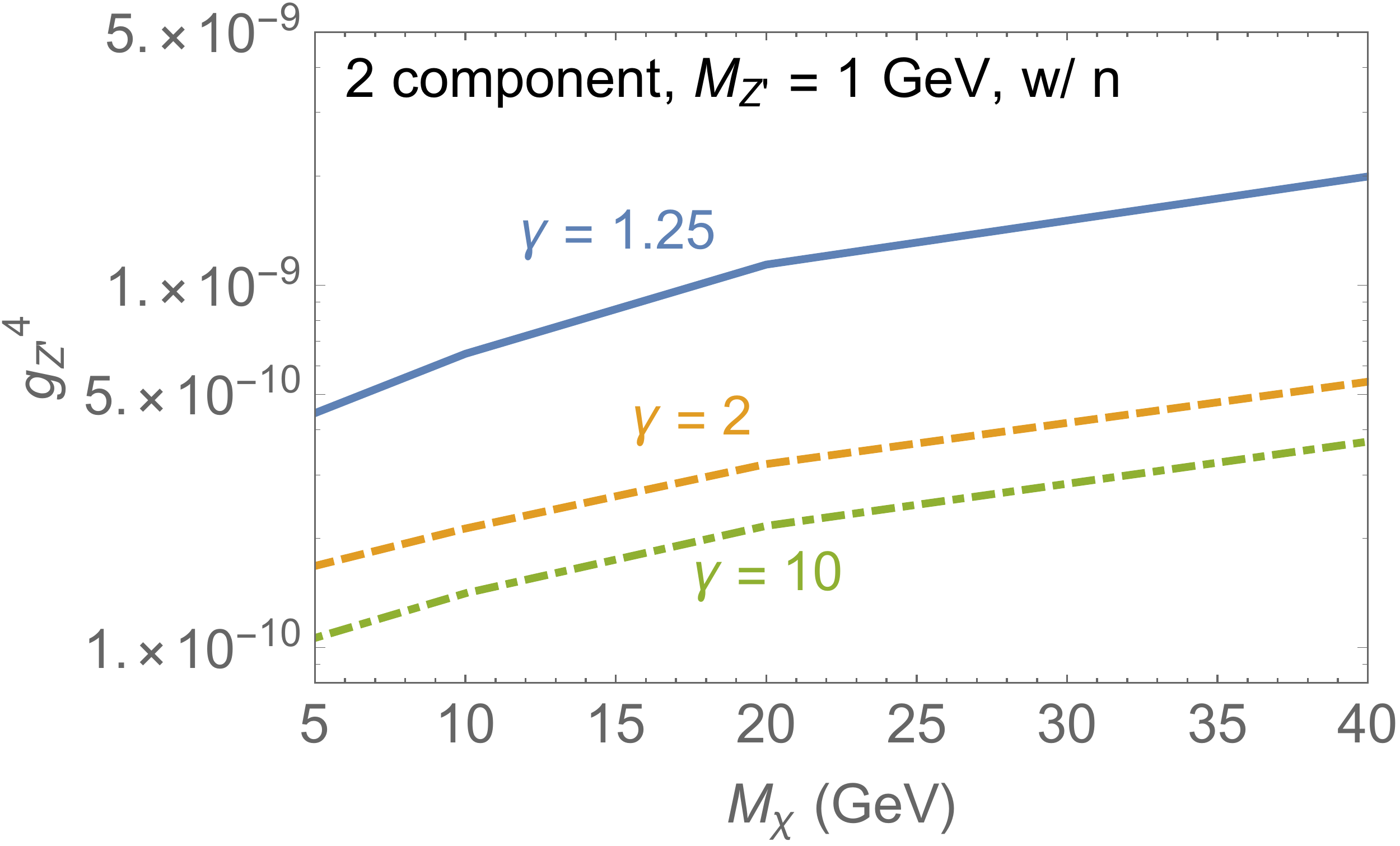}\hspace{0.05\textwidth}
\includegraphics[width=0.45\textwidth]{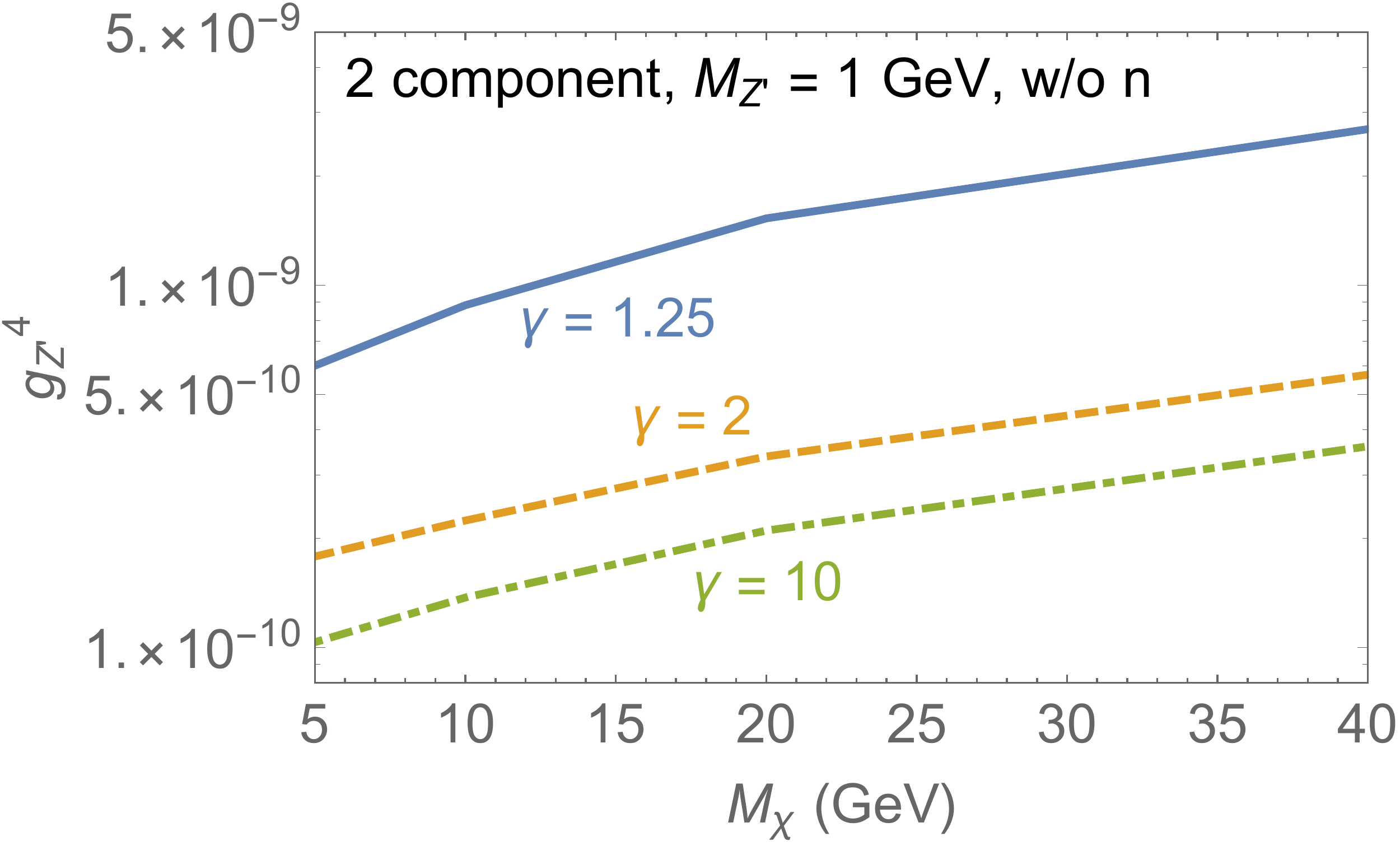}
\caption[Expected $5\sigma$ discovery reach with one year of DUNE livetime]{Expected $5\sigma$ discovery reach with one year of DUNE livetime for one \nominalmodsize module including neutrons in reconstruction (left) and excluding neutrons (right).\label{fig:significance}}
\end{figure*}
\begin{figure*}[!htb]
\centering
\includegraphics[width=0.45\textwidth]{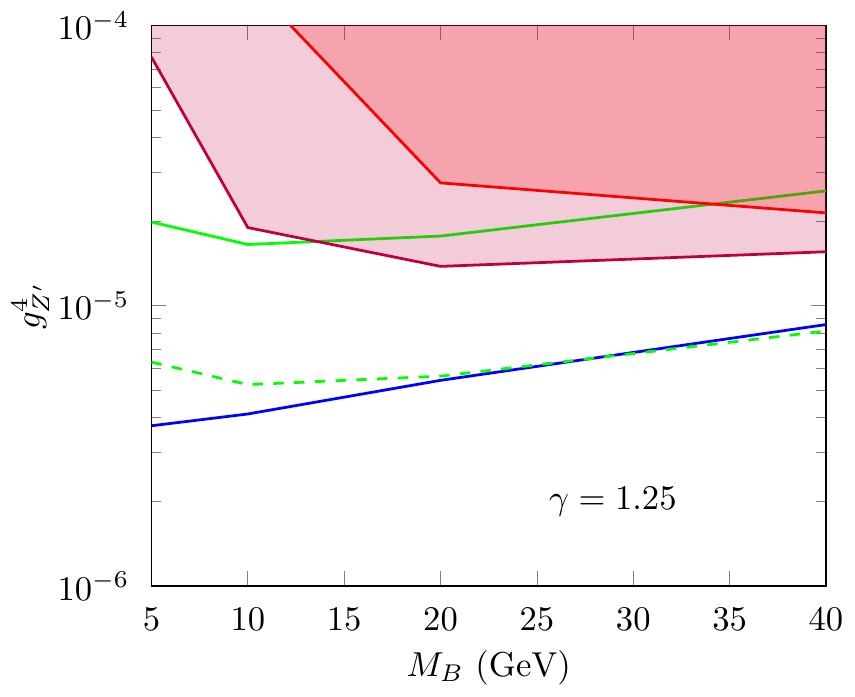}\hspace{0.05\textwidth}
\includegraphics[width=0.45\textwidth]{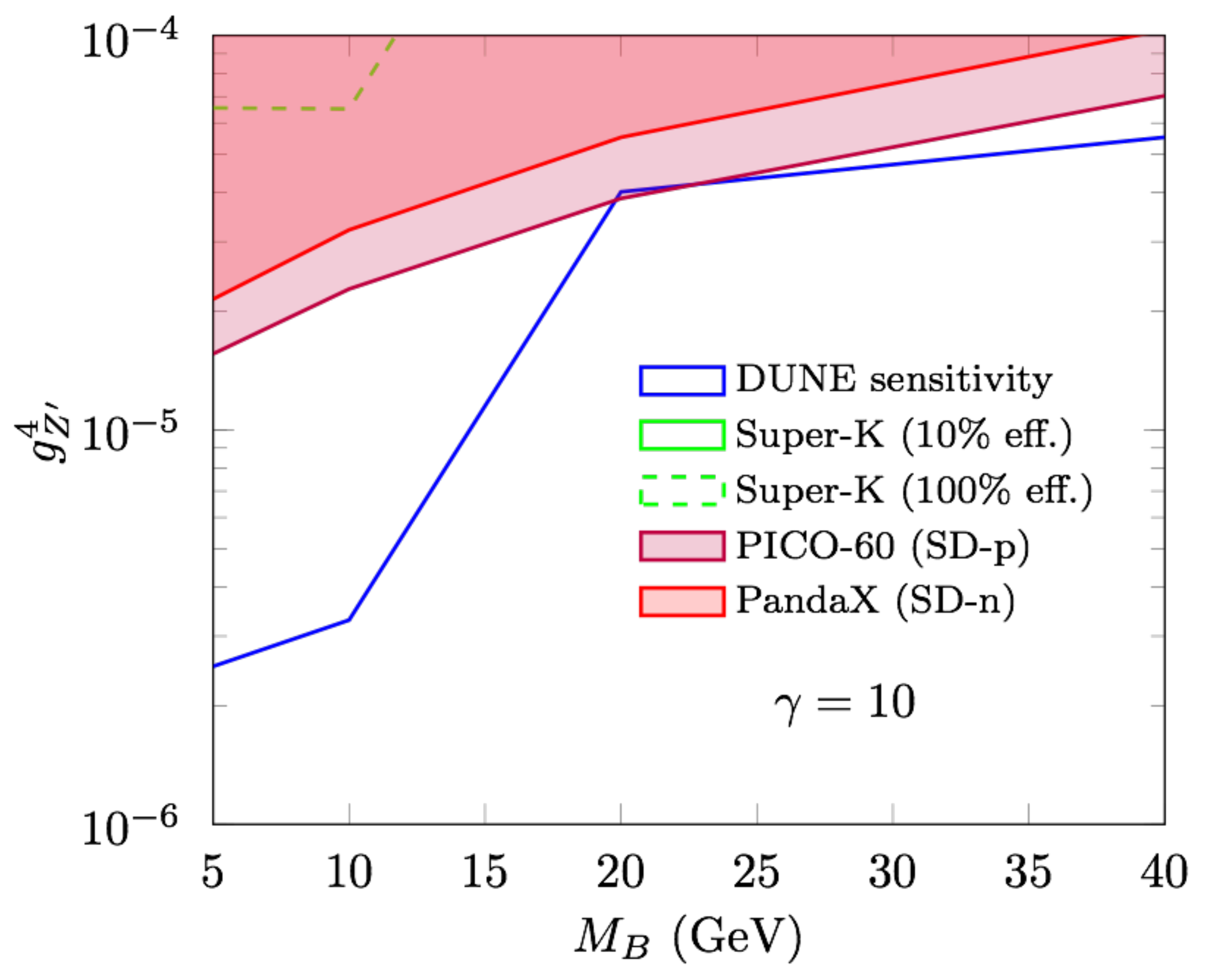}
\caption[Comparison of DUNE (10 yr) sensitivity to \superk sensitivity]{Comparison of sensitivity of DUNE for 10 years of data collection and 40~kton of detector mass with Super Kamiokande, assuming 10\% and 100\% of the selection efficiency on the atmospheric neutrino analysis in Ref.~\cite{Fechner:2009aa}, and with the reinterpretations of the current results from PICO-60~\cite{Amole:2019fdf} and PandaX~\cite{Xia:2018qgs}.  The samples with two boosted factors, $\gamma = 1.25$ (left) and $\gamma = 10$ (right), are also presented. \label{fig:bdm_sensitivity_comparison}}
\end{figure*}

The resulting expected sensitivity is presented in Figure~\ref{fig:significance} in terms of the \dword{dm} mass and the $Z^\prime$ gauge coupling for potential \dword{dm} boosts of $\gamma = 1.25,2,10$ and for a fixed mediator mass of $m_{Z^\prime} = 1~{\rm GeV}$.  We assume a DUNE livetime of one year for one \nominalmodsize module.  The models presented here are currently unconstrained by direct detection searches if the thermal relic abundance of the \dword{dm} is chosen to fit current observations.
Figure~\ref{fig:bdm_sensitivity_comparison} compares the sensitivity of 10 years of data collected in DUNE (40~kton) to re-analyses of the results from other experiments, including Super Kamiokande~\cite{Fechner:2009aa} and \dword{dm} direct detection, PICO-60~\cite{Amole:2019fdf} and PandaX~\cite{Xia:2018qgs}. 

\subsection{Discussion and Conclusions}

In this work, we have conducted simulation studies of the dark matter models described in Eqs.~\eqref{eq:lagrangian} and \eqref{eq:zprimelag} in terms of their detection prospects at the DUNE \dword{nd} and \dword{fd}. 
Thanks to its relatively low threshold and strong particle identification capabilities, DUNE presents an opportunity to significantly advance the search for \dword{ldm} and \dword{bdm} beyond what has been possible with water Cherenkov detectors.

In the case of the \dword{nd}, we assumed that the relativistic \dword{dm} is being produced directly at the target and leaves an experimental signature through an elastic electron scattering. Using two constrained parameters of the light \dword{dm} model and a range of two free parameters, a sensitivity map was produced. Within the context of the vector portal \dword{dm} model and the chosen parameter constraints along with the electron scattering as the signal event, this result sets stringent limits on \dword{dm} parameters that are comparable or even better than recent experimental bounds in the sub-GeV mass range.

By contrast, in the case of the \dword{fd} modules, we assumed that the signal events are due to \dword{dm} coming from the galactic halo and the sun with a significant boost factor. 
For the \textit{in}elastic scattering case, the \dword{dm} scatters off either an electron or proton in the detector material into a heavier unstable dark-sector state.
The heavier state, by construction, decays back to \dword{dm} and an electron-positron pair via a dark-photon exchange. 
Therefore, in the final state, a signal event comes with an electron or proton recoil plus an electron-positron pair. 
This distinctive signal feature enabled us to perform (almost) background-free analyses. 
As \dword{protodune} detectors are prototypes of DUNE \dword{fd} modules, the same study was conducted and corresponding results were compared with the ones of the DUNE \dword{fd} modules.  
We first investigated the experimental sensitivity in a dark-photon parameter space, dark-photon mass $m_V$ versus kinetic mixing parameter $\epsilon$. 
The results were shown separately for Scenario 1 and 2. 
They suggest that \dword{protodune} and DUNE \dword{fd} modules would probe a broad range of unexplored regions; they would allow for reaching $\sim 1-2$ orders of magnitude smaller $\epsilon$ values than the current limits along MeV to sub-GeV-range dark photons. 
We also examined model-independent reaches at both \dword{protodune} detectors and DUNE \dword{fd} modules, providing limits for models that assume the existence of $i$\dword{bdm} (or $i$\dword{bdm}-like) signals (i.e., a target recoil and a fermion pair). 

For the elastic scattering case, we considered the case in which \dword{bdm} comes from the sun. 
With one year of data, the $5\sigma$ sensitivity is expected to reach a coupling of $g_{Z^\prime}^4 = 9.57 \times 10^{-10}$ for a boost of 1.25 and $g_{Z^\prime}^4 = 1.49 \times 10^{-10}$ for a boost of 10 at a \dword{dm} mass of \SI{10}{GeV} without including neutrons in the reconstruction.

\section{Other \dword{bsm} Physics Opportunities}
\label{sec:otheropps}

\subsection{Tau Neutrino Appearance}

With only 19 $\nu_{\tau}$-\dword{cc} and $\bar{\nu}_{\tau}$-\dword{cc} candidates detected with high purity, we have less direct experimental knowledge of tau neutrinos than of any other \dword{sm} particle. Of these, nine $\nu_{\tau}$-\dword{cc} and $\bar{\nu}_{\tau}$-\dword{cc} candidate events with a background of 1.5 events, observed by the DONuT experiment~\cite{Kodama:2000mp, Kodama:2007aa}, were directly produced though $D_S$ meson decays.  The remaining 10 $\nu_{\tau}$-\dword{cc} candidate events with an estimated background of two events, observed by the OPERA experiment~\cite{Guler:2000bd,Agafonova:2018auq}, were produced through the oscillation of a muon neutrino beam. From this sample, a 20\% measurement of $\Delta m^{2}_{32}$ was performed under the assumption that $\sin^22\theta_{23} = 1$.  The \superk and IceCube experiments developed methods to statistically separate samples of $\nu_{\tau}$-\dword{cc} and $\bar{\nu}_{\tau}$-\dword{cc} events in atmospheric neutrinos to exclude the no-tau-neutrino appearance hypothesis at the 4.6$\sigma$ level and 3.2$\sigma$ level respectively~\cite{Abe:2012jj, Li:2017dbe, Aartsen:2019tjl}, but limitations of Cherenkov detectors constrain the ability to select a high-purity sample and perform precision measurements.

\begin{figure}[!htb]
 \centering
        \includegraphics[width=0.4\columnwidth]{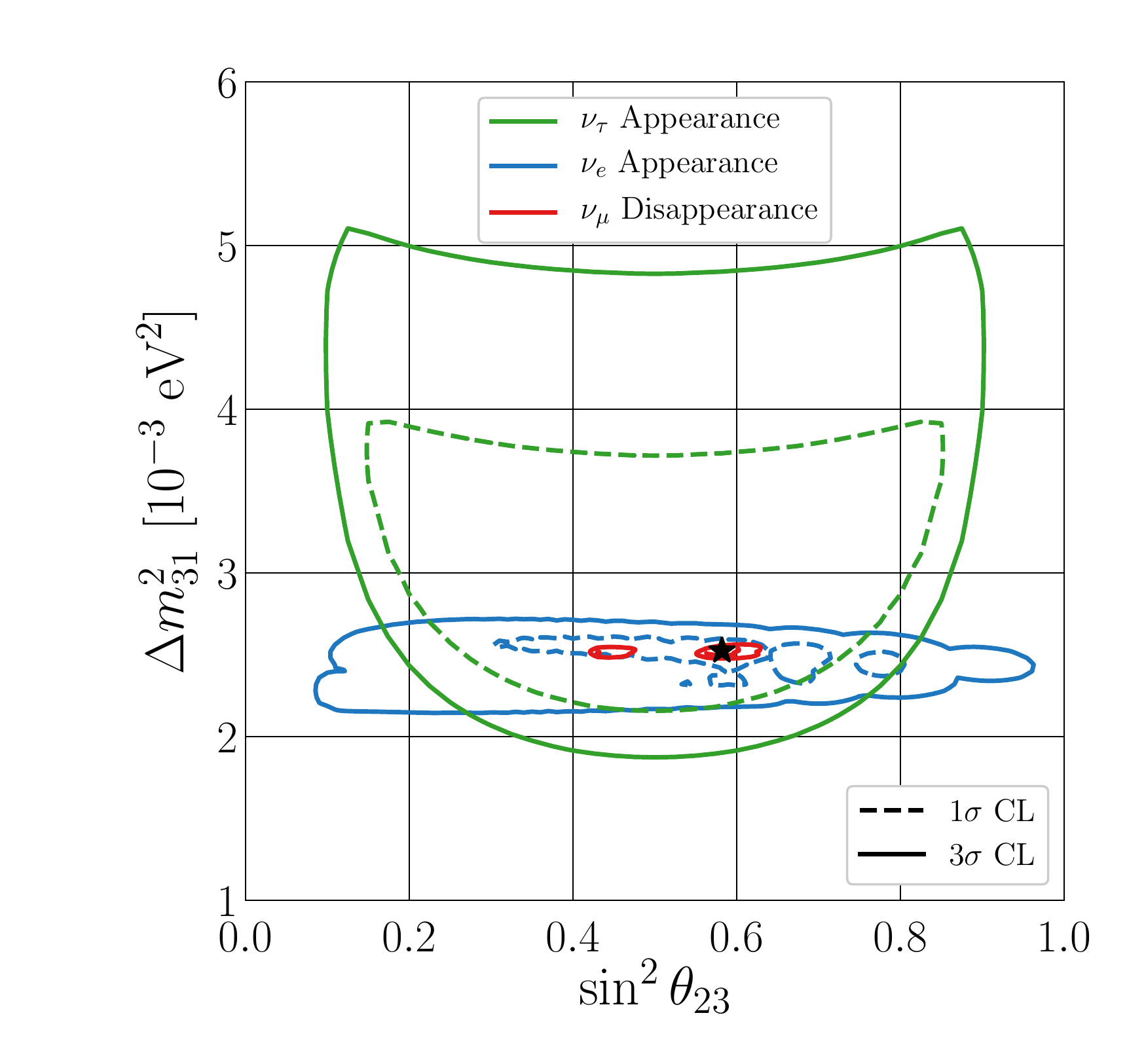}
        \includegraphics[width=0.4\columnwidth]{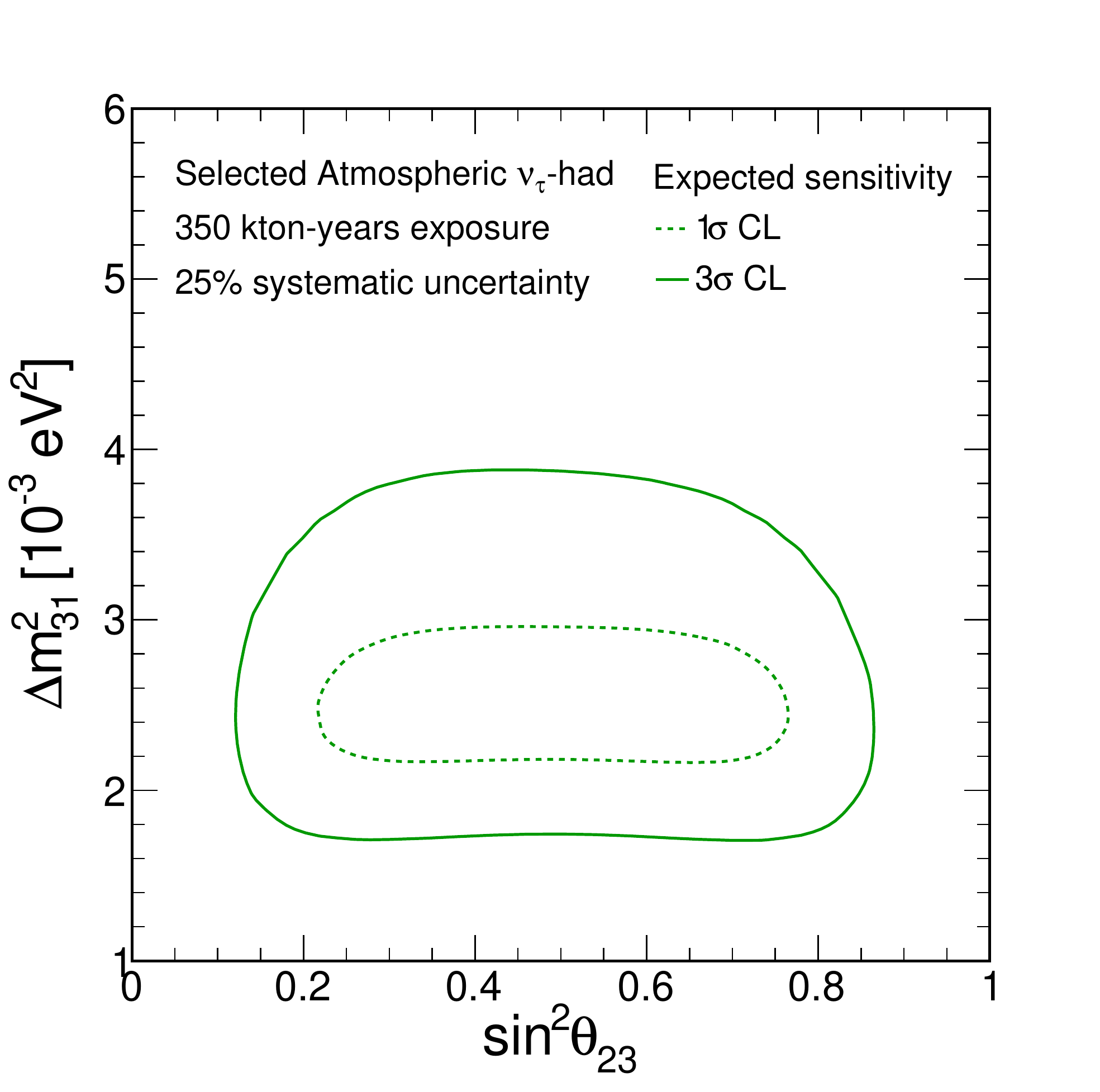}
	\caption[The 1$\sigma$ and 3$\sigma$ expected sensitivity for measuring $\Delta m^2_{31}$ and $\sin^2\theta_{23}$] 
	{The 1$\sigma$ (dashed) and 3$\sigma$ (solid) expected sensitivity for measuring $\Delta m^2_{31}$ and $\sin^2\theta_{23}$ using a variety of samples. Left: The expected sensitivity for seven years of beam data collection, assuming 3.5 years each in neutrino and antineutrino modes, measured independently using $\nu_{e}$ appearance (blue), $\nu_{\mu}$ disappearance (red), and $\nu_{\tau}$ appearance (green). Adapted from Ref.~\cite{deGouvea:2019ozk}. Right: The expected sensitivity for the $\nu_{\tau}$ appearance channel using 350 kton-years of atmospheric exposure.}
	\label{fig:nutauContours}
\end{figure}

The DUNE experiment has the possibility of significantly improving the experimental situation. Tau-neutrino appearance can potentially improve the discovery potential for sterile neutrinos, \dword{nc} \dword{nsi}, and non-unitarity.  For model independence, the first goal should be measuring the atmospheric oscillation parameters in the $\nu_{\tau}$ appearance channel and checking the consistency of this measurement with those performed using the $\nu_{\mu}$ disappearance channel.  A truth-level study of $\nu_{\tau}$ selection in atmospheric neutrinos in a large, underground LArTPC detector suggested that $\nu_{\tau}$-\dword{cc} interactions with hadronically decaying $\tau$-leptons, which make up 65\% of total $\tau$-lepton decays~\cite{Tanabashi:2018oca}, can be selected with high purity~\cite{Conrad:1008}.  This analysis suggests that it may be possible to select up to 30\% of $\nu_{\tau}$-\dword{cc} events with hadronically decaying $\tau$-leptons with minimal neutral current background.  Under these assumptions, we expect to select $\sim$25 $\nu_{\tau}$-\dword{cc} candidates per year using the \dword{cpv} optimized beam. The physics reach of this sample has been studied in Ref.~\cite{deGouvea:2019ozk}. As shown in Figure~\ref{fig:nutauContours} (left), this sample is sufficient to simultaneously constrain $\Delta m^2_{31}$ and $\sin^22\theta_{23}$. Independent measurements of $\Delta m^2_{31}$ and $\sin^22\theta_{23}$ in the $\nu_{e}$ appearance, $\nu_{\mu}$ disappearance, and $\nu_{\tau}$ appearance channels should allow DUNE to constrain $|U_{e3}|^2+|U_{\mu 3}|^2+|U_{\tau 3}|^2$ to 6\%~\cite{deGouvea:2019ozk}, a significant improvement over current constraints~\cite{Parke:2015goa}.

However, all of the events in the beam sample occur at energies higher than the first oscillation maximum due to kinematic constraints.  Only seeing the tail of the oscillation maximum creates a partial degeneracy between the measurement of $\Delta m^2_{31}$ and $\sin^22\theta_{23}$.  Atmospheric neutrinos, due to sampling a much larger $L/E$ range, allow for measuring both above and below the first oscillation maximum with $\nu_{\tau}$ appearance. Although we only expect to select $\sim$70 $\nu_{\tau}$-\dword{cc} and $\bar{\nu}_{\tau}$-\dword{cc} candidates in 350 kt-year in the atmospheric sample, as shown in Figure~\ref{fig:nutauContours} (right), a direct measurement of the oscillation maximum breaks the degeneracy seen in the beam sample. The complementary shapes of the beam and atmospheric constraints combine to reduce the uncertainty on $\sin^2\theta_{23}$, directly leading to improved unitarity constraints.  Finally, a high-energy beam option optimized for $\nu_{\tau}$ appearance should produce $\sim$150 selected  $\nu_{\tau}$-\dword{cc} candidates in one year.  These higher energy events are further in the tail of the first oscillation maximum, but they will permit a simultaneous measurement of the $\nu_{\tau}$ cross section. When analyzed within the non-unitarity framework described in Section~\ref{sec:nonUnitarity}, the high-energy beam significantly improves constraints on the parameter $\alpha_{33}$ due to increased matter effects~\cite{deGouvea:2019ozk}.

\subsection{Large Extra-Dimensions}
DUNE can search for or constrain the size of large extra-dimensions 
by looking for distortions of the oscillation pattern predicted by the three-flavor paradigm. These distortions arise through mixing between the right-handed neutrino Kaluza-Klein modes, which propagate in the compactified extra dimensions, and the active neutrinos, which exist only in the four-dimensional brane~~\cite{Dienes:1998sb,ArkaniHamed:1998vp,Davoudiasl:2002fq}. Such distortions are determined by two parameters in the model, specifically R, the radius of the circle where the extra-dimension is compactified, and $m_0$, defined as the lightest active neutrino mass ($m_1$ for normal mass ordering, and $m_3$ for inverted mass ordering). Searching for these distortions in, for instance, the $\nu_\mu$~\dword{cc} disappearance spectrum, should provide significantly enhanced sensitivity over existing results from the MINOS/MINOS+ experiment~\cite{Adamson:2016yvy}.

\begin{figure}[ht]
\centerline{
\includegraphics[width=0.7\textwidth]{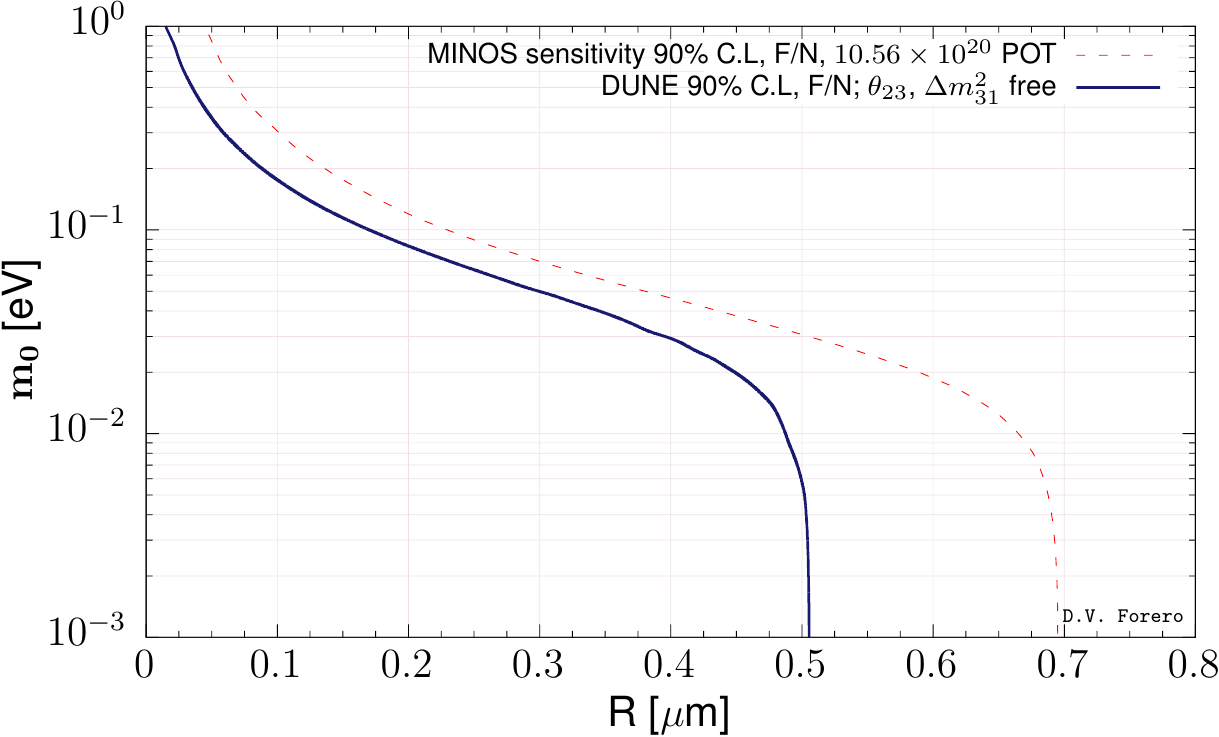}
}
\caption[DUNE sensitivity to the LED model]{Sensitivity to the LED model in Ref.~\cite{Dienes:1998sb,ArkaniHamed:1998vp,Davoudiasl:2002fq} through its impact on the neutrino oscillations expected at 
DUNE. For comparison, the MINOS sensitivity~\cite{Adamson:2016yvy} is also shown.}
\label{fig:ledsensitivity}
\end{figure}

Figure~\ref{fig:ledsensitivity} shows a comparison between the DUNE and MINOS~\cite{Adamson:2016yvy} 
sensitivities to LED at $90\%$ \dword{cl} for two degrees of freedom represented by the solid and dashed lines, respectively. 
In the case of DUNE, an exposure of $300\,\text{kt}\,\text{MW}\,\text{year}$ 
was assumed and spectral information from the four oscillation channels, (anti)neutrino 
appearance and disappearance, were included in the analysis. The muon (anti)neutrino 
fluxes, cross sections for the neutrino interactions in argon, detector energy 
resolutions, efficiencies and systematical errors were taken into account by the use of 
GLoBES files prepared for the DUNE LBL studies. In the analysis, we assumed DUNE 
simulated data as compatible with the standard three neutrino hypothesis (which corresponds to the limit $R\to 0$) and we have 
tested the LED model. The solar parameters were kept fixed, and also the reactor mixing 
angle, while the atmospheric parameters were allowed to float free. In general, DUNE 
improves over the MINOS sensitivity for all values of $m_0$ and this is more noticeable 
for $m_0\sim 10^{-3}$~eV, where the most conservative sensitivity limit to $R$ is 
obtained. 

\subsection{Heavy Neutral Leptons}
The high intensity of the LBNF neutrino beam and the production of charm and bottom mesons in the beam enables DUNE to search for a wide variety of lightweight long-lived, exotic particles, by looking for topologies of rare event interactions and decays in the fiducial volume of the DUNE \dword{nd}. These particles include weakly interacting heavy neutral leptons (HNLs), such as right-handed partners of the active neutrinos, light super-symmetric particles, or vector, scalar, and/or axion portals to a Hidden Sector containing new interactions and new particles. 
Assuming these heavy neutral leptons are the lighter particles of their hidden sector, they will only decay into \dword{sm} particles. The parameter space explored by the DUNE \dword{nd} extends into the cosmologically relevant region complementary to the LHC heavy-mass dark-matter searches through missing energy and mono-jets.

Thanks to small mixing angles, the particles can be stable enough to travel from the baseline to the detector and decay inside the active region.
It is worth noting that, differently from a light neutrino beam, an HNL beam is not polarized, due to their large mass.
The correct description of the helicity components in the beam is important for predicting the angular distributions
of HNL decays, as they might depend on the initial helicity state.
More specifically, there is a different phenomenology if the decaying HNL is a Majorana or a Dirac fermion~\cite{Balantekin:2018ukw, Ballett:2019bgd}.
Typical decay channels are two-body decays into a charged lepton and a pseudo-scalar meson, or a vector meson if
the mass allows it, two-body decays into neutral mesons, and three-body leptonic decays.

\begin{figure}[!htb]
	\begin{center}
	  	\includegraphics[width=0.73\textwidth]{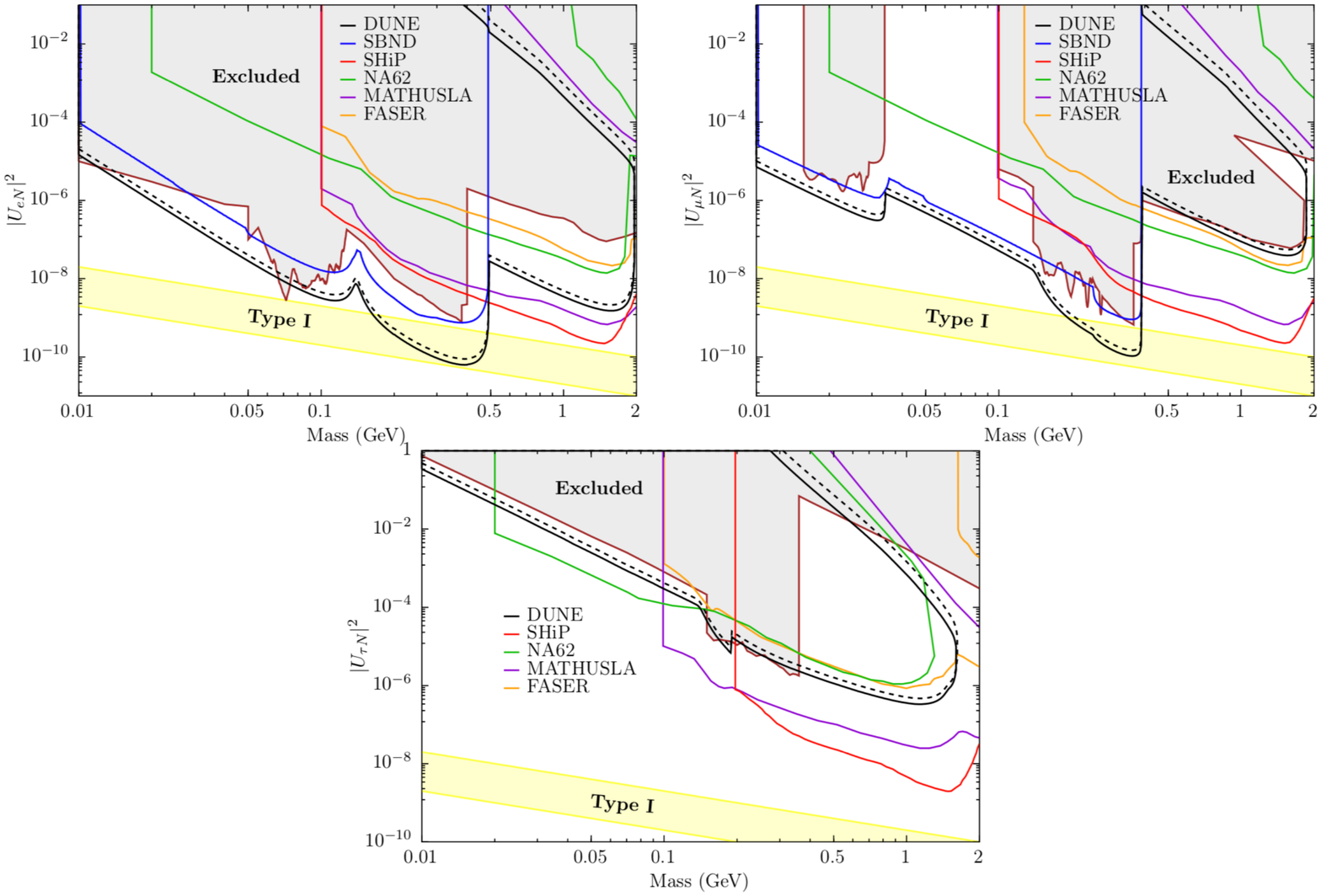}
	\end{center}
\caption[The 90\,\% \dword{cl} sensitivity regions for dominant mixings
		$|U_{\alpha N}|^2$]
		{The 90\,\% \dword{cl} sensitivity regions for dominant mixings %
		$|U_{e N}|^2$ (top~left), $|U_{\mu N}|^2$ (top right), and $|U_{\tau N}|^2$ (bottom) are presented for DUNE ND (black)~\cite{Ballett:2019bgd}.
		The regions are a combination of the sensitivity to HNL decay channels with good detection prospects.These are $N\to\nu e e$, $\nu e \mu$, $\nu \mu \mu$, $\nu \pi^0$, $e \pi$, and $\mu \pi$.The study is performed for Majorana neutrinos (solid) and Dirac neutrinos (dashed), %
		assuming no background. The region excluded by experimental constraints (grey/brown) is obtained by combining the results from PS191~\cite{Bernardi:1985ny, Bernardi:1987ek}, %
		peak searches~\cite{Artamonov:2014urb, Britton:1992pg, Britton:1992xv, Aguilar-Arevalo:2017vlf, Aguilar-Arevalo:2019owf}, %
		CHARM~\cite{Vilain:1994vg}, NuTeV~\cite{Vaitaitis:1999wq}, DELPHI~\cite{Abreu:1996pa}, and T2K~\cite{Abe:2019kgx}. The sensitivity for DUNE ND is compared to the predictions of future experiments, SBN~\cite{Ballett:2016opr} (blue), %
		SHiP~\cite{Alekhin:2015byh} (red), NA62~\cite{Drewes:2018gkc} (green), MATHUSLA~\cite{Curtin:2018mvb} (purple), and the Phase II of FASER~\cite{Kling:2018wct}.
		For reference, a band corresponding to the contribution light neutrino masses between 20~meV and 200~meV in a single generation see-saw type I model is shown (yellow).
		Larger values of the mixing angles are allowed if an extension to see-saw models is invoked,
		for instance, in an inverse or extended see-saw scheme.}
\label{fig:sensa_hnl}
\end{figure}

A recent study illustrates the potential sensitivity for  HNLs searches with the DUNE Near Detector~\cite{Ballett:2019bgd}. The sensitivity for HNL particles with masses in the range of 10 MeV to 2 GeV, from decays of mesons produced
in the proton beam dump that produces the pions for the neutrino beam production, was studied. The production
of $D_s$ mesons leads to access to high mass HNL production. The dominant HNL decay modes to SM particles
have been included, and basic detector constraints as well as the dominant background process have 
been taking into account. 

The experimental signature for these decays is a decay-in-flight event with no interaction vertex, typical of
neutrino--nucleon scattering, and a rather forward direction with respect to the beam.
The main background to this search comes from SM neutrino--nucleon scattering events in which the hadronic activity
at the vertex is below threshold.
Charged current quasi-elastic events with pion emission from resonances are background to the semi-leptonic decay channels,
whereas mis-identification of long pion tracks into muons can constitute a background to three-body leptonic decays.
Neutral pions are often emitted in neutrino scattering events and can be a challenge for decays into %
neutral meson or channels with electrons in the final state.

We report in Fig.~\ref{fig:sensa_hnl} the physics reach of the DUNE ND in its current configuration %
without backgrounds and for a Majorana and a Dirac HNL.
The sensitivity was estimated assuming a total of 1.32 x $10^{22}$ POT, i.e. for a running scenario with 6 years with a 80 GeV proton beam of 1.2 MW, followed by six years of a beam with 2.4 MW, but using only the neutrino mode configuration, which corresponds to half of the total
runtime.
As a result, HNLs with masses up to 2 GeV can be searched for in all flavor-mixing channels.

The results show that DUNE will have an improved sensitivity to small values of the
mixing parameters $|U_{\alpha N}|^2$, where $\alpha=e,\,\mu,\,\tau$, compared to the presently available experimental
limits on mixing of HNLs with the three lepton flavors. At 90\% \dword{cl} sensitivity, DUNE can probe mixing parameters as low as 
$10^{-9}-10^{-10}$ in the mass range of 300-500 MeV, for  mixing with the electron or muon neutrino flavors. In the region above 500 MeV the sensitivity
is reduced to $10^{-8}$ for $eN$ mixing and $10^{-7}$ for $\mu N$ mixing. The $\tau N$ mixing 
sensitivity is weaker but still covering a new unexplored regime. A large fraction of the covered parameter space for all neutrino flavors falls in the region that is relevant for explaining the baryon asymmetry in the universe.

Studies are ongoing with full detector simulations to validate these 
encouraging results.

\subsection{Dark Matter Annihilation in the Sun}
DUNE's large \dword{fd} LArTPC modules provide an excellent setting to conduct searches for neutrinos arising from \dword{dm} annihilation in the core of the sun. These would typically result in a high-energy neutrino signal almost always accompanied by a low-energy neutrino component, which has its origin in a hadronic cascade that develops in the dense solar medium and produces large numbers of light long-lived mesons, such as $\pi^+$ and $K^+$ that
then stop and decay at rest. The decay of each $\pi^+$ and $K^+$ will produce monoenergetic neutrinos with an energy \SI{30}{MeV} or \SI{236}{MeV}, respectively.
The  \SI{236}{MeV} flux can be measured with the DUNE \dword{fd}, thanks to its excellent energy resolution, and importantly, will benefit from directional information. By selecting neutrinos arriving from the direction of the sun, large reduction in backgrounds can be achieved.
This directional resolution for sub-GeV neutrinos will enable DUNE to be competitive with experiments with even larger fiducial masses, but less precise angular information, such as Hyper-K~\cite{ref:DMannihilation}.

\section{Conclusions and Outlook}
DUNE will be a powerful discovery tool on a variety of physics topics under very active exploration today, from the potential discovery of new particles beyond those predicted in the \dword{sm}, to precision neutrino measurements that may uncover deviations from the present three-flavor mixing paradigm and unveil new interactions and symmetries.
The \dword{nd} alone will offer excellent opportunities to search for light \dword{dm} and mixing with light sterile neutrinos, and to measure rare processes such as neutrino trident interactions. Besides looking for deviations from the three-flavor oscillation paradigm such as nonstandard interactions, DUNE's massive high-resolution \dword{fd} will probe the possible existence of \dword{bdm}. The flexibility of the LBNF beamline enables planning for high-energy beam running, providing access to probing and measuring tau neutrino physics with unprecedented precision.

DUNE will offer a long-term privileged setting for collaboration between experimentalists and theorists in the domain areas of neutrino physics, astrophysics, and cosmology, and will provide the highest potential for breakthrough discoveries among the new near-term facilities projected to start operations during the next decade.

\cleardoublepage


\cleardoublepage


\cleardoublepage
\printglossaries

\cleardoublepage
\cleardoublepage
\renewcommand{\bibname}{References}
\bibliographystyle{utphys} 
\bibliography{common/tdr-citedb}

\end{document}